\documentclass[onecolumn,useamsfonts]{pasj02}
\usepackage{natbib}
\usepackage{aas2pasj}
 
\usepackage{float}
\usepackage{amsmath}
\usepackage{tikz}
\usetikzlibrary{positioning, arrows.meta}
 
\usepackage{xr}
 
\newif\ifXRFROZEN \XRFROZENtrue
\ifXRFROZEN
  \input{xref_frozen_S2_pasj}
\else
\fi

\usepackage{xcolor}

\def\amm{NH$_3$}
 
\jyear{2026}
\Received{}
\Accepted{}

\graphicspath{{./}{figures/}}

\begin{document}
 
\title{B-fields In STar-forming Region Observations (BISTRO): \\ Population-Level Universality and Environmental Structuring of 450/850\,$\mu$m Polarization Signatures}
 
\author{%
Ray S.~\textsc{Furuya},\altaffilmark{1,2}\altemailmark\orcid{0000-0002-0786-7307} \email{rsfuruya@gmail.com}
David \textsc{Berry},\altaffilmark{3}\orcid{0000-0001-6524-2447}
Pierre \textsc{Bastien},\altaffilmark{4}\orcid{0000-0002-0794-3859}
Tetsuo \textsc{Hasegawa},\altaffilmark{5}\orcid{0000-0003-1853-0184}
Masafumi \textsc{Matsumura},\altaffilmark{6}\orcid{0000-0002-6906-0103}
Thiem \textsc{Hoang},\altaffilmark{7,8}\orcid{0000-0003-2017-0982}
Lapo \textsc{Fanciullo},\altaffilmark{9}\orcid{0000-0001-9930-9240}
Kazunari \textsc{Iwasaki},\altaffilmark{10}\orcid{0000-0002-9892-1881}
Simon \textsc{Coud\'{e}},\altaffilmark{5}\orcid{0000-0002-0859-0805}
Takashi \textsc{Onaka},\altaffilmark{11}\orcid{0000-0002-8234-6747}
Pham Ngoc \textsc{Diep},\altaffilmark{12}\orcid{0000-0002-2808-0888}
Sarah \textsc{Sadavoy},\altaffilmark{13}\orcid{0000-0001-7474-6874}
Hua-bai \textsc{Li},\altaffilmark{14}\orcid{0000-0003-2641-9240}
Koji \textsc{Kawabata},\altaffilmark{15,16,17}\orcid{0000-0001-6099-9539}
Doris \textsc{Arzoumanian},\altaffilmark{18,19}\orcid{0000-0002-1959-7201}
Yasuo \textsc{Doi},\altaffilmark{20}\orcid{0000-0001-8746-6548}
Jia-Wei \textsc{Wang},\altaffilmark{3}\orcid{0000-0002-6668-974X}
Tao-Chung \textsc{Ching},\altaffilmark{21}\orcid{0000-0001-8516-2532}
Janik \textsc{Karoly},\altaffilmark{22}\orcid{0000-0001-5996-3600}
Jihye \textsc{Hwang},\altaffilmark{18,19}\orcid{0000-0001-7866-2686}
Glenn J. \textsc{White},\altaffilmark{23}\orcid{0000-0002-7126-691X}
A-Ran \textsc{Lyo},\altaffilmark{7}
Kohji \textsc{Tomisaka},\altaffilmark{24}\orcid{0000-0003-2726-0892}
Le Ngoc \textsc{Tram},\altaffilmark{25}\orcid{0000-0002-6488-8227}
Fr\'{e}d\'{e}rick \textsc{Poidevin},\altaffilmark{26,27}\orcid{0000-0002-5391-5568}
Valentin J.~M.\ \textsc{Le Gouellec},\altaffilmark{28,29}\orcid{0000-0002-5714-799X}
Hiroko \textsc{Shinnaga},\altaffilmark{30}\orcid{0000-0001-9407-6775}
Masato I.~N. \textsc{Kobayashi},\altaffilmark{31}\orcid{0000-0003-3990-1204}
Junhao \textsc{Liu},\altaffilmark{32}\orcid{0000-0002-4774-2998}
Mehrnoosh \textsc{Tahani},\altaffilmark{33}\orcid{0000-0001-8749-1436}
Nguyen Bich \textsc{Ngoc},\altaffilmark{12}\orcid{0000-0002-5913-5554}
Jennifer \textsc{Hatchell},\altaffilmark{34}\orcid{0000-0002-4870-2760}
Archana \textsc{Soam},\altaffilmark{35}\orcid{0000-0002-6386-2906}
Hsi-Wei \textsc{Yen},\altaffilmark{21}\orcid{0000-0003-1412-893X}
Chakali \textsc{Eswaraiah},\altaffilmark{36,37,38}\orcid{0000-0003-4761-6139}
Vera \textsc{K\"{o}nyves},\altaffilmark{39}\orcid{0000-0002-3746-1498}
Jintai \textsc{Wu},\altaffilmark{40}
Jeong-Eun \textsc{Lee},\altaffilmark{41}\orcid{0000-0003-3119-2087}
Yusuke \textsc{Tsukamoto},\altaffilmark{30}\orcid{0000-0001-6273-805X}
Ekta \textsc{Sharma},\altaffilmark{36}\orcid{0000-0002-4541-0607}
Tie \textsc{Liu},\altaffilmark{42}\orcid{0000-0002-5286-2564}
Qilao \textsc{Gu},\altaffilmark{43}
Hiroyuki \textsc{Nakanishi},\altaffilmark{30}
Tsuyoshi \textsc{Inoue},\altaffilmark{44}\orcid{0000-0002-7935-8771}
Fumitaka \textsc{Nakamura},\altaffilmark{24,45}\orcid{0000-0001-5431-2294}
Rachel \textsc{Friesen},\altaffilmark{46}\orcid{0000-0001-7594-8128}
Jean-Fran\c{c}ois \textsc{Robitaille},\altaffilmark{47}\orcid{0000-0001-5079-8573}
Yoshiaki \textsc{Misugi},\altaffilmark{48}\orcid{0000-0002-4093-6925}
Raiga \textsc{Kashiwagi},\altaffilmark{48}\orcid{0000-0002-1461-3866}
Per \textsc{Friberg},\altaffilmark{3}\orcid{0000-0002-8010-8454}
Sarah \textsc{Graves},\altaffilmark{3}
Mark \textsc{Rawlings},\altaffilmark{3}\orcid{0000-0002-6529-202X}
Harriet \textsc{Parsons},\altaffilmark{3}\orcid{0000-0002-6327-3423}
David \textsc{Eden},\altaffilmark{49,50}\orcid{0000-0002-5881-3229}
Huei-Ru Vivien \textsc{Chen},\altaffilmark{51}\orcid{0000-0002-9774-1846}
Najimudeen \textsc{Bijas},\altaffilmark{52}\orcid{0000-0002-6398-7530}
Motohide \textsc{Tamura},\altaffilmark{5,11,53}\orcid{0000-0002-6510-0681}
Tyler L.\ \textsc{Bourke},\altaffilmark{54,55}\orcid{0000-0001-7491-0048}
Nagayoshi \textsc{Ohashi},\altaffilmark{21}\orcid{0000-0003-0998-5064}
Woojin \textsc{Kwon},\altaffilmark{56,57}\orcid{0000-0003-4022-4132}
Jongsoo \textsc{Kim},\altaffilmark{7,8}\orcid{0000-0002-1229-0426}
Shih-Ping \textsc{Lai},\altaffilmark{58,21}\orcid{0000-0001-5522-486X}
Keping \textsc{Qiu},\altaffilmark{40,59}\orcid{0000-0002-5093-5088}
Di \textsc{Li},\altaffilmark{36,60}\orcid{0000-0003-3010-7661}
Shu-ichiro \textsc{Inutsuka},\altaffilmark{61}\orcid{0000-0003-4366-6518}
Kate \textsc{Pattle},\altaffilmark{22}\orcid{0000-0002-8557-3582}
Derek \textsc{Ward-Thompson},\altaffilmark{39}\orcid{0000-0003-1140-2761}
and the BISTRO Collaboration}
\altaffiltext{1}{Institute of Liberal Arts and Sciences, Tokushima University, Minami Jousanjima-Cho 1-1, Tokushima, Tokushima 770-8502, Japan}
\altaffiltext{2}{Visiting Associate Professor, Center for Computational Astrophysics (CfCA), National Astronomical Observatory of Japan (FY~2023), Osawa 2-21-1, Mitaka, Tokyo 181-8588, Japan}
\altaffiltext{3}{East Asian Observatory, 660 N.\ A`oh\={o}k\={u} Place, University Park, Hilo, HI 96720, USA}
\altaffiltext{4}{Centre de recherche en astrophysique du Qu\'ebec \& D\'epartement de physique, Universit\'e de Montr\'eal, C.P.~6128, Succursale Centre-ville, Montr\'eal, QC H3C~3J7, Canada}
\altaffiltext{5}{National Astronomical Observatory of Japan, National Institutes of Natural Sciences, Osawa, Mitaka, Tokyo 181-8588, Japan}
\altaffiltext{6}{Faculty of Education \& Center for Educational Development and Support, Kagawa University, Saiwai-cho 1-1, Takamatsu, Kagawa 760-8522, Japan}
\altaffiltext{7}{Korea Astronomy and Space Science Institute, 776 Daedeokdae-ro, Yuseong-gu, Daejeon 34055, Republic of Korea}
\altaffiltext{8}{University of Science and Technology, Korea, 217 Gajeong-ro, Yuseong-gu, Daejeon 34113, Republic of Korea}
\altaffiltext{9}{Tamkang University, No.\ 151, Yingzhuan Rd., Tamsui Dist., New Taipei City 251301, Taiwan}
\altaffiltext{10}{Center for Computational Astrophysics, National Astronomical Observatory of Japan, Mitaka, Tokyo 181-8588, Japan}
\altaffiltext{11}{Department of Astronomy, Graduate School of Science, The University of Tokyo, 7-3-1 Hongo, Bunkyo-ku, Tokyo 113-0033, Japan}
\altaffiltext{12}{Vietnam National Space Center, Vietnam Academy of Science and Technology, 18 Hoang Quoc Viet, Hanoi, Vietnam}
\altaffiltext{13}{Department for Physics, Engineering Physics and Astrophysics, Queen's University, Kingston, ON K7L 3N6, Canada}
\altaffiltext{14}{Department of Physics, The Chinese University of Hong Kong, Shatin, N.T., Hong Kong}
\altaffiltext{15}{Hiroshima Astrophysical Science Center, Hiroshima University, Kagamiyama 1-3-1, Higashi-Hiroshima, Hiroshima 739-8526, Japan}
\altaffiltext{16}{Department of Physics, Hiroshima University, Kagamiyama 1-3-1, Higashi-Hiroshima, Hiroshima 739-8526, Japan}
\altaffiltext{17}{Core Research for Energetic Universe (CORE-U), Hiroshima University, Kagamiyama 1-3-1, Higashi-Hiroshima, Hiroshima 739-8526, Japan}
\altaffiltext{18}{Institute for Advanced Study, Kyushu University, Japan}
\altaffiltext{19}{Department of Earth and Planetary Sciences, Faculty of Science, Kyushu University, Nishi-ku, Fukuoka 819-0395, Japan}
\altaffiltext{20}{Department of Earth Science and Astronomy, Graduate School of Arts and Sciences, The University of Tokyo, 3-8-1 Komaba, Meguro, Tokyo 153-8902, Japan}
\altaffiltext{21}{Academia Sinica Institute of Astronomy and Astrophysics, No.1, Sec.\ 4., Roosevelt Road, Taipei 10617, Taiwan}
\altaffiltext{22}{Department of Physics and Astronomy, University College London, WC1E 6BT London, UK}
\altaffiltext{23}{School for Physical Sciences, The Open University, Walton Hall, Milton Keynes MK7 6AA, United Kingdom, and RALSpace, The Rutherford Appleton Laboratory, Chilton, Didcot, OX11 0QX, United Kingdom}
\altaffiltext{24}{Division of Theoretical Astronomy, National Astronomical Observatory of Japan, Mitaka, Tokyo 181-8588, Japan}
\altaffiltext{25}{Leiden Observatory, Leiden University, PO Box 9513, 2300 RA Leiden, The Netherlands}
\altaffiltext{26}{Instituto de Astrof\'isica de Canarias, E-38205 La Laguna, Tenerife, Canary Islands, Spain}
\altaffiltext{27}{Universidad de La Laguna, Dpto.\ Astrof\'isica, E-38206 La Laguna, Tenerife, Spain}
\altaffiltext{28}{SOFIA Science Center, Universities Space Research Association, NASA Ames Research Center, Moffett Field, California 94035, USA}
\altaffiltext{29}{Universit\'e Paris-Saclay, CNRS, CEA, Astrophysique, Instrumentation et Mod\'elisation de Paris-Saclay, 91191 Gif-sur-Yvette, France}
\altaffiltext{30}{Department of Physics and Astronomy, Graduate School of Science and Engineering, Kagoshima University, 1-21-35 Korimoto, Kagoshima, Kagoshima 890-0065, Japan}
\altaffiltext{31}{Physikalisches Institut, University of Cologne, Z\"{u}lpicher Str.\ 77, D-50937 K\"{o}ln, Germany}
\altaffiltext{32}{Division of ALMA, National Astronomical Observatory of Japan, Mitaka, Tokyo 181-8588, Japan}
\altaffiltext{33}{University of South Carolina, Columbia, SC 29208, USA}
\altaffiltext{34}{Physics and Astronomy, University of Exeter, Stocker Road, Exeter EX4 4QL, UK}
\altaffiltext{35}{Indian Institute of Astrophysics, II Block, Koramangala, Bengaluru 560034, India}
\altaffiltext{36}{CAS Key Laboratory of FAST, National Astronomical Observatories, Chinese Academy of Sciences, People's Republic of China}
\altaffiltext{37}{National Astronomical Observatories, Chinese Academy of Sciences, A20 Datun Road, Chaoyang District, Beijing 100012, People's Republic of China}
\altaffiltext{38}{Indian Institute of Science Education and Research (IISER) Tirupati, Rami Reddy Nagar, Karakambadi Road, Mangalam (P.O.), Tirupati 517507, India}
\altaffiltext{39}{Jeremiah Horrocks Institute, University of Central Lancashire, Preston PR1 2HE, UK}
\altaffiltext{40}{School of Astronomy and Space Science, Nanjing University, 163 Xianlin Avenue, Nanjing 210023, People's Republic of China}
\altaffiltext{41}{Astronomy Program, Department of Physics and Astronomy, Seoul National University, 1 Gwanak-ro, Gwanak-gu, Seoul 08826, Republic of Korea}
\altaffiltext{42}{Key Laboratory for Research in Galaxies and Cosmology, Shanghai Astronomical Observatory, Chinese Academy of Sciences, 80 Nandan Road, Shanghai 200030, People's Republic of China}
\altaffiltext{43}{Shanghai Astronomical Observatory, Chinese Academy of Sciences, 80 Nandan Road, Shanghai 200030, People's Republic of China}
\altaffiltext{44}{Department of Physics, Konan University, Okamoto 8-9-1, Higashinada-ku, Kobe 658-8501, Japan}
\altaffiltext{45}{SOKENDAI (The Graduate University for Advanced Studies), Hayama, Kanagawa 240-0193, Japan}
\altaffiltext{46}{National Radio Astronomy Observatory, 520 Edgemont Road, Charlottesville, VA 22903, USA}
\altaffiltext{47}{Univ.\ Grenoble Alpes, CNRS, IPAG, 38000 Grenoble, France}
\altaffiltext{48}{Division of Science, National Astronomical Observatory of Japan, 2-21-1 Osawa, Mitaka, Tokyo 181-8588, Japan}
\altaffiltext{49}{Department of Physics, University of Bath, Claverton Down, Bath, BA2 7AY, UK}
\altaffiltext{50}{Armagh Observatory and Planetarium, College Hill, Armagh, BT61 9DB, UK}
\altaffiltext{51}{Institute of Astronomy and Department of Physics, National Tsing Hua University, Hsinchu 300044, Taiwan}
\altaffiltext{52}{Jodrell Bank Centre for Astrophysics, School of Physics and Astronomy, The University of Manchester, Manchester M13 9PL, UK.}
\altaffiltext{53}{Astrobiology Center, National Institutes of Natural Sciences, 2-21-1 Osawa, Mitaka, Tokyo 181-8588, Japan}
\altaffiltext{54}{SKA Observatory, Jodrell Bank, Lower Withington, Macclesfield SK11 9FT, UK}
\altaffiltext{55}{Jodrell Bank Centre for Astrophysics, School of Physics and Astronomy, University of Manchester, Oxford Road, Manchester M13 9PL, UK}
\altaffiltext{56}{Department of Earth Science Education, Seoul National University, 1 Gwanak-ro, Gwanak-gu, Seoul 08826, Republic of Korea}
\altaffiltext{57}{SNU Astronomy Research Center, Seoul National University, 1 Gwanak-ro, Gwanak-gu, Seoul 08826, Republic of Korea}
\altaffiltext{58}{Institute of Astronomy and Department of Physics, National Tsing Hua University, Hsinchu 30013, Taiwan}
\altaffiltext{59}{Key Laboratory of Modern Astronomy and Astrophysics (Nanjing University), Ministry of Education, Nanjing 210023, People's Republic of China}
\altaffiltext{60}{University of Chinese Academy of Sciences, Beijing 100049, People's Republic of China}
\altaffiltext{61}{Department of Physics, Graduate School of Science, Nagoya University, Furo-cho, Chikusa-ku, Nagoya 464-8602, Japan}
 
 
\maketitle
 
\begin{abstract}
We observed 69 molecular cloud regions at 450 and 850\,\micron\ with the POL-2 polarimeter on the James Clerk Maxwell Telescope and matched the maps to a common $12\farcs6$ beam.
We tested whether their polarization ratios and spectral-index differences are universal.
Polarization fractions are typically lower at 450\,\micron.
The median of 30 measured ratios is $P_{450}/P_{850} = 0.69 \pm 0.02_{\rm sys} \pm 0.03_{\rm stat}$.
Across 35 regions with reliable spectral indices of total ($\alpha$) and polarized ($\gamma$) intensity, the medians over sources are $\langle\alpha\rangle = 3.33$ and $\langle\gamma\rangle = 2.85$--$3.04$.
The median $\gamma - \alpha$ is $-0.57$ to $-0.45$, with $\gamma < \alpha$ in 24--27 regions.
These ranges span the two extreme placements of $3\sigma$ limits for polarized intensity below the noise floor.
The position angles at the two wavelengths agree within $10\arcdeg$ in 10 of 13 regional medians, disfavoring a geometric origin that would rotate the angles.
The medians of $\gamma-\alpha$ in 9 low-mass, 19 high-mass and 7 Central Molecular Zone (CMZ) regions are not ordered low-mass $<$ high-mass $<$ CMZ (Jonckheere--Terpstra $p = 0.43$--$0.70$).
A difference among classes is not robustly detected.
The Kruskal--Wallis $p = 0.056$--$0.078$ becomes $p = 0.026$--$0.041$ when the Rosette Molecular Cloud alone is excluded, although the high-mass class has the lowest median in every treatment.
The classes follow a distance sequence (the beam spans 0.008--0.51\,pc), coupling environment and physical scale.
Within clouds, the 450\,\micron\ polarization fraction decreases with \textit{Herschel} column density in all 7 fields with sufficient dynamic range.
Coefficients from pooled pixels mix trends within clouds, differences between clouds, and the influence of OMC\,1.
They therefore do not define a single law for individual pixels.
The wavelength dependence across sources provides a benchmark for grain alignment.
We cannot order it by environment and do not propose a unique mechanism.
\end{abstract}

\section{Introduction}
\label{sec:intro}
 
\subsection{Submillimeter Polarimetry as a Probe of Interstellar Magnetism}
\label{ss:S2_intro_polarimetry}
 
Magnetic fields ($\vec{B}$-fields) pervade the interstellar medium (ISM) across nearly every scale at which molecular clouds fragment into stars.
They couple to predominantly neutral gas through ion--neutral collisions \citep{Alfven1942}.
However, in the dense nucleus of a prestellar core, the ionization fraction may fall to $\sim 10^{-9}$, where ambipolar diffusion can proceed on timescales comparable to the free-fall time \citep{Caselli2002}.
We must therefore quantify the geometry of the fields in the plane of the sky and their relative strength to understand how molecular clouds evolve and collapse to form stars \citep{Hennebelle2019,Pattle2023PPVII}.
Thermal emission from aspherical dust grains aligned with their major axes perpendicular to the local $\vec{B}$-field \citep{DavisGreenstein1951} provides the most direct observational route to this geometry in the cold, dense molecular gas ($T \lesssim 100$\,K).
The POL-2 polarimeter on the James Clerk Maxwell Telescope (JCMT) \citep{Friberg2016} operates with the Submillimetre Common-User Bolometer Array 2 (SCUBA-2) camera, which has 5120 bolometers at each wavelength \citep{Holland2013}.
At these wavelengths, dust emission remains optically thin from diffuse star-forming cloud envelopes ($N_{\rm H} \sim 10^{21}\,{\rm cm}^{-2}$) to the densest cores ($N_{\rm H} \sim 10^{24}\,{\rm cm}^{-2}$).
This statement about optical depth is based on the submillimeter dust opacities of \citet{Hildebrand1983} and \citet{Ossenkopf1994}.
The range of column density is set instead by what SCUBA-2 detects, with the lower bound corresponding to a visual extinction of $A_V \gtrsim 5$\,mag rather than to the onset of dust emission.
 
Since the advent of POL-2, the B-fields In STar-forming Region Observations (BISTRO) Large Program \citep{Ward-Thompson2017} has transformed this observational capability into a large-scale polarimetric survey of star-forming regions.
The program has produced studies of individual nearby and distant molecular clouds \citep{Pattle2023PPVII,Pattle2017, Kwon2018, Soam2018, Coude2019, Pattle2019,Doi2020, Doi2021, Pattle2021, Kwon2022, Ngoc2021, Karoly2023, Hwang2023_HHeadN, Wu2024, Sugitani2025, Choi2024, Hwang2025, Wang2024_NGC2264}.
Most of these studies analyze only the 850\,\micron\ data, where the atmospheric opacity at the JCMT site is most favorable and POL-2 systematics are best characterized \citep{Friberg2018, Mairs2021}.
POL-2 studies that also used the 450\,\micron\ data have treated individual regions \citep{Hwang2021,Lyo2021,Tram2024,Yang2025,Scheiter2026}.
In this three-paper study of submillimeter polarization across star-forming environments, we ask whether the wavelength-dependent signatures in the companion paper (\citealt{Furuya2026_PaperS1}; hereafter Paper~S1) are universal or vary systematically with environment.
 
\subsection{The Three-Paper Series}
\label{ss:S2_intro_series}
A difference between the polarization fractions at 450 and 850\,$\mu$m can reflect changes in emission from aligned dust, in magnetic geometry, or in observational response.
Interpreting that difference across molecular clouds requires both a validated comparison of the two bands and a clear account of the environments being sampled.
 
This paper presents the source census in a three-paper series on the complete reduction of the 450\,\micron\ data and analysis at both wavelengths for the BISTRO {\sc PolSrc} sample.
Tables for individual sources and an atlas of the $\vec{B}$-field at both wavelengths accompany the paper.
The technical paper of this series (\citealt{Furuya2026_PaperT}; hereafter Paper~T) presents the JCMT calibration and validates the noise model.
It establishes the correction of additive zero-level offsets in Stokes-$Q/U$, the debiasing prescription, and the Weather Band criteria (a zenith atmospheric optical depth $\tau^{\rm zenith}_{225\,{\rm GHz}} < 0.05$; \citealt[table~2]{Mairs2021}) required for reliable polarization measurements at 450\,\micron.
It also independently validates the polarimetry at 450\,\micron\ against the observations of Ophiuchus~A at 154\,\micron\ with the Stratospheric Observatory for Infrared Astronomy (SOFIA)/High-resolution Airborne Wideband Camera Plus (HAWC+).
Paper~S1 presents the astrophysical interpretation and establishes three principal findings (F1--F3) from subsamples selected for the measurement quality required by each observable.
F1 is the similarity of the projected field orientations at the two wavelengths in the sample of 21 sources with angles measured in both bands ($|\chi_{450} - \chi_{850}| < 10\degr$ for 62\% of $N_{\rm samp} = 1420$ pixels).
F2 is the systematic relation $\gamma < \alpha$ in 69\%--77\% (24--27/35) of the 35 sources in the spectral-index ({\sc SpIdx}) subsample, interpretable as a lower dust opacity spectral index $\beta$ for the grain population weighted by polarization.
F3 is the prevalence of $R_{\rm p} \equiv P_{450}/P_{850} < 1$ in the same {\sc SpIdx} subsample, which holds for the majority of sources in every environment sampled.
For a given line of sight, $\gamma < \alpha$ and $R_{\rm p} < 1$ both correspond to a polarization fraction that is higher at 850 than at 450\,\micron\ (equation~\ref{eq:Pratio_SpIdx}).
 
The present paper (Paper~S2) complements Paper~S1 by testing \emph{universality against environmental dependence}.
Paper~S1 demonstrated a survey-wide polarization difference between 450 and 850\,\micron\ within the {\sc SpIdx} subsample.
It did not establish whether the difference reflects a universal alignment behavior or varies with star-forming environment, local column density, dust temperature, and $\vec{B}$-field disorder.
This question motivates the present paper.
 
In this paper, we test whether the F1--F3 signatures are universal phenomena across the complete BISTRO {\sc PolSrc} sample and whether they vary systematically across major star-forming regimes.
Within the HGBS subsample of 9 fields from the \textit{Herschel} Gould Belt Survey, we also test whether the wavelength response diverges under local cloud conditions.
Maps of column density and dust temperature at matched resolution are available for these fields \citep{Andre2010}.
We adopt a common F1--F6 numbering scheme across the two science papers: F1--F3 (the detection of the signatures) reside in Paper~S1.
In this paper, we report F4--F6 (tests of these signatures across the source population).
We compare polarization spectra (the polarization fraction as a function of wavelength, $P(\lambda)$) and their relation to cloud conditions.
We use these comparisons to establish how consistently dust polarization can trace magnetic fields in studies of star-forming environments.
They extend the interpretation of the shared observations rather than providing an independent replication of the Paper~S1 results.
 
\subsection{Contribution of this paper}
\label{ss:S2_intro_contribution}
The main question is whether the lower polarization fraction at 450\,$\mu$m recurs under the different conditions found in low-mass star-forming clouds, regions hosting more massive young stars, and the Central Molecular Zone.
A related question is whether changes within individual clouds follow their density, heating, or $\vec{B}$-field structure.
 
We address these questions through three complementary scientific tests on the complete BISTRO {\sc PolSrc} sample.
We do not propose a unique mechanism of grain alignment.
The tests address three linked questions.
(i)~Do the F1--F3 wavelength-dependent polarization signatures hold universally across the BISTRO {\sc PolSrc} sample, rather than only within the {\sc SpIdx} subsample of Paper~S1?
(ii)~Do the wavelength-dependent observables vary systematically across major star-forming regimes (low-mass star-forming, high-mass star-forming, and Central Molecular Zone)?
(iii)~Does the polarization response at the two wavelengths diverge with changes in local cloud conditions and the degree of $\vec{B}$-field disorder?
We describe the three corresponding tests below.
 
\emph{F4 -- Population-level universality test of F1--F3.}
We measure $R_p$, $\alpha$, $\gamma$, $\mathcal{S}$, and $\mathcal{S}\times\mathcal{P}^{\rm frac}$ across the {\sc PolSrc} sample (subsections~\ref{ss:S2_Sdisp}--\ref{ss:S2_SpIdx}).
These measurements test question~(i): whether F1--F3 of Paper~S1 characterize the sample as a whole or are specific to the {\sc SpIdx} subsample.
 
\emph{F5 -- Regime-dependent wavelength response.}
We compare regions of low-mass star formation (LMSF; $N=9$), high-mass star formation (HMSF; $N=19$), and the Central Molecular Zone (CMZ; $N=7$) in our galaxy (subsection~\ref{ss:S2_EnvStrat}).
We use the $(\gamma-\alpha,\,R_{\rm EW})$ phase diagram and a one-dimensional comparison among classes.
These tests address question~(ii): whether the F2 and F3 observables vary across star-forming regimes.
 
\emph{F6 -- Wavelength-dependent polarization--environment associations}.
We correlate polarimetric observables ($P_{450}$, $P_{850}$, $\mathcal{S}$, $\mathcal{S}\times\mathcal{P}^{\rm frac}$) with \textit{Herschel}-derived $N_{\rm H}$ and $T_d$ in the HGBS subsample of 9 fields (subsection~\ref{ss:S2_HGBS}).
These tests for individual pixels address question~(iii): whether the wavelength dependence changes with local cloud conditions and $\vec{B}$-field disorder.
 
Paper~T documents the technical methods that support these three tests as three pillars.
Pillar~1 [calibration of the flux conversion factor (FCF) and the $C(T_{\rm d})$ correction of equation~(\ref{eq:planck_correction}); Paper~T~subsection~\ref{T-ss:flux}] provides the spectral indices $\alpha$, $\gamma$, $\beta$ used in F4 and F5.
Pillar~2 (noise model for Stokes\,$I$ and $Q,U$; Paper~T~subsections~\ref{T-ss:noiseI}--\ref{T-ss:noisemodel}) provides the detection masks for individual pixels used in F6.
Pillar~3 (correction of additive zero-level offsets in 450\,\micron\ Stokes-$Q/U$ and final validation of beam harmonization; Paper~T~subsections~\ref{T-ss:offset}--\ref{T-ss:beamval}) provides the Stokes~$I,Q,U$ maps at matched resolution used throughout.
Paper~T table~\ref{T-tbl:systematics} and subsection~\ref{T-sec:precisionfloor} combine the three pillars into a budget of systematic uncertainties.
We adopt its practical floors $\sigma(P_{450}/P_{850}) \lesssim 0.02$ and $\sigma(\gamma - \alpha) \lesssim 0.04$ for F4--F6.
Paper~S1 presents the physical interpretation of F4--F6 in terms of size-dependent grain alignment, evolution of dust properties, and the physics of radiative torques.
Section~\ref{sec:S2_discussion} discusses how F4--F6 jointly answer the motivating question.
 
We also define the three complementary statistical estimators of $R_p$ used throughout this series (Methods~1, 2, and 3 of subsection~\ref{ss:S2_P_ratio}).
We provide tables for individual sources in the complete {\sc PolSrc} sample, together with an atlas of $\vec{B}$-field orientations at both wavelengths.
Appendix~\ref{app:S2_atlas} presents 65 of the 69 {\sc PolSrc} sources in figure form (all but W51 and Orion~A Fields~4, 5 and 7; see the notes to table~\ref{tab:surv} in Appendix), and Flexible Image Transport System (FITS) products for individual sources are released as a Dataverse supplement.
 
\begin{figure*}[ht]
\centering
\includegraphics{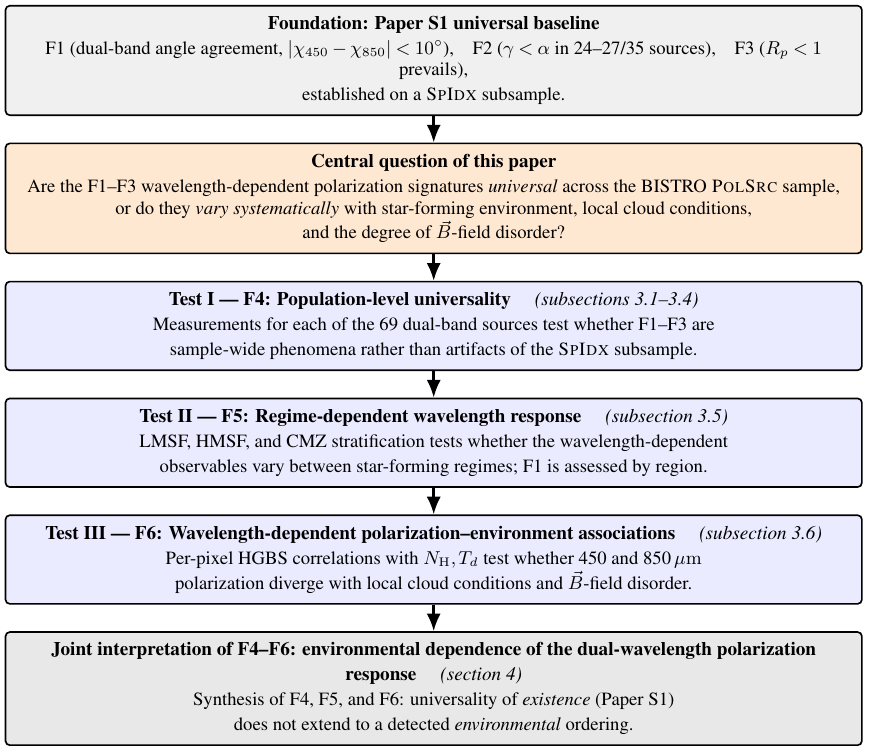}
\caption{
\textbf{Scientific questions and tests in Paper~S2.}
The diagram proceeds from the F1--F3 results of Paper~S1 (gray, top), through the question of their dependence on environment, local conditions, and $\vec{B}$-field disorder (orange), to three tests (blue).
F4 tests how representative the S1 results are across the population (subsections~\ref{ss:S2_Sdisp}--\ref{ss:S2_SpIdx}).
F5 compares star-forming environments (subsection~\ref{ss:S2_EnvStrat}).
F6 examines how the polarization fraction, $\mathcal{S}$, and the proxy $\mathcal{S}\times\mathcal{P}^{\rm frac}$ for grain alignment efficiency correlate locally with column density and dust temperature (subsection~\ref{ss:S2_HGBS}).
The three tests address questions~(i)--(iii) of subsection~\ref{ss:S2_intro_contribution}, respectively.
The bottom gray box summarizes their joint interpretation (section~\ref{sec:S2_discussion}; subsection~\ref{ss:S2_disc_topology}).
Paper~T provides the technical validation, and Paper~S1 develops the interpretation in terms of grain alignment.
\\ {Alt text: Flow diagram of text boxes linked by arrows.}
}
\label{fig:S2_test_overview}
\end{figure*}

A figure and a table summarize the structure of this paper.
Figure~\ref{fig:S2_test_overview} shows the logical structure of this paper, from the Paper~S1 baseline through the three tests to the synthesis in section~\ref{sec:S2_discussion}.
Table~\ref{tbl:S2_test_index} connects each technical topic to F4, F5, or F6 and identifies the relevant sections here and in Papers~T and~S1 (Paper~T Pillar or Paper~S1 finding).
 
\begin{longrotatetable}
\begin{deluxetable*}{lcllc}
\tablecaption{Index of the three tests in Paper~S2 and related material in Papers~T and~S1\label{tbl:S2_test_index}}
\tablewidth{0pt}
\tablehead{
  \colhead{Topic} &
  \colhead{Finding} &
  \colhead{Location in this paper series} &
  \colhead{Key result / figure} &
  \colhead{Note}
}
\startdata
\multicolumn{5}{c}{\textit{Paper~S2 --- F4: Population-level universality test of F1--F3}} \\
\hline
 \parbox[t]{5.4cm}{\raggedright Polarization-angle dispersion $\mathcal{S}$\strut}& F4 & Subsection~\ref{ss:S2_Sdisp} & table of $\mathcal{S}_{850}$ by source & 1 \\
 \parbox[t]{5.4cm}{\raggedright Grain-alignment efficiency proxy $\mathcal{S}\times\mathcal{P}^{\rm frac}$\strut}& F4 & Subsection~\ref{ss:S2_SP} & table by source + distance-bias check & 2 \\
 \parbox[t]{5.4cm}{\raggedright Polarization-fraction ratio $R_p$ (Methods 1, 2, and 3)\strut}& F4 & Subsection~\ref{ss:S2_P_ratio} & \parbox[t]{6cm}{\raggedright $R_{\rm IWM}, R_{\rm Corr}, R_{\rm EW}$ by source; median $R_{\rm EW} = 0.69$ over the 30 of the 35 {\sc SpIdx} sources with a measured ratio\strut}& 3 \\
Spectral indices $\alpha$ and $\gamma$ & F4 & Subsection~\ref{ss:S2_SpIdx} & \parbox[t]{6cm}{\raggedright $\alpha, \gamma, \beta$ by source; $\langle\alpha\rangle = 3.33$, $\langle\gamma\rangle = 2.85$--$3.04$\strut}& 4 \\
\hline
\multicolumn{5}{c}{\textit{Paper~S2 --- F5: Regime-dependent wavelength response}} \\
\hline
{Per-environment F2 and F3 breakdown; regional F1} & F5 & Sub-subsection~\ref{sss:env_breakdowns} & \parbox[t]{6cm}{\raggedright LMSF, HMSF, and CMZ medians of $\gamma{-}\alpha$ and $R_p$, together with the region-level $\widetilde{|\Delta\chi|}$ measurements\strut}& 5 \\
Environment phase diagram & F5 & Sub-subsection~\ref{sss:env_phase} & \parbox[t]{6cm}{\raggedright $(\gamma{-}\alpha,R_{\rm EW})$ plane\strut}& 6 \\
Comparison among environmental classes & F5 & Sub-subsection~\ref{sss:env_ladder} & 1-D box plots & 7 \\
\hline
\multicolumn{5}{c}{\textit{Paper~S2 --- F6: Wavelength-dependent polarization--environment associations}} \\
\hline
HGBS 9-field per-pixel correlation & F6 & Subsection~\ref{ss:S2_HGBS} & $r_s$ values with and without OMC\,1 & 8 \\
 \parbox[t]{5.4cm}{\raggedright $\mathcal{S}\times\mathcal{P}^{\rm frac}$ versus $N_{\rm H}$ at both wavelengths\strut}& F6 & Subsection~\ref{ss:S2_HGBS} & negative $r_s$ in both bands without OMC\,1 & 9 \\
\hline
\multicolumn{5}{c}{\textit{Paper~S2 --- Methodological infrastructure (supporting data products)}} \\
\hline
Three estimators of $R_p$ & --- & Subsection~\ref{ss:S2_P_ratio} & Methods 1, 2, and 3 definitions & 10 \\
Data release by source & --- & tables by source in subsections~\ref{ss:S2_P_ratio}--\ref{ss:S2_SpIdx} & \parbox[t]{6cm}{\raggedright 68-source $R_p, \alpha, \gamma, \mathcal{S}, \mathcal{S}\times\mathcal{P}^{\rm frac}$\strut}& 11 \\
 \parbox[t]{5.4cm}{\raggedright Dual-band $\vec{B}$-field orientation atlas\strut}& --- & Appendix~\ref{app:S2_atlas} & 65 sources in figure form + Dataverse supplement & 12 \\
\hline
\multicolumn{5}{c}{\textit{Material in Papers~T and~S1 (informational; cross-referenced from Paper~S2)}} \\
\hline
F1: dual-band angle agreement & --- & Paper~S1 & \parbox[t]{6cm}{\raggedright $|\chi_{450}{-}\chi_{850}|<10\degr$ on the 21-source dual-band-angle sample ($N_{\rm samp}=1420$)\strut} & 13 \\
F2: $\gamma < \alpha$ in 69\%--77\% & --- & Paper~S1 & $\gamma-\alpha = -0.57$ to $-0.45$ on 35-source subsample & 14 \\
F3: $R_p < 1$ prevails & --- & Paper~S1 & \parbox[t]{6cm}{\raggedright median $R_{\rm EW} = 0.69$ on the 30 of 35 subsample sources with a measured ratio\strut}& 15 \\
Pillar~1 (FCF + $C(T_{\rm d})$ correction) & --- & Paper~T~subsection~\ref{T-ss:flux} & basis of $\alpha$, $\gamma$ used in F4 and F5 & 16 \\
 \parbox[t]{5.4cm}{\raggedright Pillar~2 (Stokes\,$I$ and $Q,U$ noise model)\strut}& --- & Paper~T~subsections~\ref{T-ss:noiseI}--\ref{T-ss:noisemodel} & basis of per-pixel detection masks used in F6 & 17 \\
 \parbox[t]{5.4cm}{\raggedright Pillar~3 (offset corr.\ + beam harmonization)\strut}& --- & Paper~T~subsections~\ref{T-ss:offset}--\ref{T-ss:beamval} & basis of matched-resolution maps throughout & 18 \\
\enddata
\tablecomments{
The ``Finding'' column identifies F4--F6 in this paper and cross-references F1--F3 in Paper~S1.
``---'' denotes supporting material or material from Papers~T and~S1.
Topics follow their order in the text.
(1)~Polarization-angle dispersion measured with a $3\times3$ pixel ($36\arcsec\times36\arcsec$) kernel on the 12\arcsec\ grid; ordered LMSF, HMSF, CMZ.
(2)~Grain-alignment efficiency proxy [Planck/Balloon-borne Large Aperture Submillimeter Telescope for Polarimetry (BLASTPol) convention]; tested for distance bias across $0.1$--$8$\,kpc.
(3)~Comparison of the three estimators.
We adopt Method~3 for the census of individual sources.
(4)~Sample medians; Planck shape factor $C(T_{\rm d})$ cancels in the difference $\gamma-\alpha$.
(5)~Fractions of sources with $\gamma<\alpha$ at $1\sigma$ for each environment: 33\% in LMSF, 84\% in HMSF, and 57\% in CMZ.
These fractions are not monotonic in the environment sequence.
(6)~Phase plane $(\gamma-\alpha, R_{\rm EW})$ with markers for individual sources annotated by the regional $\widetilde{|\Delta\chi|}$; spectral-index consistency locus overlaid.
(7)~Two box plots ($R_{\rm EW}$, $\gamma-\alpha$) by environment, beside the individual $\widetilde{|\Delta\chi|}$ measurements.
(8)~Spearman $r_s$ correlations for individual pixels between polarization observables and $N_{\rm H}, T_d$ derived from HGBS; results for subsamples with and without OMC\,1 are reported separately.
(9)~Grain-alignment efficiency proxy $\mathcal{S}\times\mathcal{P}^{\rm frac}$ against $N_{\rm H}$, compared between 450\,\micron\ and 850\,\micron.
(10)~Method~1 $\equiv R_{\rm IWM}$ (intensity-weighted mean), Method~2 $\equiv R_{\rm Corr}$ (correlation analysis), Method~3 $\equiv R_{\rm EW}$ (equal-weighted median of pixel ratios).
The method numbers identify the subsections that define them.
(11)~Tables for individual sources include statistical and systematic uncertainties from Paper~T~subsection~\ref{T-sec:uncertainty} and table~\ref{T-tbl:systematics}.
(12)~65 of the 69 {\sc PolSrc} sources in figure form (all but W51 and Orion~A Fields~4, 5 and 7), ordered by distance from the Sun (L43 at 125\,pc to Sagittarius~B2 at 8.34\,kpc).
FITS products for individual sources (Stokes $I$/$Q$/$U$ maps, debiased $P_\lambda$ maps, $\mathcal{S}_\lambda$ and $\mathcal{S} \times \mathcal{P}^{\rm frac}$ maps, tables of statistics for individual sources) are released as a Dataverse supplement.
(13)~Agreement for individual pixels in the sample of 21 sources with angles measured in both bands (Paper~S1); basis for the scale of 10\arcdeg\ adopted as the physical tolerance and propagated to F5.
(14)~Spectral indices of individual sources in the {\sc SpIdx} subsample of Paper~S1; basis for the spectral-index quantities in F4 and F5.
(15)~$R_{\rm EW}$ of individual sources in the {\sc SpIdx} subsample of Paper~S1; basis for the polarization-fraction quantities in F4 and F5.
(16)~The choice of FCF$_{\rm arcsec}$ and the $C(T_{\rm d})$ correction, $\beta = \alpha - C(T_{\rm d})$ (equation~\ref{eq:planck_correction}), provide the basis for the definitions of $\alpha$, $\gamma$, and $\beta$ used throughout F4 and F5.
(17)~The two-component noise model and the verification of isotropy provide the basis for the detection masks for individual pixels used in the correlation analysis of F6.
(18)~Beam harmonization and the three-step correction of the 450\,\micron\ Stokes-$Q/U$ zero-level offset, based on principal component analysis (PCA), provide the basis for the Stokes~$I,Q,U$ maps at matched resolution that feed all three tests.
}
\end{deluxetable*}
\end{longrotatetable}
 
Table~\ref{tbl:S2_F4F6_master} summarizes the quantitative results for each of F4--F6, with reference rows for the F1--F3 findings of Paper~S1.
Table~\ref{tbl:S2_test_index} above gives the full index to the material in this paper.
 
For scalar medians in the quantitative summaries below, we adopt the notation $X = X_0 \pm \delta X_{\rm sys} \pm \delta X_{\rm stat}$.
Here $\delta X_{\rm sys}$ denotes the systematic floor inherited from the uncertainty budget of Paper~T (table~\ref{T-tbl:systematics}).
The term $\delta X_{\rm stat}$ denotes the empirical uncertainty of the sample: the standard error (SE) of the median evaluated by the Maritz--Jarrett estimator on the relevant sample.
The notation follows the convention of Paper~S1 (section~\ref{S1-sec:obs}).
For correlation coefficients and the diagnostics of individual clouds, we instead quote percentile or cluster block-bootstrap confidence intervals (CIs).
 
\begin{longrotatetable}
\setlength{\tabcolsep}{4pt}
\begin{deluxetable*}{clllll}
\tablecaption{Paper~S2 F4--F6 quantitative summary, with reference rows for the F1--F3 findings of Paper~S1.\label{tbl:S2_F4F6_master}}
\tablewidth{0pt}
\tablehead{
  \colhead{Label} &
  \colhead{Observable; sample} &
  \colhead{Result} &
  \colhead{Uncertainty (stat.; sys.)} &
  \colhead{Location} &
  \colhead{Limitations}
}
\startdata
\multicolumn{6}{c}{\textit{Paper~S2 findings (F4--F6)}} \\
\hline
F4 &
\parbox[t]{3.3cm}{\raggedright population-level universality of $R_{\rm EW}$, $\gamma-\alpha$, $|\Delta\chi|$; 69-source {\sc PolSrc}\strut} &
\parbox[t]{8.2cm}{\raggedright $R_{\rm EW} = 0.69 \pm 0.02_{\rm sys} \pm 0.03_{\rm stat}$ (median over the 30 of the 35 {\sc SpIdx} sources with a measured ratio; the 69-source {\sc PolSrc} census is the parent sample, not the basis of this median); $\gamma-\alpha = -0.57$ to $-0.45$ (systematic floor $0.04$); $\widetilde{|\Delta\chi|}<10\arcdeg$ in ten of the thirteen regional measurements of Paper~S1\strut} &
\parbox[t]{3.6cm}{\raggedright SE of the median (Maritz--Jarrett); Paper~T table~\ref{T-tbl:systematics}\strut} &
\parbox[t]{2.0cm}{\raggedright Subsections~\ref{ss:S2_Sdisp}--\ref{ss:S2_SpIdx}\strut} &
\parbox[t]{2.9cm}{\raggedright non-uniform regime sampling\strut} \\
F5 &
\parbox[t]{3.3cm}{\raggedright environmental stratification of $\gamma-\alpha$, $R_p$ (null result); 35-source {\sc SpIdx} (LMSF $9$, HMSF $19$, CMZ $7$)\strut} &
\parbox[t]{8.2cm}{\raggedright $\gamma-\alpha = +0.01$ (LMSF), $-0.65$ (HMSF), $-0.10$ (CMZ) at the measured medians, with 95\% percentile-bootstrap intervals $[-1.09,+0.86]$, $[-0.91,-0.41]$, $[-0.62,+0.03]$; implied $R_p = 1.90^{\gamma-\alpha}$ medians $1.01, 0.66, 0.94$ (the medians of the measured $R_{\rm EW}$ are $0.78/0.66/0.84$ on $N=6/17/7$; sub-subsection~\ref{sss:env_breakdowns}); not ordered along the LMSF--HMSF--CMZ sequence (ordered-alternative $p = 0.43$--$0.70$), and a difference among the classes not robustly detected (Kruskal--Wallis $p = 0.056$--$0.078$), although the HMSF median is the lowest in every treatment\strut} &
\parbox[t]{3.6cm}{\raggedright percentile-bootstrap 95\% CIs on the per-environment medians; Jonckheere--Terpstra permutation test ($20\,000$ permutations); Paper~T\strut} &
\parbox[t]{2.0cm}{\raggedright Subsection~\ref{ss:S2_EnvStrat}\strut} &
\parbox[t]{2.9cm}{\raggedright environment--distance coupling ($\rho = 0.828$); small $N$ per environment; two-class contrast not robust to exclusion of single sources ($p = 0.086$--$0.233$)\strut} \\
F6 &
\parbox[t]{3.3cm}{\raggedright per-sample $r_s(N_{\rm H},T_d,P_\lambda,\mathcal{S}_\lambda)$ + per-cloud analysis; HGBS-9 ($417$ matched-resolution samples)\strut} &
\parbox[t]{8.2cm}{\raggedright Hierarchical band-dependent structure: (i) $N_{\rm H}$--$P_{450}$ within-cloud universal, sign retained in 7/7 fields, pooled $r_s = -0.09$ with 95\% block-bootstrap CI $[-0.62,+0.21]$; (ii) $T_d$--$P_{850}$ OMC\,1-driven (full-sample $r_s = +0.36$, $N=383$); (iii) $I_{450}/I_{850}$--$P_{450}/P_{850}$ mixed cross-cloud + OMC\,1 origin, pooled $r_s = -0.45$ with 95\% block-bootstrap CI $[-0.74,-0.10]$\strut} &
\parbox[t]{3.6cm}{\raggedright cluster block-bootstrap 95\% CIs; Paper~T\strut} &
\parbox[t]{2.0cm}{\raggedright Subsection~\ref{ss:S2_HGBS} and sub-subsection~\ref{sss:S2_source_aware_correlation}\strut} &
\parbox[t]{2.9cm}{\raggedright OMC\,1 high-radiation regime influence; treatment of influential sources in sub-subsection~\ref{sss:S2_influential_source_policy}\strut} \\
\hline
\multicolumn{6}{c}{\textit{Paper~S1 reference findings (F1--F3); see Paper~S1 table~\ref{S1-tbl:S1_F1F3_master} for full quantitative content}} \\
\hline
F1 &
\parbox[t]{3.3cm}{\raggedright $|\chi_{450}-\chi_{850}|$; 21-source dual-band ($N_{\rm samp}=1420$)\strut} &
\parbox[t]{8.2cm}{\raggedright $\widetilde{|\Delta\chi|}<10\arcdeg$ in 10 of 13 targets (overall $6.7\arcdeg$)\strut} &
\parbox[t]{3.6cm}{\raggedright {typical per-pixel $\sigma_{\Delta\chi}\simeq 3.5\arcdeg$}; $4$--$6\arcdeg$ (Paper~T subsection~\ref{T-ss:beamval})\strut} &
\parbox[t]{2.0cm}{\raggedright Paper~S1 subsection~\ref{S1-ss:bfield_consistency}\strut} &
\parbox[t]{2.9cm}{\raggedright Mon~R2 joint test case\strut} \\
F2 &
\parbox[t]{3.3cm}{\raggedright $\gamma-\alpha$; 35-source {\sc SpIdx}\strut} &
\parbox[t]{8.2cm}{\raggedright $\gamma-\alpha = -0.57$ to $-0.45$, 24--27/35 sources $\gamma<\alpha$\strut} &
\parbox[t]{3.6cm}{\raggedright {Maritz--Jarrett SE of the median $0.10$ ($N=35$)}; $\lesssim 0.04$ (Paper~T table~\ref{T-tbl:systematics})\strut} &
\parbox[t]{2.0cm}{\raggedright Paper~S1 subsection~\ref{S1-ss:SpIdx}\strut} &
\parbox[t]{2.9cm}{\raggedright Line-of-sight temperature mixing with a polarization-efficiency contrast (Paper~S1 subsection~4.4)\strut} \\
F3 &
\parbox[t]{3.3cm}{\raggedright $R_p \equiv P_{450}/P_{850}$; {30 of the 35 {\sc SpIdx} sources + HGBS-9 (417 parent samples)}\strut} &
\parbox[t]{8.2cm}{\raggedright $R_{\rm EW} = 0.69$ sample median\strut} &
\parbox[t]{3.6cm}{\raggedright {Maritz--Jarrett SE of the median $0.03$ ($N=30$)}; $\lesssim 0.02$ (Paper~T table~\ref{T-tbl:systematics})\strut} &
\parbox[t]{2.0cm}{\raggedright Paper~S1 subsection~\ref{S1-ss:P_ratio}\strut} &
\parbox[t]{2.9cm}{\raggedright {OMC\,1 high-radiation regime; equal-factor cancellation in $R_p$ is a limiting model}\strut} \\
\enddata
\tablecomments{
The first three rows summarize F4--F6.
The last three rows give the F1--F3 inputs from Paper~S1, using the format of table~\ref{S1-tbl:S1_F1F3_master} of Paper~S1.
The uncertainty column gives the statistical estimator, then the source of the systematic floor.
For F4--F6 the floors are those of Paper~T.
Statistical errors follow the Maritz--Jarrett convention for the SE of the median, with bootstrap and block-bootstrap results for the analyses in subsections~\ref{ss:S2_EnvStrat}--\ref{ss:S2_HGBS}.
The F1 entry is the exception: it is the typical formal uncertainty of an individual angle difference between the bands, not the standard error of a regional or sample median.
The F2 and F3 entries are Maritz--Jarrett standard errors of the medians over 35 and 30 source entries, respectively.
The F2 range $-0.57$ to $-0.45$ spans the two extreme placements of the $3\sigma$ limits and is not a confidence interval.
The F2 and F3 systematic entries are the contributions of the terms quantified in Paper~T (calibration and residual zero-level offsets in Stokes $Q/U$ at 450\,\micron) to the uncertainties of these sample medians.
These entries do not include the unevaluated dependence on beam matching and spatial filtering.
The F2 and F3 systematic entries also do not include a difference in the modulation efficiency of POL-2 between the two bands (Paper~T~subsection~\ref{T-ss:flux}).
The systematic entries come from Paper~T through table~\ref{tbl:S2_dependency}.
Table~\ref{tbl:S2_test_index} additionally indexes methodological material and material from Papers~T and~S1.
}
\end{deluxetable*}
\end{longrotatetable}
 
The central scientific contribution of this paper lies in F4--F6 above and in their synthesis in section~\ref{sec:S2_discussion}.
Together, the source census and the comparisons within clouds constrain which aspects of the wavelength dependence are common and which vary with the material observed.
They do not, by themselves, select a unique mechanism of grain alignment or separate environment from physical resolution in the CMZ comparison.
 
\subsection{Structure of this Paper}
\label{ss:S2_intro_structure}
 
This paper is organized as follows.
Section~\ref{sec:S2_obs} summarizes the {\sc PolSrc} sample and directs the reader to Paper~T for the details of the JCMT/POL-2 observations and reduction.
Section~\ref{sec:S2_results} presents F4 in subsections~\ref{ss:S2_Sdisp}--\ref{ss:S2_SpIdx}, F5 in subsection~\ref{ss:S2_EnvStrat}, and F6 in subsection~\ref{ss:S2_HGBS}, following figure~\ref{fig:S2_test_overview} and table~\ref{tbl:S2_test_index}.
Section~\ref{sec:S2_discussion} combines F4--F6 to interpret how the polarization response at the two wavelengths depends on environment (subsections~\ref{ss:S2_disc_universality}--\ref{ss:S2_disc_scope}) and answers the motivating question of subsection~\ref{ss:S2_intro_series}.
Section~\ref{sec:S2_summary} summarizes the results and conclusions.
Appendices~\ref{app:S2_stacked}--\ref{app:S2_atlas} present supporting material.
The supporting material includes the stacked distribution of 450\,\micron\ polarization fractions (appendix~\ref{app:S2_stacked}), distance-bias residuals (appendix~\ref{app:S2_dist_bias}), and the reference scale of $\mathcal{S}$ for random orientations and its noise response (appendix~\ref{as:S_bounds}).
The $\vec{B}$-field atlas at both wavelengths includes figures for 65 of the 69 {\sc PolSrc} sources (appendix~\ref{app:S2_atlas}).
FITS products for individual sources are provided as a Dataverse supplement.

\section{Data and Methods}
\label{sec:S2_obs}
 
This section summarizes the POL-2 polarimetric sample analyzed at both wavelengths in this paper.
Paper~T section~\ref{T-sec:obs} through subsection~\ref{T-ss:offset} establish and validate the JCMT observational setup, reduction pipeline, noise model, correction of additive zero-level offsets in 450\,\micron\ Stokes-$Q/U$, and criteria for selecting the sample used here.
 
We processed all regions analyzed here using the reduction and calibration methods of Paper~T, with a common procedure to match the beams and spatial-frequency response.
That paper documents the reduction settings for each field and the applicability of the correction for 450\,\micron\ $Q/U$ zero-level offsets.
We use the same data products as Paper~S1, with identical beam matching, $12\arcsec$ pixel grid, and calibration.
The results in both papers do not rest on combining maps reduced under different conventions.
The use of AI tools in the analysis and in the preparation of the text is described in the Acknowledgments.
 
\subsection{Definitions and notation}
\label{ss:S2_definitions}
 
We define all quantities used here, including those shared with Papers~T and~S1.
Here $I_\lambda$ is the total intensity, and $Q_\lambda$ and $U_\lambda$ are the Stokes parameters describing linear polarization.
$PI_\lambda = \sqrt{Q_\lambda^2+U_\lambda^2}$ is the polarized intensity.
We debias it with the modified asymptotic estimator (MAS) of \citet{Plaszczynski2014} and \citet{Montier2015} to remove the positive noise bias of the quadrature sum before forming any ratio.
All $PI$ and $\mathcal{P}^{\rm frac}$ values quoted below are debiased in this way.
Polarization angles are calculated from the adopted Stokes $Q$ and $U$ maps, including the correction of 450\,\micron\ Stokes-$Q/U$ zero-level offsets where applicable.
We do not debias the angles with the MAS amplitude estimator.
 
$\mathcal{P}^{\rm frac}_\lambda \equiv PI_\lambda/I_\lambda$ is the polarization fraction.
We quote this dimensionless ratio as a percentage in table columns headed (\%) and where the text gives $P_\lambda = \dots\%$, and write it as $P_\lambda$ when unambiguous.
The alignment proxy $\mathcal{S}\times\mathcal{P}^{\rm frac}$ uses the dimensionless ratio rather than its percentage value; uppercase $P$ always denotes the fraction and $PI$ always the polarized intensity.
The polarization position angle $\chi_\lambda$ is measured east of north over $0\degr$--$180\degr$, with $1\sigma$ uncertainty $\sigma_\chi = 0.5\,\sigma_{PI}/PI$ in radians.
The field orientation in the plane of the sky is $\chi_\lambda + 90\degr$.
 
$\alpha$ and $\gamma$ are the spectral indices of total and polarized intensity,
$\log(X_{450}/X_{850})/\log(\nu_{450}/\nu_{850})$ with $X = I$ and $PI$ respectively, and $R_{\rm p} \equiv P_{450}/P_{850}$.
$\mathcal{S}$ is the polarization-angle dispersion function.
The three estimators of the polarization ratio across the source population are named for what they compute: $R_{\rm IWM}$ (intensity-weighted mean), $R_{\rm Corr}$ (correlation analysis) and $R_{\rm EW}$ (equal-weighted median of pixel ratios).
These estimators are defined in sub-subsection~\ref{sss:Methodology}.

\subsection{Sample Overview}
\label{ss:S2_sample}
 
Our full sample comprises 69 unique star-forming regions observed with the JCMT/POL-2 polarimeter at 450\,\micron\ and 850\,\micron\ simultaneously \citep{Friberg2016, Holland2013}.
Paper~T section~\ref{T-sec:obs} describes the sample selection and inventory.
We restricted the 450\,\micron\ data to Weather Band~1 observations ($\tau^{\rm zenith}_{225\,{\rm GHz}} < 0.05$), yielding 61 single fields.
Mosaicking, quality exclusions, and subdivision of wide-area observations produce 69 physically distinct regions.
We obtained the 850\,\micron\ data without selecting them by atmospheric opacity.
The sample spans 125\,pc (L43; figure~\ref{afig:Bmap_L43L43} in Appendix) to 8.3\,kpc (Sagittarius~B2; figure~\ref{afig:Bmap_GSGRB2} in Appendix), covering LMSF regions, HMSF regions, and clouds of the CMZ, the innermost ${\sim}200$\,pc of the Galaxy.
Paper~T table~\ref{T-tbl:T_sample_inventory} gives the seven-row inventory of counts for the source-based and beam-based samples.
Appendix~\ref{app:S2_survey_fields} (table~\ref{tab:surv} in Appendix) lists the sources, coordinates, adopted distances, archival references, and atlas figures in appendix~\ref{app:S2_atlas}.
 
\subsection{Subsamples Used in Paper S2}
\label{ss:S2_subsamples}
Different questions draw on different parts of the survey.
Mapping projected field structure requires usable polarization angles, whereas comparing dust spectra requires reliable measurements in both bands.
The available lines of sight and source counts, therefore, depend on the observable being examined.
 
Our analyses draw on the following nested subsamples defined in Paper~T section~\ref{T-sec:obs}.
They sample overlapping but unequal portions of the cloud population.
We therefore state conclusions about polarization ratios, spectral indices, and field orientations for the subset on which each quantity is measured.
 
\begin{itemize}
\item 69-source full sample.
All unique regions with successful 450\,\micron\ reduction (peak Stokes~$I$ signal-to-noise ratio $>100$ at 450\,\micron).
W51 satisfies this definition, but we exclude it from every table for individual sources and from the atlas, because its default 850\,\micron\ reduction is corrupted (see the note to the table~\ref{tab:surv} in Appendix).
The data products therefore cover 68 sources.
Used for the tables of $\mathcal{S}_{850}$ for individual sources (subsection~\ref{ss:S2_Sdisp}), the tables of $R_{\rm p}$ (subsection~\ref{ss:S2_P_ratio}), and the atlas of $\vec{B}$-field orientations at both wavelengths (appendix~\ref{app:S2_atlas}; Dataverse supplement).
\item 36-row $\alpha$/$\gamma$ table.
These sources have enough pixels passing the Stokes-$I$ selection of sub-subsection~\ref{sss:SpIdx_Global} to estimate spectral indices.
Of the 36 printed rows, 35 satisfy $N_{\rm samp} \geq 5$ and belong to the present reduction set; these form the statistical sample.
Three further regions yield $\alpha$ and $\gamma$ but have $N_{\rm samp} < 5$, so we do not tabulate them.
In total, 39 of the 69 regions yield spectral indices for individual sources.
Used for the $\alpha$ and $\gamma$ analysis of subsection~\ref{ss:S2_SpIdx}.
\item {\sc SpIdx} subsample.
Sources with complete paired measurements $P_{450}$, $P_{850}$, $\chi_{450}$, $\chi_{850}$, $\alpha$, $\gamma$, $\mathcal{S}$, and $\mathcal{S} \times \mathcal{P}^{\rm frac}$ that pass the consistency and signal-to-noise thresholds of Paper~T subsections~\ref{T-ss:offset}--\ref{T-ss:beamval}.
We divide them by environment into LMSF ($N=9$), HMSF ($N=19$), and CMZ ($N=7$).
This {\sc SpIdx} subsample underlies findings F2 and F3 of Paper~S1 and the comparison by environment in subsection~\ref{ss:S2_EnvStrat}.
\item HGBS 9-field subsample.
The subset of LMSF and HMSF sources with published maps of column density $N_{\rm H}$ and dust temperature $T_{\rm d}$ from the \textit{Herschel} Gould Belt Survey \citep{Andre2010} at a resolution compatible with the POL-2 12\arcsec\ grid.
The 9 fields are: Ophiuchus~A (figure~\ref{afig:Bmap_OphiuA} in Appendix), Taurus~B211/213 (figure~\ref{afig:Bmap_B211MO} in Appendix), NGC~1333 (figure~\ref{afig:Bmap_N133MO} in Appendix), Perseus~B1 (figure~\ref{afig:Bmap_PersB1} in Appendix), IC~348/HH~211 (figure~\ref{afig:Bmap_IC348H} in Appendix), OMC-1/2/3 (figures~\ref{afig:Bmap_OriAS1}--\ref{afig:Bmap_OriAS3} in Appendix), and Serpens~Main (figure~\ref{afig:Bmap_SerpMO} in Appendix).
Used for the per-pixel correlation analysis of subsection~\ref{ss:S2_HGBS}.
\end{itemize}
 
\subsection{Environment Classification}
\label{ss:S2_envcls}
 
We assign each {\sc SpIdx} source to one of three environments according to its own young stellar population.
Distance and membership in a parent complex do not determine the class; the distances below only describe the sample.
(i) LMSF: sources whose star formation is dominated by low-mass ($M \lesssim 1\,M_\odot$) core formation, typically in nearby Gould Belt molecular clouds at $d \lesssim 500$\,pc;
(ii) HMSF: sources hosting young stellar objects (YSOs) above the low-mass regime, i.e.\ intermediate-mass ($M \sim 2$--$8\,M_\odot$) or high-mass ($M \gtrsim 8\,M_\odot$) YSOs.
The label HMSF is retained as shorthand for this combined bin.
The class spans $d \approx 400$\,pc (the OMC~1, OMC~2, and OMC~3 fields of Orion~A) to $d \approx 4.8$\,kpc (G28.34$+$0.06, figure~\ref{afig:Bmap_G283MO} in Appendix).
Cepheus~A (figure~\ref{afig:Bmap_CepACe} in Appendix) is at $d \approx 0.7$\,kpc and DR21 (figure~\ref{afig:Bmap_DR21DR} in Appendix) at $d \approx 1.5$\,kpc;
(iii) CMZ: sources projected within the central $\sim$200\,pc of the Galaxy ($d \approx 8.3$\,kpc), characterized by elevated turbulence and cosmic-ray ionization rates relative to disk clouds.
Fields drawn from the same parent cloud do not need to share a class, because we classify each observed field by its own young stellar content.
The OMC~1--3 mosaic fields of Orion~A are HMSF on this basis, as is the Orion~A Field~4 pointing.
The center of the latter lies $8\farcm5$ from that of OMC~1, within one POL-2 field of view and closer than OMC~1 is to OMC~2.
The Orion~A Field~5 pointing lies more than one field of view further south, samples gas forming low-mass cores, and is LMSF.
The environment blocks of the table of spectral indices by source (sub-subsection~\ref{sss:SpIdx_Global}) give the classification of each source, and Paper~T subsection~\ref{T-ss:obs_reduction} sets out the sample definitions on which it rests.
 
\subsection{Data Products Used in This Paper}
\label{ss:S2_products}
 
All analyses in this paper use Stokes $I$, $Q$, $U$ maps at both wavelengths on a 12\arcsec\ grid, produced by the reduction of Paper~T subsections~\ref{T-ss:flux}--\ref{T-ss:noisemodel}.
We also use the polarized intensity $PI_\lambda$ and polarization fraction $P_\lambda = PI_\lambda/I_\lambda$ at 450\,\micron\ and 850\,\micron, debiased with the MAS estimator (\citealt{Plaszczynski2014, Montier2015}).
The beam sizes of the JCMT are $\theta_{450}^{\rm FWHM} = 8\farcs6 \pm 1\farcs3$ and $\theta_{850}^{\rm FWHM} = 12\farcs6 \pm 1\farcs9$ \citep{Mairs2021}.
These are the effective full widths at half maximum (FWHMs) of the two-component beams (main and secondary) in equation~12 of \citet{Mairs2021}.
We adopted them throughout this series, including the target resolution for beam harmonization.
These FWHMs differ from the main-beam widths ($\theta_{\rm MB} = 6\farcs2$ and $11\farcs0$; their table~6) and from the empirical effective widths ($10\farcs0 \pm 0\farcs6$ and $14\farcs4 \pm 0\farcs3$) inferred from the ratios of the peak FCF to the arcsecond FCF.
When we need the beam shape rather than a single width, we use both components explicitly (Paper~T~subsection~\ref{T-ss:beamval}).
Before all analyses, we apply the baseline correction for zero-level offsets in Stokes $Q/U$ at 450\,\micron\ (Paper~T subsection~\ref{T-ss:offset}).
In Paper~T, we validated this correction against SOFIA/HAWC+ 154\,\micron\ observations of Ophiuchus~A.
Throughout, we convert calibrated maps to surface-brightness units (mJy\,arcsec$^{-2}$) using the flux conversion factor for extended sources.
Paper~T subsection~\ref{T-ss:flux} discusses the rationale.
 
\subsection{Treatment of the Polarization Noise Floor}
\label{ss:S2_noise_floor}
 
We define every parent sample analyzed in this paper by conditions based on the total intensity $I$ alone, after basic validity checks that exclude entries with non-finite or unphysical values.
We do not select any parent sample on the polarized intensity $PI$, the polarization fraction $P$, or the ratio of $P$ between the bands, because a selection on weak polarization, or on the sensitivity to it, would itself bias the quantities being measured.

To define the parent samples, we apply the stated Stokes-$I$ thresholds to the intensities and their per-pixel uncertainties in the adopted map catalogs.
In these criteria, $\sigma_I$ denotes the catalogued uncertainty for each pixel, wavelength, and map product.
Paper~T subsection~\ref{T-ss:noiseI} characterizes these uncertainties and separates the additive noise from the component that depends on intensity.
To reproduce our selection, the stated thresholds must be applied with these same catalog uncertainties.
Replacing them with an independently estimated root-mean-square (rms) noise level or a single additive noise floor can change which pixels pass a given threshold.
The blank-sky noise estimates quoted in the atlas captions of appendix~\ref{ass:S2_atlas_maps} are distinct from the per-pixel uncertainties used for selecting data.

We apply the $3\sigma$ detection floor to the observables entering each statistic, not to the parent sample, because $P$ below this floor is not treated as a detection.
Polarization correlations within a single band require $P_\lambda \geq 3\,\sigma_{P_\lambda}$ in that band.
Correlations involving $\mathcal{S}_\lambda$ additionally require its raw value (equation~\ref{eq:S_def}) to exceed $3\sigma_{\mathcal{S}}$.
The ratio $R_{\rm EW}$ and correlations involving $P_{450}/P_{850}$ require $P_{450} \geq 3\,\sigma_{P_{450}}$ and $P_{850} \geq 3\,\sigma_{P_{850}}$.
Pixels that fail these requirements are represented as limits where specified and do not enter the median $R_{\rm EW}$, which uses detections only.
 
For the spectral index of polarized intensity $\gamma$ and the dispersion function $\mathcal{S}$, a pixel below the corresponding floor is kept as a $3\sigma$ limit.
The index $\gamma$ is an upper limit when $PI_{450} < 3\,\sigma_{PI_{450}}$ and a lower limit when $PI_{850} < 3\,\sigma_{PI_{850}}$.
We do not assign a value when both lie below the floor.
A value of $\mathcal{S}$ is an upper limit when $\mathcal{S} \leq 3\,\sigma_{\mathcal{S}}$.
Each limit is evaluated with the quantity below the floor set to $3\sigma$.
When the median for a source involves such limits, we quote the range between their two extreme placements.
 
Polarization angles are not subject to a floor.
An angle is determined by $Q$ and $U$, and we use its uncertainty directly.
The precision floor of the angle comparison, $4\arcdeg$--$6\arcdeg$ (Paper~T equation~\ref{T-eq:precision_floors}), is itself derived from pixels selected on $I$ alone.
At the thresholds on total intensity used here ($I/\sigma_I > 30$), the floors on $P$ and on $PI$ are nearly the same condition.
With $\sigma_P/P = [(\sigma_{PI}/PI)^2 + (\sigma_I/I)^2]^{1/2}$, $PI/\sigma_{PI} = 3$ corresponds to $P/\sigma_P \geq 2.98$.

\section{Results and Analysis}
\label{sec:S2_results}
A lower polarization fraction at one wavelength can arise because the emitting grains are less efficiently aligned or because their polarized emission is more strongly canceled by field structure.
We examine the polarization spectrum together with projected angle dispersion to distinguish the observational signatures of these possibilities.
Table~\ref{tbl:S2_dependency} summarizes the technical inputs taken from Paper~T and the F1--F3 findings of Paper~S1 on which the F4--F6 tests of this section build.
 
\begin{longrotatetable}
\begin{deluxetable*}{lll}
\tablecaption{Paper~S2: methods and findings adopted from Papers~T and~S1 (basis for the F4--F6 tests of section~\ref{sec:S2_results})\label{tbl:S2_dependency}}
\tablewidth{0pt}
\tablehead{
  \colhead{Adopted framework / finding} &
  \colhead{Source} &
  \colhead{Used in this paper}
}
\startdata
Extended-source FCF + $C(T_{\rm d})$-corrected $\beta$ (Pillar~1) &
Paper~T subsection~\ref{T-ss:flux} &
F4 spectral-index analyses (subsection~\ref{ss:S2_SpIdx}) \\
Stokes-$I$ + $Q,U$ analytic noise model (Pillar~2) &
Paper~T subsections~\ref{T-ss:noiseI}--\ref{T-ss:noisemodel} &
per-pixel detection masks for F6 (subsection~\ref{ss:S2_HGBS}) \\
{450\,$\mu$m Stokes-$Q/U$ zero-level offset correction} + beam harmonization (Pillar~3) &
Paper~T subsections~\ref{T-ss:offset}--\ref{T-ss:beamval} &
matched-resolution dual-band maps used throughout F4--F6 \\
Systematic-uncertainty floor (quantified terms) $\sigma(P_{450}/P_{850}) \lesssim 0.02$, $\sigma(\gamma{-}\alpha) \lesssim 0.04$ &
Paper~T subsection~\ref{T-sec:uncertainty}, table~\ref{T-tbl:systematics} &
$\delta X_{\rm sys}$ entries throughout section~\ref{sec:S2_results} \\
F1--F3 universal signatures of the dual-band polarization spectrum &
Paper~S1 section~\ref{S1-sec:results} &
\parbox[t]{6.5cm}{\raggedright F4 population test (subsections~\ref{ss:S2_Sdisp}--\ref{ss:S2_SpIdx});\\ F5 regime stratification (subsection~\ref{ss:S2_EnvStrat});\\ F6 per-pixel correlations (subsection~\ref{ss:S2_HGBS})\strut} \\
\enddata
\tablecomments{
The four inputs adopted from Paper~T provide maps at matched resolution and the common notation for uncertainties used throughout section~\ref{sec:S2_results}.
The F1--F3 findings from Paper~S1 provide the starting point for the F4--F6 tests of this paper.
Paper~S1 in turn uses our F4--F6 findings across the source population in its discussion of grain alignment; Paper~S1 table~\ref{S1-tbl:S1_dependency} summarizes that use.
}
\end{deluxetable*}
\end{longrotatetable}
 
This section presents the results for individual sources from the 450\,\micron\ and 850\,\micron\ polarimetry and defines the statistical estimators and diagnostic quantities used throughout the BISTRO paper series on both wavelengths.
Subsection~\ref{ss:S2_Sdisp} defines the polarization-angle dispersion function $\mathcal{S}$ as used throughout Papers~S1 and~S2.
Subsection~\ref{ss:S2_SP} defines the proxy for grain-alignment efficiency $\mathcal{S} \times \mathcal{P}^{\rm frac}$ and examines possible distance bias using the adopted reference distances in table~\ref{tab:surv} in Appendix.
Subsection~\ref{ss:S2_P_ratio} defines the three complementary estimators of the polarization-fraction ratio $R_{\rm p} \equiv P_{450}/P_{850}$ (Methods~1, 2, 3) referenced throughout Paper~T and Paper~S1.
Subsection~\ref{ss:S2_SpIdx} presents spectral indices $\alpha$ and $\gamma$ for individual sources in the {\sc SpIdx} subsample adopted by Paper~S1.
Subsections~\ref{ss:S2_EnvStrat} and~\ref{ss:S2_HGBS} compare F2 and F3 among environments (LMSF, HMSF, and CMZ; based on 35 sources), present the regional F1 measurements, and analyze correlations for individual pixels in the HGBS subsample of 9 fields.
Together, these analyses place the wavelength-dependent polarization deficit alongside the field structure and cloud conditions measured in the same survey.
The interpretation accounts for the different selections, physical resolutions, and source coverage of the individual diagnostics.

\subsection{Polarization-Angle Dispersion Function $\mathcal{S}$}
\label{ss:S2_Sdisp}
Changes in the orientation of the projected $\vec{B}$-field can lower the observed polarization even when the grains remain well aligned.
The polarization-angle dispersion function measures how steeply the observed field direction changes between neighboring lines of sight and provides a geometric context for the wavelength-dependent polarization fractions.
The measurements of $\mathcal{S}_{850}$ for individual sources characterize $\vec{B}$-field disorder in the F4 population test (figure~\ref{fig:S2_test_overview}, table~\ref{tbl:S2_test_index}).
 
To distinguish between $\vec{B}$-field structure and grain alignment efficiency as the drivers of the observed wavelength-dependent polarization, we introduce diagnostic quantities for the correlation analyses.
The polarization angle dispersion function $\mathcal{S}$ is calculated from the polarization angles alone and measures the local disorder of the projected $\vec{B}$-field on scales comparable to the POL-2 beam.
In contrast, the polarization fraction $\mathcal{P}^{\rm frac}$ reflects the product of grain alignment efficiency and field structure, as formalized by the reduction factor $\Phi = R \, F \, \cos^2\gamma$ \citep{LeeDraine1985}.
Here $\gamma$ denotes the inclination of the mean field relative to the plane of the sky, distinct from the spectral index of polarized intensity.
By combining these two quantities into the product $\mathcal{S} \times \mathcal{P}^{\rm frac}$, we can partially cancel the geometric contribution from field tangling.
The result is an empirical proxy that is more sensitive to intrinsic grain alignment efficiency (subsection~\ref{ss:S2_SP}).
 
We first define and calculate $\mathcal{S}$ in sub-subsection~\ref{sss:S_definition}, then apply it to L1688 Ophiuchus~A (figure~\ref{afig:Bmap_OphiuA} in Appendix) in sub-subsection~\ref{sss:S850_OphA}.
We summarize the properties across the survey in sub-subsection~\ref{sss:S850_global}.
Angle dispersion identifies where the projected field is relatively coherent and where its direction varies across the map.
It complements the polarization fraction without uniquely separating grain alignment from unresolved structure along the line of sight.

\subsubsection{Definition of $\mathcal{S}$}
\label{sss:S_definition}
 
The dispersion function of the polarization angle $\mathcal{S}$ \citep{Hildebrand2009,Planck2020-XII} quantifies how much the orientation of the projected $\vec{B}$-field varies between neighboring positions in the plane of the sky.
The dispersion does not resolve the structure within the beam or along the line of sight, although a large value suggests that the orientation also changes within the beam \citep{Fissel2016}.
We use this quantity to help distinguish geometric depolarization from possible variations in intrinsic grain alignment efficiency, although it does not separate them uniquely.
The dependence of the polarization fraction on $\mathcal{S}$ and column density was examined in submillimeter observations of Vela~C by \citet{Fissel2016} and of nearby molecular clouds by \citet{Sullivan2021}.
 
We adopt the standard definition of \citet{Planck2020-XII}, also applied to data from the Atacama Large Millimeter/submillimeter Array (ALMA) by \citet{LeGouellec2020}.
At each pixel $\mathbf{r}$, $\mathcal{S}$ is the rms of the angle differences to the $N$ pixels $\mathbf{r}+\boldsymbol{\delta}_i$ in an annulus around $\mathbf{r}$ with inner and outer radii $\delta/2$ and $3\delta/2$, where $\delta$ is the angular lag:
\begin{equation}
\mathcal{S}(\mathbf{r},\delta) = \sqrt{\frac{1}{N}\sum_{i=1}^{N}\left[\chi(\mathbf{r}+\boldsymbol{\delta}_i)-\chi(\mathbf{r})\right]^{2}},
\label{eq:S_def}
\end{equation}
where $\chi$ denotes the polarization angle and all angle differences are wrapped to the range $(-90\arcdeg, +90\arcdeg)$ before calculation.
Noise on $Q$ and $U$ biases $\mathcal{S}$.
Following \citet{Planck2020-XII}, we use the debiased value
\[
\mathcal{S}_{\rm db} = \sqrt{\mathcal{S}^{2}-\sigma_{\mathcal{S}}^{2}} \quad (\mathcal{S}>\sigma_{\mathcal{S}}), \qquad \mathcal{S}_{\rm db}=0 \quad \mbox{otherwise}.
\]
We propagate the variance $\sigma_{\mathcal{S}}^{2}$ from the angle uncertainties $\sigma_{\chi}$ for individual pixels using equation~8 of \citealt{Planck2020-XII}.
We obtain $\sigma_{\chi}$ from the Stokes $Q,U$ variances of the noise model in Paper~T subsection~\ref{T-ss:noisemodel}.
Throughout this paper $\mathcal{S}$ denotes $\mathcal{S}_{\rm db}$.
A pixel in which the value of equation~(\ref{eq:S_def}) does not exceed $3\sigma_{\mathcal{S}}$ is not a detection of $\mathcal{S}$, and we keep it as a $3\sigma$ upper limit (subsection~\ref{ss:S2_noise_floor}).
Appendix~\ref{as:S_bounds} describes the noise response of the raw estimator of equation~(\ref{eq:S_def}) and the reference scale of $\mathcal{S}$ for random orientations, $\approx 52\arcdeg$.
Neither is a strict limit on an individual measurement.
 
\citet{Planck2020-XII} set the lag to half the beam FWHM.
On our 12\arcsec\ grid the smallest available lag is one pixel, $\delta = 12\arcsec$ ($\approx 0.95$ times the 12\farcs6 matched beam).
The annulus for this lag ($6\arcsec \le r < 18\arcsec$) contains the eight nearest neighbors, i.e.\ a $3 \times 3$ window ($36\arcsec \times 36\arcsec$).
We use the same lag at both wavelengths, so $\mathcal{S}$ samples identical physical scales.
We also require at least five valid neighbors.

The dispersion $\mathcal{S}$ depends on the lag \citep{Hildebrand2009,Planck2020-XII,LeGouellec2020}, as expected for a statistic similar to a structure function \citep{Hildebrand2009}.
At a lag of two pixels (24\arcsec), the median $\mathcal{S}$ for an individual source is larger than at one pixel by a factor of 1.30 at 850\,$\mu$m and 1.39 at 450\,$\mu$m (median over sources); the corresponding factor for $\mathcal{S}\times\mathcal{P}^{\rm frac}$ is 1.49.
All values in this paper use the same lag of one pixel at both wavelengths.
The turbulence model of \citet{Planck2020-XII} gives a weak dependence of $\mathcal{S} \times \mathcal{P}^{\rm frac}$ on resolution, $\propto {\rm FWHM}^{0.18}$, i.e.\ $\sim 13\%$ over a factor of 2 in beam size (sub-subsection~\ref{sss:SP_definition}).
The model scaling refers to a lag tied to the beam, whereas the ratios above change the lag at a fixed beam, so the two are not the same test.
 
We calculate $\mathcal{S}$ with the same estimator and lag in both bands, whose projected field orientations are generally similar (Paper~S1 F1).
We therefore use $\mathcal{S}$ to compare the projected angle structure at the two wavelengths.

\begin{figure}[t]
\centering
\includegraphics[width=0.85\textwidth]{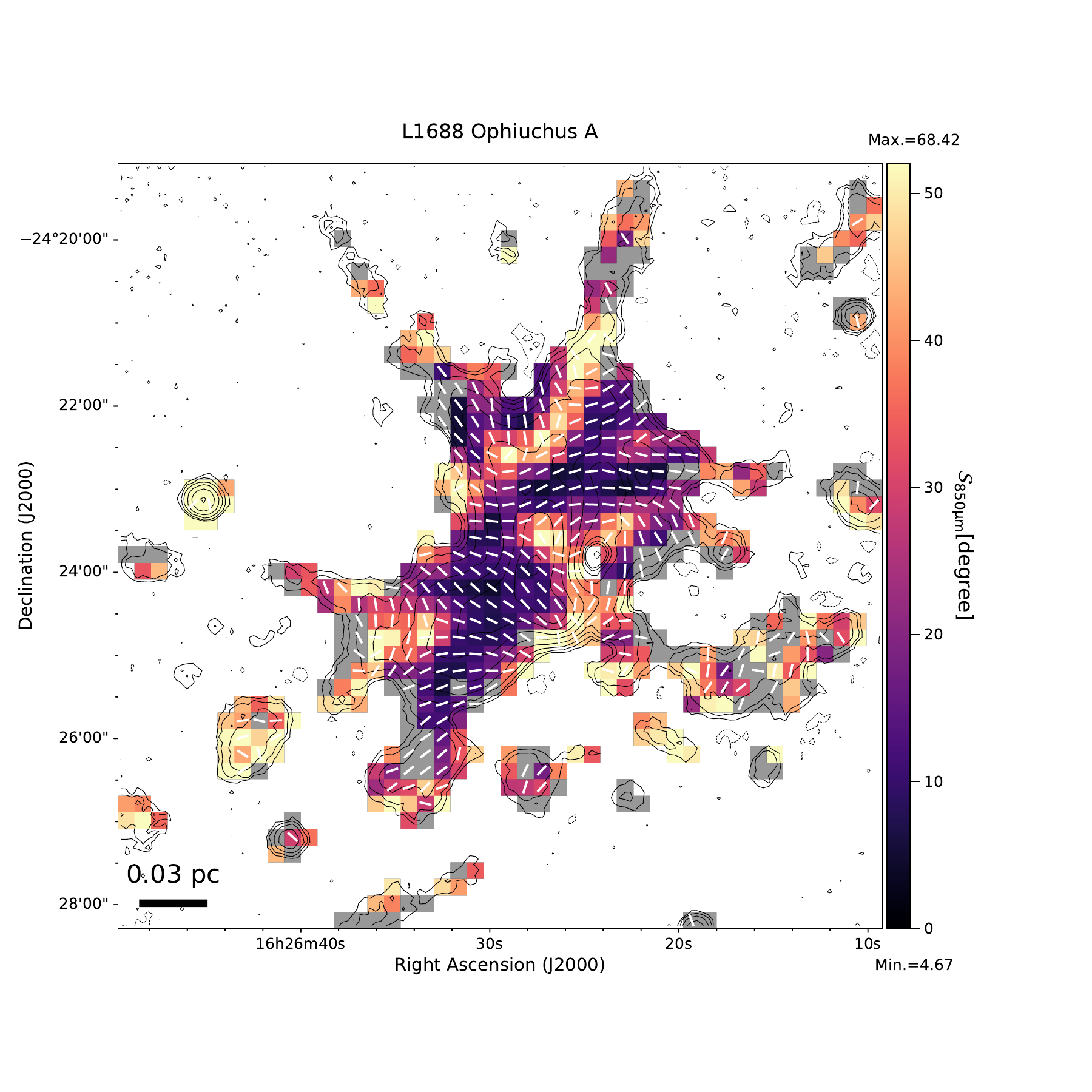}
\caption{
850\,$\mu$m polarization-angle dispersion $\mathcal{S}_{850}$ in L1688 Ophiuchus~A, computed at a lag of one pixel from the eight nearest neighbors in a $3\times3$ pixel ($36\arcsec\times36\arcsec$) window, and debiased (equation~\ref{eq:S_def}; sub-subsection~\ref{sss:S_definition}).
The pixel centers are spaced by $12\farcs0$, nearly one beam width ($12\farcs6$ FWHM).
Colors show pixels with $I_{850}/\Delta I_{850}>10$ where $\mathcal{S}_{850}$ is detected at $3\sigma_{\mathcal{S}}$.
Here, $\Delta I_{850}$ is the uncertainty of Stokes~$I$ for each pixel, derived from the variance map.
Gray pixels are $3\sigma$ upper limits (24\% of the pixels; table~\ref{tbl:S850stats}).
White $\vec{B}$-field segments require $I_{850}/\Delta I_{850}>30$, without a cut on the $S/N$ of polarized intensity, and therefore lie within the colored region.
Black Stokes-$I$ contours are at $15\sigma\times2^n$, $n=0,1,2,\ldots$.
The $0\arcdeg$--$52\arcdeg$ color scale is the common range adopted across the survey.
The upper end of this scale is the reference value of $\mathcal{S}$ for random orientations (appendix~\ref{as:S_bounds}), not an upper bound on individual measurements.
The detected values in this panel span $4\fdg7$--$68\fdg4$ and can exceed the upper end of the color scale.
The scale bar is based on the adopted reference distance of 139\,pc (table~\ref{tab:surv} in Appendix).
Table~\ref{tbl:S850stats} gives summary statistics for individual sources.
\\ {Alt text: Map with short line segments over contours.}
}
\label{fig:S850_OphiuA}
\end{figure}

\subsubsection{Application to L1688 Ophiuchus~A}
\label{sss:S850_OphA}
 
Figure~\ref{fig:S850_OphiuA} shows the spatial distribution of the polarization angle dispersion $\mathcal{S}_{850}$ at 850\,$\mu$m across L1688 Ophiuchus~A.
The map reveals a systematic variation in the coherence of the $\vec{B}$-field from the interior of the dense filament to its periphery.
Among the detected pixels, $\mathcal{S}_{850}$ is about $17\arcdeg$ along the ridge of high intensity ($I_{850}/\Delta I_{850}\geq120$) and increases toward regions of lower intensity, reaching medians of about $40\arcdeg$--$46\arcdeg$ at $I_{850}/\Delta I_{850}=10$--30.
The upper-limit fraction also increases in the lower-intensity bins (32--46\%), so the increase of $\mathcal{S}_{850}$ toward lower intensity refers to detections only.
The overlaid $\vec{B}$-field segments, selected with $I_{850}/\Delta I_{850} > 30$, display smooth and continuous orientations in these regions of low dispersion, consistent with a coherent field threading the dense filament.
 
Elevated values of $\mathcal{S}_{850}$ ($>40\arcdeg$) appear predominantly at the edges of the filament and in the vicinity of embedded substructures, where the projected field orientation varies more rapidly.
(The morphology of the filament and substructure referred to here is as mapped in the continuum by \citealt{Andre1993} and \citealt{Motte1998}.)
Rapid changes in the projected field at cloud edges can arise from the superposition of distinct field components along the line of sight, from field lines draped around the denser structure, and from general turbulent disordering of the field.
All three increase the measured angle dispersion without requiring a change in grain alignment.
Bending or convergence of field lines may also contribute to the elevated $\mathcal{S}_{850}$.
 
High $\mathcal{S}_{850}$ is spatially associated with the boundaries of the dense filament.
This association suggests that the transition from an ordered field in the coherent core to a more complex configuration in the surrounding envelope occurs over relatively small scales.
 
Table~\ref{tbl:S850stats} summarizes the statistical properties of $\mathcal{S}_{850}$ for L1688 Ophiuchus~A and the full survey sample.
For Ophiuchus~A, the median dispersion is $\mathcal{S}_{50} = 31\fdg0$ with a scatter of $19\fdg7$ based on the median absolute deviation (MAD).
The median lies between $22\fdg8$ and $35\fdg4$ when the 24\% of pixels in which $\mathcal{S}$ is a $3\sigma$ upper limit are set to zero or to their limit.
The fraction of pixels with $\mathcal{S}_{850} > 40\arcdeg$ is $f_{>40} = 30.8\%$, indicating that regions of high dispersion occupy a limited but non-negligible fraction of the mapped area.
The $40\arcdeg$ threshold is a reference value rather than a physical one.
This threshold lies just below the random-orientation reference value of $\approx 52\arcdeg$ (appendix~\ref{as:S_bounds}), so pixels above it have angle dispersions approaching the value expected for a completely random field.
None of the conclusions below depend on the precise value.
Among nearby ($d < 300$\,pc) star-forming regions, Ophiuchus~A exhibits intermediate values of both median $\mathcal{S}_{850}$ and $f_{>40}$, comparable to other well-studied clouds such as NGC~1333 (figure~\ref{afig:Bmap_N133MO} in Appendix; \citealt{Doi2020}) and Perseus~IC348 (figure~\ref{afig:Bmap_IC348H} in Appendix; \citealt{Choi2024}).
 
A previous POL-2 study of Ophiuchus~A at 850\,$\mu$m identified ten subregions with distinct polarization fractions and angles, some of which can be part of a coherent $\vec{B}$-field structure \citep{Kwon2018}.
Our $\mathcal{S}_{850}$ map quantifies this coherence, showing where the ordered field persists and where its orientation becomes more disordered within the same region.

\begin{longrotatetable}
\footnotesize
\begin{deluxetable*}{lcccccccl}
\tablecaption{Statistics of 850\,$\mu$m Magnetic Field Angle Dispersion\label{tbl:S850stats}}
\tablewidth{0pt}
\tablehead{
  \colhead{Source} &
  \colhead{$N$} &
  \colhead{$\mathcal{S}_{25}$} &
  \colhead{$\mathcal{S}_{50} \pm \sigma_{\rm MAD}$} &
  \colhead{$\mathcal{S}_{75}$} &
  \colhead{$f_{>40}$} &
  \colhead{$\mathcal{S}_{50}$ range} &
  \colhead{$f_{\rm lim}$} &
  \colhead{Note} \\
  \colhead{} &
  \colhead{} &
  \colhead{(\arcdeg)} &
  \colhead{(\arcdeg)} &
  \colhead{(\arcdeg)} &
  \colhead{(\%)} &
  \colhead{(\arcdeg)} &
  \colhead{(\%)} &
  \colhead{}
}
\startdata
\cutinhead{LMSF --- low-mass star-forming}
  Auriga                              &   553 &  24.8 & $36.4 \pm 14.9$ &  45.5 &  39.6 & 30.5--42.1 & 31.5 &  \\
  G209.29-19.65                       &   604 &  25.8 & $39.4 \pm 15.7$ &  48.1 &  48.0 & 25.9--48.5 & 42.7 &  \\
  IC 5146                             &   341 &  19.3 & $32.5 \pm 17.0$ &  42.5 &  29.9 & 19.9--41.5 & 43.7 &  \\
  L1448 IRS2                          &   288 &  25.1 & $36.0 \pm 16.0$ &  46.3 &  40.3 & 30.7--41.9 & 34.0 &  \\
  L1688 Ophiuchus A                   &   613 &  15.9 & $31.0 \pm 19.7$ &  43.2 &  30.8 & 22.8--35.4 & 24.3 &  \\
  L1688 Ophiuchus B                   &   509 &  20.9 & $35.2 \pm 18.0$ &  44.9 &  36.9 & 28.0--40.0 & 32.0 &  \\
  L1689 North                         &   319 &  13.3 & $24.5 \pm 19.2$ &  39.0 &  23.8 & 13.8--32.3 & 36.1 &  \\
  L1689 South                         &   282 &  22.3 & $33.9 \pm 17.1$ &  45.3 &  37.6 & 25.3--40.4 & 31.9 &  \\
  L43                                 &   208 &  32.5 & $40.9 \pm 12.8$ &  49.6 &  52.4 & 36.8--47.6 & 33.7 &  \\
  L483                                &    86 &  26.4 & $33.6 \pm 15.4$ &  45.9 &  37.2 & 32.1--40.1 & 27.9 &  \\
  NGC1333 North                       &   913 &  20.5 & $31.7 \pm 17.1$ &  43.4 &  31.7 & 23.6--39.9 & 33.2 &  \\
  NGC1333 South                       &   832 &  24.0 & $35.0 \pm 15.9$ &  45.2 &  37.1 & 24.9--42.1 & 37.9 &  \\
  Orion A Field 5                     &  1147 &  19.3 & $31.4 \pm 17.3$ &  42.7 &  30.9 & 23.2--37.5 & 32.9 &  \\
  Orion A Field 7                     &   440 &  27.8 & $37.1 \pm 14.6$ &  47.7 &  43.0 & 32.0--45.2 & 40.2 &  \\
  Orion-B NGC2071                     &   810 &  25.4 & $36.8 \pm 14.9$ &  45.8 &  41.4 & 31.3--42.5 & 34.6 &  \\
  Perseus B1                          &   297 &  16.4 & $27.8 \pm 17.8$ &  40.2 &  25.9 & 16.5--37.2 & 40.4 &  \\
  Perseus IC348                       &   558 &  16.4 & $29.7 \pm 19.0$ &  42.1 &  28.5 & 17.4--38.6 & 37.5 &  \\
  Serpens South                       &   699 &  17.7 & $32.4 \pm 18.9$ &  43.6 &  32.5 & 14.8--41.5 & 43.2 &  \\
\cutinhead{HMSF --- high-mass star-forming}
  Cepheus A                           &   385 &  14.0 & $28.9 \pm 20.5$ &  41.0 &  27.0 & 15.7--35.4 & 30.1 &  \\
  DR21                                &  1209 &  12.6 & $23.6 \pm 18.1$ &  37.7 &  20.8 & 14.1--28.9 & 28.2 &  \\
  G31.41+0.31                         &   504 &  16.3 & $30.4 \pm 19.5$ &  42.9 &  30.8 & 15.5--39.7 & 36.1 &  \\
  M16                                 &   533 &  23.4 & $34.4 \pm 14.8$ &  44.2 &  36.0 & 24.6--43.8 & 42.2 &  \\
  Monoceros R2                        &  1051 &  14.4 & $28.9 \pm 20.4$ &  41.4 &  27.5 & 20.6--33.2 & 22.5 &  \\
  NGC 6334 North                      &  1411 &  14.1 & $26.5 \pm 18.5$ &  39.1 &  23.6 & 22.2--28.5 & 15.7 &  \\
  NGC 6334 South                      &  1378 &  15.2 & $25.1 \pm 16.6$ &  38.3 &  21.3 & 20.0--28.5 & 19.4 &  \\
  NGC 7538 IRS4                       &  1006 &  19.8 & $32.7 \pm 17.3$ &  43.2 &  32.3 & 24.2--38.5 & 30.2 &  \\
  NGC2264 Single Field\#1              &  1005 &  19.0 & $29.7 \pm 17.5$ &  42.7 &  30.1 & 21.4--37.2 & 31.1 &  \\
  NGC2264 Single Field\#2              &   930 &  14.7 & $27.7 \pm 19.9$ &  41.3 &  27.6 & 15.7--35.4 & 33.4 &  \\
  OMC 1                               &  1352 &  10.4 & $19.3 \pm 15.9$ &  35.2 &  17.0 & 15.8--21.6 & 16.1 &  \\
  OMC 2                               &  1414 &  19.4 & $30.8 \pm 16.4$ &  41.5 &  28.5 & 26.1--35.6 & 24.8 &  \\
  OMC 3                               &  1222 &  14.6 & $26.9 \pm 18.2$ &  39.1 &  22.9 & 18.8--33.1 & 32.2 &  \\
  Onsala 2                            &   995 &  18.8 & $30.8 \pm 17.4$ &  42.2 &  29.8 & 19.9--38.5 & 33.6 &  \\
  Orion A Field 4                     &  1188 &  13.8 & $26.0 \pm 18.6$ &  38.7 &  22.1 & 22.0--28.8 & 15.7 &  \\
  S106 (mosaic component field)       &   552 &  13.9 & $27.0 \pm 19.5$ &  40.3 &  26.1 & 14.2--35.5 & 36.2 &  \\
  SDC 13 mm-2                         &  1201 &  24.8 & $36.3 \pm 15.1$ &  45.9 &  40.2 & 29.4--43.0 & 32.4 &  \\
  W3 HFS2                             &   655 &  16.4 & $29.3 \pm 19.5$ &  42.5 &  29.8 & 20.3--35.2 & 27.3 &  \\
\cutinhead{CMZ --- Central Molecular Zone}
  Brick                               &  1364 &  22.0 & $34.5 \pm 16.7$ &  44.8 &  37.2 & 26.1--41.7 & 32.5 &  \\
  Sagittarius B2                      &  1509 &  13.7 & $26.4 \pm 18.7$ &  38.9 &  22.7 & 19.9--30.0 & 21.4 &  \\
  The Gal. Cent. 20 km s$^{-1}$ cloud        &  1174 &  24.4 & $36.4 \pm 16.3$ &  47.0 &  41.8 & 29.7--43.6 & 29.9 &  \\
  The Gal. Cent. Clouds E and F       &  1372 &  24.3 & $35.9 \pm 15.5$ &  45.5 &  38.9 & 29.4--41.3 & 28.9 &  \\
  The Gal. Cent. Middle               &  1627 &  19.9 & $31.3 \pm 16.9$ &  42.6 &  30.7 & 25.6--36.3 & 23.2 &  \\
  The Gal. Cent. North                &  1564 &  22.6 & $34.9 \pm 16.5$ &  45.1 &  36.9 & 28.7--41.1 & 30.4 &  \\
  The Gal. Cent. South                &  1375 &  22.8 & $35.1 \pm 16.7$ &  45.4 &  37.8 & 29.6--39.6 & 23.3 &  \\
\cutinhead{Stacked fields}
  Orion\,A (stacked)                  &  3988 &  14.5 & $26.1 \pm 18.0$ &  38.8 &  22.9 & 19.7--31.0 & 24.1 & stacked (3 fields) \\
  NGC\,2264 (stacked)                 &  1935 &  16.9 & $29.0 \pm 18.5$ &  41.9 &  28.9 & 18.7--36.2 & 32.2 & stacked (2 fields) \\
  NGC\,6334 (stacked)                 &  2789 &  14.7 & $25.8 \pm 17.5$ &  38.7 &  22.5 & 20.9--28.5 & 17.6 & stacked (2 fields) \\
  CMZ (stacked)                       &  9985 &  21.1 & $33.4 \pm 17.1$ &  44.3 &  34.8 & 26.8--39.1 & 26.9 & stacked (7 fields) \\
\enddata
\tablecomments{
Statistics of the dispersion $\mathcal{S}$ of $\vec{B}$-field position angles at 850\,$\mu$m
computed with equation~(\ref{eq:S_def}) and debiased as in sub-subsection~\ref{sss:S_definition}, at a lag of one pixel (eight nearest neighbors on the 12\arcsec\ grid; $I_{850}/\sigma_{I} > 10$).
Columns show: source name, number of valid pixels $N$,
25th percentile $\mathcal{S}_{25}$, median $\tilde{\mathcal{S}}$ with MAD-based dispersion $\sigma_{\rm MAD}$,
75th percentile $\mathcal{S}_{75}$, the fraction of pixels with $\mathcal{S} > 40\arcdeg$,
the range of the median when the pixels in which $\mathcal{S}$ is a $3\sigma$ upper limit [equation~(\ref{eq:S_def}) not exceeding $3\sigma_{\mathcal{S}}$; subsection~\ref{ss:S2_noise_floor}] are set to zero and to $3\sigma_{\mathcal{S}}$,
and the percentage $f_{\rm lim}$ of such pixels.
The percentiles, the median, and $f_{>40}$ use the debiased values of all pixels.
We group the sources by star-forming environment (LMSF / HMSF / CMZ; subsection~\ref{ss:S2_envcls}) under printed block headings and order them alphabetically within each block.
The HMSF block collects regions hosting intermediate- to high-mass YSOs.
The grouping is by environment and not by distance.
The HMSF block spans $d \approx 400$\,pc (the OMC fields) to $d \approx 3.8$\,kpc (G31.41$+$0.31), and only the CMZ block lies at the Galactic Center.
Table~\ref{tab:surv} in Appendix lists the distances separately.
The stacked entries in the final block are aggregates of the individual fields already listed above, and are excluded from the totals quoted for each environment in sub-subsection~\ref{sss:env_synthesis}.
Only sources with $N \geq 50$ are included in this statistical sample.
}
\end{deluxetable*}
\end{longrotatetable}

\subsubsection{Global Properties of $\mathcal{S}_{850}$}
\label{sss:S850_global}
 
To provide the geometric context for the polarization comparison, we examine how the observed coherence of the projected magnetic field varies among the surveyed clouds.
Table~\ref{tbl:S850stats} gives the statistics of the angle dispersion at 850\,$\mu$m for 43 individual sources and four stacked fields, grouped by star-forming environment.
The median dispersion across all sources ranges from $19\fdg3$ (OMC~1; figure~\ref{afig:Bmap_OriAS1} in Appendix) to $40\fdg9$ (L43; figure~\ref{afig:Bmap_L43L43} in Appendix), with most sources falling between $27\arcdeg$ and $37\arcdeg$.
The MAD-based scatter within individual sources typically spans $15\arcdeg$--$20\arcdeg$, reflecting substantial internal variation in field coherence.
Between 16\% and 44\% of the pixels of each source are $3\sigma$ upper limits of $\mathcal{S}$.
The range column of table~\ref{tbl:S850stats} spans the medians obtained with the upper-limit pixels set to zero and with them set to their limits.
For the two extremes of the sample, the ranges do not overlap (OMC~1, $15\fdg8$--$21\fdg6$; L43, $36\fdg8$--$47\fdg6$).
 
The fraction of pixels with $\mathcal{S}_{850} > 40\arcdeg$, denoted $f_{>40}$, provides a measure of how much of each region exhibits high angular dispersion approaching the random-orientation reference value.
This fraction ranges from $17.0\%$ (OMC~1) to $52.4\%$ (L43), with most sources falling between $23\%$ and $41\%$.
OMC~1 has the lowest median $\mathcal{S}_{850}$ and the lowest $f_{>40}$ in the sample, while L43 has the highest values of both.
 
We did not find a systematic trend of $\mathcal{S}_{850}$ with distance in a comparison of three distance groups.
The nearby sources ($d < 300$\,pc) span median values of $24\fdg5$--$40\fdg9$, while intermediate ($300 \leq d < 1000$\,pc) and distant ($d \geq 1000$\,pc) sources show ranges of $19\fdg3$--$39\fdg4$ and $23\fdg6$--$36\fdg4$, respectively.
We do not quote a coefficient or interval for this comparison, so it has little power to exclude a difference between the groups.
The comparison supports, rather than establishes, the use of $\mathcal{S}_{850}$ as a diagnostic of intrinsic field structure.
The absence of a distance trend in $\mathcal{S}_{850}$ is consistent with previous submillimeter polarization studies, which used $\mathcal{S}$ as a measure of the small-scale disorder of the projected field \citep{Fissel2016,Sullivan2021}.
The same $12\farcs6$ beam nevertheless smooths the field structure over physical scales that increase with distance.
The value of $\mathcal{S}$ also depends on the lag (sub-subsection~\ref{sss:S_definition}).
 
The stacked fields [Orion~A, NGC~2264 (figure~\ref{afig:Bmap_N226MO} in Appendix), NGC~6334 (figure~\ref{afig:Bmap_N633MO} in Appendix), and the CMZ] show median dispersions consistent with their constituent individual fields.
For example, the Orion~A stacked field ($\mathcal{S}_{50} = 26\fdg1$) falls between OMC~1 ($19\fdg3$) and OMC~2/3 (figures~\ref{afig:Bmap_OriAS2}--\ref{afig:Bmap_OriAS3} in Appendix; $30\fdg8$ and $26\fdg9$), as expected for a combination weighted by area.
Similarly, the CMZ stacked field ($\mathcal{S}_{50} = 33\fdg4$) represents an intermediate value among the seven constituent Galactic Center fields, which range from $26\fdg4$ (Sgr~B2) to $36\fdg4$ (GC~20~km~s$^{-1}$).
At the adopted angular resolution and lag, the projected field orientations are more coherent in OMC~1 than in L43.
This contrast remains under the two extreme treatments of the pixels with upper limits.

\subsection{The Product $\mathcal{S} \times \mathcal{P}^{\rm frac}$ as a Proxy for Grain-Alignment Efficiency}
\label{ss:S2_SP}
Weak polarization in a dense cloud can indicate inefficient grain alignment, a tangled magnetic field, or both.
Combining the polarization fraction with the local angle dispersion provides an empirical way to examine whether differences remain after partially compensating for projected field disorder.
We used $\mathcal{S}\times\mathcal{P}^{\rm frac}$ at both wavelengths in the F4 population test (figure~\ref{fig:S2_test_overview}, table~\ref{tbl:S2_test_index}) and examined its distance dependence before comparing sources.
 
This subsection introduces the product $\mathcal{S} \times \mathcal{P}^{\rm frac}$ as an empirical diagnostic for grain alignment efficiency.
We first give the physical motivation and definition (sub-subsection~\ref{sss:SP_definition}), then show its spatial variation in L1688 Ophiuchus~A (sub-subsection~\ref{sss:SP450_OphA}; figure~\ref{afig:Bmap_OphiuA} in Appendix).
We next assess how much of the variation between sources can be attributed to the different physical scales sampled by a fixed angular beam (sub-subsection~\ref{sss:SP_bias}), because we use this quantity to compare sources.
We then summarize its properties across the survey sample (sub-subsection~\ref{sss:SP_global}).
We use the product to compare polarization behavior after partially compensating for the projected field disorder.
Variations in the product can motivate an alignment interpretation, but viewing geometry, sampling scale, and measurement noise remain part of the comparison.

\subsubsection{Definition of $\mathcal{S} \times \mathcal{P}^{\rm frac}$}
\label{sss:SP_definition}
 
The observed polarization fraction $\mathcal{P}^{\rm frac}$ is reduced by both poor grain alignment and tangling of the $\vec{B}$-field within the beam.
To isolate the contribution of grain alignment physics from geometric depolarization, we adopt the product $\mathcal{S} \times \mathcal{P}^{\rm frac}$ as an empirical diagnostic for alignment efficiency.
 
We base this diagnostic on the description of polarization in \citet{LeeDraine1985}.
In their formulation, the observed polarization depends on a reduction factor $\Phi = R \, F \, \cos^2\gamma$ (equation~3.12 of \citealt{LeeDraine1985}).
Here $R$ measures grain alignment with the local field, with $R = 0$ for random orientations and $R = 1$ for perfect alignment.
The factor $F$ describes tangling of the field along the line of sight, with $F = 0$ for completely disordered fields and $F = 1$ for a uniform field.
The factor $\cos^2\gamma$ accounts for the inclination of the mean field relative to the plane of the sky.
Consequently, the polarization fraction scales as $\mathcal{P}^{\rm frac} \propto R \, F \, \cos^2\gamma$, representing the product of intrinsic grain alignment efficiency ($R$) and geometric factors ($F \, \cos^2\gamma$).
 
The diagnostic relies on an inverse relation between field order, measured by $F$, and angular dispersion, measured by $\mathcal{S}$.
When field tangling dominates, $F$ decreases and $\mathcal{S}$ increases because both respond to the same disorder.
If $F \propto 1/\mathcal{S}$, which we infer from the inverse relation between $\mathcal{S}$ and $\mathcal{P}^{\rm frac}$ in the Planck turbulence model (\citealt{Planck2020-XII}), the product $\mathcal{S} \times \mathcal{P}^{\rm frac}$ cancels the first-order depolarization from field tangling:
\begin{equation}
\mathcal{S} \times \mathcal{P}^{\rm frac} \propto R \, \cos^2\gamma.
\label{eq:SP_cancellation}
\end{equation}
In this limit, $\mathcal{S} \times \mathcal{P}^{\rm frac}$ is primarily sensitive to the intrinsic grain alignment efficiency through $R$ and to the large-scale viewing geometry via $\cos^2\gamma$, while the small-scale tangling term $F$ cancels out.

The inverse relation between $\mathcal{S}$ and $\mathcal{P}^{\rm frac}$ follows from the turbulent-field model of \citet{Planck2020-XII}.
Their analytical treatment in appendix~F (\citealt{Planck2020-XII}) gives, for the case of strong turbulence,
\begin{equation}
\langle \mathcal{S}(\delta) \rangle_p \approx \frac{f_m(\delta)}{\sqrt{6N}} \frac{p_{\rm max}}{p} \propto \frac{1}{\mathcal{P}^{\rm frac}},
\label{eq:S_p_relation}
\end{equation}
where $p$ is the dimensionless polarization fraction in the notation of the cited model, $N$ is the effective number of independent polarizing layers along the line of sight, and $p_{\rm max}$ is the maximum polarization fraction that the dust can produce.
This maximum is the value that would be observed for a uniform magnetic field parallel to the plane of the sky (analogous to $R$ in the formalism of \citealt{LeeDraine1985}).
At fixed angular scale $\delta$, the model predicts $\mathcal{S} \propto 1/\mathcal{P}^{\rm frac}$, so that $\mathcal{S} \times \mathcal{P}^{\rm frac}$ is approximately constant apart from the weak dependence on scale encoded in $f_m(\delta)$.
 
The turbulence parameter $f_m(\delta)$ characterizes the magnetic field fluctuations at angular scale $\delta$.
With the signed spectral-index convention $C_\ell \propto \ell^{\alpha_M}$ for the angular power spectrum $C_\ell$ at multipole $\ell$, it scales as $f_m(\delta) \propto \delta^{-1 - \alpha_M/2}$ (appendix~F.7 of \citealt{Planck2020-XII}).
The effective index $\alpha_M = -2.36$ gives an exponent of $+0.18$ (appendix~F.9 of \citealt{Planck2020-XII}).
The numerical model of \citet{Planck2020-XII} motivates the effective scaling $\mathcal{S} \times \mathcal{P}^{\rm frac} \propto \mathrm{FWHM}^{0.18}$, corresponding to only $\sim$13\% variation over a factor of 2 in angular scale.
We use this scaling as a reference from the model, not as a resolution correction for the cloud measurements presented here.
 
Empirical studies support the validity of this proxy.
\citet{Planck2020-XII} demonstrated that while $\mathcal{P}^{\rm frac}$ decreases by a factor of 3--4 with increasing column density from $N_{\rm H} = 10^{20}$ to $2 \times 10^{22}$\,cm$^{-2}$, the product $\mathcal{S} \times \mathcal{P}^{\rm frac}$ decreases only $\sim$25\% (full sky at $10\arcmin$--$160\arcmin$ resolution).
Similarly, \citet{LeGouellec2020} found that $\mathcal{S} \times \mathcal{P}^{\rm frac}$ remains approximately constant at $0.36^{+0.10}_{-0.17}$ degrees across three orders of magnitude in column density ($10^{22}$--$10^{25}$\,cm$^{-2}$) in ALMA observations of Class~0 protostars.
These observations probe envelope radii of $\sim$10--2000\,au.
However, in their low-luminosity sources (bolometric luminosities below $10\,L_\odot$), $\mathcal{S} \times \mathcal{P}^{\rm frac}$ hints at a decrease with column density.
Their $\mathcal{S}$--$\mathcal{P}^{\rm frac}$ relation is also shallower than the one that \citet{Planck2020-XII} found in Gould Belt molecular clouds at $40\arcmin$ resolution, that is, on parsec scales.
We interpret these studies as indicating that $\mathcal{S} \times \mathcal{P}^{\rm frac}$ behaves approximately as a constant within a given environment.
Our $12\farcs6$ maps sample physical scales from 0.008 to 0.51\,pc, between the envelope scales of \citet{LeGouellec2020} and the cloud scales of \citet{Planck2020-XII}.
 
We compute $\mathcal{S} \times \mathcal{P}^{\rm frac}$ at both 450 and 850\,$\mu$m, denoted $\mathcal{S}_{450} \times \mathcal{P}_{450}^{\rm frac}$ and $\mathcal{S}_{850} \times \mathcal{P}_{850}^{\rm frac}$, respectively:
\begin{equation}
(\mathcal{S} \times \mathcal{P})_{\lambda} = \mathcal{S}_{\lambda} \times \mathcal{P}_{\lambda}^{\rm frac},
\label{eq:SP_def}
\end{equation}
where $\mathcal{S}_{\lambda}$ is the angle dispersion (in degrees) and $\mathcal{P}_{\lambda}^{\rm frac}$ is the polarization fraction (dimensionless) at wavelength $\lambda$.
The resulting quantity has units of degrees.
 
We use $\mathcal{S} \times \mathcal{P}^{\rm frac}$ as an empirical first-order diagnostic that partially compensates for geometric depolarization.
Beam averaging and integration along the line of sight complicate its interpretation in the complex, high-density environments of our star-forming regions.
This diagnostic is not a fundamental measure of alignment derived from first principles.

\begin{figure}[t]
\centering
\includegraphics[width=0.85\textwidth]{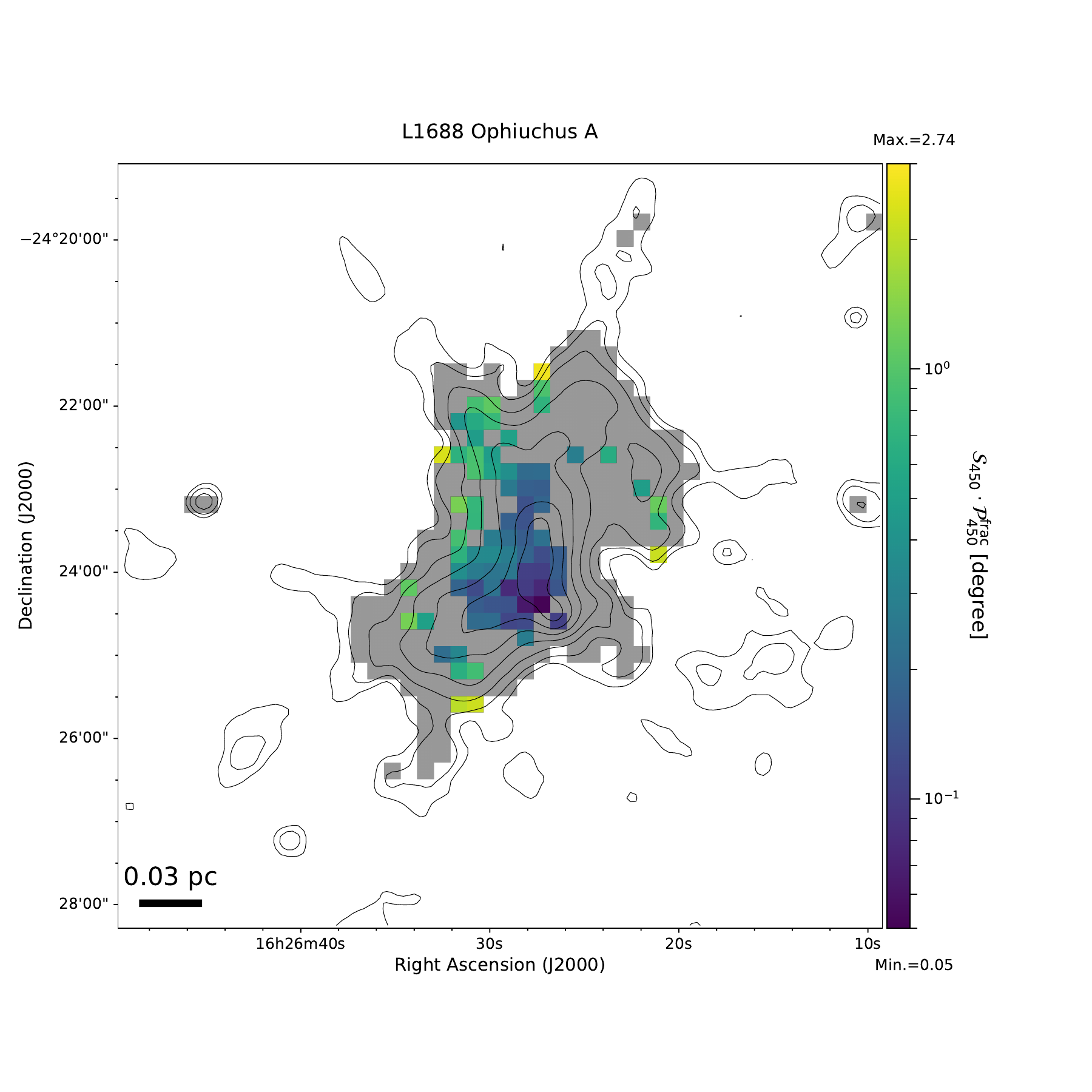}
\caption{
450\,$\mu$m grain-alignment efficiency proxy $\mathcal{S}_{450}\times\mathcal{P}_{450}^{\rm frac}$ in L1688 Ophiuchus~A (equation~\ref{eq:SP_def}; sub-subsection~\ref{sss:SP_definition}).
The product combines angle dispersion with the dimensionless polarization fraction and is expressed in degrees.
Colors show pixels with $I_{450}/\sigma_I>30$ where the product is detected ($\mathcal{S}$ detected at $3\sigma_{\mathcal{S}}$ and $P \geq 3\sigma_P$).
Gray pixels are $3\sigma$ upper limits, 73\% of the pixels (table~\ref{tbl:SP450stats} in Appendix).
The logarithmic color scale spans $0.05\arcdeg$--$3.0\arcdeg$, the display range used throughout the survey.
The detected values in this panel span $0\fdg05$--$2\fdg74$.
Black contours show 850\,$\mu$m Stokes~$I$ at $15\sigma\times2^n$, $n=0,1,2,\ldots$.
The scale bar is based on the adopted reference distance of 139\,pc (table~\ref{tab:surv} in Appendix).
Along the bright filament, detections form groups of adjacent pixels with relatively low values of $\mathcal{S}_{450}\times\mathcal{P}^{\rm frac}_{450}$.
A few higher values occur in scattered single pixels or pairs toward the cloud periphery, where the 850\,$\mu$m Stokes~$I$ intensity is lower.
The pixel centers are spaced by $12\farcs0$, nearly one beam width ($12\farcs6$ FWHM).
This contrast is conditional on detection.
The map therefore does not establish that the proxy is higher throughout the envelope.
Table~\ref{tbl:SP450stats} in Appendix gives the point estimates, percentiles, and upper-limit fractions for individual sources.
\\ {Alt text: Map with contours.}
}
\label{fig:SP450_OphiuA}
\end{figure}

\subsubsection{Application to L1688 Ophiuchus~A}
\label{sss:SP450_OphA}
 
Figure~\ref{fig:SP450_OphiuA} shows the spatial distribution of the grain-alignment efficiency proxy $\mathcal{S}_{450} \times \mathcal{P}_{450}^{\rm frac}$ across L1688 Ophiuchus~A.
The product provides an empirical diagnostic that partially mitigates geometric depolarization associated with field tangling (sub-subsection~\ref{sss:SP_definition}).
 
Binning by the 850\,$\mu$m contours, we found median detected values of $0\fdg21$--$0\fdg27$ where $I_{850}/\sigma_I \geq 60$ and $0\fdg71$ at $I_{850}/\sigma_I = 30$--60.
Only two and one pixels are detected at $I_{850}/\sigma_I = 15$--30 and 10--15, respectively.
The upper-limit fraction rises from 60\%--73\% where $I_{850}/\sigma_I \geq 60$ to 85\%--88\% at lower intensity.
Because the product contains $\mathcal{S}_{450}$, its spatial correspondence with $\mathcal{S}_{850}$ in figure~\ref{fig:S850_OphiuA} is not independent evidence for a link between field order and alignment efficiency.
 
Table~\ref{tbl:SP450stats} in Appendix gives the statistics of $\mathcal{S}_{450} \times \mathcal{P}_{450}^{\rm frac}$ for Ophiuchus~A.
The median value is $(\mathcal{S} \times \mathcal{P})_{50} = 0\fdg364$ with a scatter of $0\fdg324$ based on the MAD.
In this sample, 73\% of the pixels are $3\sigma$ upper limits (in $\mathcal{S}$, in $\mathcal{P}^{\rm frac}$, or both).
The median is therefore bounded only from above ($0\fdg000$--$0\fdg818$ with these pixels set to zero and to their limits).
The distribution is positively skewed, with a relatively small fraction of pixels forming a tail toward high values.
 
Ophiuchus~A has been observed with POL-2 at 850\,$\mu$m \citep{Kwon2018} and with SOFIA/HAWC+ at far-infrared wavelengths \citep{Santos2019}.
Along the filament traced by the peak 850\,$\mu$m Stokes~$I$ emission, the detected proxy values are lower than the few values detected at lower intensity.
Most selected pixels provide only upper limits, so the map does not establish a systematic increase of the proxy throughout the envelope.

\subsubsection{Effect of Physical Resolution on $\mathcal{S} \times \mathcal{P}^{\rm frac}$}
\label{sss:SP_bias}
 
A fixed angular beam does not necessarily sample the same $\vec{B}$-field structure in sources at different distances.
The dispersion function is sensitive to the spatial scale of magnetic fluctuations through the $f_m(\delta)$ term in equation~(\ref{eq:S_p_relation}).
The Planck turbulence model predicts only a weak dependence, $\mathcal{S} \times \mathcal{P}^{\rm frac} \propto \mathrm{FWHM}^{0.18}$, corresponding to a $\sim$13\% variation over a factor of 2 in resolution \citep{Planck2020-XII}.
Our sample spans a larger range in physical resolution.
The source distances are 125\,pc (L43) to 8.34\,kpc (Sagittarius~B2; table~\ref{tab:surv} in Appendix).
The $12\farcs6$ matched beam (subsection~\ref{ss:S2_products}) therefore corresponds to physical scales from 0.008\,pc to 0.51\,pc, a factor of 67.
The same angular resolution thus probes sub-core or core scales in the nearest clouds, increasingly the scales of clumps at distances of several kiloparsecs, and $\sim$0.5\,pc substructure in the dense gas of CMZ clouds.
Any comparison across these environments must therefore consider the different amount of spatial averaging introduced by the beam.
 
We tested this effect with the median point estimates of $\mathcal{S}_{450} \times \mathcal{P}_{450}^{\rm frac}$ for the 39 individual {\sc SP450} sources with $N \geq 50$ (table~\ref{tbl:SP450stats} in Appendix).
We excluded stacked fields.
We exclude W51 from this paper altogether (see the note to table~\ref{tab:surv} in Appendix and Paper~T subsection~\ref{T-ss:noisemodel} for the corresponding $Q,U$ diagnostics).
We did not find a significant dependence of $\mathcal{S}_{450} \times \mathcal{P}_{450}^{\rm frac}$ on physical beam size.
The Spearman coefficient is $\rho = -0.25$ ($p = 0.12$), while a log--log fit gives a slope of $-0.05 \pm 0.04$ with $R^2 = 0.03$.
Thus, over the physical scales sampled here, we did not find evidence for the positive dependence on resolution expected from the simple Planck scaling.
In particular, beam averaging does not produce a detectable systematic increase in $\mathcal{S} \times \mathcal{P}^{\rm frac}$ toward the more distant sources.
 
Distance and environment are closely coupled in the present survey.
The environmental rank is strongly correlated with distance (Spearman $\rho = +0.84$, $p = 2\times10^{-11}$).
The three samples occupy markedly different distance ranges: 0.13--0.53\,kpc for LMSF, 0.40--3.83\,kpc for HMSF, and 7.86--8.34\,kpc for the CMZ.
The distances of the CMZ sources do not overlap those of either disk population.
The present data therefore do not compare CMZ and disk environments at matched physical resolution.
The result above does not by itself demonstrate that differences among the LMSF, HMSF, and CMZ subsamples are intrinsic environmental effects.
Removing an empirical distance trend cannot resolve the coupling between distance and environment, because any genuine environmental variation that follows the same ordering would be eliminated at the same time.
We therefore cannot separate the dependences on distance and on environment uniquely for the CMZ comparison with the present sample.
 
The behavior of $\beta$ provides a useful contrast.
The inferred $\beta$, unlike $\mathcal{S} \times \mathcal{P}^{\rm frac}$, depends clearly on physical beam size.
For the {\sc SpIdx} subsample, the Spearman coefficient is $\rho = +0.61$ with $p = 1\times10^{-4}$ (appendix~\ref{app:S2_dist_bias}).
Spatial averaging that depends on distance is therefore detectable when sufficiently strong.
Appendix~\ref{app:S2_dist_bias} examines this degeneracy between resolution and environment in detail for $\gamma - \alpha$ and $\beta$.
We therefore interpret the differences among the estimates of $\mathcal{S} \times \mathcal{P}^{\rm frac}$ for each environment (sub-subsection~\ref{sss:env_synthesis}) as differences in the distribution of point estimates across the three source populations.
These do not by themselves establish differences in the underlying proxy, and we leave open whether they reflect intrinsic environment or physical resolution.

\subsubsection{Global Properties of $\mathcal{S} \times \mathcal{P}^{\rm frac}$}
\label{sss:SP_global}
If differences in polarization among clouds were controlled only by projected field disorder, the diagnostics of angle dispersion and of polarization would be expected to change together.
The distribution of $\mathcal{S} \times \mathcal{P}^{\rm frac}$ among sources tests how much additional variation remains in the combined observable.
 
Table~\ref{tbl:SP450stats} in Appendix presents the statistical properties of $\mathcal{S}_{450} \times \mathcal{P}_{450}^{\rm frac}$ for the 39 sources satisfying $N \geq 50$, including the median ranges and the fractions of pixels with upper limits.
A large fraction of the selected pixels provides only upper limits on $\mathcal{S}_{450} \times \mathcal{P}_{450}^{\rm frac}$.
For 35 of the 39 sources, more than half of the pixels fall in this category, and the lower endpoint of the corresponding median range is zero under our adopted treatment of limits.
We therefore do not rank sources by the fraction of pixels exceeding a threshold.
The tabulated point estimates summarize the debiased values of all selected pixels.
Differences between these estimates do not by themselves establish differences in the underlying proxy for alignment efficiency.
 
The median point estimates of $\mathcal{S} \times \mathcal{P}^{\rm frac}$ range from $0\fdg197$ in NGC~6334 North to $0\fdg982$ in IC~5146, a factor of $\sim$5.0.
The corresponding range for the medians of $\mathcal{S}_{850}$ alone is a factor of $\sim$2.1.
The factor of $\sim$5.0 describes how the point estimates are distributed.
Most of the pixels entering these estimates are upper limits, so this spread does not by itself establish that the underlying proxy differs among the sources by a factor of $\sim$5.
The distributions within individual sources are positively skewed, as evidenced by the scatter based on the MAD, which is typically large relative to the median.
 
As established in sub-subsection~\ref{sss:SP_bias}, the sample shows no detectable systematic increase of $\mathcal{S} \times \mathcal{P}^{\rm frac}$ with physical beam size.
The point estimates of the nearby Perseus sources (NGC~1333, L1448~IRS2) are high and comparable to those of some sources at intermediate distances (e.g., Orion~A Field~5).
The point estimates of other nearby sources [L1689~North (figure~\ref{afig:Bmap_L1689N} in Appendix), Ophiuchus~A] are more moderate and similar to those of distant sources such as NGC~6334.
The point estimates therefore show no detected ordering by distance, and we found no evidence that beam averaging alone reproduces their spread.
However, the absence of a distance ordering does not by itself establish that the differences between environments are intrinsic (sub-subsection~\ref{sss:SP_bias}).
The CMZ and disk populations are not observed at matched physical resolution.
 
The stacked fields show median point estimates consistent with area-weighted combinations of those of their constituents.
For instance, the CMZ stacked field ($\mathcal{S} \times \mathcal{P}^{\rm frac} = 0\fdg420$) lies near the middle of the range spanned by individual Galactic Center fields ($0\fdg306$--$0\fdg553$).
From this consistency we conclude that stacking does not introduce artifacts into the statistics of $\mathcal{S} \times \mathcal{P}^{\rm frac}$.
 
The quantities $\mathcal{S}$ and $\mathcal{S}\times\mathcal{P}^{\rm frac}$ complement the primary observables $P_{450}/P_{850}$ and $\gamma-\alpha$ in subsections~\ref{ss:S2_P_ratio}--\ref{ss:S2_SpIdx}.
The dispersion $\mathcal{S}$ describes the disorder of the projected field, and the product $\mathcal{S}\times\mathcal{P}^{\rm frac}$ serves as a proxy for grain-alignment efficiency.
We examine their ordering by environment in subsection~\ref{ss:S2_EnvStrat}.
The point estimates of the combined observable vary substantially among the measured clouds, and we did not detect a monotonic increase with physical beam size.
This spread does not by itself establish differences in the underlying proxy, because most of the pixels entering the point estimates are upper limits.
The present sampling also does not isolate intrinsic alignment changes from geometry and resolution.

\subsection{Three Statistical Estimators of $P_{450}/P_{850}$}
\label{ss:S2_P_ratio}
Does the reduced polarization at 450\,$\mu$m occur throughout the bright material detected in both bands, or mainly in the regions that dominate a cloud's flux?
To address this question, we compare estimators with different pixel weights, without assuming that all parts of a cloud contribute equally to the measured ratio.
 
We measure $R_p \equiv P_{450}/P_{850}$ for each source with three complementary estimators (Methods~1, 2, and 3) for the F4 population test (figure~\ref{fig:S2_test_overview}, table~\ref{tbl:S2_test_index}).
We adopt Method~3 as the standard estimator for the census of individual sources.
 
We examine how dust polarization efficiency changes between 450 and 850\,\micron, after correcting the 450\,\micron\ Stokes-$Q/U$ zero-level offset where required (Paper~T subsection~\ref{T-ss:offset}).
We begin with the observational context and previous measurements of wavelength-dependent polarization (sub-subsection~\ref{sss:P_ratio_intro}), then describe three complementary methods (sub-subsection~\ref{sss:Methodology}).
We apply these methods to L1688 Ophiuchus~A (figure~\ref{afig:Bmap_OphiuA} in Appendix) in sub-subsection~\ref{sss:P_ratio_OphA}, then present the results across the survey in sub-subsection~\ref{sss:P_ratio_Global}.
We synthesize the findings to establish robust constraints on the physics of grain alignment at submillimeter wavelengths (sub-subsection~\ref{sss:Synthesis}).
Comparing the three estimates establishes how sensitive the inferred wavelength deficit is to weighting and selection, rather than providing an independent test of each possible explanation based on grain physics.

\subsubsection{Previous Measurements of Wavelength-Dependent Polarization}
\label{sss:P_ratio_intro}
 
Observations of the submillimeter polarization spectrum have revealed complex wavelength dependencies that challenge simple dust models.
\citet{Hildebrand1999} pioneered far-infrared polarization spectroscopy and found rising spectra in dense cores ($P_{350}/P_{100} \sim 2$) but falling spectra in envelopes ($P_{350}/P_{100} \sim 0.6$).
Using polarimetric maps of OMC\,1 at 350 and 450\,\micron\ obtained with the SHARP polarimeter at the Caltech Submillimeter Observatory (CSO), \citet{Vaillancourt2008} reported $P_{450}/P_{350} \sim 1.3$ at positions more than 20\arcsec\ from the two submillimeter intensity peaks (the Kleinmann--Low nebula and Orion South), falling by more than a factor of two toward the Kleinmann--Low nebula.
They placed this ratio on a polarization spectrum normalized at 350\,\micron\ that combines earlier measurements of several bright Galactic clouds (their figure~3).
At 60 and 100\,\micron, this spectrum includes M17, W51, and OMC\,1, all with higher polarization fractions than at 350\,\micron; at 850 and 1300\,\micron, it includes only W51, DR21, and OMC\,3.
Assuming that the spectrum of OMC\,1 continues to rise beyond 450\,\micron\ as in those clouds, they concluded that the minimum of the polarization spectrum of the OMC\,1 envelope lies between 100 and 350\,\micron.
For OMC\,1 itself, \citet{Schleuning1998} compared 100\,\micron\ polarimetry from the Kuiper Airborne Observatory with 350\,\micron\ polarimetry from the CSO.
They found that the polarization fractions at the two wavelengths agree within the uncertainties at spatially coincident positions along the ridge.
The exceptions are the Kleinmann--Low nebula and the submillimeter source Orion South, where the polarization fraction increases with wavelength (their table~1).
 
Balloon-borne Large Aperture Submillimeter Telescope for Polarimetry (BLASTPol) observations \citep{Gandilo2016, Shariff2019} revealed flat polarization spectra in Vela~C and Carina, finding no pronounced minimum near 350\,\micron.
They reported $P_{250}/P_{350} \sim 0.9$--1.1 and $P_{350}/P_{500} \sim 0.8$--1.0, with variations within $\pm 15$\% relative to 350\,\micron.
At 353\,GHz, Planck provides a large-area statistical reference for polarized dust emission \citep{Planck2020-XII}.
 
Recent studies with multiple instruments have provided new insights.
\citet{Tram2024} combined SOFIA/HAWC+ and JCMT/POL-2 data for OMC\,1 and found V-shaped polarization spectra.
The polarization fraction decreases from 54 to $\sim$300\,\micron\ and increases at longer wavelengths.
Their spatially resolved spectral analysis and two-phase modeling of warm and cold dust components along the line of sight extend beyond the single-component approach adopted here.
 
These diverse results arise in part from different measurement techniques and wavelength regimes.
Our POL-2 data provide systematic measurements of $P_{450}/P_{850}$ at both wavelengths with the same instrument and three complementary methods for up to 45 individual sources.
We designed this comparison among methods to test whether the result depends on how the data are weighted.

\subsubsection{Methodology: Three Complementary Approaches}
\label{sss:Methodology}
Bright cores and fainter cloud material do not necessarily contribute the same polarization spectrum.
We therefore compare a flux-weighted ratio, a regression between the two polarization fractions, and the typical pixel ratio within a specified Stokes-$I$ mask.
 
After the basic validity checks, we define the parent samples without cuts on the polarization fraction, polarized intensity, or their signal-to-noise ratios, to avoid preferentially excluding weak polarization \citep{Pattle2019}.
Within each parent sample, the detection requirements depend on the statistic (subsection~\ref{ss:S2_noise_floor}).
The three estimators distinguish the ratio emphasized by bright emission from the ratio typical of the selected lines of sight.
We interpret agreement or disagreement within those selections, which do not provide uniform coverage of the entire molecular cloud.
 
\paragraph{Method 1: Intensity-Weighted Mean (Mass-Weighted)}
The brightest structures contribute a large fraction of a cloud's submillimeter flux and can strongly influence an integrated polarization measurement.
With an intensity-weighted estimate, we ask how the polarized fraction of that detected emission differs between 450 and 850\,$\mu$m.
In Method~1, we weight each polarization measurement by its total intensity to estimate the net polarization efficiency of the bulk dust mass.
We calculate the polarization-fraction ratio as
\begin{equation}
R_{\rm IWM} \equiv \frac{\langle P_{450}\rangle_I}{\langle P_{850}\rangle_I} = \frac{\sum PI_{450} / \sum I_{450}}{\sum PI_{850} / \sum I_{850}},
\label{eq:R1}
\end{equation}
where summations extend over pixels satisfying $I_{450}/\sigma_{I,450} > 30$ and $I_{850}/\sigma_{I,850} > 30$.
Since Stokes $I$ directly traces dust column density, this weighting emphasizes regions containing most of the cloud mass, typically dense cores where star formation occurs.
This method is therefore particularly sensitive to grain alignment in environments with high column density where depolarization effects are strongest.
\citet{Hildebrand1999} likewise wrote the polarization fraction measured in a beam as a flux-weighted sum over the contributing grain populations, although they compared ratios of median values over many beam measurements.
$R_{\rm IWM}$ characterizes the wavelength dependence of the flux-weighted polarization fraction in the selected emission.
Its interpretation as a mass-weighted alignment measure would additionally require control of temperature, opacity, and the structure along the line of sight.
Table~\ref{tbl:Pratio1} also lists, in column~(4), the median pixel ratio $\tilde{R}_1$ on the same mask, which is distinct from $R_{\rm IWM}$.

\paragraph{Method 2: Correlation Analysis (Alignment-Weighted) --- $R_{\rm Corr}$}
If the two wavelengths respond similarly to changes across a cloud, their polarization fractions should vary together from one line of sight to another.
A fitted relation between $P_{450}$ and $P_{850}$ measures this covariation while allowing the better measured pixels to contribute more strongly.
In Method~2, we fit $P_{450}$ versus $P_{850}$ for individual pixels by weighted least-squares regression through the origin, with weights $w \propto 1/(\Delta P_{450})^2$.
Since the uncertainty in polarization fraction scales as $\Delta P_{450} \propto \sigma_{PI}/I_{450}$ with approximately constant $\sigma_{PI}$, the inverse-variance weighting becomes $w \propto I_{450}^2$.
The regression therefore gives higher statistical weight to bright regions.
Brighter regions in star-forming clouds tend also to have stronger polarization.
Pixels with both high intensity and high polarization fraction therefore have a greater influence on the fitted slope $R_{\rm Corr}$, even though the formal statistical weight depends only on $I_{450}^2$.
This correlation approach for individual pixels is similar in method to the BLASTPol analyses of \citet{Gandilo2016} and \citet{Shariff2019}.
 
$R_{\rm Corr}$ summarizes how the two polarization fractions scale together in the fitted lines of sight.
It emphasizes bright structures with precise measurements, which is useful for comparison with $R_{\rm EW}$, but the fitted slope is not a direct measure of grain-alignment efficiency.
 
\paragraph{Method 3: Statistical Distribution (Equal-Weighted) --- $R_{\rm EW}$}
An integrated ratio can be dominated by a small number of bright structures, even when most detected lines of sight behave differently.
With the median of individual pixel ratios, we instead ask what wavelength dependence is typical of the material reliably detected in Stokes $I$ at both bands.
To estimate the typical polarization ratio across the sampled physical conditions, we give equal weight to all qualifying pixels in Method~3.
We compute $R_{\rm EW}$ as the median of individual pixel ratios:
\begin{equation}
R_{\rm EW} = {\rm median}\left\{\frac{P_{450,i}}{P_{850,i}}\right\}_{i \in S_{\rm det}},
\label{eq:R3}
\end{equation}
where $S$ denotes pixels that satisfy the common dual-band Stokes-$I$ detection mask: $I_{450}/\sigma_{I,450} > 50$ and $I_{850}/\sigma_{I,850} > 50$, Stokes~$I$ above the 95th percentile of the source at both wavelengths, and Stokes~$I$ above an absolute floor at both wavelengths ($10\%$ of the mean intensity of the pixels at or above the 99th percentile).
We do not select pixels by the signal-to-noise ratio of polarized intensity when defining $S$, so that intrinsically low-polarization pixels are retained in this parent mask.
The median runs over $S_{\rm det} \subset S$, the members of $S$ with $P_{450} \geq 3\,\sigma_{P_{450}}$ and $P_{850} \geq 3\,\sigma_{P_{850}}$ (subsection~\ref{ss:S2_noise_floor}).
It describes the detected ratios within $S$ and does not include ratios represented only by limits.
Selecting the parent mask on $PI/\sigma_{PI}$ would preferentially exclude the weak-$P$ tail of the pixel distribution from every analysis that uses this mask.
Unlike Methods~1 and 2 that emphasize bright or well-polarized regions, Method~3 provides equal statistical weight to every qualifying pixel regardless of its intensity or polarization strength.
 
However, the equal weighting within the mask is not the same as that of an unbiased sample of the cloud.
The requirements that define the mask are themselves brightness cuts, so the pixel ensemble is weighted toward dense, star-forming material and away from the diffuse envelope.
We therefore read $R_{\rm EW}$ as the typical ratio across the conditions POL-2 detects in both bands, not across the cloud as a whole.
Within this mask the ensemble still spans the bright interiors of the clouds (about $10^{22}$--$10^{24}$\,cm$^{-2}$ in $N_{\rm H_2}$ in the nine HGBS fields; Paper~S1), so the method tests whether the wavelength dependence varies with local conditions among the detected lines of sight.
The unweighted statistical approach corresponds most closely to the Planck all-sky analyses \citep{Planck2020-XII}.
$R_{\rm EW}$ describes the typical ratio among these lines of sight selected by brightness.
It retains weakly polarized lines of sight within the Method~3 mask when their polarization is detected in both bands, but it does not represent diffuse material that fails the intensity requirement imposed at both wavelengths.
 
These estimators emphasize different parts of the detected cloud material.
Comparing them tests whether the inferred polarization deficit strongly depends on that emphasis.
Systematic differences between the weighted ($R_{\rm IWM}$, $R_{\rm Corr}$) and unweighted ($R_{\rm EW}$) estimators would indicate wavelength effects that depend on density.
For example, $R_{\rm EW} \gg R_{\rm IWM}$ would suggest stronger depolarization at 450\,\micron\ in the brightest regions than on fainter lines of sight.
If the three estimators give consistent values of $P_{450}/P_{850}$ below unity for the same source, we would infer that the higher polarization fraction at 850 than at 450\,\micron\ does not depend strongly on the choice of estimator.
This agreement alone does not establish uniform grain properties or exclude calibration errors shared by the three methods.

\begin{figure}
\centering
\includegraphics[width=0.7\textwidth]{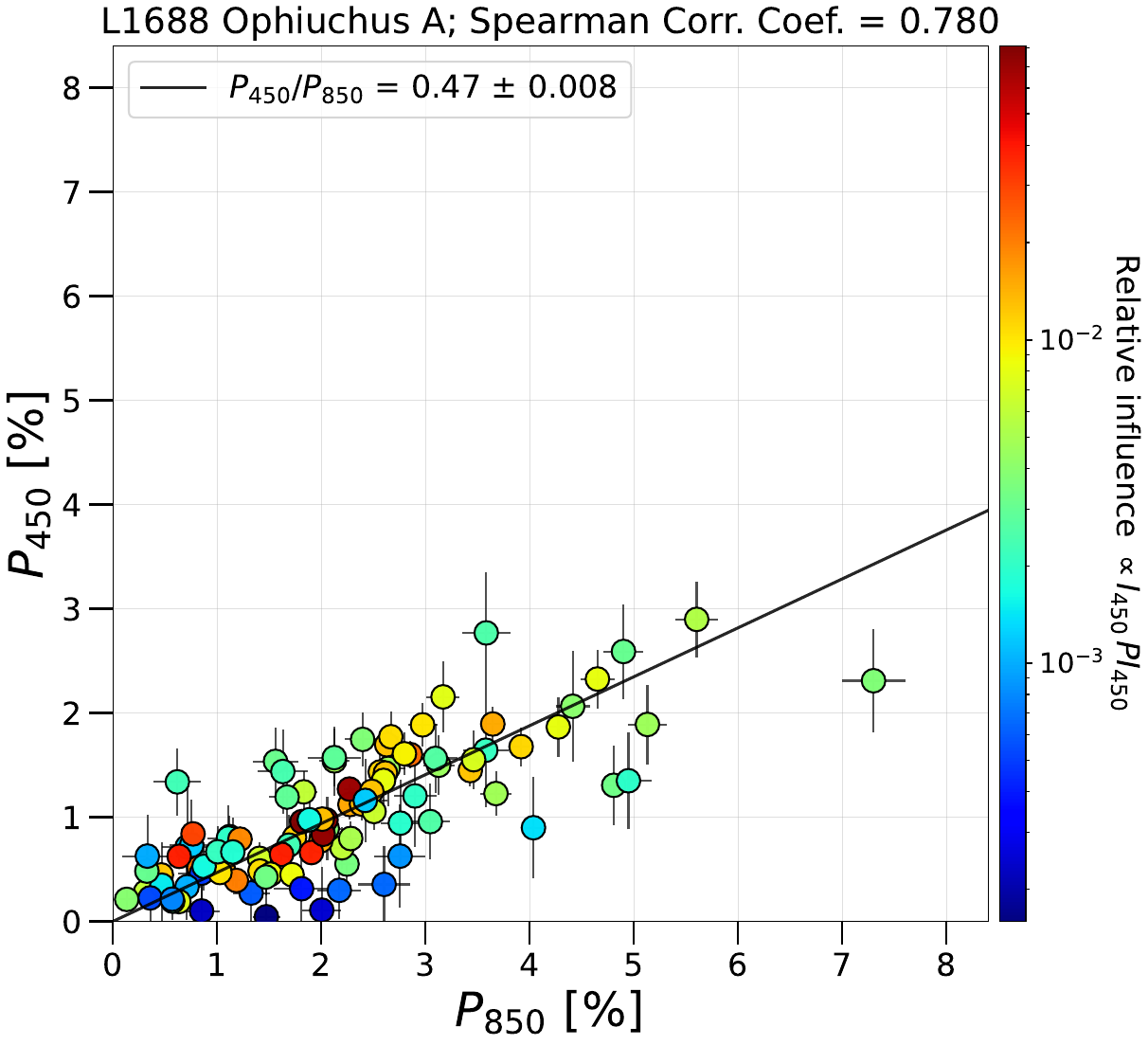}
\caption{
Illustration of Method~2 (weighted least-squares regression of $P_{450}$ versus $P_{850}$ through the origin) for L1688 Ophiuchus~A (sub-subsection~\ref{sss:P_ratio_OphA}).
The statistical weight is $w \propto I_{450}^2$ (inverse-variance weighting assuming constant $\sigma_{PI}$).
The color scale shows the combined effect of intensity and polarization on the influence each pixel has on the fitted slope.
The slope $R_{\rm Corr} = 0.47 \pm 0.01$ represents the polarization fraction ratio.
Error bars indicate the measurement uncertainties.
Paper~S1 (figure~\ref{S1-fig:Pratio_OphA_method3}) shows the corresponding Method~3 plot of ratios for individual pixels in the same source.
\\ {Alt text: Scatter plot with error bars and a fitted line.}
}
\label{fig:Pratio_OphA}
\end{figure}

\subsubsection{Application to L1688 Ophiuchus~A}
\label{sss:P_ratio_OphA}
 
We illustrate the three methods using L1688 Ophiuchus~A, the source with the highest signal-to-noise ratio in our sample.
Figure~\ref{fig:Pratio_OphA} shows Method~2.
Paper~S1 (figure~\ref{S1-fig:Pratio_OphA_method3}) shows the corresponding distribution of ratios for individual pixels in Method~3.
 
The intensity-weighted analysis (Method~1) yields $\langle P_{850} \rangle_I = 1.82 \pm 0.01$\% and $\langle P_{450} \rangle_I = 0.90 \pm 0.02$\%, giving a ratio $R_{\rm IWM} = 0.49 \pm 0.02$.
This value indicates that the polarization efficiency at 450\,\micron, weighted by intensity, is approximately half that at 850\,\micron.
 
The weighted least-squares regression for individual pixels (Method~2; figure~\ref{fig:Pratio_OphA}) yields $R_{\rm Corr} = 0.47 \pm 0.01$ with a Spearman correlation coefficient of 0.78.
We interpret the tight correlation and the color sequence from regions of low intensity and high polarization (blue) to regions of high intensity and low polarization (red) as evidence that depolarization affects both bands similarly.
 
Method~3 gives a median $R_{\rm EW} = 0.47 \pm 0.07$.
Paper~S1 figure~\ref{S1-fig:Pratio_OphA_method3} shows the distribution of the individual pixel ratios.
From the flat distribution across nearly two orders of magnitude in intensity, we infer that the polarization fraction ratio in this source does not depend measurably on the local intensity.

For L1688 Ophiuchus~A, the three methods agree ($R_{\rm IWM}=0.49$, $R_{\rm Corr}=0.47$, $R_{\rm EW}=0.47$).
This agreement shows that the inferred polarization fraction ratio does not depend sensitively on whether the analysis emphasizes regions containing most of the mass, grains that are well aligned, or the pixel ensemble as a whole.
We examine the physical interpretation of the wavelength dependence in subsection~\ref{ss:S2_SpIdx} and Paper~S1 subsection~4.2 (Size-Dependent Grain Alignment).

\begin{longrotatetable}
\footnotesize
\begin{deluxetable*}{lcccc}
\tablecaption{Intensity-Weighted Polarization Fractions and Median Pixel Ratios on the Method 1 Mask\label{tbl:Pratio1}}
\tablewidth{0pt}
\tablehead{
  \colhead{Source} &
  \colhead{$\langle P_{850} \rangle_I$} &
  \colhead{$\langle P_{450} \rangle_I$} &
  \colhead{$\tilde{R}_1$} &
  \colhead{$N$} \\
  \colhead{} &
  \colhead{(\%)} &
  \colhead{(\%)} &
  \colhead{} &
  \colhead{}
}
\startdata
  L43                                 & $0.77 \pm 0.05$      & $1.31 \pm 0.08$      & $1.63 \pm 0.17$      &   59 \\
  Taurus B211/213 (mosaicked field)   & $0.93 \pm 0.09$      & $1.67 \pm 0.16$      & $1.69 \pm 0.44$      &   15 \\
  L1689 South                         & $1.14 \pm 0.17$      & $3.69 \pm 0.84$      & $1.97 \pm 0.58$      &    3 \\
  L1688 Ophiuchus A                   & $1.82 \pm 0.01$      & $0.90 \pm 0.02$      & $0.48 \pm 0.02$      &  100 \\
  L1688 Ophiuchus B                   & $1.07 \pm 0.03$      & $1.16 \pm 0.11$      & $0.87 \pm 0.08$      &   63 \\
  L1689 North                         & $0.56 \pm 0.01$      & $0.33 \pm 0.02$      & $0.74 \pm 0.14$      &   19 \\
  L1448 IRS2                          & $0.75 \pm 0.04$      & $1.04 \pm 0.13$      & $1.11 \pm 0.23$      &   26 \\
  L483                                & $0.92 \pm 0.09$      & $1.38 \pm 0.29$      & $1.87 \pm 0.18$      &    9 \\
  Perseus B1                          & $0.98 \pm 0.05$      & $2.70 \pm 0.41$      & $4.27 \pm 0.66$      &   11 \\
  NGC1333 North                       & $1.28 \pm 0.04$      & $1.44 \pm 0.15$      & $0.99 \pm 0.14$      &   22 \\
  NGC1333 South                       & $0.98 \pm 0.03$      & $1.34 \pm 0.10$      & $1.26 \pm 0.16$      &   37 \\
  Perseus IC348                       & $1.95 \pm 0.03$      & $1.78 \pm 0.06$      & $0.85 \pm 0.04$      &   83 \\
  OMC 1                               & $4.55 \pm 0.01$      & $2.19 \pm 0.01$      & $0.58 \pm 0.04$      &   96 \\
  OMC 2                               & $1.07 \pm 0.01$      & $0.61 \pm 0.01$      & $0.55 \pm 0.03$      &  114 \\
  OMC 3                               & $2.05 \pm 0.01$      & $1.14 \pm 0.02$      & $0.57 \pm 0.02$      &  116 \\
  Orion-B NGC2071                     & $0.64 \pm 0.03$      & $1.13 \pm 0.18$      & $1.56 \pm 0.21$      &    9 \\
  IC 5146                             & $1.12 \pm 0.07$      & $0.67 \pm 0.22$      & $0.40 \pm 0.14$      &    8 \\
  Auriga                              & $1.44 \pm 0.03$      & $2.50 \pm 0.11$      & $1.56 \pm 0.09$      &   54 \\
  Serpens South                       & $1.77 \pm 0.04$      & $1.02 \pm 0.08$      & $0.68 \pm 0.08$      &   29 \\
  Cepheus A                           & $1.54 \pm 0.01$      & $0.96 \pm 0.03$      & $0.58 \pm 0.04$      &   54 \\
  NGC2264 Single Field\#1              & $1.14 \pm 0.01$      & $0.67 \pm 0.02$      & $0.63 \pm 0.03$      &   96 \\
  NGC2264 Single Field\#2              & $1.20 \pm 0.02$      & $0.74 \pm 0.03$      & $0.60 \pm 0.03$      &   96 \\
  SDC 13 mm-2                         & $1.03 \pm 0.02$      & $0.82 \pm 0.04$      & $0.78 \pm 0.05$      &   93 \\
  Monoceros R2                        & $2.15 \pm 0.01$      & $1.08 \pm 0.00$      & $0.45 \pm 0.01$      &  123 \\
  S106 (mosaic component field)       & $1.72 \pm 0.02$      & $1.38 \pm 0.04$      & $0.79 \pm 0.05$      &   85 \\
  Onsala 2                            & $1.42 \pm 0.01$      & $1.29 \pm 0.04$      & $0.80 \pm 0.06$      &   80 \\
  DR21                                & $1.91 \pm 0.00$      & $1.07 \pm 0.00$      & $0.55 \pm 0.01$      &  128 \\
  M16                                 & $1.47 \pm 0.10$      & $1.96 \pm 0.37$      & $1.54 \pm 0.56$      &    4 \\
  NGC 6334 North                      & $1.68 \pm 0.00$      & $0.82 \pm 0.01$      & $0.44 \pm 0.01$      &  113 \\
  NGC 6334 South                      & $1.24 \pm 0.00$      & $0.91 \pm 0.01$      & $0.73 \pm 0.02$      &  119 \\
  W3 HFS2                             & $1.05 \pm 0.01$      & $0.73 \pm 0.02$      & $0.67 \pm 0.02$      &   35 \\
  NGC 7538 IRS4                       & $1.80 \pm 0.01$      & $1.25 \pm 0.02$      & $0.74 \pm 0.03$      &   96 \\
  G31.41+0.31                         & $0.43 \pm 0.01$      & $0.31 \pm 0.01$      & $0.65 \pm 0.11$      &   17 \\
  G28.34+0.06  (mosaicked field)      & $0.96 \pm 0.02$      & $0.93 \pm 0.01$      & $0.97 \pm 0.05$      &  141 \\
  The Gal. Cent. 20 km s$^{-1}$ cloud        & $1.28 \pm 0.01$      & $0.86 \pm 0.02$      & $0.67 \pm 0.03$      &   94 \\
  Brick                               & $1.20 \pm 0.01$      & $0.86 \pm 0.01$      & $0.70 \pm 0.04$      &  107 \\
  The Gal. Cent. Clouds E and F       & $1.50 \pm 0.01$      & $1.44 \pm 0.02$      & $0.95 \pm 0.04$      &  113 \\
  The Gal. Cent. Middle               & $1.16 \pm 0.01$      & $1.10 \pm 0.02$      & $0.96 \pm 0.04$      &  110 \\
  The Gal. Cent. North                & $1.38 \pm 0.02$      & $1.34 \pm 0.03$      & $0.89 \pm 0.04$      &   78 \\
  The Gal. Cent. South                & $1.15 \pm 0.01$      & $1.08 \pm 0.02$      & $0.98 \pm 0.06$      &  74 \\
  CMZ (stacked)                       & $1.21 \pm 0.00$      & $1.01 \pm 0.00$      & $0.84 \pm 0.01$      &  701 \\
  Sagittarius B2                      & $1.18 \pm 0.00$      & $0.93 \pm 0.00$      & $0.72 \pm 0.03$      &  125 \\
  Orion\,A (stacked)                  & $4.01 \pm 0.00$      & $1.95 \pm 0.01$      & $0.56 \pm 0.02$      &  326 \\
  G209.29-19.65            & $2.03 \pm 0.08$      & $1.18 \pm 0.31$      & $0.71 \pm 0.12$      &   12 \\
  Orion A Field 7                     & $1.43 \pm 0.21$      & $2.57 \pm 0.55$      & $1.99 \pm 0.29$      &    8 \\
  Orion A Field 4                     & $3.91 \pm 0.01$      & $2.06 \pm 0.01$      & $0.60 \pm 0.03$      &  100 \\
  Orion A Field 5                     & $2.25 \pm 0.04$      & $1.81 \pm 0.06$      & $0.90 \pm 0.08$      &   99 \\
  NGC\,2264 (stacked)                 & $1.17 \pm 0.01$      & $0.71 \pm 0.02$      & $0.62 \pm 0.02$      &  192 \\
  NGC\,6334 (stacked)                 & $1.52 \pm 0.00$      & $0.85 \pm 0.01$      & $0.60 \pm 0.02$      &  232 \\
\enddata
\tablecomments{
Intensity-weighted mean polarization fractions $\langle P_{\lambda} \rangle_I = \sum PI_{\lambda} / \sum I_{\lambda}$,
computed on the Method~1 mask (see sub-subsection~\ref{sss:Methodology}).
Column~(4) is not the quotient of columns~(3) and~(2).
The ratio $\tilde{R}_1$ is the median of $P_{450}/P_{850}$ over individual pixels, with uncertainty
$\sigma_{\tilde{R}_1}$ given by the standard error of the median, based on the MAD.
Of the 45 individual sources, 34 show $\tilde{R}_1 < 1$ and 11 show $\tilde{R}_1 > 1$.
Four stacked entries present statistics computed from the combined pixels
of multiple single-field observations:
Orion\,A (stacked) from OMC\,1, OMC\,2, and OMC\,3;
NGC\,2264 (stacked) from Single Field\,1 and Single Field\,2;
NGC\,6334 (stacked) from North and South;
and the CMZ (stacked) from seven individual fields
(Sgr\,B2, Clouds~E and~F, the 20\,km\,s$^{-1}$ cloud, the Brick, North, Middle, and South).
}
\end{deluxetable*}
\end{longrotatetable}

\begin{longrotatetable}
\footnotesize
\begin{deluxetable}{lcc}
\tablecaption{Summary of Polarization-fraction Ratios, $R_{\rm p}$, Derived from Methods 2 and 3\label{tbl:Pratios23}}
\tablewidth{0pt}
\tablehead{
  \colhead{Source} &
  \colhead{$R_{\rm Corr} \pm \delta R_{\rm Corr}$} &
  \colhead{$R_{\rm EW} \pm \delta R_{\rm EW}$} \\
  \colhead{} &
  \colhead{(Method 2)} &
  \colhead{(Method 3)}
}
\startdata
  L43                                 & $1.25 \pm 0.08$      & \nodata              \\
  Taurus B211/213 (mosaicked field)   & $1.16 \pm 0.14$      & \nodata              \\
  L1688 Ophiuchus A                   & $0.47 \pm 0.01$      & $0.47 \pm 0.07$      \\
  L1688 Ophiuchus B                   & $0.96 \pm 0.08$      & \nodata              \\
  L1689 North                         & $0.53 \pm 0.03$      & $0.70 \pm 0.38$      \\
  L1448 IRS2                          & $0.95 \pm 0.15$      & \nodata              \\
  L483                                & $1.48 \pm 0.29$      & \nodata              \\
  Perseus B1                          & $2.07 \pm 0.27$      & \nodata              \\
  NGC1333 North                       & $0.83 \pm 0.07$      & \nodata              \\
  NGC1333 South                       & $0.96 \pm 0.06$      & \nodata              \\
  Perseus IC348                       & $0.78 \pm 0.02$      & $0.85 \pm 0.16$      \\
  OMC 1                               & $0.36 \pm 0.00$      & $0.48 \pm 0.07$      \\
  OMC 2                               & $0.47 \pm 0.01$      & $0.55 \pm 0.10$      \\
  OMC 3                               & $0.47 \pm 0.01$      & $0.48 \pm 0.04$      \\
  Orion\,A (stacked)                  & $0.36 \pm 0.00$      & $0.49 \pm 0.04$      \\
  Orion-B NGC2071                     & $1.40 \pm 0.23$      & \nodata              \\
  G209.29-19.65            & $0.55 \pm 0.14$      & \nodata              \\
  Orion A Field 7                     & $1.34 \pm 0.30$      & \nodata              \\
  Orion A Field 4                     & $0.43 \pm 0.00$      & $0.46 \pm 0.04$      \\
  Orion A Field 5                     & $0.60 \pm 0.02$      & $1.27 \pm 0.16$      \\
  G28.34+0.06  (mosaicked field)      & $0.70 \pm 0.01$      & $0.73 \pm 0.08$      \\
  IC 5146                             & $0.57 \pm 0.17$      & \nodata              \\
  Serpens South                       & $0.52 \pm 0.03$      & $0.59 \pm 0.07$      \\
  Auriga                              & $1.39 \pm 0.06$      & $1.28 \pm 0.10$      \\
  Cepheus A                           & $0.57 \pm 0.01$      & $0.72 \pm 0.10$      \\
  NGC2264 Single Field\#1              & $0.55 \pm 0.02$      & $0.78 \pm 0.31$      \\
  NGC2264 Single Field\#2              & $0.56 \pm 0.02$      & $0.65 \pm 0.15$      \\
  NGC\,2264 (stacked)                 & $0.56 \pm 0.01$      & $0.67 \pm 0.20$      \\
  SDC 13 mm-2                         & $0.78 \pm 0.03$      & $0.91 \pm 0.10$      \\
  DR21                                & $0.56 \pm 0.00$      & $0.51 \pm 0.03$      \\
  Monoceros R2                        & $0.45 \pm 0.00$      & $0.62 \pm 0.04$      \\
  S106 (mosaic component field)       & $0.68 \pm 0.02$      & $0.92 \pm 0.14$      \\
  NGC 6334 North                      & $0.42 \pm 0.00$      & $0.46 \pm 0.04$      \\
  NGC 6334 South                      & $0.70 \pm 0.00$      & $0.73 \pm 0.06$      \\
  NGC\,6334 (stacked)                 & $0.48 \pm 0.00$      & $0.64 \pm 0.04$      \\
  W3 HFS2                             & $0.68 \pm 0.01$      & $0.66 \pm 0.04$      \\
  NGC 7538 IRS4                       & $0.62 \pm 0.01$      & $0.65 \pm 0.05$      \\
  Onsala 2                            & $0.78 \pm 0.01$      & $0.78 \pm 0.04$      \\
  G31.41+0.31                         & $0.68 \pm 0.02$      & $0.67 \pm 0.09$      \\
  The Gal. Cent. 20 km s$^{-1}$ cloud        & $0.68 \pm 0.01$      & $0.68 \pm 0.09$      \\
  Brick                               & $0.70 \pm 0.01$      & $0.71 \pm 0.05$      \\
  The Gal. Cent. Clouds E and F       & $0.85 \pm 0.01$      & $0.84 \pm 0.23$      \\
  The Gal. Cent. Middle               & $0.96 \pm 0.01$      & $0.87 \pm 0.08$      \\
  The Gal. Cent. North                & $0.85 \pm 0.02$      & $0.86 \pm 0.07$      \\
  The Gal. Cent. South                & $0.83 \pm 0.01$      & $0.97 \pm 0.22$      \\
  CMZ (stacked)                       & $0.68 \pm 0.00$      & $0.82 \pm 0.05$      \\
  Sagittarius B2                      & $0.67 \pm 0.00$      & $0.68 \pm 0.08$      \\
\enddata
\tablecomments{
Method~2 gives the weighted regression slope $R_{\rm Corr}$ of $P_{450}$ versus $P_{850}$; Method~3 gives the median pixel ratio $R_{\rm EW}$ in bright regions (sub-subsection~\ref{sss:Methodology}).
Of the individual sources analyzed with Method~2, 36 show $R_{\rm Corr} < 1$ and 7 show $R_{\rm Corr} > 1$.
Of the individual sources analyzed with Method~3, 29 show $R_{\rm EW} < 1$ and 2 show $R_{\rm EW} > 1$.
Method~3 uses only pixels in which both $P_{450}$ and $P_{850}$ exceed three times their uncertainties, and we do not tabulate the 12 sources that retain fewer than three such pixels.
Table~\ref{tbl:Pratio1} describes the composition of the four stacked entries.
We apply additional selection criteria to the spectral-index analysis (subsection~\ref{ss:S2_SpIdx}), so the usable $12\arcsec$ pixels of a given region differ in number and in sky position between this table and that analysis.
For each source, we use its maps to name the entry for the field or mosaic where those pixels concentrate.
}
\end{deluxetable}
\end{longrotatetable}

\begin{figure}[htbp]
\centering
\includegraphics[width=0.75\textwidth]{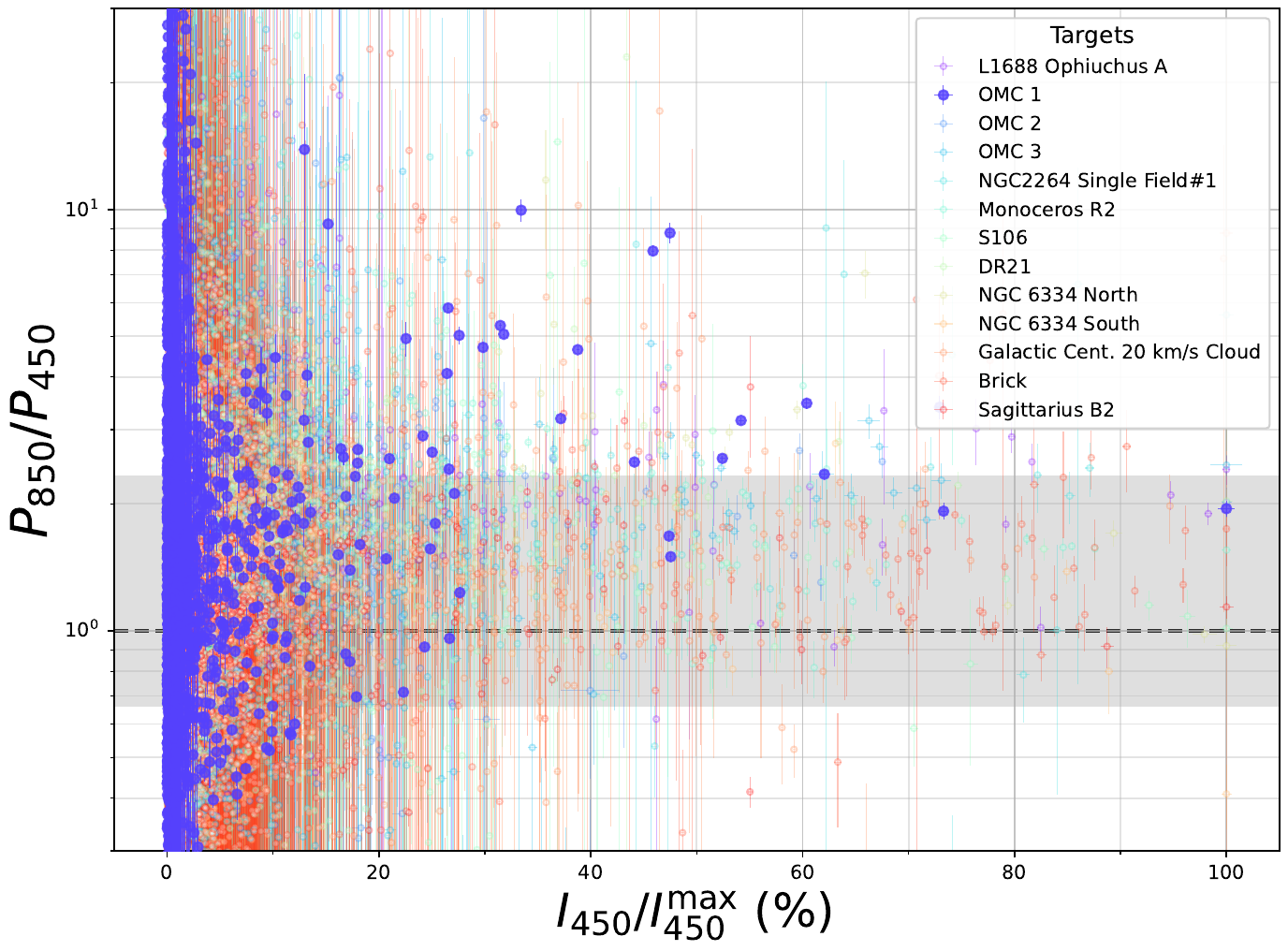}
\caption{
Inverted polarization-fraction ratio $P_{850}/P_{450}$ versus $I_{450}$ normalized to each source's peak.
The plot includes all pixels with $I_{450}/\sigma_{I,450} > 3$ in 13 regions with high-$S/N$ data in both bands.
As in Method~3, the 450\,\micron\ polarization is taken from the map corrected for the 450\,\micron\ Stokes-$Q/U$ zero-level offsets where required (Paper~T subsection~\ref{T-ss:offset}).
The inverted ratio follows the format of \citet{Vaillancourt2008}, whose figure~4 compares $P_{450}/P_{350}$ with normalized 350\,\micron\ intensity.
The shaded band spans the pooled 25th--75th percentiles of 15,708 pixels: $P_{850}/P_{450}=0.66$--2.33, with median 1.27, equivalent to $P_{450}/P_{850}=0.43$--1.52 with median 0.79.
It describes observed pixel-to-pixel and source-to-source scatter, not uncertainty on the median.
In 4255 of the plotted pixels (27\%), both $P_{450}$ and $P_{850}$ exceed three times their uncertainties.
For these pixels alone, the quartiles are $P_{850}/P_{450}=0.91$--1.90 with median 1.33.
The dashed line marks unity; values above it have $P_{450}<P_{850}$.
The distribution straddles unity, while its median indicates lower polarization at 450\,\micron.
Over all plotted pixels the ratio rises only weakly with normalized intensity (Spearman $r_s=+0.13$), and the trend differs between sources.
For OMC~1 (filled circles), the ratio rises with normalized intensity: its median is 1.5 at 5\%--20\% of peak and 2.6 at or above 20\% (Spearman $r_s=+0.40$ over its 174 pixels at or above 5\% of peak).
\\ {Alt text: Scatter plot of many points with a shaded band and a horizontal line.}
}
\label{fig:Vaillancourt_comparison}
\end{figure}

\subsubsection{Global Results from All Sources}
\label{sss:P_ratio_Global}
The three methods agree closely in L1688 Ophiuchus~A ($\tilde{R}_1 = 0.48$, $R_{\rm Corr} = 0.47$, $R_{\rm EW} = 0.47$; sub-subsection~\ref{sss:P_ratio_OphA}).
We now examine whether this consistency and the finding of $P_{450}/P_{850} < 1$ extend to the full sample, including clouds with different luminosities, distances, and star-forming activity.
We compare the ratios for individual sources under each estimator, stating the available number of sources separately for each method.
 
Tables~\ref{tbl:Pratio1} and \ref{tbl:Pratios23} show the polarization fraction ratios derived from Methods~1, 2 and 3 for all sources with sufficient signal-to-noise ratio for analysis in both bands.
The sample contains 45 individual sources from $125$\,pc (L43) to $\sim 8$\,kpc (Sagittarius~B2), including low-mass cloud cores, high-mass star-forming regions, and the CMZ.
Table~\ref{tab:surv} in Appendix lists their distances.
We also combine fields into four stacked entries: Orion\,A (three fields), NGC\,2264 (two fields), NGC\,6334 (two fields), and the CMZ (seven fields).
Ratios below unity predominate among the sources measured with each estimator, extending the polarization deficit at the shorter wavelength beyond the representative Ophiuchus A field.
The methods select different sources, which limits the precision of a single ratio for the survey as a whole.

Table~\ref{tbl:Pratio1} lists the intensity-weighted mean polarization fractions in columns~(2) and~(3) and the median pixel ratio $\tilde{R}_1$ on the Method~1 mask in column~(4).
The ratio $\tilde{R}_1$ is distinct from the quotient $R_{\rm IWM}$ of the two mean fractions, and it is below unity in the majority of sources.
Of the 45 individual sources, 34 (76\%) show $\tilde{R}_1 < 1$ and 11 (24\%) show $\tilde{R}_1 > 1$.
The median ratio among the 45 individual sources is $\tilde{R}_1 = 0.78$, with values between 0.40 (IC \, 5146) and 4.27 (Perseus \, B1).
 
Several of the most massive and luminous regions have $\tilde{R}_1 < 1$: OMC\,1 ($\tilde{R}_1 = 0.58 \pm 0.04$), Monoceros~R2 (figure~\ref{afig:Bmap_MonoR2} in Appendix; $\tilde{R}_1 = 0.45 \pm 0.01$), DR\,21 ($\tilde{R}_1 = 0.55 \pm 0.01$), and NGC\,6334 North ($\tilde{R}_1 = 0.44 \pm 0.01$).
L1688 Ophiuchus~A, our representative example from sub-subsection~\ref{sss:P_ratio_OphA}, yields $\tilde{R}_1 = 0.48 \pm 0.02$, placing it among the sources with the lowest ratios and demonstrating that the reduced polarization efficiency at 450\,\micron\ is not restricted to high-mass environments.
 
Sources with $\tilde{R}_1 > 1$ tend to have low signal-to-noise ratios.
Most have $N < 30$, including L1689~South (figure~\ref{afig:Bmap_L1689S} in Appendix; $N = 3$), Perseus~B1 (figure~\ref{afig:Bmap_PersB1} in Appendix; $N = 11$), and L483 (figure~\ref{afig:Bmap_L483L4} in Appendix; $N = 9$).
The notable exception is Auriga (figure~\ref{afig:Bmap_Auriga} in Appendix; $\tilde{R}_1 = 1.56 \pm 0.09$, $N = 54$), which warrants further investigation.
Paper~S1 subsection~4.2 (Size-Dependent Grain Alignment) discusses these outliers.
 
The intensity-weighted mean polarization fractions themselves show substantial variation across the sample.
At 850\,\micron, $\langle P_{850} \rangle_I$ ranges from 0.43\% (G31.41+0.31; figure~\ref{afig:Bmap_G3141G} in Appendix) to 4.55\% (OMC\,1), with a median of $\sim 1.3$\%.
At 450\,\micron, $\langle P_{450} \rangle_I$ ranges from 0.31\% (G31.41+0.31) to 3.69\% (L1689~South), with a median of $\sim 1.1$\%.

A polarization deficit confined to the brightest structures could appear in a weighted measurement without characterizing a typical detected line of sight.
Comparing the regression slope with the median pixel ratio tests whether the prevalence of $P_{450}<P_{850}$ depends strongly on that emphasis.
Table~\ref{tbl:Pratios23} presents the results based on correlation (Method~2) and statistical distribution (Method~3).
Ratios below unity are even more common in these methods.
In Method~2, 36 of 43 sources (84\%) have $R_{\rm Corr} < 1$, with a median of $0.68$.
In Method~3, 29 of 31 sources (94\%) have $R_{\rm EW} < 1$.
The median over those 31 individual sources is $R_{\rm EW} = 0.70$.
The value $R_{\rm EW} = 0.70$ differs from that quoted in the abstract and in Paper~S1, $R_{\rm EW} = 0.69$.
The value $R_{\rm EW} = 0.69$ is the median over the 30 sources of the 35-source {\sc SpIdx} subsample that have a measured ratio (subsection~\ref{ss:S2_SpIdx}).
The two values come from different samples.
 
For the 31 sources with all three estimators, the three ratios are below unity in 29, above unity in Auriga, and mixed in Orion~A Field~5 ($R_{\rm EW} = 1.27$).
For L1688 Ophiuchus~A, the three agree closely ($\tilde{R}_1 = 0.48$, $R_{\rm Corr} = 0.47$, and $R_{\rm EW} = 0.47$).
DR\,21 ($\tilde{R}_1 = 0.55$, $R_{\rm Corr} = 0.56$, $R_{\rm EW} = 0.51$) and NGC\,6334 North ($\tilde{R}_1 = 0.44$, $R_{\rm Corr} = 0.42$, $R_{\rm EW} = 0.46$) show similar agreement.
 
Systematic differences between methods appear primarily in sources with a low signal-to-noise ratio.
Of the 14 individual sources without an $R_{\rm EW}$ entry, 12 have an $R_{\rm Corr}$ entry in table~\ref{tbl:Pratios23}, whereas L1689~South and M16 have no Method~2 entry either.
Among these 14 sources, $\tilde{R}_1$ exceeds unity in 10, and $R_{\rm Corr}$ exceeds unity in 6 of the 12.
For example, L43 (figure~\ref{afig:Bmap_L43L43} in Appendix) yields $\tilde{R}_1 = 1.63 \pm 0.17$ but $R_{\rm Corr} = 1.25 \pm 0.08$.
We attribute this difference to the sensitivity of different weighting schemes to noise fluctuations.
Its six Method~3 pixels lie in two groups: four contiguous pixels at the 850\,\micron\ intensity peak, on the starless core of L43 \citep{Karoly2023}, and two adjacent pixels about $80\arcsec$ to the west, near RNO~91.
RNO~91 is a YSO in transition from Class~I to Class~II \citep{Koyamatsu2014}, with a bolometric luminosity of $2.5\,L_\odot$ for a distance of 130\,pc \citep{Chen2009L43}.
None of the six has both $P_{450}$ and $P_{850}$ above three times their uncertainties, so table~\ref{tbl:Pratios23} gives no $R_{\rm EW}$ for L43.
Their pixel ratios range from 0.12 to 7.41 (median 2.56, interquartile range 0.47--3.77), and the ratios within each group differ by more than a factor of 20.
 
The Method~2 correlation coefficient provides an additional diagnostic.
Sources with ordered fields show strong correlations, such as DR \, 21 with Spearman $\rho = 0.95$, while sources with complex fields or measurements dominated by noise show weaker correlations.
The predominance of ratios below unity in both summaries shows that the detected deficit is not restricted to the particular weighting used by one estimator.
We first consider measurement and sampling issues when interpreting larger discrepancies in poorly measured sources, before assigning them an astrophysical origin.

The four stacked entries provide improved statistics by combining pixels from multiple fields.
All four stacked entries have $\tilde{R}_1 < 1$.
Orion\,A gives $\tilde{R}_1 = 0.56 \pm 0.02$ from 326 pixels.
NGC\,2264 gives $\tilde{R}_1 = 0.62 \pm 0.02$ from 192 pixels.
NGC\,6334 gives $\tilde{R}_1 = 0.60 \pm 0.02$ from 232 pixels.
The CMZ gives $\tilde{R}_1 = 0.84 \pm 0.01$ from 701 pixels.
The CMZ stacked ratios are higher than those of the local star-forming regions, a difference also visible in the individual CMZ fields (table~\ref{tbl:Pratio1}).
However, it is not statistically supported as an ordered environmental response (subsection~\ref{ss:S2_EnvStrat}).
Subsection~\ref{ss:S2_SpIdx} examines whether this environmental difference persists after accounting for effects related to distance.
We defer to future work a dedicated analysis of the CMZ fields that separates the dust emission from other contributions to the 850\,$\mu$m continuum, such as synchrotron emission toward Sgr~A$^{*}$ \citep{Sato2024}.

\subsubsection{Synthesis of Polarization Fraction Ratios}
\label{sss:Synthesis}
The key observational result is that polarized dust emission constitutes a smaller fraction of the total emission at 450\,$\mu$m than at 850\,$\mu$m in most measured sources.
Comparing the three summaries for individual sources shows how this conclusion depends on the parts of each cloud emphasized by the analysis.
 
We consider the existence of a wavelength-dependent polarization deficit secure across the source population, because the sample medians of $P_{450}/P_{850}$ are below unity under all three weighting schemes.
We test formally whether its amplitude in individual sources follows an ordered pattern with environment in subsection~\ref{ss:S2_EnvStrat}.
The recurring deficit constrains the relative spectra of polarized and total dust emission.
Its physical origin must be assessed with the field geometry and cloud conditions, because agreement among summaries of the same observations does not remove shared systematic effects.

Across the three methods, approximately 76--94\% of sources have $R_{\rm p} < 1$, with sample medians of $R_{\rm p} = 0.68$--0.78.
This behavior is seen across diverse environments: nearby low-mass cores (L1688 Ophiuchus~A at 139\,pc), intermediate-distance high-mass regions (OMC\,1 at 399\,pc, DR\,21 at 1500\,pc), and distant Galactic Center clouds (Sagittarius~B2 at 8.3\,kpc).
Table~\ref{tab:surv} in Appendix lists the adopted reference distances cited here and below.

Dense bright structures and the wider detected material can weight a cloud's polarization spectrum differently.
Agreement between their corresponding estimators is therefore informative about whether the observed wavelength deficit is confined to one part of the measured emission.
For several sources with high signal-to-noise ratios, we find agreement among the median pixel ratio $\tilde{R}_1$ of Method~1, the alignment-weighted slope $R_{\rm Corr}$, and the equal-weighted median $R_{\rm EW}$.
We therefore conclude that, in these sources, wavelength-dependent polarization is not an artifact of the particular statistical treatment adopted.
For L1688 Ophiuchus~A, the three methods yield $\tilde{R}_1 = 0.48$, $R_{\rm Corr} = 0.47$, and $R_{\rm EW} = 0.47$.
The spread of only 0.01 is far smaller than the measurement uncertainties.
Similar behavior is found for DR\,21 and NGC\,6334 North.
 
The budget of systematic uncertainties in Paper~T subsection~\ref{T-sec:uncertainty} provides the relevant comparison.
The dominant term is the residual zero-level offset in 450\,$\mu$m Stokes-$Q/U$.
It contributes $\lesssim 0.05$ per source to $P_{450}/P_{850}$, larger than the $0.01$ spread between methods.
The three weighting schemes therefore agree well within the systematic floor.
Their consistency shows that the result does not depend on the statistical treatment.
The systematic floor sets the precision of the measurement, rather than the spread between methods.
For these sources with precise measurements, the polarization deficit at the shorter wavelength remains whether we emphasize bright emission or give equal weight to pixels within the selected mask.
However, this persistence leaves unresolved how grain alignment and geometry contribute to the deficit.
 
We interpret the measured levels of $P_{450}/P_{850}$ and $\gamma-\alpha$ on the assumption that the modulation efficiency of POL-2 is equal in the two bands.
The ranking of the sources and the associations of $P_{450}/P_{850}$ and $\gamma-\alpha$ with cloud conditions do not depend on this assumption.
The budget of systematic uncertainties in Paper~T does not include a difference in the modulation efficiency between the two bands.
The modulation efficiency has not been calibrated in either band to the best of our knowledge.
A difference in the modulation efficiency between the two bands would scale $P_{450}/P_{850}$ by a common factor and shift $\gamma-\alpha$ by a common amount for all sources (Paper~T~subsection~\ref{T-ss:flux}).

Figure~\ref{fig:Vaillancourt_comparison} presents the polarization fraction ratio for individual pixels as a function of normalized intensity in 13 sources with high signal-to-noise polarimetry at both wavelengths.
We plot $P_{850}/P_{450}$ against $I_{450}/I_{450}^{\rm max}$, following the format of \citet{Vaillancourt2008}, who reported systematic variations of $P_{450}/P_{350}$ with intensity in OMC\,1.
Across all plotted pixels, the ratio shows only a weak positive correlation with normalized intensity (Spearman $r_s=+0.13$), and the trend differs between sources.
For pixels at or above 5\% of each source's peak intensity, six of the 13 sources have positive correlations with nominal $p<0.01$ calculated from individual pixels ($r_s=+0.10$ to $+0.40$).
Monoceros~R2 has a negative correlation meeting the same criterion ($r_s=-0.34$).
The remaining six sources do not meet this significance criterion.
 
For OMC\,1, the median $P_{850}/P_{450}$ rises from 1.5 at 5\%--20\% of peak intensity to 2.6 at or above 20\% ($r_s=+0.40$).
The Method~3 value $P_{450}/P_{850} \approx 0.48 \pm 0.07$, corresponding to an inverse ratio of approximately $2.1 \pm 0.3$, summarizes a different set of pixels: those that meet the stricter criteria for intensity and polarization detection in both bands defined in sub-subsection~\ref{sss:Methodology}.
The 13 sources span distances from 139\,pc to 8.3\,kpc (table~\ref{tab:surv} in Appendix).

We can compare qualitative trends with earlier studies, although differences in wavelength coverage and measurement methods limit a direct comparison.
\citet{Vaillancourt2008} reported, from CSO/SHARP polarimetric maps of OMC\,1 at 350 and 450\,\micron, $P_{450}/P_{350} \sim 1.3$ at positions more than 20\arcsec\ from the two submillimeter intensity peaks (the Kleinmann--Low nebula and Orion South).
In their maps, the ratio falls by more than a factor of two toward the Kleinmann--Low nebula.
For OMC\,1, the median pixel ratio in the Method~1 mask is $\tilde{R}_1 = 0.58 \pm 0.04$ (table~\ref{tbl:Pratio1}).
These measurements do not locate the minimum of the V-shaped polarization spectrum reported for OMC\,1 by \citet{Tram2024}.
\citet{Lyo2021} found median polarization fractions of 3.0\% at 450\,\micron\ and 1.2\% at 850\,\micron\ in the central $3\arcmin$ of NGC~2071IR, using vectors with $P/\delta P \geq 3$ selected separately in each band.
Applied to our maps of this region, at the native 450\,\micron\ resolution rather than their $14\farcs1$, the same selection gives medians of 2.6\% and 1.2\%.
We define our masks on Stokes~$I$ (sub-subsection~\ref{sss:Methodology}), because selecting on polarization preferentially retains pixels with high polarization fractions.
On the Method~1 mask, our ratio for NGC~2071 is also above unity ($\tilde{R}_1 = 1.56 \pm 0.21$; table~\ref{tbl:Pratio1}).
\citet{Scheiter2026} combined SOFIA/HAWC+ 154\,\micron\ data with POL-2 data of L43 and found that the polarization fraction decreases from 450 to 850\,\micron.
Our ratio for L43 is also above unity ($\tilde{R}_1 = 1.63 \pm 0.17$; table~\ref{tbl:Pratio1}).
 
The polarization fraction compares the polarized emission with the total dust emission at each wavelength.
Its wavelength dependence can therefore be expressed as a difference between the spectral slopes of total and polarized intensity.
In subsection~\ref{ss:S2_SpIdx}, we express this ratio through $(\alpha,\gamma)$ using the identity $P_{450}/P_{850} = (\nu_{450}/\nu_{850})^{\gamma-\alpha}$.
The sample median $R_{\rm EW} = 0.70$ implies $\gamma - \alpha \approx -0.56$.
Equation~\ref{eq:spectral_indices} defines $\alpha$ and $\gamma$, and equation~\ref{eq:Pratio_SpIdx} gives this relation explicitly.
This decomposition connects the observed polarization deficit to the shallower spectrum of polarized emission.
Interpreting that spectral difference as a difference in dust opacity indices requires additional assumptions about the emitting material.
 
In summary, our analysis establishes that $P_{450}/P_{850} < 1$ is the dominant behavior across star-forming regions, with sample medians of 0.68--0.78 under the three methods.
We stress that the result is not affected by the systematic terms quantified in Paper~T (calibration and residual $Q/U$ zero-level offsets; $\sigma(P_{450}/P_{850}) \lesssim 0.02$; Paper~T subsection~\ref{T-sec:precisionfloor}).
Within individual sources, the dependence of the ratio on local intensity differs from source to source (figure~\ref{fig:Vaillancourt_comparison}).
Subsection~\ref{ss:S2_SpIdx} and Paper~S1 subsections~4.2--4.4 examine whether the wavelength dependence reflects grain sizes, alignment efficiency, optical depth, or temperature structure along the line of sight.

\subsection{Spectral Indices of Total and Polarized Intensity}
\label{ss:S2_SpIdx}
A polarization fraction that depends on wavelength means that polarized dust emission and total dust emission have different spectral slopes.
Measuring those slopes separately allows us to ask how the dust that contributes to polarization differs from the full emitting column.
 
We complete the F4 test of universality across the population in our three-test framework (figure~\ref{fig:S2_test_overview}, table~\ref{tbl:S2_test_index}) by measuring $\alpha$ and $\gamma$ for each source in the {\sc SpIdx} subsample.
With $\langle\alpha\rangle = 3.33$, $\langle\gamma\rangle = 2.85$--$3.04$, and $\gamma<\alpha$ in 24--27 out of 35 sources, this test confirms the F2 finding of Paper~S1 across the population.
 
The analysis of the polarization fraction ratio (subsection~\ref{ss:S2_P_ratio}) established that $P_{450}/P_{850} < 1$ holds for the majority of sources in every environment class, with a median $R_{\rm EW} = 0.69$ over the 30 {\sc SpIdx} sources with a measured ratio.
The spectral indices provide a complementary description by measuring the frequency dependence of total and polarized intensity separately.
The indices $\alpha$ and $\gamma$ therefore connect the polarization-fraction ratio to models of dust emissivity and the physics of grain alignment.
We first describe the framework and data selection (sub-subsection~\ref{sss:SpIdxFramework}) and apply it to L1688 Ophiuchus~A (figure~\ref{afig:Bmap_OphiuA} in Appendix; sub-subsection~\ref{sss:SpIdx_OphA}).
We then report the global results (sub-subsection~\ref{sss:SpIdx_Global}) and discuss their implications (sub-subsection~\ref{sss:SpIdx_Synthesis}).
Relating this wavelength contrast to models of dust emission is conditional on the temperature structure and on the correspondence between the polarized and total emitting components.

\subsubsection{Spectral Index Framework and Data Selection}
\label{sss:SpIdxFramework}
The polarized emission is fainter than the total dust emission, so noise and pixel selection can mimic a difference between their spectral slopes.
Therefore, a physical interpretation of $\gamma-\alpha$ begins with a precise definition of the measured intensities and the lines of sight admitted to the comparison.
 
We define the spectral indices for total and polarized intensity as
\begin{equation}
\alpha \equiv \frac{\log(I_{450}/I_{850})}{\log(\nu_{450}/\nu_{850})}
\quad \text{and} \quad
\gamma \equiv \frac{\log(PI_{450}/PI_{850})}{\log(\nu_{450}/\nu_{850})},
\label{eq:spectral_indices}
\end{equation}
where $\nu_{450} = 665.0$\,GHz and $\nu_{850} = 349.5$\,GHz are the mid-frequencies of the SCUBA-2/POL-2 passbands.
 
The polarization-fraction ratio $P_{450}/P_{850}$ analyzed in subsection~\ref{ss:S2_P_ratio} follows from equation~(\ref{eq:spectral_indices}) directly as the algebraic identity
\begin{equation}
\frac{P_{450}}{P_{850}} = \left(\frac{\nu_{450}}{\nu_{850}}\right)^{\gamma - \alpha},
\label{eq:Pratio_SpIdx}
\end{equation}
which decomposes the polarization-fraction ratio of subsection~\ref{ss:S2_P_ratio} into a dust-emission spectrum ($\alpha$) and a polarized-emission spectrum ($\gamma$).
Thus, $\gamma < \alpha$ corresponds to a polarization fraction that is higher at 850 than at 450\,\micron.
For the sample medians $P_{450}/P_{850} = 0.68$--0.78 obtained in subsection~\ref{ss:S2_P_ratio}, equation~(\ref{eq:Pratio_SpIdx}) implies a prediction $\gamma - \alpha \approx -0.60$ to $-0.39$ (since $\nu_{450}/\nu_{850} \approx 1.90$).
The measurements of $(\alpha,\gamma)$ for individual sources presented below in this subsection test this prediction.
These measurements also discriminate between scenarios where the wavelength dependence arises from variations in dust opacity versus changes in grain alignment efficiency.
Paper~S1 subsection~4.2 develops the physical interpretation in terms of size-dependent grain alignment.
 
For modified blackbody emission $S_\nu \propto \nu^\beta B_\nu(T_{\rm d})$, the observed spectral index between two frequencies is
\begin{equation}
\alpha = \beta + C(T_{\rm d}), \qquad C(T_{\rm d}) \equiv \frac{\log\left[B_{\nu_{450}}(T_{\rm d})/B_{\nu_{850}}(T_{\rm d})\right]}{\log(\nu_{450}/\nu_{850})},
\label{eq:planck_correction}
\end{equation}
where $C(T_{\rm d})$ is the shape factor of the Planck function between the two POL-2 bands.
The observed slope of the intensity spectrum, $\alpha$, is thus the opacity slope, $\beta$, plus the slope of the Planck function, $C(T_{\rm d})$.
The POL-2 band centers correspond to $h\nu/k_{\rm B} = 31.9$~K and $16.8$~K, comparable to the molecular cloud dust temperatures of 10--30~K probed by our sample.
We therefore retain the full Planck form throughout.
The indices $\alpha$ and $\gamma$ are treated as the primary observables, and whenever $\beta$ is quoted, it is computed directly via equation~(\ref{eq:planck_correction}).
We refer to obtaining $\beta$ by subtracting $C(T_{\rm d})$ from $\alpha$ as the $C(T_{\rm d})$ correction.
 
If total and polarized emission are described by a single common dust temperature, the shape factor $C(T_{\rm d})$ of the Planck function enters $\alpha$ and $\gamma$ as the same additive term.
It therefore cancels \emph{exactly} in their difference, $\gamma - \alpha = \beta_{\rm pol} - \beta$.
Here $\beta_{\rm pol}$ is the effective opacity index of the aligned grain population.
The cancellation is an algebraic identity of the modified-blackbody representation with a single temperature, rather than a Rayleigh--Jeans approximation.
This identity holds for the full Planck form at any finite $T_{\rm d}$.
As a result, the central finding $\gamma < \alpha$ and all derived values $P_{450}/P_{850}$ reported below are independent of the $C(T_{\rm d})$ correction at the adopted common single temperature, however uncertain that correction may be.
Only absolute $\beta$ values (table~\ref{tbl:SpIdx_Global}) depend directly on it.
 
However, the identity fails where components at different temperatures along the line of sight also differ in polarization efficiency.
In that case, $\alpha$ and $\gamma$ are weighted by different moments of the temperature distribution ($\gamma$ by the polarized emission and $\alpha$ by the total emission), and the residual does not vanish.
This residual is an $\mathcal{O}(1)$ function of the component temperatures, the warm mass fraction, and the contrast in polarization efficiency between the components, not a small higher-order correction.
Uncertainty in a single adopted $T_{\rm d}$ therefore cannot overturn $\gamma < \alpha$, whatever the size of that uncertainty.
However, temperature mixing with multiple physically distinct components along the line of sight is not covered by that statement and remains the one temperature-related effect that could.
We tested that case in Paper~S1 (subsection~4.4 and appendix~\ref{S1-app:Tmixing}).
For a common polarization angle, such mixing reproduces the observed median $\gamma-\alpha$ only if the warm dust is polarized less efficiently than the cold dust.
For cold dust at 10\,K or warmer and warm dust at 40\,K or cooler, the ratio of the warm to the cold effective polarization efficiency must be about 0.18 or less.
 
Where $\beta$ is reported, the same shape factor propagates the uncertainty in $T_{\rm d}$ into an uncertainty in $\beta$ for individual sources.
We adopt representative ranges for each environment: $T_{\rm d} = 15 \pm 5$~K for LMSF and cold infrared dark clouds (IRDCs), and $T_{\rm d} = 25 \pm 10$~K for HMSF and CMZ sources.
Over these ranges, $C \approx 0.39$ at $T_{\rm d} = 10$~K, $1.01$ at $15$~K, $1.30$ at $20$~K, $1.45$ at $25$~K, and $1.56$ at $30$~K.
The corresponding half-width, $\sigma_{C(T_{\rm d})} \approx 0.35$, gives $\sigma_{\beta, T} \approx 0.35$ through $\beta = \alpha - C(T_{\rm d})$.
This uncertainty of temperature broadens the absolute $\beta$ values in table~\ref{tbl:SpIdx_Global}, in addition to their empirical statistical errors.
 
For the Orion~A sub-fields and L1688 Ophiuchus~A, we adopt the median dust temperature derived from HGBS for each field.
For the other sources, including six LMSF regions covered by the HGBS maps of dust temperature, we use the defaults for each environment given above.
For the two cold infrared dark clouds in the sample (G28.34$+$0.06, mosaicked field, figure~\ref{afig:Bmap_G283MO} in Appendix; SDC~13 mm-2, figure~\ref{afig:Bmap_SDC13M} in Appendix), we adopt $T_{\rm d} = 15$~K.
For the Brick (figure~\ref{afig:Bmap_GCBRCK} in Appendix), we adopt $T_{\rm d} = 20$~K.
These values reflect their published dust temperatures.
We do not attempt spectral energy distribution (SED) fitting for individual sources because the two POL-2 bands alone do not constrain $T_{\rm d}$ robustly.
We regard the adopted values as practical parameters for the correction rather than independently measured dust temperatures.
Uncertainties in these adopted temperatures broaden the absolute $\beta$ values but do not propagate to $\gamma - \alpha$ or any $P_{450}/P_{850}$ result.
 
The polarization fraction ratio is directly related to these indices through equation~(\ref{eq:Pratio_SpIdx}).
With $\nu_{450}/\nu_{850} = 1.90$, $P_{450}/P_{850} < 1$ for matched measurements requires $\gamma < \alpha$.
The polarized intensity must rise less steeply with frequency than the total intensity.
A constant polarization-fraction ratio $P_{450}/P_{850}$ corresponds to a fixed value of $\gamma - \alpha$, but corresponds to a flat polarization spectrum only when the ratio is unity, that is, when $\gamma = \alpha$.
Deviations of $\gamma$ from $\alpha$ therefore directly diagnose a dependence of the polarization efficiency on wavelength.
This framework connects the empirical measurements of the ratio in subsection~\ref{ss:S2_P_ratio} to the underlying spectral behavior of total and polarized emission.
 
The polarized intensity is intrinsically weaker and noisier than the total intensity, so $\gamma$ is more susceptible than $\alpha$ to biases that depend on the $S/N$ ratio.
We assess the dependence of both indices on their formal $S/N$ ratios source by source in Paper~T (sub-subsections~\ref{T-ss:spidx_snrbias}--\ref{T-ss:spidx_snrbias_sens}), with running medians in logarithmic $S/N$ bins.
We use the running median as a diagnostic, not as a selection cut.
We do not apply a $S/N(\gamma)$ cut when computing the values for individual sources in table~\ref{tbl:SpIdx_Global} (sub-subsection~\ref{sss:SpIdx_Global}).
 
The spatially resolved maps (figure~\ref{fig:SpIdx_maps_OphA} for Ophiuchus~A) apply a display threshold common to all fields, $S/N(\alpha) > 50$ and $S/N(\gamma) > 7$, and are therefore sparsely populated for many sources.
We compute the values for individual sources using the pixel selection for the statistical analysis in sub-subsection~\ref{sss:SpIdx_Global}.
The number of selected pixels also varies significantly among targets.
The strict data selection required for spectral index analyses motivates the use of multiple complementary approaches for polarization fraction ratios (sub-subsection~\ref{sss:Methodology}), which remain robust under less restrictive selection criteria.
The present results establish baseline values for future polarimetric surveys with broader wavelength coverage.
However, measurements of spectral indices at two wavelengths inherently provide fewer constraints than analyses at multiple wavelengths.
 
We apply the same analysis procedure to all sources.
First, we assess potential biases that depend on $S/N$, using running medians of the spectral indices as a function of $S/N$ (Paper~T sub-subsection~\ref{T-ss:spidx_snrbias}).
Second, we do not use $S/N(\gamma)$ to define the parent sample.
As a sensitivity test, we impose and vary a common $S/N(\gamma)$ threshold to examine how the population result responds (Paper~T sub-subsection~\ref{T-ss:spidx_snrbias_sens}).
Third, we examine whether any residual trends remain after controlling for observational effects such as distance (appendix~\ref{app:S2_dist_bias}).
The selected measurements constrain the relative spectral behavior of the polarized and total emission in the detected cloud material.
A common temperature correction cancels from their difference, but interpreting that difference as a contrast in opacity still requires control of temperature mixing along the line of sight.

\subsubsection{Application to L1688 Ophiuchus~A}
\label{sss:SpIdx_OphA}
 
To examine whether polarized and total dust emission have different spectra near the SM1 core \citep{Kwon2018}, we compare their spectral indices in L1688 Ophiuchus~A (figure~\ref{afig:Bmap_OphiuA} in Appendix).
This resolved example complements the comparison of ratios of polarization fractions in sub-subsection~\ref{sss:P_ratio_OphA}.
\citet{Kwon2018} studied this source at 850\,\micron\ using the same POL-2 observations.
We applied the data-reduction procedure described in Paper~T and extended the analysis to spectral indices between the two wavelengths.
We assessed the bias for this source by quantifying the dependence of the measured spectral indices on their formal $S/N$ ratios (Paper~T figure~\ref{T-fig:SpIdx_BiasCheck_OphA}).
 
A total of $N = 238$ pixels satisfy the basic selection criteria ($I/\sigma_I > 30$ at both wavelengths and $\alpha > 0$).
Of these, 175 pixels have $S/N(\alpha) > 50$, whereas only 8 pixels have $S/N(\gamma) > 7$.
For Ophiuchus~A, the running median of $\gamma$ levels off over its two bins with the highest $S/N$ (Paper~T figure~\ref{T-fig:SpIdx_BiasCheck_OphA}).
The resolved maps below use the common display threshold $S/N(\gamma)>7$.
Three pixels yield $\gamma < 0$, likely because of noise fluctuations or residual systematic effects at low $S/N$.
We excluded these pixels from the spatially resolved maps.
 
\begin{figure*}
\centering
\includegraphics{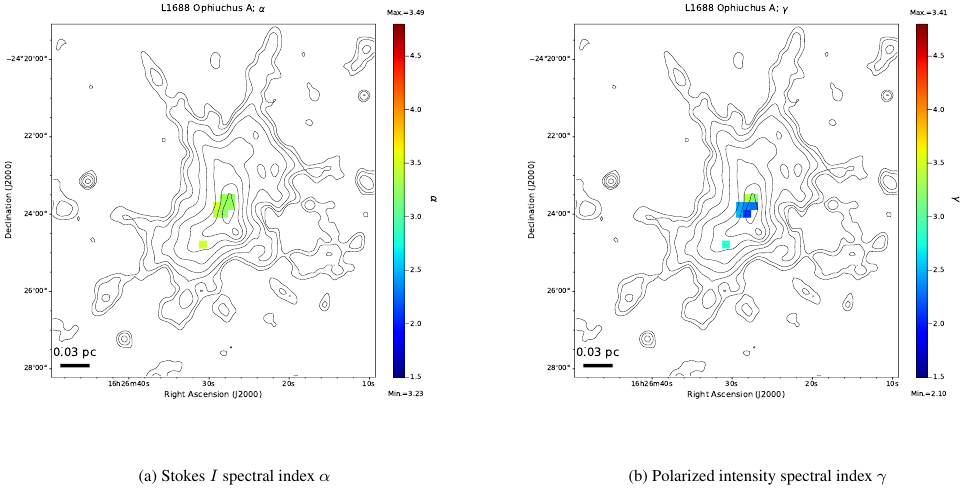}
\caption{Spatially resolved maps of the spectral indices for L1688 Ophiuchus~A.
(a)~The effective two-band frequency index $\alpha$ of Stokes~$I$ and
(b)~the corresponding index $\gamma$ of polarized intensity, defined in equation~(\ref{eq:spectral_indices}).
The pixel centers in both panels are spaced by $12\farcs0$, nearly one beam width ($12\farcs6$ FWHM).
Black contours show the 850\,\micron\ Stokes~$I$ distribution.
The scale bar indicates the physical size at the adopted reference distance of Ophiuchus (139\,pc; table~\ref{tab:surv} in Appendix).
Only pixels satisfying $S/N(I) > 30$ at both wavelengths, $S/N(\alpha) > 50$, and $S/N(\gamma) > 7$ are displayed, yielding $N = 8$ pixels located near the SM1 core.
The small number of retained pixels reflects the $S/N(\gamma) > 7$ display threshold common to all fields, which limits but does not by itself remove $S/N$-dependent effects.
The color scale ranges from 1.5 to 4.8 for both panels, and the displayed values span $\alpha = 3.23$--$3.49$ and $\gamma = 2.10$--$3.41$.
In 7 of the 8 displayed pixels, $\gamma<\alpha$, meaning that the polarization fraction is higher at 850 than at 450\,$\mu$m.
\\ {Alt text: Two maps with contours.}
}
\label{fig:SpIdx_maps_OphA}
\end{figure*}

The $S/N(\gamma) > 7$ display threshold leaves $N = 8$ pixels for the spatially resolved spectral-index analysis.
This threshold limits but does not remove $S/N$-dependent effects by itself.
Figure~\ref{fig:SpIdx_maps_OphA} shows the corresponding maps of $\alpha$ and $\gamma$ overlaid on the 850\,\micron\ Stokes~$I$ map (contours).
The few valid pixels are concentrated near the intensity peak (the SM1 core region), where both $I$ and $PI$ achieve the highest $S/N$ ratios.
 
Across these 8 pixels, the spectral index of Stokes~$I$ spans $\alpha = 3.23$--$3.49$, while the spectral index of polarized intensity spans $\gamma = 2.10$--$3.41$.
The median values are $\langle \alpha \rangle = 3.27 \pm 0.02$ and $\langle \gamma \rangle = 2.48 \pm 0.17$.
In contrast, $\gamma$ exhibits a larger pixel-to-pixel scatter but is lower than $\alpha$ in 7 of the 8 pixels, producing a median difference of $\gamma - \alpha = -0.79 \pm 0.17$.
We evaluated $C(T_{\rm d}) \approx 1.26$ with the full Planck function (equation~\ref{eq:planck_correction}), using the median dust temperature derived from HGBS for this source ($T_{\rm d} \approx 19.2$~K).
The measured $\alpha = 3.27$ then gives $\beta = \alpha - C(T_{\rm d}) \approx 2.0$, comparable to the $\beta \approx 1.6$--2 found toward the dense cores of Perseus~B1 \citep{Sadavoy2013}.
This estimate depends on the opacity index assumed in the HGBS fits, $\beta = 2$ \citep{Ladjelate2020}, through the adopted $T_{\rm d}$.
 Therefore, we do not regard this value as an independent confirmation of $\beta = 2$.
 
The inequality $\gamma < \alpha$ directly implies $P_{450} < P_{850}$ through equation~(\ref{eq:Pratio_SpIdx}).
Using the median spectral indices and propagating uncertainties, we obtained the following.
\begin{equation}
\frac{P_{450}}{P_{850}}
= \left( \frac{\nu_{450}}{\nu_{850}} \right)^{\gamma - \alpha}
= 1.90^{-0.79 \pm 0.17}
= 0.60 \pm 0.07.
\label{eq:pratio_indices_applied}
\end{equation}
This value is consistent, within the uncertainties, with those derived from the analysis of polarization fraction ratios ($R_{\rm IWM} = 0.49 \pm 0.02$, $R_{\rm Corr} = 0.47 \pm 0.01$, $R_{\rm EW} = 0.47 \pm 0.07$; sub-subsection~\ref{sss:P_ratio_OphA}).
This agreement confirms that the spectral index approach and the direct ratio approach yield mutually compatible constraints on the wavelength dependence of polarization.
Different spatial selections can contribute to the modest difference (the spectral index method yields a $\sim$25\% higher ratio).
We have not isolated this contribution by comparing measurements selected with the same mask.
The spatially resolved spectral-index analysis is restricted to the 8 highest-$S/N$ pixels near the SM1 core.

 The polarization-ratio methods use different selections in Ophiuchus~A.
The Method~1 table contains 100 pixels, while the parent set selected on intensity for Method~3 contains 18 pixels, of which 16 have polarization detections in both bands and enter $R_{\rm EW}$ (Paper~S1 figure~\ref{S1-fig:Pratio_OphA_method3}).
The much smaller number of pixels in the spatially resolved spectral-index analysis ($N = 8$) reflects the adopted display threshold $S/N(\gamma) > 7$.
The two approaches are complementary.
Ratios of polarization fractions provide statistics over a broader area and across diverse environments, while spectral indices connect the measurements to models of dust emissivity and grain alignment \citep[e.g.,][]{Bethell2007, Draine2009}.
Seven of the eight pixels displayed near SM1 have $\gamma<\alpha$, showing that the polarization fraction is higher at 850 than at 450\,$\mu$m along these lines of sight.
This result applies to the selected region near the core and does not establish the same behavior throughout Ophiuchus~A.

\begin{figure}
\centering
\includegraphics[width=0.85\columnwidth]{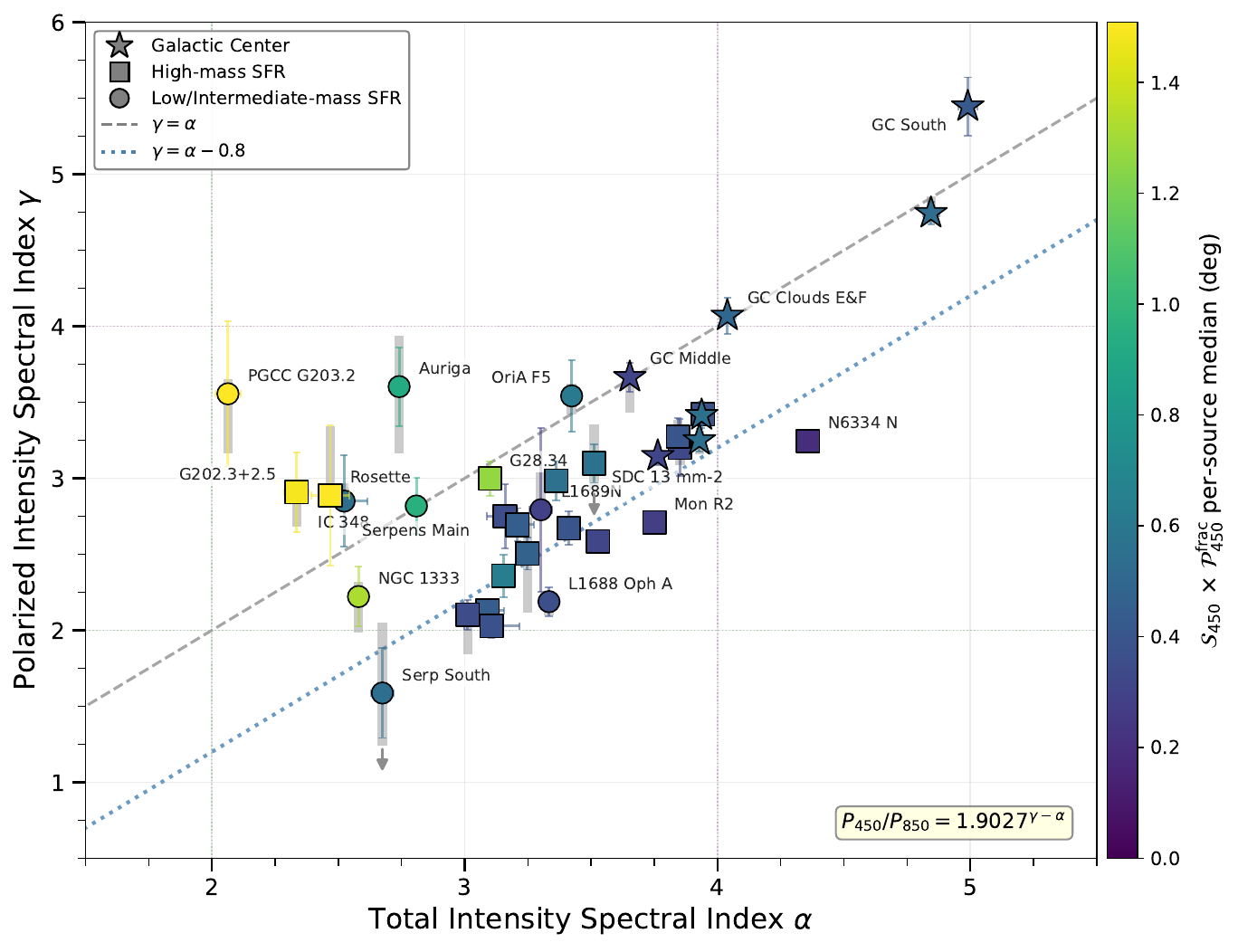}
\caption{
Correlation between the spectral index $\langle\alpha\rangle$ of Stokes~$I$ and the spectral index $\langle\gamma\rangle$ of polarized intensity for the 35 sources in the statistical sample ($N_{\rm samp} \geq 5$).
Symbols show the measured medians of individual sources, and error bars represent the Maritz--Jarrett standard error of each median (sub-subsection~\ref{sss:SpIdxFramework}).
Symbol shapes encode environment: stars for Central Molecular Zone (CMZ) sources, squares for intermediate- to high-mass star-forming regions (HMSF), and circles for low-mass star-forming regions (LMSF).
Marker fill colors show, for individual sources, the point estimates of the proxy $\mathcal{S}_{450} \times \mathcal{P}^{\rm frac}_{450}$ for grain alignment efficiency (subsection~\ref{ss:S2_SP}), calculated as the median of the debiased values over all pixels selected for this proxy.
For most sources, more than half of these pixels provide only upper limits, so the colors rank the point estimates without establishing differences in the underlying proxy.
The linear color scale is a display choice.
The proxy uses the broader 450\,$\mu$m mask ($I/\sigma_I>30$), whereas $\alpha$ and $\gamma$ use the spectral-index mask.
The colors therefore describe each source as a whole and do not necessarily represent the same pixels as the spectral indices, especially where spectral-index coverage is sparse.
The seven mosaicked fields are not listed in table~\ref{tbl:SP450stats} in Appendix.
We calculate their colors from their mosaic maps using the same proxy estimator and selection.
The light-gray vertical bars show the range of each source's median $\gamma$ when measurements of polarized intensity below three times their uncertainty are treated as $3\sigma$ limits.
A nondetection at 450\,$\mu$m gives an upper limit on $\gamma$, whereas one at 850\,$\mu$m gives a lower limit.
For every source, the measured median lies within the corresponding range.
The gray bars and arrows indicate this range, not a correction to the measured median.
For Serpens~South (figure~\ref{afig:Bmap_SerpS3} in Appendix) and SDC~13 mm-2 (figure~\ref{afig:Bmap_SDC13M} in Appendix), the upper end of the gray bar gives an upper limit on the median $\gamma$.
For these two sources, the downward arrow indicates that the range extends without a lower bound.
The downward lengths of these two bars and arrows are chosen for display only and do not indicate lower bounds.
The dashed line shows $\gamma = \alpha$ ($P_{450}/P_{850} = 1$).
The dotted line shows $\gamma = \alpha - 0.8$ ($P_{450}/P_{850} \approx 0.6$).
A majority of sources lie below the $\gamma = \alpha$ line: 25 of 35 at the measured medians, and 24--27 of 35 under the two extreme placements of the $3\sigma$ limits.
For matched measurements, $\gamma<\alpha$ corresponds to $P_{450}<P_{850}$, so a higher polarization fraction at the longer wavelength is the dominant behavior in the sample.
The five sources with $\gamma > \alpha$ at the $1\sigma$ level are Auriga (figure~\ref{afig:Bmap_Auriga} in Appendix),
Perseus~IC348 (figure~\ref{afig:Bmap_IC348H} in Appendix), and PGCC~G203.21$-$11.20 (figure~\ref{afig:Bmap_OrBPMO} in Appendix), all LMSF; G202.3$+$2.5 (HMSF; figure~\ref{afig:Bmap_G202MO} in Appendix); and GC~South (CMZ; figure~\ref{afig:Bmap_GCSOUT} in Appendix).
The following sources are consistent with $\gamma = \alpha$ at the $1\sigma$ level: L1689~North (figure~\ref{afig:Bmap_L1689N} in Appendix), Orion~A~Field~5, and Serpens~Main (figure~\ref{afig:Bmap_SerpMO} in Appendix), all LMSF; G28.34$+$0.06 (figure~\ref{afig:Bmap_G283MO} in Appendix) and the Rosette
Molecular Cloud (figure~\ref{afig:Bmap_RoseMO} in Appendix), both HMSF; and GC~Clouds~E~and~F (figure~\ref{afig:Bmap_GCCLEF} in Appendix) and GC~Middle (figure~\ref{afig:Bmap_GCMIDD} in Appendix), both CMZ.
Subsection~\ref{ss:S2_EnvStrat} develops the environmental comparison.
The HMSF class has the lowest median point estimate of the proxy among the three classes (table~\ref{tbl:env_4metric}), and 17 of its 19 sources lie below the $\gamma = \alpha$ line.
The CMZ class lies near $\gamma = \alpha$ with an intermediate median point estimate, while the LMSF class spans both regimes and has the highest median point estimate.
The three classes are not ordered along the LMSF--HMSF--CMZ sequence (sub-subsection~\ref{sss:env_synthesis}).
Paper~S1 subsection~\ref{S1-ss:Discussion_GammaAlpha} develops the interpretation in terms of grain alignment.
\\ {Alt text: Scatter plot with error bars and two lines.}
}
\label{fig:alpha_vs_gamma}
\end{figure}

\begin{figure}
\centering
\includegraphics[width=0.85\columnwidth]{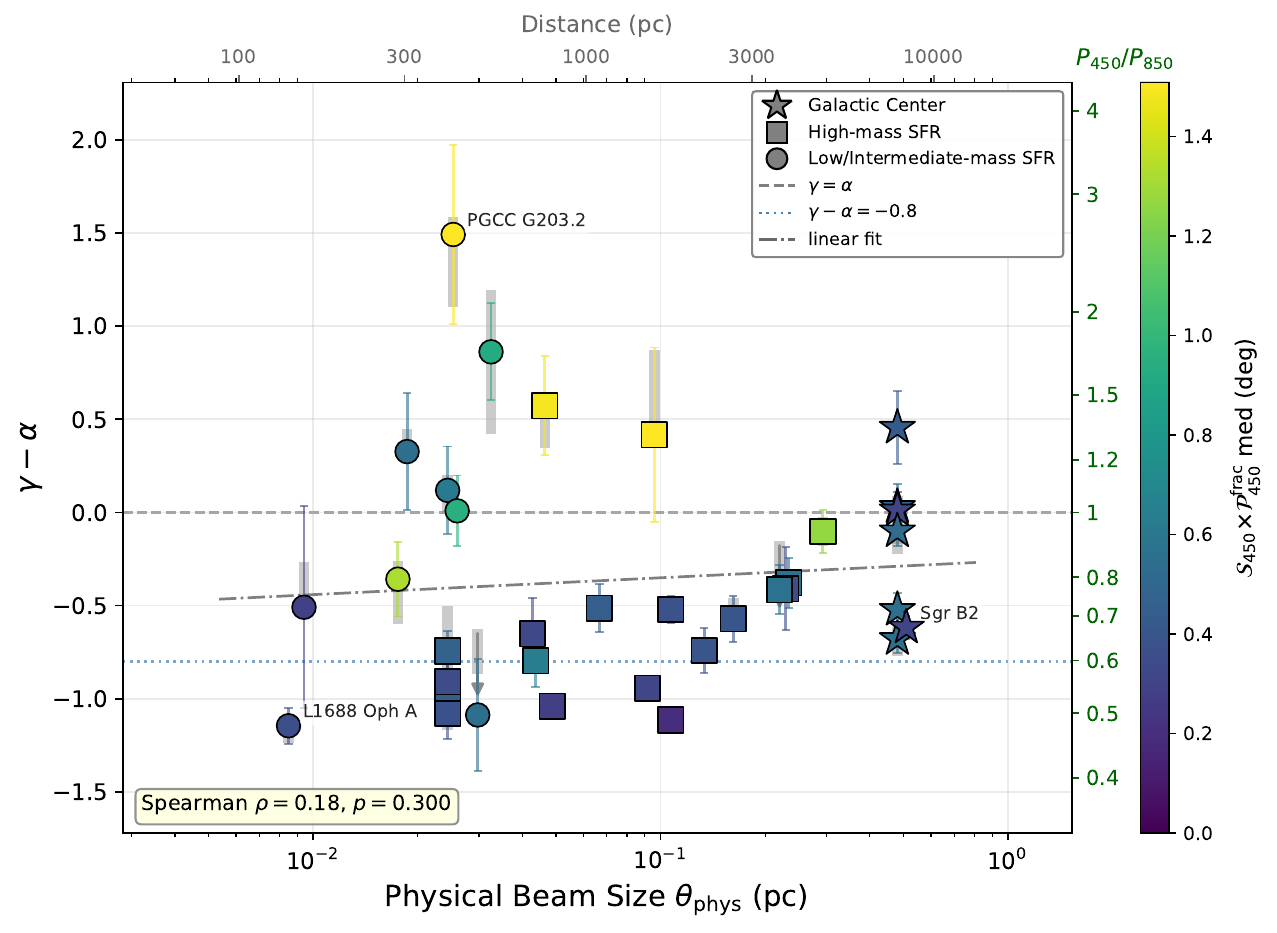}
\caption{Spectral index difference $\gamma - \alpha$ as a function of physical beam size for the 35 sources in the statistical sample.
The physical beam size is computed as $\theta_{\rm beam} \times d / 206265$, where $\theta_{\rm beam} = 12\farcs6$ is the POL-2 beam at 850\,\micron\ and $d$ is the adopted reference distance of each source in parsecs (table~\ref{tab:surv} in Appendix).
Symbol conventions and the $\mathcal{S}_{450} \times \mathcal{P}^{\rm frac}_{450}$ color axis follow figure~\ref{fig:alpha_vs_gamma}, including the distinction between the spatial masks described there.
Vertical error bars are the standard error of the median for the difference, $\sigma(\gamma{-}\alpha) = \sqrt{\sigma_\alpha^2 + \sigma_\gamma^2}$, where each $\sigma$ is the SE of the median for an individual source (Maritz--Jarrett estimator; subsection~\ref{ss:S2_intro_contribution}).
Symbols are the measured medians of individual sources.
The light-gray vertical bars show the range of each median when pixels whose polarized intensity lies below three times its uncertainty are treated as $3\sigma$ limits on $\gamma$ (upper limits for $PI_{450}$, lower limits for $PI_{850}$).
An arrow marks an unbounded end (two sources).
The horizontal dashed line marks $\gamma - \alpha = 0$ ($P_{450}/P_{850} = 1$).
The dotted line marks $\gamma - \alpha = -0.8$ ($P_{450}/P_{850} \approx 0.6$).
The right-hand axis shows the corresponding $P_{450}/P_{850}$ scale.
We do not detect a significant correlation among the measured medians (Spearman $\rho = 0.18$, $p = 0.30$; $R^2 = 0.007$), so physical beam size accounts for less than 1\% of the total variance.
The scatter among sources observed at broadly similar physical scales is far larger than the measurement uncertainties.
The present data neither assign the scatter to intrinsic differences among sources nor bound the residual contribution of spatial resolution, because even the sources observed at broadly similar physical scales span a factor of 2.6 in physical beam size.
Appendix~\ref{app:S2_dist_bias} presents the residual analysis.
\\ {Alt text: Scatter plot with error bars and lines.}
}
\label{fig:beamSize_gammaMalpha}
\end{figure}

\begin{figure*}[htbp]
\centering
\includegraphics[width=0.85\textwidth]{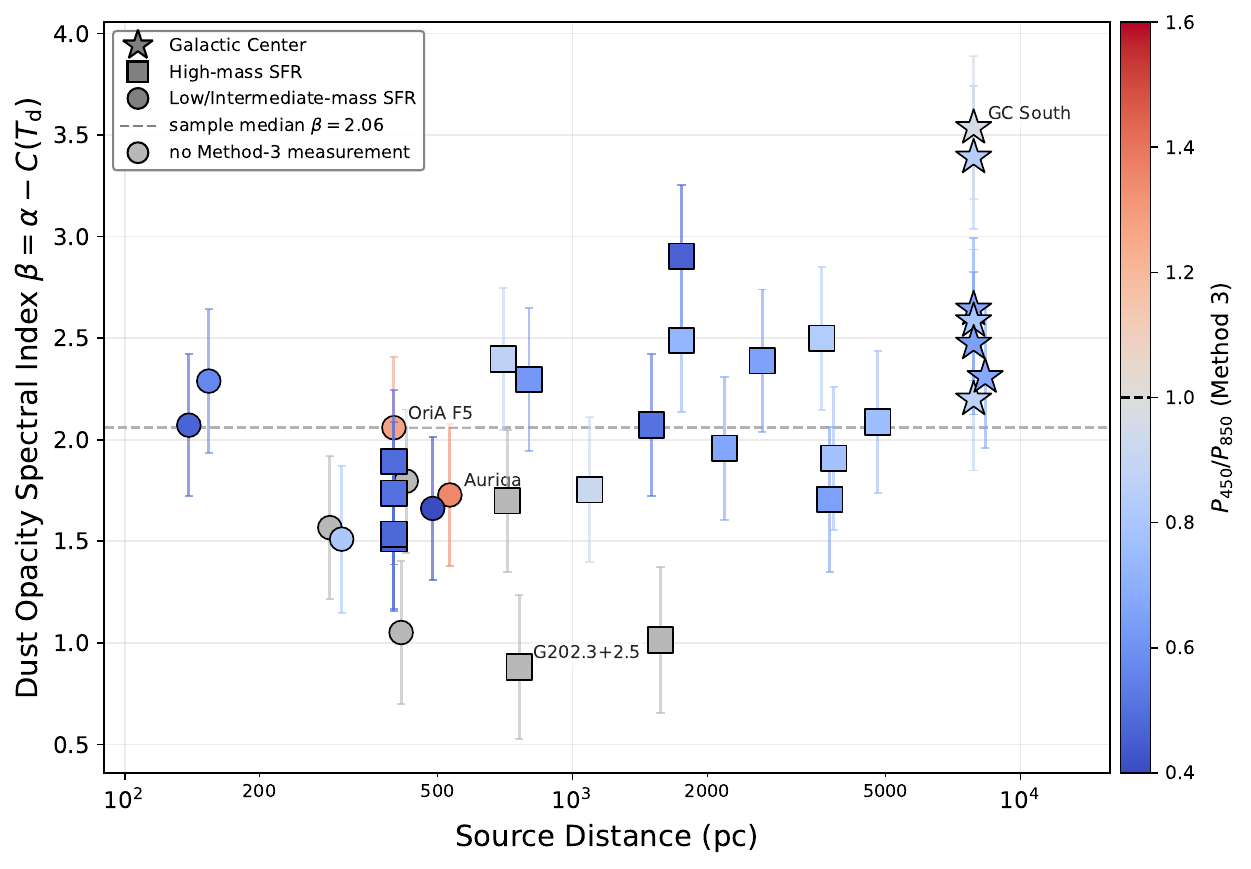}
\caption{Dust opacity spectral index $\beta = \alpha - C(T_{\rm d})$ (equation~\ref{eq:planck_correction}, sub-subsection~\ref{sss:SpIdxFramework}) as a function of source distance for the 35 sources in the statistical sample, color-coded by the polarization ratio $P_{450}/P_{850}$.
Symbol shapes encode environment as in figure~\ref{fig:alpha_vs_gamma}.
Vertical error bars are $\sigma_\beta = \sqrt{\sigma_\alpha^2 + \sigma_{\beta,T}^2}$.
The systematic contribution from the uncertainty in $T_{\rm d}$, $\sigma_{\beta,T} \approx 0.35$ (sub-subsection~\ref{sss:SpIdxFramework}), dominates for all sources.
The values for individual sources are tabulated in table~\ref{tbl:SpIdx_Global} column~(7).
The horizontal dashed line marks the sample median $\beta$.
The dust temperatures for individual sources are the median of the Herschel Gould Belt Survey (HGBS) dust-temperature maps for the Orion~A sub-fields and L1688 Ophiuchus~A, and environment-based defaults for the other sources (LMSF: $T_{\rm d} = 15$~K; HMSF: $T_{\rm d} = 25$~K; CMZ: $T_{\rm d} = 25$~K, with cold IRDCs at $T_{\rm d} = 15$~K and the Brick at $T_{\rm d} = 20$~K).
Most sources have $P_{450}/P_{850} < 1$ (blue colors), consistent with wavelength-dependent depolarization.
The dashed horizontal mark on the colorbar indicates $P_{450}/P_{850} = 1$.
The two reddest symbols, with $P_{450}/P_{850} > 1.2$ (Method~3, subsection~\ref{ss:S2_P_ratio}), are Auriga ($d = 532$~pc) and Orion~A Field~5 ($d = 399$~pc), both in the LMSF environment.
In the sample of 35 sources, the lowest $\beta$ is that of G202.3$+$2.5 ($\beta = 0.88$; figure~\ref{afig:Bmap_G202MO} in Appendix), followed by the Rosette Molecular Cloud ($\beta = 1.01$; figure~\ref{afig:Bmap_RoseMO} in Appendix).
\\ {Alt text: Scatter plot with error bars.}
}
\label{fig:distance_vs_beta}
\end{figure*}

\subsubsection{Global Results from All Sources}
\label{sss:SpIdx_Global}
The source census tests whether a shallower spectrum of polarized emission is confined to a few bright regions or recurs across the observed cloud population.
We compare the source summaries for the 35 regions that meet the spectral-index selection.
 
The preceding analysis of L1688 Ophiuchus~A (sub-subsection~\ref{sss:SpIdx_OphA}) demonstrated that spectral index measurements provide constraints on grain alignment physics that are consistent with, yet complementary to, the polarization fraction ratio approach.
We now extend this investigation to all sources in our sample with sufficient Stokes-$I$ signal-to-noise for spectral index determination.
A negative spectral-index difference occurs in a majority of the selected regions, so the wavelength contrast is not unique to the illustrative Ophiuchus A field.
Its source-to-source variation and its environmental interpretation require the more specific comparisons that follow.

Spectral index analysis requires a stricter data selection than polarization fraction ratio measurements, as discussed in sub-subsection~\ref{sss:SpIdxFramework}.
The pixel selection proceeds in two stages.
A preliminary selection retains pixels with $S/N(I) > 10$ at both wavelengths, $0 < P < 100\%$ at both wavelengths, and a defined uncertainty in polarization angle at both wavelengths.
We evaluate the percentiles quoted below on this preliminary set.
The selection for the scientific analysis then requires, at both wavelengths simultaneously, $S/N(I) > 50$, an intensity exceeding $10\%$ of the mean intensity of the pixels at or above the 99th percentile, and membership of the top decile of the intensity distribution.
We apply no cut on $\alpha$ or $S/N(\gamma)$, and we do not require a minimum $PI/\sigma_{PI}$ at either wavelength.
Avoiding this requirement prevents the selection bias that it would introduce at low signal-to-noise ratios.
We examine how the running medians depend on $S/N(\gamma)$ source by source as a diagnostic (Paper~T sub-subsection~\ref{T-ss:spidx_snrbias}).
Paper~T sub-subsection~\ref{T-ss:spidx_snrbias_sens} quantifies the effect of imposing an $S/N(\gamma)$ cut.
 
For each source, we report $\langle\alpha\rangle$ and $\langle\gamma\rangle$ as the median spectral indices of all selected pixels.
The uncertainties printed with them in table~\ref{tbl:SpIdx_Global} are the Maritz--Jarrett standard errors of the medians, evaluated using the selected pixels in each source.
We use the same estimator when combining these values for individual sources into medians for each environment (table~\ref{tbl:env_4metric}), so that one convention applies throughout.
 
Applying these criteria, we list 36 fields in table~\ref{tbl:SpIdx_Global}, with the number of valid matched-resolution samples in each field ranging from $N_{\rm samp} = 13$ to $N_{\rm samp} = 121$.
To ensure statistically meaningful comparisons, we define the statistical sample as the 35 of the 36 fields that satisfy $N_{\rm samp} \geq 5$ and belong to the reduction set from which this analysis is drawn.
We therefore exclude one tabulated row from the statistical sample: W3~HFS1, whose reduction is published separately by \citet{Kumar2025} and is not part of that set.
Three further fields return $\alpha$ and $\gamma$ but fall below the $N_{\rm samp} \geq 5$ floor and are not tabulated here.
Including them, 39 of the 69 regions yield spectral indices for individual sources.

If the grains contributing to polarization have the same effective spectrum as the full dust column, the spectral indices of total and polarized intensity would coincide.
The distribution of the source values relative to $\gamma=\alpha$ therefore summarizes the observed wavelength contrast in the cloud sample.
 
Figure~\ref{fig:alpha_vs_gamma} shows the correlation between the spectral index $\langle\alpha\rangle$ of Stokes~$I$ and the spectral index $\langle\gamma\rangle$ of polarized intensity for the 35 sources in the statistical sample.
The marker fill colors encode the proxy $\mathcal{S}_{450} \times \mathcal{P}^{\rm frac}_{450}$ for grain alignment efficiency in individual sources (subsection~\ref{ss:S2_SP}).
The plot combines the spectral contrast with this proxy and summarizes the environmental comparisons developed in subsection~\ref{ss:S2_EnvStrat}.
We use this plot to compare broad trends among sources rather than to measure a one-to-one quantitative correlation.
We use a broader mask to estimate $\mathcal{S}\times\mathcal{P}^{\rm frac}$ than to derive the spectral indices.
The two samples are not pixel-matched at the current sensitivity of POL-2.
Across this sample, the median of the Stokes~$I$ indices of individual sources is $\langle\alpha\rangle = 3.33$ (range: 2.06--4.99).
The corresponding median index of polarized intensity is $\langle\gamma\rangle = 2.91$ (range: 1.59--5.45) at the measured medians and $2.85$--$3.04$ with the $3\sigma$ limits at their two extremes.
In table~\ref{tbl:SpIdx_Global}, $\langle\alpha\rangle$ and $\langle\gamma\rangle$ denote medians over the $N_{\rm samp}$ matched pixels within each source.
The two sample values quoted here are the medians of these medians over the 35 sources.
Both levels of aggregation use medians, not means.
Where the brackets do denote a mean, the surrounding text identifies it as one, as for the intensity-weighted mean polarization fractions $\langle P_{\lambda}\rangle_I$ of subsection~\ref{ss:S2_P_ratio} and the arithmetic mean $\langle P \rangle$ of the distribution of pixel values across the survey.
 
With the $C(T_{\rm d})$ correction (sub-subsection~\ref{sss:SpIdxFramework}, equation~\ref{eq:planck_correction}), we obtained a sample median $\beta \approx 2.0$ (range: $\sim$0.9--3.5 across the 35 sources of the statistical sample).
This median is comparable to the $\beta \approx 1.6$--2 found toward dense cores in Perseus~B1 \citep{Sadavoy2013} and the $\beta$ of up to $\sim$2.2 found in individual cold clumps \citep{Juvela2015}.
As discussed above, the $C(T_{\rm d})$ correction cancels from $\gamma - \alpha = \beta_{\rm pol} - \beta$ for a common dust temperature.
The results below on the wavelength dependence of the polarization are therefore independent of the adopted $T_{\rm d}$ within that representation.
 
A majority of sources (24--27 of 35; 69\%--77\%) lie below the $\gamma = \alpha$ line, indicating $\gamma < \alpha$ and therefore $P_{450}/P_{850} < 1$.
A one-sided Wilcoxon signed-rank test on the differences for individual sources rejects a non-negative median at $p = 5.9\times10^{-5}$ and $5.1\times10^{-3}$ at the two ends of the limit range (appendix~\ref{app:S2_secondary}).
The median spectral index difference is $\gamma - \alpha = -0.51$ at the measured medians (range: $-1.14$ to $+1.49$) and $-0.57$ to $-0.45$ with the $3\sigma$ limits at their two extremes.
These values correspond to a median polarization fraction ratio of $P_{450}/P_{850} = 1.90^{-0.51} = 0.72$ (range: 0.48--2.61; $0.69$--$0.75$ with the limits).
This median value is consistent with the polarization-fraction ratios derived in subsection~\ref{ss:S2_P_ratio}.
This consistency confirms that the approaches based on the spectral indices and on the direct ratios yield mutually compatible constraints on the wavelength dependence of the polarization.
 
The wavelength dependence of the polarization we observed, with $P$ decreasing toward shorter wavelengths, is consistent with prior studies of submillimeter polarization spectra in star-forming regions.
\citet{Hildebrand1999} found far-infrared polarization spectra that rise with wavelength in dense cloud cores, in the same sense as ours, and fall in cloud envelopes.
For the cloud envelopes, they proposed a heterogeneous cloud model in which the warmer domains contain the aligned grains.
Subsequently, \citet{VaillancourtMatthews2012} directly compared 350\,\micron\ polarimetry with Hertz at the CSO and 850\,\micron\ polarimetry with the SCUBA polarimeter at the JCMT in 17 Galactic clouds.
These authors did not apply corrections for spatial resolution because the effective beams of the two instruments were both about 20\arcsec.
For positions with $P \geq 3\sigma_P$ at both wavelengths and differences of at most $10\arcdeg$ in angle between the bands, they found a median ratio $P_{850}/P_{350} = 1.7$ with a median absolute deviation of 0.6.
This ratio corresponds to $P_{350}/P_{850} \approx 0.6$, consistent with our $P_{450}/P_{850} = 0.69$--$0.75$.
Our analysis of POL-2 data at two wavelengths extends the earlier measurements of the ratio to a substantially larger and more diverse sample, in which the polarization fraction is lower at the shorter wavelength in most, but not all, sources.
Negative differences between the medians reported for each source predominate, indicating a tendency toward a shallower spectrum of polarized emission in these summaries of individual sources.
The sources with a positive offset show that a single polarization spectrum does not describe every cloud.

The higher polarization fraction at 850 than at 450\,\micron\ in most of our sources is consistent with the rise of a V-shaped polarization spectrum toward long wavelengths.
Earlier measurements of several molecular clouds, taken together, suggested such a spectrum with a minimum near 350\,\micron\ \citep{Vaillancourt2008,Gandilo2016}.
However, our two-band measurements alone do not establish the presence or wavelength of a minimum.
\citet{Gandilo2016} noted that the minimum had been seen directly only in OMC\,1.
They found a relatively flat spectrum without a pronounced minimum in Vela~C, so the V shape cannot be regarded as universal.

Five sources in the statistical sample show $\gamma > \alpha$ at $1\sigma$, corresponding to $P_{450}/P_{850} > 1$.
These are Auriga, Perseus~IC348, and PGCC~G203.21$-$11.20 (LMSF); G202.3$+$2.5 (HMSF); and GC~South (CMZ).
They are the entries flagged with $\oplus$ in table~\ref{tbl:SpIdx_Global} that fall inside the statistical sample.
They span environments from low-mass star-forming regions to the Galactic Center.
Paper~S1 (subsection~\ref{S1-ss:Discussion_GammaAlpha}) discusses the physical interpretation of these exceptions.

To assess potential distance-dependent biases, we examine the relationship between $\gamma - \alpha$ and the physical size of the beam in figure~\ref{fig:beamSize_gammaMalpha} (adopted reference distances in table~\ref{tab:surv} in Appendix).
We quantify monotonic trends using the Spearman rank coefficient and estimate the fraction of variance explained by a simple linear regression of $\gamma - \alpha$ against $\log_{10}(\mathrm{beam~size})$.
For the 35 sources of the statistical sample, we did not detect a significant monotonic association between $\gamma - \alpha$ and the physical size of the beam (Spearman $\rho = 0.18$, $p = 0.30$).
A linear regression against $\log_{10}(\mathrm{beam~size})$ gives a slope of $0.09 \pm 0.19$ per dex with $R^2 = 0.007$.
The source-to-source variation is therefore not well described by a simple systematic distance trend.
The absence of a distance trend is a non-detection, not a demonstration that the observable is insensitive to resolution.
The $95\%$ interval on the slope runs from $-0.30$ to $+0.48$ per dex, so a trend of $0.5$ per dex is not excluded by these data.
 
The scatter among sources observed at comparable physical scales is large.
The 14 sources between 300 and 800~pc span $\gamma - \alpha$ from $-1.09$ to $+1.49$, corresponding to $P_{450}/P_{850}$ from 0.50 to 2.6, a factor of $\sim$5.
These sources are not at a common distance: they run from 305 to 799~pc, a factor of 2.6 in physical beam size.
The comparison therefore shows that a large spread persists among sources observed at broadly similar physical scales.
This comparison does not eliminate resolution as a contributor.
Environment and distance remain confounded across the sample (sub-subsection~\ref{sss:SP_bias}; appendix~\ref{app:S2_dist_bias}).

At a fixed angular resolution, more distant clouds are averaged over larger physical structures.
A trend in the inferred opacity index with distance could therefore reflect changing mixtures of dust temperatures and emitting material rather than an intrinsic variation of grain properties with distance.
 
Figure~\ref{fig:distance_vs_beta} shows the dust opacity spectral index $\beta = \alpha - C(T_{\rm d})$ as a function of source distance for the 35 sources in the statistical sample.
Symbol color indicates the polarization fraction ratio $P_{450}/P_{850}$.
We found a significant correlation between the inferred $\beta$ and source distance (Spearman $\rho = 0.61$, $p = 1.2\times10^{-4}$ for the 35-source statistical sample).
The corresponding correlation for $\gamma - \alpha$ is not significant (figure~\ref{fig:beamSize_gammaMalpha}), and appendix~\ref{app:S2_dist_bias} (figure~\ref{fig:beamSize_beta} in Appendix) compares the two.
We do not interpret this trend as an intrinsic dependence of grain properties on distance.
Rather, increasing distance corresponds to coarser physical resolution and hence more beam averaging and mixing along the line of sight, because the observations have approximately fixed angular resolution.
The same $12\farcs6$ beam subtends $\approx 0.008$~pc in the nearest LMSF fields (core scales and below), $\approx 0.10$~pc in the median HMSF field (the core-to-clump transition), and $\approx 0.48$~pc in the CMZ (clump scales), a span of a factor of $\sim$60.
The contrast is between a detection and a non-detection, and does not establish that $\gamma - \alpha$ is unaffected by spatial averaging.
We therefore account for this distance-dependent resolution trend before assessing environmental differences (appendix~\ref{app:S2_dist_bias}).
The $C(T_{\rm d})$ correction for a common single temperature cancels from $\gamma - \alpha$ (sub-subsection~\ref{sss:SpIdxFramework}), whereas $\beta$ depends on the adopted $T_{\rm d}$.
Our use of the full Planck function therefore does not by itself exclude a residual contribution of temperature to the distance trend.
 
The sample median is $\beta \approx 2.0$ (range: $\sim$0.9--3.5).
This distribution is broadly consistent with expectations for dense molecular cloud material ($\beta \simeq 1.5$--$2.0$; \citealt{Ossenkopf1994, Juvela2015}).
The median lies at the upper end of that standard range, which we attribute to the temperature dependence of the $C(T_{\rm d})$ correction rather than to an anomaly of the dust population itself.
The distribution has a low tail reaching $\beta \approx 0.9$.
Some sources have $\beta > 3$, formally above typical values for interstellar dust.
These high values may result from temperature mixing along the line of sight, calibration uncertainties, or departures from a single-temperature modified blackbody, rather than intrinsic grain properties.
Sources at similar distances scatter: at 300--800~pc, $\beta$ lies between $\sim$0.9 and $\sim$2.4, a spread that exceeds the measurement uncertainties.
These sources are not at a common distance (305--799\,pc, a factor of 2.6 in physical beam size), and $\beta$ is the one quantity in this analysis for which a trend with distance \emph{is} detected.
We therefore do not interpret the spread as purely intrinsic.
 
We did not find a significant environmental dependence of $\beta$ (Kruskal--Wallis $p \approx 0.43$; appendix~\ref{app:S2_dist_bias}).
The inferred opacity slope of the bulk dust changes systematically with the physical scale sampled by the beam, while its attribution to resolution, environment, and temperature structure remains degenerate.
The residual comparison does not establish identical dust properties in the three environments.

The statistical sample spans three broad environment categories as defined in subsection~\ref{ss:S2_envcls}: LMSF regions (9 sources), HMSF regions (19 sources), and the Galactic Center / CMZ (7 sources).
The fraction of sources satisfying $\gamma < \alpha$ at $1\sigma$ varies by environment: 3 of 9 (33\%) for LMSF, 16 of 19 (84\%) for HMSF, and 4 of 7 (57\%) for CMZ.
These fractions express, for each environment, the pooled finding F2 of Paper~S1.
The fractions are not monotonic in the sequence of environments.
Subsection~\ref{ss:S2_EnvStrat} shows that the classes are not ordered along this sequence and that a difference among the classes is not robustly detected, although the HMSF class has the lowest median.
Subsection~\ref{ss:S2_EnvStrat} presents a full synthesis of $\gamma - \alpha$, $R_{\rm p}$, and $\mathcal{S}$ statistics for each environment.
 
The Galactic Center sources show higher $\gamma - \alpha$ values than the sample median, with one of the seven (GC~South) having $\gamma > \alpha$ at $1\sigma$.
The present sample does not establish a difference between the CMZ and disk clouds, with one-sided $p = 0.13$--$0.28$ depending on how the Orion~A and CMZ fields are grouped (sub-subsection~\ref{sss:env_synthesis}).
We caution that the CMZ sources also occupy a narrow range of systematically large beam sizes ($\sim$0.5~pc at $d \approx 8$~kpc), with no overlap in distance with either the LMSF or the HMSF sample.
Environment rank and distance are consequently strongly correlated across the sample ($\rho = 0.83$; appendix~\ref{app:S2_dist_bias}).
We therefore cannot presently distinguish whether the higher CMZ values of $\gamma - \alpha$ reflect environmental differences in grain alignment or the larger physical beam size.
Paper~S1 discusses this question in subsection~\ref{S1-ss:Discussion_GammaAlpha}.

The census of spectral indices examines whether the material producing polarized emission follows the same submillimeter spectrum as the bulk dust.
Across the selected sources, polarized intensity predominantly rises less steeply with frequency than total intensity, with substantial differences among individual clouds.
 
The spectral index analysis of the 35-source statistical sample confirms and extends the results from the approach based on the polarization fraction ratio.
First, $\gamma < \alpha$ is observed in 24--27 of 35 sources (69\%--77\%), with a median $\gamma - \alpha$ between $-0.57$ and $-0.45$ corresponding to $P_{450}/P_{850} = 0.69$--$0.75$.
A one-sided Wilcoxon signed-rank test rejects a non-negative median at $p \le 5.1\times10^{-3}$ at either end of that range.
Second, we do not detect a significant correlation between physical beam size and $\gamma - \alpha$ ($\rho = 0.18$, $p = 0.30$; 35 sources), so we do not find evidence that the result is an artifact of beam dilution.
However, with a fitted slope of $0.09 \pm 0.19$ per dex, these data do not exclude one.
Third, $\beta$ correlates with source distance, and hence with physical beam size ($\rho = 0.61$, $p = 1.2\times10^{-4}$).
In appendix~\ref{app:S2_dist_bias}, we treat this correlation as an effect of distance to be controlled rather than as an intrinsic dependence.
We do not find residual environmental structure in $\beta$ once the trend is removed (Kruskal--Wallis $p \approx 0.43$).
Fourth, the five sources showing $\gamma > \alpha$ at $1\sigma$ are drawn from all three environment classes.
The measured contrast is a constraint on polarized emissivity across the sample.
This contrast motivates models of grain alignment while leaving temperature mixtures, viewing geometry, and the origin of the cloud-to-cloud scatter to be assessed separately.

\begin{longrotatetable}
\begingroup\footnotesize
\begin{deluxetable*}{lrccccccl}
\tablecaption{Spectral Index Statistics for the 35-Source Statistical Sample\label{tbl:SpIdx_Global}}
\tablewidth{0pt}
\tablehead{
\colhead{Source} & \colhead{$N_{\rm samp}$} & \colhead{$\langle\alpha\rangle \pm \mathrm{SE}$} & \colhead{$\langle\gamma\rangle \pm \mathrm{SE}$} & \colhead{$\gamma-\alpha \pm \sigma_{\gamma-\alpha}$} & \colhead{$P_{450}/P_{850}$} & \colhead{$\beta \pm \sigma_\beta$} & \colhead{$\gamma-\alpha$ with} & \colhead{Note} \\
\colhead{} & \colhead{} & \colhead{} & \colhead{} & \colhead{} & \colhead{} & \colhead{} & \colhead{$3\sigma$ limits} & \colhead{}
}
\startdata
\cutinhead{LMSF --- low-mass star-forming}
Auriga                                   &   22 & $2.74 \pm 0.02$ & $3.60 \pm 0.26$ & $+0.86 \pm 0.26$ & 1.74 & $1.73 \pm 0.35$ & $+0.42$ to $+1.20$ & $\oplus$ \\
L1688 Ophiuchus A                        &   39 & $3.33 \pm 0.02$ & $2.19 \pm 0.09$ & $-1.14 \pm 0.10$ & 0.48 & $2.07 \pm 0.35$ & $-1.24$ to $-1.13$ &  \\
L1689 North                              &   13 & $3.30 \pm 0.05$ & $2.79 \pm 0.54$ & $-0.51 \pm 0.54$ & 0.72 & $2.29 \pm 0.35$ & $-0.57$ to $-0.27$ & $\odot$ \\
NGC 1333 (mosaicked field)               &   78 & $2.58 \pm 0.04$ & $2.22 \pm 0.20$ & $-0.36 \pm 0.20$ & 0.79 & $1.57 \pm 0.35$ & $-0.60$ to $-0.26$ & $\ast$ \\
Orion A Field 5                          &   55 & $3.42 \pm 0.01$ & $3.54 \pm 0.24$ & $+0.12 \pm 0.24$ & 1.08 & $2.06 \pm 0.35$ & $-0.01$ to $+0.20$ & $\odot$, $\diamond$ \\
Perseus IC348                            &   25 & $2.52 \pm 0.09$ & $2.85 \pm 0.30$ & $+0.33 \pm 0.31$ & 1.23 & $1.51 \pm 0.36$ & $+0.33$ to $+0.45$ & $\oplus$ \\
PGCC G203.21-11.20 (mosaicked field)     &   16 & $2.06 \pm 0.05$ & $3.56 \pm 0.48$ & $+1.49 \pm 0.48$ & 2.61 & $1.05 \pm 0.35$ & $+1.10$ to $+1.59$ & $\ast$, $\oplus$ \\
Serpens Main (mosaicked field)           &   58 & $2.81 \pm 0.04$ & $2.82 \pm 0.18$ & $+0.01 \pm 0.19$ & 1.01 & $1.80 \pm 0.35$ & $+0.01$ to $+0.01$ & $\ast$, $\odot$ \\
Serpens South 3                          &   14 & $2.67 \pm 0.04$ & $1.59 \pm 0.30$ & $-1.09 \pm 0.30$ & 0.50 & $1.66 \pm 0.35$ & $<-0.63$ & $\ast$ \\
\cutinhead{HMSF --- high-mass star-forming}
Cepheus A                                &   18 & $3.85 \pm 0.03$ & $3.20 \pm 0.19$ & $-0.65 \pm 0.19$ & 0.66 & $2.40 \pm 0.35$ & $-0.76$ to $-0.65$ &  \\
Onsala 2                                 &   28 & $3.36 \pm 0.04$ & $2.98 \pm 0.13$ & $-0.38 \pm 0.14$ & 0.78 & $1.91 \pm 0.35$ & $-0.45$ to $-0.34$ &  \\
DR21                                     &   86 & $3.53 \pm 0.02$ & $2.58 \pm 0.03$ & $-0.94 \pm 0.04$ & 0.55 & $2.07 \pm 0.35$ & $-0.96$ to $-0.94$ &  \\
G202.3+2.5 (mosaicked field)             &   38 & $2.34 \pm 0.05$ & $2.91 \pm 0.26$ & $+0.57 \pm 0.27$ & 1.45 & $0.88 \pm 0.35$ & $+0.35$ to $+0.61$ & $\ast$, $\oplus$ \\
G28.34+0.06  (mosaicked field)           &  121 & $3.10 \pm 0.03$ & $3.00 \pm 0.11$ & $-0.10 \pm 0.12$ & 0.94 & $2.09 \pm 0.35$ & $-0.18$ to $-0.03$ & $\ast$, $\S$, $\odot$ \\
G31.41+0.31                              &   16 & $3.16 \pm 0.07$ & $2.75 \pm 0.21$ & $-0.41 \pm 0.22$ & 0.77 & $1.70 \pm 0.36$ & $-0.43$ to $-0.41$ &  \\
Monoceros R2                             &   73 & $3.75 \pm 0.02$ & $2.71 \pm 0.06$ & $-1.04 \pm 0.07$ & 0.51 & $2.29 \pm 0.35$ & $-1.04$ to $-1.04$ &  \\
NGC 2264 (mosaicked field)               &   95 & $3.15 \pm 0.02$ & $2.36 \pm 0.14$ & $-0.80 \pm 0.14$ & 0.60 & $1.70 \pm 0.35$ & $-0.80$ to $-0.80$ & $\ast$ \\
NGC 6334 North                           &   77 & $4.36 \pm 0.02$ & $3.24 \pm 0.07$ & $-1.11 \pm 0.07$ & 0.49 & $2.91 \pm 0.35$ & $-1.11$ to $-1.09$ &  \\
NGC 6334 South (field 2)                 &   87 & $3.94 \pm 0.02$ & $3.42 \pm 0.07$ & $-0.52 \pm 0.07$ & 0.72 & $2.49 \pm 0.35$ & $-0.57$ to $-0.49$ &  \\
NGC 7538 IRS4                            &   52 & $3.84 \pm 0.01$ & $3.27 \pm 0.12$ & $-0.57 \pm 0.12$ & 0.69 & $2.39 \pm 0.35$ & $-0.57$ to $-0.46$ &  \\
Orion A Field 4                          &   79 & $3.09 \pm 0.06$ & $2.13 \pm 0.04$ & $-0.96 \pm 0.07$ & 0.54 & $1.51 \pm 0.36$ & $-0.96$ to $-0.96$ & $\dagger$ \\
OMC 1 (Orion A Mosaic S1)                &   77 & $3.11 \pm 0.11$ & $2.03 \pm 0.08$ & $-1.08 \pm 0.14$ & 0.50 & $1.53 \pm 0.37$ & $-1.08$ to $-1.08$ & $\dagger$ \\
OMC 2 (Orion A Mosaic S2)                &   83 & $3.25 \pm 0.01$ & $2.50 \pm 0.11$ & $-0.74 \pm 0.11$ & 0.62 & $1.90 \pm 0.35$ & $-1.13$ to $-0.50$ & $\dagger$ \\
OMC 3 (Orion A Mosaic S3)                &   72 & $3.01 \pm 0.01$ & $2.10 \pm 0.10$ & $-0.91 \pm 0.10$ & 0.56 & $1.74 \pm 0.35$ & $-1.17$ to $-0.84$ & $\dagger$ \\
Rosette Molecular Cloud (mosaicked field) &   28 & $2.47 \pm 0.07$ & $2.89 \pm 0.46$ & $+0.42 \pm 0.47$ & 1.31 & $1.01 \pm 0.36$ & $+0.42$ to $+0.87$ & $\ast$, $\odot$ \\
S106                                     &   18 & $3.21 \pm 0.07$ & $2.69 \pm 0.11$ & $-0.51 \pm 0.13$ & 0.72 & $1.75 \pm 0.36$ & $-0.57$ to $-0.45$ &  \\
SDC 13 mm-2                              &   51 & $3.51 \pm 0.03$ & $3.10 \pm 0.13$ & $-0.41 \pm 0.13$ & 0.77 & $2.50 \pm 0.35$ & $<-0.16$ & $\S$ \\
W3 HFS1                                  &   41 & $3.21 \pm 0.04$ & $3.01 \pm 0.05$ & $-0.19 \pm 0.06$ & 0.88 & $1.75 \pm 0.35$ & \nodata & $\natural$ \\
W3 HFS2                                  &   30 & $3.41 \pm 0.04$ & $2.67 \pm 0.11$ & $-0.74 \pm 0.12$ & 0.62 & $1.96 \pm 0.35$ & $-0.78$ to $-0.74$ &  \\
\cutinhead{CMZ --- Central Molecular Zone}
Brick (G0.253+0.016)                     &   72 & $3.94 \pm 0.02$ & $3.42 \pm 0.09$ & $-0.52 \pm 0.09$ & 0.71 & $2.64 \pm 0.35$ & $-0.56$ to $-0.50$ & $\P$ \\
GC Clouds E and F                        &   70 & $4.04 \pm 0.03$ & $4.07 \pm 0.12$ & $+0.03 \pm 0.12$ & 1.02 & $2.58 \pm 0.35$ & $+0.03$ to $+0.09$ & $\odot$ \\
GC Middle (molecular ridge)              &   82 & $3.65 \pm 0.01$ & $3.67 \pm 0.10$ & $+0.01 \pm 0.10$ & 1.01 & $2.20 \pm 0.35$ & $-0.22$ to $+0.05$ & $\odot$, $\diamond$ \\
GC North                                 &   73 & $4.84 \pm 0.03$ & $4.74 \pm 0.07$ & $-0.10 \pm 0.08$ & 0.94 & $3.39 \pm 0.35$ & $-0.17$ to $+0.01$ & $\diamond$ \\
GC South                                 &   63 & $4.99 \pm 0.04$ & $5.45 \pm 0.19$ & $+0.46 \pm 0.20$ & 1.34 & $3.54 \pm 0.35$ & $+0.45$ to $+0.46$ & $\oplus$ \\
Galactic Center 20 km/s cloud            &   61 & $3.93 \pm 0.01$ & $3.25 \pm 0.07$ & $-0.68 \pm 0.08$ & 0.65 & $2.47 \pm 0.35$ & $-0.77$ to $-0.62$ &  \\
Sagittarius B2                           &  101 & $3.76 \pm 0.02$ & $3.15 \pm 0.06$ & $-0.62 \pm 0.06$ & 0.67 & $2.31 \pm 0.35$ & $-0.62$ to $-0.62$ &  \\
\enddata
\tablecomments{
The SE printed in this table is the Maritz--Jarrett standard error of the median, evaluated on the sample of pixels of each source.
Table~\ref{tbl:env_4metric} aggregates these values for individual sources into medians for each environment with the same estimator.
Column~(1): Source name. The entries do not correspond one for one to those of table~\ref{tbl:Pratios23}, because the spectral-index analysis applies additional selection criteria (notes to table~\ref{tbl:Pratios23}). For example, NGC~1333 appears in table~\ref{tbl:Pratios23} as its North and South pointings and in this table as one mosaicked field. Several sources also appear under different forms of the same name, such as ``The Gal. Cent. Clouds E and F'' and ``GC Clouds E and F'' in tables~\ref{tbl:Pratios23} and~\ref{tbl:SpIdx_Global}, respectively.
Column~(2): Number of matched-resolution samples ($12\arcsec$ pairs after harmonization; see Paper~T~subsection~\ref{T-ss:beamval}) passing the quality cuts for the spectral indices in sub-subsection~\ref{sss:SpIdx_Global}.
The quality cuts are applied in two stages, a preliminary selection followed by a science selection.
We do not apply cuts on $\alpha$ or $S/N(\gamma)$, and we do not require a minimum signal-to-noise ratio for the polarized intensities themselves.
Avoiding a threshold on the signal-to-noise ratio of $PI$ prevents the selection bias that it would introduce at low $S/N$.
Columns~(3)--(4): Median spectral indices of Stokes~$I$ and polarized intensity for individual sources.
Column~(5): Spectral-index difference $\gamma-\alpha$ (median minus median) for individual sources, with the propagated 1$\sigma$ joint SE, $\sigma(\gamma-\alpha)=\sqrt{\mathrm{SE}_\alpha^2+\mathrm{SE}_\gamma^2}$.
Column~(6): Implied polarization-fraction ratio $P_{450}/P_{850} = (\nu_{450}/\nu_{850})^{\gamma-\alpha}$, with $\nu_{450}/\nu_{850} = 665.0/349.5 = 1.9027$, quoted as 1.90 in the text.
Column~(7): Dust opacity spectral index $\beta = \alpha - C(T_{\rm d})$ (equation~\ref{eq:planck_correction}) with combined uncertainty $\sigma_\beta = \sqrt{\sigma_\alpha^2 + \sigma_{\beta,T}^2}$, where $\sigma_{\beta,T} \approx 0.35$ from the $T_{\rm d}$-driven systematic (sub-subsection~\ref{sss:SpIdxFramework}) dominates for all sources.
The dust temperature $T_{\rm d}$ is the HGBS median for the Orion~A sub-fields and L1688 Ophiuchus~A and an environment-based default for the other sources (LMSF: 15~K; HMSF: 25~K; CMZ: 25~K), with the exceptions marked in Column~(9).
Column~(8): Range of the median of $\gamma-\alpha$ for each source when pixels below the polarized-intensity noise floor are treated as $3\sigma$ limits.
Pixels in which $PI_{450}$ or $PI_{850}$ lies below three times its uncertainty give only a limit on $\gamma$: an upper limit when $PI_{450}$ is below the floor and a lower limit when $PI_{850}$ is.
The limit is evaluated with the polarized intensity set to $3\sigma$, and a pixel below the floor in both bands does not give a value.
The range brackets the median for each source between the two extreme placements of these limits (upper limits at $-\infty$ or at their $3\sigma$ value, lower limits at their $3\sigma$ value or at $+\infty$).
An entry of the form $<x$ has an unbounded lower end.
$\ast$~Mosaicked field.
$\oplus$~$\gamma>\alpha$ at $1\sigma$ (i.e., $\gamma-\alpha-\sigma>0$, robustly $P_{450}/P_{850}>1$).
$\diamond$~The sign of $\gamma-\alpha$ is not determined by the $3\sigma$ limits (Column~8).
$\odot$~$\gamma\approx\alpha$ at $1\sigma$ (the 1$\sigma$ band crosses $\gamma=\alpha$).
Sources without $\oplus$/$\odot$ markers satisfy $\gamma<\alpha$ at $1\sigma$.
Marker counts quoted in the text are over the statistical sample.
$\natural$~Reduction published separately by \citet{Kumar2025}.
$\S$~Cold infrared dark cloud (the $C(T_{\rm d})$ correction of equation~\ref{eq:planck_correction} uses $T_{\rm d}=15$~K rather than the HMSF default).
$\P$~The Brick (G0.253+0.016) uses $T_{\rm d}=20$~K rather than the CMZ default.
$\dagger$~Orion~A field classified HMSF on the intermediate- to high-mass young stellar content of the field itself (subsection~\ref{ss:S2_envcls}).
Membership of the same parent cloud is not in itself a criterion for classification.
Orion~A Field~4 has this mark because it observes essentially the same gas as OMC~1.
The two centers lie $8\farcm5$ apart within the $\sim$$12\arcmin$ POL-2 field of view, closer than OMC~1 is to OMC~2 ($9\farcm2$), itself HMSF.
The two rows return $\alpha$, $\gamma$, $T_{\rm d}$, and $\beta$ that agree within their uncertainties.
We retain the two fields as separate rows because they are separate pointings, but they are not independent samples of distinct gas.
The HMSF class of table~\ref{tbl:env_4metric} therefore contains both members of a largely co-spatial pair.
Excluding Field~4 moves the sample medians to $\langle\alpha\rangle = 3.35$, $\langle\gamma\rangle = 2.95$ and $\gamma-\alpha = -0.51$ at the measured medians ($-0.57$ to $-0.43$ with the $3\sigma$ limits at their two extremes), with $\gamma<\alpha$ in $23$--$26$ of $34$ sources ($68\%$--$76\%$), leaving the conclusions of subsection~\ref{ss:S2_EnvStrat} unchanged.
The environmental tests there are quoted with and without Field~4 for that reason.
Orion~A Field~5 lies $14\farcm6$ from OMC~1 and remains in LMSF.
CMZ sources are at $d = 7.86$--$8.34$~kpc.
Of the 36 rows printed here, the 35 that do not carry $\natural$ constitute the statistical sample.
The other row is printed for reference and does not enter any statistical analysis.
Fields with $N_{\rm samp}<5$ are not tabulated.
The entry names do not necessarily correspond one for one to tables~\ref{tbl:Pratio1} and~\ref{tbl:Pratios23}: the NGC~6334 complex appears here as NGC~6334 North and NGC~6334 South (field~2) and there as a single entry.
Elsewhere the difference is one of form only: GC~North for The Gal.\ Cent.\ North, Brick (G0.253$+$0.016) for Brick.
The comments to table~\ref{tbl:Pratios23} describe the naming.
}
\end{deluxetable*}
\endgroup
\end{longrotatetable}

\subsubsection{Synthesis: Spectral Index Constraints on Grain Alignment}
\label{sss:SpIdx_Synthesis}
The polarization-fraction ratio $P_{450}/P_{850}$ of subsection~\ref{ss:S2_P_ratio} and the spectral-index difference $\gamma-\alpha$ of this section describe the same wavelength dependence, because the same four measured intensities determine both quantities at a given line of sight.
We found $\gamma<\alpha$ in 24--27 of the 35 sources, with $\widetilde{\gamma-\alpha}$ between $-0.57$ and $-0.45$, and $\widetilde{R_{\rm EW}}=0.69$ for the 30 sample members with an entry in table~\ref{tbl:Pratios23}.
From the relation $P_{450}/P_{850} = (\nu_{450}/\nu_{850})^{\gamma - \alpha} = 1.90^{\gamma - \alpha}$ (equation~\ref{eq:Pratio_SpIdx}), this median $\gamma - \alpha$ predicts $P_{450}/P_{850} = 0.69$--$0.75$.
The predicted ratio agrees, within the statistical uncertainties, with the directly measured polarization fraction ratios, $\tilde{R}_1 \approx R_{\rm Corr} \approx R_{\rm EW} \approx 0.68$--$0.78$ across the three weighting schemes, depending on which sample the median is taken over (sub-subsection~\ref{sss:P_ratio_Global}).
Both summaries indicate that polarized emission generally rises less steeply with frequency than total emission in the selected clouds across the submillimeter wavelength range probed by POL-2.
In wavelength terms, the polarization fraction is generally higher at 850 than at 450\,$\mu$m.
The shallower frequency dependence of the polarized emission is the observational requirement for dust models.
It does not uniquely identify the grain population or alignment process responsible.

We take the agreement between the ratio and spectral-index summaries to show that the observed $P_{450} < P_{850}$ does not depend strongly on how the pixels are selected and aggregated.
Monte Carlo simulations of the dominant quantified source of systematic error (residual zero-level offsets in 450\,$\mu$m Stokes-$Q/U$; Paper~T~subsection~\ref{T-sec:uncertainty}) yield $\sigma(\gamma - \alpha) \lesssim 0.04$, an order of magnitude smaller than the observed signal.
The numerical difference between the two summaries cannot by itself be used as evidence for temperature mixing.
A physical interpretation of any offset first requires the two selections and aggregation rules to be reconciled.

We examine whether the spectral index $\alpha$ of Stokes~$I$ correlates linearly with the polarization-fraction ratio $P_{450}/P_{850}$ implied by $\gamma-\alpha$.
The index $\alpha$ is a direct observable that does not require the $C(T_{\rm d})$ correction.
Any correlation of $\alpha$ with $P_{450}/P_{850}$ directly maps to a correlation of $\beta$ with $P_{450}/P_{850}$ at fixed temperature, because $C(T_{\rm d})$ enters $\beta = \alpha - C(T_{\rm d})$ as an offset that depends only on temperature.
Across 35 sources, the raw implied ratio $1.90^{\gamma-\alpha}$ has a negative Pearson association with $\alpha$ ($r=-0.37$, $p=0.028$), whereas the correlation is weaker on the log-ratio scale ($r=-0.30$, $p=0.077$).
The Spearman rank correlation is not significant ($r_s=-0.22$, $p=0.21$).
This association is not an independent test of a physical link, because the implied ratio is derived from $\gamma-\alpha$, which itself contains $-\alpha$.
For the 29 sources with a measured $R_{\rm EW}$ that are plotted in figure~\ref{fig:env_phase}, we did not detect a Pearson association with $\alpha$ ($r=0.06$, $p=0.75$).
 
The dust opacity spectral index $\beta$ is primarily controlled by the size distribution and composition of the bulk grain population \citep[e.g.,][]{Hildebrand1983, Draine2006}.
In contrast, $P_{450}/P_{850}$ probes the wavelength-dependent properties of the subset of grains that contribute to polarized emission.
The absence of a detected association with the measured ratio is compatible with scenarios in which the grains that are most efficiently aligned may not have exactly the same effective optical properties as the bulk.
Such a difference is expected in radiative alignment torque (RAT) models, in which alignment efficiency depends on grain size relative to the radiation wavelength \citep{Lazarian2007, Andersson2015, HoangLazarian2016}.
A quantitative test of these scenarios requires forward modeling of the grain populations and radiation fields.
We defer this analysis to Paper~S1 (subsection~\ref{S1-ss:Discussion_theory}).

Combining the $\alpha$--$\gamma$ correlation and the beam-size analysis, we identify $\gamma < \alpha$ for the majority of sources as the most robust empirical result.
Of the 35 sources in the statistical sample, 24--27 (69\%--77\%) show $\gamma < \alpha$.
The median difference $\gamma - \alpha$, between $-0.57$ and $-0.45$, corresponds to $P_{450}/P_{850} = 0.69$--$0.75$.
This finding holds across the observational parameters and physical environments of the sample.
The significance of this finding is unaffected by the floor of systematic uncertainty, $\sigma(\gamma - \alpha) \lesssim 0.04$ (Paper~T~subsection~\ref{T-sec:precisionfloor} and table~\ref{T-tbl:systematics}).
 
The sample spans distances from 139~pc (L1688 Ophiuchus~A) to 8340~pc (Sagittarius~B2; figure~\ref{afig:Bmap_GSGRB2} in Appendix), corresponding to physical beam sizes from 0.008~pc to 0.5~pc.
The dust opacity spectral index $\beta = \alpha - C(T_{\rm d})$ ranges from $\sim$0.9 to $\sim$3.5 across the 35 sources of the statistical sample.
This range encompasses values characteristic of both grain growth in dense cores and more pristine ISM-like dust.
The star-forming environments represented include quiescent low-mass cores, intermediate-mass protoclusters, high-mass star-forming regions, and the extreme conditions of the Galactic Center.
Despite this diversity, $\gamma < \alpha$ remains the dominant behavior.
 
The principal finding $\gamma < \alpha$ is robust across distance subsamples (adopted reference distances in table~\ref{tab:surv} in Appendix).
Among nearby sources ($d < 500$\,pc), 8 of 12 sources (67\%) satisfy $\gamma < \alpha$.
Among sources at intermediate distances (500\,pc $< d <$ 2000\,pc), 7 of 10 sources (70\%) satisfy $\gamma < \alpha$.
Among distant sources ($d > 2000$\,pc), 10 of 13 sources (77\%) satisfy $\gamma < \alpha$.
We do not detect a significant correlation between $\gamma - \alpha$ and distance in this sample ($\rho = 0.18$, $p = 0.30$; appendix~\ref{app:S2_dist_bias}).
The relation $\gamma < \alpha$ prevails in every distance bin, which is not what a beam-averaging artifact scaling with distance would naturally produce.
The absence of a dependence on distance weighs against such an artifact without excluding it.
 
The inequality $\gamma < \alpha$ requires that aligned grains have a different spectral energy distribution than the bulk dust population.
Specifically, the polarized intensity must rise less steeply with frequency than the total intensity.
Within the framework of modified blackbody emission, this requirement can be expressed as:
\begin{equation}
\gamma - \alpha = (\beta_{\rm pol} + C(T_{\rm d})) - (\beta + C(T_{\rm d})) = \beta_{\rm pol} - \beta,
\label{eq:gamma_alpha_beta}
\end{equation}
where $\beta_{\rm pol}$ is the effective opacity index of the aligned grain population.
The observed $\gamma - \alpha \approx -0.5$ therefore implies $\beta_{\rm pol} \approx \beta - 0.5$.
In other words, the aligned grains have a shallower opacity spectrum than the bulk population.
Within the common single-temperature representation, $C(T_{\rm d})$ enters $\alpha$ and $\gamma$ as the same additive term and cancels exactly from their difference.
The relation is therefore independent of the adopted $C(T_{\rm d})$ correction.
Paper~S1 (subsection~\ref{S1-ss:Discussion_GammaAlpha}) gives the full treatment of this relation.
 
This simple relation assumes isothermal dust along each line of sight.
In reality, temperature gradients exist within the beam, and RAT models predict that alignment efficiency itself depends on the intensity of the local radiation field \citep{Lazarian2007, HoangLazarian2016}.
Paper~S1 (subsection~4.4) tests a two-temperature model of the line of sight against the HGBS dust temperatures and the \amm\ kinetic temperatures of four fields.
In subsection~\ref{ss:S2_HGBS}, we examine correlations between polarization properties and dust temperature derived from Herschel observations.
 
The shallower opacity spectrum of the aligned grains ($\beta_{\rm pol} < \beta$) could arise from several physical mechanisms: size-dependent alignment efficiency that preferentially aligns larger grains, which have lower $\beta$; compositional differences between aligned and unaligned populations; or wavelength-dependent alignment efficiency.
Distinguishing among these possibilities requires polarimetric observations at multiple wavelengths extending beyond the 450--850\,\micron\ range, combined with detailed modeling of grain alignment.
Paper~S1 (subsection~\ref{S1-ss:Discussion_theory}) develops these physical interpretations.
 
The comparison across sources examines whether the short-wavelength polarization deficit is a recurring property of star-forming clouds.
\textit{This subsection completes the F4 population-level universality test of F1--F3.}
Across the BISTRO {\sc PolSrc} sample, the principal observables of Paper~S1 are recovered as phenomena in the whole sample rather than artifacts from representative sources.
The polarization-fraction ratio $R_p \equiv P_{450}/P_{850}$ has a sample median of $R_{\rm EW} = 0.69$ among the 30 members of the spectral sample with a ratio measurement (subsection~\ref{ss:S2_P_ratio}).
The spectral indices give median $\langle\alpha\rangle = 3.33$ and $\langle\gamma\rangle = 2.85$--$3.04$ with $\gamma<\alpha$ in 24--27/35 of the sources in the statistical subsample (subsection~\ref{ss:S2_SpIdx}).
The polarization-angle dispersion $\mathcal{S}_{850}$ together with the alignment proxy $\mathcal{S}\times\mathcal{P}^{\rm frac}$ are well-defined observables for individual sources across the survey (subsections~\ref{ss:S2_Sdisp}--\ref{ss:S2_SP}).
We thus conclude from the F4 test that F1--F3 are results for the population rather than anecdotal ones.
The F4 test provides the population-scale baseline against which we interpret the F5 test of regime-dependent wavelength response (subsection~\ref{ss:S2_EnvStrat}) and the F6 test of wavelength-dependent polarization--environment associations (subsection~\ref{ss:S2_HGBS}).
The deficit occurs in a majority of the measured sources.
These results set the population constraint that models of aligned dust emission must reproduce.

\subsection{Environmental Stratification: LMSF / HMSF / CMZ Breakdown}
\label{ss:S2_EnvStrat}
Low-mass star-forming clouds, regions hosting more massive young stars, and CMZ clouds differ in radiation fields, dynamical conditions, and the physical scales resolved by these observations.
We ask whether the observed polarization spectrum follows a systematic progression across these environments or instead distinguishes a particular source population.
 
We test the F5 hypothesis of a regime-dependent wavelength response (figure~\ref{fig:S2_test_overview}, table~\ref{tbl:S2_test_index}) in three ways.
We compare the summaries of LMSF, HMSF, and CMZ sources, plot a $(\gamma-\alpha,\,R_{\rm EW})$ phase diagram, and contrast the one-dimensional distributions of the environmental classes.
Together they test whether the F2 and F3 observables vary systematically across star-forming regimes.
F1 enters only through the regional measurements.
 
We use the geometric and alignment proxies $\mathcal{S}$ and $\mathcal{S} \times \mathcal{P}^{\rm frac}$ (subsections~\ref{ss:S2_Sdisp}, \ref{ss:S2_SP}), the three estimators of $P_{450}/P_{850}$ (subsection~\ref{ss:S2_P_ratio}), and the spectral indices $(\alpha, \gamma)$ (subsection~\ref{ss:S2_SpIdx}).
This stratification provides the F5 test of regime-dependent wavelength response in the three-test framework introduced in subsection~\ref{ss:S2_intro_contribution} (figure~\ref{fig:S2_test_overview}, table~\ref{tbl:S2_test_index}), complementing the F4 population-level universality test of subsections~\ref{ss:S2_Sdisp}--\ref{ss:S2_SpIdx} and the F6 test of wavelength-dependent polarization--environment associations in subsection~\ref{ss:S2_HGBS}.
For consistency with the spectral-index analysis, the comparison among environments in this subsection uses the {\sc SpIdx} subsample of table~\ref{tbl:SpIdx_Global} pooled into three broad environment categories: LMSF regions ($N=9$), HMSF regions ($N=19$), and CMZ sources ($N=7$).
The two-dimensional environment-phase diagram (sub-subsection~\ref{sss:env_phase}) and the one-dimensional comparison among environmental classes (sub-subsection~\ref{sss:env_ladder}) below are drawn on this same 35-source sample wherever the quantity displayed permits it.
The median $|\chi_{450}-\chi_{850}|$ within a source is not available for every member.
It therefore sets the sample of the one panel that displays it (figure~\ref{fig:env_ladder}c) and the symbol size in figure~\ref{fig:env_phase}.
We do not use it to restrict $R_{\rm EW}$ or $\gamma-\alpha$.
The difference $\gamma-\alpha$ is defined for all 35, and $R_{\rm EW}$ is tabulated for 30 of them (6 LMSF, 17 HMSF, and 7 CMZ; table~\ref{tbl:Pratios23}).
The five without an entry are mosaicked fields, for which no mosaic-level ratio is available.
Figures~\ref{fig:env_phase} and~\ref{fig:env_ladder}a show 29 of these 30 (6, 16, and 7) and exclude the stacked NGC~2264 ratio, whose input maps differ from the mosaic used for its spectral-index entry.
Of the 29 sources plotted in figure~\ref{fig:env_phase}, 20 carry a $|\chi_{450}-\chi_{850}|$ of their own and nine do not.
The thirteen measurements for regions and groups shown in figure~\ref{fig:env_ladder}c together cover 21 of the 35 sources.
Subsection~\ref{ss:S2_HGBS} separately presents the pixel-by-pixel analysis of HGBS correlations on the HGBS-matched subsample of 9 fields.
 
This subsection is organized as follows.
Sub-subsection~\ref{sss:env_breakdowns} first partitions the {\sc SpIdx} subsample into LMSF, HMSF, and CMZ classes and reports the breakdown by environment of F2, F3, and the $\mathcal{S}$ and $\mathcal{S}\times\mathcal{P}^{\rm frac}$ proxies, together with the regional F1 measurements.
Sub-subsection~\ref{sss:env_phase} projects the measurements for individual sources onto the environment-phase plane $(\gamma-\alpha,\,R_{\rm EW})$, to provide a compact graphical summary of how the environments are distributed.
Sub-subsection~\ref{sss:env_ladder} complements that two-dimensional view with a one-dimensional comparison among the environmental classes in three side-by-side panels mapped onto the F3, F2, and F1 findings of Paper~S1, respectively.
The first two panels show the distributions of $R_{\rm EW}$ and $\gamma-\alpha$ for each environment over the sample.
The third panel does not show a distribution by environment.
Several targets share one regional measurement of $\widetilde{|\Delta\chi|}$.
We therefore treat the angle diagnostic region by region and do not count those shared values as independent measurements of individual sources in the environmental comparison.
After sub-subsection~\ref{sss:env_ladder}, sub-subsection~\ref{sss:env_synthesis} summarizes the results that Paper~S1 uses in its theoretical discussion.
The environmental comparison can identify differences among the sampled populations, but a difference does not necessarily form a monotonic sequence.
In particular, the separation of CMZ and disk sources in distance limits any attribution to environment alone.
 
\subsubsection{Per-Environment F2 and F3 Breakdowns and Regional F1}
\label{sss:env_breakdowns}

For the disk clouds, the relevant distinction is the young stellar population associated with the observed field, while CMZ membership identifies the environment of the Galactic Center.
In Orion A, neighboring pointings sample different parts of the cloud and sometimes overlap substantially, so the local field must be distinguished from the parent complex.
 
The {\sc SpIdx} subsample is the set of sources with $N_{\rm samp} \geq 5$ matched-resolution samples above the Stokes-$I$ $S/N$ threshold used for the spectral-index analysis (table~\ref{tbl:SpIdx_Global}).
The LMSF class ($N=9$) comprises Auriga (figure~\ref{afig:Bmap_Auriga} in Appendix), L1688 Ophiuchus~A (figure~\ref{afig:Bmap_OphiuA} in Appendix), L1689 North (figure~\ref{afig:Bmap_L1689N} in Appendix), NGC~1333 (figure~\ref{afig:Bmap_N133MO} in Appendix; mosaicked field), Orion~A Field~5, Perseus~IC348 (figure~\ref{afig:Bmap_IC348H} in Appendix), PGCC~G203.21$-$11.20 (figure~\ref{afig:Bmap_OrBPMO} in Appendix; mosaicked field), Serpens~Main (figure~\ref{afig:Bmap_SerpMO} in Appendix; mosaicked field), and Serpens South~3 (figure~\ref{afig:Bmap_SerpS3} in Appendix).
The HMSF class ($N=19$) comprises Cepheus~A (figure~\ref{afig:Bmap_CepACe} in Appendix), Onsala~2 (figure~\ref{afig:Bmap_DR15DR} in Appendix), DR21 (figure~\ref{afig:Bmap_DR21DR} in Appendix), G202.3$+$2.5 (figure~\ref{afig:Bmap_G202MO} in Appendix; mosaicked field), G28.34$+$0.06 (figure~\ref{afig:Bmap_G283MO} in Appendix; mosaicked field), G31.41$+$0.31 (figure~\ref{afig:Bmap_G3141G} in Appendix), Monoceros~R2 (figure~\ref{afig:Bmap_MonoR2} in Appendix), NGC~2264 (figure~\ref{afig:Bmap_N226MO} in Appendix; mosaicked field), NGC~6334 North and NGC~6334 South (field~2) (figure~\ref{afig:Bmap_N633MO} in Appendix), NGC~7538~IRS4 (figure~\ref{afig:Bmap_N753I4} in Appendix), OMC~1, OMC~2, and OMC~3 (the Orion~A mosaic-component fields; figures~\ref{afig:Bmap_OriAS1}--\ref{afig:Bmap_OriAS3} in Appendix), Orion~A Field~4, the Rosette Molecular Cloud (figure~\ref{afig:Bmap_RoseMO} in Appendix; mosaicked field), S106 (figure~\ref{afig:Bmap_S106S1} in Appendix), SDC~13 mm-2 (figure~\ref{afig:Bmap_SDC13M} in Appendix), and W3~HFS2 (figure~\ref{afig:Bmap_W3HFS2} in Appendix).
W3~HFS1 (figure~\ref{afig:Bmap_W3HFS1} in Appendix) is printed in table~\ref{tbl:SpIdx_Global} but does not enter the statistical sample, and so is not counted in this class.
We assign the three OMC fields to HMSF on the basis of the intermediate- to high-mass young stellar content of the fields themselves (subsection~\ref{ss:S2_envcls}), which places them with the other HMSF sources despite their proximity.
We assign Orion~A Field~4 to HMSF with them.
Its center lies $8\farcm5$ from that of OMC~1, within one POL-2 field of view.
The two rows return $\alpha$, $\gamma$, $T_{\rm d}$, and $\beta$ that agree within their uncertainties (table~\ref{tbl:SpIdx_Global}).
The Orion~A Field~5 pointing lies further south, does not share pixels with OMC~1, samples low-mass core-forming gas, and remains in LMSF.
The CMZ class ($N=7$) comprises the Galactic Center 20~km\,s$^{-1}$ cloud (figure~\ref{afig:Bmap_GC20KC} in Appendix), the Brick (figure~\ref{afig:Bmap_GCBRCK} in Appendix; G0.253$+$0.016), GC~Clouds~E and~F (figure~\ref{afig:Bmap_GCCLEF} in Appendix), GC~Middle (figure~\ref{afig:Bmap_GCMIDD} in Appendix), GC~North, GC~South, and Sagittarius~B2 (figure~\ref{afig:Bmap_GSGRB2} in Appendix).
With these groups, we compare the observed fields in their local star-forming contexts.
Overlapping pointings and unequal coverage of the ancillary observables remain explicit limitations on the independence and scope of the environmental comparison.

The fraction of sources satisfying $\gamma < \alpha$ at the 1$\sigma$ level (table~\ref{tbl:SpIdx_Global} classification flag absent), the per-environment expression of the pooled finding F2 of Paper~S1, is 3/9 (33\%) for LMSF, 16/19 (84\%) for HMSF, and 4/7 (57\%) for CMZ.
The HMSF subsample exhibits the most consistent behavior.
Only G202.3$+$2.5, which has the highest formal positive $\gamma-\alpha$ in the class, lies on the $\gamma > \alpha$ side at $1\sigma$ within HMSF.
The LMSF subsample is the broadest of the three, spanning $\gamma-\alpha$ from $-1.14$ (L1688 Ophiuchus~A) to $+1.49$ (PGCC~G203.21$-$11.20).
It includes three sources with $\gamma > \alpha$ at $1\sigma$ (Perseus~IC348, Auriga, and PGCC~G203.21$-$11.20), three with $\gamma \approx \alpha$ (L1689 North, Serpens~Main, and Orion~A Field~5), and three with $\gamma < \alpha$ at $1\sigma$.
One of the seven CMZ sources, GC~South, lies on the $\gamma > \alpha$ side at $1\sigma$.
GC~Clouds~E and~F and GC~Middle are consistent with $\gamma \approx \alpha$.
The medians for each environment are $\gamma - \alpha = +0.01$ (LMSF), $-0.65$ (HMSF), and $-0.10$ (CMZ) at the measured medians ($-0.01$ to $+0.01$, $-0.78$ to $-0.50$, and $-0.22$ to $+0.01$ with the $3\sigma$ limits at their two extremes).
The HMSF class, not the LMSF class, carries the most negative median, so the three medians do not order monotonically along the LMSF--HMSF--CMZ sequence.
Sub-subsection~\ref{sss:env_synthesis} shows that the ordered alternative is not detected and that a difference among the classes is not robustly detected.
Figure~\ref{fig:alpha_vs_gamma} compares the distributions directly, and figure~\ref{fig:env_ladder} below shows them as box plots for each environment.

We computed the fraction of sources satisfying $R_{\rm p} = P_{450}/P_{850} < 1$ from table~\ref{tbl:Pratios23} restricted to the {\sc SpIdx} subsample.
All three environment classes are dominated by sources with $R_{\rm p} < 1$, but not every source satisfies this condition.
In the LMSF class, six of the nine sources carry an $R_{\rm EW}$ entry.
L1688 Ophiuchus~A is lowest at $0.47$, and Serpens South~3 follows at $0.59$.
L1689 North and Perseus~IC348 lie at $0.70$ and $0.85$.
Auriga and Orion~A Field~5 lie above unity at $1.28$ and $1.27$ (consistent with the $\gamma>\alpha$ flag for Auriga and the $\gamma\approx\alpha$ flag for Orion~A Field~5 in table~\ref{tbl:SpIdx_Global}).
NGC~1333, PGCC~G203.21$-$11.20 and Serpens~Main (all mosaicked fields) have no entry in table~\ref{tbl:Pratios23}.
In the HMSF class, seventeen of the nineteen sources carry an entry, spanning $R_{\rm EW} = 0.46$ (NGC~6334 North and Orion~A Field~4) to $0.92$ (S106).
The three OMC fields lie at $0.48$--$0.55$, and DR21, Monoceros~R2, NGC~2264, G31.41$+$0.31 and NGC~7538~IRS4 lie between $0.51$ and $0.67$.
G202.3$+$2.5 and the Rosette Molecular Cloud (both mosaicked fields) have no entry.
In the CMZ class all seven sources carry an entry, spanning $0.68$ (Galactic Center 20~km\,s$^{-1}$ cloud) to $0.97$ (GC~South).
Sgr~B2 itself has $R_{\rm EW} = 0.68$.
The medians of the measured $R_{\rm EW}$ are $0.78$ (LMSF, $N=6$), $0.66$ (HMSF, $N=17$) and $0.84$ (CMZ, $N=7$).
The HMSF median includes the stacked NGC~2264 ratio, which figures~\ref{fig:env_phase} and~\ref{fig:env_ladder}a exclude.
These medians are based on unequal and incomplete subsets of each class.
They are not the quantity tested in sub-subsection~\ref{sss:env_synthesis}, which uses the implied $R_p = 1.90^{\gamma-\alpha}$ available for all 35 sources.
Neither quantity establishes an environmental ordering.

Environmental differences in polarization could accompany changes in projected field disorder, grain alignment, or both.
Comparing $\mathcal{S}_{850}$ with $\mathcal{S}_{450} \times \mathcal{P}_{450}^{\rm frac}$ provides a descriptive test of whether these aspects of the observed emission vary together.
 
The $\mathcal{S}_{850}$ statistics for individual sources in table~\ref{tbl:S850stats} (subsection~\ref{ss:S2_Sdisp}) show that HMSF sources with embedded internal heating (the NGC~6334 complex) and the Orion~A complex (Field~4/5 and Mosaic~S1/S2/S3) jointly contribute the pixels with the lowest $\mathcal{S}_{850}$.
This result is consistent with the ordered magnetic fields expected in actively star-forming cores.
LMSF sources occupy an intermediate range, while the CMZ class populates the high-$\mathcal{S}_{850}$ tail.
We attribute the CMZ tail either to increased beam-diluted field tangling at the $\sim 0.5$\,pc physical beam size characteristic of the CMZ or to genuinely enhanced complexity of the field structure along the long lines of sight toward the Galactic Center.
 
The point estimates of $\mathcal{S}_{450} \times \mathcal{P}_{450}^{\rm frac}$, the proxy for grain alignment efficiency (subsection~\ref{ss:S2_SP}), follow $\mathcal{S}_{850}$ only in placing the HMSF class lowest.
The HMSF class has the lowest median point estimate (table~\ref{tbl:env_4metric}), below both the CMZ and the LMSF classes.
The differences among the classes do not by themselves establish differences in the underlying proxy, because most of the pixels entering these point estimates are upper limits (sub-subsection~\ref{sss:SP_global}).
In this tabulation, HMSF sources have the most ordered fields and the lowest point estimate of the alignment proxy, and CMZ sources have the most disordered fields and an intermediate point estimate.
We do not apply a test to either quantity.
The entries for each environment are columns 5 and 6 of half~(a) of table~\ref{tbl:env_4metric}, evaluated on class members with entries in table~\ref{tbl:S850stats} and in table~\ref{tbl:SP450stats} in Appendix.
These quantities provide context for any environmental differences in polarization.
They do not by themselves separate intrinsic alignment efficiency from unresolved structure or from the different physical scales sampled in disk and CMZ clouds.

A change in polarization amplitude does not necessarily imply that the two wavelengths trace different projected $\vec{B}$-field directions.
The angle measurements in the two bands provide a separate constraint on how much the observed field orientation changes while the polarization spectrum varies.
 
Finding F1 of Paper~S1 (similar projected field orientations at the two wavelengths; $N_{\rm samp}=1420$ dual-wavelength pixels, drawn from the 21-source dual-band-angle sample, with typical $|\chi_{450}-\chi_{850}|\lesssim10\degr$) is a pooled statement over the subset with angle measurements at both wavelengths, whereas the environmental stratification below uses the 35-source {\sc SpIdx} sample.
The regional median $\widetilde{|\Delta\chi|}$ is within the adopted $10\degr$ tolerance of F1 for 10 of the 13 regional measurements, which between them cover 21 of the 35 {\sc SpIdx} sources (figure~\ref{fig:env_ladder}c).
Whatever wavelength-dependent mechanism of grain alignment is invoked to explain F2 and F3, it must be compatible with $\vec{B}$-field orientation being set on scales shared by both wavelengths.
The generally small regional angle differences show that a large difference in polarization amplitude can occur without a comparably large change in projected field direction.
One regional measurement (OMC~2/3 $+$ Orion~A Field~5) spans two environment classes.
Given this grouping and the limited coverage of each class, we do not compare the angle distributions between environments with a statistical test.
 
\subsubsection{Environment-Phase Diagram}
\label{sss:env_phase}
\begin{figure}[htbp]
\centering
\includegraphics[width=0.85\textwidth]{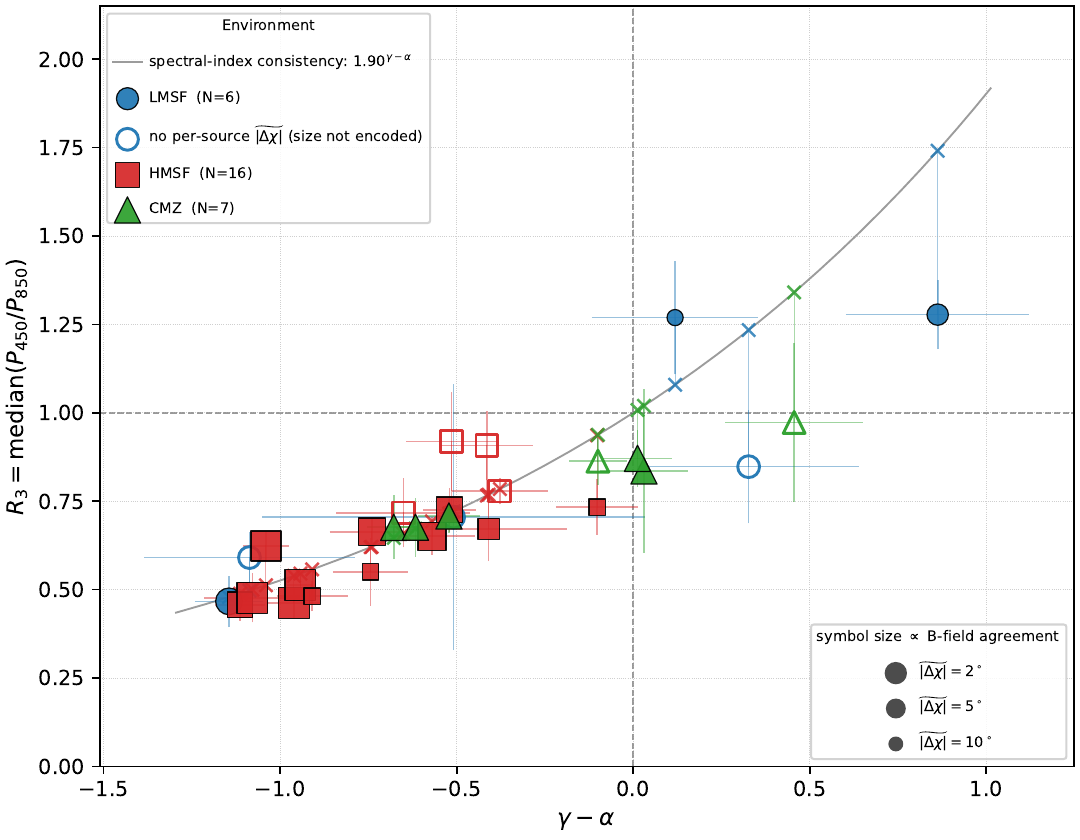}
\caption{
Environment phase diagram in the $(\gamma-\alpha,\,R_{\rm EW})$ plane.
For matched measurements, $\gamma-\alpha<0$ and $R_{\rm EW}<1$ both correspond to $P_{450}<P_{850}$, or a higher polarization fraction at the longer wavelength.
Blue circles (LMSF, $N=6$), red squares (HMSF, $N=16$), and green triangles (CMZ, $N=7$) show 29 of the 30 sources in the 35-source {\sc SpIdx} statistical sample of table~\ref{tbl:SpIdx_Global} that have an $R_{\rm EW}$ entry in table~\ref{tbl:Pratios23}.
The difference $\gamma-\alpha$ is defined for all 35, but the five mosaicked fields without a Method-3 $R_{\rm EW}$ have no position in this plane, and the stacked NGC~2264 ratio is not plotted (sub-subsection~\ref{sss:env_phase}).
The size of filled symbols is inversely proportional to the regional $\widetilde{|\Delta\chi|}$ of Paper~S1 figure~\ref{S1-fig:PA450vsPA850}.
Sources sharing a regional measurement share its size (figure~\ref{fig:env_ladder}c).
Nine sources lack a reliable angle statistic and use open symbols of fixed size.
The availability of an angle statistic changes only the annotation, not inclusion in the plot.
Error bars show only the statistical uncertainties of individual sources: the joint standard error of $\gamma-\alpha$ [table~\ref{tbl:SpIdx_Global}, column~(5)] and the uncertainty of $R_{\rm EW}$ (table~\ref{tbl:Pratios23}).
The systematic floors of Paper~T ($0.02$ in $P_{450}/P_{850}$ and $0.04$ in $\gamma-\alpha$) limit sample medians and are not added.
For an individual source at $PI_{450}=0.68$\,mJy\,arcsec$^{-2}$, the corresponding contributions of the residual zero-level offset in 450\,\micron\ Stokes-$Q/U$ are $\lesssim0.05$ and $\lesssim0.11$ (Paper~T table~\ref{T-tbl:systematics}, note~b).
The gray curve is the relation for individual pixels, $P_{450}/P_{850}=(\nu_{450}/\nu_{850})^{\gamma-\alpha}=1.90^{\gamma-\alpha}$, which follows from the definitions without a Rayleigh--Jeans or isothermal approximation.
It serves as a reference for the estimators used for individual sources, whose different masks and aggregation procedures do not necessarily place them exactly on this curve.
Cross markers ($\times$) show the corresponding derived value, connected to the measured $R_{\rm EW}$ by thin vertical segments (sub-subsection~\ref{sss:env_phase}).
\label{fig:env_phase}
\\ {Alt text: Scatter plot with error bars and a curve.}
}
\end{figure}

We next ask whether clouds forming low-mass stars, clouds hosting intermediate- or high-mass young stars, and CMZ clouds exhibit different short-wavelength polarization deficits.
To compare the polarization properties at the two wavelengths across Galactic environments, we plot the directly measured ratio $R_{\rm EW}$ against the spectral-index difference $\gamma-\alpha$ (figure~\ref{fig:env_phase}).
Of the 30 {\sc SpIdx} sources with an $R_{\rm EW}$ entry in table~\ref{tbl:Pratios23}, 29 are plotted.
The stacked NGC~2264 ratio is omitted (subsection~\ref{ss:S2_EnvStrat}), and the five mosaicked fields without a Method-3 ratio have no ordinate in this plane.
Sources without a $\widetilde{|\Delta\chi|}$ measurement are shown as open symbols.
We exclude them only from the symbol-size encoding, not from the figure.
We plot them because $R_{\rm EW}$ and $\gamma-\alpha$ do not depend on $\widetilde{|\Delta\chi|}$, and because the availability criterion is a threshold on the number of matched pixels that is unrelated to any environmental signal.
The medians for each environment on the full sample are $\widetilde{\gamma-\alpha} = +0.01$ (LMSF), $-0.65$ (HMSF), and $-0.10$ (CMZ).
These medians do not vary monotonically along the LMSF--HMSF--CMZ sequence.
The tests reported in sub-subsection~\ref{sss:env_synthesis} detect neither the ordered alternative nor, robustly, a difference among the classes, although the HMSF class has the lowest median in every treatment of the noise floor.
 
Each point represents a single source, with $R_{\rm EW}$ defined as the median of individual pixel ratios $P_{450}/P_{850}$ and $\gamma-\alpha$ computed from the spectral indices at each pixel (table~\ref{tbl:Pratios23}; table~\ref{tbl:SpIdx_Global}).
Both quantities are ratios or differences of observables measured at the same frequency pair, so the scaling factors that depend on distance and enter absolute quantities do not affect them.
Whether a residual dependence on distance nevertheless survives is an empirical question, addressed in appendix~\ref{app:S2_dist_bias}, where none is detected for $\gamma-\alpha$ although one is detected for $\beta$.
The ratio $R_{\rm EW}$ and the difference $\gamma-\alpha$ both trace the wavelength dependence of polarization analyzed in Paper~S1.
 
The environments differ more in the width of their $\gamma-\alpha$ distributions than in their centers.
The LMSF sources span the widest interval ($\gamma-\alpha \sim -1.1$ to $+1.5$), while the HMSF sample occupies a narrower one ($\gamma-\alpha \sim -1.1$ to $+0.6$) displaced to negative values.
The CMZ sources ($\gamma-\alpha \sim -0.7$ to $+0.4$) cluster near $\gamma-\alpha \approx 0$, with some cases showing $\gamma \gtrsim \alpha$.
However, the medians for each environment are not ordered along the environment sequence.
The Kruskal--Wallis test does not robustly detect a difference among the classes, although the HMSF class has the lowest median (sub-subsection~\ref{sss:env_synthesis}).
 
Despite this variation in spectral-index behavior, 27 out of the 29 plotted sources (28 of the 30 tabulated entries) have $R_{\rm EW} < 1$ when measured directly via the Method~3 estimator (sub-subsection~\ref{sss:Methodology}).
The two exceptions are Auriga ($R_{\rm EW} = 1.28$) and Orion~A Field~5 (1.27), which show $\gamma > \alpha$ ($\oplus$) and $\gamma \approx \alpha$ ($\odot$) at $1\sigma$, respectively (table~\ref{tbl:SpIdx_Global}).
We phrase this in terms of the specific Method~3 quantity $R_{\rm EW}$ here rather than the generic notation $R_{\rm p}$ to avoid mixing, for individual sources, values defined under different weighting schemes.
The fundamental wavelength-dependent polarization signature is therefore preserved across all three environments in most individual sources.
The source-to-source variation in its amplitude and spectral expression is not resolved into an ordered environmental sequence (subsection~\ref{ss:S2_EnvStrat}).
 
In addition, we found offsets between the directly measured $R_{\rm EW}$ values of individual sources and those inferred from $\gamma-\alpha$ via the spectral-index consistency relation, $R_{\rm p} = 1.90^{\gamma-\alpha}$ (equation~\ref{eq:Pratio_SpIdx}).
This relation follows from the definitions of $\alpha$ and $\gamma$ at each pixel (equation~\ref{eq:spectral_indices}) and requires neither a Rayleigh--Jeans nor an isothermal approximation.
The two routes select different pixels: the brightest 5\% with $P \geq 3\,\sigma_P$ in both bands for $R_{\rm EW}$ and the brightest 10\% without that cut for $\gamma-\alpha$ (sub-subsections~\ref{sss:Methodology} and~\ref{sss:SpIdx_Global}).
We also summarize the selected pixels differently in the two routes: we take the median of the pixel ratios in one and the difference between the median indices in the other.
Across the 30 sources with both quantities, $R_{\rm EW}$ lies below the corresponding derived value $1.90^{\gamma-\alpha}$ for 19, and above it for 11.
The largest positive excursions are S106 ($+0.20$), Orion~A Field~5 ($+0.19$), SDC~13 mm-2 ($+0.14$) and Monoceros~R2 ($+0.11$).
The remaining seven (Serpens~South~3, NGC~2264, Cepheus~A, W3~HFS2, the Galactic Center 20~km\,s$^{-1}$ cloud, NGC~6334 South and Sagittarius~B2) lie within $+0.09$ of the locus.
In the Paper~S1 F1 test, the median angle difference $|\chi_{450}-\chi_{850}|$ of Mon~R2 is $12\degr$ before the Stokes-$Q/U$ zero-level offset correction (Paper~S1 appendix~\ref{S1-app:F1_uncorrected}) and $4.0\degr$ after it, below the adopted $10\degr$ tolerance.
This change reflects the sensitivity of the angles to the correction.
The comparison here is made source by source, so the relevant uncertainty is that for an individual source due to the residual 450\,\micron\ Stokes-$Q/U$ zero-level offset (Paper~T subsection~\ref{T-sec:uncertainty}), $\sigma(R_{\rm p}) \lesssim 0.05$ at the sample median of polarized intensity.
The $\lesssim 0.02$ of the Paper~T table~\ref{T-tbl:systematics} limits the survey-level median and is not the appropriate comparison at this level.
The excursions quoted above exceed the uncertainty of 0.05 for an individual source.
We interpret them as a consequence of the different pixel selections and aggregation rules, because equation~(\ref{eq:Pratio_SpIdx}) holds at each pixel for any temperature structure along the line of sight.
The two routes give consistent results for the sample as a whole.
The sample medians $R_{\rm EW} = 0.69 \pm 0.02_{\rm sys} \pm 0.03_{\rm stat}$ and $1.90^{-0.51} = 0.72 \pm 0.02_{\rm sys} \pm 0.05_{\rm stat}$ ($0.69$--$0.75$ with the $3\sigma$ limits at their two extremes) differ by $0.03$, within their combined uncertainty of $0.06$ (Paper~S1 subsection~\ref{S1-ss:SpIdx}).
We therefore use the derived $P_{450}/P_{850}$ values as a complementary systematic check rather than as the primary estimator.
 
The CMZ sources provide a particularly instructive example: even in cases where $\gamma-\alpha \gtrsim 0$, the directly measured ratios remain at or below unity.
At a single pixel, equation~(\ref{eq:Pratio_SpIdx}) gives $R_{\rm p} \geq 1$ whenever $\gamma-\alpha \geq 0$.
The CMZ values below unity can therefore arise from the different selections and aggregation rules of the two routes and from the uncertainties of individual sources.
 
The phase diagram projects the polarization trends at the two wavelengths onto a space in which sources are distinguished by environment.
It does not test the existence of these trends, which are already established for the sample as a whole in sub-subsections~\ref{sss:P_ratio_Global} and \ref{sss:SpIdx_Synthesis}.
The primary result of this projection is therefore the offset between measured and derived ratios, which we interpret as a consequence of how the two routes select and combine pixels.
Taking these results together, we conclude that the fundamental polarization behavior is universal across the sample, and that the present data do not resolve its source-to-source variation into an ordered sequence across the environments sampled here.
 
The Orion~A entries populate both the LMSF class (Field~5) and the HMSF class (OMC~1--3 and Field~4).
The comparison among environmental classes below quantifies how far the high-mass class lies from the other two.
The phase diagram places the HMSF sample at the most negative median spectral-index displacement.
The tests reported below assess both an unordered population difference and the proposed environmental sequence.
Neither is robustly detected, although the HMSF class has the lowest median.
Differences between directly measured and spectrally inferred ratios still require their selections and aggregation rules to be reconciled before a physical interpretation.
 
\subsubsection{Comparison among Environmental Classes}
\label{sss:env_ladder}
The source distributions show whether the polarization spectrum changes steadily from nearby low-mass star-forming gas to the CMZ, or whether one class departs from the other two.
Displaying the ratios and spectral-index differences separately makes that distinction visible without requiring a two-dimensional interpretation.
 
\begin{figure*}[htbp]
\centering
\includegraphics[width=0.95\textwidth]{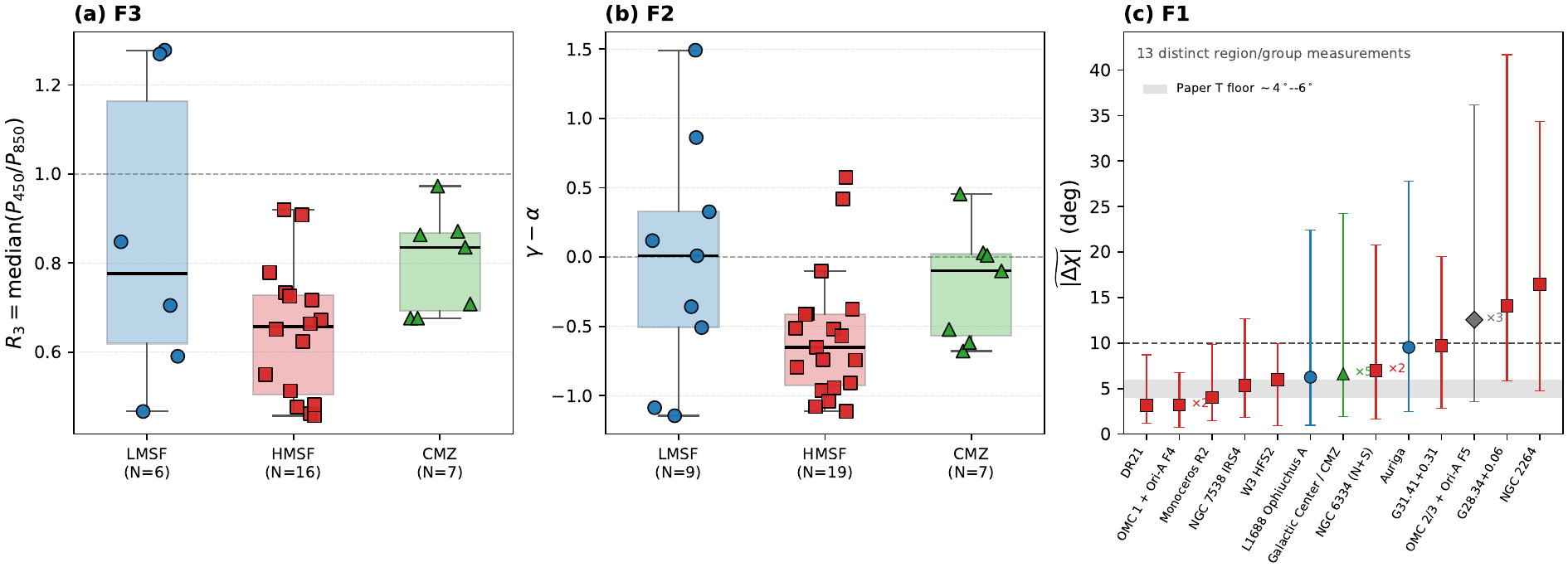}
\caption{
Environmental comparison for the {\sc SpIdx} sample of table~\ref{tbl:SpIdx_Global}: \textbf{(a)} $R_{\rm EW}=\mathrm{median}(P_{450}/P_{850})$ (Paper~S1 F3), \textbf{(b)} $\gamma-\alpha$ (F2), and \textbf{(c)} $\widetilde{|\Delta\chi|}$ (F1).
Panel~(b) uses the full 35-source sample (LMSF $N=9$, HMSF $N=19$, CMZ $N=7$) and panel~(a) 29 of the 30 sources that carry an entry of $R_{\rm EW}$ in table~\ref{tbl:Pratios23} (LMSF $N=6$, HMSF $N=16$, CMZ $N=7$), without the stacked NGC~2264 ratio.
The $N$ in each panel is printed beneath each box.
Panel~(c) shows the thirteen distinct measurements of $\widetilde{|\Delta\chi|}$ for individual regions or groups.
Repeated assignments of the same regional measurement to multiple constituent sources are not counted as separate data points.
The thirteen are distinct measurements, not statistically independent ones.
These measurements share a survey, a calibration, and, for the two Orion entries, a parent cloud.
All thirteen measurements of Paper~S1 cover sources in this sample and are plotted, together accounting for $21$ of the 35.
In panel~(c), light gray shading marks the Paper~T precision floor for region-level median angle differences, $\widetilde{|\Delta\chi|}\sim4\arcdeg$--$6\arcdeg$ (Paper~T equation~\ref{T-eq:precision_floors}), which is the same estimator as the plotted points.
Vertical bars span the 16th--84th percentiles of the distribution of $|\Delta\chi|$ among individual pixels.
These bars are not confidence intervals on the median.
The typical formal error for individual pixels is about $3.5\arcdeg$ (Paper~S1 subsection~\ref{S1-ss:bfield_consistency}).
The $\times n$ annotation gives the number of sources sharing a group measurement.
The gray diamond marks the one Orion measurement whose constituent sources fall in different environment classes (OMC~2/3 $+$ Orion~A Field~5), so that we cannot assign a single class to it.
In panels~(a) and~(b), boxes span the 25th--75th percentiles for each environment, the heavy horizontal bar marks the median, and whiskers extend to the data range of each environment.
Jittered markers for individual sources use blue circles for LMSF, red squares for HMSF, and green triangles for CMZ, matching the convention of figure~\ref{fig:env_phase}.
Panel~(c) uses the same colors where a measurement lies wholly within one class.
Reference lines mark $R_{\rm EW}=1$, $\gamma-\alpha=0$, and the scale $|\Delta\chi|=10\arcdeg$ adopted in Paper~S1 as the physical tolerance (subsection~\ref{S1-ss:bfield_consistency}).
The figure summarizes the central environmental result of this paper.
Panels (a) and (b) show that the three classes overlap in $R_{\rm EW}$ and in $\gamma-\alpha$, so that a significant environmental ordering is not resolved and a difference among the classes is not robustly detected (sub-subsection~\ref{sss:env_synthesis}).
Panel~(c) is not an environmental comparison.
Ten of its thirteen measurements stay below the scale of $10\arcdeg$ adopted as the physical tolerance.
\label{fig:env_ladder}
\\ {Alt text: Three panels, two box plots with points and a plot of values with error bars.}
}
\end{figure*}

The two-dimensional environment-phase diagram of sub-subsection~\ref{sss:env_phase} (figure~\ref{fig:env_phase}) shows where each environment populates the $(\gamma-\alpha,\,R_{\rm EW})$ plane.
Figure~\ref{fig:env_ladder} summarizes the central values and spreads in three one-dimensional panels, corresponding to the principal findings of Paper~S1:
\textbf{(a)}~$R_{\rm EW} = \mathrm{median}(P_{450}/P_{850})$ (Paper~S1 F3);
\textbf{(b)}~$\gamma - \alpha$ (Paper~S1 F2); and
\textbf{(c)}~the individual $\widetilde{|\Delta\chi|}$ measurements (Paper~S1 F1).
In panels~(a) and~(b), box widths span the 25th--75th percentiles for each environment, the heavy horizontal bar marks the median, and whiskers extend to the data range of each environment.
Individual sources are overlaid as jittered markers using the same color and shape conventions as figure~\ref{fig:env_phase} (blue circles for LMSF, red squares for HMSF, green triangles for CMZ).
Panel~(c) shows once each measurement of $\widetilde{|\Delta\chi|}$ for a region or a group, ordered by value.
It does not show boxes or statistics for each environment.
A measurement retains an environment color only if all of its constituent sources belong to the same class.
The reference lines mark $R_{\rm EW} = 1$, $\gamma-\alpha = 0$, and the scale $|\Delta\chi| = 10\arcdeg$ adopted in Paper~S1 as the physical tolerance (subsection~\ref{S1-ss:bfield_consistency}).
 
Three results follow directly from the panels.
First, panel~(c) shows that, in ten of the thirteen measurements that cover this sample, $\widetilde{|\Delta\chi|}$ stays below the scale of $10\arcdeg$ adopted in Paper~S1 as the physical tolerance.
The exceptions are OMC\,2/3 + Orion~A Field~5 ($12.6\arcdeg$), G28.34+0.06 ($14.1\arcdeg$), and NGC~2264 ($16.5\arcdeg$).
The three exceptions occur where polarized intensities are weak and the per-pixel angle uncertainties are larger ($6\arcdeg$--$8\arcdeg$; Paper~S1 subsection~\ref{S1-ss:bfield_consistency}).
The two bands therefore generally show similar projected field orientations where the angles are well constrained.
These angles constrain changes in the net plane-of-sky polarization direction at the adopted resolution but do not establish identical weights along the line of sight or within the beam, so wavelength-dependent depolarization can remain.
Second, panels~(a) and~(b) show that the three environment classes overlap heavily in both $R_{\rm EW}$ and $\gamma-\alpha$.
The HMSF class is compact and displaced to the corner of negative $\gamma-\alpha$ and low $R_{\rm EW}$.
The CMZ class lies at $\gamma-\alpha \approx 0$ with $R_{\rm EW}$ closer to (but still below) unity, and the LMSF class spans both regimes.
The medians for each environment are not ordered, and a difference among the classes is not robustly detected, although the HMSF class has the lowest median (sub-subsection~\ref{sss:env_synthesis}).
Third, the panels together summarize the central environmental result of this paper.
The polarization observables $R_{\rm EW}$ and $\gamma-\alpha$ based on the two wavelengths are present in every class, and this sample does not resolve them into an ordered environmental sequence or robustly into different classes.
 
The paper's constraint on grain-alignment physics rests on the pooled findings F2 and F3 rather than on any environmental contrast, because the ordered environmental comparison returns a null result and a difference among the classes is not robustly detected.
The F1 comparison in panel~(c) is descriptive and does not test environmental independence.
One regional measurement spans two environment classes, and the coverage of each class is limited.
We therefore do not compare the angle distributions between environments with a statistical test, and the absence of an environmental dependence is not established.
The pooled interpretation is unchanged.
We interpret the F2 and F3 trends most simply as an intrinsic modulation by dust physics (size-dependent grain alignment efficiency) rather than as having a geometric origin that would rotate the angles.
The angle panel supplies a regional geometric constraint and is not a separate comparison of box plots among environments.
 
\subsubsection{Synthesis: Environment-Stratified View}
\label{sss:env_synthesis}
Neither an environmental difference among the sampled cloud populations nor the proposed ordered sequence is robustly detected.
The HMSF sources have the most negative median spectral-index difference, whereas the LMSF and CMZ medians lie close to zero.
 
The observational conclusion of section~\ref{sec:S2_results} is that the polarization discrepancy between the two wavelengths exists for the population as a whole, and that the present sample does not resolve it into an ordered sequence across star-forming environments.
Taken together, the environment-stratified view yields three results.
First, the measured $R_{\rm EW}$ is below unity for a majority of the tabulated entries in each class, but the pooled spectral-index finding F2 is not equally expressed in every class.
The LMSF median $\gamma-\alpha$ is close to zero, whereas the HMSF median is the most negative.
Ten of the thirteen region-level angle medians covering this sample lie within the adopted $10\arcdeg$ tolerance [panel~(c) of figure~\ref{fig:env_ladder}].
The relation $\gamma < \alpha$ at $1\sigma$ holds for 3 of 9 LMSF and 16 of 19 HMSF sources and for 4 of 7 CMZ sources.
The relation $R_{\rm EW} < 1$ is preserved for a majority of sources in every class (LMSF 4 of 6, HMSF 17 of 17, CMZ 7 of 7; 28 of the 30 sample members carrying an entry of $R_{\rm EW}$).
Second, the medians for each environment do not vary monotonically along the environment sequence.
The ordered alternative is not detected, and the Kruskal--Wallis test does not robustly detect a difference among the classes (see below).
The interquartile widths are comparable: $0.51$ in $\gamma-\alpha$ for the HMSF class against $0.84$ for the LMSF class, and $0.21$ against $0.54$ in $R_{\rm EW}$ over the 17 and 6 sources of those classes that carry an $R_{\rm EW}$ entry.
However, with $N = 9$ and $7$ in the LMSF and CMZ classes, we do not apply a dispersion test or draw an environmental conclusion from the widths.
Both the most negative individual $\gamma-\alpha$ in the sample ($-1.14$, L1688 Ophiuchus~A) and the most positive ($+1.49$, PGCC~G203.21$-$11.20) lie in the LMSF class.
 
Third, the two diagnostics are not measured at the same level, and the environmental comparison is confined to the level at which it is defined.
The 35-source census carries the $\gamma-\alpha$ and implied $R_p$ statistics tested above, and the 30 sources with an entry in table~\ref{tbl:Pratios23} carry the measured $R_{\rm EW}$ counts.
Panel~(b) of figure~\ref{fig:env_ladder} shows all 35 values of $\gamma-\alpha$, while the phase diagram and panel~(a) show 29 of the 30 ratios, without the NGC~2264 stacked entry.
In contrast, $\widetilde{|\Delta\chi|}$ is measured for regions rather than for individual sources.
Because one of the regional measurements spans two environment classes and the class-specific coverage is limited, we do not define an environment-stratified angle-distribution test and do not quote a distribution of $\widetilde{|\Delta\chi|}$ for each environment.
The F1 diagnostic enters the environmental discussion only through the statement that it stays within the Paper~S1 tolerance in ten of the thirteen regional measurements.
 
In both views, the HMSF median ($-0.65$; $-0.78$ to $-0.50$ with the $3\sigma$ limits at their two extremes) is the lowest, and the medians of LMSF ($+0.01$) and CMZ ($-0.10$) agree well with each other, well within the interquartile width of either class.
The data therefore suggest a lower value in one class rather than a progression across the sequence.
We do not offer a physical interpretation of this difference because we did not robustly detect it.
Paper~S1 (subsection~\ref{S1-ss:Discussion_theory}) develops the physical interpretation of the pooled findings themselves---size-dependent grain alignment in the context of radiative alignment torque theory.
In the exploratory comparison between HMSF and the other two classes, the HMSF sample has the lowest median.
This contrast is not a measurement of the effect of star-formation regime alone, since the populations also differ in physical resolution.

Two different physical pictures are at issue: a steady change of the polarization spectrum along the LMSF--HMSF--CMZ sequence, or a difference associated with one of the sampled populations.
We test these alternatives separately and examine how much the conclusions depend on overlapping or influential fields.
The environmental medians of $\gamma-\alpha$ and $R_p$ summarized above are paired with non-parametric statistical tests on the {\sc SpIdx} subsample (35 sources; LMSF $N=9$, HMSF $N=19$, CMZ $N=7$).
Percentile-bootstrap 95\% confidence intervals on the medians of $\gamma-\alpha$ for each environment are $\widetilde{\gamma-\alpha} = +0.01\,[-1.09,\,+0.86]$ (LMSF), $-0.65\,[-0.91,\,-0.41]$ (HMSF), and $-0.10\,[-0.62,\,+0.03]$ (CMZ) at the measured medians.
The corresponding values for $R_p$ are $\widetilde{R_p} = 1.01\,[0.50,\,1.74]$ (LMSF), $0.66\,[0.56,\,0.77]$ (HMSF), and $0.94\,[0.67,\,1.02]$ (CMZ).
We bootstrapped each quantity separately.
Because all three classes have odd sizes and $R_p = 1.90^{\gamma-\alpha}$ is strictly increasing, the $R_p$ medians and intervals agree with the transforms of the $\gamma-\alpha$ values to the printed precision.
The medians are not ordered along the LMSF--HMSF--CMZ sequence.
The HMSF class has the most negative median, and the LMSF and CMZ classes on either side of it both lie near zero.
 
Two statements follow from this comparison, and they are distinct.
The ordered alternative posed in advance, LMSF $<$ HMSF $<$ CMZ, is not detected.
The Jonckheere--Terpstra test, using 20\,000 permutations of the environment labels on the values of individual sources, returns a one-sided Monte-Carlo permutation $p = 0.57$ at the measured medians and $0.43$--$0.70$ with the $3\sigma$ limits at their two extremes.
Because $R_p = 1.90^{\gamma-\alpha}$ is a strictly increasing transformation of $\gamma-\alpha$, the rank-based test yields the identical test statistic for $R_p$.
The tests on the two quantities are therefore one result on two scales rather than two results.

The Kruskal--Wallis test posed alongside it, which assumes no ordering, does not robustly detect a difference among the three classes.
With pixels below the polarized-intensity noise floor entered as $3\sigma$ limits, it returns $p = 0.056$--$0.078$ with the pointings treated as independent.
When the largely co-spatial Orion~A Field~4 / OMC~1 pair is counted once by excluding Field~4, 34 sources remain and $p = 0.077$--$0.101$.
This is the same two-way accounting applied to the disk-versus-CMZ contrast below.
When the Rosette Molecular Cloud alone is excluded, we obtain $p = 0.026$--$0.041$.
What lowers $p$ is one class rather than a sequence.
The HMSF class lies at $\widetilde{\gamma-\alpha} = -0.65 \pm 0.12$ while LMSF and CMZ lie at $+0.01 \pm 0.36$ and $-0.10 \pm 0.26$ and agree with each other.
The three regimes therefore do not fall along the low-mass-to-CMZ axis for which the ordered test was posed.
The HMSF class retains the lowest median under the stricter accounting.
Counting the overlapping Orion~A gas once therefore neither creates nor removes a detection.
We do not attribute this contrast to star-formation regime alone, because the environmental classes are also a distance sequence (subsection~\ref{ss:S2_EnvStrat}).

We also compared clouds in the Galactic disk with clouds in the Central Molecular Zone, a two-class contrast that does not assume an environmental order.
Treating the 35 sources as independent draws gives a one-sided Mann--Whitney $p = 0.132$.
This result gives an upper bound on the evidence because the sources are not independent draws: the five Orion~A fields are separate pointings within one cloud.
Entering them once as their median gives $p = 0.18$.
A cluster permutation that keeps Orion~A intact gives $p = 0.16$, and keeping the CMZ intact also gives $p = 0.280$.
Excluding one source at a time moves the two-class $p$ between $0.086$ and $0.233$.
Excluding the two Galactic-Center fields with the steepest total-intensity slopes (GC~North and GC~South, $\alpha = 4.84$ and $4.99$), which is where a residual-background or spatial-filtering mismatch between the bands would show up first, moves the CMZ median from $-0.10$ to $-0.52$ and the one-sided $p$ to $0.322$.
Of the contrasts posed in advance, the ordered LMSF--HMSF--CMZ alternative and the disk-versus-CMZ split do not reach significance in the dual-wavelength polarization observables, and the Kruskal--Wallis test across the three classes does so only when the Rosette Molecular Cloud is excluded.
 
The disk-versus-CMZ split is the one two-class contrast whose definition does not depend on star-formation state, which is why it is the one posed.
However, it is not the split with the smallest $p$.
Contrasting the HMSF class against the other two gives a one-sided Mann--Whitney $p = 0.015$ on $\gamma-\alpha$ ($p = 0.045$ after a correction by a factor of three for the three possible splits of one class against the other two).
We chose this split by inspecting the medians and did not pose it in advance.
A one-sided Fisher exact test on the $1\sigma$ fractions (16/19 against 7/16) gives $p = 1.5\times10^{-2}$.
Both tests treat the 35 fields as independent draws, an assumption the disk-versus-CMZ contrast above was shown not to satisfy.
We report these comparisons as exploratory.
The split was not motivated by a physical hypothesis or a prediction of F5, and the $1\sigma$ fractions depend on the signal-to-noise ratio of individual sources as well as on $\gamma-\alpha$.
The Kruskal--Wallis test across all three classes ($p = 0.056$--$0.101$, quoted above; appendix~\ref{app:S2_secondary}) remains the appropriate summary of ``any difference among the three'', and it is the one posed in advance.
We report the HMSF-against-others split only because it names which class carries that difference.
 
The pooled findings are robust under the same checks that the environmental comparison fails.
Excluding one source at a time from the full sample moves the median $\gamma-\alpha$ by at most $0.003$ and the implied $\widetilde{R_p}$ by at most $0.001$, and the fraction of sources with $\gamma<\alpha$ stays between $71\%$ and $74\%$.
The population result is stable under these changes in sample membership, whereas the environmental comparison is sensitive to them.
The ordered environmental hypothesis is not detected, and the Kruskal--Wallis test does not robustly detect a difference among the three classes.
An exploratory HMSF-versus-others comparison points to the HMSF sample as the main contributor.
The disk--CMZ contrast and the physical attribution remain sensitive to source grouping and the confounding of distance with environment.
 
The proposed progression in polarization spectrum from low-mass star-forming gas through massive star-forming regions to the Galactic Center is not detected.
In the measured source summaries, the HMSF population instead has a more negative spectral-index displacement than the LMSF and CMZ populations.
\textit{This subsection completes the F5 test of regime-dependent wavelength response.}
The medians of $\gamma-\alpha$ and $R_p$ for each environment are not ordered along the LMSF--HMSF--CMZ sequence, and the test for the ordered alternative returns $p = 0.43$--$0.70$.
The Kruskal--Wallis test posed alongside it does not robustly separate the classes.
It returns $p = 0.056$--$0.078$ with the pointings treated as independent, $p = 0.077$--$0.101$ with the largely co-spatial Orion~A Field~4 / OMC~1 pair counted once, and $p = 0.026$--$0.041$ without the Rosette Molecular Cloud.
In every treatment the HMSF class has the lowest median ($\widetilde{\gamma-\alpha}$ from $-0.78$ to $-0.50$ at the two ends of the limit range), against LMSF ($-0.01$ to $+0.01$) and CMZ ($-0.22$ to $+0.01$).
A lower median in one class, with mutually consistent medians in the other two, does not support the progression from quiescent low-mass gas to the Galactic Center proposed by F5.
The one two-class contrast defined without reference to star-formation state --- clouds in the Galactic disk against clouds in the Central Molecular Zone --- gives a one-sided $p = 0.132$ with the 35 sources treated as independent, $p = 0.18$ once the five Orion~A pointings are counted as a single cloud, and a value between $0.086$ and $0.23$ when any single source is excluded.
The fractions of sources with $\gamma<\alpha$ at $1\sigma$ in each environment (LMSF 3/9, HMSF 16/19, CMZ 4/7) count $1\sigma$ detections and therefore depend on the signal-to-noise ratio of individual sources as well as on $\gamma-\alpha$ itself.
The only contrast reported on them is the post-hoc HMSF-versus-others comparison noted above, which is exploratory.
The regional median $\widetilde{|\Delta\chi|}$ is within the adopted $10\degr$ tolerance of F1 for 10 of the 13 regional measurements covering the sample.
Because one of them spans two classes and the class-specific coverage is limited, we do not stratify F1 by environment.
 
F5 therefore returns a null result on the ordered hypothesis it posed, and it does not robustly detect an unordered difference among the classes.
The HMSF class has the lowest median in every treatment, but we cannot establish a monotonic trend of the F2 and F3 observables across star-forming environments.
Determining whether the HMSF displacement arises from star-forming conditions, spatial averaging, or differences in the sampled material requires a comparison at matched physical scales.
Our constraint on the physics of grain alignment therefore rests on the pooled findings.
In the F6 test of subsection~\ref{ss:S2_HGBS}, we examine whether cloud conditions and the degree of $\vec{B}$-field disorder drive the local wavelength-dependent response.
We test this at the level of individual pixels, where the number of samples is far larger.

The spectrum of polarized emission and the coherence of the projected $\vec{B}$-field do not necessarily respond to the environment of a cloud in the same way.
A joint tabulation of the spectral diagnostics, the angle dispersion, and the proxy for grain alignment efficiency makes clear which aspects of the environmental comparison are directly measured and which remain incomplete.
Table~\ref{tbl:env_4metric} consolidates the F5 census of the four metrics.
The table presents the medians for each environment on both the {\sc SpIdx} subsample shared with Paper~S1 and the broader 43-source SP450 census scope of this paper.
None of the four diagnostics orders monotonically across the three environments.
Among the environments, the spectral observables $\gamma-\alpha$ and $R_p$ reach their most negative and lowest medians in the HMSF class, not at either end of the LMSF--HMSF--CMZ sequence.
Neither an ordered progression nor, robustly, a difference among the classes is detected.
The angle dispersion $\mathcal{S}_{850}$ and the alignment proxy $\mathcal{S}_{450} \times \mathcal{P}_{450}^{\rm frac}$ do not order along the sequence either.
The \emph{Integrated observational signature} column therefore describes each class in its own row without asserting a relation between them.
The physical interpretation of the pooled findings is developed in Paper~S1 (subsection~\ref{S1-ss:Discussion_theory}).
The angle dispersion and the alignment proxy provide complementary context, but this census alone does not identify a single environmental parameter that controls the polarization response.

\begin{longrotatetable}
\begin{deluxetable*}{lcccccl}
\tablecaption{Environmental Census of Dual-Wavelength Polarization Observables across LMSF, HMSF, and CMZ Regimes\label{tbl:env_4metric}}
\tablewidth{0pt}
\tablehead{
  \colhead{Environment} &
  \colhead{$N_{\rm src}$} &
  \colhead{$\widetilde{\gamma-\alpha}$} &
  \colhead{$\widetilde{R_p}$} &
  \colhead{$\widetilde{\mathcal{S}_{850}}$} &
  \colhead{$\widetilde{\mathcal{S}_{450} \times \mathcal{P}_{450}^{\rm frac}}$} &
  \colhead{Integrated observational signature} \\
  \colhead{} & \colhead{} & \colhead{} & \colhead{} & \colhead{(\arcdeg)} & \colhead{(\arcdeg)} & \colhead{}
}
\startdata
\multicolumn{7}{c}{\textit{(a) {\sc SpIdx} subsample (Paper~S1-common; table~\ref{tbl:SpIdx_Global})}} \\
\hline
  LMSF & 9\tablenotemark{a} & $+0.01 \pm 0.36$ & $1.01 \pm 0.23$ & $31.4 \pm 1.28$ & $0.54 \pm 0.09$ & \parbox[t]{8.0cm}{\raggedright Widest spread of the three classes in $\gamma-\alpha$, spanning both signs, with a median indistinguishable from zero, that is, consistent with equal polarization fractions at 450 and 850\,\micron\ at the level of the class median.\strut} \\
  HMSF & 19\tablenotemark{a} & $-0.65 \pm 0.12$ & $0.66 \pm 0.05$ & $28.9 \pm 1.27$ & $0.38 \pm 0.03$ & \parbox[t]{8.0cm}{\raggedright Most homogeneous class, with the most negative median spectral-index difference (corresponding to a higher polarization fraction at 850 than at 450\,\micron) and the lowest median $\mathcal{S}_{850}$.\strut} \\
  CMZ & 7 & $-0.10 \pm 0.26$ & $0.94 \pm 0.14$ & $34.9 \pm 1.42$ & $0.50 \pm 0.07$ & \parbox[t]{8.0cm}{\raggedright Median spectral-index difference near zero ($\gamma-\alpha \approx 0$, $R_p \approx 1$), consistent with equal polarization fractions at 450 and 850\,\micron\ at the level of the class median; elevated median $\mathcal{S}_{850}$ and an intermediate point estimate of $\mathcal{S} \times \mathcal{P}^{\rm frac}$.\strut} \\
\hline
\multicolumn{7}{c}{\textit{(b) 43-source SP450 broader survey scope (this paper)}} \\
\hline
  LMSF & 18 & \nodata & \nodata & $33.7 \pm 1.02$ & $0.62 \pm 0.04$ & \parbox[t]{8.0cm}{\raggedright Same rank of the class medians as in half~(a)\strut} \\
  HMSF & 18 & \nodata & \nodata & $28.9 \pm 1.08$ & $0.38 \pm 0.02$ & \parbox[t]{8.0cm}{\raggedright Same rank of the class medians as in half~(a)\strut} \\
  CMZ & 7 & \nodata & \nodata & $34.9 \pm 1.42$ & $0.50 \pm 0.07$ & \parbox[t]{8.0cm}{\raggedright Same rank of the class medians as in half~(a)\strut} \\
\enddata
\tablecomments{
Medians of the four principal dual-wavelength polarization observables in each environment of the LMSF, HMSF, and CMZ stratification.
Half~(a) reports the {\sc SpIdx} subsample of table~\ref{tbl:SpIdx_Global} (LMSF~$N=9$, HMSF~$N=19$, CMZ~$N=7$), the basis on which $\gamma-\alpha$ and $R_p$ are defined for each source.
Columns 5 and 6 of half~(a) are evaluated on the subset of sources in each class that have a matched entry in the $\mathcal{S}_{850}$ and $\mathcal{S}\times\mathcal{P}^{\rm frac}$ tables for the whole survey ($N=7$ for LMSF, $N=15$ for HMSF, $N=7$ for CMZ).
Unlike columns 3 and 4, these columns cannot be evaluated on all 35 sources, because the survey-wide tables do not contain an entry for a mosaicked field.
Of the seven mosaicked fields in the sample, only NGC~1333 and NGC~2264 have component pointings from which a mean weighted by pixel count could be formed (NGC~1333 is so represented here, by $33.277$ and $0.658$).
Half~(b) reports the broader 43-source SP450 census scope (LMSF~$N=18$, HMSF~$N=18$, CMZ~$N=7$).
This sample underlies table~\ref{tbl:S850stats} and table~\ref{tbl:SP450stats} in Appendix, and includes four sources not printed in the latter.
L1689~South, L483, Perseus~B1, and M16 have 450\,\micron\ pixel counts of 22, 27, 41, and 48 and so fall below the $N \geq 50$ printing threshold of table~\ref{tbl:SP450stats} in Appendix, while each of them still has a valid median.
Their 850\,\micron\ counts are larger and all four are printed in table~\ref{tbl:S850stats}.
The medians of $\mathcal{S}_{850}$ and $\mathcal{S}_{450} \times \mathcal{P}_{450}^{\rm frac}$ for individual sources on this broader basis rank the three classes in the same order as in half~(a).
Values are reported as median~$\pm$~SE-of-median, the SE evaluated by the Maritz--Jarrett estimator.
Column~3 ($\widetilde{\gamma-\alpha}$): spectral-index differences for individual sources from table~\ref{tbl:SpIdx_Global} column~(5).
Column~4 ($\widetilde{R_p}$): implied $P_{450}/P_{850} = 1.90^{\gamma-\alpha}$ for each source.
Column~5 ($\widetilde{\mathcal{S}_{850}}$): median dispersion of polarization angles at 850\,\micron\ for individual sources, from table~\ref{tbl:S850stats} (subsection~\ref{ss:S2_Sdisp}).
Column~6 ($\widetilde{\mathcal{S}_{450} \times \mathcal{P}_{450}^{\rm frac}}$): median of the proxy for grain alignment efficiency at 450\,\micron\ for individual sources, from table~\ref{tbl:SP450stats} in Appendix.
We use the same $I/\sigma_I > 30$ pixel selection as elsewhere in this paper, so the medians for each environment and the tabulated medians for individual sources are directly comparable.
These medians for individual sources are point estimates over all selected debiased values.
For 35 of the 39 tabulated sources, more than half of those pixels are upper limits.
Medians of the point estimates for individual sources therefore do not by themselves establish differences in the underlying proxy.
\tablenotetext{a}{$N_{\rm src}$ is the class size of columns 3 and 4.
Columns 5 and 6 rest on the smaller matched subsets stated in the note above.}
Column~7 (Integrated observational signature) empirically summarizes the observed pattern across columns~3--6 within each environment.
None of the four diagnostics orders monotonically across LMSF, HMSF, and CMZ on this sample.
Among the environments, the spectral observables $\gamma-\alpha$ and $R_p$ take their most negative and lowest median, respectively, in the HMSF class rather than in the LMSF class.
Neither an ordered progression across the sequence nor, robustly, a difference among the classes is detected (sub-subsection~\ref{sss:env_synthesis}).
The angle dispersion $\mathcal{S}_{850}$ and the alignment proxy $\mathcal{S}_{450} \times \mathcal{P}_{450}^{\rm frac}$ likewise do not order along the sequence.
The table is therefore a census of the three classes rather than evidence for an environmental control parameter.
}
\end{deluxetable*}
\end{longrotatetable}

\subsection{HGBS 9-Field Per-Pixel Correlation Analysis}
\label{ss:S2_HGBS}
Dust shielding and internal heating can change the conditions for grain alignment within a molecular cloud.
Herschel maps of column density and temperature allow us to compare these conditions with the two polarization bands and ask whether the same local behavior recurs in different clouds.
 
This subsection presents the F6 test of wavelength-dependent polarization--environment associations in the three-test framework (figure~\ref{fig:S2_test_overview}, table~\ref{tbl:S2_test_index}).
The test correlates polarization observables with \textit{Herschel}-derived $N_{\rm H}$ and $T_d$ for individual pixels in 9 fields with HGBS coverage.
We compare results with and without OMC\,1 and within individual clouds (sub-subsection~\ref{sss:S2_source_aware_correlation}).
These analyses reveal a hierarchical structure that depends on the band, rather than a single universal law at the scale of individual pixels.
We calculate these correlations for individual pixels using polarization fractions and angle-dispersion functions measured at the matched POL-2 resolution, starting from the common parent mask defined by Stokes~$I$ in both bands.
We interpret them using the Stokes-$Q,U$ noise model and the limits on map comparability established in Paper~T, and assess the environmental trends relative to those technical limits.
 
To investigate the physical drivers of the observed polarization properties in individual pixels, we require independent constraints on dust temperature and column density that cannot be reliably determined from the POL-2 observations in the two bands alone.
We therefore restrict this subsection to the nine HGBS-matched star-forming regions [L1688 Ophiuchus~A (figure~\ref{afig:Bmap_OphiuA} in Appendix), Taurus~B211/213 (figure~\ref{afig:Bmap_B211MO} in Appendix), NGC~1333 (figure~\ref{afig:Bmap_N133MO} in Appendix), Perseus~B1 (figure~\ref{afig:Bmap_PersB1} in Appendix), Perseus~IC348 (figure~\ref{afig:Bmap_IC348H} in Appendix), OMC~1, OMC~2, and OMC~3 (figures~\ref{afig:Bmap_OriAS1}--\ref{afig:Bmap_OriAS3} in Appendix), and Serpens~Main (figure~\ref{afig:Bmap_SerpMO} in Appendix)].
For these regions we report the $T_{\rm d}$--$P$, $N_{\rm H}$--$P$, and $P$--$\mathcal{S}$ correlation analysis for individual pixels.
This analysis provides the basis for the pooled synthesis for each environment in subsection~\ref{ss:S2_EnvStrat}.
This subsample overlaps with that of Paper~S1 (subsection~\ref{S1-ss:samples}).
Here we extend the analysis by tabulating the full matrix of Spearman correlations for each band and reporting results with and without OMC~1.
The comparison separates trends within individual clouds from differences between cloud populations and from the strongly irradiated OMC\,1 field.
It constrains the local expression of the polarization spectrum without treating every pooled correlation as a universal alignment law.

\subsubsection{Dust Temperature and Column Density from Herschel Observations}
\label{sss:S2_HGBSparams}
 
The two POL-2 bands at 450 and 850\,$\mu$m do not adequately separate temperature from column density in a fit to the spectral energy distribution.
Ground-based bolometer arrays also filter structures larger than $\sim$300\arcsec\ when removing atmospheric emission (Paper~T section~\ref{T-sec:obs}).
We therefore use maps of column density ($N_{\rm H}^{\rm HGBS}$) and dust temperature ($T_{\rm d}^{\rm HGBS}$) from the HGBS \citep{Andre2010}, which were derived from Herschel images at multiple wavelengths with a large spatial dynamic range.
Hereafter, $N_{\rm H}^{\rm HGBS}$ denotes the hydrogen-equivalent column density inferred from Herschel dust emission, assuming a standard gas-to-dust ratio.
The HGBS provides uniform coverage of nearby ($d < 500$\,pc) star-forming regions with angular resolutions of 18\farcs2 for column density and 36\arcsec\ for dust temperature.
The maps of temperature and column density were derived by fitting a modified blackbody to the spectral energy distribution at each pixel, assuming a dust opacity law $\kappa_{\nu} \propto \nu^{\beta}$ with $\beta = 2$ \citep{Hildebrand1983}.
This assumption is broadly consistent with the distribution of $\beta$ derived from our 450/850\,$\mu$m data (median $\beta \approx 2.0$ for the 35-source statistical sample; table~\ref{tbl:SpIdx_Global}).
This consistency supports the use of parameters derived from the HGBS in our correlation analysis.
Our $\beta$ values depend on the adopted $T_{\rm d}$, which is the HGBS temperature for the Orion~A sub-fields and L1688 Ophiuchus~A and a representative value for the other sources (sub-subsection~\ref{sss:SpIdxFramework}).
We therefore do not regard this consistency as an independent confirmation of $\beta = 2$.
For the correlation analysis presented below, we use nine POL-2 fields with HGBS coverage, comprising approximately $4.5 \times 10^5$ pixels with valid measurements.
The central 90\% of the pixels with HGBS coverage span $T_{\rm d}^{\rm HGBS} = 13$--$30$\,K and $N_{\rm H}^{\rm HGBS} = 10^{21.0}$--$10^{22.4}$\,cm$^{-2}$.
The full ranges are 10--52\,K and $10^{20.3}$--$10^{23.9}$\,cm$^{-2}$.
 
We reprojected the HGBS maps onto the POL-2 pixel grid by bilinear interpolation.
Regridding does not change their angular resolutions ($18\farcs2$ for $N_{\rm H}^{\rm HGBS}$ and $36\arcsec$ for $T_{\rm d}^{\rm HGBS}$), which remain coarser than the $12\farcs6$ resolution common to the two POL-2 bands.
We did not convolve the maps to a common beam or correct their different spatial frequency coverage.
Herschel preserves all angular scales, while POL-2 filters structures larger than $\sim$300\arcsec.
This procedure avoids artifacts from poorly constrained corrections, but the comparison remains subject to the differences in resolution and spatial filtering.
 
The extended emission filtered out by POL-2 likely originates from foreground material or diffuse envelopes surrounding the molecular clouds.
In contrast, the correlation analysis in this work focuses on the innermost, densest regions where the emission is dominated by compact structures well within the spatial filtering scale of POL-2.
We expect the impact of this mismatch to be small, because the polarization diagnostics analyzed here are dominated by structures on scales comparable to or smaller than the POL-2 beam.
The principal polarimetric quantities in this work---polarization fraction $P = PI/I$ and spectral-index differences---are ratio quantities.
However, the filtered emission does not necessarily affect their numerators and denominators in proportion, so we do not assume that its effect cancels in them.
For our most distant sources ($d \sim 5$--8\,kpc), the maximum recoverable scale of $\sim$300\arcsec\ corresponds to 7--12\,pc, far exceeding the physical extent of the dense structures analyzed.
For nearby sources ($d \sim 140$\,pc), the filtering scale of $\sim$0.2\,pc primarily removes diffuse envelope emission that lies outside the high-column-density regions selected by our Stokes-$I$ signal-to-noise criteria.
However, we note that we have not quantitatively evaluated this effect.
 
The HGBS map of dust temperature in the L1688 Ophiuchus complex reveals a clear gradient from $T_{\rm d}^{\rm HGBS} \sim 28$\,K at the externally heated western edge of the complex to $\sim$12 \, K in the shielded interiors of the dense cores.
This range of $\sim$15\,K within a single cloud complex demonstrates that our observations sample diverse thermal environments, from regions dominated by external radiation to well-shielded interiors of filaments.
The temperature gradient is relevant to RAT theory, which predicts that alignment efficiency depends on the local radiation field strength relative to gas density (see Paper~S1, subsection~\ref{S1-ss:Discussion_theory}).
The anticorrelation between temperature and column density reflects dust cooling in shielded regions, as the interstellar radiation field is attenuated in denser material.
Across the fields with HGBS coverage in our sample, a power-law fit yields $T_{\rm d}^{\rm HGBS} \propto (N_{\rm H}^{\rm HGBS})^{-0.4}$, consistent with previous HGBS studies of nearby star-forming regions \citep[e.g.,][]{Konyves2015,Marsh2016}.

\subsubsection{Construction of the Correlation Data Set}
\label{sss:S2_correlation_sample}
A trend across many cloud pixels can reflect either changes within a cloud or differences between the clouds themselves.
To distinguish these possibilities, we retain the field identity of each matched line of sight while comparing its polarization with the Herschel-derived column density and temperature.
 
We combine high-quality measurements for individual pixels from the nine fields with HGBS coverage to compare polarization properties across these environments.
This approach follows earlier multi-region comparisons of molecular clouds \citep{Lada2010,Dib2019,Sullivan2021}.
 
We select only pixels satisfying the Method~3 Stokes-$I$ detection mask (sub-subsection~\ref{sss:Methodology}): $I/\sigma_I > 50$, Stokes~$I$ above the 95th percentile of the field, and the absolute brightness floor, all at both wavelengths.
As in Method~3, we do not impose a threshold on the signal-to-noise ratio of polarized intensity when defining the parent sample.
This retains intrinsically low-polarization pixels in the parent sample.
Each correlation then uses the subset that meets the detection requirements of its observables (subsection~\ref{ss:S2_noise_floor}).
For the subset with HGBS coverage, we additionally require valid values of $T_{\rm d}^{\rm HGBS}$ and $N_{\rm H}^{\rm HGBS}$ at each pixel position.
The resulting correlation data set comprises the {\sc DualBand} sample (introduced in Paper~S1 subsection~\ref{S1-ss:samples}) of $N_{\rm total} = 417$ sampled pixel pairs spanning column densities from $N_{\rm H}^{\rm HGBS} \sim 10^{22}$ to $10^{24}$\,cm$^{-2}$ and dust temperatures from $T_{\rm d}^{\rm HGBS} \sim 10$ to 40\,K.
Both bands are sampled on the $12\arcsec$ grid at the common $12\farcs6$ effective resolution (Paper~T~subsection~\ref{T-ss:beamval}), so adjacent samples can remain correlated.
The significance levels calculated from individual pixels are therefore nominal.
The per-cloud tests and cluster block-bootstrap analyses address some, but not all, of these dependencies.
For each pixel, the diagnostic quantities available are: $\mathcal{P}_{450}^{\rm frac}$, $\mathcal{P}_{850}^{\rm frac}$, $\mathcal{S}_{450}$, $\mathcal{S}_{850}$, $\mathcal{S}_{450} \times \mathcal{P}_{450}^{\rm frac}$, $\mathcal{S}_{850} \times \mathcal{P}_{850}^{\rm frac}$, $T_{\rm d}^{\rm HGBS}$, and $N_{\rm H}^{\rm HGBS}$.
All correlation coefficients reported below are Spearman rank coefficients ($r_s$), which are robust against outliers and do not assume linear relationships.
We adopt a significance threshold of $p < 0.01$ for claiming a statistically significant correlation.
 
\begin{figure*}[htbp]
\centering
\includegraphics{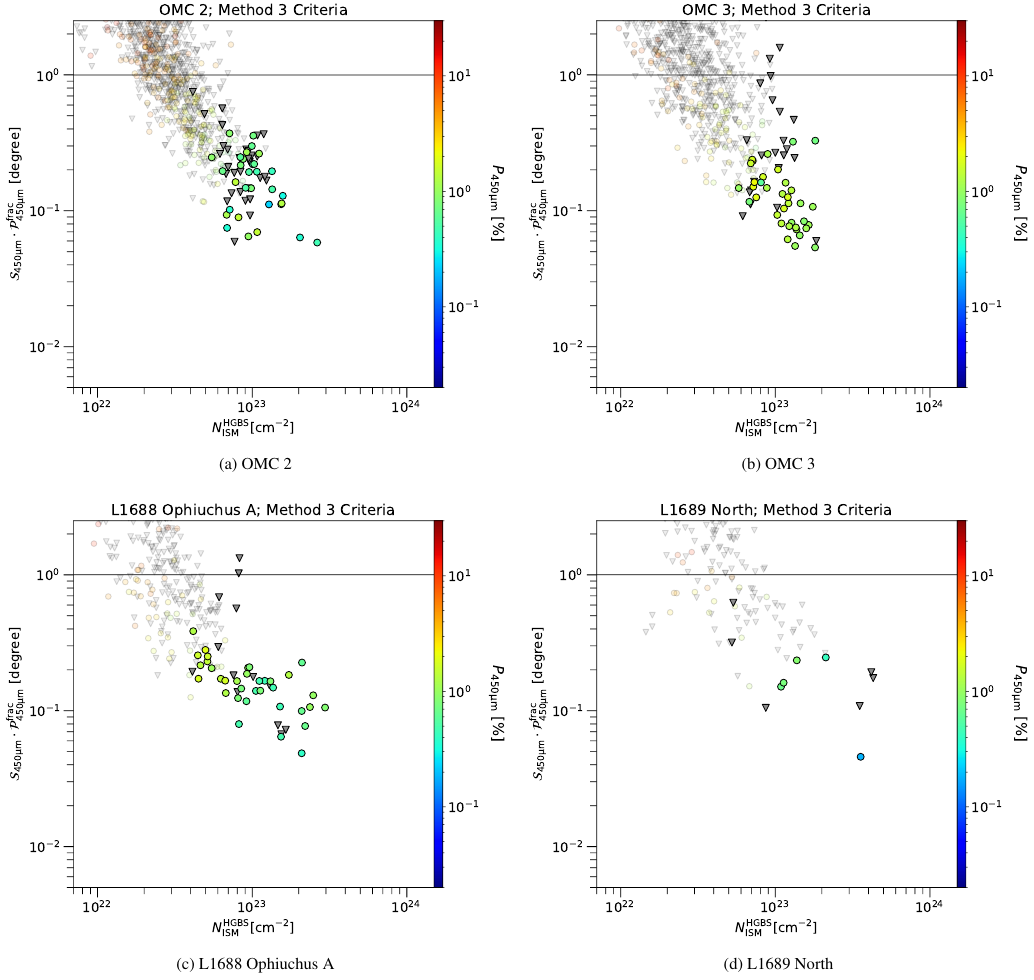}
\caption{
The $\mathcal{S}_{450} \times \mathcal{P}_{450}^{\rm frac}$ diagnostic versus column density for four representative star-forming regions from Orion~A (a, b) and Ophiuchus (c, d), color-coded by $P_{450}$.
Opaque symbols satisfy the Method~3 selection (sub-subsection~\ref{sss:Methodology}).
Translucent symbols show the remaining pixels for context.
Gray downward triangles mark $3\sigma$ upper limits on $\mathcal{S}\times\mathcal{P}^{\rm frac}$, defined when $\mathcal{S}$ is undetected or $P_{450}<3\sigma_P$.
Each triangle is placed with the undetected quantity at its $3\sigma$ value.
The limits of most translucent pixels lie above the plotted range.
Among selected pixels with a detected product, the product decreases with $N_{\rm H}$ in OMC~3 (Spearman $r_s=-0.54$, $p=8\times10^{-4}$) and Ophiuchus~A ($r_s=-0.65$, $p=2\times10^{-5}$).
We do not detect a corresponding rank trend in OMC~2 ($r_s=-0.26$, $p=0.16$).
Only five selected pixels have a detected product in L1689 North.
These correlations describe the detected subsets and do not include the upper limits.
\\ {Alt text: Four scatter plots in two rows.}
}
\label{fig:SPvsN_individual}
\end{figure*}

We first examine $\mathcal{S}_{450} \times \mathcal{P}_{450}^{\rm frac}$ within individual clouds, before combining the fields.
Figure~\ref{fig:SPvsN_individual} illustrates this diagnostic for four representative regions spanning two molecular cloud complexes.
Among the Method~3 pixels with a detected product, $\mathcal{S} \times \mathcal{P}^{\rm frac}$ decreases with column density in OMC~3 and Ophiuchus~A.
We do not detect a corresponding rank trend in OMC~2, and only five selected pixels have a detected product in L1689 North.
These rank correlations describe the detected subsets and do not include the upper limits.
Because the product contains $\mathcal{P}^{\rm frac}$ by definition, its correspondence with the $P_{450}$ color coding is not independent evidence.
Four regions cannot establish whether such a decrease is general, because of their limited sample sizes and potential biases specific to each source.
The combined data set allows us to quantify these correlations across the sampled star-forming environments.
This data set samples the bright material detected in both POL-2 bands and covered by the Herschel maps.
The field labels and the selection requirements for each observable are essential for deciding whether a pooled relation describes the interiors of the clouds or differences among the sampled regions.

\begin{figure}[htbp]
\centering
\includegraphics[width=0.95\textwidth]{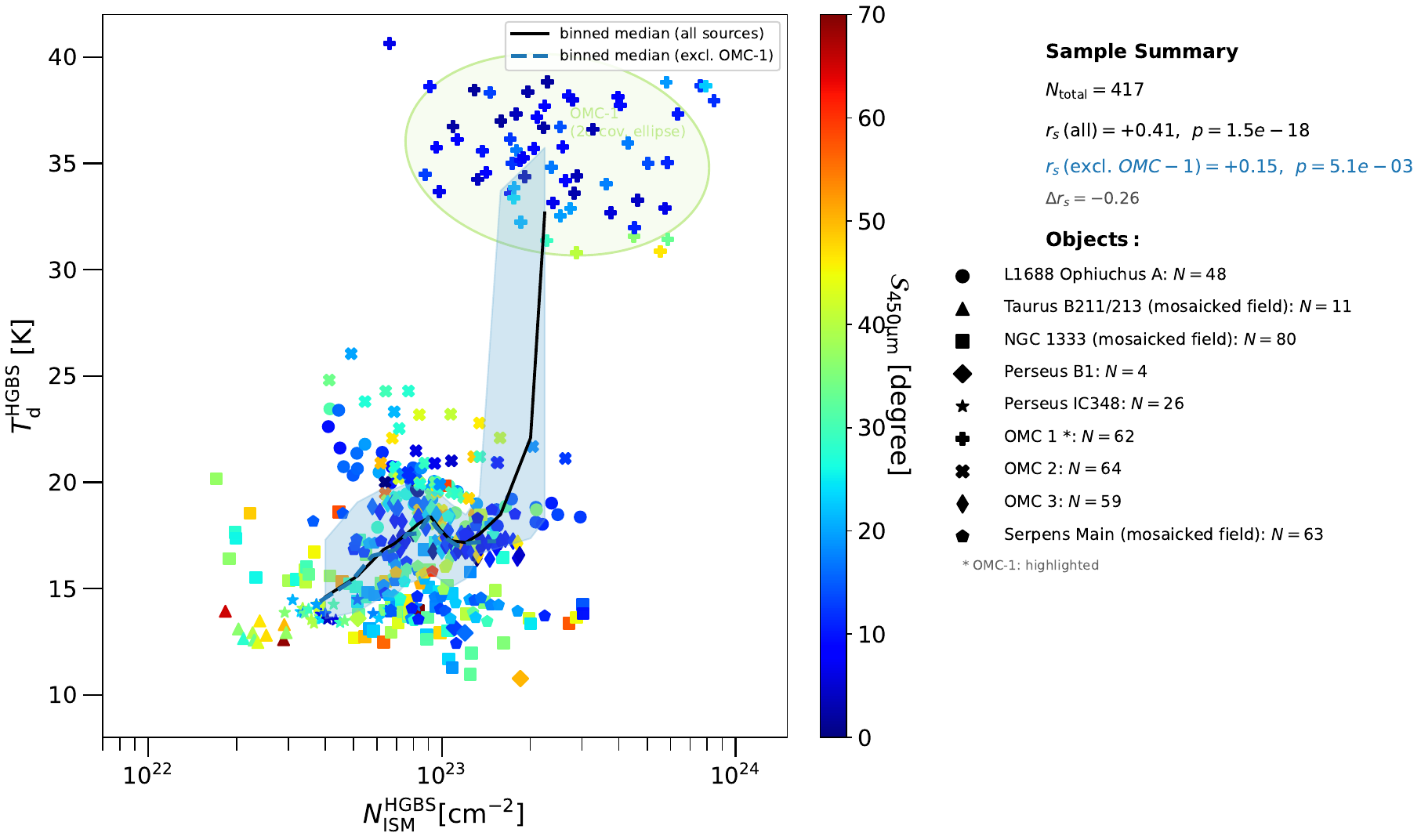}
\caption{
Distribution of the Method~3 sample in the $N_{\rm H}^{\rm HGBS}$--$T_{\rm d}^{\rm HGBS}$ parameter space, color-coded by the dispersion $\mathcal{S}_{450}$ of polarization angles at 450\,$\mu$m.
The positive correlation across the full HGBS subsample ($r_s = +0.41$, $N = 417$, $p = 1.5 \times 10^{-18}$) reflects the dominance of internally heated sources (OMC regions) in the high-column-density regime.
Excluding OMC\,1 weakens the trend to $r_s = +0.15$ ($N = 355$, $p = 5.1 \times 10^{-3}$).
Pixels with low $|\mathcal{S}|$ (coherent $\vec{B}$-field) preferentially occupy the region of high $N_{\rm H}$ and high $T_{\rm d}$.
This non-independence of $N_{\rm H}$ and $T_{\rm d}$ must be considered when interpreting correlations with either parameter.
\\ {Alt text: Scatter plot with a line and a shaded band.}
}
\label{fig:NvsT}
\end{figure}

\subsubsection{The $N_{\rm H}$--$T_{\rm d}$ Parameter Space}
\label{sss:S2_N_T_space}
Shielded material within a cloud can be colder at higher column density, whereas internally heated clouds can contain both warm and dense regions.
The relation between $N_{\rm H}$ and $T_{\rm d}$ therefore helps distinguish local shielding from differences in the heating environments sampled by the survey.
 
Before examining polarization correlations, we characterize the environmental parameter space sampled by our data.
Figure~\ref{fig:NvsT} shows the distribution of pixels in the $N_{\rm H}^{\rm HGBS}$--$T_{\rm d}^{\rm HGBS}$ plane, color-coded by the 450\,$\mu$m polarization angle dispersion $\mathcal{S}_{450}$.
 
The data exhibit a statistically significant positive correlation between column density and temperature over the full HGBS subsample: $r_s(N_{\rm H}, T_{\rm d}) = +0.41$ ($N = 417$, $p = 1.5 \times 10^{-18}$).
We attribute this positive correlation to the dominance of internally heated sources in the high-column-density regime.
The OMC~1, 2, and~3 regions, which contain embedded YSOs, contribute the majority of pixels with high $N_{\rm H}$ and high $T_{\rm d}$.
Excluding OMC\,1 weakens the correlation to $r_s = +0.15$ ($N = 355$, $p = 5.1 \times 10^{-3}$).
OMC\,1 therefore contributes most of the tail toward high $N_{\rm H}$ and high $T_{\rm d}$, while the remaining eight HGBS regions show a much weaker pooled correlation.
The dynamic range spans approximately two orders of magnitude in $N_{\rm H}^{\rm HGBS}$ and a factor of four in $T_{\rm d}^{\rm HGBS}$.
 
Because $N_{\rm H}^{\rm HGBS}$ and $T_{\rm d}^{\rm HGBS}$ are not independent in this sample, a correlation with one may partly reflect a correlation with the other.
We therefore cannot readily separate their effects using this data set.
This limitation applies throughout the following analysis, particularly to the $P$--$T_{\rm d}$ relation (sub-subsection~\ref{sss:S2_P_vs_T}).
 
The color coding in figure~\ref{fig:NvsT} reveals that pixels with low $\mathcal{S}$ (coherent $\vec{B}$-field) preferentially occupy the corner of the parameter space at high $N_{\rm H}$ and high $T_{\rm d}$.
This corner corresponds to actively star-forming cores with internal heating sources.
The OMC~1, 2, and~3 regions dominate the high-temperature regime ($T_{\rm d}^{\rm HGBS} \gtrsim 25$\,K), while Perseus, Taurus, and parts of Ophiuchus occupy the low-temperature regime ($T_{\rm d}^{\rm HGBS} \lesssim 20$\,K).
This environmental diversity, driven primarily by the presence or absence of embedded heating sources, enables meaningful tests of the predicted dependence of grain alignment on temperature, although the degeneracy between $N_{\rm H}$ and $T_{\rm d}$ must be carefully considered.
The positive pooled relation is not the behavior typical of the individual clouds, whose valid correlations within each field are negative.
The pooled relation reflects the combination of regions with different ranges of heating and column density, so correlations of polarization with either variable require interpretation cloud by cloud.

\begin{figure}[htbp]
\centering
\includegraphics{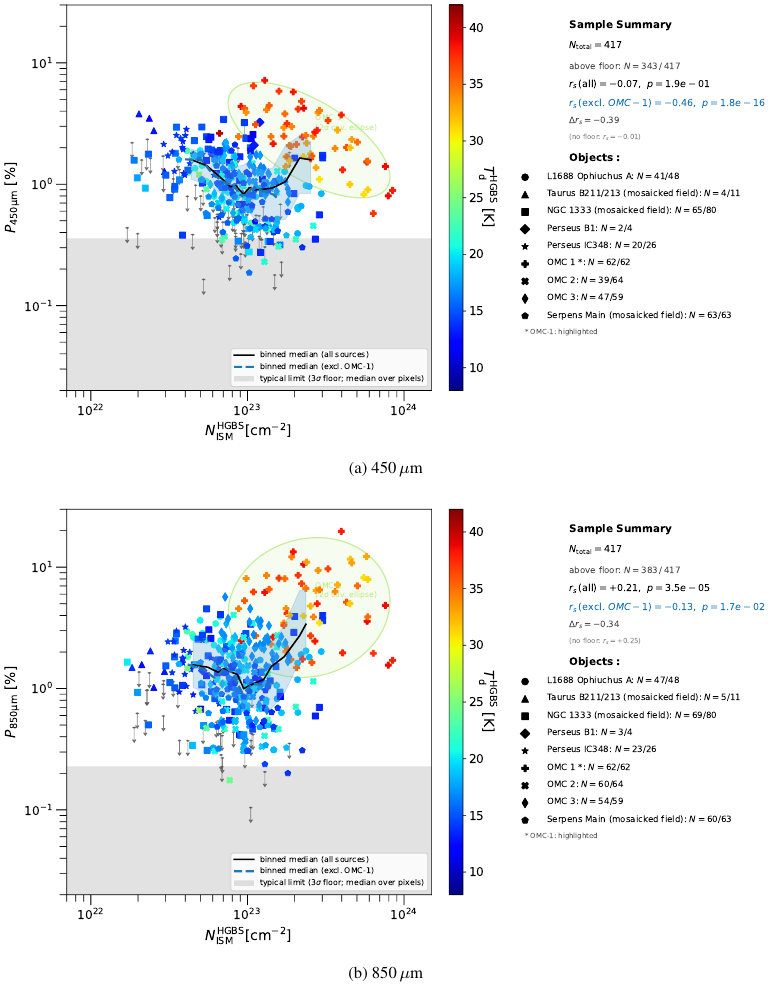}
\caption{
Polarization fraction versus column density at 450\,$\mu$m (a) and 850\,$\mu$m (b), color-coded by dust temperature.
Light gray shading marks the typical $3\sigma$ detection limit, the median over pixels of $3\,\delta P_i$ (Paper~T~subsection~\ref{T-ss:noisemodel}): $P_{450} < 0.36\%$ (a) and $P_{850} < 0.23\%$ (b).
Each pixel is tested against its own floor, so individual limits scatter about these medians.
Pixels below their own floor (74 at 450\,$\mu$m and 34 at 850\,$\mu$m) are plotted at $3\,\delta P_i$ as $3\sigma$ upper limits (downward arrows).
The quoted Spearman statistics are computed on the above-floor sample.
The light-blue band shows the interquartile range of the binned median.
At 450\,$\mu$m, we do not detect a significant correlation in the full above-floor sample ($r_s = -0.07$, $N = 343$, $p = 0.19$).
Excluding OMC\,1 yields $r_s = -0.46$ ($N = 281$, $p = 1.8 \times 10^{-16}$).
At 850\,$\mu$m, a modest positive correlation is present in the full sample ($r_s = +0.21$, $N = 383$, $p = 3.5 \times 10^{-5}$) but flips to a weak negative value when OMC\,1 is excluded ($r_s = -0.13$, $N = 321$, $p = 1.7 \times 10^{-2}$).
High-temperature pixels from OMC~1 (and secondarily from OMC~2), which dominate the high-$N_{\rm H}$ regime, drive the positive full-sample trend at 850\,$\mu$m.
Once OMC\,1 is excluded, the classical polarization-hole behavior is recovered at 450\,$\mu$m, and the 850\,$\mu$m coefficient becomes negative without reaching $p < 0.01$.
Annotations indicate the two physical regimes: the classical ``polarization hole'' regime at low $N_{\rm H}$ and the ``internal heating'' regime at high $N_{\rm H}$ where embedded sources are inferred to enhance grain alignment.
\\ {Alt text: Two scatter plots stacked vertically, each with a line and a shaded band.}
}
\label{fig:NvsP}
\end{figure}

\subsubsection{Polarization Fraction versus Column Density}
\label{sss:S2_P_vs_N}
 
Submillimeter emission from the starless core Pipe-109 shows a polarization hole, with a lower polarization fraction toward its center, where the column density is high \citep{Alves2014}.
To examine how this behavior appears in a sample of star-forming clouds that includes regions with internal heating sources, we compare polarization fraction with column density in both bands and assess the influence of OMC\,1 on the pooled relations.
Figure~\ref{fig:NvsP} shows the correlation between column density and polarization fraction at both wavelengths.
The two wavelengths show distinctly different behaviors.
 
At 450\,$\mu$m, we did not detect a significant correlation between polarization fraction and column density in the above-floor sample, which contains the pixels with $P_i \geq 3\,\delta P_i$ (the per-pixel detection floor of Paper~T~subsection~\ref{T-ss:noisemodel}):
\begin{equation}
r_s(N_{\rm H}, P_{450}) = -0.07 \quad (N = 343, \; p = 0.19).
\label{eq:rs_NH_P450}
\end{equation}
Excluding OMC\,1 from the above-floor sample yields a strong negative correlation, $r_s(N_{\rm H}, P_{450}) = -0.46$ ($N = 281$, $p = 1.8 \times 10^{-16}$).
The polarization hole behavior is recovered at 450\,$\mu$m once OMC\,1 is excluded.
The binned median trend (black line with light-blue interquartile band in figure~\ref{fig:NvsP}a) remains approximately flat over two orders of magnitude in column density when OMC\,1 is retained, because OMC\,1 pixels populate the high-$N_{\rm H}$ corner at elevated $P_{450}$ and cancel the underlying negative slope.
 
At 850\,$\mu$m, we found a statistically significant positive correlation in the full above-floor sample:
\begin{equation}
r_s(N_{\rm H}, P_{850}) = +0.21 \quad (N = 383, \; p = 3.5 \times 10^{-5}),
\label{eq:rs_NH_P850}
\end{equation}
but excluding OMC\,1 reverses the sign and reduces the magnitude to a weak negative correlation, $r_s(N_{\rm H}, P_{850}) = -0.13$ ($N = 321$, $p = 1.7 \times 10^{-2}$).
The OMC\,1 pixels combine the highest column densities in this subsample with elevated 850\,$\mu$m polarization fractions and drive the positive pooled trend.
This result therefore does not contradict the polarization-hole expectation across the survey \citep{Matthews2001,Ward-Thompson2017,Goodman1995,Whittet2008,Alves2014}.
 
The temperature colors in figure~\ref{fig:NvsP} help explain this result.
High-temperature pixels show systematically higher polarization fractions at both wavelengths and drive the positive $P_{850}$--$N_{\rm H}$ correlation in the full sample.
These pixels come predominantly from OMC~1 and secondarily from OMC~2, both of which harbor multiple embedded YSOs.
The positive $N_{\rm H}$--$T_{\rm d}$ correlation established in sub-subsection~\ref{sss:S2_N_T_space} means that the pixels with the highest column density in our sample are also the warmest, internally heated pixels.
Once those pixels are excluded (OMC\,1 exclusion), the relation between $P$ and $N_{\rm H}$ is negative at both wavelengths, as in the classical polarization hole, but reaches $p < 0.01$ only at 450\,$\mu$m.
At fixed column density, cooler pixels ($T_{\rm d}^{\rm HGBS} \lesssim 20$\,K) show lower polarization fractions with larger scatter, a pattern more consistent with the classical polarization hole behavior.
 
In the full sample, we found no significant trend at 450\,$\mu$m and a modest positive trend at 850\,$\mu$m.
Excluding OMC\,1 recovers the classical polarization-hole behavior at 450\,$\mu$m and turns the 850\,$\mu$m coefficient negative.
The latter does not reach $p < 0.01$.
These changes demonstrate the importance of OMC\,1 to the contrast between the two wavelengths.
Paper~S1 (subsection~\ref{S1-ss:Discussion_theory}) discusses the physical interpretation of this wavelength dependence, including the role of grain size distributions and the distinction between internally heated and externally illuminated environments.

\begin{figure}[htbp]
\centering
\includegraphics{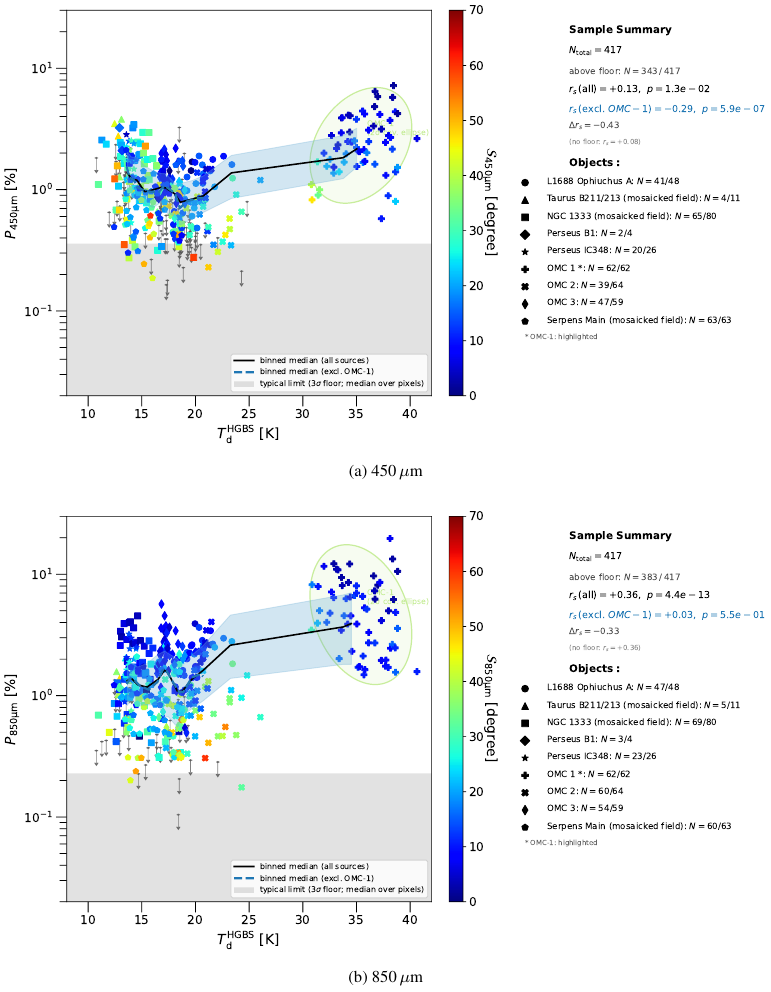}
\caption{
Polarization fraction versus dust temperature at 450\,$\mu$m (a) and 850\,$\mu$m (b), color-coded by polarization angle dispersion $|\mathcal{S}|$.
Light gray shading marks the typical $3\sigma$ detection limit, the median over pixels of $3\,\delta P_i$ (Paper~T~subsection~\ref{T-ss:noisemodel}): $P_{450} < 0.36\%$ (a) and $P_{850} < 0.23\%$ (b).
Each pixel is tested against its own floor, so individual limits scatter about these medians.
Pixels below their own floor (74 at 450\,$\mu$m and 34 at 850\,$\mu$m) are plotted at $3\,\delta P_i$ as $3\sigma$ upper limits (downward arrows).
The quoted Spearman statistics are computed on the above-floor sample.
The light-blue band shows the interquartile range of the binned median.
At 450\,$\mu$m, we do not detect a statistically significant correlation in the full above-floor sample ($r_s = +0.13$, $N = 343$, $p = 1.3 \times 10^{-2}$).
Excluding OMC\,1 yields a negative pooled correlation, $r_s = -0.29$ ($N = 281$, $p = 5.9 \times 10^{-7}$).
This negative correlation indicates that temperature does not enhance 450\,$\mu$m polarization outside the OMC\,1 radiation-dominated regime.
The value $r_s = -0.29$ is a pooled, per-pixel coefficient.
Sub-subsection~\ref{sss:S2_source_aware_correlation} shows that this coefficient is not reproduced cloud by cloud.
At 850\,$\mu$m, we detect a statistically significant positive correlation in the full sample ($r_s = +0.36$, $N = 383$, $p = 4.4 \times 10^{-13}$), consistent with RAT predictions that large grains respond to enhanced radiation fields.
Excluding OMC\,1 removes the significance of this correlation ($r_s = +0.03$, $N = 321$, $p = 0.55$).
This loss of significance shows that the RAT-consistent positive $T$--$P_{850}$ trend is driven entirely by the OMC\,1 high-radiation regime in this subsample.
The wavelength contrast is consistent with size-dependent grain alignment that is enhanced only within the OMC\,1 high-radiation regime at 850\,$\mu$m, although other factors such as differences in grain composition may also contribute.
Annotations mark the ``efficient alignment'' regime (high $T_{\rm d}$, high $P$) attributed to internal radiation, and the ``geometric depolarization'' regime (low $P$) where field tangling reduces observed polarization.
\\ {Alt text: Two scatter plots stacked vertically, each with a line and a shaded band.}
}
\label{fig:TvsP}
\end{figure}

\subsubsection{Polarization Fraction versus Temperature}
\label{sss:S2_P_vs_T}
 
Stronger radiation can enhance grain alignment by RATs when other local conditions are held fixed \citep{Lazarian2007,Andersson2015}.
For grains in radiative equilibrium with far-infrared opacity $\kappa_\nu \propto \nu^\beta$, a radiation field of fixed spectral shape gives $U_{\rm rad} \propto T_{\rm d}^{4+\beta}$, where $U_{\rm rad} \equiv u_{\rm rad}/u_{\rm ISRF}$ is the radiation energy density relative to that of the interstellar radiation field in the solar neighborhood.
For $\beta \simeq 2$, as assumed in deriving $T_{\rm d}^{\rm HGBS}$, this yields the approximate relation $U_{\rm rad} \simeq (T_{\rm d}/16.4\,{\rm K})^6$ adopted by \citet{Hoang2021}.
A $T_{\rm d}^4$ scaling would apply only in the limit of gray emissivity ($\beta = 0$), including blackbody emission.
However, radiative torque disruption can reduce the polarization fraction by disrupting large grains \citep{Hoang2021}.
We examine the resulting polarization response by correlating dust temperature, used as an approximate proxy for radiation strength, with polarization fraction at both wavelengths.
Because $T_{\rm d} > 0$, any power-law scaling $U_{\rm rad} \propto T_{\rm d}^{n}$ with $n > 0$ preserves the rank order, so the Spearman coefficients and $p$-values below do not depend on the exponent.
 
Figure~\ref{fig:TvsP}a shows $P_{450}$ versus $T_{\rm d}^{\rm HGBS}$, color-coded by the polarization angle dispersion $|\mathcal{S}_{450}|$.
At 450\,$\mu$m, we do not detect a statistically significant correlation in the full above-floor sample:
\begin{equation}
r_s(T_{\rm d}, P_{450}) = +0.13 \quad (N = 343, \; p = 1.3 \times 10^{-2}).
\label{eq:rs_Td_P450}
\end{equation}
However, excluding OMC\,1 reverses the sign of the correlation and produces a strong, highly significant negative trend,
\begin{equation}
r_s(T_{\rm d}, P_{450})_{\rm excl\,OMC\,1} = -0.29 \quad (N = 281, \; p = 5.9 \times 10^{-7}),
\label{eq:rs_Td_P450_exclOMC1}
\end{equation}
indicating that the RAT-predicted positive response of 450\,$\mu$m polarization to temperature is not present outside the OMC\,1 radiation-dominated regime.
If anything, the eight non-OMC\,1 HGBS fields show the opposite behavior.
 
At 850\,$\mu$m, we found a statistically significant positive correlation in the full above-floor sample:
\begin{equation}
r_s(T_{\rm d}, P_{850}) = +0.36 \quad (N = 383, \; p = 4.4 \times 10^{-13}),
\label{eq:rs_Td_P850}
\end{equation}
but this correlation is no longer significant when OMC\,1 is excluded:
\begin{equation}
r_s(T_{\rm d}, P_{850})_{\rm excl\,OMC\,1} = +0.03 \quad (N = 321, \; p = 0.55).
\label{eq:rs_Td_P850_exclOMC1}
\end{equation}
The positive $T_{\rm d}$--$P_{850}$ correlation in the full sample, which is consistent with RAT predictions, is therefore driven entirely by the OMC\,1 pixels.
We do not detect a temperature dependence of $P_{850}$ within the remaining eight HGBS fields.
 
The temperature correlations in the full sample parallel the results for column density in sub-subsection~\ref{sss:S2_P_vs_N}.
With OMC\,1 included, the 850\,$\mu$m polarization appears to respond to environmental parameters in a manner consistent with RAT theory, whereas the 450\,$\mu$m polarization does not.
At 450\,$\mu$m, the full-sample null result is visually apparent in figure~\ref{fig:TvsP}a: the data scatter broadly at all temperatures without a systematic trend.
The color coding reveals that the scatter in $P_{450}$ at fixed temperature correlates strongly with $|\mathcal{S}_{450}|$: pixels with low angle dispersion (coherent $\vec{B}$-field) tend to have higher polarization fractions regardless of temperature.
The correlation of the scatter with $|\mathcal{S}_{450}|$ suggests that at 450\,$\mu$m, $\vec{B}$-field structure, rather than alignment efficiency that depends on temperature, is the primary driver of variations in polarization fraction.
 
At 850\,$\mu$m, the full-sample positive correlation is consistent with RAT predictions, in which higher temperatures correspond to stronger radiation fields that more efficiently align the large grains that dominate the polarized emission.
Because the correlation vanishes once OMC\,1 is excluded, its interpretation is local to the OMC\,1 regime rather than to the survey as a whole.
The contrast between the two wavelengths in the full sample is consistent with size-dependent grain alignment rather than with either band selecting a grain size directly.
In this picture, the aligned population that dominates the polarized emission has a flatter effective opacity spectrum than the bulk dust (Paper~S1 subsection~\ref{S1-ss:Discussion_theory}).
The sensitivity of both bands to the inclusion of OMC\,1 indicates that the effect is concentrated in its high-radiation environment.
 
We note three observational effects that can modify or mimic intrinsic $P$--$T_{\rm d}$ trends.
First, fits of far-infrared spectral energy distributions (e.g., from Herschel) give an average of $T_{\rm d}$ along the line of sight, weighted toward warmer dust.
This weighting may bias comparisons with submillimeter polarization if the far-infrared emission and submillimeter polarization trace different grain populations.
Second, optical depth effects may become non-negligible at 450\,$\mu$m in the regions of highest column density, reducing the effective depth from which polarized emission emerges.
Such effects could suppress any intrinsic $P$--$T_{\rm d}$ correlation at 450\,$\mu$m while leaving the optically thinner 850\,$\mu$m emission less affected.
Third, mixing of temperatures along the line of sight can dilute correlations by combining regions with different physical conditions within a single beam, particularly at shorter wavelengths where the Planck function is more sensitive to temperature variations.
Despite these caveats, the systematic nature of the observed behavior across the full sample---a statistically significant positive correlation at 850\,$\mu$m and a null correlation at 450\,$\mu$m---suggests that the three effects alone are unlikely to fully account for the wavelength dependence.
Paper~S1 (subsection~\ref{S1-ss:Discussion_theory}) discusses the implications for RAT theory further.

\subsubsection{Polarization Angle Dispersion versus Environment}
\label{sss:S2_S_vs_env}
If heating or increasing column density is accompanied by a more coherent projected field, polarization can change even without a corresponding change in grain alignment.
We therefore examine how the magnitude of the measured angle dispersion varies with $N_{\rm H}$ and $T_{\rm d}$ at 450\,$\mu$m.
 
The polarization angle dispersion $\mathcal{S}$ \citep{PlanckIntXIX2015,Planck2020-XII} measures the coherence of the projected $\vec{B}$-field.
Because it uses the polarization angles alone, it does not depend directly on the grain alignment efficiency.
We use the debiased, non-negative $\mathcal{S}$ defined in sub-subsection~\ref{sss:S_definition}, written $|\mathcal{S}|$ here and in table~\ref{tab:correlation_summary}.
As in Paper~S1, we compute the correlations with column density and temperature over the pixels that meet the detection requirement for this observable, $\mathcal{S}_{450}$ detected at $3\sigma_{\mathcal{S}}$.
 
The 450\,$\mu$m angle dispersion decreases with column density over the full above-floor sample:
\begin{equation}
r_s(N_{\rm H}, |\mathcal{S}_{450}|) = -0.34 \quad (N = 363, \; p = 2.4 \times 10^{-11}),
\label{eq:rs_NH_S450}
\end{equation}
and the trend weakens but remains significant when OMC\,1 is excluded:
\begin{equation}
r_s(N_{\rm H}, |\mathcal{S}_{450}|)_{\rm excl\,OMC\,1} = -0.19 \quad (N = 301, \; p = 1.2 \times 10^{-3}).
\label{eq:rs_NH_S450_exclOMC1}
\end{equation}
As defined in equation~(\ref{eq:S_def}), $\mathcal{S}$ is a root-mean-square angle difference and is therefore non-negative by construction.
This pooled trend is not reproduced within individual clouds (per-cloud median $r_s = +0.04$, with 3 of 8 fields sharing the negative sign), so we attribute it to differences between clouds rather than to a systematic change with column density within them.
 
We found a statistically significant negative correlation between dust temperature and $|\mathcal{S}_{450}|$ in the full above-floor sample:
\begin{equation}
r_s(T_{\rm d}, |\mathcal{S}_{450}|) = -0.40 \quad (N = 363, \; p = 1.0 \times 10^{-15}),
\label{eq:rs_Td_S450}
\end{equation}
and the correlation weakens but remains significant once OMC\,1 is excluded:
\begin{equation}
r_s(T_{\rm d}, |\mathcal{S}_{450}|)_{\rm excl\,OMC\,1} = -0.19 \quad (N = 301, \; p = 8.1 \times 10^{-4}).
\label{eq:rs_Td_S450_exclOMC1}
\end{equation}
OMC\,1 strengthens the full-sample $T_{\rm d}$--$|\mathcal{S}_{450}|$ anticorrelation but does not produce it alone.
The median of the coefficients for individual clouds is $r_s = -0.23$, with 5 of 8 fields sharing the negative sign.
Warm regions thus tend to have more coherent field directions at 450\,$\mu$m, most clearly in OMC\,1, but the trend is not confined to it.
 
We do not analyze the corresponding 850\,$\mu$m relations ($N_{\rm H}$--$|\mathcal{S}_{850}|$ and $T_{\rm d}$--$|\mathcal{S}_{850}|$) here and defer a dedicated 850\,$\mu$m analysis to a future study of the angle dispersion in both bands.
The pooled dependences of the measured 450\,$\mu$m angle dispersion on column density and temperature are both significant, but neither is established as a trend within individual clouds.

\subsubsection{Per-Pixel Relations Reported in Paper~S1}
\label{sss:S2_PvsS_IratioRp_pointer}
Two related comparisons help place the wavelength-dependent polarization in context: polarization fraction versus projected angle dispersion, and polarization ratio versus total-intensity ratio.
The first tests whether reduced polarization accompanies field disorder.
The second describes how the wavelength contrast varies across the selected emission.
 
Paper~S1 presents and physically interprets both relations for individual pixels, derived from the same HGBS Method~3 subsample used here.
The first is the anticorrelation between polarization fraction and angle dispersion, $r_s(P_\lambda, \mathcal{S}_\lambda)$, at each wavelength (Paper~S1 subsection~\ref{S1-ss:S_dispersion} and figure~\ref{S1-fig:PvsS}), used to check geometric depolarization in F3.
The second is the anticorrelation between the intensity ratio and the polarization ratio, $r_s(I_{450}/I_{850}, R_{\rm p})$ (Paper~S1 subsection~\ref{S1-ss:Iratio} and figure~\ref{S1-fig:IratioRp}), the counterpart, pixel by pixel, of the $\gamma - \alpha$ structure in F2.
To avoid redundancy between Papers~S1 and~S2, we do not reproduce those figures here.
Briefly, the Spearman rank coefficients reported in Paper~S1 for the pixels above the detection floors are $r_s(P_{450}, \mathcal{S}_{450}) = -0.45$ ($N=308$, $p=1.1 \times 10^{-16}$) and $r_s(P_{850}, \mathcal{S}_{850}) = -0.73$ ($N=358$, $p=9.8 \times 10^{-61}$).
Both remain significant when OMC\,1 is excluded ($-0.27$ and $-0.70$), so the $P$--$\mathcal{S}$ relation tests for the presence of geometric depolarization in both bands rather than measuring its magnitude.
The $I_{450}/I_{850}$--$R_{\rm p}$ anticorrelation, $r_s = -0.49$ ($N=327$, $p=1.9 \times 10^{-21}$), persists at $r_s = -0.35$ ($N=265$, $p=5.9 \times 10^{-9}$) after OMC\,1 is excluded.
We therefore regard this anticorrelation as a general property of the HGBS subsample rather than a pattern specific to OMC\,1.
We return to both relations when comparing results with and without OMC\,1 in sub-subsections~\ref{sss:S2_correlation_summary} and~\ref{sss:S2_source_aware_correlation}.
If geometric depolarization acts as the same multiplicative factor at both wavelengths, it cancels in $R_{\rm p}$.
The $P$--$\mathcal{S}$ associations in each band do not establish that this cancellation occurs.
The correlations with field disorder persist when OMC\,1 is excluded, as does the pooled ratio relation.
However, the ratio relation does not recur uniformly within the other clouds.
These comparisons constrain the observed covariance of the diagnostics rather than identifying geometry or grain alignment as a unique cause.

\begin{figure}[htbp]
\centering
\includegraphics{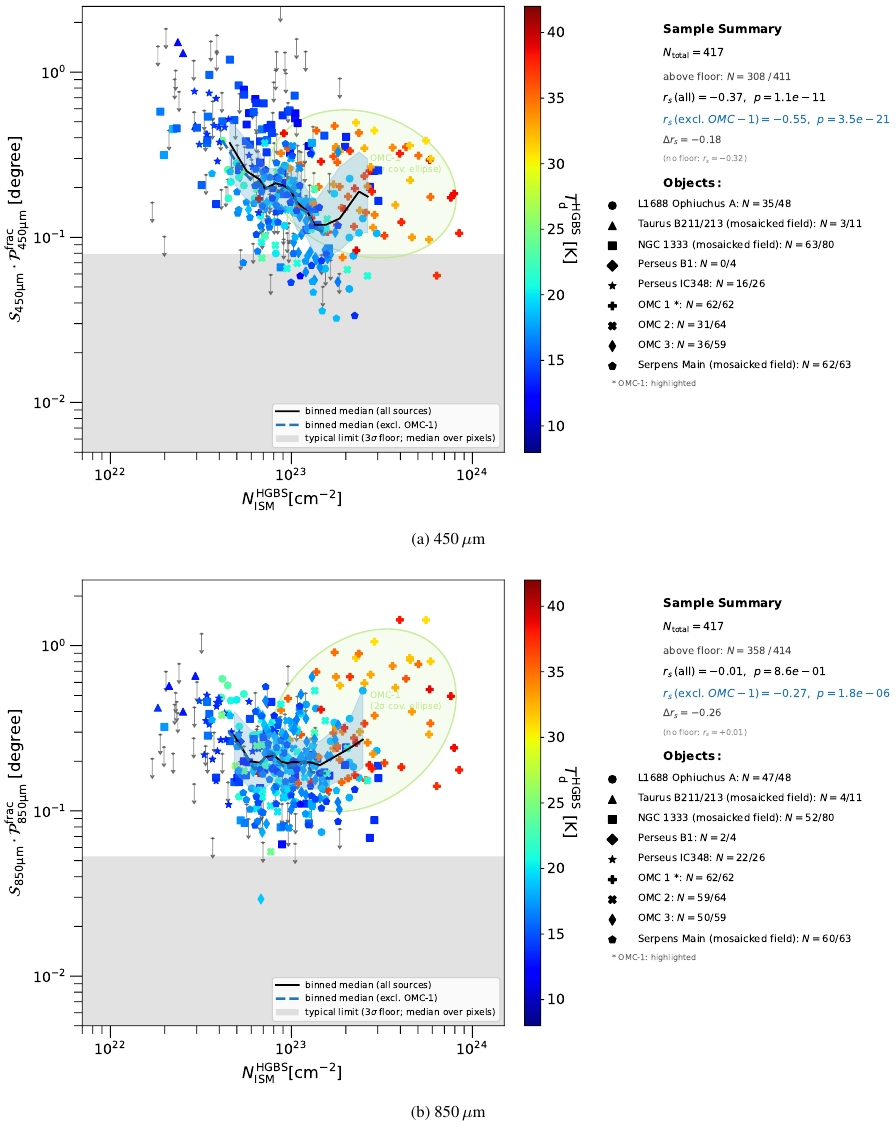}
\caption{
The $|\mathcal{S}| \times \mathcal{P}^{\rm frac}$ diagnostic versus column density at 450\,$\mu$m (a) and 850\,$\mu$m (b), color-coded by temperature.
We computed the Spearman rank coefficients over the subset that satisfies $P_i \ge 3\,\sigma_{P_i}$ with $\mathcal{S}$ detected at $3\sigma_{\mathcal{S}}$ at each wavelength.
The coefficients are negative at 450\,$\mu$m and null at 850\,$\mu$m over the full above-floor sample ($r_s = -0.37$, $N = 308$, $p = 1.1 \times 10^{-11}$; $r_s = -0.01$, $N = 358$, $p = 0.86$).
The coefficients are negative in both bands after the OMC\,1 pixels are excluded ($r_s = -0.55$, $N = 246$, $p = 3.5 \times 10^{-21}$ at 450\,$\mu$m; $r_s = -0.27$, $N = 296$, $p = 1.8 \times 10^{-6}$ at 850\,$\mu$m).
Light gray shading marks the typical $3\sigma$ limit of the product, the median over pixels of $3\sigma_{\mathcal{S}}\,\mathcal{P}^{\rm frac}$: $0.079\arcdeg$ (a) and $0.053\arcdeg$ (b).
Pixels with $\mathcal{S}$ undetected or $P$ below its floor (109 at 450\,$\mu$m and 59 at 850\,$\mu$m) are plotted with the below-floor factor set to its $3\sigma$ value, as upper limits (downward arrows), and are excluded from the Spearman statistics.
One of the 450\,$\mu$m limits lies above the plotted range.
The light-blue band shows the interquartile range of the binned median.
Annotations mark the ``alignment maintained'' (high $|\mathcal{S}| \times \mathcal{P}^{\rm frac}$) and ``alignment suppressed'' (low $|\mathcal{S}| \times \mathcal{P}^{\rm frac}$) regimes.
The horizontal line at $1\arcdeg$ separates the two regimes.
\\ {Alt text: Two scatter plots stacked vertically, each with a line and a shaded band.}
}
\label{fig:SPvsN}
\end{figure}

\subsubsection{The $\mathcal{S} \times \mathcal{P}^{\rm frac}$ Diagnostic: Wavelength Dependence}
\label{sss:S2_SP_vs_N}
A decline in polarization toward dense material may result partly from cancellation by a disordered field.
The product $|\mathcal{S}| \times \mathcal{P}^{\rm frac}$ provides an empirical check on whether a trend with column density remains after partly compensating for that geometric contribution.
 
As introduced in sub-subsection~\ref{sss:SP_definition}, the product $|\mathcal{S}| \times \mathcal{P}^{\rm frac}$ is designed to partially remove the geometric depolarization effect, so that it serves as a proxy for grain alignment efficiency \citep{Planck2020-XII,LeGouellec2020}.
We now examine how this diagnostic correlates with column density at both wavelengths, applying the per-pixel detection floors used in Paper~S1 and throughout subsection~\ref{ss:S2_HGBS}.
 
Figure~\ref{fig:SPvsN} presents $|\mathcal{S}| \times \mathcal{P}^{\rm frac}$ versus $N_{\rm H}^{\rm HGBS}$ at 450\,$\mu$m (panel a) and 850\,$\mu$m (panel b).
Over the full above-floor sample, the 450\,$\mu$m product decreases with column density, whereas the 850\,$\mu$m product shows no trend:
\begin{equation}
r_s(N_{\rm H}, |\mathcal{S}_{450}| \times P_{450}^{\rm frac}) = -0.37 \quad (N = 308, \; p = 1.1 \times 10^{-11}),
\label{eq:rs_NH_SP450}
\end{equation}
\begin{equation}
r_s(N_{\rm H}, |\mathcal{S}_{850}| \times P_{850}^{\rm frac}) = -0.01 \quad (N = 358, \; p = 0.86).
\label{eq:rs_NH_SP850}
\end{equation}
Under the floors adopted here for individual pixels, the dependence of $|\mathcal{S}| \times \mathcal{P}^{\rm frac}$ on column density is consistent with zero only at 850\,$\mu$m.
 
After OMC\,1 is excluded, both bands show a negative trend:
\begin{equation}
r_s(N_{\rm H}, |\mathcal{S}_{450}| \times P_{450}^{\rm frac})_{\rm excl\,OMC\,1} = -0.55 \quad (N = 246, \; p = 3.5 \times 10^{-21}),
\label{eq:rs_NH_SP450_exclOMC1}
\end{equation}
\begin{equation}
r_s(N_{\rm H}, |\mathcal{S}_{850}| \times P_{850}^{\rm frac})_{\rm excl\,OMC\,1} = -0.27 \quad (N = 296, \; p = 1.8 \times 10^{-6}).
\label{eq:rs_NH_SP850_exclOMC1}
\end{equation}
Both trends are statistically significant after OMC\,1 is excluded.
The 450\,$\mu$m trend is the stronger ($|r_s| = 0.55$ against $0.27$).
Over the eight HGBS fields other than OMC\,1 the two bands therefore agree in sign: the pooled proxy $|\mathcal{S}| \times \mathcal{P}^{\rm frac}$ declines toward higher column densities.
At 850\,$\mu$m, OMC\,1 contributes pixels with high $|\mathcal{S}| \times \mathcal{P}^{\rm frac}$ at high $N_{\rm H}$, canceling the negative trend in the full HGBS-9 sample.
At 450\,$\mu$m, these pixels weaken the trend without canceling it.
 
Several factors may underlie the negative trend observed outside OMC\,1 and its difference between the bands.
First, the 450 and 850\,$\mu$m emission may be dominated by different grain size populations.
If smaller grains contribute more at 450\,$\mu$m, the 450\,$\mu$m polarization could be more affected by the loss of alignment of small grains in dense gas, where the minimum aligned grain size increases with gas density \citep{Hoang2021}.
Second, the different optical depths at the two wavelengths mean that 450\,$\mu$m emission is more weighted toward warmer surface layers, while 850\,$\mu$m probes deeper into the cold interiors of the clouds.
Third, we suggest that grain growth in dense regions may affect the polarization differently at 850 and 450\,$\mu$m.
Grain growth raises the far-infrared dust opacity \citep{Ysard2013} and could modify the polarizing efficiency of the grains \citep{Guillet2018}.
Paper~S1 (subsection~\ref{S1-ss:Discussion_theory}) develops the physical interpretation of these alignment diagnostics.
In the full HGBS-9 sample the product declines with column density at 450\,$\mu$m ($r_s = -0.37$) but not at 850\,$\mu$m ($r_s = -0.01$).
After OMC\,1 is excluded, both bands show a negative association ($r_s = -0.55$ and $-0.27$).
The result is consistent with reduced grain alignment efficiency in denser material outside that field, but it does not establish a wavelength difference or a causal role for magnetic disorder.

\subsubsection{Summary of Correlation Results}
\label{sss:S2_correlation_summary}
The HGBS fields do not sample a single combination of density and irradiation.
We therefore interpret their polarization correlations by comparing the ordinary interiors of the clouds with the warm, dense material strongly represented by OMC\,1.
 
Table~\ref{tab:correlation_summary} compiles all Spearman correlation coefficients measured in this analysis.
 
\begin{deluxetable*}{llcccccc}
\tablecaption{Summary of Correlations above the Detection Floors of Individual Pixels\label{tab:correlation_summary}}
\tablewidth{0pt}
\tablehead{
\colhead{Correlation} & \colhead{Physical Interpretation} & \multicolumn{3}{c}{Above-floor sample} & \multicolumn{3}{c}{Excl.\ OMC\,1} \\
\colhead{} & \colhead{} & \colhead{$N$} & \colhead{$r_s$} & \colhead{$p$} & \colhead{$N$} & \colhead{$r_s$} & \colhead{$p$}
}
\startdata
\multicolumn{8}{c}{\textit{Environmental Parameter Correlation}} \\
$N_{\rm H}$ vs.\ $T_{\rm d}$ & Internal heating in dense cores & 417 & $+0.41$ & $1.5 \times 10^{-18}$ & 355 & $+0.15$ & $5.1 \times 10^{-3}$ \\
\hline
\multicolumn{8}{c}{\textit{Polarization Fraction Correlations}} \\
$N_{\rm H}$ vs.\ $P_{450}$ & Polarization-hole signature after OMC\,1 excluded & 343 & $-0.07$ & $0.19$ & 281 & $-0.46$ & $1.8 \times 10^{-16}$ \\
$N_{\rm H}$ vs.\ $P_{850}$ & Full-sample positive; flips excl.\ OMC\,1 & 383 & $+0.21$ & $3.5 \times 10^{-5}$ & 321 & $-0.13$ & $1.7 \times 10^{-2}$ \\
$T_{\rm d}$ vs.\ $P_{450}$ & Null full; negative outside OMC\,1 & 343 & $+0.13$ & $1.3 \times 10^{-2}$ & 281 & $-0.29$ & $5.9 \times 10^{-7}$ \\
$T_{\rm d}$ vs.\ $P_{850}$ & OMC\,1-driven RAT-consistent positive & 383 & $+0.36$ & $4.4 \times 10^{-13}$ & 321 & $+0.03$ & $0.55$ \\
$P_{450}$ vs.\ $|\mathcal{S}_{450}|$\tablenotemark{a} & Geometric depolarization & 308 & $-0.45$ & $1.1 \times 10^{-16}$ & 246 & $-0.27$ & $2.4 \times 10^{-5}$ \\
$P_{850}$ vs.\ $|\mathcal{S}_{850}|$\tablenotemark{a} & Geometric depolarization & 358 & $-0.73$ & $9.8 \times 10^{-61}$ & 296 & $-0.70$ & $6.5 \times 10^{-45}$ \\
\hline
\multicolumn{8}{c}{\textit{Angle Dispersion at 450\,$\mu$m}\tablenotemark{b}} \\
$N_{\rm H}$ vs.\ $|\mathcal{S}_{450}|$ & Negative; weaker excl.\ OMC\,1 & 363 & $-0.34$ & $2.4 \times 10^{-11}$ & 301 & $-0.19$ & $1.2 \times 10^{-3}$ \\
$T_{\rm d}$ vs.\ $|\mathcal{S}_{450}|$ & Negative; weaker excl.\ OMC\,1 & 363 & $-0.40$ & $1.0 \times 10^{-15}$ & 301 & $-0.19$ & $8.1 \times 10^{-4}$ \\
\hline
\multicolumn{8}{c}{\textit{Alignment Diagnostics}} \\
$I_{450}/I_{850}$ vs.\ $P_{450}/P_{850}$\tablenotemark{a} & Spectral index relationship & 327 & $-0.49$ & $1.9 \times 10^{-21}$ & 265 & $-0.35$ & $5.9 \times 10^{-9}$ \\
$N_{\rm H}$ vs.\ $|\mathcal{S}_{450}| \times P_{450}^{\rm frac}$ & Alignment diagnostic (450\,$\mu$m) & 308 & $-0.37$ & $1.1 \times 10^{-11}$ & 246 & $-0.55$ & $3.5 \times 10^{-21}$ \\
$N_{\rm H}$ vs.\ $|\mathcal{S}_{850}| \times P_{850}^{\rm frac}$ & Alignment diagnostic (850\,$\mu$m) & 358 & $-0.01$ & $0.86$ & 296 & $-0.27$ & $1.8 \times 10^{-6}$ \\
\enddata
\tablenotetext{a}{Figure and physical interpretation provided in Paper~S1; see sub-subsection~\ref{sss:S2_PvsS_IratioRp_pointer}.}
\tablenotetext{b}{The corresponding 850\,$\mu$m relations ($N_{\rm H}$ vs.\ $|\mathcal{S}_{850}|$ and $T_{\rm d}$ vs.\ $|\mathcal{S}_{850}|$) are not evaluated here and are deferred to a future study of angle dispersion in both bands.}
\tablecomments{Spearman rank correlation coefficients computed for the HGBS-9 sample of individual Method~3 pixels.
The ``above-floor sample'' is the subset of the $N_{\rm total} = 417$ pixels satisfying $P_i \ge 3\,\sigma_{P_i}$ at the relevant wavelength (floors differ per figure because the significance criterion depends on which $P$-bands enter the correlation).
Rows involving $\mathcal{S}$ further require $\mathcal{S}$ to be detected at $3\sigma_{\mathcal{S}}$ (sub-subsection~\ref{sss:S_definition}).
The ``excl.\ OMC\,1'' column excludes the OMC\,1 pixels from the above-floor sample.
Angle dispersions are the debiased $\mathcal{S}$ of sub-subsection~\ref{sss:S_definition}, which is non-negative by construction, so that $|\mathcal{S}| = \mathcal{S}$.
For $P_{450}$ in the four fields to which Paper~T applies the 450\,\micron\ Stokes-$Q/U$ zero-level offset correction, we use the value after that correction.
$p$-values $<0.01$ are considered statistically significant.
Figure references and full-sample discussion appear in the cited subsections of subsection~\ref{ss:S2_HGBS}.}
\end{deluxetable*}
 
The entries in table~\ref{tab:correlation_summary} fall into two broad groups.
The first group consists of the relations that remain stable with and without OMC\,1.
The second group consists of the relations whose sign or magnitude depends strongly on the inclusion of OMC\,1, and we describe it in sub-subsection~\ref{sss:S2_correlation_ex_OMC1}.
The first stable relation is the $N_{\rm H}$--$T_{\rm d}$ positive correlation (a pooled coefficient shown in sub-subsection~\ref{sss:S2_source_aware_correlation} to arise from offsets between clouds rather than from a trend within individual clouds).
The second is the $P$--$|\mathcal{S}|$ anticorrelation from geometric depolarization.
This anticorrelation is strongest over the above-floor sample and, as reported in Paper~S1, weakens but does not change sign when OMC\,1 is excluded.
The third is the $I_{450}/I_{850}$--$R_{\rm p}$ anticorrelation, which persists with and without OMC\,1 and is interpreted in Paper~S1 as the per-pixel counterpart of the $\gamma < \alpha$ structure underlying F2.

\subsubsection{OMC~1 Inclusion/Exclusion Summary}
\label{sss:S2_correlation_ex_OMC1}
OMC\,1 occupies the part of the HGBS sample with warm dust and high column density, and can strongly influence correlations made by pooling fields.
Comparing the same diagnostics with and without this region tests whether they describe typical cloud material or conditions concentrated in an intensely irradiated source.
 
OMC~1 contributes $\approx$62 Method~3 pixels in the high-$N_{\rm H}$, high-$T_{\rm d}$ corner (sub-subsection~\ref{sss:S2_N_T_space}) and strongly affects several pooled correlations.
We summarize here the results with and without OMC\,1 reported above.
Following the convention of Paper~S1 subsection~4.3, the OMC\,1-excluded subsample retains all above-floor pixels from L1688 Ophiuchus~A, Taurus~B211/213, NGC~1333, Perseus~B1, Perseus~IC348, OMC~2, OMC~3, and Serpens~Main.
The resulting sample size $N$ ranges from 246 to 355 pixels depending on the wavelength-dependent floor applied (table~\ref{tab:correlation_summary}).
 
Three patterns emerge.
First, \emph{sign-flip} relations: $N_{\rm H}$--$P_{850}$ changes from $r_s = +0.21$ in the full sample to $r_s = -0.13$ when OMC\,1 is excluded.
The $T_{\rm d}$--$P_{450}$ relation changes from $r_s = +0.13$ ($p = 0.013$), which is not statistically significant in the full sample, to $r_s = -0.29$ ($p = 5.9 \times 10^{-7}$) when OMC\,1 is excluded.
Second, \emph{weakening} relations: the positive $T_{\rm d}$--$P_{850}$ correlation falls from $r_s = +0.36$ ($p = 4.4 \times 10^{-13}$) in the full sample to $r_s = +0.03$ ($p = 0.55$) when OMC\,1 is excluded.
The $N_{\rm H}$-- and $T_{\rm d}$--$|\mathcal{S}_{450}|$ anticorrelations weaken (from $r_s = -0.34$ and $-0.40$ to $-0.19$) but do not vanish.
Third, \emph{strengthening} relations: the $N_{\rm H}$--$P_{450}$ relation is not detected in the above-floor full sample ($r_s = -0.07$, $p = 0.19$) but shows a strong classical polarization-hole signature when OMC\,1 is excluded ($r_s = -0.46$, $p = 1.8 \times 10^{-16}$).
Both $|\mathcal{S}| \times \mathcal{P}^{\rm frac}$--$N_{\rm H}$ relations strengthen, from $r_s = -0.37$ to $-0.55$ at 450\,$\mu$m and from no detected correlation to $-0.27$ at 850\,$\mu$m.
Both rank correlations are significant at $p < 0.01$ after OMC\,1 is excluded.
The two wavelengths therefore agree in sign outside OMC\,1 but differ in the full sample.
Paper~S1 develops the physical interpretation of these correlations (subsections~\ref{S1-ss:Discussion_GammaAlpha}~and~\ref{S1-ss:Discussion_theory}).

The physical picture that accounts for all three patterns is that the HGBS-9 sample contains \emph{two competing regimes}.
In the regime of strong radiation represented by OMC\,1, the association of warm dust with high $P$, consistent with RAT alignment, flattens or reverses the classical polarization-hole signature.
The background regime represented by the other eight fields shows the classical $P$-vs-$N_{\rm H}$ anticorrelation, a modest negative $T_{\rm d}$-vs-$P_{450}$ trend, and a decline of $|\mathcal{S}| \times \mathcal{P}^{\rm frac}$ with $N_{\rm H}$ at both wavelengths.
The competing regimes partially cancel in the full-sample averages.
The OMC\,1 pixels control the full-sample sign or significance because of their strong clustering at the high-$N_{\rm H}$, high-$T_{\rm d}$ corner.
Paper~S1 develops the physical interpretation of the two-regime picture (subsection~\ref{S1-ss:Discussion_theory}).
The analysis of individual clouds in sub-subsection~\ref{sss:S2_source_aware_correlation} refines this two-regime picture into a hierarchical structure that differs between the bands (table~\ref{tbl:source_aware_hgbs_correlations}).
The association of warm dust with high polarization at 850\,$\mu$m is concentrated in the sample containing OMC\,1, while at 450\,$\mu$m the anticorrelation between polarization fraction and column density becomes clearer without OMC\,1.
These contrasts identify where the pooled signals originate, but do not establish a common local law for all eight remaining fields.
 
\subsubsection{Per-Cloud Correlation Tests}
\label{sss:S2_source_aware_correlation}
A polarization trend between clouds does not necessarily describe the changes encountered when moving from the envelope to the interior of a single cloud.
We therefore ask which correlations recur within individual fields and which arise because the fields occupy different ranges of density, temperature, and polarized emission.
 
We describe the map entries as pixels, but calculate correlations on the $12\arcsec$ grid after harmonizing both bands to the common $12\farcs6$ effective resolution (Paper~T~subsection~\ref{T-ss:beamval}).
This grid provides one sample per $12\farcs6$ beam.
We therefore report counts as matched-resolution samples, or simply samples, here and in table~\ref{tbl:source_aware_hgbs_correlations}.
The beam autocorrelation and residual map-making covariance from {\tt pol2map} mean that adjacent samples are not strictly independent.
We use the tests for individual clouds below to distinguish trends within clouds from differences between clouds and from structure within a source that strongly affects the pooled statistic.
 
The full-sample correlations of table~\ref{tab:correlation_summary} aggregate pixels across all nine HGBS fields and therefore conflate structure \emph{within clouds} with structure \emph{between clouds}.
A non-zero full-sample $r_s$ may arise from covariation of the observables among pixels within typical fields, from systematic differences in the mean values between fields, or from a single source dominating one corner of the joint distribution.
To separate trends within clouds from differences between clouds, we recomputed each entry of table~\ref{tab:correlation_summary} under two complementary diagnostics for individual clouds.
First, we calculated the Spearman rank coefficient for each HGBS field using the pixels that meet the detection requirements.
We report the distribution of the nine $r_s$ values for individual clouds and the number of fields with the same sign as the full sample.
This \emph{concordance} count tests the influence of OMC\,1 and whether the trend recurs within individual clouds.
Second, we applied a cluster block-bootstrap in which the resampling unit is the HGBS field rather than the pixel.
At each of $10^4$ replicates we drew nine fields with replacement, pooled the resampled pixels, and recomputed the pooled $r_s$.
We took 95\% CIs from the 2.5--97.5 percentiles of the bootstrap distribution.
The cluster block-bootstrap accounts for the grouping of pixels within clouds.
Field-to-field differences broaden the uncertainty of the pooled $r_s$ relative to a standard per-pixel CI, complementing the per-pixel $p$-values in table~\ref{tab:correlation_summary}.
 
The following observations for individual clouds refine the full-sample picture without overturning the two-regime structure from the comparison with and without OMC\,1 in sub-subsection~\ref{sss:S2_correlation_ex_OMC1}.
Among the environmental correlations of table~\ref{tab:correlation_summary}, the most robust result for individual clouds is the anti-correlation between $N_{\rm H}$ and $P_{450}$.
Although the pooled above-floor coefficient is not significant ($r_s = -0.07$, $N=343$, $p = 0.19$), all seven HGBS fields with a sufficient dynamic range above the detection floor show the same negative sign, with a per-cloud median coefficient $r_s = -0.44$ and interquartile range $[-0.59,\,-0.31]$.
The pooled coefficient is diluted by offsets between clouds in mean column density and polarization fraction and by differences in dynamic range among fields.
The full-sample value strengthens to $r_s = -0.46$ ($N=281$, $p = 1.8\times 10^{-16}$) once the high-radiation samples of OMC\,1 that occupy the high-$N_{\rm H}$ corner of the joint distribution are excluded.
The pooled $r_s = -0.09$ from the cluster block-bootstrap has a 95\% CI of $[-0.62,\,+0.21]$, with the lower quartile firmly negative.
We therefore interpret the $450\,\mu$m polarization fraction as tracing the classical polarization-hole behavior within clouds more robustly than any other HGBS environmental trend in our sample.
 
By contrast, $P_{850}$ does not show a uniform response to column density within clouds.
The full-sample $r_s = +0.21$ ($N=383$) flips sign to $r_s = -0.13$ when OMC\,1 is excluded.
The distribution over individual sources shows the sign changing among the eight valid HGBS fields (per-cloud median $r_s = -0.21$, with only 3 of 8 fields concordant with the full-sample positive sign).
The cluster block-bootstrap CI on the pooled $r_s$ ($[-0.34,\,+0.54]$) includes zero.
This difference between $P_{450}$ and $P_{850}$ is itself informative.
The 450\,$\mu$m polarization fraction shows the more classical trend of depolarization with column density within clouds.
In contrast, the 850\,$\mu$m response is more sensitive to radiation and structural conditions specific to each source and does not constitute a universal law within clouds.
 
The $T_{\rm d}$--$P_{450}$ relation is not a robust trend within individual clouds.
The median per-cloud coefficient of this relation is close to zero ($r_s = -0.02$, with the full-sample value $r_s = -0.29$ when OMC\,1 is excluded but only 3 of 7 valid fields concordant with the full-sample positive sign).
The sign varies among individual fields.
The stronger result for individual clouds at 450\,$\mu$m is therefore the dependence on column density reported above rather than a direct dependence on temperature.
 
The apparent positive $T_{\rm d}$--$P_{850}$ correlation in the pooled sample is not a universal relation within clouds.
Excluding OMC\,1 reduces the coefficient to nearly zero ($r_s = +0.03$, $N=321$, $p = 0.55$), and the median per-cloud coefficient is also close to zero ($r_s = +0.04$, with only 4 of 8 valid fields concordant with the full-sample positive sign and the cluster block-bootstrap CI $[-0.21,\,+0.64]$ overlapping zero).
We therefore interpret the association of warm dust with high $P_{850}$ as a signature of irradiated conditions like those in OMC\,1, which strongly affect the pooled coefficient.
It is not a general law across the HGBS sample.
 
The anti-correlation between $I_{450}/I_{850}$ and $P_{450}/P_{850}$ remains negative in the pooled sample ($r_s = -0.49$, $N=327$) and in the cluster block-bootstrap estimate ($r_s = -0.45$, 95\% CI $[-0.74,\,-0.10]$), including after OMC\,1 exclusion ($r_s = -0.35$, $N=265$).
However, the median of the coefficients for individual clouds is close to zero ($r_s = -0.03$ with only 4 of 7 valid fields concordant with the negative full-sample sign).
The strongest signature among individual clouds lies within OMC\,1 itself ($r_s = -0.93$ on $N=62$ matched-resolution samples).
We therefore do not interpret this relation as a universal law for individual pixels within clouds.
This relation appears instead to encode a mixture of environmental offsets between clouds and internal structure within OMC\,1, which strongly influences the pooled coefficient.
 
The $N_{\rm H}$--$T_{\rm d}$ relation provides a useful control case.
In the pooled sample the correlation is positive ($r_s = +0.41$ for the full above-floor sample of $N = 417$ matched-resolution samples).
However, the coefficients for individual clouds show the opposite behavior: 8 of 8 valid HGBS fields exhibit $r_s < 0$ (per-cloud median $r_s = -0.31$, with the strongest signal $r_s = -0.78$ in L1688\,Ophiuchus\,A), so individual clouds preferentially follow the classical cold-core trend.
We therefore interpret the pooled positive coefficient as an inter-cloud effect, driven by the fact that the high-radiation fields occupy both higher-temperature and higher-column regimes.
The cluster block-bootstrap CI on the pooled $r_s$ ($[-0.12,\,+0.69]$) spans zero.
This result confirms that the uncertainty from the cluster bootstrap on the positive full-sample value is substantially larger than the CI based on individual pixels suggests.
This reversal is similar to Simpson's paradox.
It is consistent with HGBS studies that find higher column densities at lower dust temperatures within individual clouds \citep[e.g.,][]{Konyves2015,Marsh2016}, and shows why we must examine the statistics for each cloud alongside pooled correlations.
 
\begin{table*}
\centering
\caption{
Tests for individual clouds of the HGBS environmental correlations among pixels.
The Full and OMC\,1-excluded columns give pooled Spearman coefficients on the above-floor sample.
The per-cloud median and same-sign fraction test whether the pooled trend is reproduced within individual clouds.
The same-sign fraction gives the number of valid HGBS fields whose individual coefficient has the same sign as the full pooled coefficient (e.g.\ for $N_{\rm H}$--$T_{\rm d}$ the full pooled sign is positive and 0 of 8 valid fields have $r_s>0$).
The block-bootstrap 95\% CI resamples HGBS fields rather than pixels ($N_{\rm boot}=10^4$, random-number generator seed $20260508$).
\label{tbl:source_aware_hgbs_correlations}
}
\begin{tabular}{lccccc}
\hline
Correlation & Full $r_s$ & Excl.\ OMC\,1 & Per-cloud median & Same-sign fraction & Block-bootstrap 95\% CI \\
\hline
$N_{\rm H}$--$T_{\rm d}$                  & $+0.41$ & $+0.15$ & $-0.31$ & $0/8$ & $+0.35~[-0.12,+0.69]$ \\
$N_{\rm H}$--$P_{450}$                    & $-0.07$ & $-0.46$ & $-0.44$ & $7/7$ & $-0.09~[-0.62,+0.21]$ \\
$N_{\rm H}$--$P_{850}$                    & $+0.21$ & $-0.13$ & $-0.21$ & $3/8$ & $+0.18~[-0.34,+0.54]$ \\
$T_{\rm d}$--$P_{450}$                    & $+0.13$ & $-0.29$ & $-0.02$ & $3/7$ & $+0.12~[-0.48,+0.56]$ \\
$T_{\rm d}$--$P_{850}$                    & $+0.36$ & $+0.03$ & $+0.04$ & $4/8$ & $+0.34~[-0.21,+0.64]$ \\
$I_{450}/I_{850}$--$P_{450}/P_{850}$ & $-0.49$ & $-0.35$ & $-0.03$ & $4/7$ & $-0.45~[-0.74,-0.10]$ \\
\hline
\end{tabular}
\end{table*}
 
These tests for individual clouds change our interpretation of the HGBS correlations.
Several pooled coefficients are not universal relations within clouds, but mixtures of trends inside clouds, offsets between clouds, and the disproportionate influence of OMC\,1.
The $N_{\rm H}$--$T_{\rm d}$ relation provides the clearest control case.
The median of its coefficients for individual clouds shows the classical negative cold-core behavior, even though the pooled coefficient is positive.
This contrast demonstrates an effect across clouds similar to Simpson's paradox.
The most robust within-cloud polarization trend is instead the anti-correlation between $N_{\rm H}$ and $P_{450}$, which has the same sign in all seven fields with a sufficient dynamic range above the detection floor.
By contrast, the association of warm dust with high $P_{850}$ is not a universal relation across the HGBS sample.
It is concentrated in irradiated conditions like those in OMC\,1.
Finally, the anti-correlation between $I_{450}/I_{850}$ and $P_{450}/P_{850}$ remains negative in the pooled and block-bootstrap estimates.
However, the median of its coefficients for individual clouds is near zero.
This median indicates that the anti-correlation reflects a combination of environmental offsets between clouds and internal structure within one influential field rather than a single universal law on the scale of pixels.
 
The tests for individual clouds show three aspects of the F6 wavelength-dependent response.
$P_{450}$ has a robust anticorrelation with column density within individual clouds.
The coefficients are negative in all seven fields with sufficient usable range.
This sign pattern is the most consistent among the polarization relations examined.
The association of warm dust with high $P_{850}$ is concentrated in irradiated conditions like those in OMC\,1.
The trends in the ratios between bands combine structure within clouds, differences between clouds, and structure within one influential field.
At the population scale, F4 (existence of the polarization discrepancy between the bands across the BISTRO {\sc PolSrc} sample) is unaffected by the analysis of individual clouds, because F4 is a statement about individual sources rather than about individual pixels.
At the environment scale, the recognition that several pooled HGBS trends arise from cross-cloud environmental offsets does not translate into the ordered LMSF/HMSF/CMZ response tested by F5, which the bootstrap CIs and Jonckheere--Terpstra (JT) permutation $p$-values of subsection~\ref{ss:S2_EnvStrat} do not support.
Section~\ref{sec:S2_discussion} and Paper~S1 (subsection~\ref{S1-ss:Discussion_theory}) take up the implications for the interpretation in terms of size-dependent grain alignment.
Several other pooled relations combine cloud-to-cloud offsets with source-specific structure, so their interpretation as trends through the interiors of individual clouds requires the separate within-field results.
 
\subsubsection{Treatment of Influential Sources}
\label{sss:S2_influential_source_policy}
OMC\,1, Monoceros~R2, and Sgr~B2 probe distinct extremes: intense internal irradiation, an unusually large inter-band angle difference before the 450\,\micron\ Stokes-$Q/U$ zero-level offset correction, and the CMZ environment, respectively.
Three sources in the BISTRO dual-band sample exert a disproportionate influence on subsample medians: OMC\,1 in the HGBS regime with strong radiation, Mon\,R2 at the lower edge of the F2 distribution, and Sgr\,B2 within the stack of seven CMZ sources in subsection~\ref{ss:S2_EnvStrat}.
We treat them uniformly, as summarized in table~\ref{tbl:influential_source_policy}.
Each source remains in its primary sample, and a complementary verification test characterizes its influence on the statistics across the survey.
 
\begin{longrotatetable}
\begin{deluxetable*}{llll}
\tablecaption{Treatment of influential sources in the BISTRO dual-band sample\label{tbl:influential_source_policy}}
\tablewidth{0pt}
\tablehead{
\colhead{Source} &
\colhead{Sample-membership decision} &
\colhead{Verification test} &
\colhead{Reporting in this paper}
}
\startdata
OMC\,1 &
\parbox[t]{5.0cm}{\raggedright Retained in the full HGBS-9 per-pixel sample as the highest-significance high-radiation regime exemplar (subsection~\ref{ss:S2_HGBS}).\strut} &
\parbox[t]{6.0cm}{\raggedright OMC\,1-excluded comparison computed for every per-pixel correlation (table~\ref{tab:correlation_summary}); contrasts summarized in sub-subsection~\ref{sss:S2_correlation_ex_OMC1}.\strut} &
\parbox[t]{5.5cm}{\raggedright Inclusion/exclusion sign and amplitude shifts reported as a result on the high-radiation regime, with full and excluded values quoted side by side.\strut} \\
Mon\,R2 &
\parbox[t]{5.0cm}{\raggedright Retained in the {\sc PolSrc} and {\sc SpIdx} subsamples; bracketed against the broader sample in F1 and F2 reporting.\strut} &
\parbox[t]{6.0cm}{\raggedright {Test case for the sensitivity of the F1 angle comparison to the 450\,\micron\ Stokes-$Q/U$ zero-level offset correction and for the lower edge of the F2 distribution; flagged in Paper~S1 angle and spectral-index analyses.}\strut} &
\parbox[t]{5.5cm}{\raggedright Source-specific behavior reported as a single-source flag within F1 and F2 discussions, not as a sample-wide finding.\strut} \\
Sgr\,B2 &
\parbox[t]{5.0cm}{\raggedright Retained as part of the CMZ environment stack ($N=7$; subsection~\ref{ss:S2_EnvStrat}).\strut} &
\parbox[t]{6.0cm}{\raggedright Single-source leave-one-out (LOO) check on $\mathcal{S}_{850}$, $\gamma-\alpha$, and $R_p$ (sub-subsection~\ref{sss:env_synthesis}).\strut} &
\parbox[t]{5.5cm}{\raggedright Covered by the range of the disk--CMZ contrast under single-source exclusion (sub-subsection~\ref{sss:env_synthesis}); no Sgr\,B2-specific value is quoted.\strut} \\
\enddata
\tablecomments{
All three sources remain in their primary samples.
Their influence is assessed by comparing results with and without OMC~1 pixels, examining Mon~R2 as a diagnostic of the sensitivity of the inter-band angles to the 450\,\micron\ Stokes-$Q/U$ zero-level offset correction (Paper~S1 subsection~\ref{S1-ss:Discussion_GammaAlpha}), and excluding Sgr~B2 in a source-level sensitivity check (subsection~\ref{ss:S2_EnvStrat}).
}
\end{deluxetable*}
\end{longrotatetable}
 
The verification tests collectively indicate that none of the three sources alters the qualitative conclusions reported in section~\ref{sec:S2_results} for the survey as a whole.
We regard the contrasts between including and excluding OMC\,1 (sub-subsection~\ref{sss:S2_correlation_ex_OMC1}) themselves as a finding about the high-radiation regime rather than as evidence against the robustness of F4--F6.
The Mon\,R2 signature is captured within the conventions of Paper~S1 for flagging individual sources.
The Sgr\,B2 leave-one-out check confirms that the CMZ placement in the F5 comparison among environmental classes is qualitatively unchanged.
Keeping these regions in the census preserves its astrophysical range.
The accompanying checks on individual sources are needed to distinguish results shared across the sampled clouds from those driven by a particular irradiated or geometrically complex region.
 
Within clouds, column density and irradiation provide two routes by which the observed polarization can change.
Shielding can accompany lower polarization in dense material, while internal heating can favor a different component of polarized emission.
\textit{This subsection completes the F6 test of wavelength-dependent polarization--environment associations}.
Per-pixel correlations in the HGBS 9-field subsample, refined by the per-cloud analysis of sub-subsection~\ref{sss:S2_source_aware_correlation} (table~\ref{tbl:source_aware_hgbs_correlations}), establish that the wavelength dependence of polarization at the per-pixel level is hierarchical rather than a single universal relation with local physical conditions.
The most robust within-cloud trend is the classical 450\,\micron\ polarization hole.
The $N_{\rm H}$--$P_{450}$ relation is negative in 7 of 7 valid HGBS fields with a sufficient dynamic range above the detection floor, and the full-sample value strengthens to $r_s = -0.46$ when OMC\,1 is excluded.
The association of warm dust with high $P_{850}$ is concentrated in irradiated conditions like those in OMC\,1 rather than holding across typical HGBS fields.
The same dependence on OMC\,1 carries the positive $T_d$--$P_{850}$ trend in the full sample ($r_s = +0.36$, falling to $+0.03$ when OMC\,1 is excluded).
The anti-correlation of the ratios between bands, $r_s(I_{450}/I_{850}, P_{450}/P_{850})$, remains negative in the pooled and cluster block-bootstrap estimates.
However, this anti-correlation combines environmental offsets between clouds with internal structure within OMC\,1, which strongly influences the pooled coefficient.
Separately, the associations of $\mathcal{S}\times\mathcal{P}^{\rm frac}$ with column density also change when OMC~1 is excluded (sub-subsection~\ref{sss:S2_SP_vs_N}).
The HGBS-9 sample therefore reveals a layered structure---an OMC\,1-like high-radiation regime, a robust within-cloud column-density polarization hole at 450\,\micron, and a cross-cloud component that drives the dual-band ratio---rather than a single universal per-pixel law.
 
The F4--F6 results form the joint observational benchmark for the size-dependent grain-alignment interpretation developed in Paper~S1.
F4 establishes universality across the population and is unaffected by the analysis of individual clouds because it concerns individual sources.
F5 tests the ordered regime-dependent wavelength response and does not detect it.
Section~\ref{sec:S2_discussion} gives the joint interpretation of F4--F6.
The coefficients between 450\,$\mu$m polarization and column density are negative in all seven fields with sufficient usable range, whereas the warm, highly polarized lines of sight in the 850\,$\mu$m analysis are concentrated in OMC\,1.
The correlations of $\mathcal{S}\times\mathcal{P}^{\rm frac}$, the proxy for grain alignment efficiency, are complementary to the results for column density and temperature.
These correlations do not establish that magnetic disorder mediates the wavelength difference.
 
The survey separates a common deficit in polarization at the shorter wavelength from the ways that this deficit is expressed in individual molecular clouds.
The three tests of section~\ref{sec:S2_results} therefore return a coherent observational answer to the motivating question of subsection~\ref{ss:S2_intro_series}.
At the population scale, F4 shows that the polarization discrepancy between the two wavelengths, first established as F1--F3 in Paper~S1, is present across the BISTRO 450/850\,\micron\ {\sc PolSrc} sample ($\gamma<\alpha$ in 24--27 of 35 sources, median $\gamma-\alpha$ between $-0.57$ and $-0.45$).
At the regime scale, F5 returns a null result: none of $\gamma-\alpha$, $R_p$, $\mathcal{S}_{850}$ or $\mathcal{S} \times \mathcal{P}^{\rm frac}$ orders monotonically along the LMSF--HMSF--CMZ sequence (table~\ref{tbl:env_4metric}).
For $\gamma-\alpha$ and $R_p$ the classes are neither ordered along the sequence nor robustly different under the Kruskal--Wallis test.
We apply no test to either $\mathcal{S}_{850}$ or $\mathcal{S} \times \mathcal{P}^{\rm frac}$.
At the scale of individual pixels, F6 shows three aspects of the response (table~\ref{tbl:source_aware_hgbs_correlations}).
The $N_{\rm H}$--$P_{450}$ anticorrelation within clouds holds in 7 of 7 valid HGBS fields.
The association of warm dust with high $P_{850}$ is driven by OMC\,1.
The pooled anticorrelation of the ratios between bands combines structure within clouds, differences between clouds, and the internal physics of OMC\,1.
Taking F4--F6 together, we conclude that the dual-wavelength polarization discrepancy is a population-level property of the survey that this sample does not resolve into an environmental sequence, while its per-pixel coupling to local cloud conditions is band-dependent and hierarchically structured.
These results form a single coherent benchmark that any candidate model of grain alignment must reproduce.
Models of polarized dust emission can be assessed against the widespread measured deficit, the lowest median in the HMSF class (not robustly detected as a population difference), and the contrast between the within-cloud polarization relations and the irradiated conditions in OMC\,1.
The proposed ordered environmental sequence is not detected, and distance and environment remain confounded in the CMZ comparison.

\section{Discussion}
\label{sec:S2_discussion}
 
\subsection{Survey-Wide Trends and Local Wavelength Dependence}
\label{ss:S2_disc_universality}
The two bands usually trace similar projected field orientations even when their polarization fractions differ substantially.
This combination shifts attention from a change in the mean projected direction to the spectral weighting of polarized emission and to unresolved structure within the emitting material.
 
We now consider the physical implications of the F4--F6 results in section~\ref{sec:S2_results}.
The result is asymmetric: the F1 inter-band angle agreement set in Paper~S1 generally holds where the angles are well constrained, whereas the wavelength-amplitude observables ($R_p$, $\gamma-\alpha$, and the per-pixel response of $P$, $\mathcal{S}$, and $\mathcal{S}\times\mathcal{P}^{\rm frac}$ to $N_{\rm H}$ and $T_d$) respond to local cloud conditions in a band-dependent way.
However, table~\ref{tbl:env_4metric} shows that none of these quantities orders along the LMSF--HMSF--CMZ sequence, so the asymmetry that the data establish is one between scales rather than between star-forming regimes.
The central physical result is that the two bands trace similar projected field orientations, while the polarized dust emissivity varies from cloud to cloud.
The present sample does not resolve those variations into an environmental sequence.
 
Any candidate mechanism of grain alignment for the submillimeter polarization spectrum must therefore reproduce both constraints simultaneously: the signatures across the survey established by F1--F3 in Paper~S1 and the locally conditioned, band-dependent response documented here as F4--F6.
The remainder of section~\ref{sec:S2_discussion} examines this asymmetry.
Subsection~\ref{ss:S2_disc_topology} organizes F4--F6 as three observational tests of environmental dependence.
Subsection~\ref{ss:S2_disc_grains} traces the implications of this asymmetry for size-dependent grain alignment, and subsection~\ref{ss:S2_disc_scope} discusses the limits of the present analysis.
The survey therefore constrains both the relative spectra of polarized and total emission and the cloud conditions under which their local behavior changes.
Similar projected angles do not exclude unresolved geometric depolarization, and the environmental comparisons do not isolate a unique alignment mechanism.

\subsection{Three Observational Tests of Environmental Dependence}
\label{ss:S2_disc_topology}
\begin{figure*}[ht]
\centering
\includegraphics{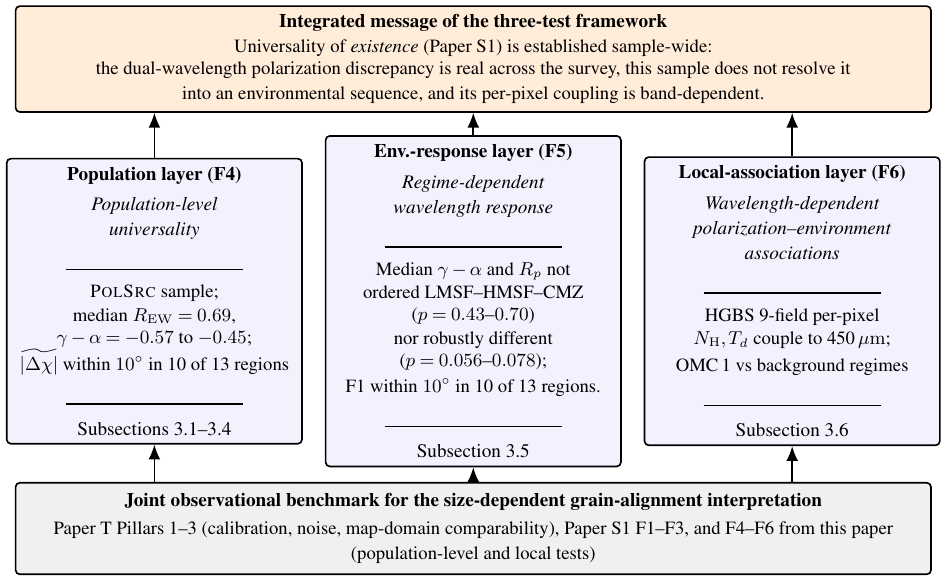}
\caption{
\textbf{Population, environment, and local-cloud comparisons of dual-band polarization.}
F4 examines the prevalence of the Paper~S1 results on the polarization ratio, the spectral indices, and the inter-band angle in the samples available for each observable.
F5 compares LMSF, HMSF, and CMZ populations.
We do not detect an ordered progression, and tests that do not assume an ordering do not robustly detect a difference, although the HMSF class has the lowest median (sub-subsection~\ref{sss:env_synthesis}).
F6 examines wavelength-dependent associations with $N_{\rm H}$ and $T_{\rm d}$, including those of the proxy $\mathcal{S}\times\mathcal{P}^{\rm frac}$ for grain alignment efficiency, and tests the influence of individual clouds.
This synthesis complements the question-to-test sequence in figure~\ref{fig:S2_test_overview}.
\\ {Alt text: Diagram of text boxes linked by arrows.}
}
\label{fig:S2_topology_summary}
\end{figure*}

The observations support three complementary comparisons: the prevalence of the wavelength contrast across the cloud sample, differences between the sampled environments, and correlations within individual clouds.
The three tests are not independent anomalies; they are nested comparisons, from the whole sample through the environmental classes to individual clouds.
We summarize F4--F6 in three layers (figure~\ref{fig:S2_topology_summary}).
The \emph{population layer} tests whether the F1--F3 signatures recur across the sample.
The \emph{environmental-response layer} tests their variation across LMSF, HMSF, and CMZ.
The \emph{local-association layer} compares the 450 and 850\,\micron\ responses to local cloud conditions and to the disorder of the projected $\vec{B}$-field.
Neither an unordered population difference nor the proposed environmental sequence is robustly detected.
Read together with the correlations that vary from source to source, this result constrains models of polarized dust emission without isolating the alignment mechanism or removing the dependence on sampled physical scale.
 
\paragraph*{Population layer (F4).}
A difference between the spectra of polarized and total dust emission recurs across the measured sources.
The F4 test (subsections~\ref{ss:S2_Sdisp}--\ref{ss:S2_SpIdx}) establishes that the Paper~S1 signatures of $R_p<1$ and $\gamma<\alpha$ are recovered across the BISTRO dual-band sample, and that inter-band $\vec{B}$-field angle agreement generally holds where the angles are well constrained.
We interpret this result as evidence that the wavelength-dependent polarization discrepancy is not an artifact of representative sources or of the smaller statistical subset of Paper~S1, but a property of the survey population.
The ratios and spectral-index differences measured for individual sources therefore constrain the effective emissivity of aligned dust relative to the total emitting material.
These population statements apply to the subset available for each diagnostic, rather than to every member of the 69-source parent census.
 
\paragraph*{Environmental-response layer (F5).}
The environmental comparison examines whether the short-wavelength polarization deficit follows a steady progression with star-forming regime.
The observed spectral-index medians are instead lowest in the HMSF population, with LMSF and CMZ close to zero.
The F5 test (subsection~\ref{ss:S2_EnvStrat}) does not detect the ordered variation it posed.
The medians of $\gamma-\alpha$ and $R_p$ for each environment are not ordered along the LMSF--HMSF--CMZ sequence.
The ordered-alternative test returns $p = 0.43$--$0.70$.
The Kruskal--Wallis test across the same three classes does not robustly detect a difference ($p = 0.056$--$0.078$ with the pointings treated as independent, $0.077$--$0.101$ with the co-spatial Orion~A Field~4 / OMC~1 pair counted once, and $0.026$--$0.041$ without the Rosette Molecular Cloud), although the HMSF class has the lowest median in every treatment.
The data therefore show neither a sequence nor a robust separation.
The one two-class contrast that does not assume an order, Galactic disk against CMZ, gives one-sided $p = 0.132$ with the sources treated as independent, $p = 0.18$ with the five Orion~A pointings entered as one cloud, and between $0.086$ and $0.23$ when individual sources are excluded one at a time.
The interquartile widths are comparable ($0.51$ in $\gamma-\alpha$ for HMSF against $0.84$ for LMSF and $0.59$ for CMZ).
With $N = 9$ and $7$ in the LMSF and CMZ classes, we do not test the dispersions or draw a conclusion from these widths.
Ten of the thirteen regional medians $\widetilde{|\Delta\chi|}$ lie within the adopted $10\arcdeg$ tolerance, and the larger medians in the other three regions occur where polarized intensities are weak and the angle uncertainties for individual pixels are higher (sub-subsection~\ref{sss:env_ladder}).
We do not test the significance of the individual regional medians, and these regional measurements do not form an environment-stratified angle distribution.
The three classes also form a distance sequence over which the matched beam samples 0.008--0.51\,pc.
Environment and physical resolution therefore remain confounded for the CMZ contrast, because the distance ranges do not overlap at all (sub-subsection~\ref{sss:SP_bias}; appendix~\ref{app:S2_dist_bias}).
The medians for the HMSF sources are displaced from those of the other two classes, while the proposed ordered environmental sequence is not detected.
The lower HMSF median cannot be assigned to star-formation conditions alone, because the classes also sample different physical scales, particularly for the CMZ.
 
\paragraph*{Local-association layer (F6).}
The interior of the clouds and the strongly irradiated regions do not need to show the same relation between polarized emission, column density, and temperature.
The HGBS analysis of individual clouds (sub-subsection~\ref{sss:S2_source_aware_correlation}, table~\ref{tbl:source_aware_hgbs_correlations}) shows that the coupling to local cloud conditions is hierarchical rather than a single universal law at the pixel scale.
The $450\,\mu$m polarization fraction shows a robust decrease with column density within clouds, while the association of warm dust with high polarization at $850\,\mu$m is concentrated in irradiated conditions like those in OMC\,1.
The trend in the ratio between the bands persists in the pooled and cluster block-bootstrap analyses but combines environmental offsets between clouds with internal structure within one influential field.
Outside OMC\,1, the proxy $\mathcal{S}\times\mathcal{P}^{\rm frac}$ for grain alignment efficiency declines with column density in both bands (sub-subsection~\ref{sss:S2_SP_vs_N}).
 
These three layers form a single observational picture rather than three independent census results.
F4 establishes the displacement across the population.
F5 tests whether this displacement varies systematically across star-forming regimes and detects neither an ordered trend nor a robust difference among the classes.
F6 reveals that the response of individual pixels within those regimes has several components rather than following a single universal law (table~\ref{tbl:source_aware_hgbs_correlations}).
These components are a robust anticorrelation between $N_{\rm H}$ and $P_{450}$ within clouds, an association of warm dust with high $P_{850}$ driven by OMC\,1, and a combination of contributions to the ratio between bands.
Together they define the observational benchmark that a viable grain-alignment model must reproduce: sample-wide existence, an amplitude that this sample does not resolve into an environmental sequence, and a hierarchical local coupling revealed by the per-cloud tests.
The correlation coefficients between polarization at 450\,$\mu$m and column density are negative in all seven fields with sufficient usable range, while the warm, highly polarized 850\,$\mu$m signal is concentrated in OMC\,1.
The two patterns are distinct observational constraints.
The present $\mathcal{S}\times\mathcal{P}^{\rm frac}$ analyses do not establish a wavelength difference or identify magnetic disorder as its cause.
 
\subsection{Implications for Size-Dependent Grain Alignment Across Star-Forming Regimes}
\label{ss:S2_disc_grains}
A lower polarization fraction at 450\,$\mu$m constrains how aligned dust contributes to the submillimeter emission relative to the full dust column.
Grain-size-dependent alignment is one possible explanation, but temperature structure and magnetic geometry can also alter the relative weighting of polarized emission.
 
The three-layer view of subsection~\ref{ss:S2_disc_topology} has a natural physical reading within the framework developed in Paper~S1 for size-dependent grain alignment.
In this interpretation, the 450 and 850\,\micron\ polarization channels do not respond identically to the same cloud environment because they weight the aligned grain population, the local radiation field, and the degree of $\vec{B}$-field disorder differently.
The negative spectral-index differences and ratios below unity describe a short-wavelength polarization deficit.
The HMSF class carries the most negative median $\gamma-\alpha$ of the three (table~\ref{tbl:env_4metric}), while the LMSF and CMZ medians lie closer to zero.
This descriptive contrast is not a robustly established environmental separation, and the classes also differ in physical resolution, so it does not by itself show where radiative-torque alignment \citep{Lazarian2007,Andersson2015,HoangLazarian2016} is stronger or weaker.
 
The wavelength dichotomy of F6, the absence of a detected ordering across regimes in F5, and the displacement of F4 across the population are therefore three observational constraints on the same underlying class of mechanisms:
models in which the contribution of aligned grains to polarized emissivity responds differently at 450 and 850\,\micron\ as the local radiation field, density structure, and magnetic disorder change.
Paper~S1 develops the detailed physical interpretation in terms of grain-size distributions, evolution of dust properties across regimes, and RAT alignment efficiencies.
We do not identify a unique microscopic mechanism of grain alignment.
We instead specify the observations that a viable model must reproduce: a widespread discrepancy between the bands in polarization amplitude and local environmental response, with no detected ordering across the environmental classes.
The warm, highly polarized 850\,$\mu$m behavior is concentrated in OMC\,1.
The observations do not determine whether grain-size-dependent alignment, temperature mixtures, or unresolved geometry dominates the wavelength contrast.

\subsection{Scope and Limits of the Present Environmental Census}
\label{ss:S2_disc_scope}
Our physical interpretation is limited by the way cloud conditions and spatial scales are sampled.
The CMZ clouds are observed on much larger linear scales than nearby disk clouds, while the HGBS correlations combine regions with unequal ranges of density, temperature, and polarized emission.
 
Several limitations apply to this analysis.
First, the F6 correlations use only the 9-field HGBS subsample.
OMC\,1 contributes many of its high-$P_{850}$ pixels and therefore has a strong influence on the pooled results.
We have therefore reported all F6 results both with and without OMC\,1 and draw our conclusion on the band dependence of these correlations from the sample that excludes OMC\,1.

Second, the \textit{Herschel}-derived $N_{\rm H}$ and $T_d$ for each pixel describe material integrated along the line of sight and do not resolve the three-dimensional structure of the molecular clouds.
The F6 correlations therefore measure the projected coupling of polarization to large-scale physical conditions rather than to the microscopic environment of individual aligned grains.

Third, the interpretation in terms of size-dependent grain alignment, outlined in subsection~\ref{ss:S2_disc_grains} and developed in Paper~S1, does not involve full radiative-transfer inversion of the dust emission spectrum or polarization spectrum.
The F4--F6 benchmark constrains \emph{what} changes across environments in the observations rather than the microscopic mechanism behind it.

Fourth, we performed the F6 per-pixel correlation analysis on the matched $12\arcsec$ pixel grid after harmonization to the common $12\farcs6$ effective resolution (Paper~T~subsection~\ref{T-ss:beamval}).
These samples are comparable in size to the beam, but adjacent samples are not strictly independent.
Beam autocorrelation, residual {\tt pol2map} covariance, common calibration within a source, and shared cloud conditions all introduce correlations.
We used the tests for individual clouds in sub-subsection~\ref{sss:S2_source_aware_correlation} (table~\ref{tbl:source_aware_hgbs_correlations}) to separate trends within clouds from environmental offsets between clouds and from internal structure within a single influential source.
These tests provide the cluster-bootstrap uncertainties that the formal per-pixel $p$-values of table~\ref{tab:correlation_summary} alone do not supply.

Fifth, the LMSF, HMSF, and CMZ classes are also a distance sequence (environment rank versus distance, Spearman $\rho = 0.828$, $p = 8\times10^{-10}$), over which the $12\farcs6$ matched beam subtends 0.008--0.51\,pc.
The CMZ class does not overlap in distance with either disk class.
Any environmental difference in these observables, and the CMZ median $\gamma-\alpha = -0.10$ ($-0.22$ to $+0.01$ for the two extreme placements of the $3\sigma$ limits) in particular, therefore, could not be uniquely attributed to the environment rather than to the physical resolution (sub-subsection~\ref{sss:SP_bias}; appendix~\ref{app:S2_dist_bias}).
In that appendix, removing the fitted distance trend leaves $JT = 169.0$, $p \approx 0.68$, compared with $JT = 178.0$, $p \approx 0.57$ on the raw values.
We use the data on the raw $4\arcsec$ pixel grid only for Stokes-$I$ contours in the $\vec{B}$-field orientation atlas (appendix~\ref{app:S2_atlas}).
These data improve the display of significant intensity structure and do not enter the quantitative correlation analysis.

Sixth, to our knowledge, the modulation efficiency of POL-2 has not been calibrated in either band (Paper~T~subsection~\ref{T-ss:flux}).
A difference in the modulation efficiency between the two bands would scale $P_{450}/P_{850}$ by a common factor and shift $\gamma-\alpha$ by a common amount for all sources.
 
The first five limitations bound our causal interpretation of the present results but do not weaken our central observational conclusion.
The sixth limitation concerns the level of the wavelength contrast but not its variation between sources.
We interpret the level of the contrast on the assumption that the modulation efficiency of POL-2 is equal in the two bands (sub-subsection~\ref{sss:Synthesis}).
The polarization response at the two wavelengths is a property of the surveyed population, and its variation between sources is not resolved into an environmental sequence by this sample.
The data establish a prevalent wavelength contrast in the selected material and identify source-dependent associations with cloud conditions.
The data do not uniquely separate environment from resolution or partition the contrast among grain properties, alignment, and unresolved structure along the line of sight.

\section{Summary}
\label{sec:S2_summary}
 
We have presented a dual-wavelength polarimetric analysis of the BISTRO 450/850\,\micron\ survey of 69 star-forming regions.
We use three complementary tests to ask whether the Paper~S1 signatures F1--F3 recur across the environments accessible from the JCMT or vary with environment and local cloud conditions.
Figure~\ref{fig:S2_test_overview} and table~\ref{tbl:S2_test_index} summarize the tests.
Section~\ref{sec:S2_discussion} and figure~\ref{fig:S2_topology_summary} give their joint interpretation:
 
\begin{enumerate}
  \item \textit{F4 --- Population-level universality test of F1--F3.}
  We conclude that F1--F3 are population-level phenomena rather than artifacts of representative sources.
This conclusion follows from measurements of $\mathcal{S}$ (subsection~\ref{ss:S2_Sdisp}), $\mathcal{S}\times\mathcal{P}^{\rm frac}$ (subsection~\ref{ss:S2_SP}), $R_p\equiv P_{450}/P_{850}$ via three complementary statistical estimators (Methods~1--3; subsection~\ref{ss:S2_P_ratio}), and the dust spectral indices $\alpha$ and $\gamma$ (subsection~\ref{ss:S2_SpIdx}) for individual sources across the BISTRO {\sc PolSrc} sample.
The median is $R_{\rm EW} = 0.69$ over the 30 {\sc SpIdx} sources carrying a measured ratio.
For the spectral indices, $\langle\alpha\rangle = 3.33$, $\langle\gamma\rangle = 2.85$--$3.04$, and the median $\gamma - \alpha$ lies between $-0.57$ and $-0.45$ (69\,\%--77\,\% of the {\sc SpIdx} subsample have $\gamma<\alpha$).
We find $\widetilde{|\Delta\chi|}<10\degr$ in ten of the thirteen regional measurements.
 
  \item \textit{F5 --- Test for a regime-dependent wavelength response: no ordered sequence and no robust difference among the classes.}
  Stratification of the {\sc SpIdx} subsample across LMSF ($N=9$), HMSF ($N=19$), and CMZ ($N=7$) sub-environments (subsection~\ref{ss:S2_EnvStrat}) does not order the classes.
The medians of $\gamma-\alpha$ and $R_p$ for each environment do not follow the LMSF--HMSF--CMZ sequence (figures~\ref{fig:env_phase} and~\ref{fig:env_ladder}), and the ordered-alternative test returns $p = 0.43$--$0.70$.
The Kruskal--Wallis test, which does not assume an ordering, returns $p = 0.056$--$0.078$, and $p = 0.026$--$0.041$ without the Rosette Molecular Cloud.
We therefore do not robustly detect a difference among the classes, although the HMSF class has the lowest median in every treatment.
The one two-class contrast defined without reference to star-formation state (Galactic disk against Central Molecular Zone) gives one-sided $p = 0.132$ with the sources treated as independent, $p = 0.18$ with the five Orion~A pointings entered as one cloud, and between $0.086$ and $0.23$ when individual sources are excluded one at a time.
The fractions of sources with $\gamma<\alpha$ at $1\sigma$ in each environment (LMSF 3/9, HMSF 16/19, CMZ 4/7) count $1\sigma$ detections.
We compare these fractions only in the exploratory contrast of HMSF against the other two classes.
The regional median $\widetilde{|\Delta\chi|}$ lies within the adopted $10\degr$ tolerance for 10 of the 13 regional measurements covering the sample.
 
  \item \textit{F6 --- Wavelength-dependent polarization--environment associations}.
  Per-pixel correlations between the polarimetric observables and \textit{Herschel}-derived $N_{\rm H}, T_d$ in the HGBS 9-field subsample (subsection~\ref{ss:S2_HGBS}) show that the coupling of polarization to local cloud conditions differs between 450 and 850\,\micron\ and depends on influential regions.
At 450\,\micron\ the classical polarization-hole signature is recovered ($r_s = -0.46$ for $N_{\rm H}$--$P_{450}$ outside OMC\,1).
The same pixel set excluding OMC\,1 gives a pooled negative $T_d$--$P_{450}$ correlation ($r_s = -0.29$).
However, the median of the coefficients for individual clouds is $-0.02$, with 3 of 7 fields sharing the positive sign of the full-sample coefficient, so the temperature relation is not a trend shared within individual clouds.
At 850\,\micron, in contrast, the warm, highly polarized lines of sight, consistent with RAT expectations, are concentrated in OMC\,1, and their inclusion flattens or reverses the pooled polarization-hole signature (table~\ref{tab:correlation_summary}, sub-subsection~\ref{sss:S2_correlation_ex_OMC1}).
The accompanying measurements of the grain-alignment efficiency proxy $\mathcal{S}\times\mathcal{P}^{\rm frac}$ do not identify magnetic disorder as the cause of the wavelength-dependent correlation patterns.
\end{enumerate}
 
Together, these tests show a widespread discrepancy in polarization between the bands: a polarization spectrum that rises with wavelength between 450 and 850\,\micron.
The present sample does not resolve this discrepancy into an environmental sequence.
The relation of this discrepancy to local cloud conditions in individual pixels differs between the bands.
The 450\,\micron\ polarization fraction decreases with column density in all seven fields with sufficient usable range, whereas the warm, highly polarized 850\,\micron\ lines of sight are concentrated in OMC\,1.
These results and F1--F3 of Paper~S1 define the observational requirements for candidate mechanisms of grain alignment.
Section~\ref{sec:S2_discussion} discusses how F4--F6 constrain the interpretation in terms of size-dependent grain alignment developed in Paper~S1 subsection~\ref{S1-ss:Discussion_theory}.
 
Paper~T provides the technical basis of our scientific analysis, and Paper~S1 develops the full physical interpretation of F1--F6 in terms of grain-size distributions, evolution of dust properties, and alignment by radiative torques.
In addition to the F4--F6 scientific tests, we provide the canonical definitions of the three $R_p$ estimators (Methods~1, 2, and 3 of subsection~\ref{ss:S2_P_ratio}) used throughout this series, tables for individual sources in the complete {\sc PolSrc} sample, and the $\vec{B}$-field orientation atlas at both wavelengths.
Appendix~\ref{app:S2_atlas} presents 65 of the 69 {\sc PolSrc} sources in figure form (all but W51 and Orion~A Fields~4, 5 and 7; see the notes to table~\ref{tab:surv} in Appendix), with the FITS products for individual sources in Dataverse.
 
The BISTRO census in both bands therefore places measurements of $P_{450}/P_{850}$ on a firm statistical footing across the diversity of star-forming regions that are bright in the submillimeter and accessible from the JCMT site.
Together with Paper~T and Paper~S1, it establishes the observational benchmark that future grain-alignment calculations and multi-wavelength polarimetric surveys \citep{Gandilo2016, Shariff2019, Fanciullo2022b, NguyenBichNgoc2024, Hull2019} must reproduce: sample-wide existence, an amplitude not resolved into an environmental sequence, and wavelength-dependent local environmental coupling.

The principal results of this paper fall into four levels of claim strength, ordered by the degree to which they rest on direct observation.
\textit{Directly measured.}
Using the matched-resolution dual-band BISTRO {\sc PolSrc} data, we measure the polarization-angle dispersion $\mathcal{S}_{850}$, the grain-alignment efficiency proxy $\mathcal{S}\times\mathcal{P}^{\rm frac}$, and $R_p$ via Methods~1--3 for individual sources.
We also measure the spectral indices $\alpha$ and $\gamma$ for individual sources in the {\sc SpIdx} subsample.
For the HGBS-9 fields, we calculate per-pixel correlations of polarization observables with $N_{\rm H}$ and $T_d$.

\textit{Empirically inferred.}
The F4--F6 census statements follow from these measurements: we infer that the F1--F3 medians match the {\sc PolSrc} census (population-level universality), with the regional median $\widetilde{|\Delta\chi|}$ within the adopted $10\degr$ tolerance for 10 of the 13 regional measurements that cover the sample.
We do not detect an ordering of $\gamma-\alpha$ and $R_p$ along the LMSF--HMSF--CMZ sequence and do not robustly detect a difference among the classes.
The per-pixel correlation patterns indicate that the response of polarization to local cloud conditions differs between bands and depends on influential regions.

\textit{Physically interpreted.}
We interpret the asymmetry between the F1 angle agreement in ten of the thirteen regional measurements and the band-dependent per-pixel coupling documented by F6 as a variation in polarized emissivity rather than a breakdown of inter-band field-angle agreement.
A causal role for magnetic disorder in that coupling remains a physical interpretation rather than a measured proportionality or mediation effect.

\textit{Not uniquely determined.}
These observations do not uniquely determine the dust composition within each cloud, for which a full radiative-transfer inversion remains necessary.
The influence of individual sources beyond the checks in table~\ref{tbl:influential_source_policy} and the relative contributions of size-dependent alignment and field geometry to F6 are also not uniquely determined.
These observations also do not uniquely determine how much a difference in the modulation efficiency of POL-2 between the two bands contributes to the levels of $R_p$ and $\gamma-\alpha$.
We discuss these limitations in subsection~\ref{ss:S2_disc_scope}.
All data products for individual sources described in this paper will be made available electronically as supplementary material to this paper.
These products include the $\vec{B}$-field orientation atlas at both wavelengths (appendix~\ref{app:S2_atlas}), the tables of polarization fractions and spectral indices for individual sources, the environment-stratified summaries of table~\ref{tbl:env_4metric}, and the HGBS-9 per-pixel correlation outputs of table~\ref{tbl:source_aware_hgbs_correlations}.
 
\begin{acknowledgments}
RSF is deeply grateful to the medical staff who have continuously supported his recovery from cancer.
Their dedicated care made this work possible.
The authors thank Paul Ho, the former Director of the East Asian Observatory (EAO), for his full and unwavering support for this project, as well as the EAO staff for their continued assistance over a decade.
We also express our sincere gratitude to Dr.~Jessica Dempsey, former Deputy Director of the EAO, for her continuous and decisive leadership of the international instrument team throughout the commissioning and validation phase of POL-2.
Dr.~Dempsey sustained institutional support for the extensive test observations and performance assessments needed to establish POL-2 as a reliable scientific instrument, particularly to characterize instrumental polarization (IP).
She also enabled the participating teams to contribute their full capabilities.
In particular, she approved and supported dedicated IP validation experiments conducted with the JCMT dome cover removed, thereby enabling critical verification of the IP correction models developed independently by DB and PB.
We acknowledge Takeshi Inagaki, a UNIX systems engineer at the Subaru Telescope, who visited RSF for approximately two weeks each year from 2015 to 2020 and made substantial contributions to improving the efficiency and reliability of the data-analysis workflow.
We also thank the successive Directors of the Subaru Telescope, who not only approved Mr.~Inagaki's long-term annual visits, but also provided extensive in-kind support during JCMT observing runs from 2015 to 2019.
We gratefully acknowledge the Astronomy Data Center (ADC) of the National Astronomical Observatory of Japan (NAOJ) for long-term support and access to shared computing resources, including 32 UNIX machines and more than 100~TB of storage.
In 2022, Mr.~Inagaki also made \texttt{pol2map} run efficiently on all 32 of these machines, including the handling of jobs under the security settings of the ADC.
The authors are grateful to Drs.~Dan Bintley and Shaoliang Li (EAO), who generously shared their in-depth knowledge of SCUBA-2 instrumentation and the JCMT optics through close collaboration as members of the SCUBA-3 development team.
They provided critical insights into instrumental characteristics and systematic effects that informed the uncertainty analysis presented in this work.
We thank Prof.\ Doug Johnstone for helpful advice on this manuscript.
Discussions of observational uncertainties associated with submillimeter telescopes and bolometer cameras greatly benefited from interactions with Prof.~Ryohei~Kawabe and Dr.~Tai~Oshima (NAOJ).
We thank Prof.~Yoshito~Shimajiri (Kyushu Kyoritsu University, Japan) for discussions of bolometer imaging algorithms, Dr.~Hiroyuki~Hirashita (ASIAA, Taiwan) for discussions of dust properties, and Dr.~Riccardo~Cesaroni (INAF Arcetri, Italy) for advice on \emph{Herschel} data handling.
Discussions with Profs.~Hideki Takami and Yutaka Hayano (Subaru Telescope), drawing on their experience in adaptive optics development and operations, helped us understand atmospheric fluctuations at Maunakea relevant to submillimeter observations.
RSF expresses sincere gratitude to Prof.~Akira~Obuchi (Tokushima University, Japan) for extensive guidance and invaluable discussions on mathematical aspects relevant to this work.
Finally, we thank Mike Chen, Jungmi Kwon, Chi-Yan Law, Tetsuya Zenko, and Yapeng Zhang
for helpful discussions.
 
These observations were obtained by the James Clerk Maxwell Telescope, operated by the East Asian Observatory on behalf of The National Astronomical Observatory of Japan; Academia Sinica Institute of Astronomy and Astrophysics; the Korea Astronomy and Space Science Institute; the National Astronomical Research Institute of Thailand; Center for Astronomical Mega-Science (as well as the National Key R\&D Program of China with No.~2017YFA0402700).
Additional funding support is provided by the Science and Technology Facilities Council of the United Kingdom and participating universities and organizations in the United Kingdom and Canada.
The POL-2 data analyzed here were obtained under JCMT programs M16AL004, M17BL011, M20AL006 and M20AL018 (BISTRO Large Programs), together with E21AC002, E21BJ001, E21BJ002, M16AD003, M17AP019, M17AP067, M17AP073, M17AP074, M17BP037, M17BP041, M17BP046, M17BP050, M17BP070, M18AH27C, M18BD002, M18BP019, M18BP030, M18BP041, M18BP055, M19AP003, M19AP009, M19AP014, M19AP016, M19AP032, M19AP034, M19AP038, M19BP015, M19BP058, M19BP061, M20AP022, M20AP023, M20AP040, M20BP007, M20BP022, M20BP043, M20BP046, M21AH20A, M21AP016, M21AP021, M21BF004, M21BF005, M21BP004, M21BP045, M21BP083, M22AP018, M23BP011, R20XP002.
Programs M21BF004 and M21BF005, which observed fields in Orion~A, used observing time purchased through the EAO Individually Funded Program with JSPS KAKENHI Grant Numbers JP19H01938 and JP21H00045.
 
\textit{Use of AI tools.}
This study began in early 2016, and AI tools took part in it during roughly the last two years, from the autumn of 2024.
Their role changed as both the research and the tools developed.
We describe it here in some detail, because AI tools contributed substantially to the analysis code and to the text of all three papers of this series.
Before that period, until about 2023, RSF used the AI-based language tools Grammarly and DeepL under paid subscriptions.
He corrected the English of his drafts with Grammarly and checked their meaning by translating them back into Japanese with DeepL.
Some of the wording corrected with Grammarly probably remains in the present text, but it can no longer be identified.
 
Before AI tools were introduced, the observations and the basic data reduction had been completed by people.
All raw data were transferred from the JCMT Science Archive between 2017 March and 2024 February and reduced to Stokes $I$, $Q$, and $U$ maps (steps~1--3 of \texttt{pol2map}) by RSF on the computers of the NAOJ Astronomy Data Center.
No command or parameter of these steps was produced by an AI tool.
This reduction rests on the research of DB, who implemented \texttt{pol2map} in the Starlink SMURF package and advised on its use.
RSF produced the polarization-vector catalogs using his procedure; their scripts were later tidied with ChatGPT and refined with Claude Code.
The core analysis code was developed by RSF from 2015, together with Takeshi Inagaki until 2022, and RSF's notebooks of 2018--2024 underlie much of the present analysis.
From the autumn of 2024, AI tools took over part of the role that Mr.~Inagaki had played, helping RSF to debug this code and to run it in new computing environments.
Development of the analysis code proceeded iteratively between RSF and the AI tools, including substantial rewriting or re-implementation of parts of RSF's earlier code.
 
The principal tools were ChatGPT (OpenAI; GPT-4o from 2024 October, later GPT-5 and, from 2026 September, GPT-6 Astra) and Claude (Anthropic).
Claude was used through the AI assistant of the Warp terminal from the autumn of 2024 (Claude 3.5-series and other models; the model that answered cannot always be identified), through Claude Cowork, and through Claude Code.
The session logs of Claude Code, which survive from 2026 June onward, record the models Claude Opus 4.8, Claude Fable 5, Claude Opus 5, Claude Fable 5.1, and Claude Opus 5.5, with Claude Haiku 4.5 for auxiliary tasks.
The models used with Claude Code before 2026 June were not recorded.
The periods of use of these tools overlapped, and we do not assert the order in which they were adopted.
OpenAI Codex was used in 2026 September to review the work records and to draft working instructions.
In 2026 October, Codex also checked independently whether the cited papers support the statements for which we cite them (the model was not recorded in either period).
Gemini (Google) was used occasionally, including the summaries shown automatically with search results and Gemini Code Assist in a code editor.
Perplexity was used occasionally until 2026 May; it suggested points to check in the manuscripts, and text or arguments from it may have entered the papers.
 
The AI tools carried out work packages that could not have been completed by people in a reasonable time, and some work packages were led by them.
They shared the management of the large survey data set and the evaluation of calibration observations, for example of quasars, used as references when inspecting the science data.
They helped RSF to read and interpret DB's implementation in the Starlink software, the Starlink source code, and the accompanying documents, including e-mail correspondence.
They converted RSF's notebooks into stand-alone scripts and packages, wrote new analysis and plotting code to RSF's specifications, and modified and accelerated RSF's existing code.
They transcribed RSF's handwritten notes, articulated, organized, and consolidated RSF's ideas, and drafted and rewrote much of the English text.
In choosing the English language, RSF adopted many of their recommendations to match the strength of each claim to the statistical result.
The three papers were first planned as one, and their division was considered from late 2024 onward.
During this period the AI tools consolidated the many comments received in several rounds from more researchers than there are co-authors.
They were instructed to weigh each comment on its scientific merit alone, irrespective of who had made it, and this also eased communication in English among non-native speakers.
They kept the source counts and the numbers consistent among the text, tables, and figures of the three papers, a task that could not have been done by hand within realistic limits.
In the final stage they read the three manuscripts adversarially as simulated referees set up by RSF.
They also checked the cited statements against the cited papers, which RSF had obtained.
In 2026 October, they revised the English toward RSF's own style, as learned from his hand edits and his published papers.
The AI tools mainly handled short-term writing plans, while RSF set the medium- and long-term plans.
 
RSF set the astronomical motivation and the questions, specified the analyses, and verified the results from the astronomical point of view.
The verification from the data-science point of view was done mainly by the AI tools.
The statistical results were interpreted by RSF together with the AI tools.
RSF decided the outline of the astronomical conclusions, the AI tools then tested this outline adversarially, and RSF made the final decision on what to adopt.
RSF wrote the original versions of much of the analysis code, examined substantial parts of the code written or modified by the AI tools, and checked their outputs.
All visual checks, including separate displays of the original data, were made by RSF.
These checks found errors both in AI-written code and in RSF's earlier code, and the errors were corrected.
For example, RSF noticed that the statistics of one comparison of polarization angles were more tightly aligned than the maps suggested, which revealed a cut on the polarization-fraction ratio inherited from code written in 2021.
RSF also identified the cause of an error in which polarization-angle maps had been read in place of angle-dispersion maps.
 
The physical interpretation of the difference between the spectral indices of polarized and total intensity developed over several years in discussions among several of the authors, in the course of which RSF also used ChatGPT.
The contributions of the individuals and of the tool to these parts cannot now be separated, and we have not tried to reconstruct them from memory.
We assess the reliability of these parts from the analyses presented here and the authors' judgment, rather than their origin.
Whether, and how far, the co-authors used AI tools in preparing their own contributions and comments is not known to the first author.
The handling of unpublished data, manuscripts, and correspondence with these tools was judged comprehensively in light of the regulations of Tokushima University and the laws of Japan.
Under this judgment, RSF shared this material with the providers of these tools, mainly Anthropic, OpenAI, and Google.
The data were astronomical maps and catalogs that RSF had derived from raw data that are publicly available in the JCMT Science Archive.
RSF regards such sharing as a premise of using AI tools in research, and this study may have contributed modestly to the development of their models.
 
The figures in this paper were produced with Matplotlib or TikZ between 2026 April and 2026 October, that is, after AI tools had entered this work.
By the origin of their plotting code they fall into three groups.
Figures~\ref{fig:SpIdx_maps_OphA} and \ref{fig:SPvsN_individual} and the atlas of figures~\ref{afig:Bmap_L43L43}--\ref{afig:SgrB2_450um_StokesI} in Appendix were made with map-plotting code that RSF wrote between 2020 and 2024, before AI tools were used, and extended with AI tools from late 2025 onward.
Figures~\ref{fig:S850_OphiuA} and \ref{fig:SP450_OphiuA} were drawn with the same code from masked maps computed with code developed with AI tools.
Figures~\ref{fig:Pratio_OphA} and~\ref{fig:Vaillancourt_comparison} in the main text and figure~\ref{afig:StackedHist} in Appendix were made with notebooks and scripts that RSF developed while using AI tools interactively.
The contributions of RSF and of the tools to them cannot be separated.
Figures~\ref{fig:S2_test_overview}, \ref{fig:alpha_vs_gamma}--\ref{fig:env_ladder}, and \ref{fig:NvsT}--\ref{fig:S2_topology_summary} in the main text and figures~\ref{fig:gammaMalpha_residuals} and~\ref{fig:beamSize_beta} in Appendix were made with scripts written by Claude-based coding tools to RSF's specifications, in several cases re-implementing RSF's notebook code.
Figures~\ref{fig:S2_test_overview} and \ref{fig:S2_topology_summary} are TikZ diagrams drafted in the same way.
For submission, Claude Code typeset the TikZ diagrams and the figures with several panels as single figure files, without changing their content.
RSF inspected every figure against the underlying data.
 
No AI tool is an author of this paper.
The authors take full responsibility for its content, including every part prepared with the help of AI tools.
No output of these tools entered the paper without RSF's judgment, and RSF has read the whole text.
Without the commercially available models of OpenAI and Anthropic that took part in this work, these survey papers could not have been brought to submission before RSF's retirement.
\end{acknowledgments}
 
\section*{Funding}
RSF was supported by the Visiting Scholars Program provided by the NAOJ Research Coordination Committee, NINS (\mbox{NAOJ-RCC-23DS-050}).
RSF was supported by Grants-in-Aid for Scientific Research of Japan (JSPS KAKENHI Grant Numbers JP19H01938 and JP21H00045).
Team BISTRO-J gratefully acknowledges partial financial support from 260 individual contributors through a crowdfunding campaign.
 
\section*{Data Availability}
The raw SCUBA-2/POL-2 observations analyzed in this paper are available from the JCMT Science Archive hosted at the Canadian Astronomy Data Centre (CADC).
The \textit{Herschel} Gould Belt Survey column-density and dust-temperature maps used here are public.
The data products for individual sources described in appendix~\ref{app:S2_atlas} (Stokes $I$/$Q$/$U$ maps, polarization tables, and the dual-wavelength $\vec{B}$-field orientation atlas) are released as a Dataverse supplement to this paper.
 
\facilities{JCMT (SCUBA-2, POL-2), Herschel (SPIRE, PACS)}
 
\software{Starlink \citep{Currie2014}, SMURF \citep{Chapin2013}, \texttt{pol2map} \citep{Parsons2021}, \textsc{kappa}, Astropy \citep{Astropy2013,Astropy2018,Astropy2022}, NumPy \citep{Harris2020}, SciPy \citep{Virtanen2020}, Matplotlib \citep{Hunter2007}, \texttt{reproject}}
 
\appendix
 
\section{Stacked Distribution of 450\,\micron\ Polarization Fractions}
\label{app:S2_stacked}
The polarization fraction varies widely among the detected lines of sight toward star-forming clouds, so a single mean does not necessarily describe the typical 450\,$\mu$m emission.
The pooled distribution shows both the prevalent polarization level and the extent of the tail toward high polarization.
 
To characterize the overall distribution of the 450 \,\micron\ polarization fractions in the entire sample, we combine the debiased $P_{450}$ measurements from all observed fields with valid 450\, \micron \ polarimetric detections into a single stacked histogram.
We use this distribution to choose the summary statistics in the main text and tables.
 
\begin{figure}[ht]
\gridline{
    \fig{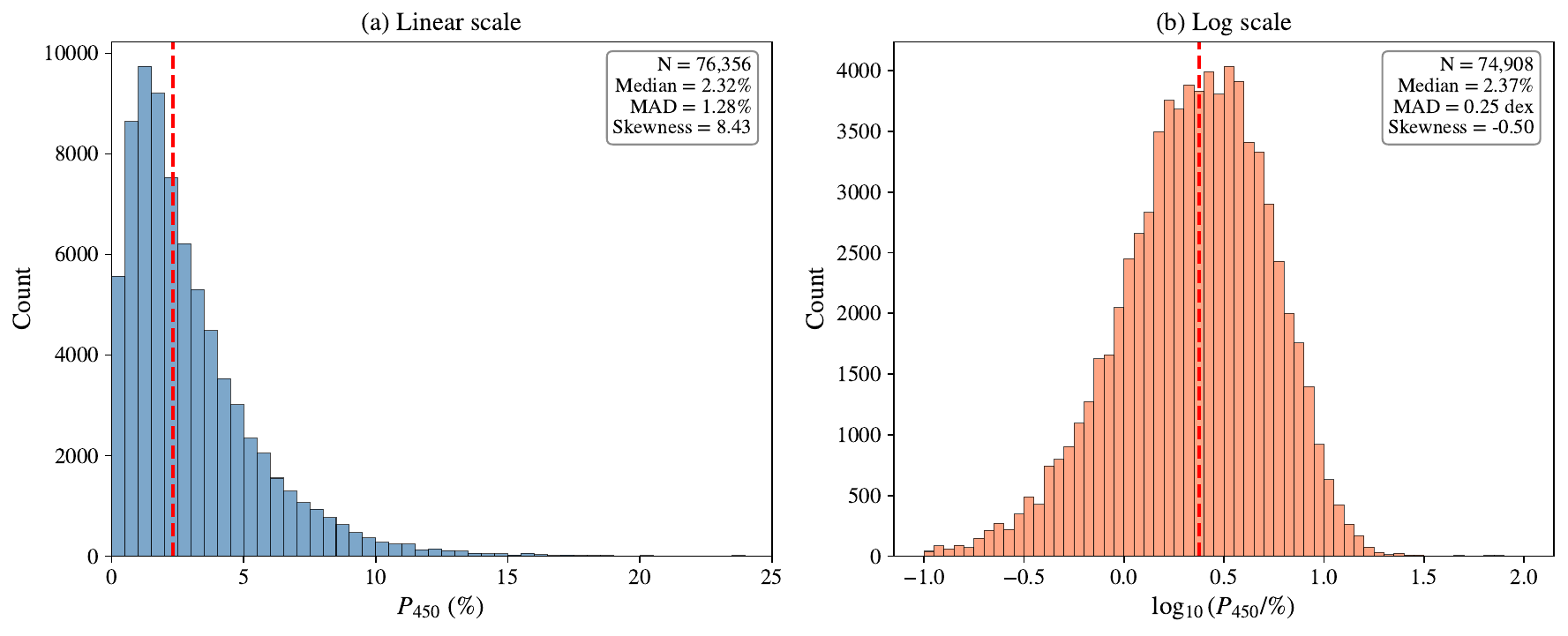}{0.95\textwidth}{}
}
\caption{Stacked histogram of debiased 450\,\micron\ polarization fractions from all observed fields with valid 450\,\micron\ polarimetric detections.
(a)~Distribution in linear space, showing strong positive skewness (skewness $= 8.43$).
(b)~Distribution in logarithmic space, showing nearly symmetric behavior (skewness $= -0.50$) consistent with a log-normal distribution.
The vertical dashed lines indicate the median values.
The total sample comprises 76{,}356 pixels satisfying $I/\sigma_I > 30$.
The transition from highly skewed to approximately symmetric behavior upon logarithmic transformation motivates the use of percentile statistics rather than Gaussian summary measures.
\\ {Alt text: Two histograms side by side.}
}
\label{afig:StackedHist}
\end{figure}

Figure~\ref{afig:StackedHist} presents the stacked distribution of $P_{450}$ values from 76{,}356 pixels that meet the signal-to-noise criterion $I/\sigma_I > 30$.
The sample includes sources with pixel counts of $<20$ to $>16{,}000$, as listed in appendix~\ref{app:S2_P450stats}.
Panel~(a) shows the distribution in a linear space, while panel~(b) displays the same data transformed into a logarithmic space.
The two representations show substantially different distributional properties.
 
In linear space, the distribution exhibits strong positive skewness (skewness $= 8.43$), with a pronounced tail that extends to high polarization fractions.
The median ($P_{50} = 2.32$\%) lies below the mean ($\langle P \rangle = 3.05$\%), a characteristic signature of right-skewed distributions.
The median absolute deviation is MAD $= 1.28$\%, indicating substantial dispersion around the central value.
 
In contrast, the logarithmic transformation yields an approximately symmetric distribution with a skewness near zero ($-0.50$).
This behavior is characteristic of a log-normal distribution, where the logarithm of the variable is normally distributed.
For log-normal data, the geometric mean provides a more representative measure of central tendency than the arithmetic mean.
For such data, multiplicative (rather than additive) measures of dispersion are appropriate.
 
The log-normal shape of the observed distribution of polarization fractions is consistent with theoretical expectations for turbulence-dominated molecular clouds.
However, this agreement alone does not establish a turbulent origin.
The polarization fraction $\mathcal{P}^{\rm frac}$ also depends nonlinearly on alignment efficiency, the radiation field, and column density.
Part of its log-normal distribution may therefore be inherited from the distribution of column density.
Our use of medians and percentiles depends only on the observed shape and does not require a particular physical origin.
A quantity that is the product of many independent random factors approaches a log-normal distribution, because the central limit theorem applies to its logarithm \citep{Vazquez-Semadeni1994}.
Conversely, an approximately log-normal distribution is consistent with, but does not establish, such a multiplicative origin.
Numerical simulations of supersonic isothermal turbulence give approximately log-normal distributions of volume density \citep{Padoan1997}, and the width of these distributions depends on the Mach number and the type of forcing \citep{Federrath2008}.
We suggest that distributions of column density may inherit a similar shape.
Similar multiplicative effects may propagate into the observed $P$ distribution because the polarization fraction depends on averaging grain alignment and magnetic field properties with weights set by column density.
 
These results justify our adoption of non-parametric percentile statistics ($P_{25}$, $P_{50}$, $P_{75}$) and the median absolute deviation in appendix~\ref{app:S2_P450stats}, rather than Gaussian summary statistics such as mean and standard deviation.
For highly skewed distributions, percentiles provide a more robust and interpretable characterization of the data.
The shape of the distribution alone does not distinguish the turbulent field structure from variations in alignment, column density, or the weighting of different cloud regions.
 
\section{Distance-Bias Residual Analysis}
\label{app:S2_dist_bias}
 
This appendix provides additional details on the analysis of how the spectral indices depend on distance, presented in sub-subsection~\ref{sss:SpIdx_Global}.
We quantify the relationship between the physical beam size, estimated using the adopted reference distances, and the spectral index difference $\gamma - \alpha$.
We then examine whether residuals from this relationship show systematic differences among star-forming environments.

As in Paper~S1 section~\ref{S1-sec:obs}, we use these distances as reference scaling factors to convert angular beam sizes to representative physical scales.
We do not treat them as precise distances to all of the material within each field because the clouds are extended along the line of sight.
Appendix~\ref{app:S2_survey_fields} lists the adopted reference distances and the corresponding atlas figures.

\subsection{Regression Setup}
\label{as:S2_SpIdxDistanceBias_Setup}
 
\subsubsection{Physical Beam Size}
 
We compute the physical beam size at each source distance as
\begin{equation}
\theta_{\rm phys} = \frac{\theta_{\rm beam} \times d}{206265} \quad [\mathrm{pc}],
\label{eq:beam_size}
\end{equation}
where $\theta_{\rm beam} = 12\farcs6$ is the POL-2 beam FWHM at 850\,\micron\ and $d$ is the adopted reference distance in parsecs.
For our sample, $\theta_{\rm phys}$ lies between 0.008\,pc (L1688 Ophiuchus~A at 139\,pc) and 0.51\,pc (Sagittarius~B2 at 8340\,pc).
 
\subsubsection{Linear Regression Model}
 
To assess trends with distance, we fit linear regression models of the form
\begin{equation}
Y = a \times \log_{10}(\theta_{\rm phys}) + b + \epsilon,
\label{eq:regression_model}
\end{equation}
where $Y$ is either $\gamma - \alpha$ or the dust opacity index $\beta = \alpha - C(T_{\rm d})$ (equation~\ref{eq:planck_correction}), $a$ is the slope, $b$ is the intercept, and $\epsilon$ represents residual scatter.
The coefficient of determination $R^2$ quantifies the fraction of variance in $Y$ explained by the linear relationship with $\log_{10}(\theta_{\rm phys})$:
\begin{equation}
R^2 = 1 - \frac{\sum_i (Y_i - \hat{Y}_i)^2}{\sum_i (Y_i - \bar{Y})^2},
\end{equation}
where $\hat{Y}_i$ is the predicted value from the regression and $\bar{Y}$ is the sample mean.
 
We also compute the Spearman rank correlation coefficient $\rho$ to assess monotonic trends without assuming linearity.
We evaluate the significance of $\rho$ using a two-sided $p$-value for the null hypothesis of no correlation.
 
\subsection{Results for $\gamma - \alpha$}
\label{as:S2_SpIdxDistanceBias_gammaMalpha}
 
\subsubsection{Correlation with Beam Size}
 
Figure~\ref{fig:beamSize_gammaMalpha} in sub-subsection~\ref{sss:SpIdx_Global} shows $\gamma - \alpha$ as a function of the physical beam size for the 35 sources in the statistical sample.
The linear regression yields:
\begin{equation}
\gamma - \alpha = (0.09 \pm 0.19) \times \log_{10}(\theta_{\rm phys}/\mathrm{pc}) - (0.26 \pm 0.24),
\end{equation}
with $R^2 = 0.007$ and Spearman $\rho = 0.18$ ($p = 0.30$).
 
The slope is consistent with zero and the regression accounts for $1\%$ of the variance in $\gamma - \alpha$, so we did not detect an ordering of $\gamma - \alpha$ by physical beam size.
We therefore do not assign the remaining variance to intrinsic differences among sources: the $95\%$ interval on the slope, $-0.30$ to $+0.48$ per dex, neither requires nor excludes a resolution-driven contribution.
 
\subsubsection{Residual Analysis by Environment}
Clouds observed with the same angular beam are sampled on different physical scales, which can complicate an environmental comparison of their polarization spectra.
Subtracting a fitted trend with beam size tests whether the environmental comparison changes after this particular dependence is accounted for.
 
To test for systematic environmental differences after removing the fitted trend with distance, we calculate the residuals
\begin{equation}
\Delta(\gamma - \alpha) = (\gamma - \alpha)_{\rm observed} - (\gamma - \alpha)_{\rm predicted}
\end{equation}
for each source, where the predicted value comes from the linear regression.
 
\begin{figure*}[htbp]
\centering
\includegraphics[width=0.95\textwidth]{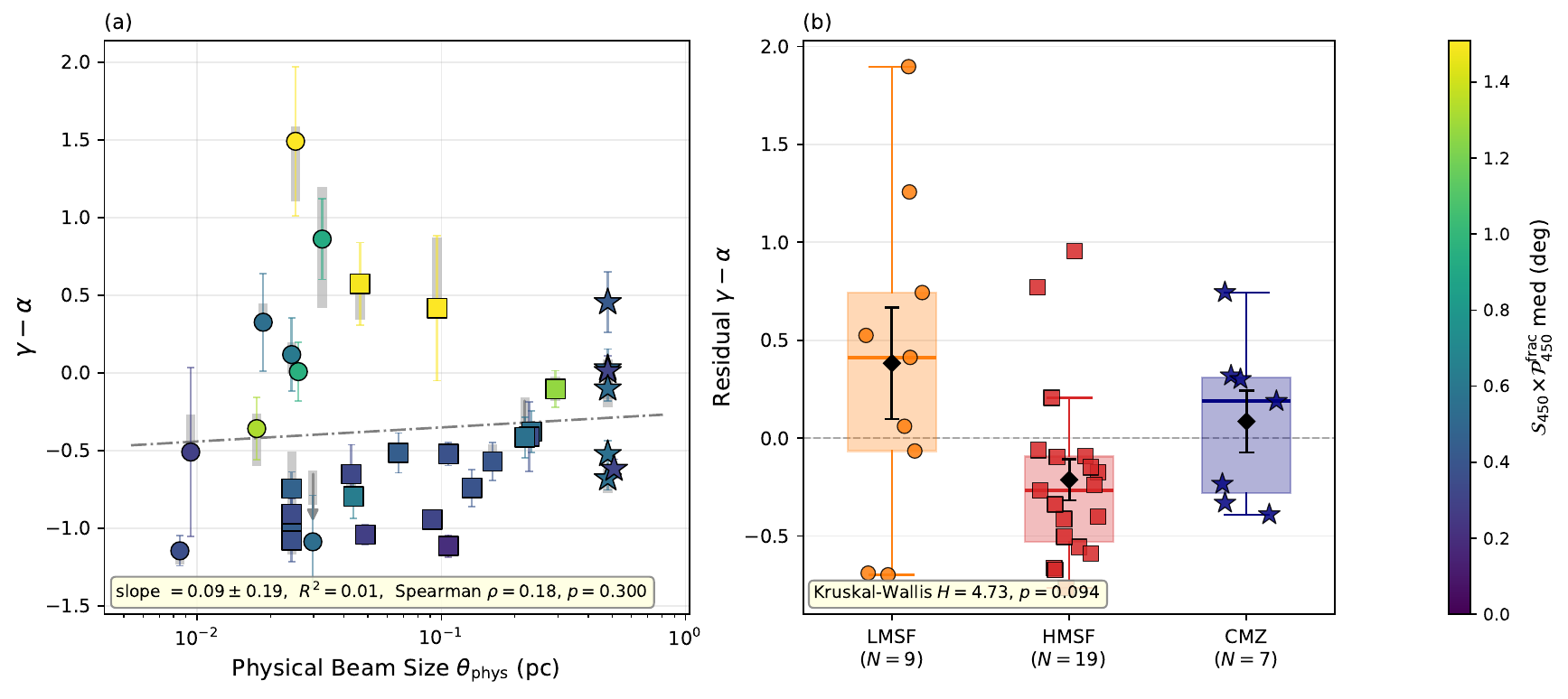}
\caption{
Distance-bias analysis for the spectral index difference $\gamma - \alpha$.
(a)~$\gamma - \alpha$ versus physical beam size for the 35 sources in the statistical sample.
Symbol conventions and the $\mathcal{S}_{450} \times \mathcal{P}^{\rm frac}_{450}$ color axis follow figure~\ref{fig:alpha_vs_gamma}, including the caveat about the spatial mask given there.
Vertical error bars are the SE of the median for $\gamma - \alpha$ ($\sigma(\gamma{-}\alpha) = \sqrt{\sigma_\alpha^2 + \sigma_\gamma^2}$).
The light-gray vertical bars and arrows show the $3\sigma$-limit range of each median, as in figure~\ref{fig:beamSize_gammaMalpha}.
The fit and the residuals in panel~(b) use the measured medians.
The dash--dotted line shows the best-fit linear trend (slope $= 0.09 \pm 0.19$, $R^2 = 0.007$).
(b)~Residuals after removing the distance trend, grouped by environment (LMSF: $N = 9$; HMSF: $N = 19$; CMZ: $N = 7$).
Box plots for each environment span the 25th--75th percentiles, with a bar marking the median.
Jittered markers show individual sources, and black diamonds show the mean $\pm$ SE for each class.
The Kruskal--Wallis test ($H=4.73$, $p=0.094$) does not detect a distributional difference after subtracting the fitted distance trend.
Environmental effects are not separated from the effects of spatial resolution in this sample.
\\ {Alt text: Two panels, a scatter plot with a fitted line and box plots.}
}
\label{fig:gammaMalpha_residuals}
\end{figure*}

Figure~\ref{fig:gammaMalpha_residuals} shows the distribution of residuals grouped by environment type: LMSF regions ($N = 9$), HMSF regions ($N = 19$), and CMZ sources ($N = 7$).
The mean residuals are $+0.38 \pm 0.28$ for LMSF, $-0.21 \pm 0.10$ for HMSF, and $+0.09 \pm 0.16$ for CMZ.
 
We use the Kruskal--Wallis $H$-test to compare the distributions of residuals among environments without assuming normality.
The test statistic is
\begin{equation}
H = \frac{12}{N(N+1)} \sum_{j=1}^{k} \frac{R_j^2}{n_j} - 3(N+1),
\end{equation}
where $N$ is the total sample size, $k$ is the number of groups, $n_j$ is the sample size of group $j$, and $R_j$ is the sum of ranks in group $j$.
 
For the $\gamma - \alpha$ residuals, we obtained $H = 4.73$ with $p = 0.094$ (2 degrees of freedom).
We also obtained $p = 0.091$ from $10^{5}$ label permutations, so the result does not rest on the $\chi^2$ approximation at these group sizes.
Removing the fitted trend with distance leaves the environmental comparison of sub-subsection~\ref{sss:env_synthesis} qualitatively unchanged.
For the measured medians of individual sources, the detrended $p$-values range from $0.056$ (Anderson--Darling, sensitive to any distributional difference) to $0.126$ (van der Waerden normal scores), compared with $0.056$--$0.145$ for the raw $\gamma-\alpha$ values.
As in sub-subsection~\ref{sss:env_synthesis}, the HMSF class retains the lowest median.
We do not infer an ordered environmental sequence.
Applying the same ordered Jonckheere--Terpstra test used in sub-subsection~\ref{sss:env_synthesis} to the residuals returns $JT = 169.0$, $p \approx 0.68$, against $JT = 178.0$, $p \approx 0.57$ on the raw $\gamma-\alpha$ values.
The ordered alternative is not supported either before or after the distance trend is removed.

However, environment and distance remain strongly confounded in the present sample.
The environmental classification itself is strongly correlated with the source distance (Spearman $\rho = 0.828$, $p = 8\times10^{-10}$).
In particular, the CMZ sources occupy 7.86--8.34~kpc, whereas the LMSF and HMSF samples span 0.139--0.532 and 0.399--4.80~kpc, respectively.
Thus, the CMZ sample does not overlap in distance with either disk population.
Subtracting the distance trend would also remove any genuine environmental component that followed the same ordering.
The HMSF class is intermediate in distance but has the most negative median $\gamma-\alpha$, so this pattern is not aligned with the distance ordering.
The coupling of environment and distance still prevents us from separating the effects of environment and physical resolution for the CMZ comparison.
 
\subsubsection{Scatter at Comparable Distances}
 
The scatter among sources observed at comparable physical scales is substantial.
Among the 14 sources at 300--800\,pc, $\gamma - \alpha$ ranges from $-1.09$ to $+1.49$, a spread of 2.58 that far exceeds the measurement uncertainties.
These sources span 305 to 799\,pc, a factor of 2.6 in physical beam size rather than a common distance.
Among the 7 sources at 1--3\,kpc, $\gamma - \alpha$ ranges from $-1.11$ to $+0.42$, a spread of 1.53.
 
This scatter shows that sources observed at broadly similar physical scales differ from each other by far more than their measurement uncertainties.
It limits neither the size nor the direction of any residual resolution effect.
The comparison spans a factor of 2.6 in beam size, and environment and distance remain confounded throughout the sample (environment rank versus distance, Spearman $\rho = 0.83$, with no distance overlap between the CMZ and either disk class).
 
\subsection{Results for $\beta$}
\label{as:S2_SpIdxDistanceBias_beta}
 
\subsubsection{Correlation with Beam Size}
 
The dust opacity spectral index $\beta = \alpha - C(T_{\rm d})$ correlates with beam size, unlike $\gamma - \alpha$, for which no such trend is detected.
For the 35 sources of the statistical sample, the linear regression yields:
\begin{equation}
\beta = (0.60 \pm 0.15) \times \log_{10}(\theta_{\rm phys}/\mathrm{pc}) + (2.71 \pm 0.19),
\label{eq:beta_regression}
\end{equation}
with $R^2 = 0.32$ and Spearman $\rho = 0.61$ ($p = 1.2\times10^{-4}$).
 
The regression accounts for about $32\%$ of the variance in $\beta$, against $1\%$ for $\gamma - \alpha$.
The observational contrast --- a distance-related change detected in $\beta$, none detected in $\gamma - \alpha$ --- is consistent with the two quantities having different sensitivities to spatial averaging.
$\beta$ is an absolute spectral slope and depends on the mix of dust temperatures and grain populations within the beam, which becomes more heterogeneous at larger physical scales.
By contrast, $\gamma - \alpha$ is a difference of two slopes measured at the same frequency pair, from which the common single-temperature Planck shape factor cancels exactly.
However, algebraic cancellation is not a statement about mixing within the beam.
It does not by itself establish that a mixture of temperatures, grain populations, or field geometries within the beam cancels in the \emph{ratio} of polarization efficiencies.
A cancellation in the ratio would require the polarized and total emission to be averaged over the same material with the same weights.
We therefore report the contrast as an observational one and do not claim that the wavelength dependence of polarization is unaffected by resolution.
 
\subsubsection{Residual Analysis by Environment}
The inferred opacity index changes with the physical scale sampled by the beam, so we cannot immediately assign its raw environmental differences to the grain properties.
The comparison of the residuals examines whether additional differences are detected after subtracting the fitted distance-related trend.
 
After subtracting the fitted distance trend from $\beta$, we applied the Kruskal--Wallis test to the residuals grouped by environment (LMSF: $N = 9$; HMSF: $N = 19$; CMZ: $N = 7$).
The residual distributions are consistent with a single parent population: the Kruskal--Wallis test yields $H = 1.68$, $p \approx 0.43$, and does not detect a significant difference among environments after removing the distance trend.
 
\begin{figure*}[htbp]
\centering
\includegraphics[width=0.95\textwidth]{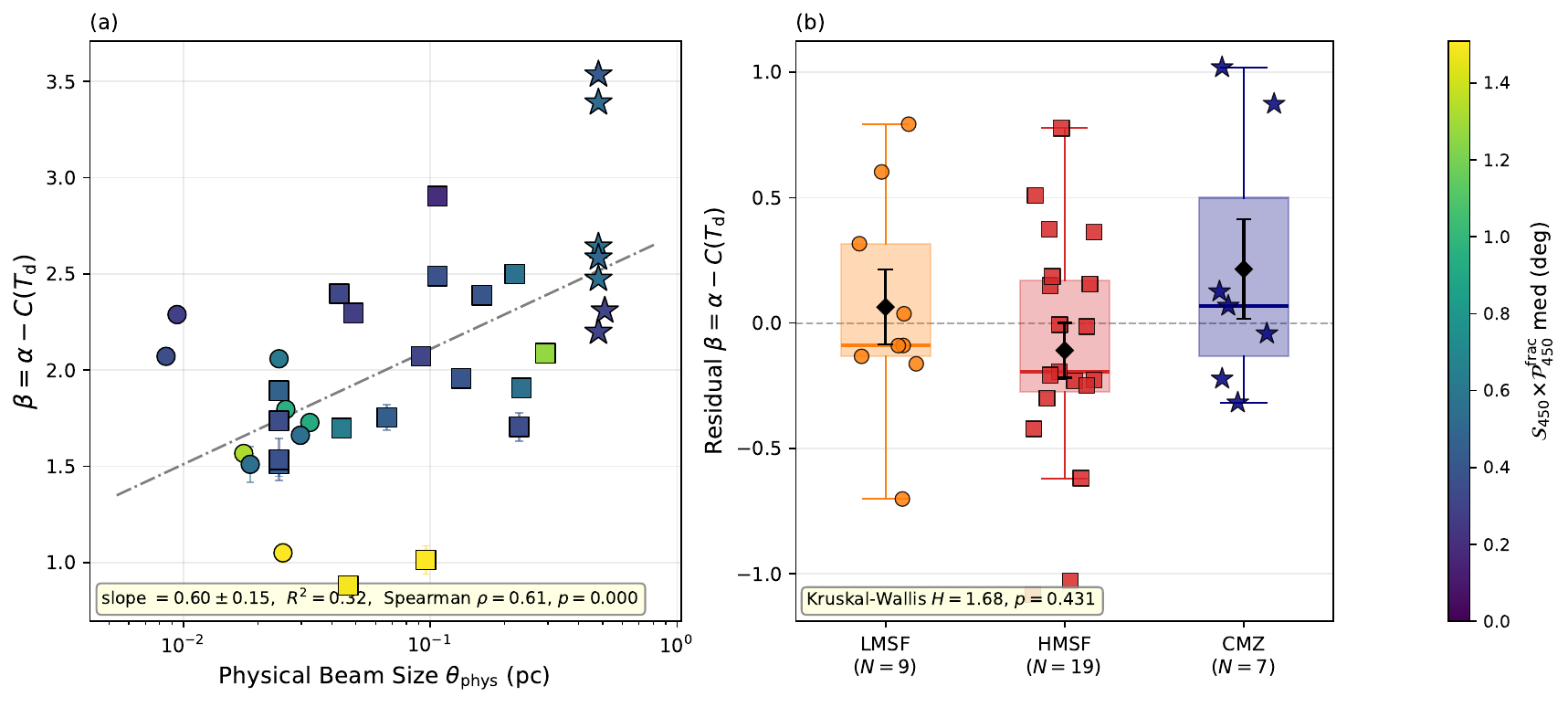}
\caption{
Distance-bias analysis for the dust opacity spectral index $\beta = \alpha - C(T_{\rm d})$ (equation~\ref{eq:planck_correction}).
(a)~$\beta$ versus physical beam size for the 35 sources in the statistical sample.
Symbol conventions and the $\mathcal{S}_{450} \times \mathcal{P}^{\rm frac}_{450}$ color axis follow figure~\ref{fig:alpha_vs_gamma}, including the caveat about the spatial mask given there.
Vertical bars show the statistical SE of the median at fixed adopted $T_{\rm d}$, $\sigma_{\beta,\rm stat}=\sigma_\alpha$.
They exclude the temperature-related contribution included in figure~\ref{fig:distance_vs_beta}.
The dash--dotted line shows the best-fit linear trend (slope $= 0.60 \pm 0.15$, $R^2 = 0.32$).
A distance-related change is detected for $\beta$ and none is detected for $\gamma - \alpha$.
This is a contrast between a detection and a non-detection, not a demonstration that $\gamma - \alpha$ is immune to beam averaging.
(b)~Residuals after removing the distance trend, grouped by environment (LMSF: $N = 9$; HMSF: $N = 19$; CMZ: $N = 7$).
Box plots for each environment span the 25th--75th percentiles, with a bar marking the median.
Jittered markers show individual sources, and black diamonds show the mean $\pm$ SE for each class.
The Kruskal--Wallis test ($H = 1.68$, $p = 0.431$) does not detect a difference among environments after subtracting the fitted distance trend.
\\ {Alt text: Two panels, a scatter plot with a fitted line and box plots.}
}
\label{fig:beamSize_beta}
\end{figure*}

A substantial component of the variation in $\beta$ observed in figure~\ref{fig:beamSize_beta}(a) is therefore associated with the physical resolution that depends on distance, and secondarily with the temperature dependence of the $C(T_{\rm d})$ correction.
However, the confounding described above for $\gamma - \alpha$ applies equally here, so these data do not establish resolution as the dominant cause in preference to an environmental difference that follows the same ordering.
This non-detection does not demonstrate that the dust populations are identical nor does it establish spatial resolution as the cause of the original trend.
 
\subsubsection{Comparison of $\gamma - \alpha$ and $\beta$}
 
We consider the contrast between $\gamma - \alpha$ and $\beta$ the clearest single result of this appendix.
For $\gamma - \alpha$, we do not detect a statistically significant trend with distance.
The fitted trend accounts for only $1\%$ of the variance.
The remaining variance is thus not shown to be intrinsic: the slope is bounded only to the interval $-0.30$ to $+0.48$ per dex.
For $\beta$, the distance trend explains $32\%$ of the variance ($R^2 = 0.32$) and is highly significant.
We do not detect additional environmental structure in the residuals.
A distance-related trend is therefore detected in the slope of the bulk dust emission and not in the wavelength dependence of polarization, which probes the aligned grain population specifically.
The contrast is between a detection and a non-detection and does not establish that the wavelength dependence of polarization is free of beam averaging.
 
\subsection{Robustness Checks}
\label{as:S2_SpIdxDistanceBias_Robustness}
 
\subsubsection{Outlier Sensitivity}
 
Mon~R2 shows a large negative $\gamma - \alpha$ value ($-1.04$), corresponding to strong wavelength-dependent depolarization.
We repeated the analysis without Mon~R2 to test whether it drives the correlations.
 
Table~\ref{tbl:robustness_outliers} summarizes the results.
The Spearman correlation coefficient $\rho$ changes by 0.0001 when Mon~R2 is excluded, and the correlation remains non-significant with or without it ($p = 0.30$ and $p = 0.31$).
The $R^2$ value moves from 0.007 to 0.005, confirming that the non-detection is not an artifact of a single source.
 
\begin{deluxetable}{lccc}
\tablecaption{Robustness of Distance Correlations to Outlier Exclusion
\label{tbl:robustness_outliers}}
\tablewidth{0pt}
\tablehead{
\colhead{Sample} &
\colhead{$N$} &
\colhead{$\rho$ ($\gamma - \alpha$)} &
\colhead{$R^2$ ($\gamma - \alpha$)}
}
\startdata
Full sample              & 35 & $0.18$ ($p = 0.30$) & 0.007 \\
Excluding Mon R2         & 34 & $0.18$ ($p = 0.31$) & 0.005 \\
\enddata
\tablecomments{
Spearman rank correlation coefficient $\rho$ and coefficient of determination $R^2$ for the relationship between $\gamma - \alpha$ and $\log_{10}(\theta_{\rm phys})$.
We identified Mon~R2 ($\gamma - \alpha = -1.04$) as a potential outlier because of its large negative value.
The correlation is not statistically significant with or without this source.
}
\end{deluxetable}
 
\subsubsection{Threshold Sensitivity}
 
We derive the values for individual sources without applying an $S/N(\gamma)$ threshold to their pixels (sub-subsection~\ref{sss:SpIdx_Global}).
We repeated the analysis after requiring $S/N(\gamma) > 5$ for pixels in every source to test whether this selection choice affects the distance result.
 
Recomputing both medians for each source using the pixels with $S/N(\gamma) > 5$ retains 1087 of the 1971 pixels.
Two sources---PGCC~G203.21$-$11.20 and Serpens~South~3---do not retain any qualifying pixels.
On the remaining 33 sources the association is $\rho = 0.08$ ($p = 0.66$, $R^2 = 0.001$), consistent with the full-sample value of table~\ref{tbl:robustness_outliers}.
The non-detection is therefore not an artifact of the absence of an $S/N(\gamma)$ cut.

\section{Secondary Tests of the Statistical Methods Used in the Main Text}
\label{app:S2_secondary}
The environmental interpretation should not depend on a single choice of statistical test, especially when the cloud populations are small and have unequal distributions.
Alternative procedures assess how consistently the data support a population difference, an ordered sequence, or a predominance of negative spectral-index offsets.
 
We repeat each primary comparison with the alternative method that we judge most suitable for the distribution of the data.
 
The tests reported in section~\ref{sec:S2_results} are those posed in advance of the analysis.
In this appendix, we examine how the same comparisons change under alternative statistical procedures, using the same samples and quantities.
We do not add new measurements.
The permutation $p$-values use $10^{5}$ relabelings of the environment assignment with the observed values kept fixed.
We report effect sizes without $p$-values because effect sizes do not depend on a test.
The alternative tests describe the sensitivity of the reported comparisons to their statistical assumptions.
Their combined role is to delimit the environmental and population claims, rather than to turn a marginal or exploratory contrast into a new detection.
 
\subsection{Comparison of $\gamma-\alpha$ among the three environmental classes}
\label{as:S2_secondary_omnibus}
 
The classes are LMSF ($N=9$), HMSF ($N=19$) and CMZ ($N=7$), with the 35 fields entered as independent draws, as in subsection~\ref{ss:S2_EnvStrat} (table~\ref{tbl:secondary_omnibus}).
Effect sizes for the same comparison are $\varepsilon^2 = 0.087$ and, pairwise, Cliff's $\delta = +0.39$ (LMSF against HMSF), $-0.49$ (HMSF against CMZ) and $+0.14$ (LMSF against CMZ).
A positive Cliff's $\delta$ means that the first-named group tends to have larger $\gamma-\alpha$ values.
 
\begin{deluxetable}{lcc}
\tablecaption{Secondary tests of $\gamma-\alpha$ across LMSF, HMSF and CMZ\label{tbl:secondary_omnibus}}
\tablewidth{0pt}
\tablehead{\colhead{Procedure} & \colhead{Statistic} & \colhead{$p$}}
\startdata
Kruskal--Wallis, $\chi^2$ approximation   & $H = 4.78$ & 0.092 \\
Kruskal--Wallis, permutation              & $H = 4.78$ & 0.090 \\
Mood median test, permutation             & $5.48$     & 0.060 \\
van der Waerden normal scores, permutation& $3.87$     & 0.145 \\
Anderson--Darling $k$-sample              & $1.88$     & 0.056 \\
\hline
Fligner--Killeen, equality of scale       & \nodata    & 0.15 \\
Brown--Forsythe, equality of scale        & \nodata    & 0.09 \\
\enddata
\tablecomments{The first block tests equality of the three distributions.
The second block tests whether their spreads are equal.
The null hypothesis of the Kruskal--Wallis test is that the three distributions are identical, so a rejection admits a difference in scale as well as in location.
We report the second block so that this alternative can be assessed directly.
The entries use the measured medians of individual sources.
With pixels below the polarized-intensity noise floor entered as $3\sigma$ limits, the Kruskal--Wallis $p$ ($\chi^2$ approximation) is $0.056$--$0.078$ for the two extreme placements of the limits.}
\end{deluxetable}

We also compare the three environmental classes using residuals about the linear fit of $\gamma-\alpha$ against $\log_{10}\theta_{\rm phys}$ of appendix~\ref{app:S2_dist_bias} (table~\ref{tbl:secondary_residual}).
 
\begin{deluxetable}{lcc}
\tablecaption{Secondary tests of the distance-detrended $\gamma-\alpha$ residuals\label{tbl:secondary_residual}}
\tablewidth{0pt}
\tablehead{\colhead{Procedure} & \colhead{Statistic} & \colhead{$p$}}
\startdata
Kruskal--Wallis, $\chi^2$ approximation    & $H = 4.73$ & 0.094 \\
Kruskal--Wallis, permutation               & $H = 4.73$ & 0.091 \\
Mood median test, permutation              & $5.48$     & 0.060 \\
van der Waerden normal scores, permutation & $4.13$     & 0.126 \\
Anderson--Darling $k$-sample               & $1.87$     & 0.056 \\
\hline
Fligner--Killeen, equality of scale        & \nodata    & 0.15 \\
Brown--Forsythe, equality of scale         & \nodata    & 0.09 \\
\enddata
\tablecomments{$\varepsilon^2 = 0.085$.
Cliff's $\delta = +0.43$ (LMSF against HMSF), $-0.43$ (HMSF against CMZ) and $+0.21$ (LMSF against CMZ).}
\end{deluxetable}
 
\subsection{Two-sample contrasts}
\label{as:S2_secondary_twosample}
 
We compared $\gamma-\alpha$ for two partitions of the 35 sources: the disk (LMSF$+$HMSF) against the CMZ, and HMSF against the other two classes (table~\ref{tbl:secondary_twosample}).
 
\begin{deluxetable}{lccccc}
\tablecaption{Two-sample contrasts on $\gamma-\alpha$\label{tbl:secondary_twosample}}
\tablewidth{0pt}
\tablehead{\colhead{Contrast} & \colhead{MWU, one-sided} & \colhead{MWU, two-sided} &
\colhead{Brunner--Munzel} & \colhead{Permutation} & \colhead{Cliff's $\delta$}}
\startdata
Disk (LMSF$+$HMSF) against CMZ & 0.132   & 0.263   & 0.186   & 0.131   & $-0.29$ \\
HMSF against the other two     & 0.0150  & 0.0301  & 0.0298  & 0.0144  & $-0.43$ \\
\enddata
\tablecomments{The contrast of the disk against the CMZ is the one posed without reference to the state of star formation (subsection~\ref{ss:S2_EnvStrat}).
The contrast of HMSF against the other two classes is the post-hoc one, reported there as exploratory and quoted here unchanged.
Brunner--Munzel does not assume equal scales.
Correcting by a factor of three for the three possible splits of one class against the other two gives 0.045 and 0.090 for the two Mann--Whitney $U$ (MWU) entries of the second row.}
\end{deluxetable}
 
Of the 19 HMSF fields, 16 satisfy $\gamma<\alpha$ at $1\sigma$.
Seven of the remaining 16 fields meet the same criterion (subsection~\ref{ss:S2_EnvStrat}; table~\ref{tbl:secondary_fisher}).
 
\begin{deluxetable}{lc}
\tablecaption{Secondary tests of the $2\times2$ counts $(16,3;7,9)$: rows are HMSF and the other environments combined; columns indicate whether $\gamma<\alpha$ is established at $1\sigma$ or not.\label{tbl:secondary_fisher}}
\tablewidth{0pt}
\tablehead{\colhead{Procedure} & \colhead{$p$}}
\startdata
Fisher exact, one-sided  & $1.5\times10^{-2}$ \\
Fisher exact, two-sided  & $3.0\times10^{-2}$ \\
Barnard exact            & $1.3\times10^{-2}$ \\
Boschloo exact           & $1.6\times10^{-2}$ \\
\enddata
\tablecomments{Odds ratio 6.9.
The HMSF-versus-others split is the post-hoc split of subsection~\ref{ss:S2_EnvStrat} and carries the caveats stated there.
The $1\sigma$ fractions depend on the signal-to-noise ratio for each source as well as on $\gamma-\alpha$.}
\end{deluxetable}
 
\subsection{The pooled $\gamma<\alpha$ result}
\label{as:S2_secondary_pooled}
 
The main text reports a one-sided Wilcoxon signed-rank test on the differences for the 35 individual sources (sub-subsection~\ref{sss:SpIdx_Global}).
We test the same claim using the following alternative procedures (table~\ref{tbl:secondary_pooled}).
 
\begin{deluxetable}{lcc}
\tablecaption{Secondary tests of $\gamma<\alpha$ on the 35-source sample\label{tbl:secondary_pooled}}
\tablewidth{0pt}
\tablehead{\colhead{Procedure} & \colhead{Quantity tested} & \colhead{$p$}}
\startdata
Wilcoxon signed-rank, one-sided & median $\gamma-\alpha$  & $5.9\times10^{-5}$ -- $5.1\times10^{-3}$ \\
Sign test, one-sided            & incidence $27/35$ -- $24/35$ & $9.4\times10^{-4}$ -- $2.0\times10^{-2}$ \\
Binomial against $1/2$, two-sided& incidence $27/35$ -- $24/35$ & $1.9\times10^{-3}$ -- $4.1\times10^{-2}$ \\
Student $t$, one-sided          & mean $\gamma-\alpha$    & \nodata\ -- $1.0\times10^{-2}$ \\
\enddata
\tablecomments{Each entry gives the two extreme placements of the $3\sigma$ limits of pixels below the polarized-intensity noise floor. 
The bootstrap $95\%$ confidence interval on the median ($2\times10^{5}$ resamples) is $[-0.78,-0.22]$ at the lower end and $[-0.62,-0.03]$ at the upper end of the limit range.
The mean is undefined at the lower end, where the values for two sources (Serpens~South and SDC~13 mm-2) have upper bounds but no lower bounds.
The two families test different statements.
The first and last procedures address the magnitude of the offset.
The middle two address only its incidence, which is the weaker of the two properties.}
\end{deluxetable}
 
\subsection{Monotonic association with physical beam size}
\label{as:S2_secondary_monotone}
 
Spearman $\rho$ is quoted in appendix~\ref{app:S2_dist_bias}.
We give Kendall's $\tau_b$ here for the same two pairs.
For $\beta$ against $\log_{10}\theta_{\rm phys}$, $\rho = 0.61$ ($p = 1.2\times10^{-4}$) and $\tau_b = 0.42$ ($p = 4.7\times10^{-4}$).
For $\gamma-\alpha$ against the same abscissa, $\rho = 0.18$ ($p = 0.30$) and $\tau_b = 0.15$ ($p = 0.20$).

\section{Reference Scales and Noise Response of $\mathcal{S}$}
\label{as:S_bounds}
 
We examine the random-orientation reference scale and the noise response of the polarization-angle dispersion $\mathcal{S}$.
These scales guide the interpretation of low and high dispersions and our choice of a common color scale for the maps.
 
\subsection{Random-Orientation Reference Scale}
\label{ass:S_upper}
 
For a completely random magnetic field, the angular differences $\chi_k - \chi_{\mathrm{center}}$ follow a uniform distribution over the interval $(-90\arcdeg, +90\arcdeg)$.
The standard deviation of this uniform distribution is
\begin{equation}
\sigma_{\rm rand} = \frac{180\arcdeg}{\sqrt{12}} \approx 52\arcdeg,
\label{eq:S_max_std}
\end{equation}
which is the random-orientation reference value for the ``dispersion about mean'' definition, i.e., the dispersion about the global mean field with which \citet{Hildebrand2009} compared their angular dispersion function (see also \citealt{Houde2009}).
We adopt the pixel-centered formulation of \citet{Planck2020-XII} (equation~\ref{eq:S_def}).
For a completely random field, the angle differences from the reference pixel, wrapped to $(-90\arcdeg, +90\arcdeg)$, are uniformly distributed over that interval.
Thus, $\langle\mathcal{S}^{2}\rangle = (90\arcdeg)^{2}/3$ and $\langle\mathcal{S}^{2}\rangle^{1/2} \approx 52\arcdeg$ (section~3.3 of \citealt{Planck2020-XII}).
The value $52\arcdeg$ provides a reference scale for random orientations.
It is not an upper bound: an individual measurement based on a finite number of neighbors can exceed this value.
Table~\ref{tab:S_definitions} summarizes the reference values for these two definitions.
 
\begin{deluxetable}{lcc}
\tablecaption{Random-Orientation Reference Scales of $\mathcal{S}$ for Different Definitions\label{tab:S_definitions}}
\tablewidth{0pt}
\tablehead{
  \colhead{Definition} &
  \colhead{Reference} &
  \colhead{Reference rms}
}
\startdata
Dispersion about mean & cf.\ \citet{Hildebrand2009} & $\approx 52\arcdeg$ \\
Pixel-centered rms    & \citet{Planck2020-XII}           & $\approx 52\arcdeg$ \\
\enddata
\tablecomments{Values assume a completely random magnetic field orientation uniformly distributed over $(-90\arcdeg, +90\arcdeg)$.
The standard deviation of such a distribution is $180\arcdeg/\sqrt{12} \approx 52\arcdeg$.
The difference between two such angles, wrapped to the same interval, is again uniformly distributed, so the expected pixel-centered rms takes the same value.
Individual values based on a finite number of neighbors can exceed it.}
\end{deluxetable}
 
\subsection{Noise Response of the Raw Estimator}
\label{ass:S_lower}
 
In the limit of a perfectly uniform magnetic field, where all true polarization angles are identical, the mathematical lower bound is $\mathcal{S} = 0\arcdeg$.
However, any measured non-zero dispersion in this idealized case arises entirely from measurement noise in $\chi$.
Here we describe the noise response of the raw estimator $\mathcal{S}_{\rm raw}$, the value of equation~(\ref{eq:S_def}) before debiasing, using the noise properties established in Paper~T subsection~\ref{T-ss:noisemodel}.
 
\subsubsection{Linking $\sigma_\chi$ to the Observed Dispersion Function}
 
Paper~T subsection~\ref{T-ss:noisemodel} finds closely matched noise scales in Stokes $Q$ and $U$ and a correlation $|\rho_{QU}|\!\lesssim\!0.02$:
\begin{equation}
\sigma_Q \simeq \sigma_U \equiv \sigma, \qquad \mathrm{Cov}(Q,U) \approx 0.
\label{eq:isotropic_noise}
\end{equation}
Under these conditions, the uncertainty of the polarization angle follows from the standard approximation:
\begin{equation}
\sigma_\chi \simeq \frac{1}{2}\frac{\sigma_P}{P},
\label{eq:sigma_phi_repeat}
\end{equation}
where $\chi$ is measured in radians and this expression is valid in the high signal-to-noise regime ($P/\sigma_P \gtrsim 3$).
 
The polarization-angle dispersion function $\mathcal{S}$ measures the rms variation of $\chi$ within a local kernel.
In the limit of a perfectly uniform magnetic field and of small angle errors, the raw estimator responds to the angle error $e_0$ of the central pixel and the errors $e_i$ of its $N$ neighbors, including their covariance:
\begin{equation}
\langle\mathcal{S}_{\rm raw}^{2}\rangle = \frac{1}{N}\sum_{i=1}^{N}\left[\sigma_{\chi,i}^{2}+\sigma_{\chi,0}^{2}-2\,\mathrm{Cov}(e_i,e_0)\right].
\label{eq:S_min_sigma}
\end{equation}
For equal independent angle errors, this gives the rms scale $\langle\mathcal{S}_{\rm raw}^{2}\rangle^{1/2} = \sqrt{2}\,\sigma_\chi$.
Positive covariance between the angle errors reduces this rms scale.
However, the scale does not impose a positive lower bound on measured or debiased $\mathcal{S}$.
We assess detections with the propagated uncertainty $\sigma_{\mathcal{S}}$ of sub-subsection~\ref{sss:S_definition}, which includes the shared central-pixel term but treats the errors of different pixels as independent.
 
\subsubsection{Expression in Terms of Signal-to-Noise Ratio}
 
Substituting equation~(\ref{eq:sigma_phi_repeat}) into the rms scale for equal independent errors yields
\begin{equation}
\langle\mathcal{S}_{\rm raw}^{2}\rangle^{1/2} \simeq \frac{\sqrt{2}}{2}\frac{\sigma_P}{P} = \frac{\sqrt{2}}{2}\frac{1}{(S/N)_P} \quad \mathrm{(radians)},
\label{eq:S_min_snr_rad}
\end{equation}
where $(S/N)_P \equiv P/\sigma_P$ is the polarization signal-to-noise ratio.
Converting to degrees:
\begin{equation}
\langle\mathcal{S}_{\rm raw}^{2}\rangle^{1/2}\;[\mathrm{deg}] \simeq \sqrt{2}\,\frac{57\fdg30}{2\,(S/N)_P} = \sqrt{2}\,\frac{28\fdg65}{(S/N)_P}.
\label{eq:S_min_snr_deg}
\end{equation}
This expression is the rms noise scale of the raw estimator for equal independent errors.
Even an intrinsically perfectly ordered magnetic field yields a non-zero raw $\mathcal{S}$ because of noise in $Q$ and $U$.
 
\subsubsection{Illustrative Numerical Values}
 
The survey imposes a uniform total-intensity requirement, $I/\sigma_I > 30$.
By design, we do not impose a threshold on the signal-to-noise ratio of polarized intensity (subsection~\ref{ss:S2_P_ratio}).
The polarization floors that we apply are the per-pixel detection floor $P_\lambda \geq 3\,\sigma_{P_\lambda}$, required in each band that enters a correlation and in both bands for $R_{\rm EW}$, and the $3\sigma$ limits that enter $\gamma$ and $\mathcal{S}$ (subsection~\ref{ss:S2_noise_floor}).
For illustration, a polarization signal-to-noise ratio of 30 in equation~(\ref{eq:S_min_snr_deg}) gives
\begin{equation}
\langle\mathcal{S}_{\rm raw}^{2}\rangle^{1/2} \simeq \sqrt{2}\,\frac{28\fdg65}{30} \approx 1\fdg35 .
\label{eq:S_min_numerical}
\end{equation}
A polarization signal-to-noise ratio of 30 is not the survey selection threshold: the Stokes-$I$ requirement does not impose the same signal-to-noise ratio on polarization.
The scale varies as $1/(S/N)_P$ and is $\approx 13.5\arcdeg$ at the per-pixel detection floor $P \geq 3\,\sigma_P$.
At this floor, the high signal-to-noise approximation of equation~(\ref{eq:sigma_phi_repeat}) is only marginal.
Pixels where we do not detect $\mathcal{S}$ with $3\sigma_{\mathcal{S}}$ (16\%--44\% of the pixels per source at 850\,$\mu$m) are kept as upper limits (subsection~\ref{ss:S2_noise_floor}).
The noise response of $\mathcal{S}$ therefore depends on the polarization signal-to-noise ratios and the error covariance of the pixels involved.
It cannot be represented by a single number for the entire survey.
 
\subsubsection{Interpretation}
Field disorder can be inferred only where the measured dispersion is distinguishable from the noise response of the estimator.
The relevant comparison is between the observed dispersion and the noise contribution expected from equation~(\ref{eq:S_min_sigma}) for the same polarization uncertainties, neighboring lines of sight, and estimator.
A selection threshold on intensity alone does not set a universal angular floor.

\section{Distance Determination}
\label{app:distance}
 
Distance estimates are required to place the survey fields on a common physical scale, although the principal dual-band diagnostics presented in this paper are largely insensitive to distance uncertainties (Paper~S1 section~\ref{S1-sec:obs}).
We performed the distance compilation described here for the complete set of 81 POL-2 survey fields, including fields not used in the statistical analysis of individual sources in Paper~S1, to maintain a uniform reference table throughout the survey.
We used a hierarchical approach based on two complementary catalogs derived from Gaia, supplemented by curated literature values.
All catalog coordinates were first transformed into the International Celestial Reference System (ICRS) to ensure a consistent reference frame for cross-matching.

We first queried the catalog of cloud distances derived from extinction and Gaia data, presented by \citet{Zucker2020}.
This catalog provides median distances ($d_{50}$) along with the 16th and 84th percentile values ($d_{16}$, $d_{84}$) for 326 sightlines toward about 60 star-forming regions within 2.5\,kpc.
These distances were derived by Bayesian inference from stellar photometry combined with Gaia Data Release 2 (DR2) parallaxes.
For each POL-2 field, we matched the field center (ICRS) to the nearest \citet{Zucker2020} entry within a maximum angular separation of $4\fdg0$ using Astropy's \texttt{SkyCoord.match\_to\_catalog\_sky} function.
When a match was found, we adopted the catalog median distance $d_{50}$ as the target distance.
We defined the asymmetric $1\sigma$ uncertainties as $d_{50} - d_{16}$ on the lower side and $d_{84} - d_{50}$ on the upper side.

For targets lacking a \citet{Zucker2020} match, typically due to coverage limits or boundaries of the extinction maps, we cross-matched against the cluster catalog compiled by \citet{Prisinzano2022} from Gaia Early Data Release 3 (EDR3) proper motions.
This catalog provides robust median distances for young stellar clusters and associations, determined from the proper motions and parallaxes of their member stars.
We applied the same $4\fdg0$ matching radius and nearest-neighbor criterion, adopting the \citet{Prisinzano2022} median distance $d$ for matched targets.
The \citet{Prisinzano2022} catalog does not provide uniform uncertainties based on percentiles.

In rare cases where both catalogs returned matches within $4\fdg0$, we selected the value from the catalog with the smaller angular separation to the field center.
For a few well-studied targets, we adopt a more precise non-Gaia literature distance, such as a very long baseline interferometry (VLBI) maser parallax, in place of the catalog value and record its reference.
Each target's final distance was therefore set to (i) the \citet{Zucker2020} value when available, (ii) otherwise the \citet{Prisinzano2022} value, subject to the smaller-separation tie-break above, and (iii) superseded by a curated non-Gaia literature value where one existed.
We prioritized \citet{Zucker2020} distances because they are tied directly to Gaia parallaxes and come with uncertainties based on percentiles.
 
We assigned distances to 38 of the 81 fields from \citet{Zucker2020} and 12 from \citet{Prisinzano2022}.
The remaining 31 fields take individually vetted literature values drawn from 23 further sources, most often a VLBI maser parallax or a dedicated distance study of the region concerned.
Examples are \citet{Ando2011} for Onsala~2, \citet{Omodaka2020} for Auriga, \citet{Peretto2014} for SDC~13, and \citet{Boehle2016} and \citet{Reid2014} for the Galactic Center fields.
Table~\ref{tab:surv} (appendix~\ref{app:S2_survey_fields}) lists the adopted value and its reference for each field.
The 16th and 84th percentiles of \citet{Zucker2020} describe statistical uncertainties, usually of order 1\%--2\%.
The authors estimate an additional systematic uncertainty of about 5\%, likely larger beyond $\approx 1.5$\,kpc.
We calculate the principal dual-band diagnostics of this paper, in particular $R_p$ and $\gamma-\alpha$, without using distances, as discussed in Paper~S1 section~\ref{S1-sec:obs}.

\section{Survey Field List for Individual Sources}
\label{app:S2_survey_fields}
 
This appendix lists the 81 POL-2 fields analyzed in the present BISTRO 450/850\,$\mu$m survey, ordered approximately by adopted reference distance $d_{\rm ref}$.
Entries with a figure reference are mapped in the atlas of appendix~\ref{app:S2_atlas} which presents 65 of the 69 unique star-forming regions of the {\sc PolSrc} sample discussed in the main text (all but W51 and Orion~A Fields~4, 5 and~7).
Table entries and distinct regions are not in one-to-one correspondence, because adjacent fields of the same complex are mosaicked with differing degrees of overlap.
Apart from Orion~A Fields~4, 5 and~7, the remaining entries (marked with $\dagger$ or carrying \nodata\ figure references, together with W51, marked $^\S$) are fields observed at 850\,$\mu$m only, non-detected fields, or, in the case of W51, a field excluded for a corrupted 850\,\micron\ reduction.
We retain these entries to document the full survey footprint.
The rightmost column identifies the corresponding map of $\vec{B}$-field orientations at both wavelengths in appendix~\ref{app:S2_atlas}.
 
\begin{longrotatetable}
\footnotesize
\begin{deluxetable}{lllllllll}
\tablecolumns{9}
\tablewidth{0pt}
\tablecaption{POL-2 Survey Fields \label{tab:surv}}
\tablehead{
\colhead{No.} & \colhead{Name} & \colhead{R.A.} & \colhead{Dec.} & \colhead{$d_{\rm ref}$} & \colhead{Sep.} & \colhead{Distance Ref.} & \colhead{POL-2 Ref.} & \colhead{Figure} \\
& & \colhead{(J2000)} & \colhead{(J2000)} & \colhead{(pc)} & \colhead{(deg)} & & &
}
\startdata
1 & L43 & 16:34:33.0 & -15:47:01 & 125 & \nodata & \citet{Geus1990} & \citet{Karoly2023} & \ref{afig:Bmap_L43L43} \\
2 & Taurus B211/213 & 4:19:50.4 & +27:11:46 & 130 & 1.65 & \citet{Zucker2020} & \citet{Chakali2021} & \ref{afig:Bmap_B211MO} \\
3 & L1688 H$^\dagger$ & 16:28:59.0 & -24:22:00 & 139 & 0.06 & \citet{Zucker2020} & \nodata & \nodata \\
4 & L1689 South & 16:31:38.6 & -24:49:49 & 139 & 0.88 & \citet{Zucker2020} & \citet{Pattle2021} & \ref{afig:Bmap_L1689S} \\
5 & Ophiuchus A & 16:26:28.1 & -24:23:29 & 139 & 0.33 & \citet{Zucker2020} & \parbox[t]{3.0cm}{\raggedright \citet{Kwon2018, Pattle2019}\strut}& \ref{afig:Bmap_OphiuA} \\
6 & Ophiuchus B & 16:27:26.2 & -24:27:07 & 139 & 0.10 & \citet{Zucker2020} & \parbox[t]{3.0cm}{\raggedright \citet{Soam2018, Pattle2019}\strut}& \ref{afig:Bmap_OphiuB} \\
7 & Ophiuchus C & 16:26:58.2 & -24:34:23 & 139 & 0.20 & \citet{Zucker2020} & \parbox[t]{3.0cm}{\raggedright \citet{JLiu2019, Pattle2019}\strut}& \ref{afig:Bmap_OphiuC} \\
8 & L183$^\dagger$ & 15:54:12.2 & -2:49:42 & 140 & \nodata & \citet{Hilton1995} & \citet{Karoly2020} & \nodata \\
9 & IRAS 15398-3359 & 15:43:01.0 & -34:09:04 & 146 & \nodata & \citet{Prisinzano2022} & \citet{Yen2024_eDisk} & \ref{afig:Bmap_I15398} \\
10 & L1527 & 4:39:53.8 & +26:03:08 & 147 & 0.75 & \citet{Zucker2020} & \parbox[t]{3.0cm}{\raggedright \citet{Yen2021, Nam2026}\strut} & \ref{afig:Bmap_L1527L} \\
11 & TMC-1A$^\dagger$ & 4:39:34.5 & +25:41:45 & 147 & 0.67 & \citet{Zucker2020} & \citet{Yen2024_eDisk} & \nodata \\
12 & L1689 North & 16:32:24.5 & -24:28:57 & 154 & 0.99 & \citet{Zucker2020} & \citet{Pattle2021} & \ref{afig:Bmap_L1689N} \\
13 & L1689B$^\dagger$ & 16:34:48.3 & -24:38:04 & 154 & 0.44 & \citet{Zucker2020} & \citet{Pattle2021} & \nodata \\
14 & B335 & 19:37:00.4 & +7:34:10 & 165 & \nodata & \citet{Watson2020} & \citet{Yen2019_B335} & \ref{afig:Bmap_B335B3} \\
15 & L1448-IRS2 & 3:25:22.4 & +30:44:49 & 240 & 0.28 & \citet{Zucker2020} & \citet{Yen2021} & \ref{afig:Bmap_L1448I} \\
16 & L483 & 18:17:34.7 & -4:40:02 & 250 & \nodata & \citet{Hilton1995} & \nodata & \ref{afig:Bmap_L483L4} \\
17 & GF 9-2$^\dagger$ & 20:51:29.8 & +60:18:38 & 270 & \nodata & \citet{Clemens2018} & \nodata & \nodata \\
18 & Perseus B1 & 3:33:17.5 & +31:08:26 & 276 & 0.50 & \citet{Zucker2020} & \citet{Coude2019} & \ref{afig:Bmap_PersB1} \\
19 & NGC 1333 & 3:29:05.6 & +31:17:29 & 287 & 0.36 & \citet{Zucker2020} & \parbox[t]{3.0cm}{\raggedright \citet{Doi2020, Doi2021}\strut} & \ref{afig:Bmap_N133MO} \\
20 & SSTc2d J033032.7$^\dagger$ & 3:30:32.0 & +30:26:26 & 291 & 0.10 & \citet{Zucker2020} & \nodata & \nodata \\
21 & IC 348 / HH 211 & 3:43:57.7 & +32:01:24 & 305 & 0.56 & \citet{Zucker2020} & \citet{Choi2024} & \ref{afig:Bmap_IC348H} \\
22 & L1157 & 20:39:11.5 & +68:02:08 & 341 & 1.83 & \citet{Zucker2020} & \citet{Yen2021} & \ref{afig:Bmap_L1157L} \\
23 & OMC 1 & 5:35:16.5 & -5:18:50 & 399 & 0.57 & \citet{Zucker2020} & $^\ddagger$ & \ref{afig:Bmap_OriAS1} \\
24 & OMC 2 & 5:35:25.5 & -5:09:52 & 399 & 0.53 & \citet{Zucker2020} & \citet{Wu2024} & \ref{afig:Bmap_OriAS2} \\
25 & OMC 3 & 5:35:22.5 & -5:01:08 & 399 & 0.49 & \citet{Zucker2020} & \citet{Wu2024} & \ref{afig:Bmap_OriAS3} \\
26 & HH 111 & 5:51:46.5 & +2:48:56 & 414 & 0.09 & \citet{Zucker2020} & \citet{Yen2021} & \ref{afig:Bmap_HH111H} \\
27 & PGCC G203.21-11.20 & 5:53:43.1 & +3:22:46 & 414 & 0.88 & \citet{Zucker2020} & \citet{Mallick2025} & \ref{afig:Bmap_OrBPMO} \\
28 & Orion B / NGC 2071 & 5:47:04.2 & +0:22:02 & 415 & 0.86 & \citet{Zucker2020} & \parbox[t]{3.0cm}{\raggedright \citet{Lyo2021, Fanciullo2022b}\strut} & \ref{afig:Bmap_NG2068} \\
29 & Serpens Main & 18:29:53.4 & +1:14:36 & 425 & 0.39 & \citet{Zucker2020} & \citet{Kwon2022} & \ref{afig:Bmap_SerpMO} \\
30 & SDC 24.013+0.488 & 18:33:33.4 & -7:39:02 & 430 & \nodata & \citet{Prisinzano2022} & \nodata & \ref{afig:Bmap_SDC240} \\
31 & SDC 24.489 & 18:38:26.6 & -7:49:10 & 430 & \nodata & \citet{Prisinzano2022} & \citet{Ramkumar2026} & \ref{afig:Bmap_SD24P4} \\
32 & SDC 25.166 & 18:38:13.7 & -7:02:35 & 430 & \nodata & \citet{Prisinzano2022} & \citet{Ramkumar2026} & \ref{afig:Bmap_SD25MO} \\
33 & HH 212$^\dagger$ & 5:43:51.4 & -1:02:53 & 433 & 0.39 & \citet{Zucker2020} & \citet{Yen2021} & \nodata \\
34 & Orion A Field 7 & 5:35:01.0 & -5:55:40 & 399$^\P$ & \nodata & \citet{Zucker2020} & \nodata & \nodata \\
35 & Orion A Field 4 & 5:35:13.3 & -5:27:19 & 399$^\P$ & \nodata & \citet{Zucker2020} & $^\ddagger$ & \nodata \\
36 & Orion A Field 5 & 5:35:05.0 & -5:33:06 & 399$^\P$ & \nodata & \citet{Zucker2020} & \nodata & \nodata \\
37 & Horsehead Nebula$^\dagger$ & 5:40:59.0 & -2:27:30 & 451 & 0.88 & \citet{Zucker2020} & \citet{Hwang2023_HHeadN} & \nodata \\
38 & IRAS 17527-2439 & 17:55:48.2 & -24:40:03 & 451 & \nodata & \citet{Prisinzano2022} & \nodata & \ref{afig:Bmap_I17527} \\
39 & IC 5146 & 21:47:25.9 & +47:32:00 & 460 & \nodata & \citet{Lada1999} & \citet{Wang2019_IC5146} & \ref{afig:Bmap_IC5146} \\
40 & IC 5146 Western Hub & 21:44:56.8 & +47:40:17 & 460 & \nodata & \citet{Lada1999} & \citet{Chung2022} & \ref{afig:Bmap_IC51WH} \\
41 & Serpens South & 18:29:52.3 & -1:57:29 & 487 & 0.40 & \citet{Zucker2020} & \nodata & \ref{afig:Bmap_SerpS3} \\
42 & L1688 (3-field mosaic) & 16:27:12.8 & -24:31:59 & 139 & \nodata & \citet{Zucker2020} & \citet{Pattle2019} & \ref{afig:Bmap_OPHIMO} \\
43 & Auriga & 4:30:17.0 & +35:16:36 & 532 & \nodata & \citet{Omodaka2020} & \citet{Ngoc2021} & \ref{afig:Bmap_Auriga} \\
44 & L1478 & 4:21:39.3 & +37:34:16 & 544 & \nodata & \citet{Prisinzano2022} & \citet{Chung2023} & \ref{afig:Bmap_L1478L} \\
45 & Cepheus A & 22:56:20.2 & +62:02:11 & 700 & \nodata & \citet{Moscadelli2009} & \citet{Sandhyarani2025} & \ref{afig:Bmap_CepACe} \\
46 & NGC 2264 & 6:41:07.1 & +9:32:50 & 715 & \nodata & \citet{Wang2024_NGC2264} & \citet{Wang2024_NGC2264} & \ref{afig:Bmap_N226MO} \\
47 & G202.3+2.5 & 6:41:01.0 & +10:36:33 & 760 & \nodata & \citet{Montillaud2019} & \citet{Gu2024} & \ref{afig:Bmap_G202MO} \\
48 & Monoceros R2 & 6:07:46.0 & -6:22:54 & 799 & 0.70 & \citet{Zucker2020} & \citet{Hwang2022} & \ref{afig:Bmap_MonoR2} \\
49 & G35.2 North$^\dagger$ & 18:58:13.0 & +1:40:36 & 838 & 0.53 & \citet{Prisinzano2022} & \nodata & \nodata \\
50 & G16$^\|$ & 18:20:42.5 & -14:05:29 & 849 & \nodata & \citet{Prisinzano2022} & \citet{Gu2024_G16} & \ref{afig:Bmap_G16G16} \\
51 & SDC 18.624-0.070 & 18:25:09.0 & -12:42:34 & 856 & \nodata & \citet{Prisinzano2022} & \citet{Lee2025_SDC18} & \ref{afig:Bmap_SDC186} \\
52 & IC 1396E & 21:40:45.3 & +58:16:47 & 905 & 0.12 & \citet{Zucker2020} & \citet{Sugitani2025} & \ref{afig:Bmap_IC139E} \\
53 & IC 1396A & 21:37:06.5 & +57:30:49 & 916 & 0.23 & \citet{Zucker2020} & \nodata & \ref{afig:Bmap_IC139A} \\
54 & IRAS 18360 & 18:38:40.6 & -5:34:38 & 957 & \nodata & \citet{Prisinzano2022} & \nodata & \ref{afig:Bmap_I18360} \\
55 & S106 & 20:27:36.2 & +37:18:49 & 1091 & 0.41 & \citet{Zucker2020} & \citet{Kumar2025} & \ref{afig:Bmap_S106S1} \\
56 & G34.257+0.155 & 18:53:17.8 & +1:19:56 & 1092 & 0.33 & \citet{Prisinzano2022} & \parbox[t]{3.0cm}{\raggedright \citet{Khan2024, Ramkumar2026}\strut} & \ref{afig:Bmap_G342MO} \\
57 & CB 54$^\dagger$ & 7:04:21.5 & -16:23:19 & 1100 & \nodata & \citet{Brand1993} & \citet{Pattle2022_CB54} & \nodata \\
58 & DR21 & 20:39:02.7 & +42:21:25 & 1500 & \nodata & \citet{Rygl2012} & \citet{Ching2022} & \ref{afig:Bmap_DR21DR} \\
59 & G192.16--3.84 & 5:58:14.6 & +16:32:02 & 1520 & \nodata & \citet{Shiozaki2011} & \nodata & \ref{afig:Bmap_G19216} \\
60 & Rosette & 6:34:11.5 & +4:22:59 & 1571 & 0.71 & \citet{Prisinzano2022} & \citet{Konyves2021} & \ref{afig:Bmap_RoseMO} \\
61 & M16 & 18:18:51.6 & -13:49:45 & 1739 & 0.71 & \citet{Zucker2020} & \citet{Pattle2018} & \ref{afig:Bmap_M16M16} \\
62 & NGC 6334 & 17:20:35.8 & -35:50:08 & 1750 & \nodata & \citet{Russeil2020} & \parbox[t]{3.0cm}{\raggedright \citet{Arzoumanian2021, Tahani2023}\strut}& \ref{afig:Bmap_N633MO} \\
63 & SH2-201 & 3:03:23.9 & +60:27:35 & 1962 & 0.71 & \citet{Zucker2020} & \citet{Eswaraiah2020} & \ref{afig:Bmap_SH2F1S} \\
64 & W3 HFS1 & 2:25:39.9 & +62:05:58 & 2184 & 0.08 & \citet{Zucker2020} & \citet{Kumar2025} & \ref{afig:Bmap_W3HFS1} \\
65 & W3 HFS2 & 2:27:10.2 & +61:52:52 & 2184 & 0.33 & \citet{Zucker2020} & \citet{Kumar2025} & \ref{afig:Bmap_W3HFS2} \\
66 & NGC 7538 & 23:13:33.3 & +61:29:35 & 2650 & \nodata & \citet{Sharma2017} & \nodata & \ref{afig:Bmap_N753I4} \\
67 & SDC 13 & 18:14:27.4 & -17:33:23 & 3600 & \nodata & \citet{Peretto2014} & \citet{Wang2022_SDC13} & \ref{afig:Bmap_SDC13M} \\
68 & G31.41+0.31 & 18:47:33.2 & -1:12:38 & 3750 & \nodata & \citet{Immer2019} & \citet{Law2025} & \ref{afig:Bmap_G3141G} \\
69 & Onsala 2 & 20:21:45.7 & +37:26:14 & 3830 & \nodata & \citet{Ando2011} & \nodata & \ref{afig:Bmap_DR15DR} \\
70 & G28.34+0.06 & 18:42:52.0 & -4:00:08 & 4800 & \nodata & \citet{Carey1998} & \citet{Hwang2025} & \ref{afig:Bmap_G2834G}, \ref{afig:Bmap_G283MO} \\
71 & G240.310 & 7:44:50.8 & -24:07:33 & 5320 & \nodata & \citet{Choi2014} & \nodata & \ref{afig:Bmap_G24031} \\
72 & W51$^\S$ & 19:23:43.9 & +14:30:31 & 5419 & \nodata & \citet{Sato2010} & \nodata & \nodata \\
73 & Gal. Cent. 20 km/s & 17:45:36.6 & -29:04:20 & 7860 & \nodata & \citet{Boehle2016} & \citet{Karoly2025} & \ref{afig:Bmap_GC20KC} \\
74 & Gal. Cent. Brick & 17:46:08.5 & -28:42:50 & 7860 & \nodata & \citet{Boehle2016} & \citet{Karoly2025} & \ref{afig:Bmap_GCBRCK} \\
75 & Gal. Cent. Clouds E/F & 17:46:44.2 & -28:30:54 & 7860 & \nodata & \citet{Boehle2016} & \citet{Karoly2025} & \ref{afig:Bmap_GCCLEF} \\
76 & Gal. Cent. Middle & 17:45:38.6 & -29:00:18 & 7860 & \nodata & \citet{Boehle2016} & \parbox[t]{3.0cm}{\raggedright \citet{Sato2024, Yang2025, Karoly2025}\strut} & \ref{afig:Bmap_GCMIDD} \\
77 & Gal. Cent. North & 17:45:44.6 & -28:54:07 & 7860 & \nodata & \citet{Boehle2016} & \parbox[t]{3.0cm}{\raggedright \citet{Yang2025, Karoly2025}\strut} & \ref{afig:Bmap_GCNORT} \\
78 & Gal. Cent. South & 17:45:36.1 & -29:03:39 & 7860 & \nodata & \citet{Boehle2016} & \parbox[t]{3.0cm}{\raggedright \citet{Yang2025, Karoly2025}\strut} & \ref{afig:Bmap_GCSOUT} \\
79 & Gal.~Cent.~(7-field mosaic) & 17:45:38.6 & -29:00:18 & 7860 & \nodata & \citet{Boehle2016} & \citet{Karoly2025} & \ref{afig:Bmap_GC20KC}--\ref{afig:Bmap_GCSOUT} \\
80 & G33.92+0.119 & 18:52:49.7 & +0:55:25 & 8200 & \nodata & \citet{Wink1982} & \citet{Wang2020_G3392} & \ref{afig:Bmap_G3392G} \\
81 & Sagittarius B2 & 17:47:19.0 & -28:22:55 & 8340 & \nodata & \citet{Reid2014} & \citet{Karoly2025} & \ref{afig:Bmap_GSGRB2} \\
\enddata
\tablecomments{
Columns:
(1) sequential number;
(2) source name;
(3)--(4) J2000 equatorial coordinates of the map center;
(5) adopted reference distance $d_{\rm ref}$ in parsecs;
(6) angular separation between the POL-2 field center and the reference position from the distance catalog;
(7) distance reference;
(8) representative POL-2 publication for this field, if available;
(9) figure number in appendix~\ref{app:S2_atlas} showing the map of $\vec{B}$-field orientations at both wavelengths.
The listed distances $d_{\rm ref}$ are reference scaling factors adopted for angular-to-physical conversion, rather than uniquely defined physical distances to point-like objects.
Appendix~\ref{app:distance} describes how we determine these reference distances.
When the separation column shows ``\nodata'', the distance was derived from maser proper motion or literature values without catalog cross-matching.
$^\dagger$Sources observed at 850\,$\mu$m only; 450\,$\mu$m data are either non-detection or of insufficient quality for polarization analysis.
We include these sources in the statistical analyses presented in the main text where applicable.
We applied the data-quality checks to L1689B, but the signal is too weak for any position to pass them.
We therefore exclude this field from the analyses.
$^\S$W51 was observed but is excluded from all quantitative analyses in this paper and in Paper~S1: its default 850\,\micron\ reduction is affected by systematics in the filtering (FLT) and principal component analysis (PCA) stages of the iterative map maker \texttt{makemap}~\citep{Chapin2013} documented in Paper~T appendix~\ref{T-app:QU_diagnostics}, which corrupt the reconstructed Stokes parameters toward the bright central emission.
None of the alternative reductions produced under the survey's default masking parameters is free of this artifact.
We therefore exclude the entire field rather than mask only part of it.
$^\ddagger$For OMC~1, multiple BISTRO/POL-2 publications exist; see \citet{Ward-Thompson2017,Pattle2017,Hwang2021}. Orion~A Field~4 (entry~35) carries the same mark because most of its pixels with $I/\sigma_I > 30$ lie within the OMC~1 field.
$^\P$Orion~A Fields~4, 5 and 7 lie within the Orion~A cloud and are assigned the OMC~1 distance adopted for the mosaicked Orion~A fields.
The \citet{Zucker2020} solutions at their own pointing centers would give $445$, $445$ and $438$\,pc, respectively.
These three fields are retained in the statistical samples, although their maps are not reproduced in the paper.
$^\|$The 450\,$\mu$m data of G16 are of low quality, so we exclude this field from the dual-band analyses.
We retain it for the data-quality checks of Paper~T.
Mosaics are survey products rather than additional independent sources.
Entry~42 combines entries~5--7 (Ophiuchus~A, B, and C), and entry~79 combines the seven CMZ fields listed below.
The seven Central Molecular Zone (CMZ) sources used in the statistical analysis of this paper are entries 73--78 and 81: the Gal.~Cent.~20~km~s$^{-1}$ cloud, Brick, Clouds~E/F, Middle, North, and South, together with Sagittarius~B2. The figure column of entry~79 lists only the maps of entries~73--78. We give the map of Sagittarius~B2 with entry~81 alone, so that the atlas ends with Sagittarius~B2 and its extended 450\,$\mu$m map (appendix~\ref{ass:S2_SgrB2_450note}).
Entry~80, G33.92+0.119, lies at a comparable adopted distance but at Galactic longitude $l = 33\fdg9$, well outside the CMZ, and does not enter the spectral-index analysis.
See \citet{Karoly2025} for an overview of the BISTRO CMZ survey.
For G28.34+0.06 (row~70), the atlas presents both the 450\,$\mu$m observation of the single field (figure~\ref{afig:Bmap_G2834G}) and the wider map of the mosaicked field (figure~\ref{afig:Bmap_G283MO}).
The row for this survey field points to both figures.
}
\end{deluxetable}
\end{longrotatetable}

\section{Dual-Wavelength Magnetic Field Orientation Maps}
\label{app:S2_atlas}
 
This appendix presents an atlas of maps of $\vec{B}$-field orientations
derived from POL-2 polarimetry at 450 and 850\,$\mu$m.
Appendix~\ref{ass:S2_atlas_maps} provides the maps for 65 of the 69 sources of the {\sc PolSrc} sample with detected 450\,$\mu$m polarization---all but W51, whose 850\,$\mu$m map is corrupted, and Orion~A Fields~4, 5 and 7 (see the notes to table~\ref{tab:surv})--- while appendix~\ref{ass:S2_SgrB2_450note}
addresses the extended 450\,$\mu$m polarization coverage observed toward Sagittarius~B2.
The listing of individual survey fields in appendix~\ref{app:S2_survey_fields} (table~\ref{tab:surv}) provides the master index linking each POL-2 field to its corresponding atlas figure below.
 
\subsection{Map Atlas}
\label{ass:S2_atlas_maps}
 
\begin{figure}[H]
\centering
\includegraphics[width=0.95\textwidth]{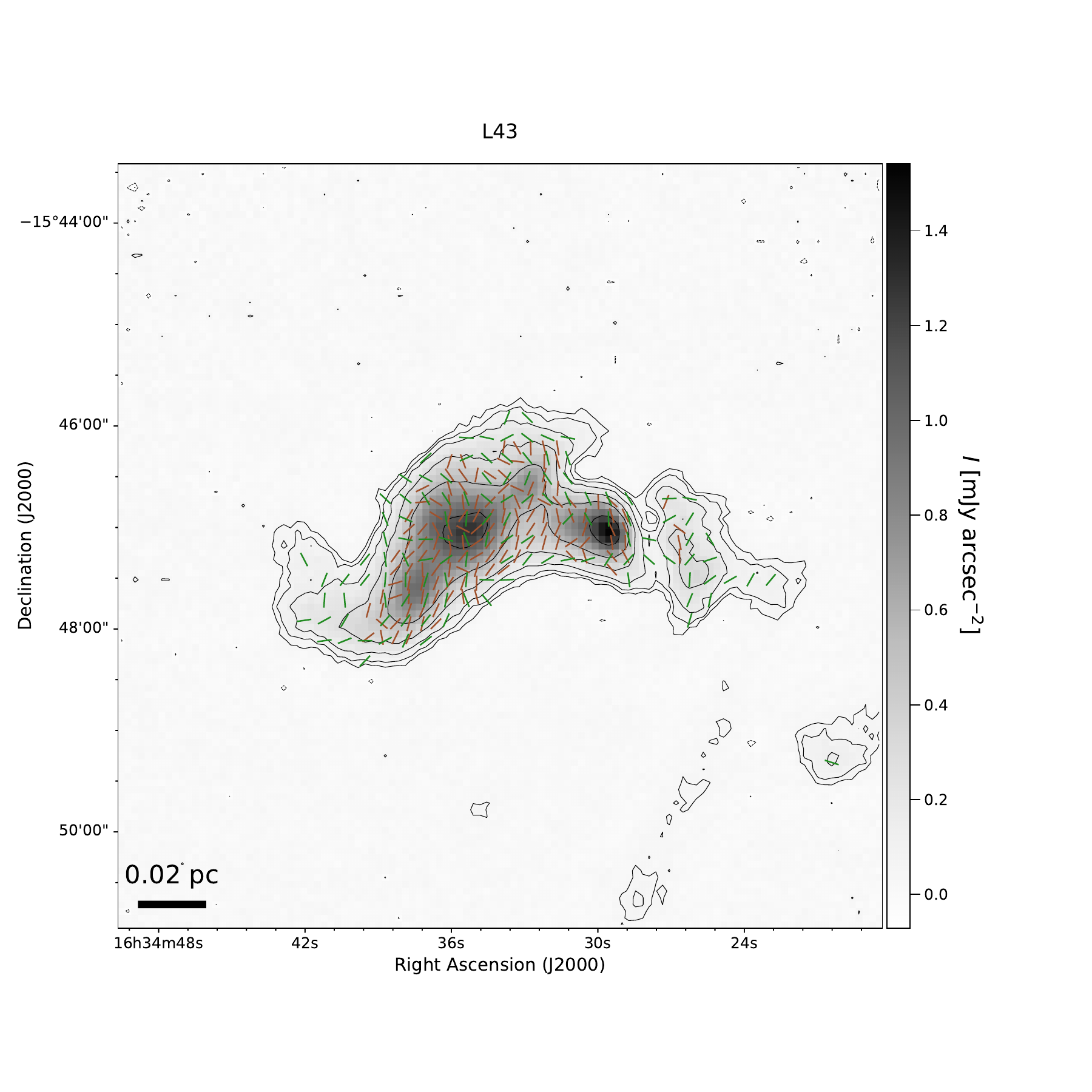}
\caption{Magnetic field orientation map of L43 observed with JCMT POL-2.
Green and sienna segments show the $\vec{B}$-field orientations in the plane of the sky, inferred from 850 and 450\,$\mu$m polarization, respectively, with the polarization angles rotated by $90\degr$.
Contours trace 850\,$\mu$m Stokes $I$ intensity with levels at $15\sigma \times 2^n$
($n = 0, 1, 2, \ldots$), where $\sigma_{850} = 0.0023$\,mJy\,arcsec$^{-2}$.
The 450\,$\mu$m noise level (not shown as contours) is $\sigma_{450} = 0.035$\,mJy\,arcsec$^{-2}$.
These noise levels are blank-sky estimates rescaled to a $4\arcsec$ pixel scale, following the procedure in Paper~T subsection~\ref{T-ss:noiseI}.
The quoted 850\,$\mu$m value sets the contour levels.
We select magnetic field segments independently at each wavelength using $I/\Delta I > 30$, where $\Delta I$ is the uncertainty of Stokes~$I$ for each pixel in the polarization catalog.
These per-pixel uncertainties are distinct from the quoted blank-sky noise levels.
The segments are sampled on $12\arcsec$ and $8\arcsec$ grids at 850 and 450\,$\mu$m, respectively.
We do not apply an additional selection based on the signal-to-noise ratio of polarized intensity.
The horizontal bar at lower left shows the physical scale.
Table~\ref{tab:surv} (appendix~\ref{app:S2_survey_fields}) lists the adopted distance and reference for every atlas field.
Appendix~\ref{app:distance} describes how we assign these distances.
\\ {Alt text: Map with short line segments over contours.}
}
\label{afig:Bmap_L43L43}
\end{figure}

\begin{figure}[H]
\centering
\includegraphics[width=0.95\textwidth]{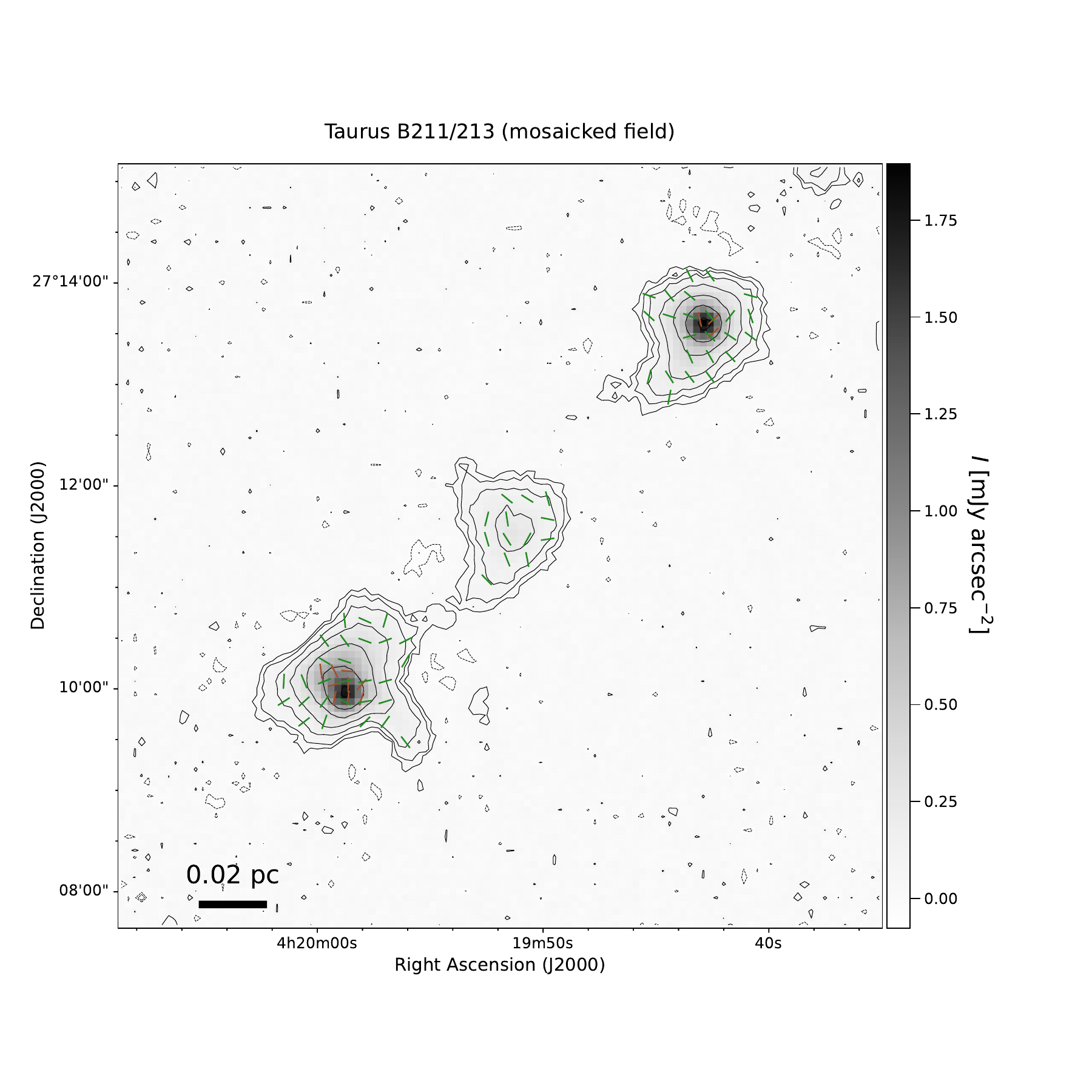}
\caption{Same as figure~\ref{afig:Bmap_L43L43}, but for Taurus B211/213 (mosaicked field).
The noise levels are $\sigma_{850} = 0.0017$\,mJy\,arcsec$^{-2}$ and $\sigma_{450} = 0.033$\,mJy\,arcsec$^{-2}$.
\\ {Alt text: Map with short line segments over contours.}
}
\label{afig:Bmap_B211MO}
\end{figure}

\begin{figure}[H]
\centering
\includegraphics[width=0.95\textwidth]{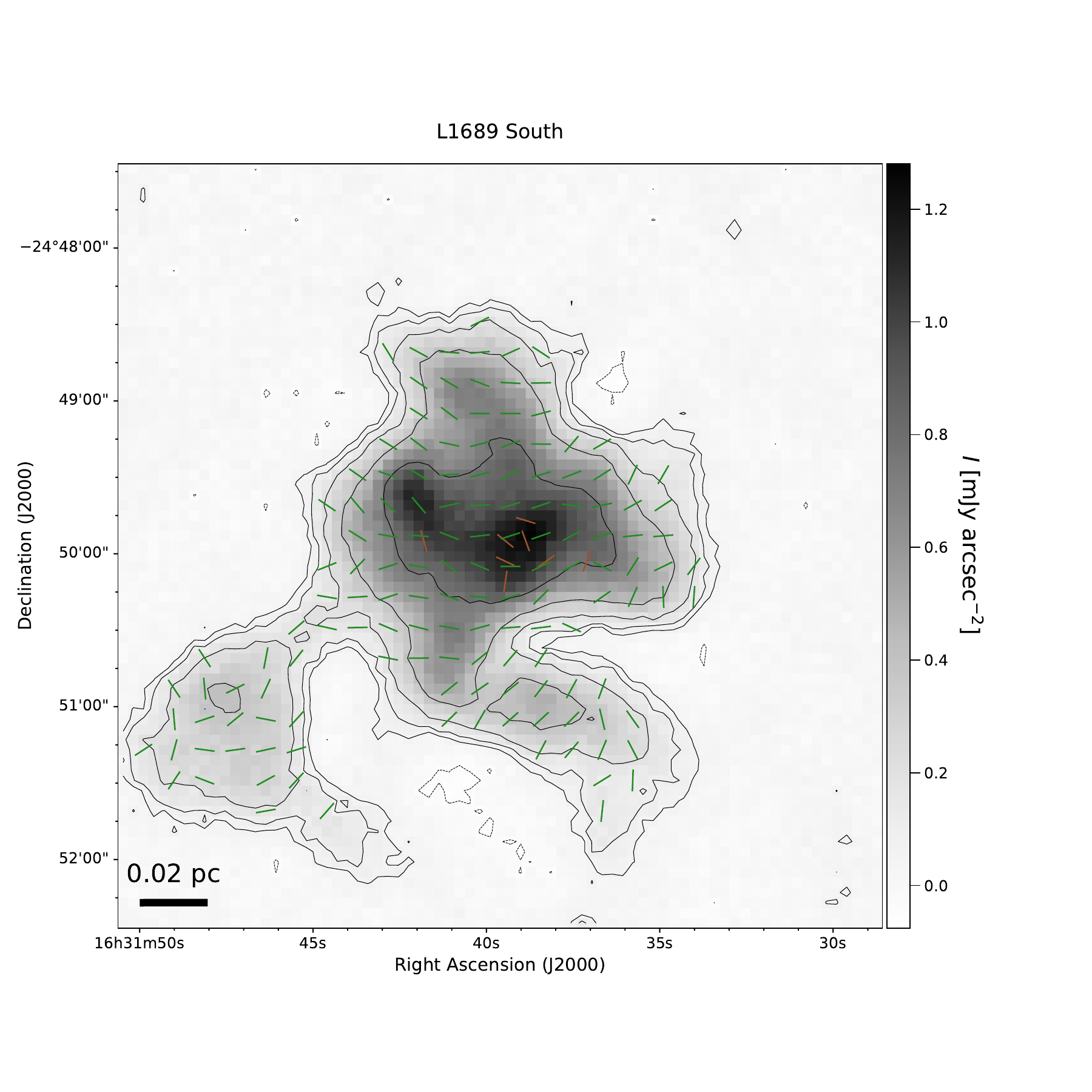}
\caption{Same as figure~\ref{afig:Bmap_L43L43}, but for L1689 South.
The noise levels are $\sigma_{850} = 0.0032$\,mJy\,arcsec$^{-2}$ and $\sigma_{450} = 0.18$\,mJy\,arcsec$^{-2}$.
\\ {Alt text: Map with short line segments over contours.}
}
\label{afig:Bmap_L1689S}
\end{figure}

\begin{figure}[H]
\centering
\includegraphics[width=0.95\textwidth]{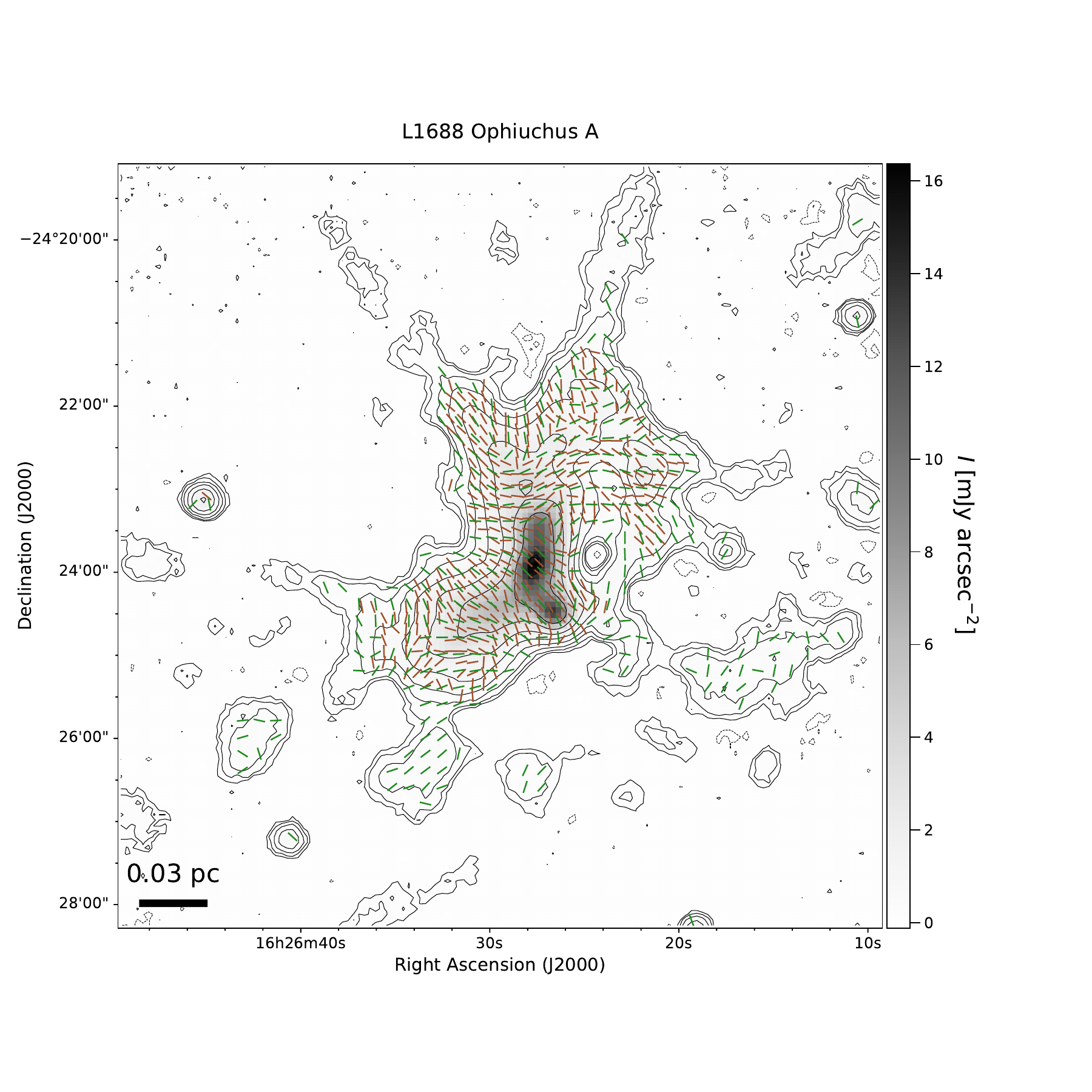}
\caption{Same as figure~\ref{afig:Bmap_L43L43}, but for L1688 Ophiuchus A.
The noise levels are $\sigma_{850} = 0.0038$\,mJy\,arcsec$^{-2}$ and $\sigma_{450} = 0.071$\,mJy\,arcsec$^{-2}$.
\\ {Alt text: Map with short line segments over contours.}
}
\label{afig:Bmap_OphiuA}
\end{figure}

\begin{figure}[H]
\centering
\includegraphics[width=0.95\textwidth]{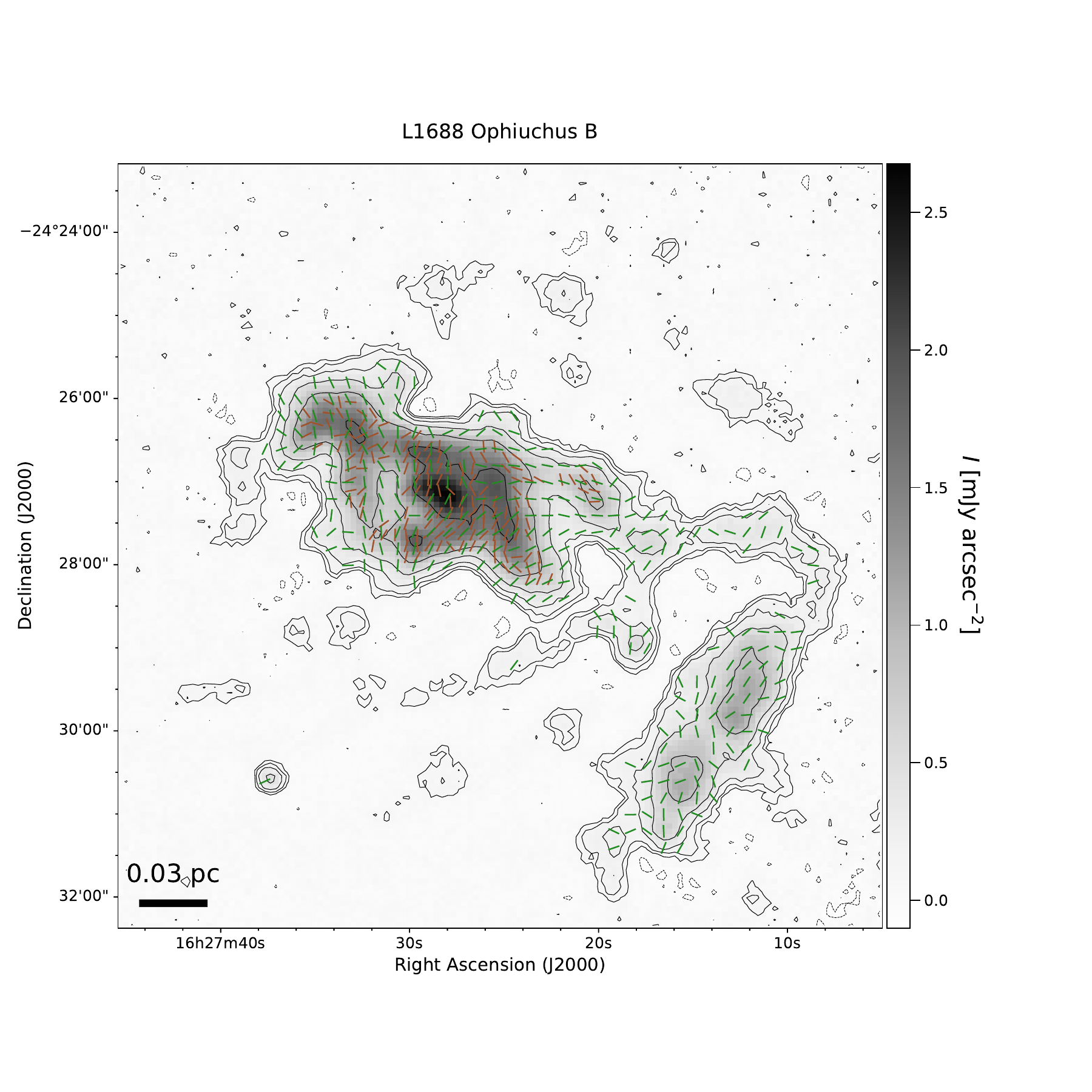}
\caption{Same as figure~\ref{afig:Bmap_L43L43}, but for L1688 Ophiuchus B.
The noise levels are $\sigma_{850} = 0.0037$\,mJy\,arcsec$^{-2}$ and $\sigma_{450} = 0.079$\,mJy\,arcsec$^{-2}$.
\\ {Alt text: Map with short line segments over contours.}
}
\label{afig:Bmap_OphiuB}
\end{figure}

\begin{figure}[H]
\centering
\includegraphics[width=0.95\textwidth]{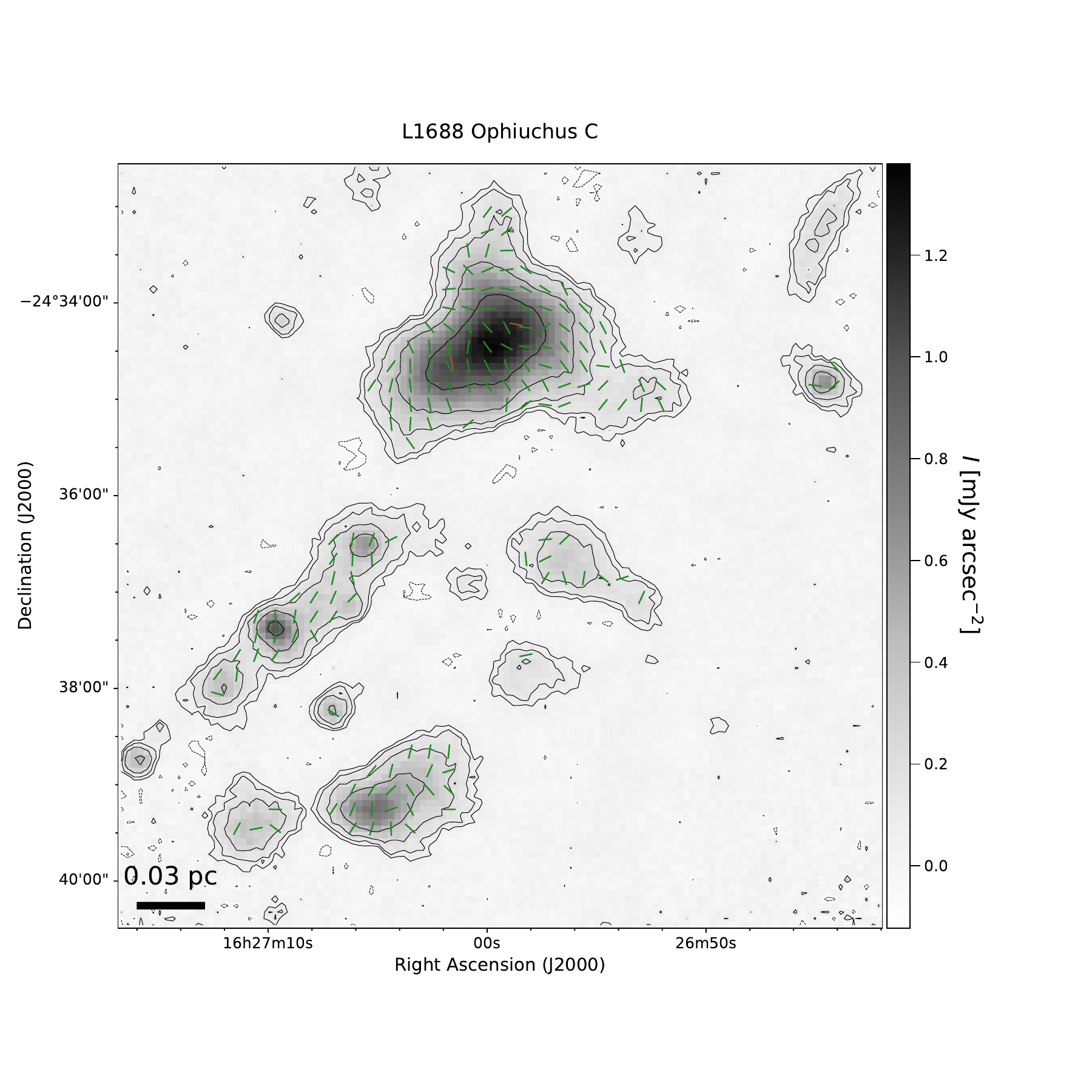}
\caption{Same as figure~\ref{afig:Bmap_L43L43}, but for L1688 Ophiuchus C.
The noise levels are $\sigma_{850} = 0.0036$\,mJy\,arcsec$^{-2}$ and $\sigma_{450} = 0.085$\,mJy\,arcsec$^{-2}$.
\\ {Alt text: Map with short line segments over contours.}
}
\label{afig:Bmap_OphiuC}
\end{figure}

\begin{figure}[H]
\centering
\includegraphics[width=0.95\textwidth]{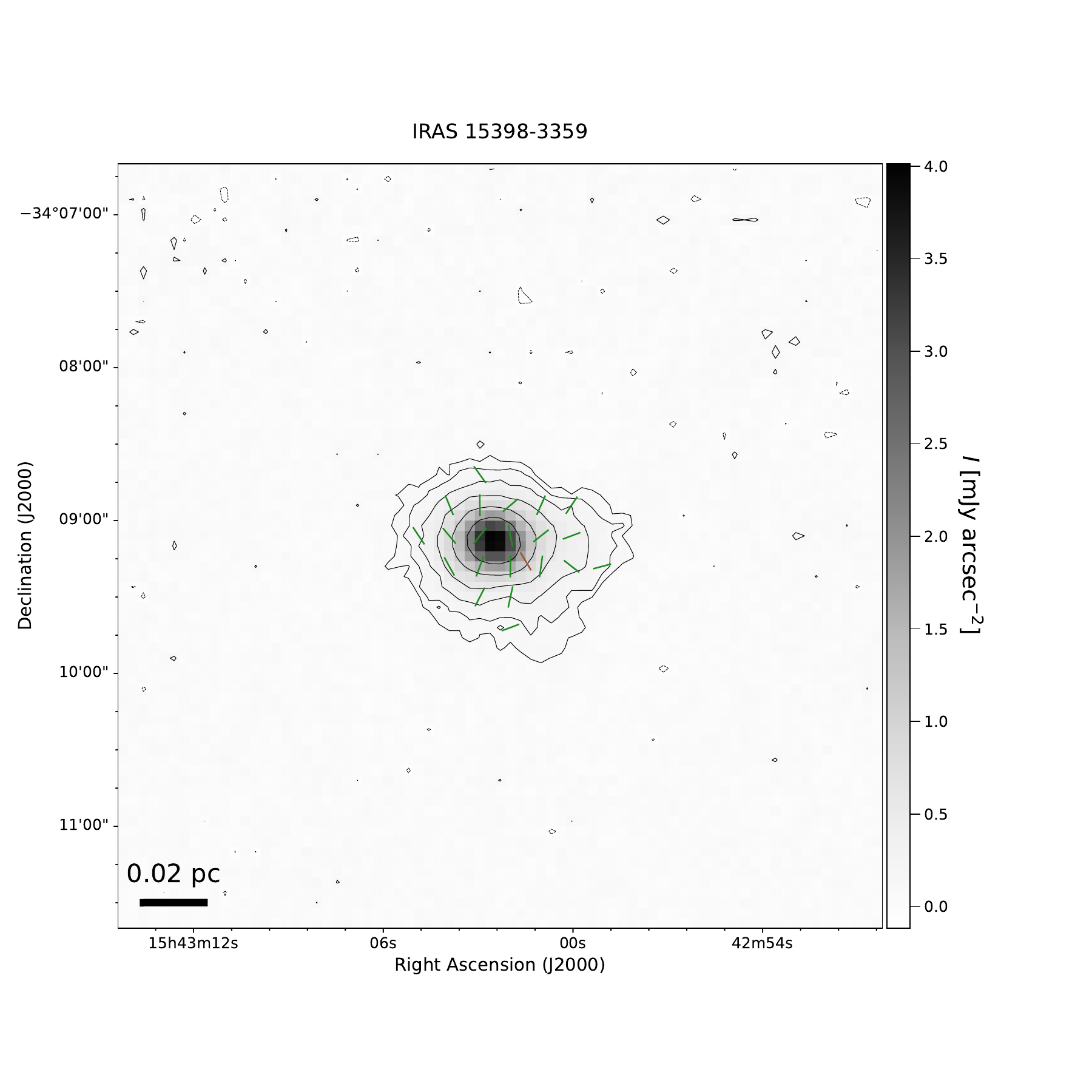}
\caption{Same as figure~\ref{afig:Bmap_L43L43}, but for IRAS 15398-3359.
The noise levels are $\sigma_{850} = 0.0044$\,mJy\,arcsec$^{-2}$ and $\sigma_{450} = 0.25$\,mJy\,arcsec$^{-2}$.
\\ {Alt text: Map with short line segments over contours.}
}
\label{afig:Bmap_I15398}
\end{figure}

\begin{figure}[H]
\centering
\includegraphics[width=0.95\textwidth]{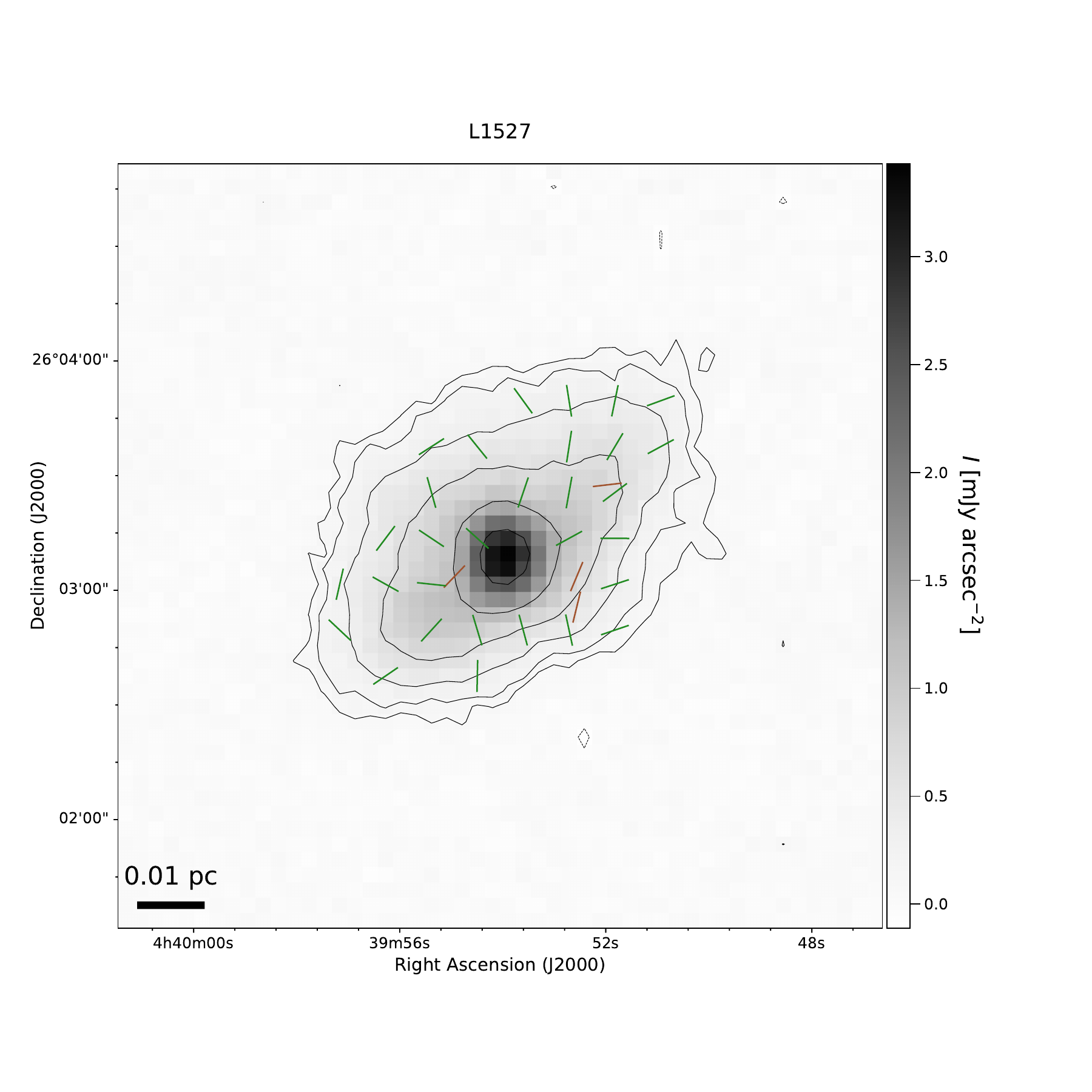}
\caption{Same as figure~\ref{afig:Bmap_L43L43}, but for L1527.
The noise levels are $\sigma_{850} = 0.0053$\,mJy\,arcsec$^{-2}$ and $\sigma_{450} = 0.18$\,mJy\,arcsec$^{-2}$.
\\ {Alt text: Map with short line segments over contours.}
}
\label{afig:Bmap_L1527L}
\end{figure}

\begin{figure}[H]
\centering
\includegraphics[width=0.95\textwidth]{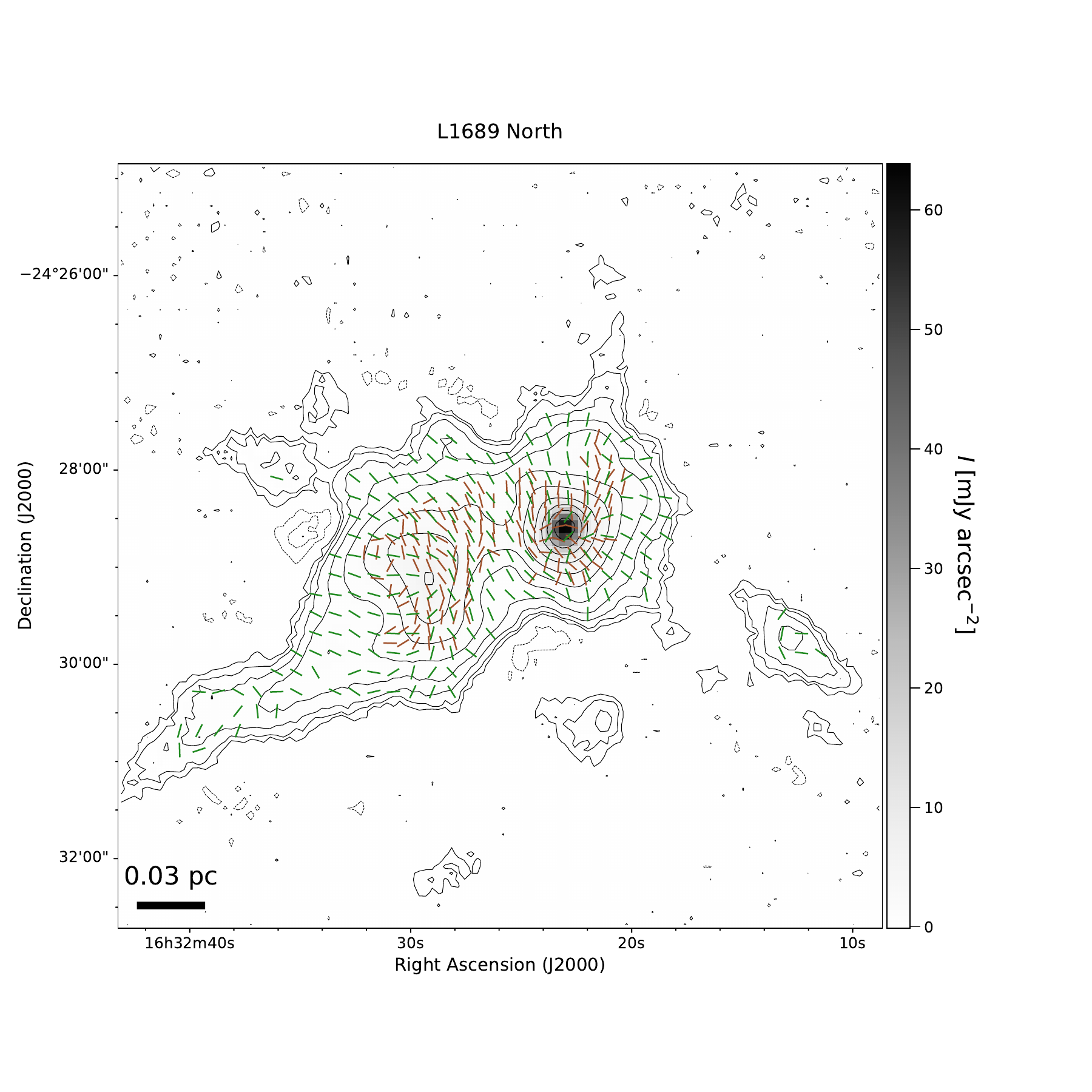}
\caption{Same as figure~\ref{afig:Bmap_L43L43}, but for L1689 North.
The noise levels are $\sigma_{850} = 0.0029$\,mJy\,arcsec$^{-2}$ and $\sigma_{450} = 0.098$\,mJy\,arcsec$^{-2}$.
\\ {Alt text: Map with short line segments over contours.}
}
\label{afig:Bmap_L1689N}
\end{figure}

\begin{figure}[H]
\centering
\includegraphics[width=0.95\textwidth]{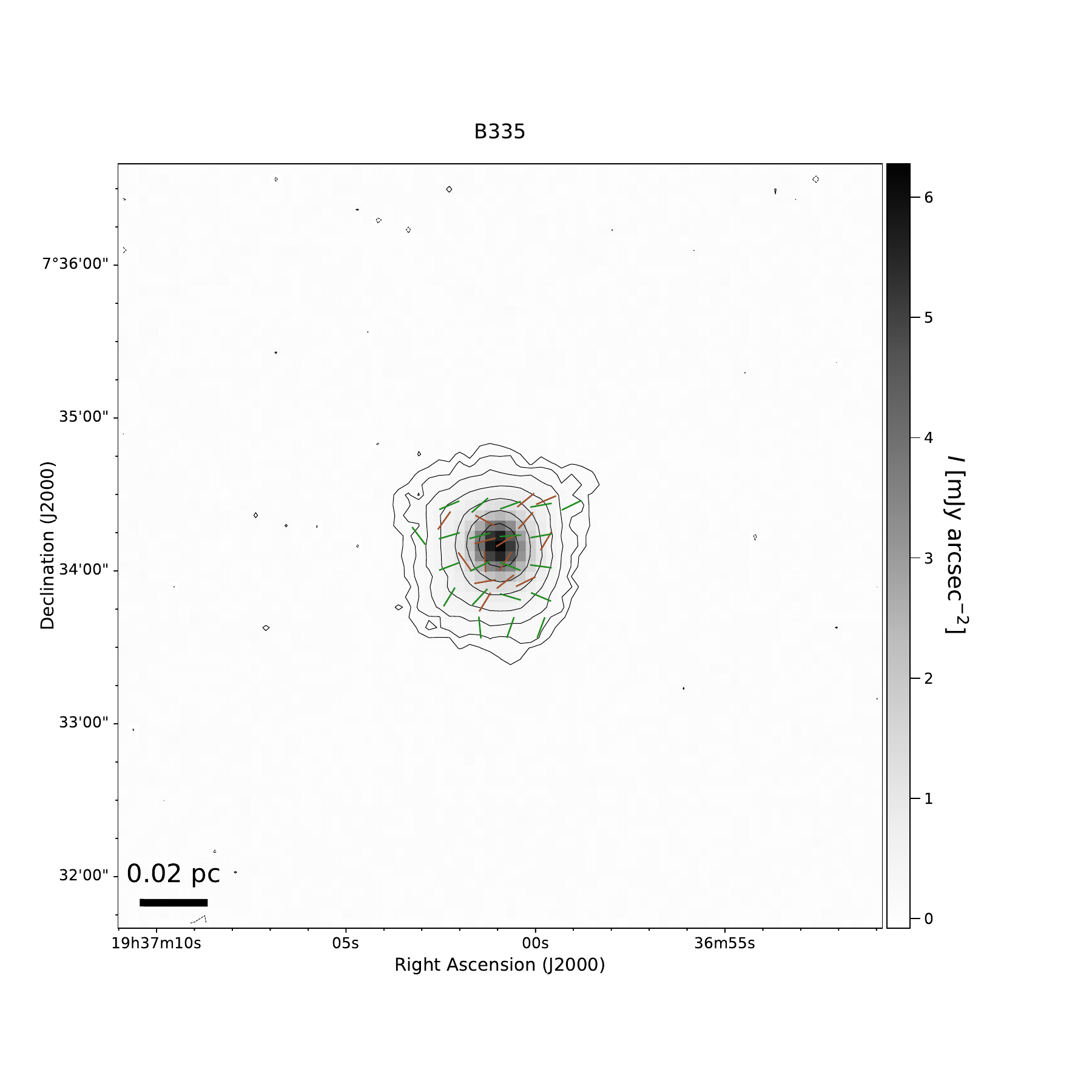}
\caption{Same as figure~\ref{afig:Bmap_L43L43}, but for B335.
The noise levels are $\sigma_{850} = 0.0039$\,mJy\,arcsec$^{-2}$ and $\sigma_{450} = 0.058$\,mJy\,arcsec$^{-2}$.
\\ {Alt text: Map with short line segments over contours.}
}
\label{afig:Bmap_B335B3}
\end{figure}

\begin{figure}[H]
\centering
\includegraphics[width=0.95\textwidth]{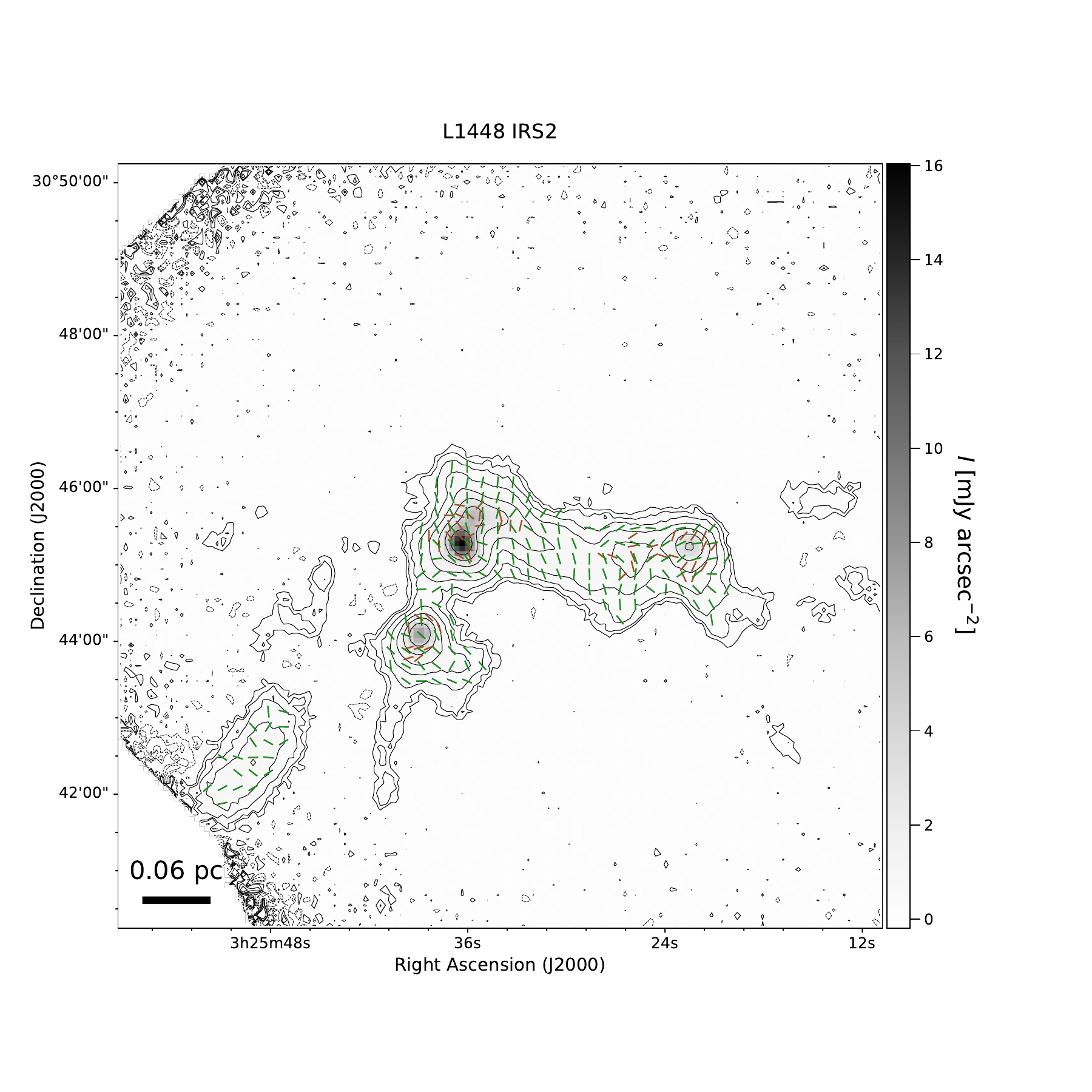}
\caption{Same as figure~\ref{afig:Bmap_L43L43}, but for L1448 IRS2.
The noise levels are $\sigma_{850} = 0.0042$\,mJy\,arcsec$^{-2}$ and $\sigma_{450} = 0.066$\,mJy\,arcsec$^{-2}$.
\\ {Alt text: Map with short line segments over contours.}
}
\label{afig:Bmap_L1448I}
\end{figure}

\begin{figure}[H]
\centering
\includegraphics[width=0.95\textwidth]{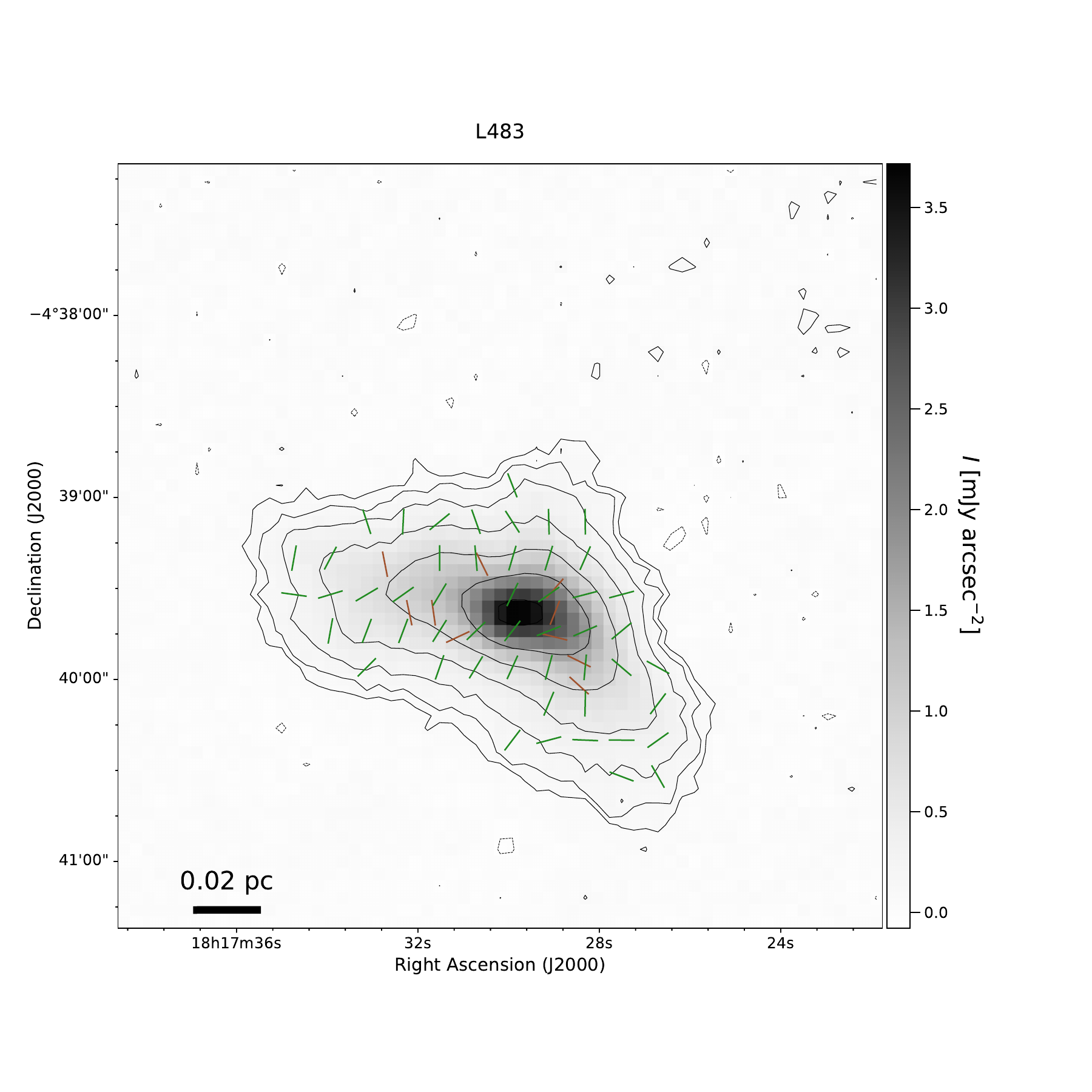}
\caption{Same as figure~\ref{afig:Bmap_L43L43}, but for L483.
The noise levels are $\sigma_{850} = 0.0035$\,mJy\,arcsec$^{-2}$ and $\sigma_{450} = 0.11$\,mJy\,arcsec$^{-2}$.
\\ {Alt text: Map with short line segments over contours.}
}
\label{afig:Bmap_L483L4}
\end{figure}

\begin{figure}[H]
\centering
\includegraphics[width=0.95\textwidth]{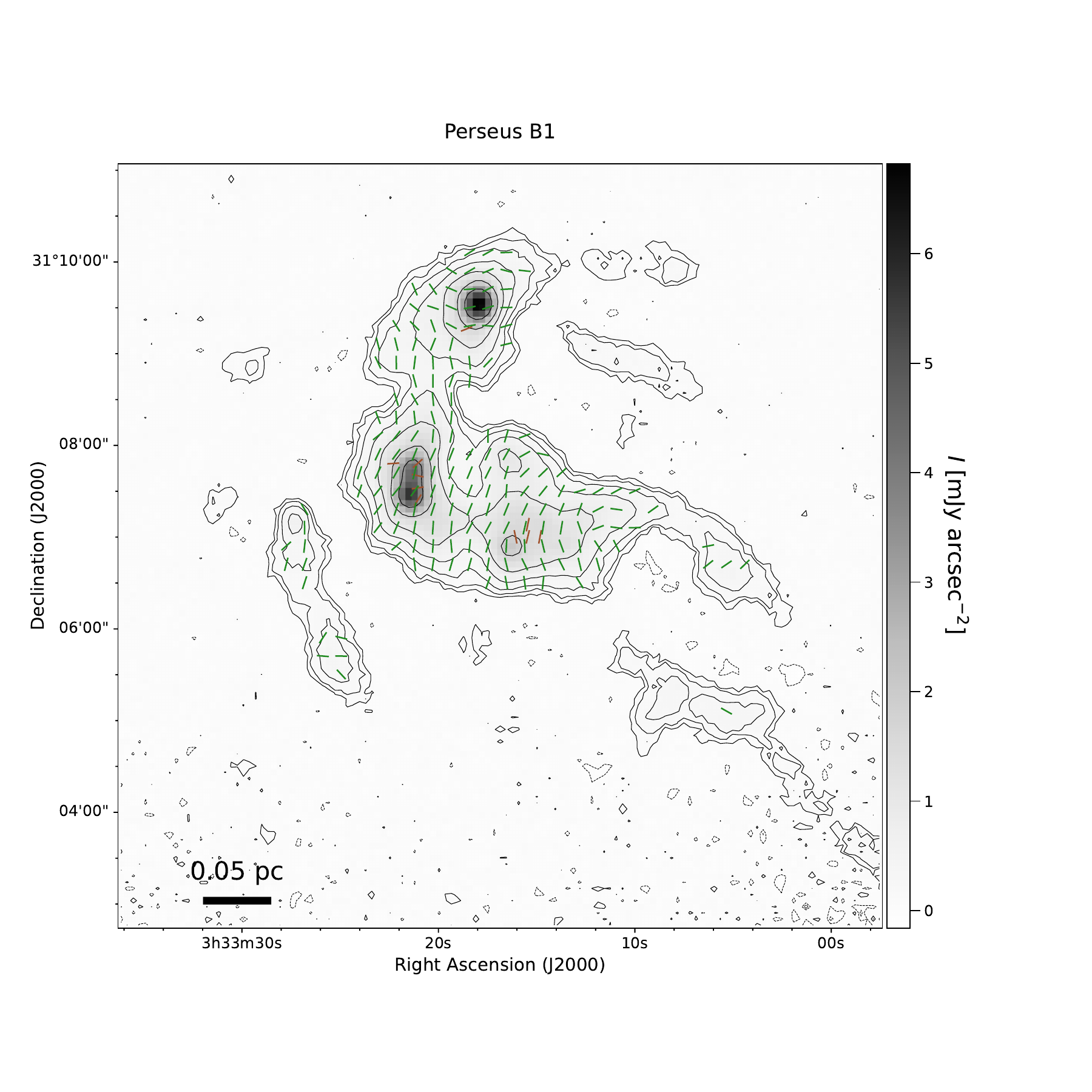}
\caption{Same as figure~\ref{afig:Bmap_L43L43}, but for Perseus B1.
The noise levels are $\sigma_{850} = 0.0036$\,mJy\,arcsec$^{-2}$ and $\sigma_{450} = 0.10$\,mJy\,arcsec$^{-2}$.
\\ {Alt text: Map with short line segments over contours.}
}
\label{afig:Bmap_PersB1}
\end{figure}

\begin{figure}[H]
\centering
\includegraphics[width=0.95\textwidth]{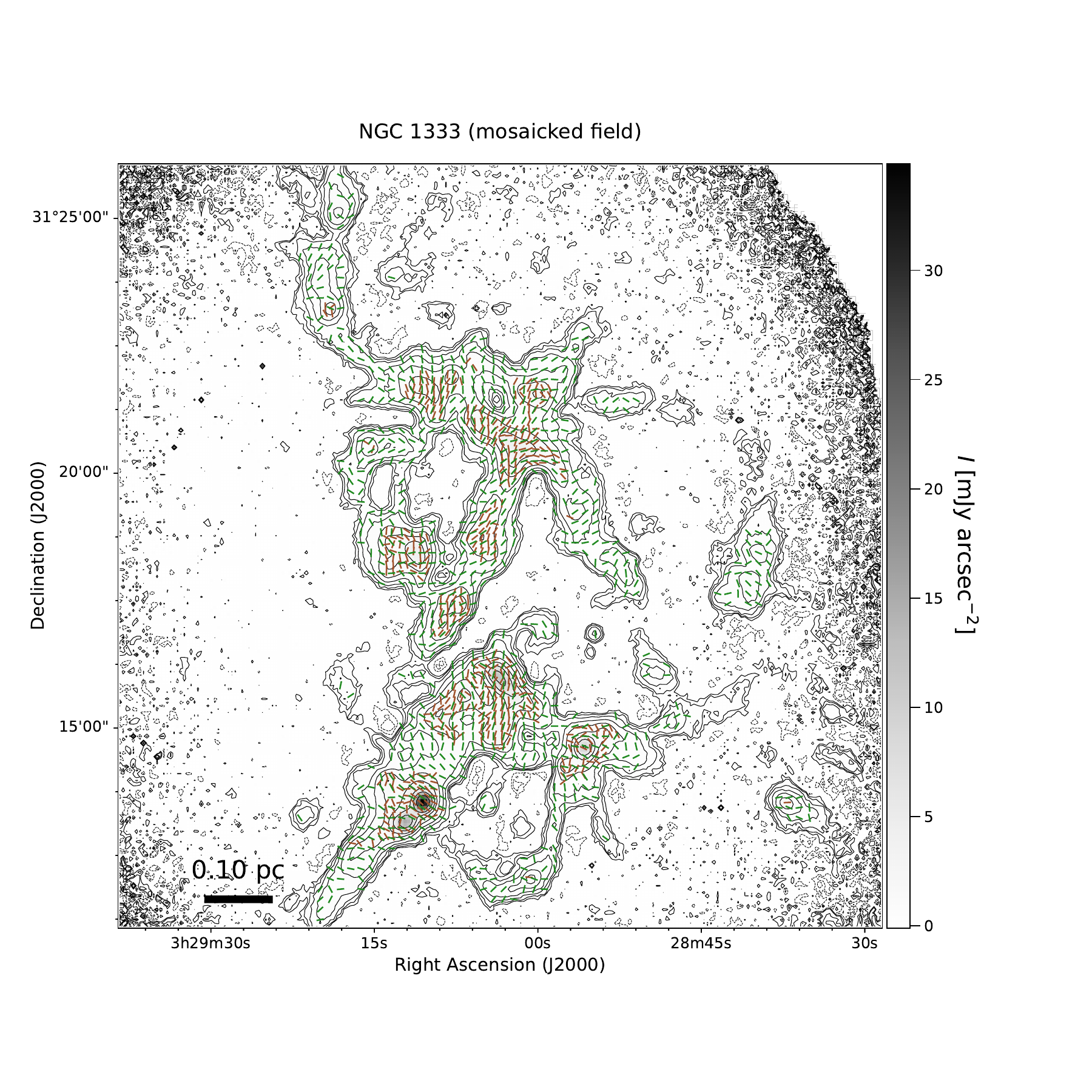}
\caption{Same as figure~\ref{afig:Bmap_L43L43}, but for NGC 1333 (mosaicked field).
The noise levels are $\sigma_{850} = 0.0024$\,mJy\,arcsec$^{-2}$ and $\sigma_{450} = 0.032$\,mJy\,arcsec$^{-2}$.
\\ {Alt text: Map with short line segments over contours.}
}
\label{afig:Bmap_N133MO}
\end{figure}

\begin{figure}[H]
\centering
\includegraphics[width=0.95\textwidth]{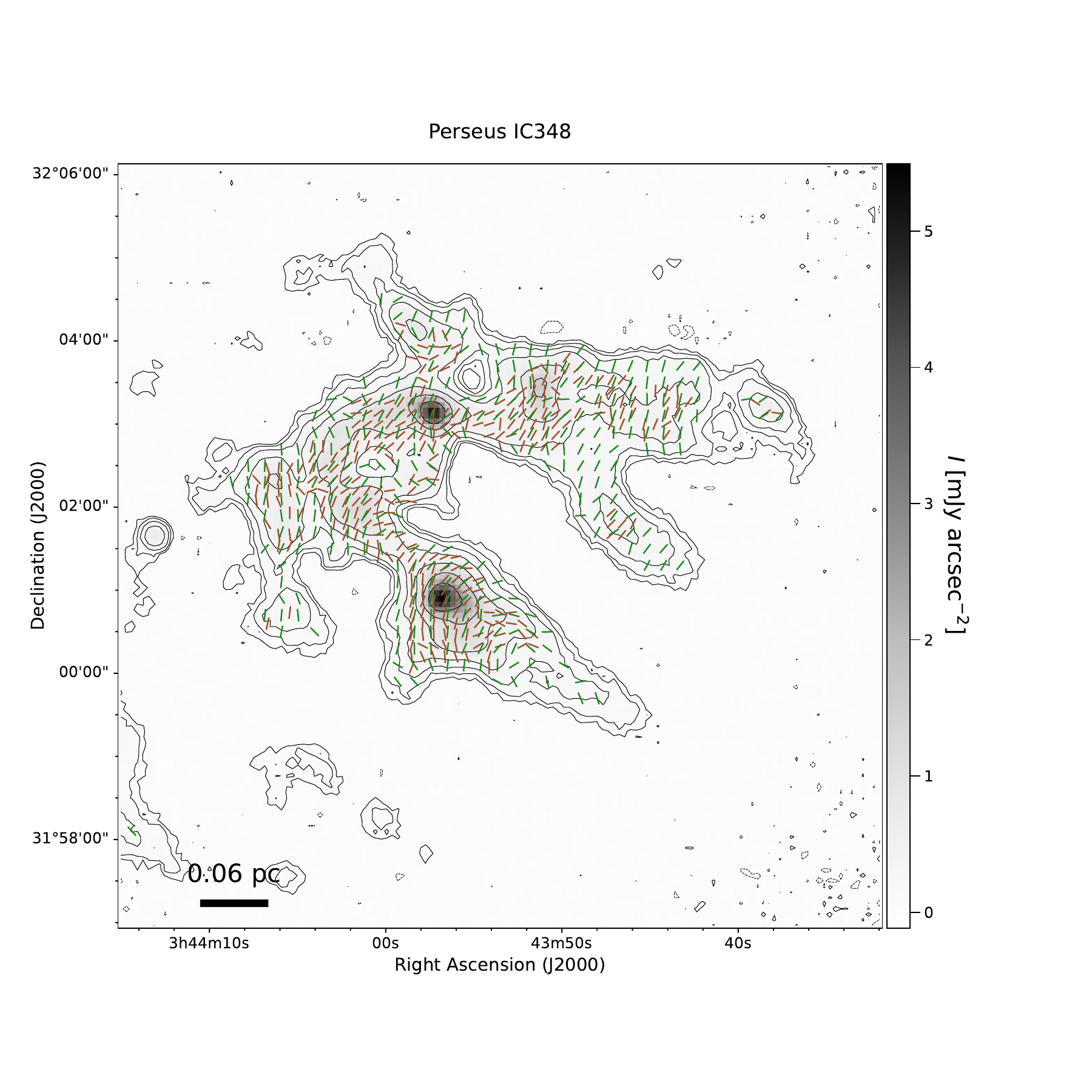}
\caption{Same as figure~\ref{afig:Bmap_L43L43}, but for Perseus IC348.
The noise levels are $\sigma_{850} = 0.0029$\,mJy\,arcsec$^{-2}$ and $\sigma_{450} = 0.043$\,mJy\,arcsec$^{-2}$.
\\ {Alt text: Map with short line segments over contours.}
}
\label{afig:Bmap_IC348H}
\end{figure}

\begin{figure}[H]
\centering
\includegraphics[width=0.95\textwidth]{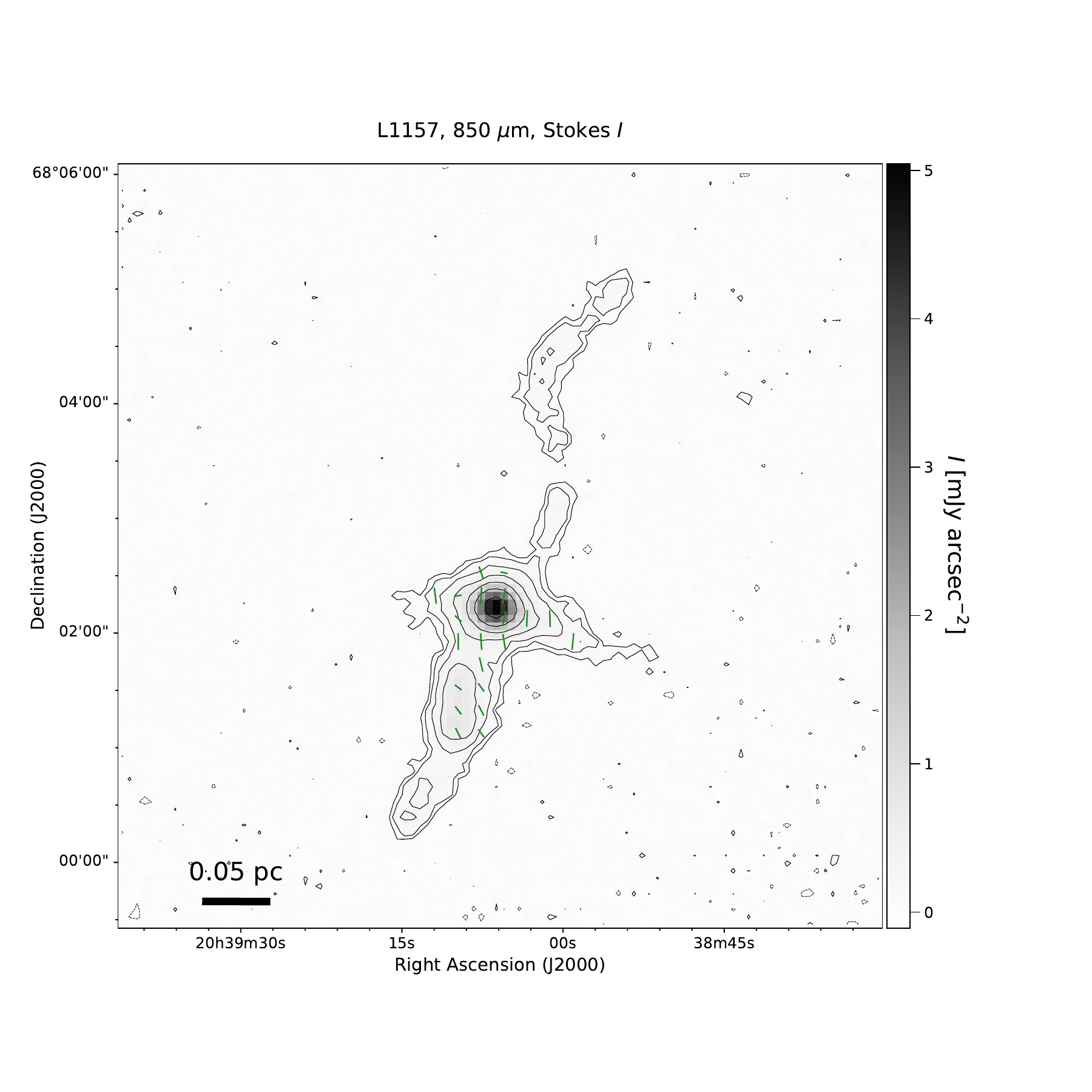}
\caption{Same as figure~\ref{afig:Bmap_L43L43}, but for L1157.
The noise levels are $\sigma_{850} = 0.0039$\,mJy\,arcsec$^{-2}$ and $\sigma_{450} = 0.19$\,mJy\,arcsec$^{-2}$.
\\ {Alt text: Map with short line segments over contours.}
}
\label{afig:Bmap_L1157L}
\end{figure}

\begin{figure}[H]
\centering
\includegraphics[width=0.95\textwidth]{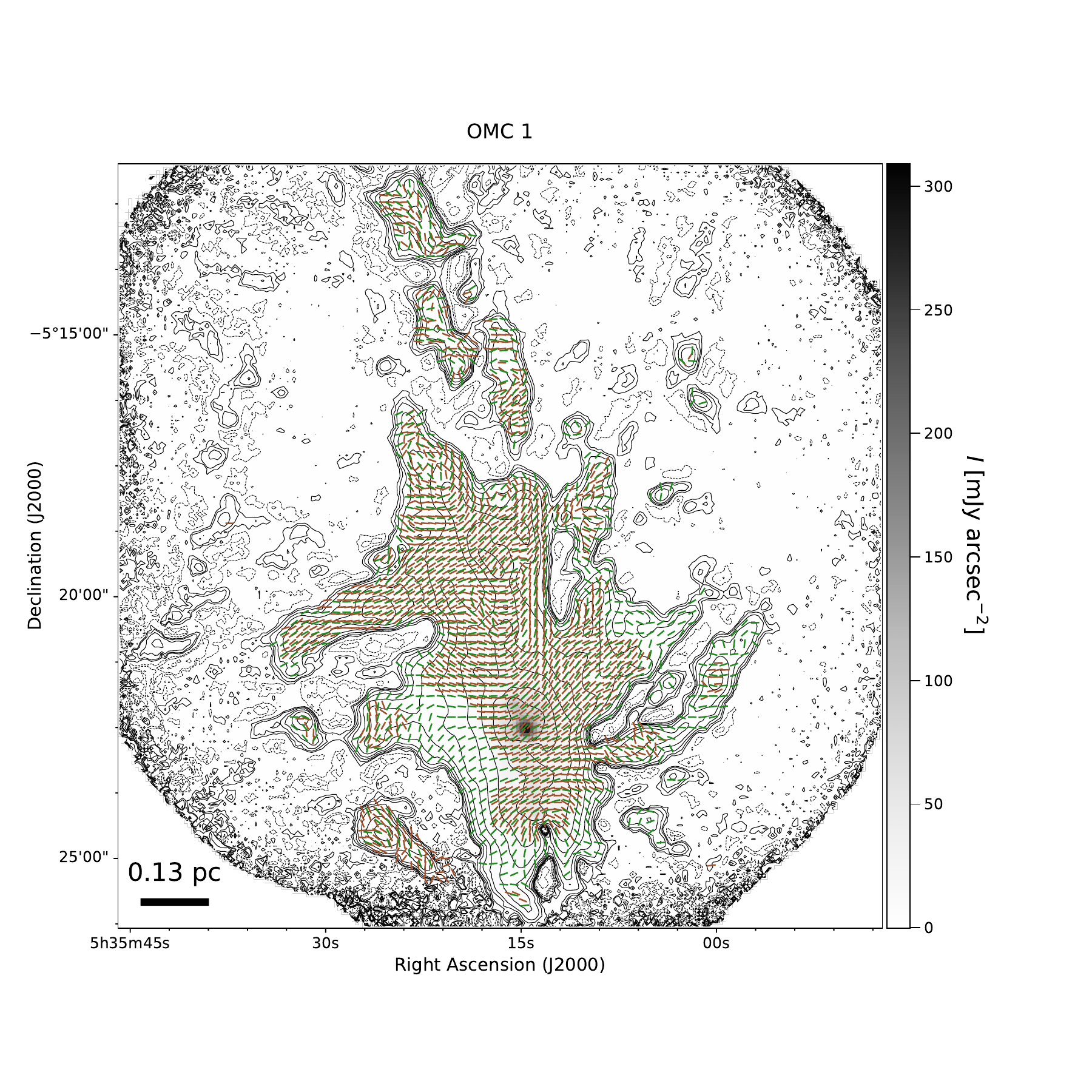}
\caption{Same as figure~\ref{afig:Bmap_L43L43}, but for OMC 1.
The noise levels are $\sigma_{850} = 0.0045$\,mJy\,arcsec$^{-2}$ and $\sigma_{450} = 0.13$\,mJy\,arcsec$^{-2}$.
\\ {Alt text: Map with short line segments over contours.}
}
\label{afig:Bmap_OriAS1}
\end{figure}

\begin{figure}[H]
\centering
\includegraphics[width=0.95\textwidth]{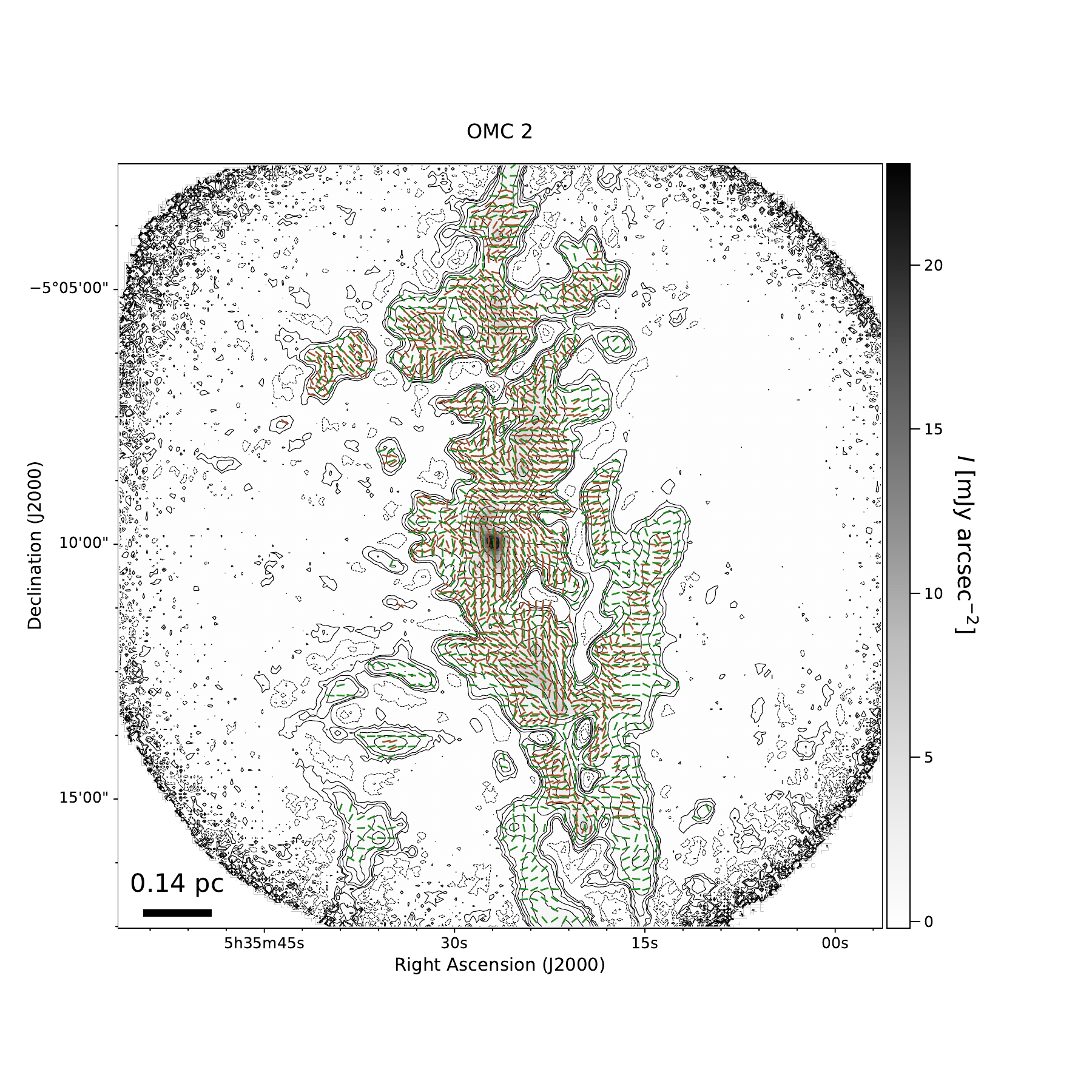}
\caption{Same as figure~\ref{afig:Bmap_L43L43}, but for OMC 2.
The noise levels are $\sigma_{850} = 0.0046$\,mJy\,arcsec$^{-2}$ and $\sigma_{450} = 0.18$\,mJy\,arcsec$^{-2}$.
\\ {Alt text: Map with short line segments over contours.}
}
\label{afig:Bmap_OriAS2}
\end{figure}

\begin{figure}[H]
\centering
\includegraphics[width=0.95\textwidth]{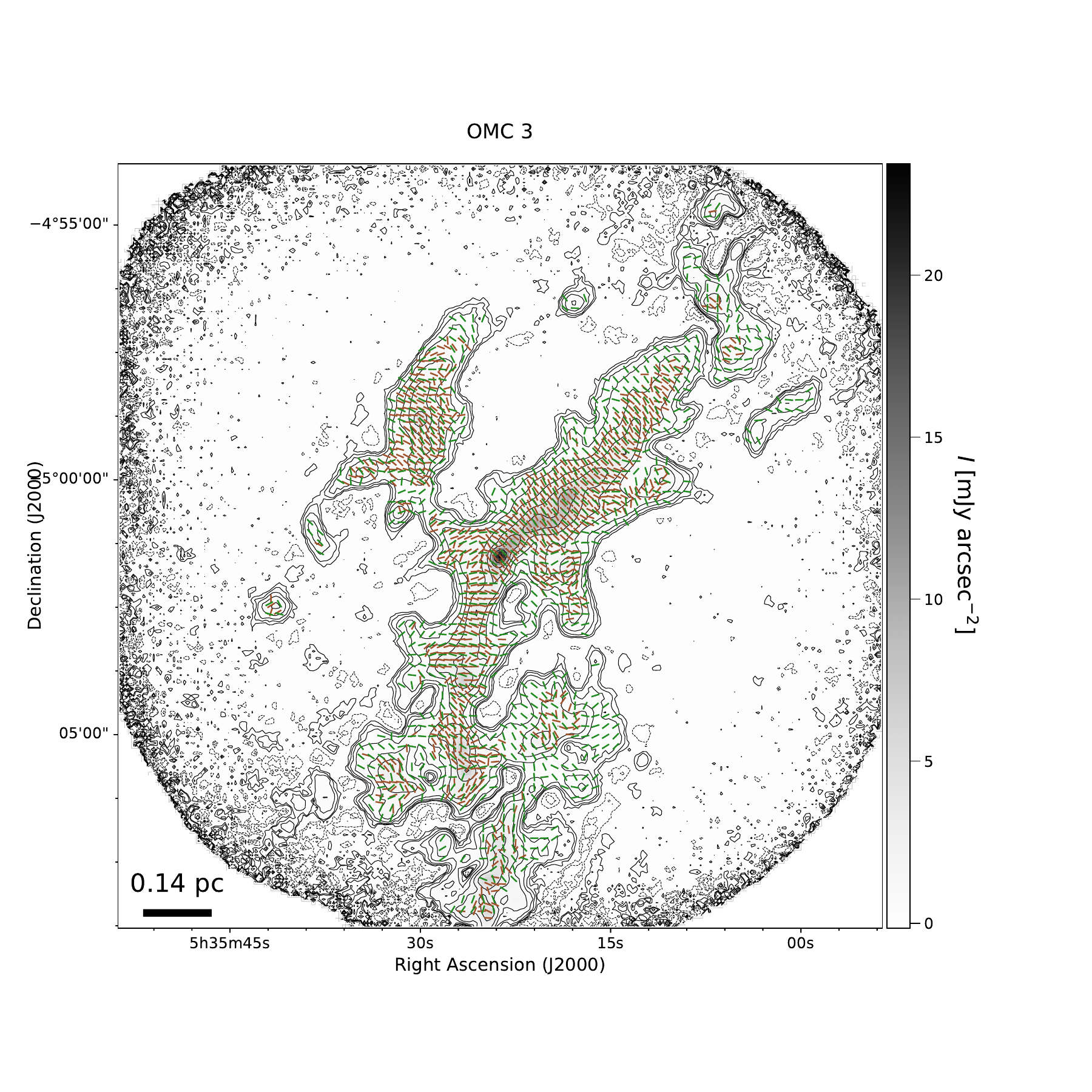}
\caption{Same as figure~\ref{afig:Bmap_L43L43}, but for OMC 3.
The noise levels are $\sigma_{850} = 0.0035$\,mJy\,arcsec$^{-2}$ and $\sigma_{450} = 0.12$\,mJy\,arcsec$^{-2}$.
\\ {Alt text: Map with short line segments over contours.}
}
\label{afig:Bmap_OriAS3}
\end{figure}

\begin{figure}[H]
\centering
\includegraphics[width=0.95\textwidth]{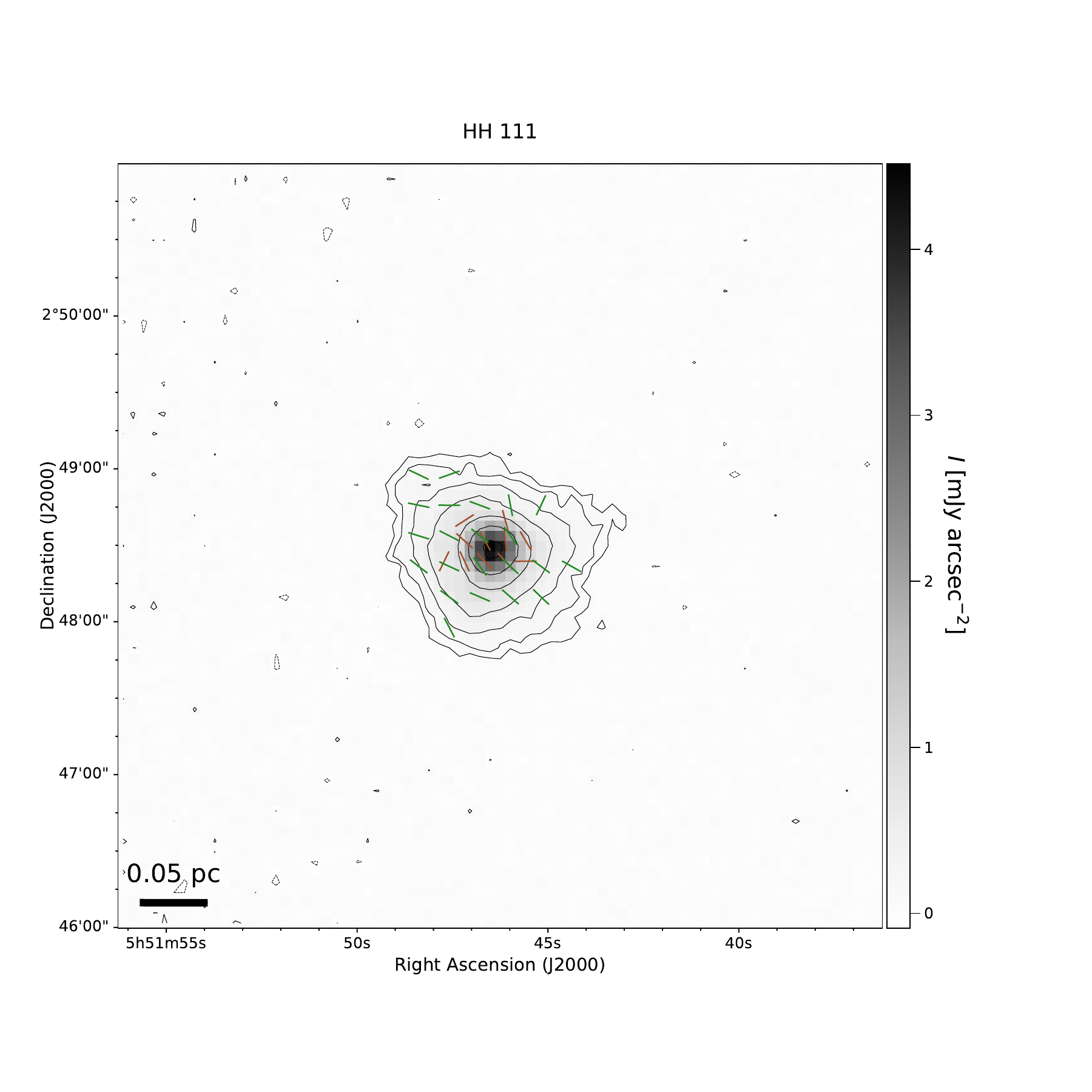}
\caption{Same as figure~\ref{afig:Bmap_L43L43}, but for HH 111.
The noise levels are $\sigma_{850} = 0.0040$\,mJy\,arcsec$^{-2}$ and $\sigma_{450} = 0.072$\,mJy\,arcsec$^{-2}$.
\\ {Alt text: Map with short line segments over contours.}
}
\label{afig:Bmap_HH111H}
\end{figure}

\begin{figure}[H]
\centering
\includegraphics[width=0.95\textwidth]{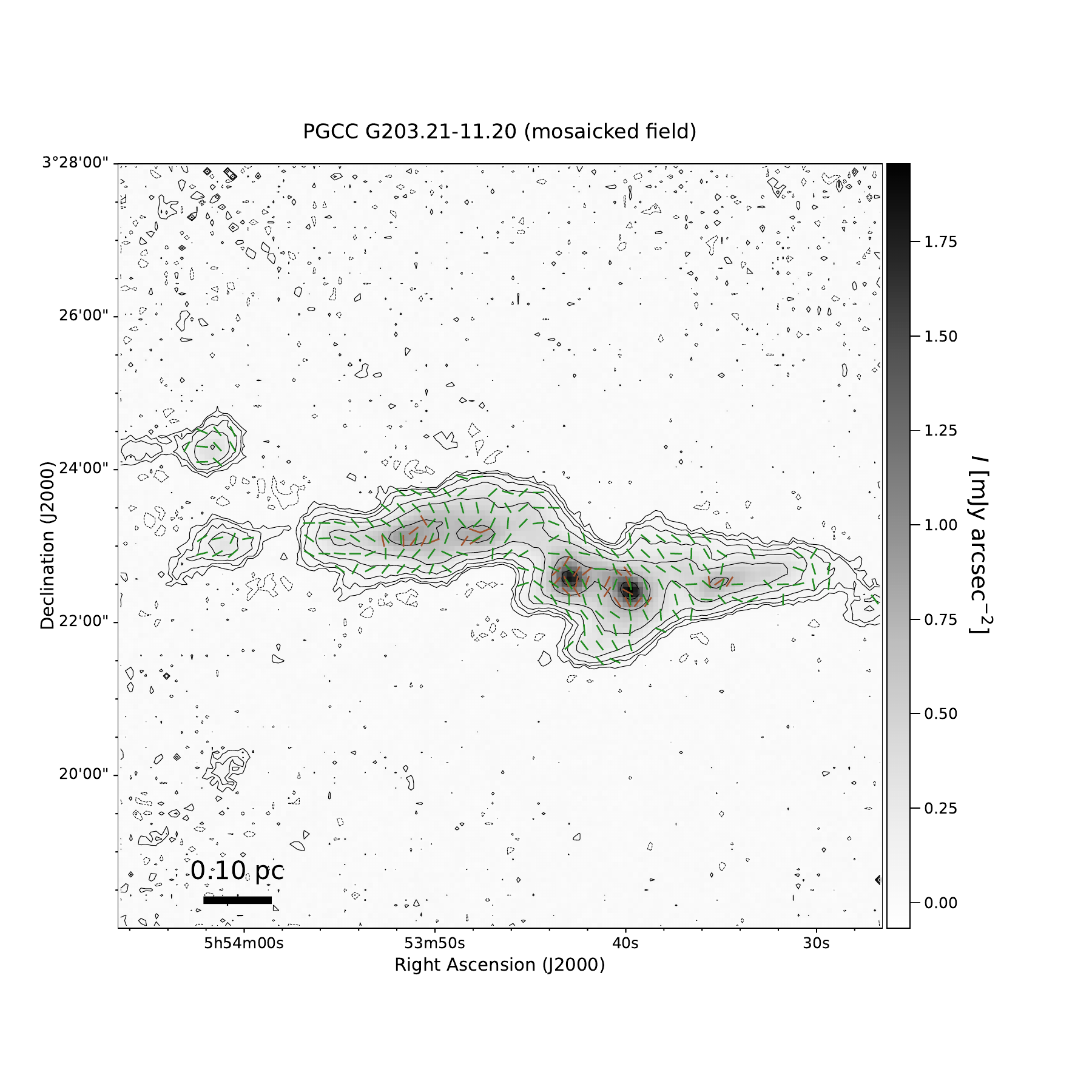}
\caption{Same as figure~\ref{afig:Bmap_L43L43}, but for PGCC G203.21-11.20 (mosaicked field).
The noise levels are $\sigma_{850} = 0.0015$\,mJy\,arcsec$^{-2}$ and $\sigma_{450} = 0.036$\,mJy\,arcsec$^{-2}$.
\\ {Alt text: Map with short line segments over contours.}
}
\label{afig:Bmap_OrBPMO}
\end{figure}

\begin{figure}[H]
\centering
\includegraphics[width=0.95\textwidth]{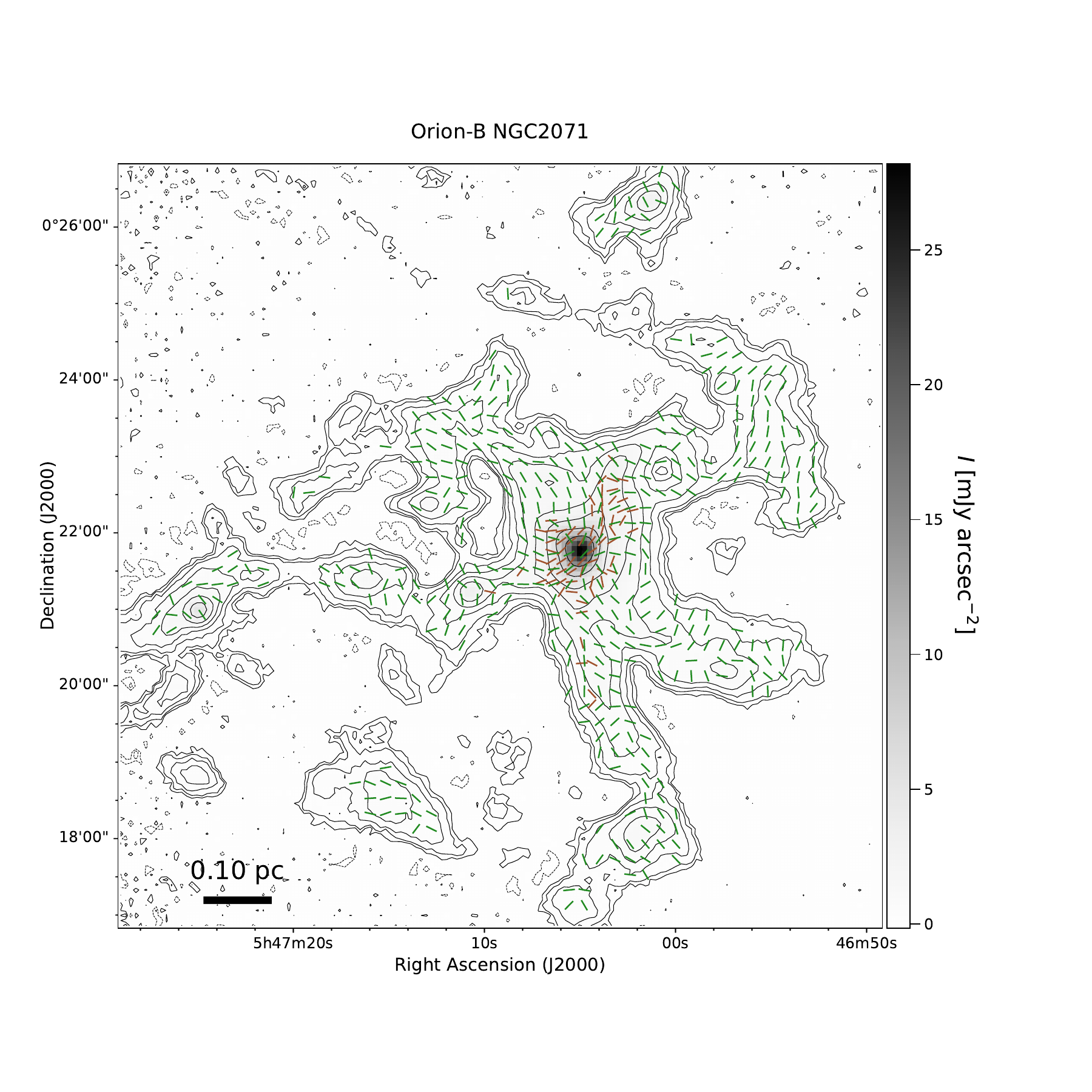}
\caption{Same as figure~\ref{afig:Bmap_L43L43}, but for Orion-B NGC2071.
The noise levels are $\sigma_{850} = 0.0036$\,mJy\,arcsec$^{-2}$ and $\sigma_{450} = 0.10$\,mJy\,arcsec$^{-2}$.
\\ {Alt text: Map with short line segments over contours.}
}
\label{afig:Bmap_NG2068}
\end{figure}

\begin{figure}[H]
\centering
\includegraphics[width=0.95\textwidth]{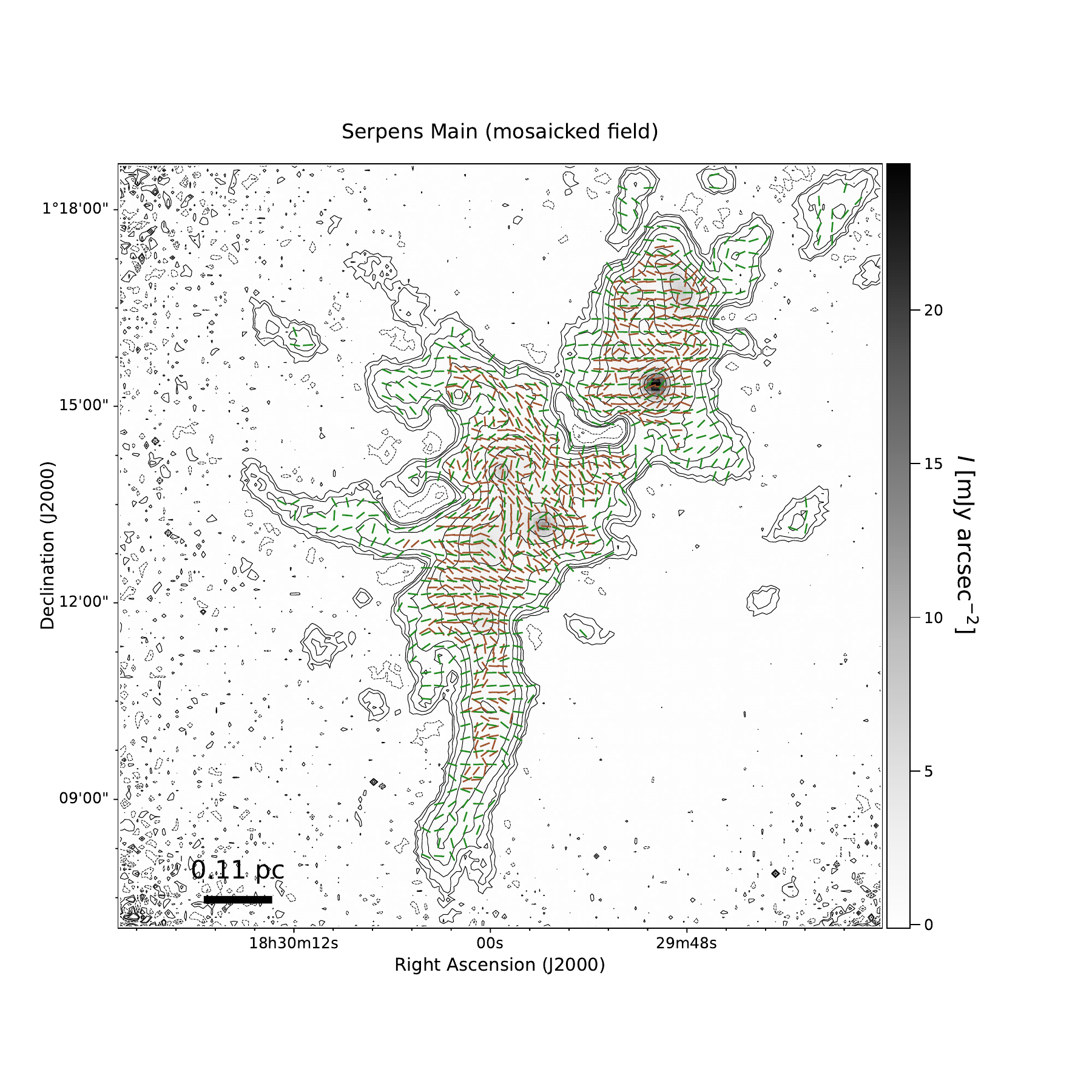}
\caption{Same as figure~\ref{afig:Bmap_L43L43}, but for Serpens Main (mosaicked field).
The noise levels are $\sigma_{850} = 0.0025$\,mJy\,arcsec$^{-2}$ and $\sigma_{450} = 0.022$\,mJy\,arcsec$^{-2}$.
\\ {Alt text: Map with short line segments over contours.}
}
\label{afig:Bmap_SerpMO}
\end{figure}

\begin{figure}[H]
\centering
\includegraphics[width=0.95\textwidth]{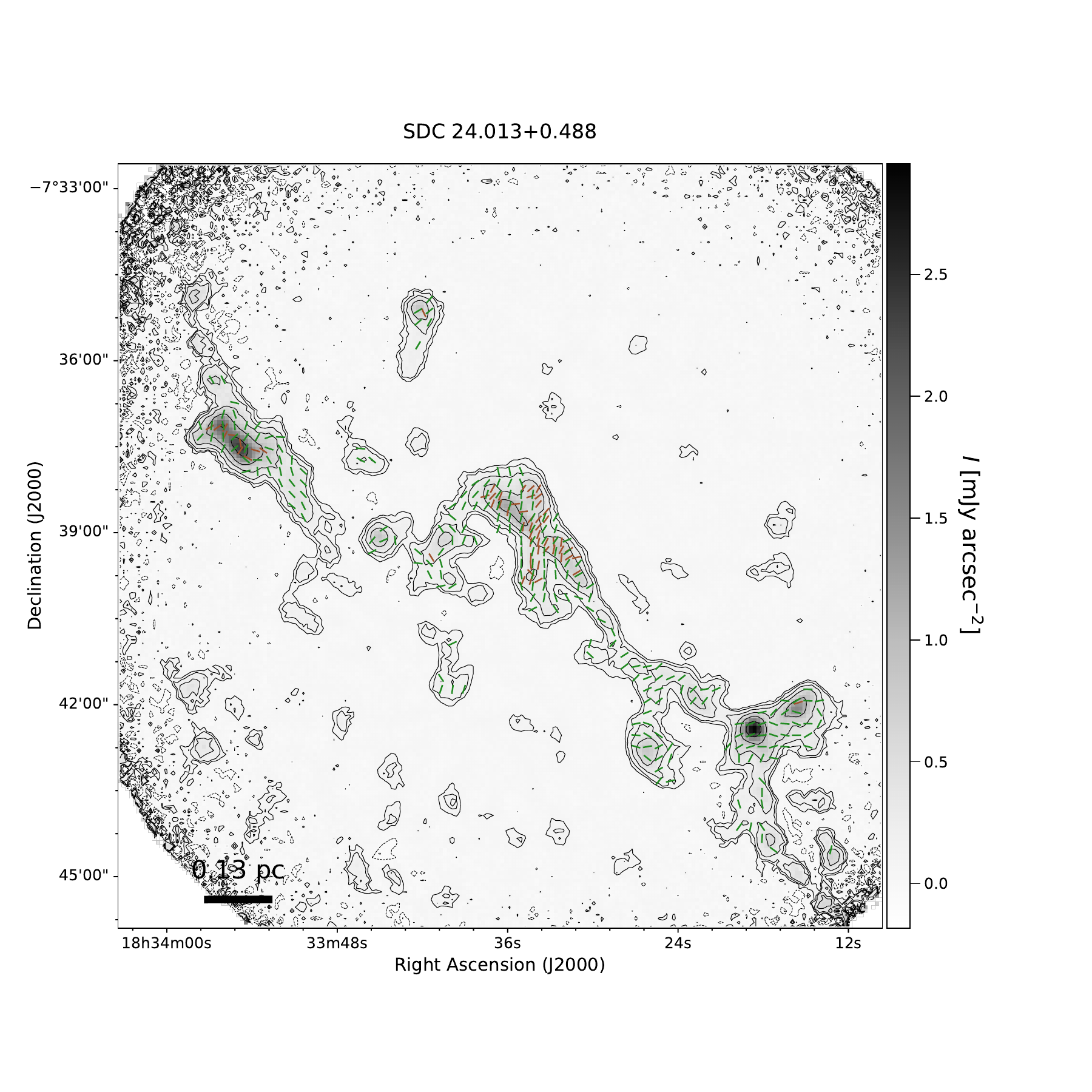}
\caption{Same as figure~\ref{afig:Bmap_L43L43}, but for SDC 24.013+0.488.
The noise levels are $\sigma_{850} = 0.0041$\,mJy\,arcsec$^{-2}$ and $\sigma_{450} = 0.12$\,mJy\,arcsec$^{-2}$.
\\ {Alt text: Map with short line segments over contours.}
}
\label{afig:Bmap_SDC240}
\end{figure}

\begin{figure}[H]
\centering
\includegraphics[width=0.95\textwidth]{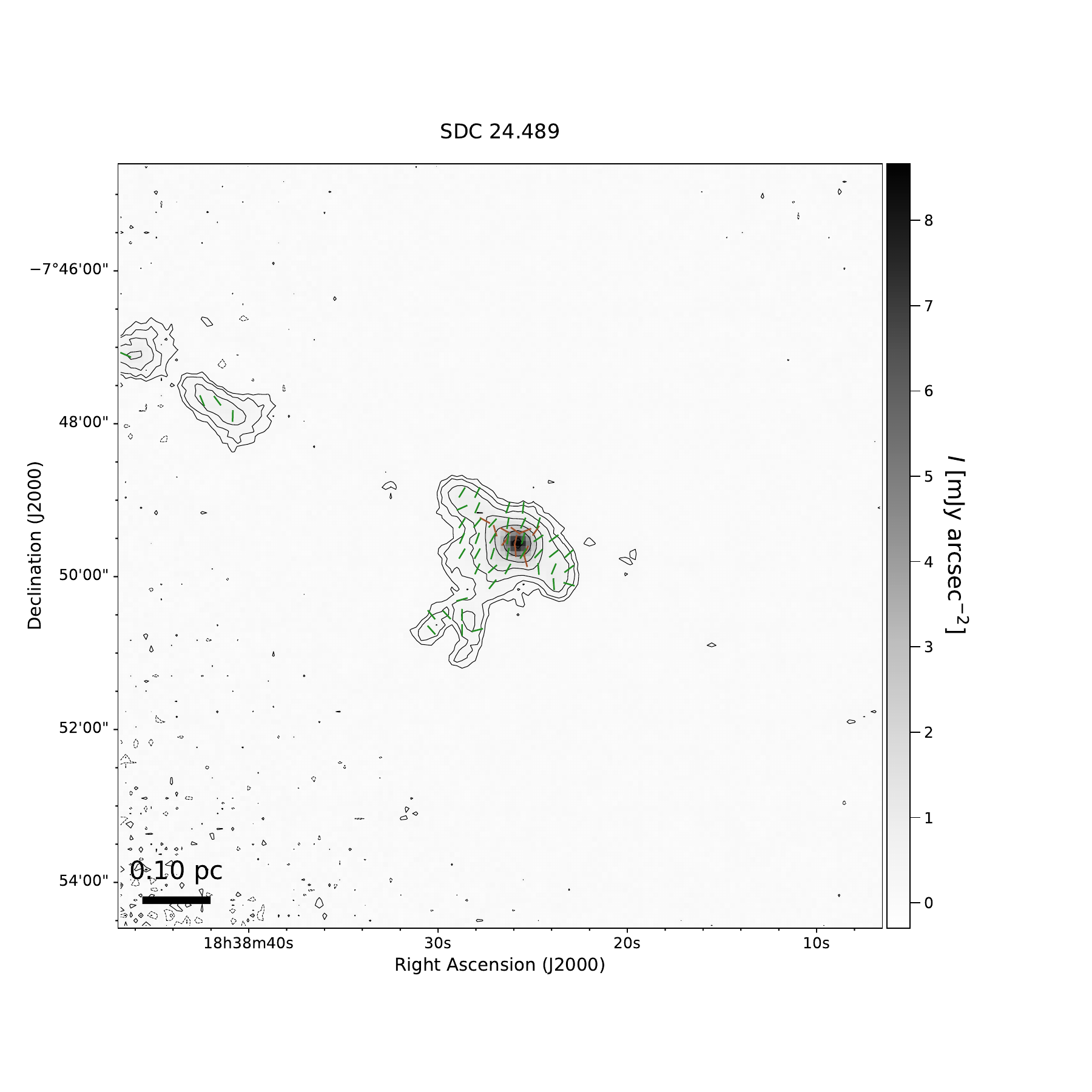}
\caption{Same as figure~\ref{afig:Bmap_L43L43}, but for SDC 24.489.
The noise levels are $\sigma_{850} = 0.0077$\,mJy\,arcsec$^{-2}$ and $\sigma_{450} = 0.21$\,mJy\,arcsec$^{-2}$.
\\ {Alt text: Map with short line segments over contours.}
}
\label{afig:Bmap_SD24P4}
\end{figure}

\begin{figure}[H]
\centering
\includegraphics[width=0.95\textwidth]{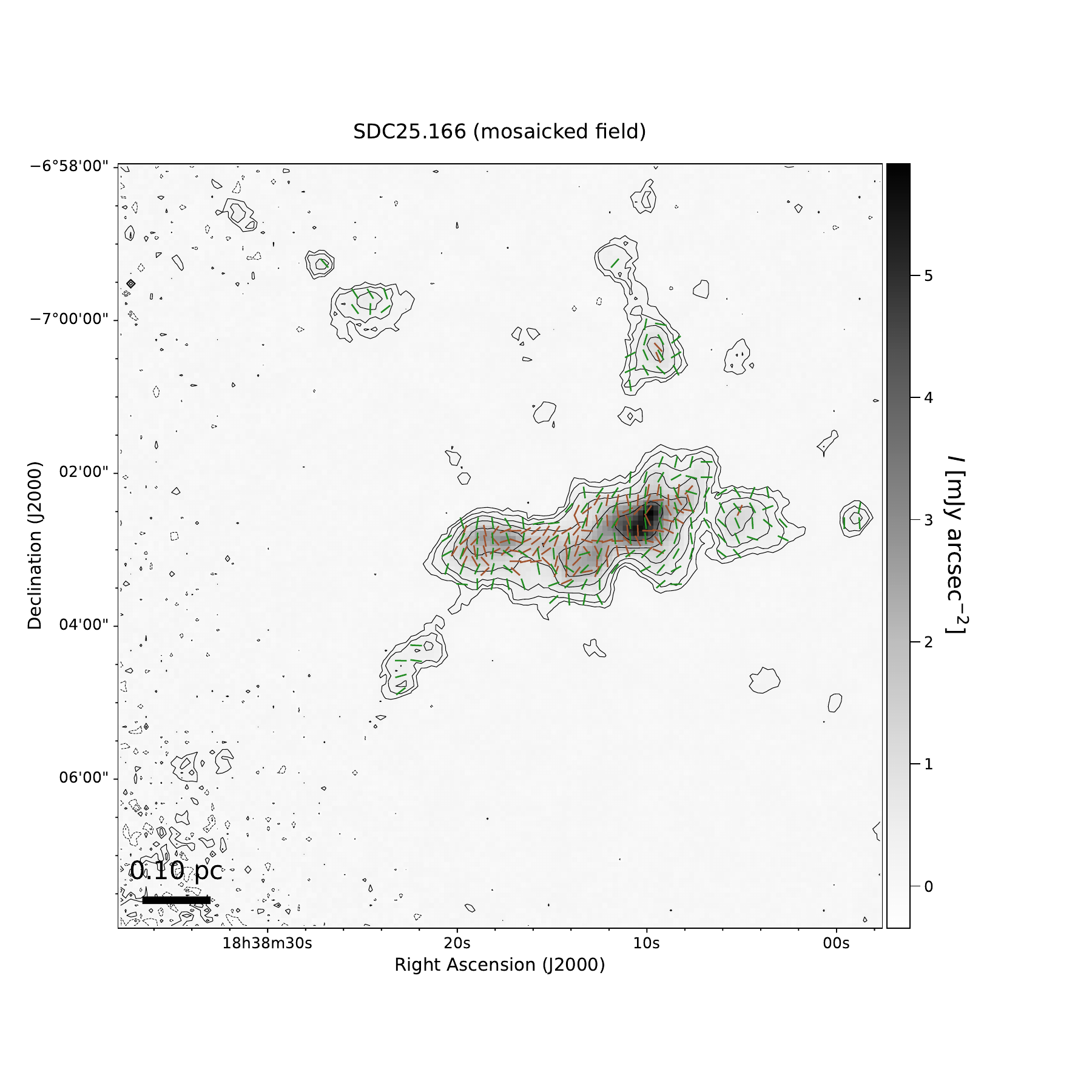}
\caption{Same as figure~\ref{afig:Bmap_L43L43}, but for SDC 25.166 (mosaicked field).
The noise levels are $\sigma_{850} = 0.0077$\,mJy\,arcsec$^{-2}$ and $\sigma_{450} = 0.088$\,mJy\,arcsec$^{-2}$.
\\ {Alt text: Map with short line segments over contours.}
}
\label{afig:Bmap_SD25MO}
\end{figure}

\begin{figure}[H]
\centering
\includegraphics[width=0.95\textwidth]{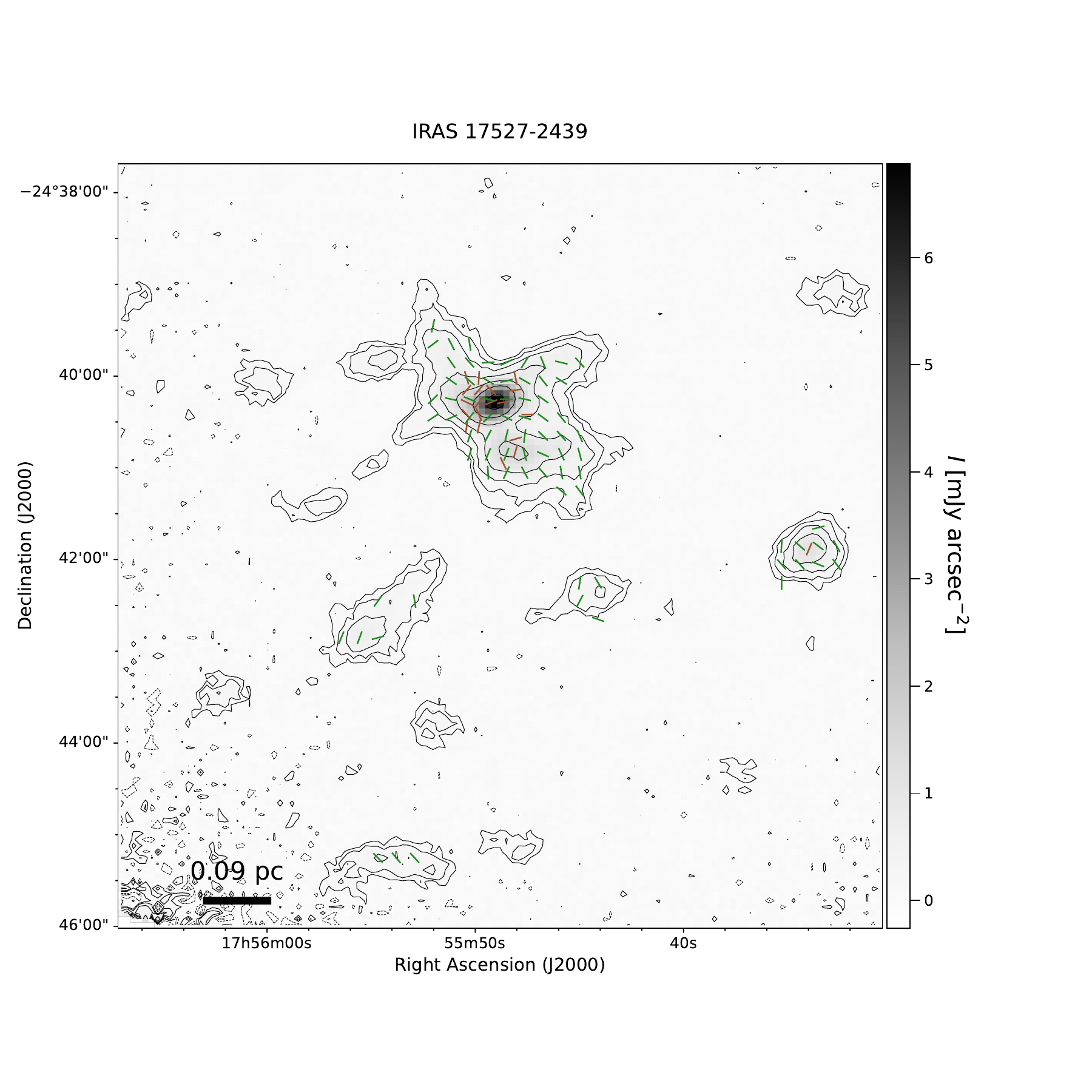}
\caption{Same as figure~\ref{afig:Bmap_L43L43}, but for IRAS 17527-2439.
The noise levels are $\sigma_{850} = 0.0058$\,mJy\,arcsec$^{-2}$ and $\sigma_{450} = 0.20$\,mJy\,arcsec$^{-2}$.
\\ {Alt text: Map with short line segments over contours.}
}
\label{afig:Bmap_I17527}
\end{figure}

\begin{figure}[H]
\centering
\includegraphics[width=0.95\textwidth]{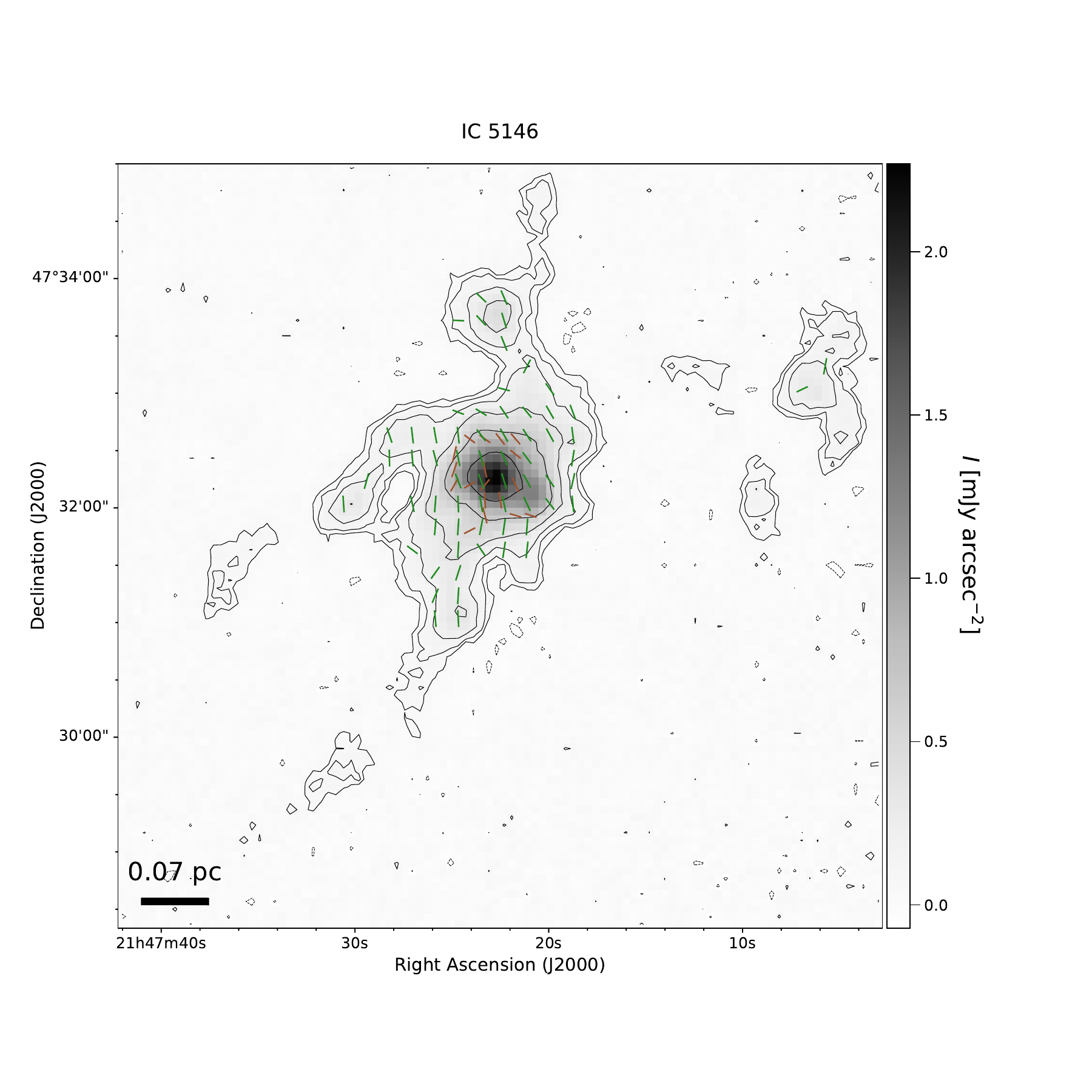}
\caption{Same as figure~\ref{afig:Bmap_L43L43}, but for IC 5146.
The noise levels are $\sigma_{850} = 0.0027$\,mJy\,arcsec$^{-2}$ and $\sigma_{450} = 0.056$\,mJy\,arcsec$^{-2}$.
\\ {Alt text: Map with short line segments over contours.}
}
\label{afig:Bmap_IC5146}
\end{figure}

\begin{figure}[H]
\centering
\includegraphics[width=0.95\textwidth]{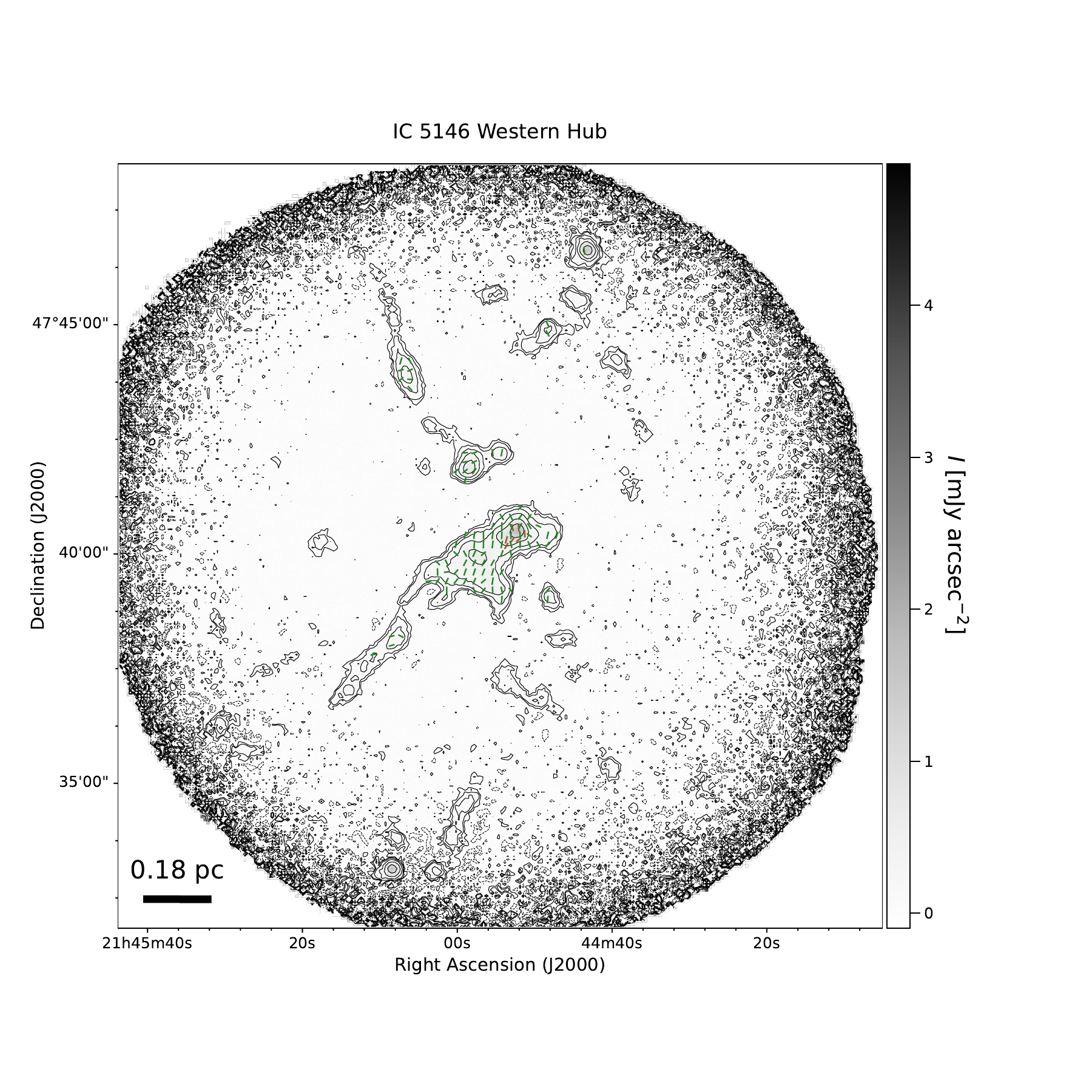}
\caption{Same as figure~\ref{afig:Bmap_L43L43}, but for IC 5146 Western Hub.
The noise levels are $\sigma_{850} = 0.0022$\,mJy\,arcsec$^{-2}$ and $\sigma_{450} = 0.053$\,mJy\,arcsec$^{-2}$.
\\ {Alt text: Map with short line segments over contours.}
}
\label{afig:Bmap_IC51WH}
\end{figure}

\begin{figure}[H]
\centering
\includegraphics[width=0.95\textwidth]{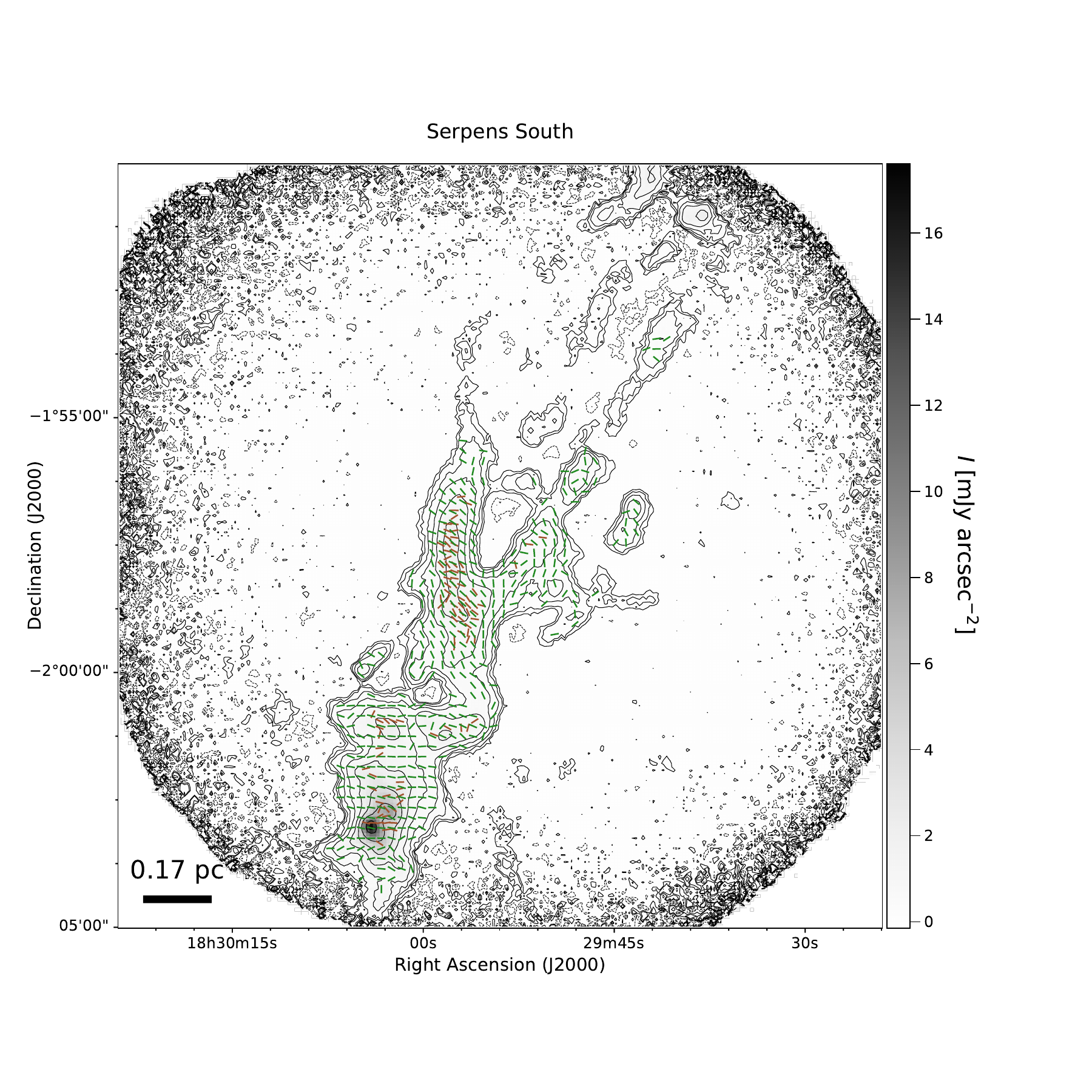}
\caption{
Same as figure~\ref{afig:Bmap_L43L43}, but for Serpens South, with $\sigma_{850}=0.0033$ and $\sigma_{450}=0.051$\,mJy\,arcsec$^{-2}$.
We exclude the upper-right rectangle after visual inspection because large Stokes-$I$ uncertainties near the edge of the bolometer array produce spurious positive emission.
We cannot reliably remove those artifacts without also discarding real emission near the opposite edge.
Paper~T appendix~\ref{T-app:QU_diagnostics} describes how anomalous Stokes-$Q,U$ behavior provides a complementary check on the reliability of the total intensity.
\\ {Alt text: Map with short line segments over contours.}
}
\label{afig:Bmap_SerpS3}
\end{figure}

\begin{figure}[H]
\centering
\includegraphics[width=0.95\textwidth]{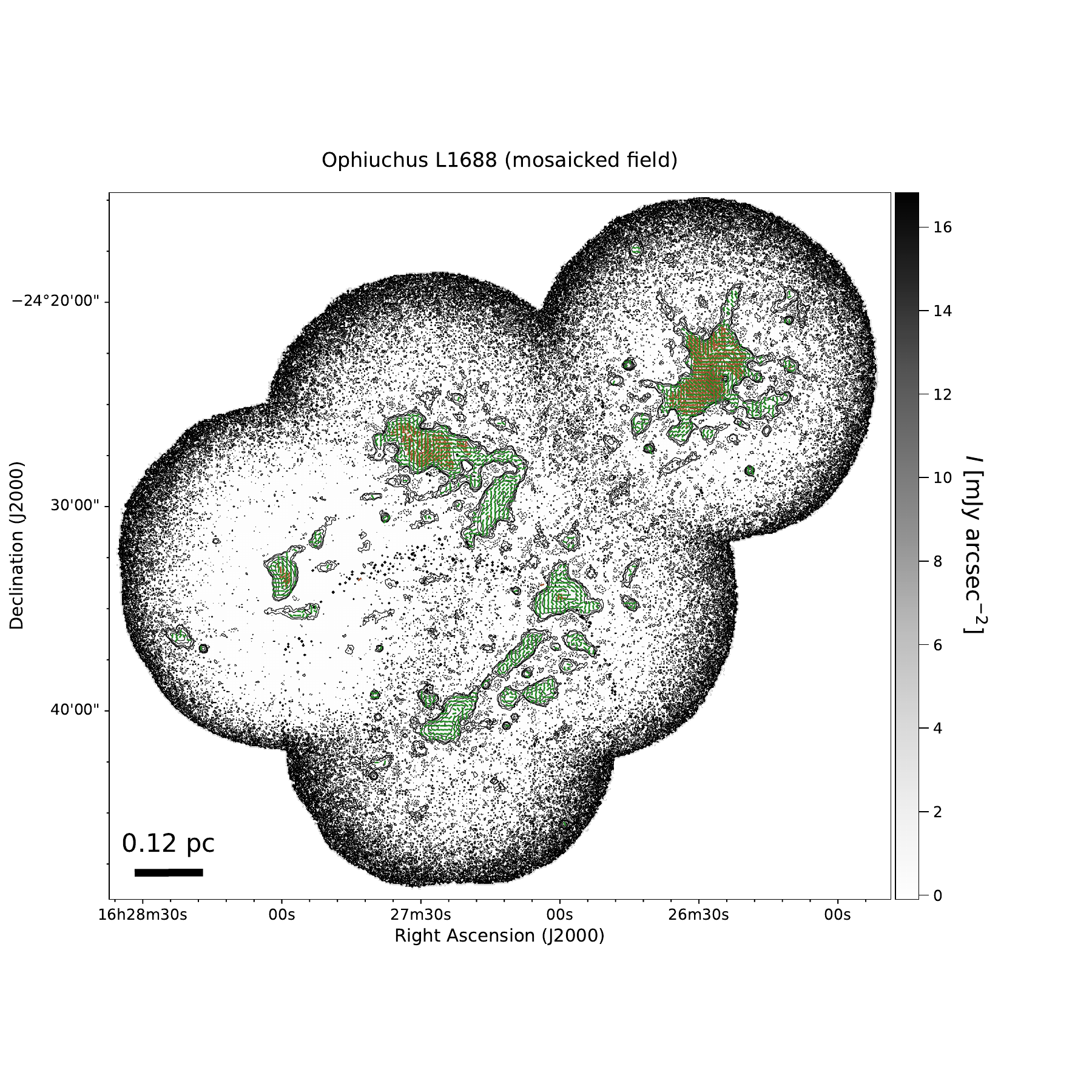}
\caption{Same as figure~\ref{afig:Bmap_L43L43}, but for L1688 Ophiuchus mosaicked field.
The noise levels are $\sigma_{850} = 0.0022$\,mJy\,arcsec$^{-2}$ and $\sigma_{450} = 0.034$\,mJy\,arcsec$^{-2}$.
\\ {Alt text: Map with short line segments over contours.}
}
\label{afig:Bmap_OPHIMO}
\end{figure}

\begin{figure}[H]
\centering
\includegraphics[width=0.95\textwidth]{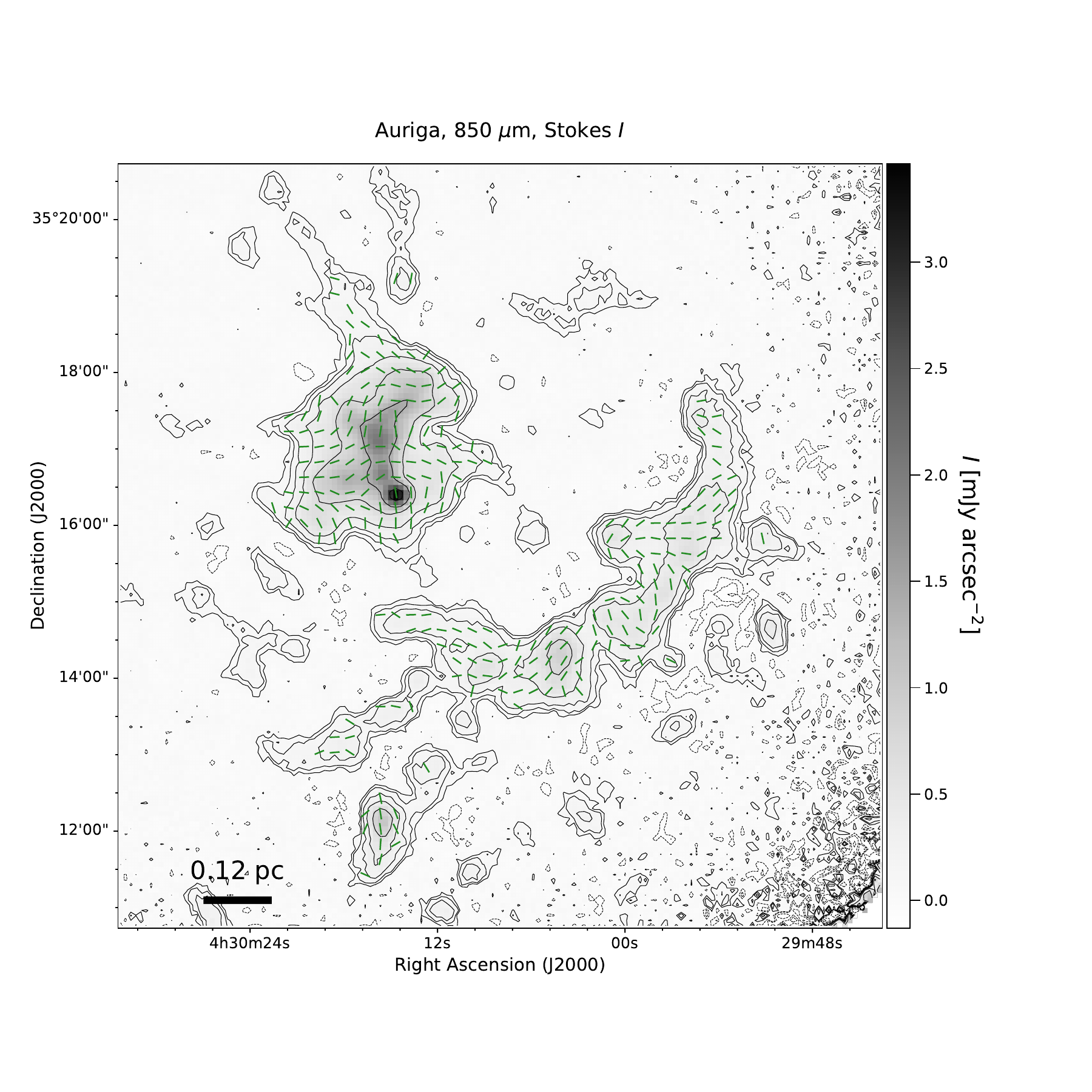}
\caption{Same as figure~\ref{afig:Bmap_L43L43}, but for Auriga.
The noise levels are $\sigma_{850} = 0.0029$\,mJy\,arcsec$^{-2}$ and $\sigma_{450} = 0.070$\,mJy\,arcsec$^{-2}$.
We adopt $d = 532 \pm 28$\,pc from the H$_2$O maser parallax of LkH$\alpha$\,101 measured with VLBI Exploration of Radio Astrometry (VERA) \citep[$1.879 \pm 0.096$\,mas;][]{Omodaka2020}.
This value replaces the $466$\,pc used by \citet{Ngoc2021} in the BISTRO study of Auriga and agrees with the independent value of $535 \pm 29$\,pc from the Very Long Baseline Array (VLBA) \citep{Dzib2018_LkHa101}.
We use the adopted distance for the scale bar.
\\ {Alt text: Map with short line segments over contours.}
}
\label{afig:Bmap_Auriga}
\end{figure}

\begin{figure}[H]
\centering
\includegraphics[width=0.95\textwidth]{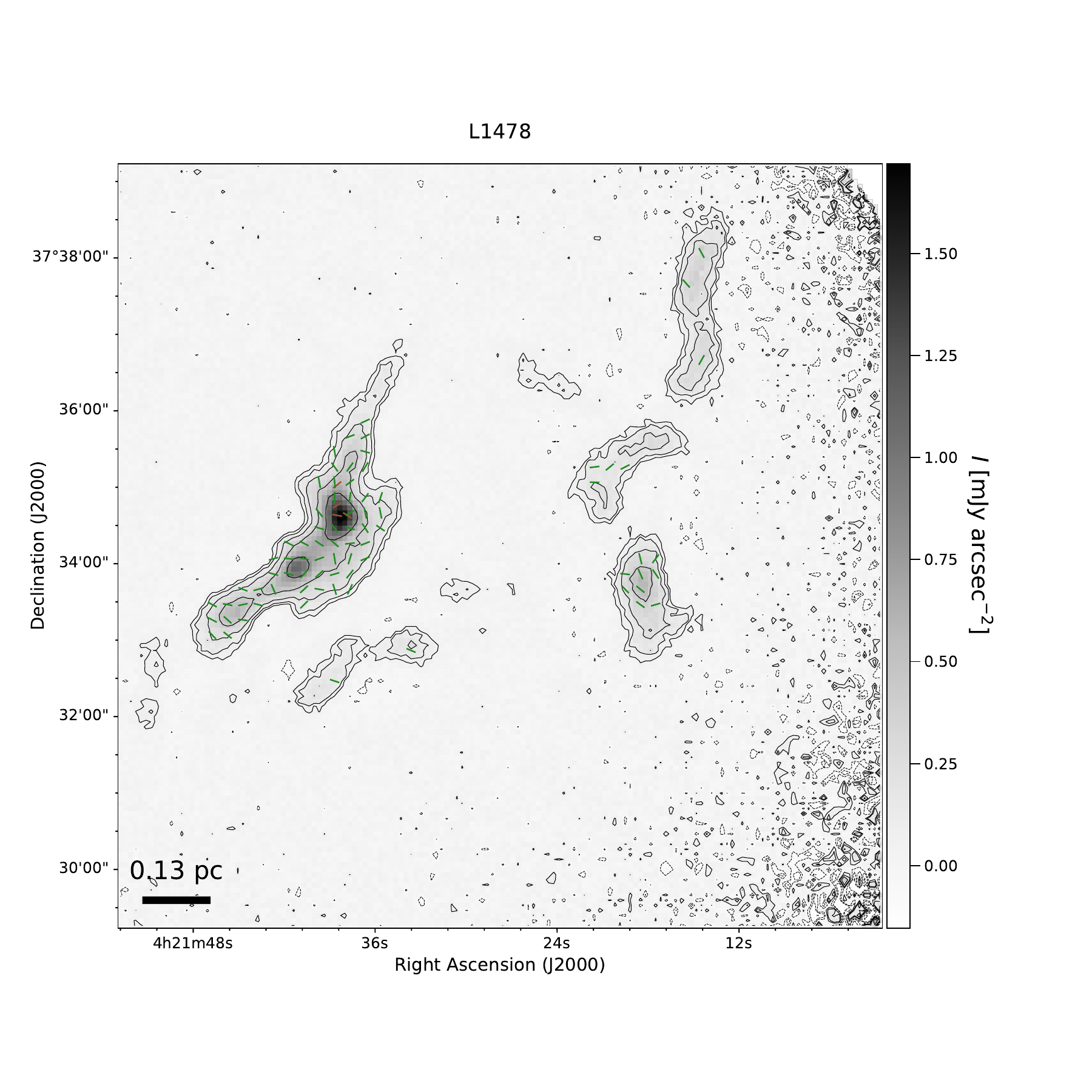}
\caption{Same as figure~\ref{afig:Bmap_L43L43}, but for L1478.
The noise levels are $\sigma_{850} = 0.0034$\,mJy\,arcsec$^{-2}$ and $\sigma_{450} = 0.14$\,mJy\,arcsec$^{-2}$.
\\ {Alt text: Map with short line segments over contours.}
}
\label{afig:Bmap_L1478L}
\end{figure}

\begin{figure}[H]
\centering
\includegraphics[width=0.95\textwidth]{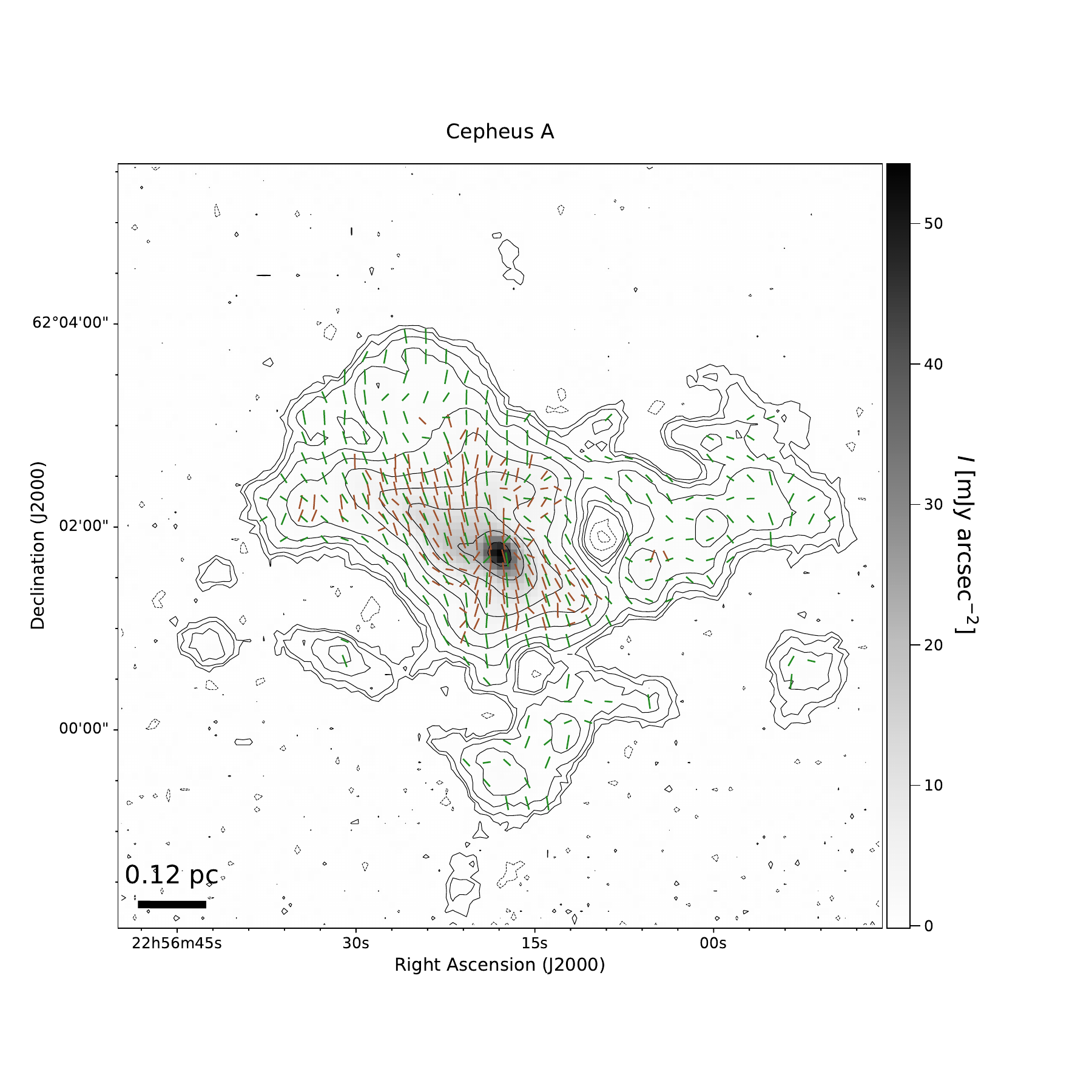}
\caption{Same as figure~\ref{afig:Bmap_L43L43}, but for Cepheus A.
The noise levels are $\sigma_{850} = 0.0051$\,mJy\,arcsec$^{-2}$ and $\sigma_{450} = 0.30$\,mJy\,arcsec$^{-2}$.
\\ {Alt text: Map with short line segments over contours.}
}
\label{afig:Bmap_CepACe}
\end{figure}

\begin{figure}[H]
\centering
\includegraphics[width=0.95\textwidth]{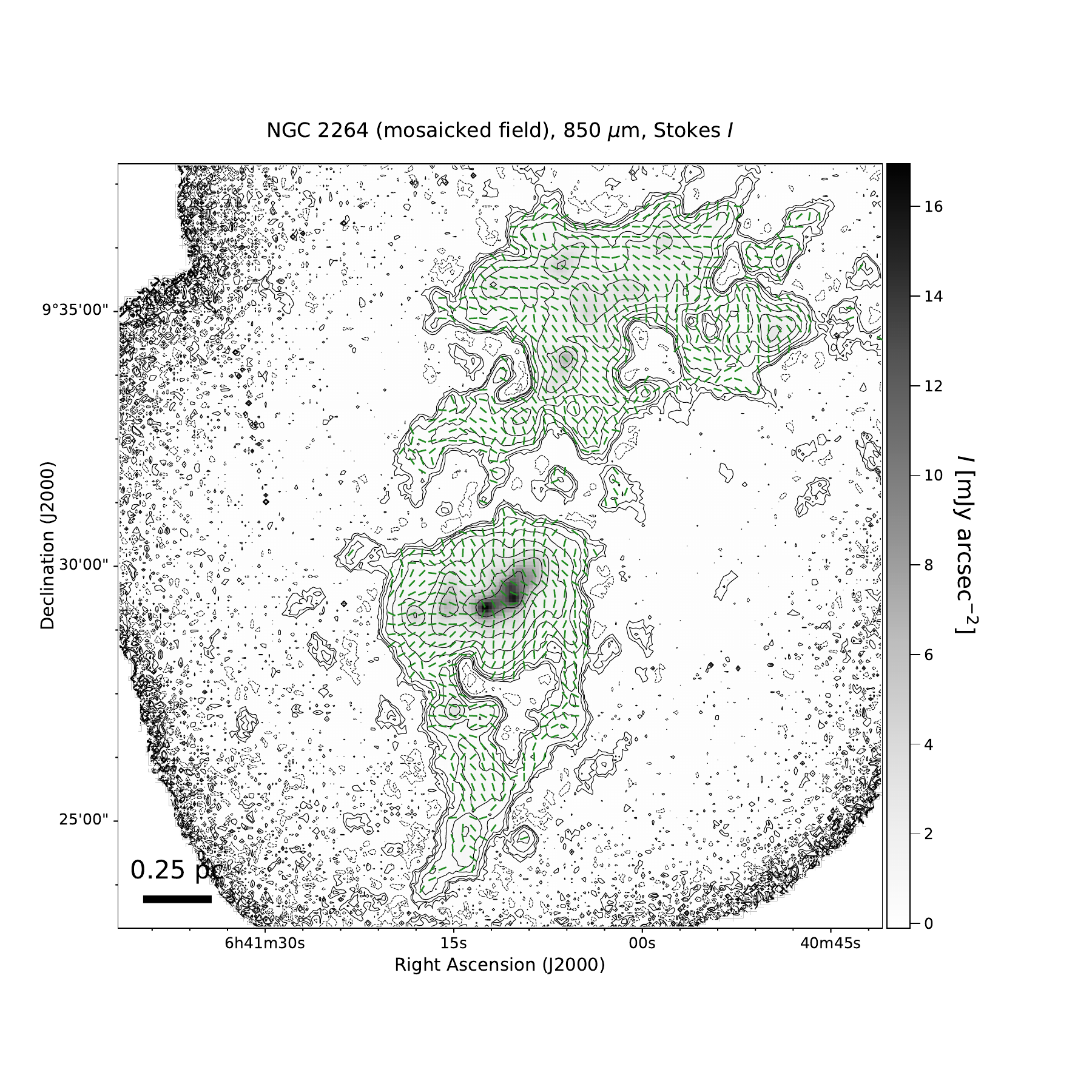}
\caption{Same as figure~\ref{afig:Bmap_L43L43}, but for NGC 2264 (mosaicked field).
The noise levels are $\sigma_{850} = 0.0024$\,mJy\,arcsec$^{-2}$ and $\sigma_{450} = 0.035$\,mJy\,arcsec$^{-2}$.
\\ {Alt text: Map with short line segments over contours.}
}
\label{afig:Bmap_N226MO}
\end{figure}

\begin{figure}[H]
\centering
\includegraphics[width=0.95\textwidth]{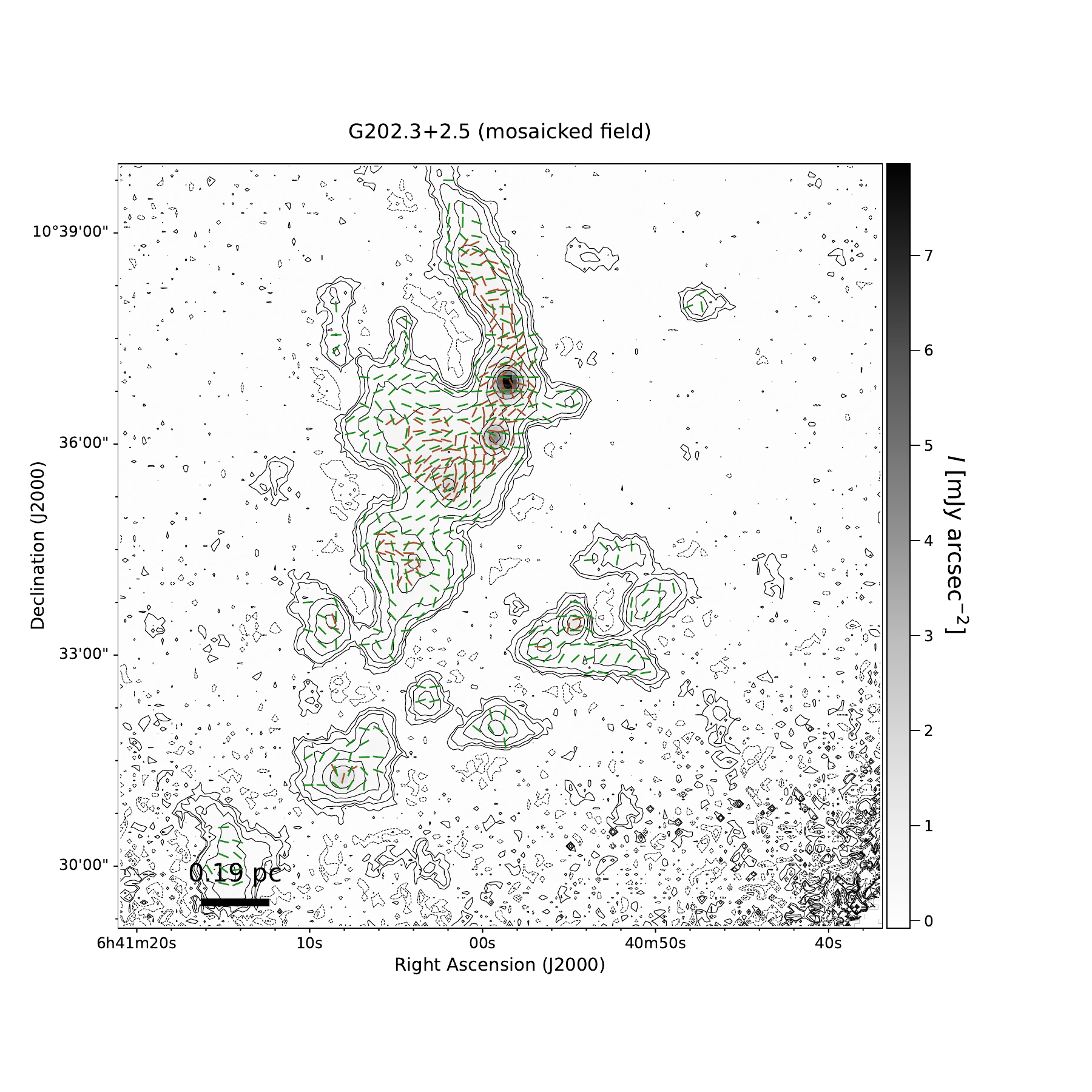}
\caption{Same as figure~\ref{afig:Bmap_L43L43}, but for G202.3+2.5 (mosaicked field).
The noise levels are $\sigma_{850} = 0.0018$\,mJy\,arcsec$^{-2}$ and $\sigma_{450} = 0.020$\,mJy\,arcsec$^{-2}$.
\\ {Alt text: Map with short line segments over contours.}
}
\label{afig:Bmap_G202MO}
\end{figure}

\begin{figure}[H]
\centering
\includegraphics[width=0.95\textwidth]{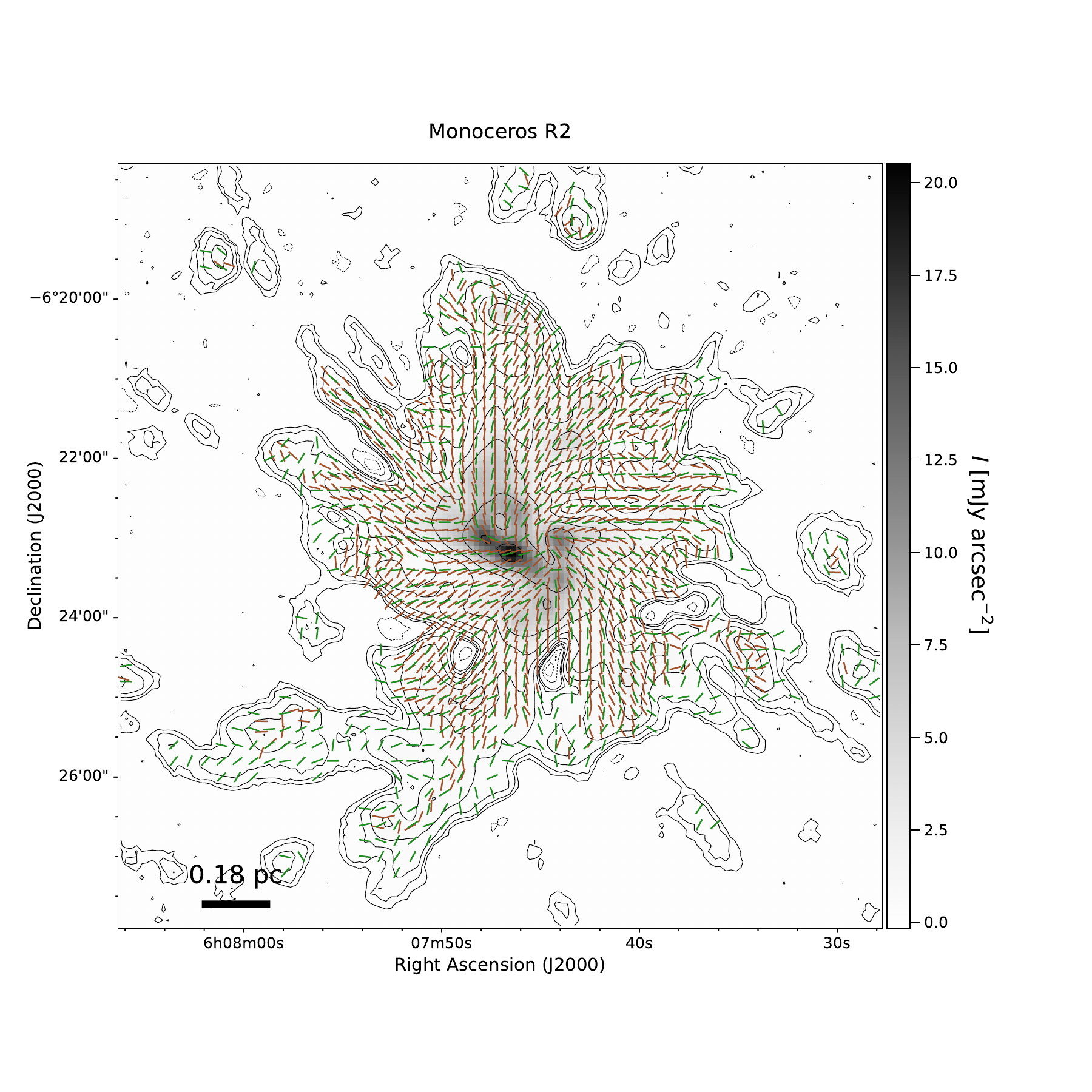}
\caption{Same as figure~\ref{afig:Bmap_L43L43}, but for Monoceros R2.
The noise levels are $\sigma_{850} = 0.0040$\,mJy\,arcsec$^{-2}$ and $\sigma_{450} = 0.078$\,mJy\,arcsec$^{-2}$.
\\ {Alt text: Map with short line segments over contours.}
}
\label{afig:Bmap_MonoR2}
\end{figure}

\begin{figure}[H]
\centering
\includegraphics[width=0.95\textwidth]{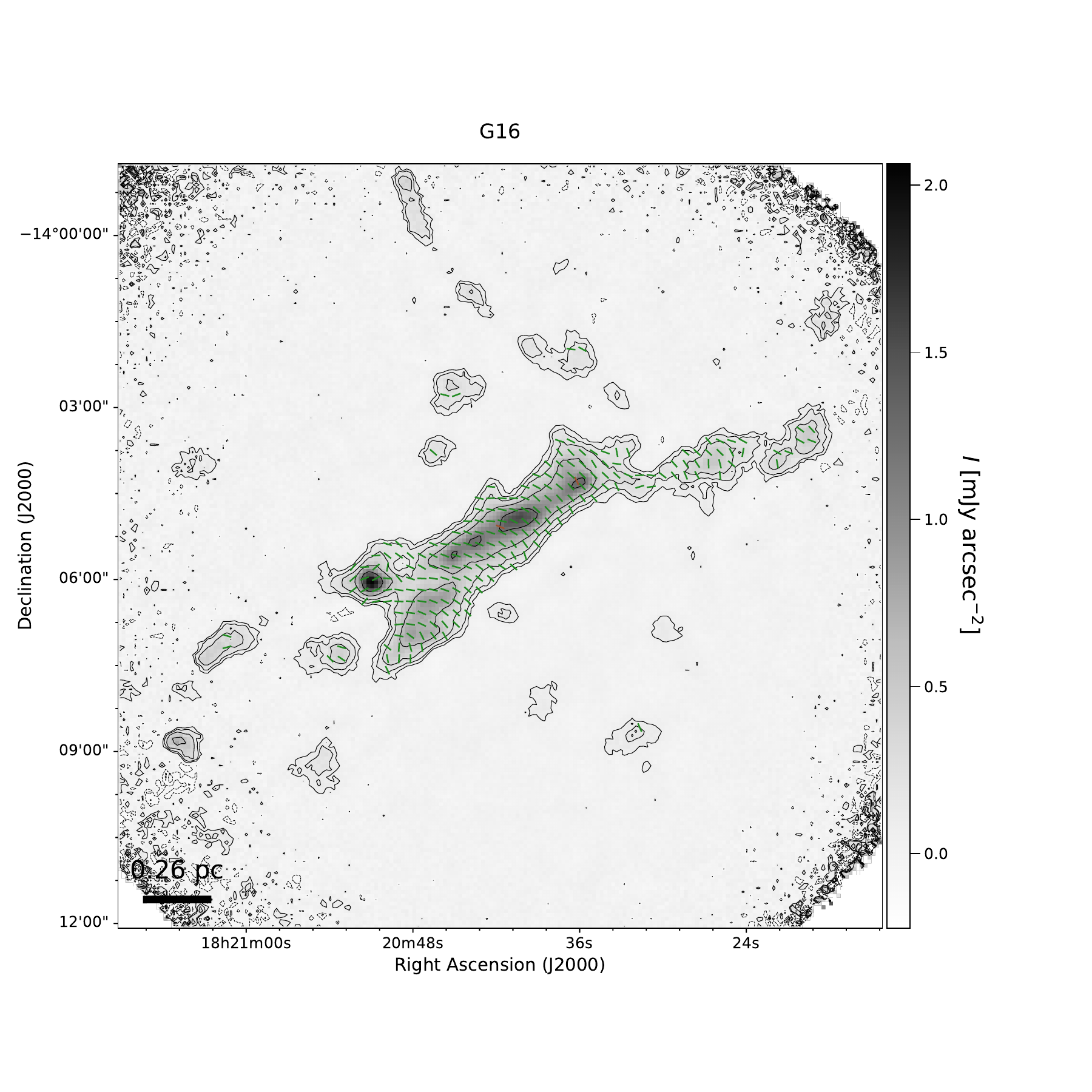}
\caption{Same as figure~\ref{afig:Bmap_L43L43}, but for G16.
The noise levels are $\sigma_{850} = 0.0050$\,mJy\,arcsec$^{-2}$ and $\sigma_{450} = 0.29$\,mJy\,arcsec$^{-2}$.
\\ {Alt text: Map with short line segments over contours.}
}
\label{afig:Bmap_G16G16}
\end{figure}

\begin{figure}[H]
\centering
\includegraphics[width=0.95\textwidth]{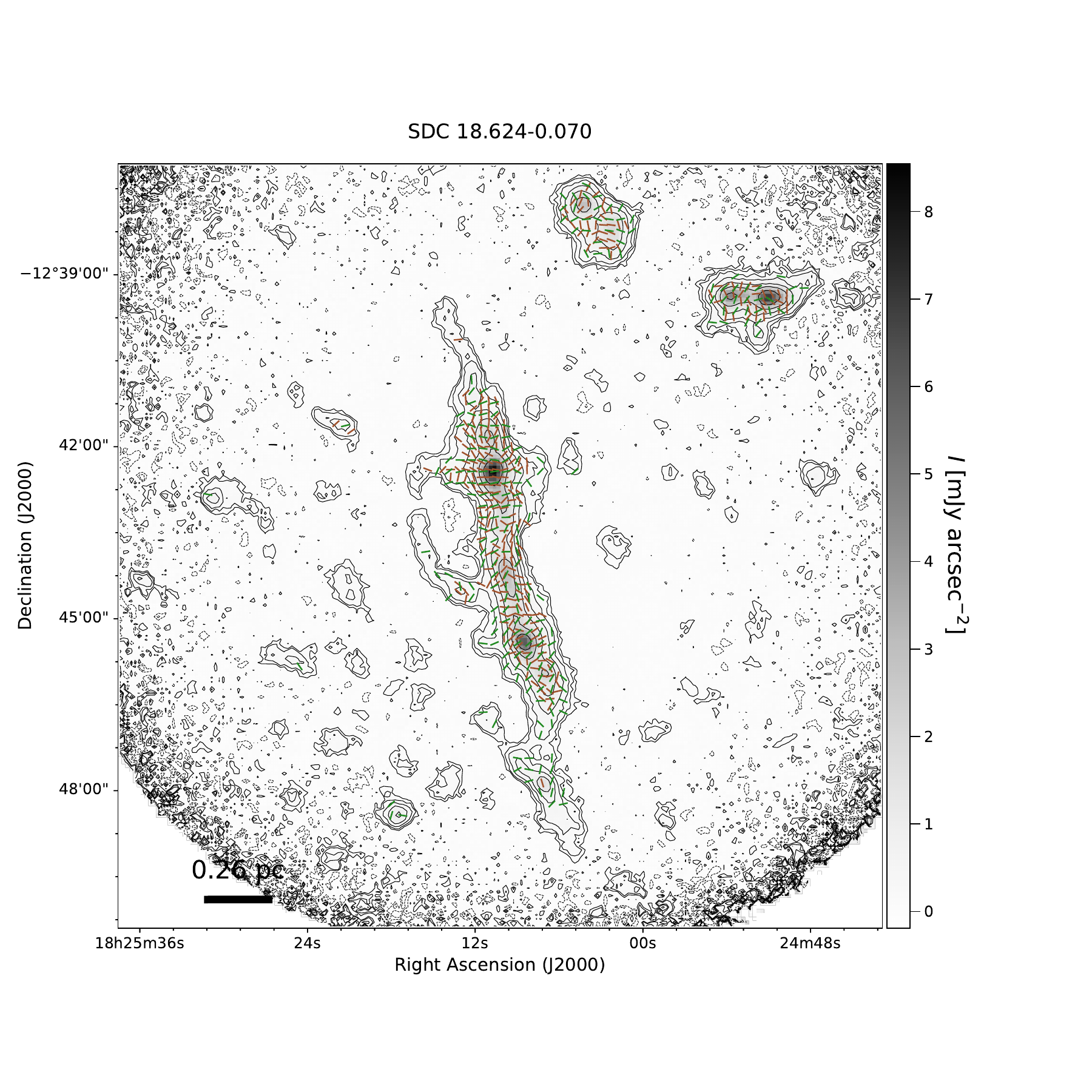}
\caption{Same as figure~\ref{afig:Bmap_L43L43}, but for SDC 18.624-0.070.
The noise levels are $\sigma_{850} = 0.0042$\,mJy\,arcsec$^{-2}$ and $\sigma_{450} = 0.14$\,mJy\,arcsec$^{-2}$.
\\ {Alt text: Map with short line segments over contours.}
}
\label{afig:Bmap_SDC186}
\end{figure}

\begin{figure}[H]
\centering
\includegraphics[width=0.95\textwidth]{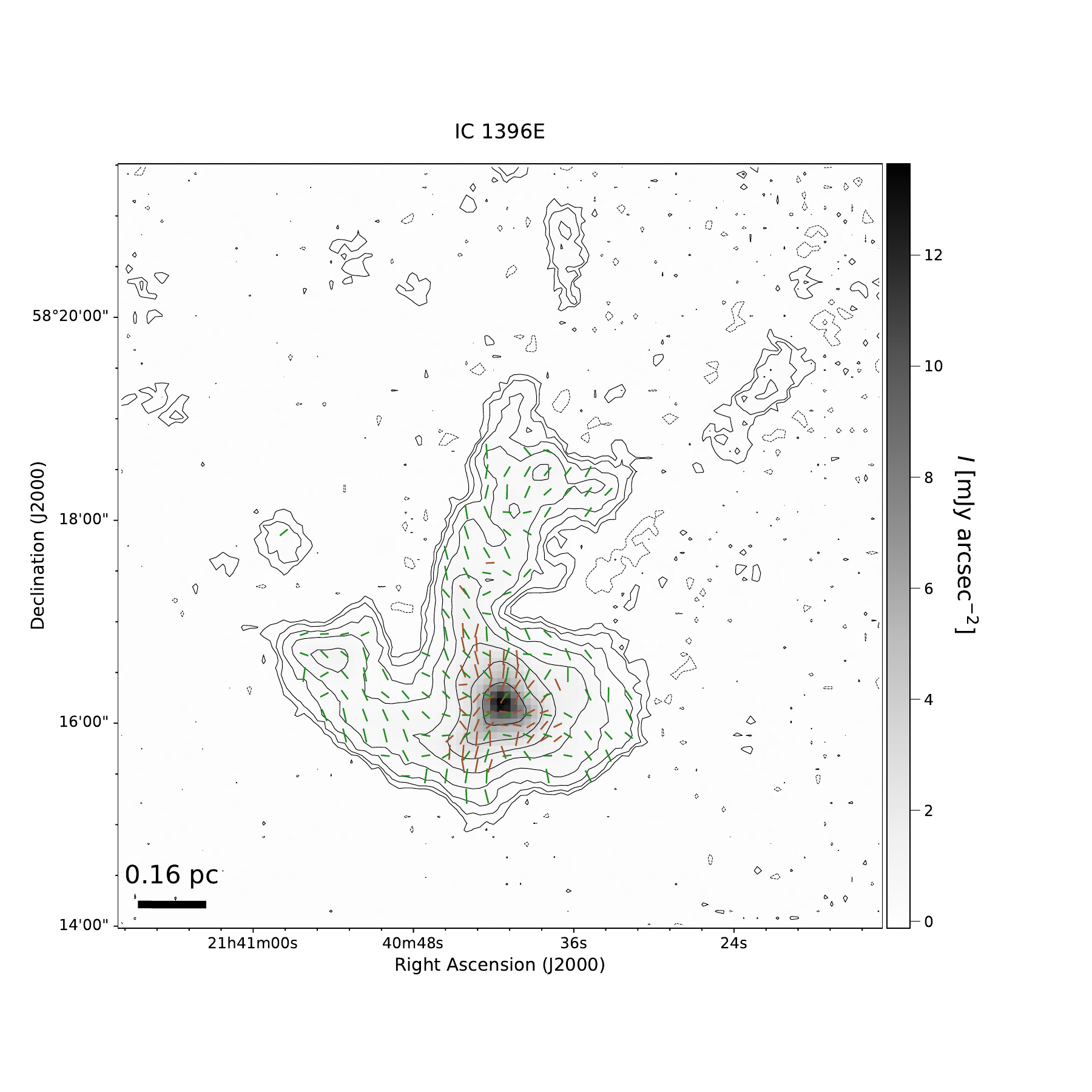}
\caption{Same as figure~\ref{afig:Bmap_L43L43}, but for IC 1396E.
The noise levels are $\sigma_{850} = 0.0033$\,mJy\,arcsec$^{-2}$ and $\sigma_{450} = 0.14$\,mJy\,arcsec$^{-2}$.
\\ {Alt text: Map with short line segments over contours.}
}
\label{afig:Bmap_IC139E}
\end{figure}

\begin{figure}[H]
\centering
\includegraphics[width=0.95\textwidth]{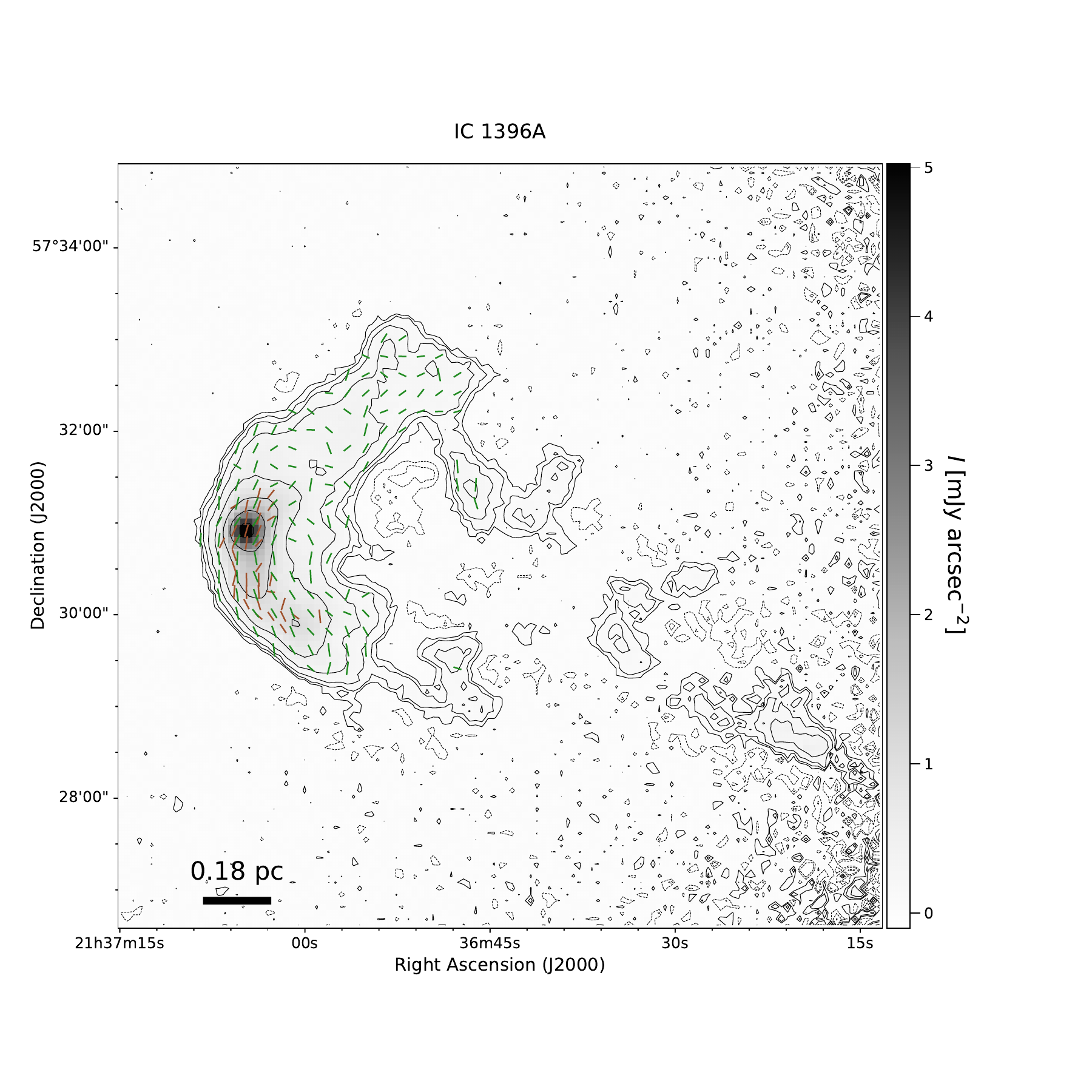}
\caption{Same as figure~\ref{afig:Bmap_L43L43}, but for IC 1396A.
The noise levels are $\sigma_{850} = 0.0023$\,mJy\,arcsec$^{-2}$ and $\sigma_{450} = 0.089$\,mJy\,arcsec$^{-2}$.
\\ {Alt text: Map with short line segments over contours.}
}
\label{afig:Bmap_IC139A}
\end{figure}

\begin{figure}[H]
\centering
\includegraphics[width=0.95\textwidth]{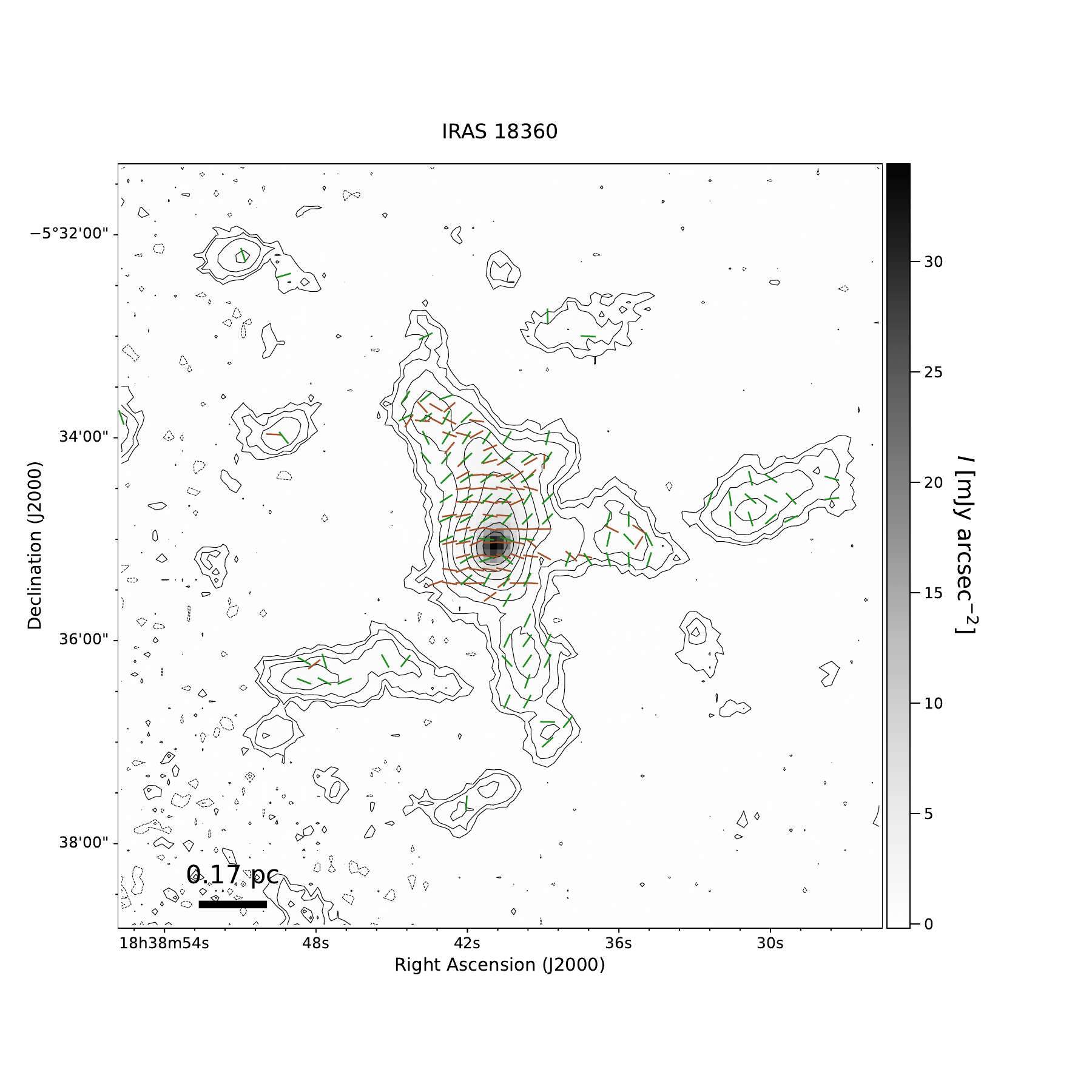}
\caption{Same as figure~\ref{afig:Bmap_L43L43}, but for IRAS 18360.
The noise levels are $\sigma_{850} = 0.0051$\,mJy\,arcsec$^{-2}$ and $\sigma_{450} = 0.11$\,mJy\,arcsec$^{-2}$.
\\ {Alt text: Map with short line segments over contours.}
}
\label{afig:Bmap_I18360}
\end{figure}

\begin{figure}[H]
\centering
\includegraphics[width=0.95\textwidth]{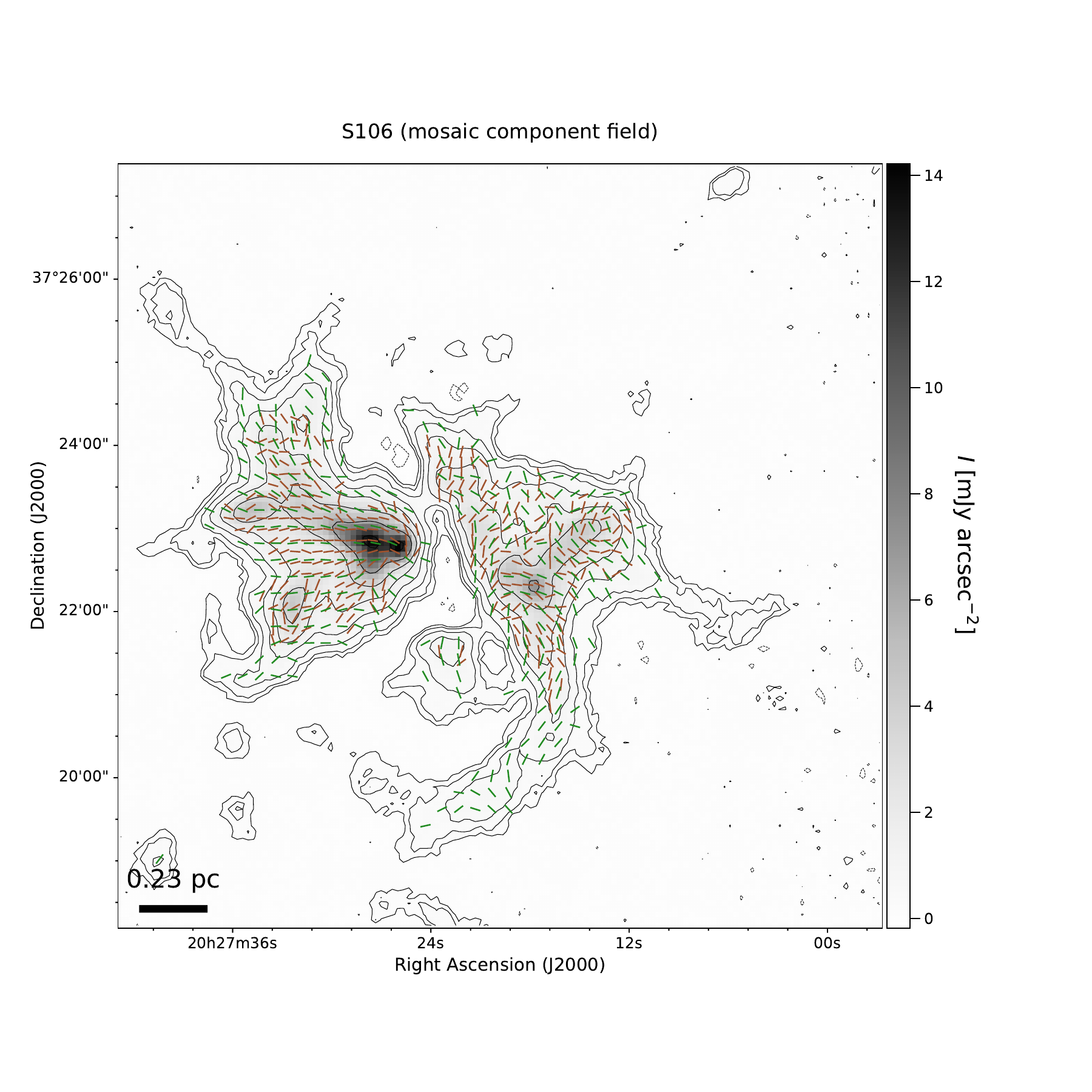}
\caption{Same as figure~\ref{afig:Bmap_L43L43}, but for S106 (mosaic component field).
The noise levels are $\sigma_{850} = 0.0067$\,mJy\,arcsec$^{-2}$ and $\sigma_{450} = 0.16$\,mJy\,arcsec$^{-2}$.
\\ {Alt text: Map with short line segments over contours.}
}
\label{afig:Bmap_S106S1}
\end{figure}

\begin{figure}[H]
\centering
\includegraphics[width=0.95\textwidth]{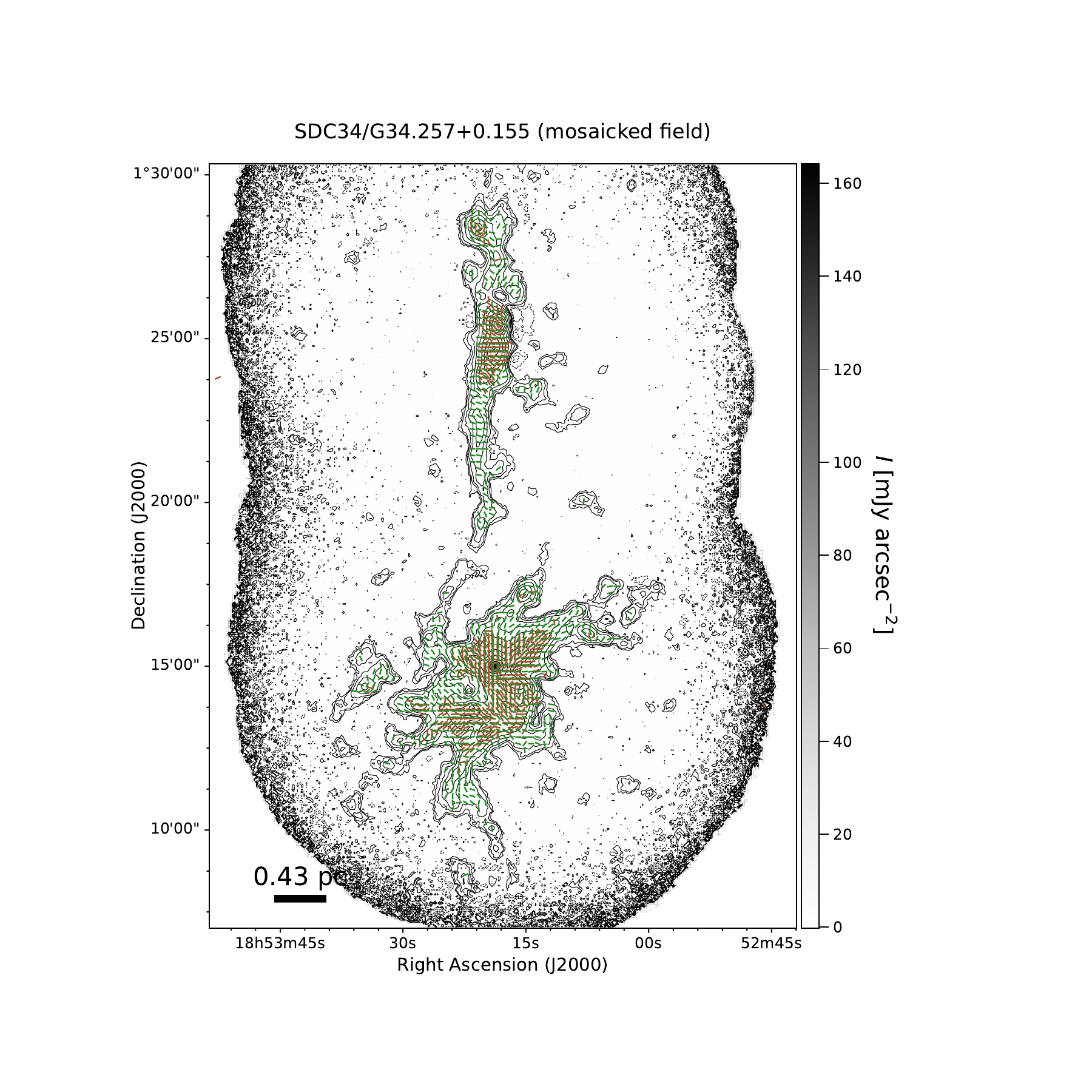}
\caption{Same as figure~\ref{afig:Bmap_L43L43}, but for SDC34/G34.257+0.155 (mosaicked field).
The noise levels are $\sigma_{850} = 0.0049$\,mJy\,arcsec$^{-2}$ and $\sigma_{450} = 0.16$\,mJy\,arcsec$^{-2}$.
\\ {Alt text: Map with short line segments over contours.}
}
\label{afig:Bmap_G342MO}
\end{figure}

\begin{figure}[H]
\centering
\includegraphics[width=0.95\textwidth]{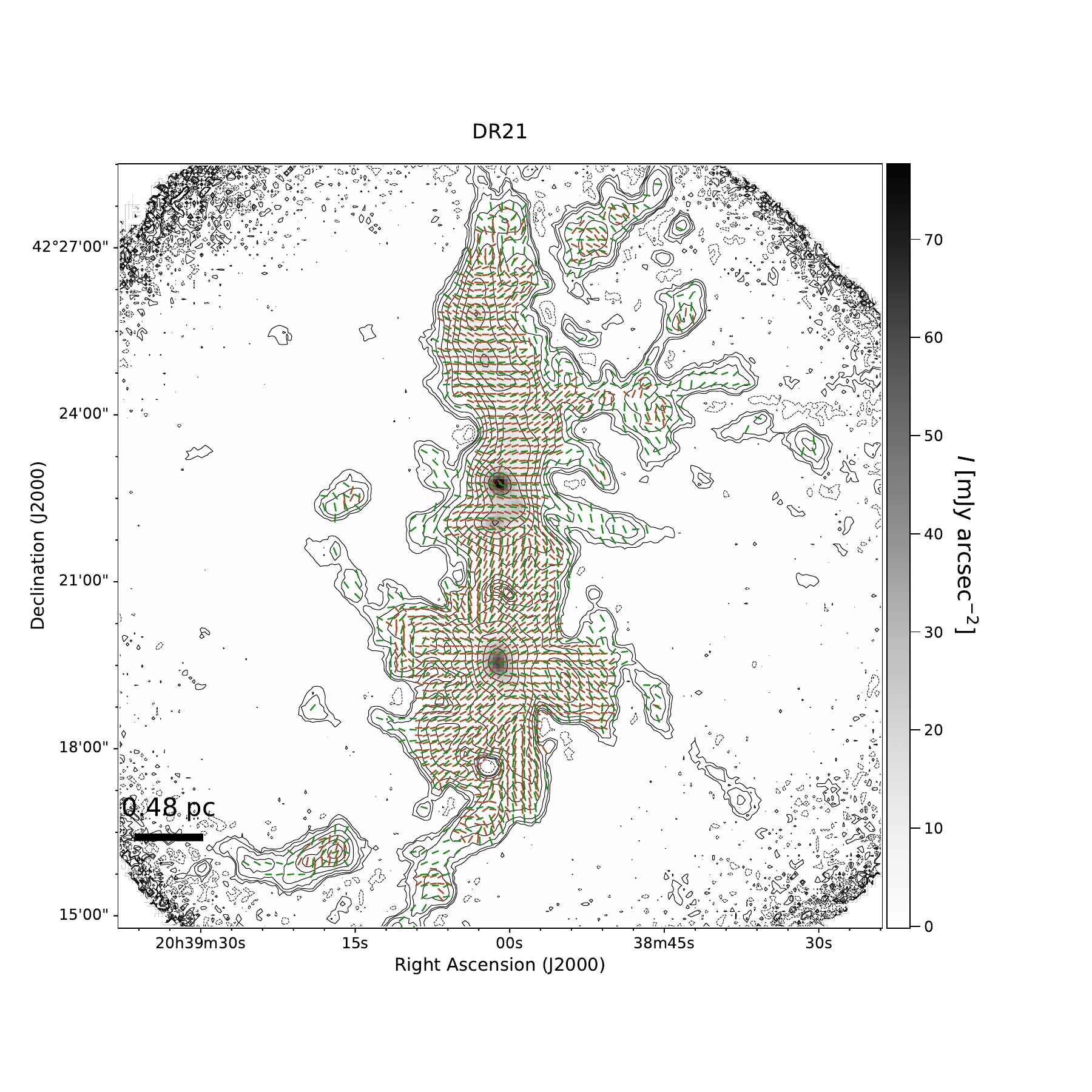}
\caption{
Same as figure~\ref{afig:Bmap_L43L43}, but for DR21.
The noise levels are $\sigma_{850} = 0.0044$\,mJy\,arcsec$^{-2}$ and
$\sigma_{450} = 0.062$\,mJy\,arcsec$^{-2}$.
In this field, systematics in the map making affect the extremely bright central region.
They lead to anomalous Stokes~$I$ behavior and elevated Stokes~$Q,U$ uncertainties.
We exclude the affected pixels from the quantitative polarization analysis.
Paper~T appendix~\ref{T-app:QU_diagnostics} ($Q,U$ diagnostics) describes how we diagnose these effects and discusses them in detail.
\\ {Alt text: Map with short line segments over contours.}
}
\label{afig:Bmap_DR21DR}
\end{figure}

\begin{figure}[H]
\centering
\includegraphics[width=0.95\textwidth]{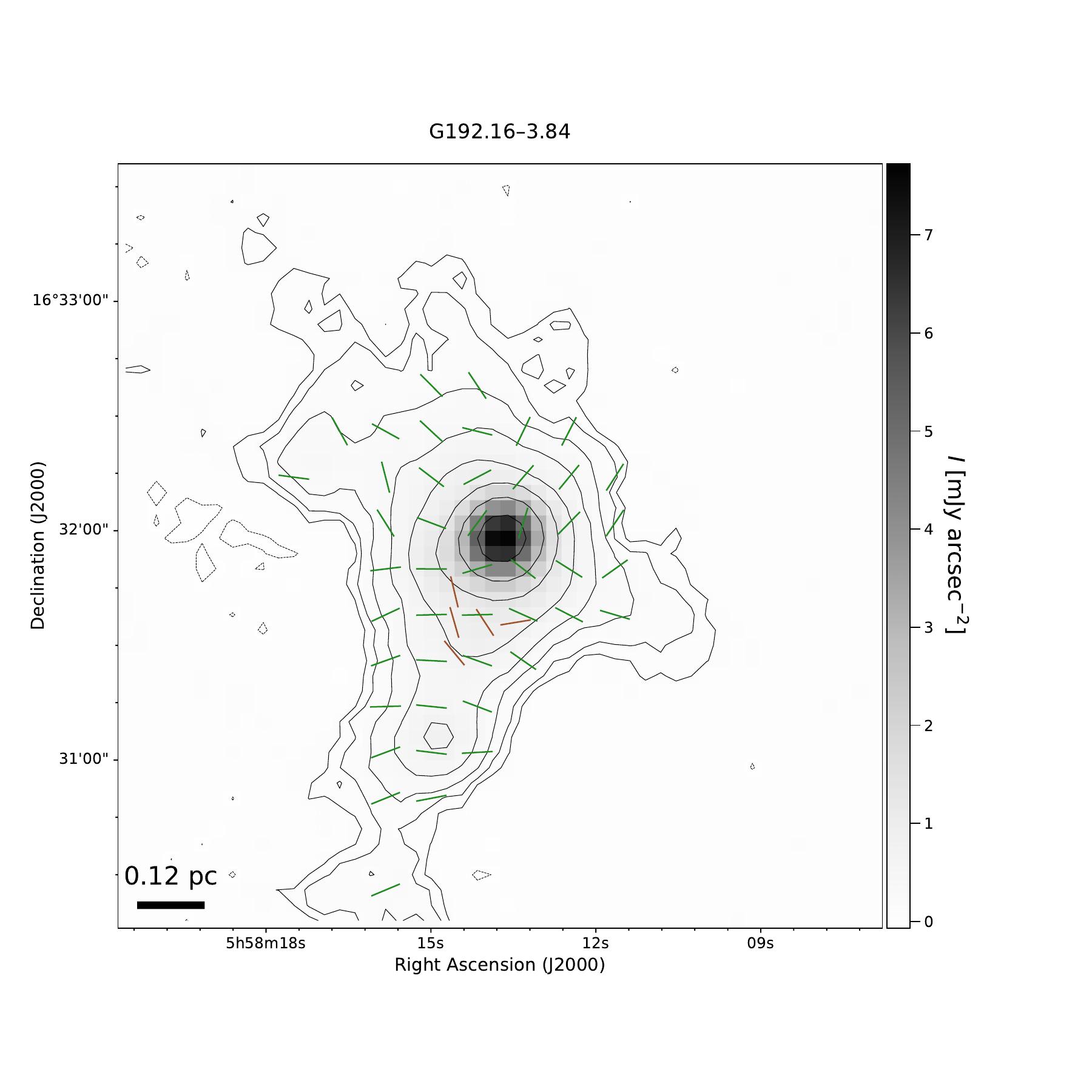}
\caption{Same as figure~\ref{afig:Bmap_L43L43}, but for G192.16--3.84.
The noise levels are $\sigma_{850} = 0.0028$\,mJy\,arcsec$^{-2}$ and $\sigma_{450} = 0.17$\,mJy\,arcsec$^{-2}$.
\\ {Alt text: Map with short line segments over contours.}
}
\label{afig:Bmap_G19216}
\end{figure}

\begin{figure}[H]
\centering
\includegraphics[width=0.95\textwidth]{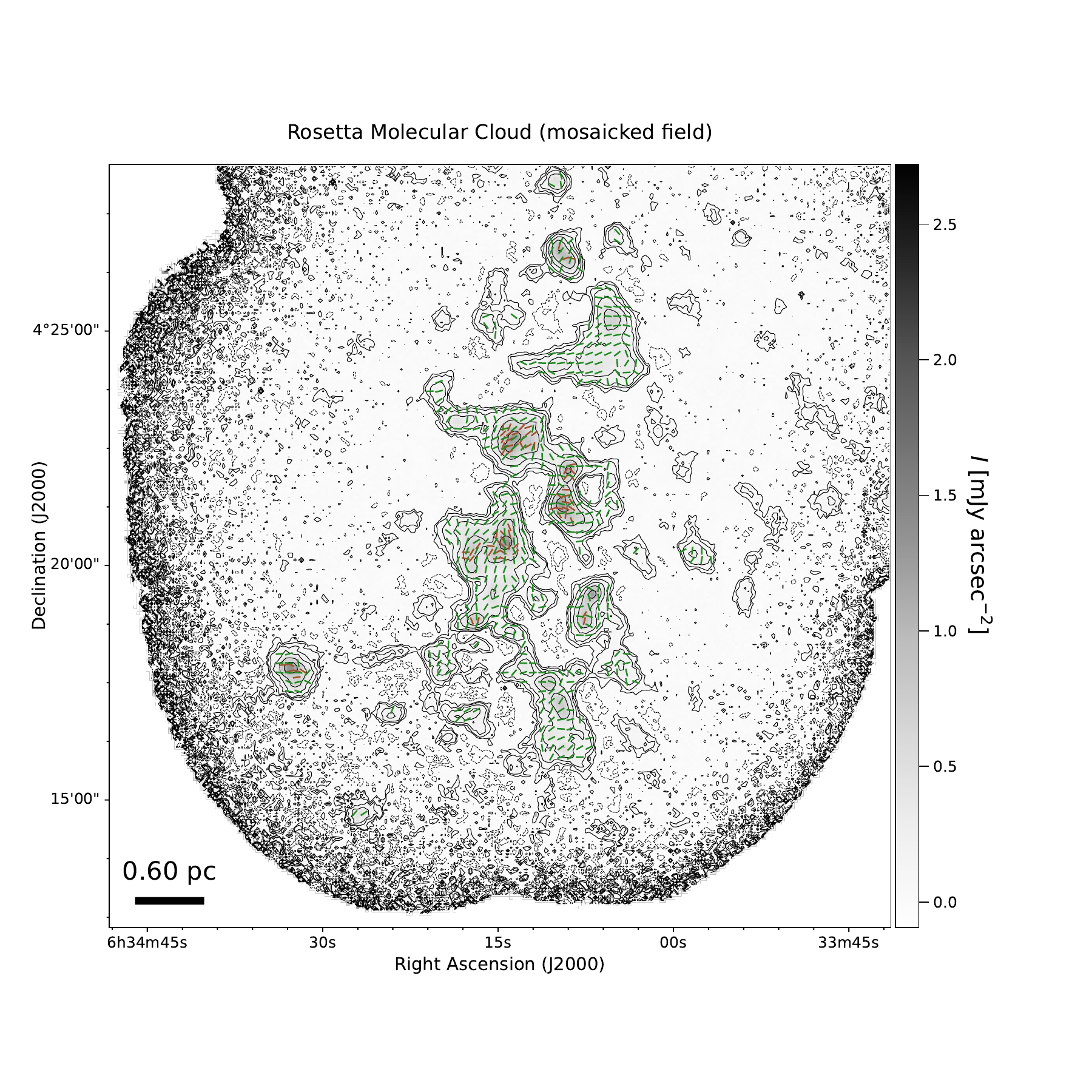}
\caption{Same as figure~\ref{afig:Bmap_L43L43}, but for Rosette Molecular Cloud (mosaicked field).
The noise levels are $\sigma_{850} = 0.0021$\,mJy\,arcsec$^{-2}$ and $\sigma_{450} = 0.042$\,mJy\,arcsec$^{-2}$.
\\ {Alt text: Map with short line segments over contours.}
}
\label{afig:Bmap_RoseMO}
\end{figure}

\begin{figure}[H]
\centering
\includegraphics[width=0.95\textwidth]{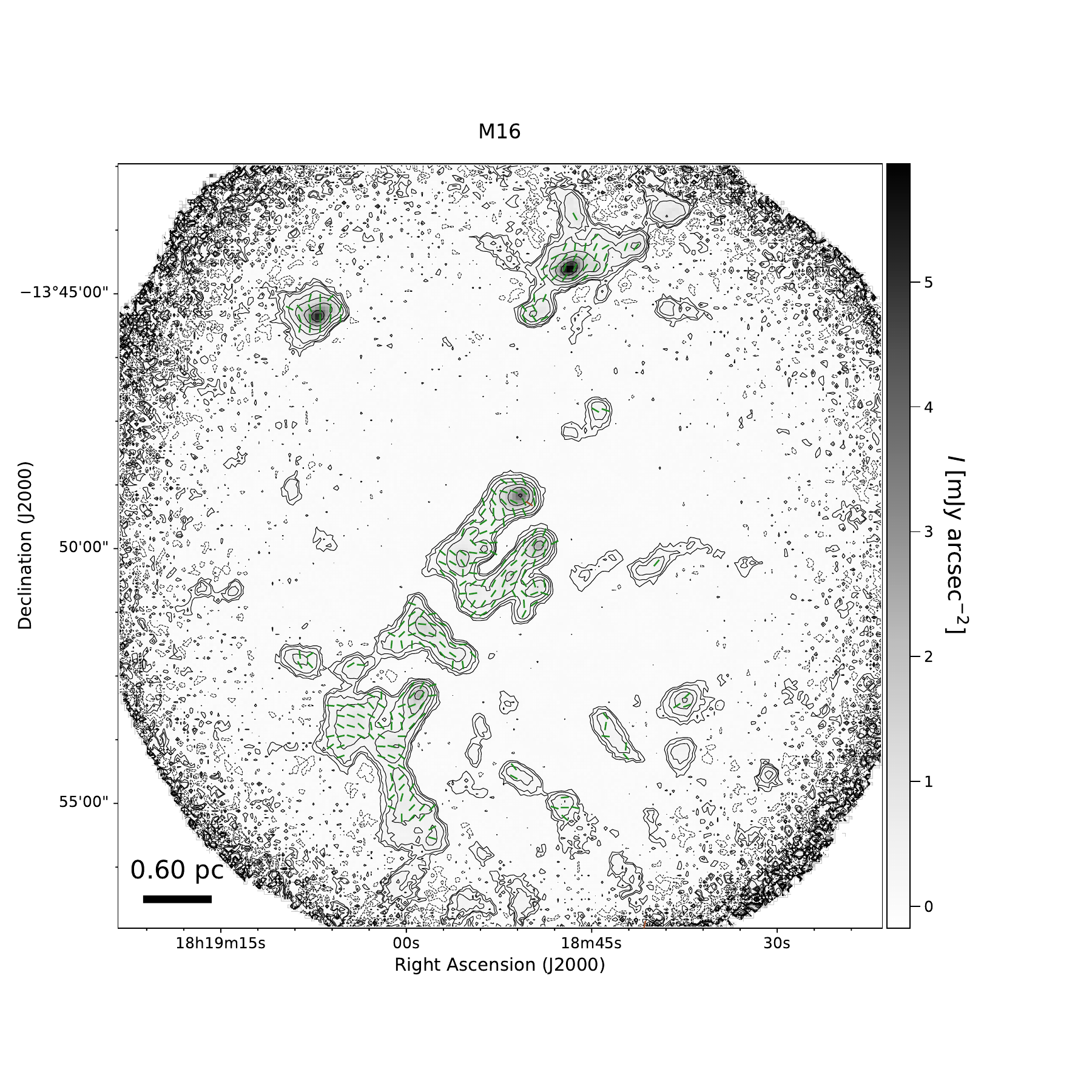}
\caption{Same as figure~\ref{afig:Bmap_L43L43}, but for M16.
The noise levels are $\sigma_{850} = 0.0039$\,mJy\,arcsec$^{-2}$ and $\sigma_{450} = 0.13$\,mJy\,arcsec$^{-2}$.
\\ {Alt text: Map with short line segments over contours.}
}
\label{afig:Bmap_M16M16}
\end{figure}

\begin{figure}[H]
\centering
\includegraphics[width=0.95\textwidth]{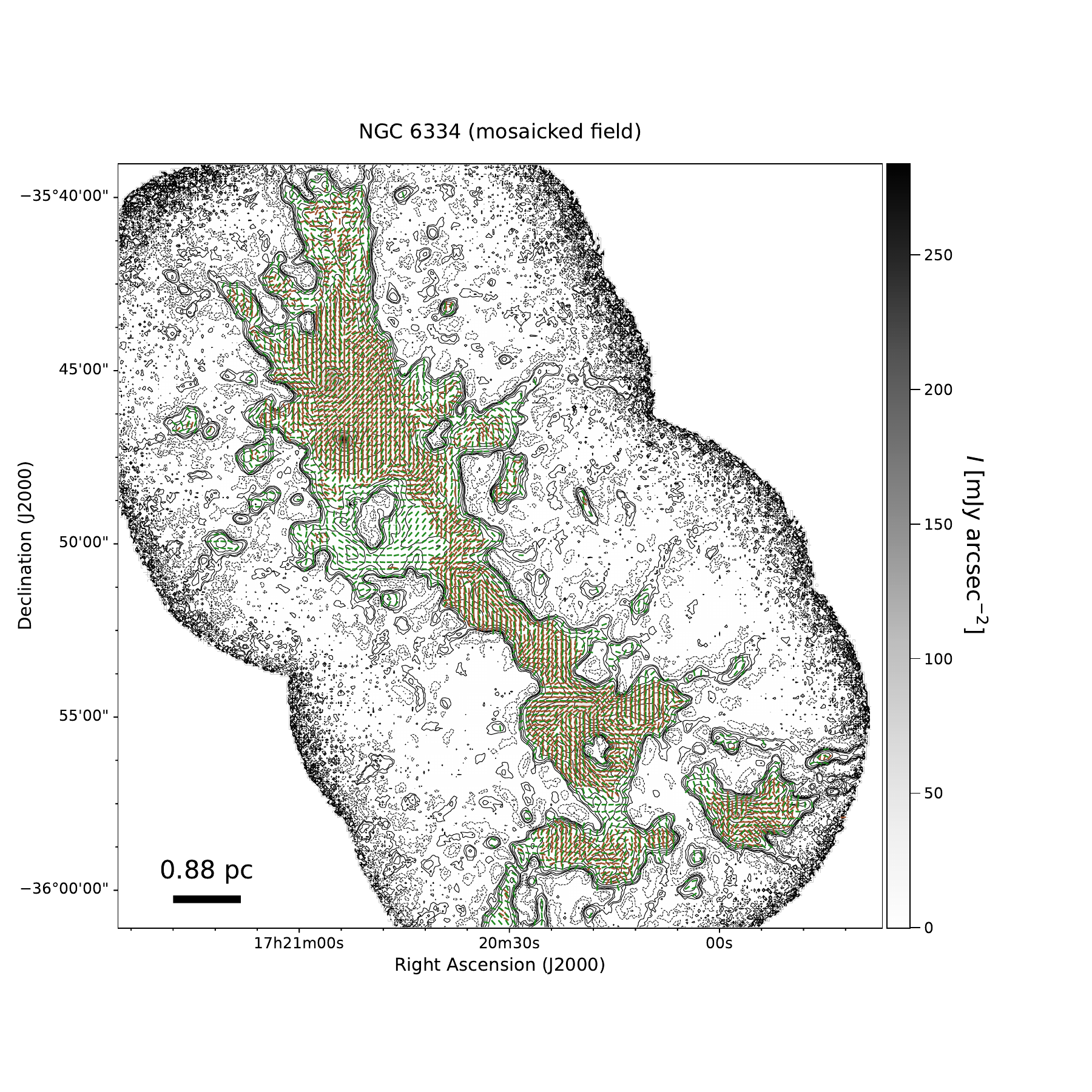}
\caption{Same as figure~\ref{afig:Bmap_L43L43}, but for NGC 6334 (mosaicked field).
The noise levels are $\sigma_{850} = 0.0038$\,mJy\,arcsec$^{-2}$ and $\sigma_{450} = 0.12$\,mJy\,arcsec$^{-2}$.
\\ {Alt text: Map with short line segments over contours.}
}
\label{afig:Bmap_N633MO}
\end{figure}

\begin{figure}[H]
\centering
\includegraphics[width=0.95\textwidth]{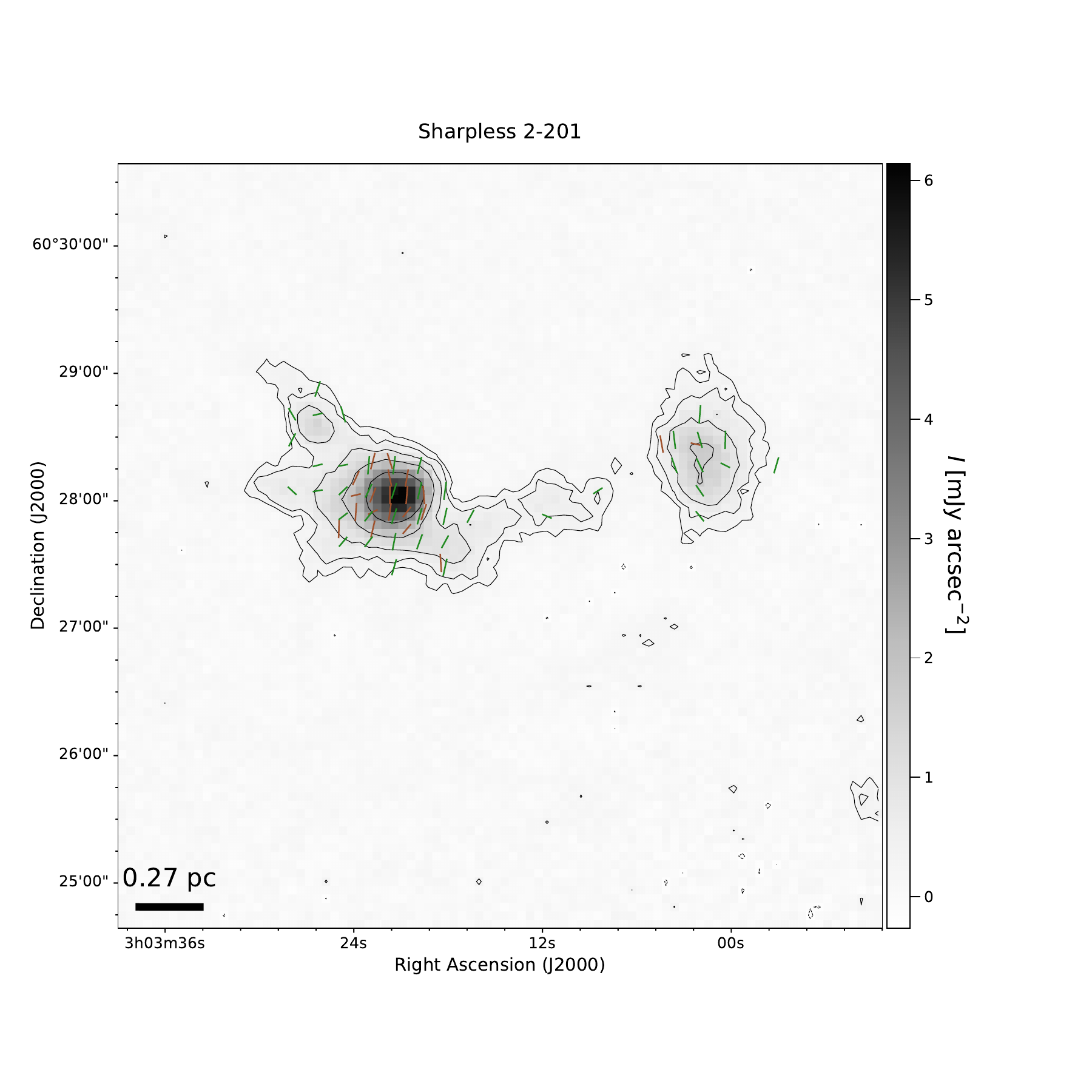}
\caption{Same as figure~\ref{afig:Bmap_L43L43}, but for Sharpless 2-201.
The noise levels are $\sigma_{850} = 0.013$\,mJy\,arcsec$^{-2}$ and $\sigma_{450} = 0.26$\,mJy\,arcsec$^{-2}$.
\\ {Alt text: Map with short line segments over contours.}
}
\label{afig:Bmap_SH2F1S}
\end{figure}

\begin{figure}[H]
\centering
\includegraphics[width=0.95\textwidth]{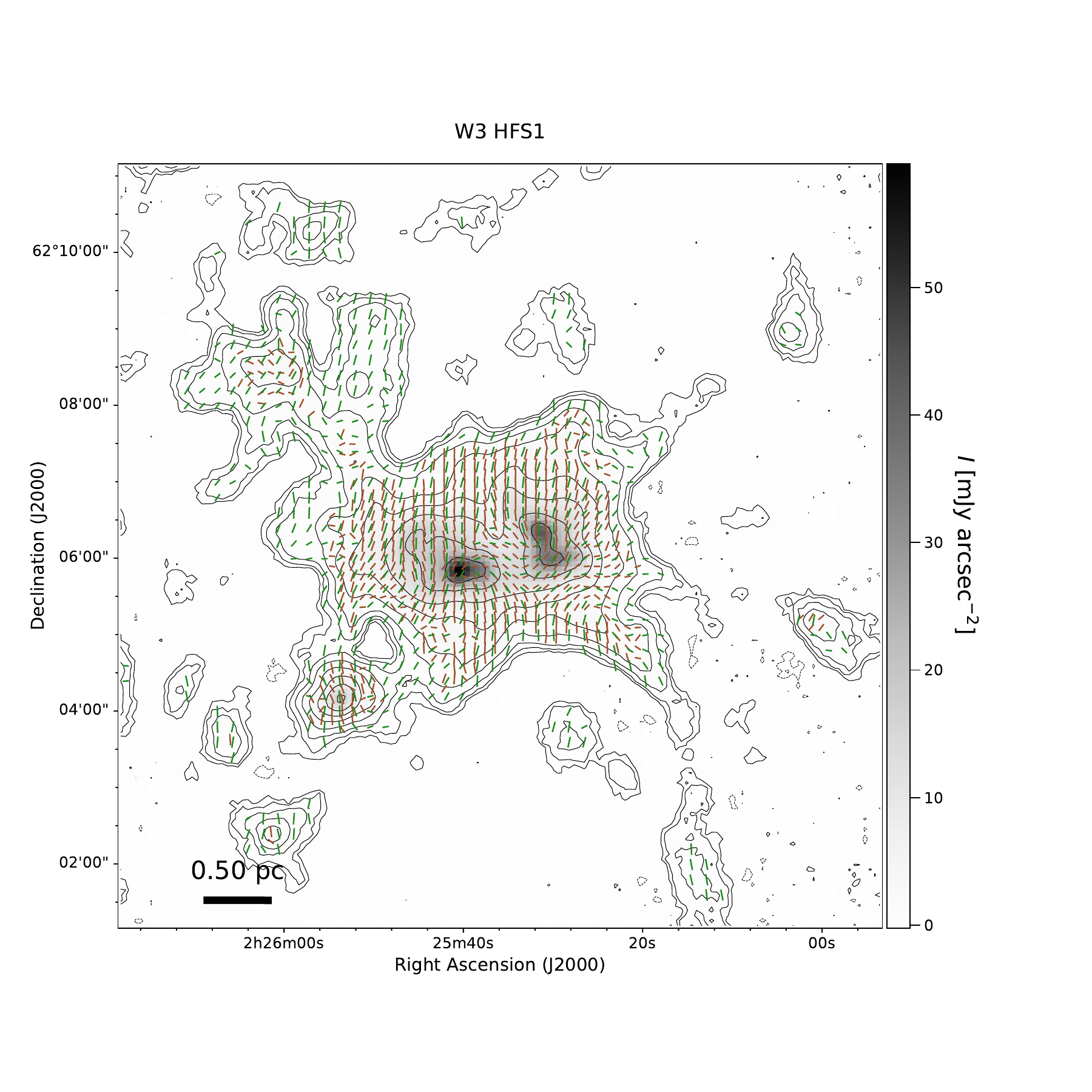}
\caption{Same as figure~\ref{afig:Bmap_L43L43}, but for W3 HFS1.
The noise levels are $\sigma_{850} = 0.0084$\,mJy\,arcsec$^{-2}$ and $\sigma_{450} = 0.21$\,mJy\,arcsec$^{-2}$.
\\ {Alt text: Map with short line segments over contours.}
}
\label{afig:Bmap_W3HFS1}
\end{figure}

Figures~\ref{afig:Bmap_L43L43}--\ref{afig:Bmap_GSGRB2} present maps of the
$\vec{B}$-field orientation in the plane of the sky at both wavelengths for 65 of the 69 sources in which at least one 450\,$\mu$m polarization segment was detected.
The maps are ordered approximately by distance from the Sun, ranging from L43 at 125\,pc to Sagittarius~B2 at 8.3\,kpc.
Each figure shows the $\vec{B}$-field orientation in the plane of the sky, inferred from
POL-2 polarization observations at 850\,$\mu$m (green segments) and 450\,$\mu$m (sienna segments), with contours tracing the 850\,$\mu$m
Stokes~$I$ intensity.
The first figure (figure~\ref{afig:Bmap_L43L43}) provides a complete description of the map elements and selection criteria.
Subsequent figures follow the same format, with the noise levels for each source given in the corresponding caption.
We exclude sources without 450\,$\mu$m polarization detections from this appendix but include them in the statistical analyses presented in the main text where applicable.
 
\clearpage
\clearpage
\clearpage
\clearpage
\clearpage
\clearpage
\clearpage
\clearpage
\clearpage
\clearpage
\clearpage
\clearpage
\clearpage
\clearpage
\clearpage
\clearpage
\clearpage
\clearpage
\clearpage
\clearpage
\clearpage
\clearpage
\clearpage
\clearpage
\clearpage
\clearpage
\clearpage
\clearpage
\clearpage
\clearpage
\clearpage
\clearpage
\clearpage
\clearpage
\clearpage
\clearpage
\clearpage
\clearpage
\clearpage
\clearpage
\clearpage
\clearpage
\clearpage
\clearpage
\clearpage
\clearpage
\clearpage
\begin{figure}[H]
\centering
\includegraphics[width=0.95\textwidth]{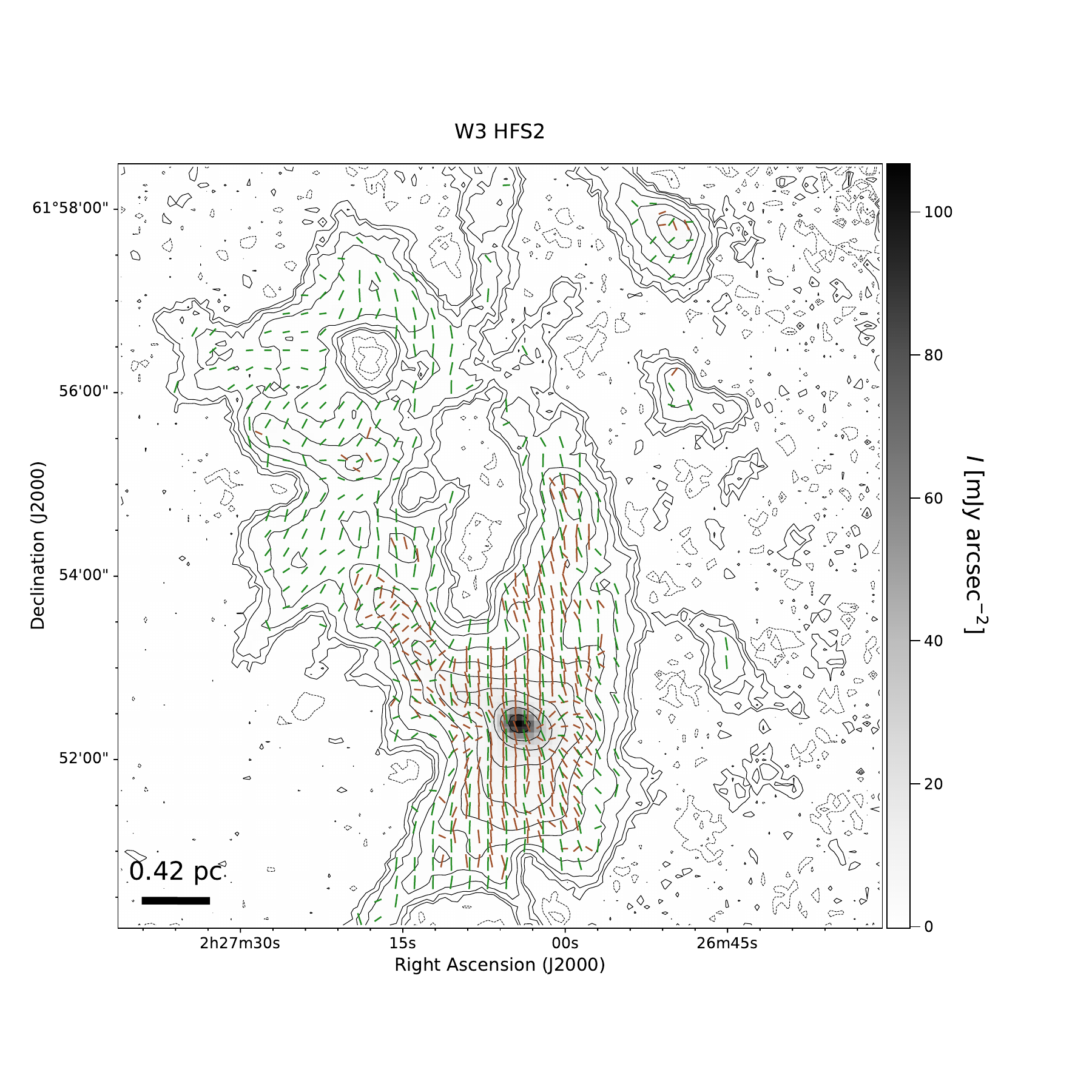}
\caption{Same as figure~\ref{afig:Bmap_L43L43}, but for W3 HFS2.
The noise levels are $\sigma_{850} = 0.0048$\,mJy\,arcsec$^{-2}$ and $\sigma_{450} = 0.22$\,mJy\,arcsec$^{-2}$.
\\ {Alt text: Map with short line segments over contours.}
}
\label{afig:Bmap_W3HFS2}
\end{figure}
\clearpage
\begin{figure}[H]
\centering
\includegraphics[width=0.95\textwidth]{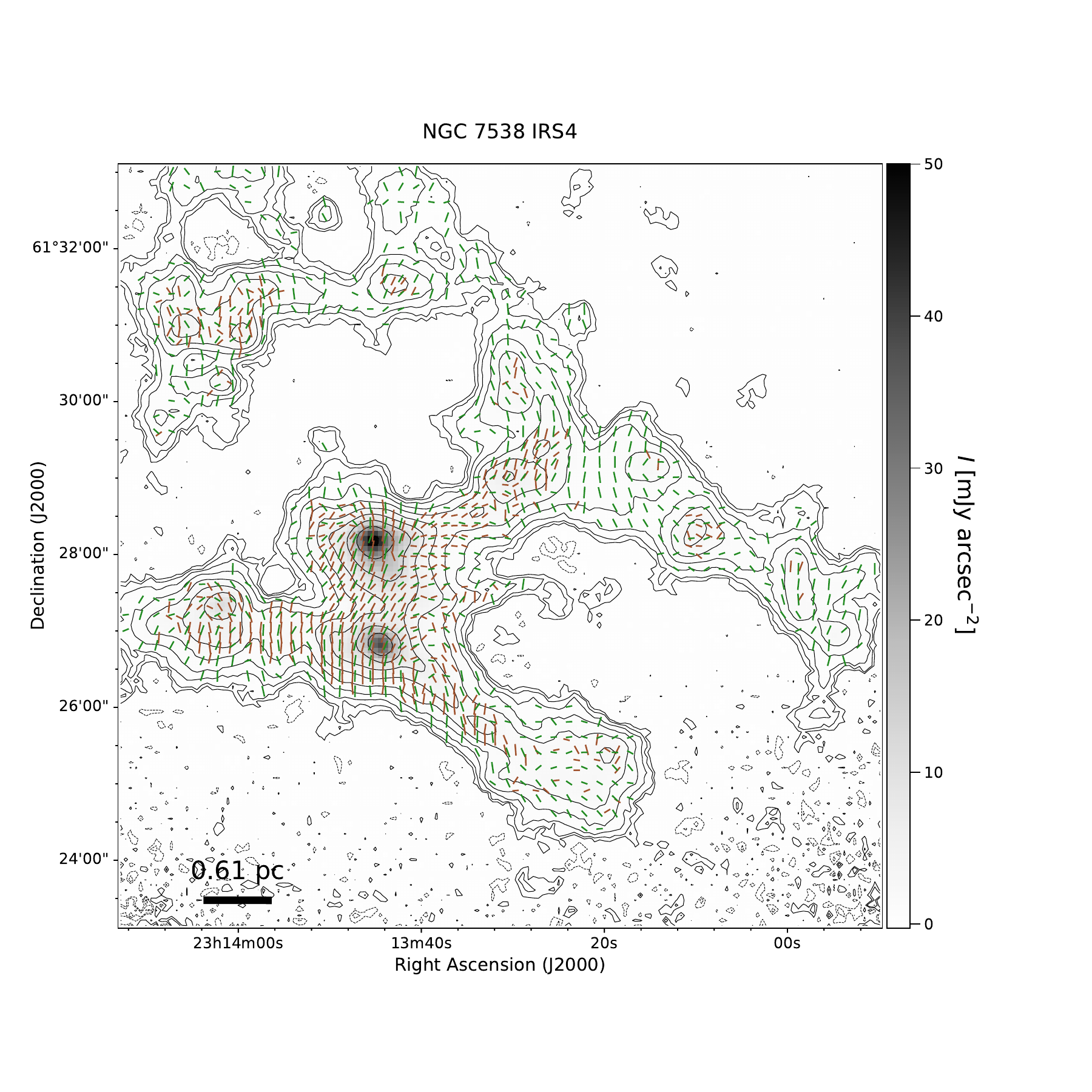}
\caption{Same as figure~\ref{afig:Bmap_L43L43}, but for NGC 7538 IRS4.
The noise levels are $\sigma_{850} = 0.0059$\,mJy\,arcsec$^{-2}$ and $\sigma_{450} = 0.25$\,mJy\,arcsec$^{-2}$.
\\ {Alt text: Map with short line segments over contours.}
}
\label{afig:Bmap_N753I4}
\end{figure}
\clearpage
\begin{figure}[H]
\centering
\includegraphics[width=0.95\textwidth]{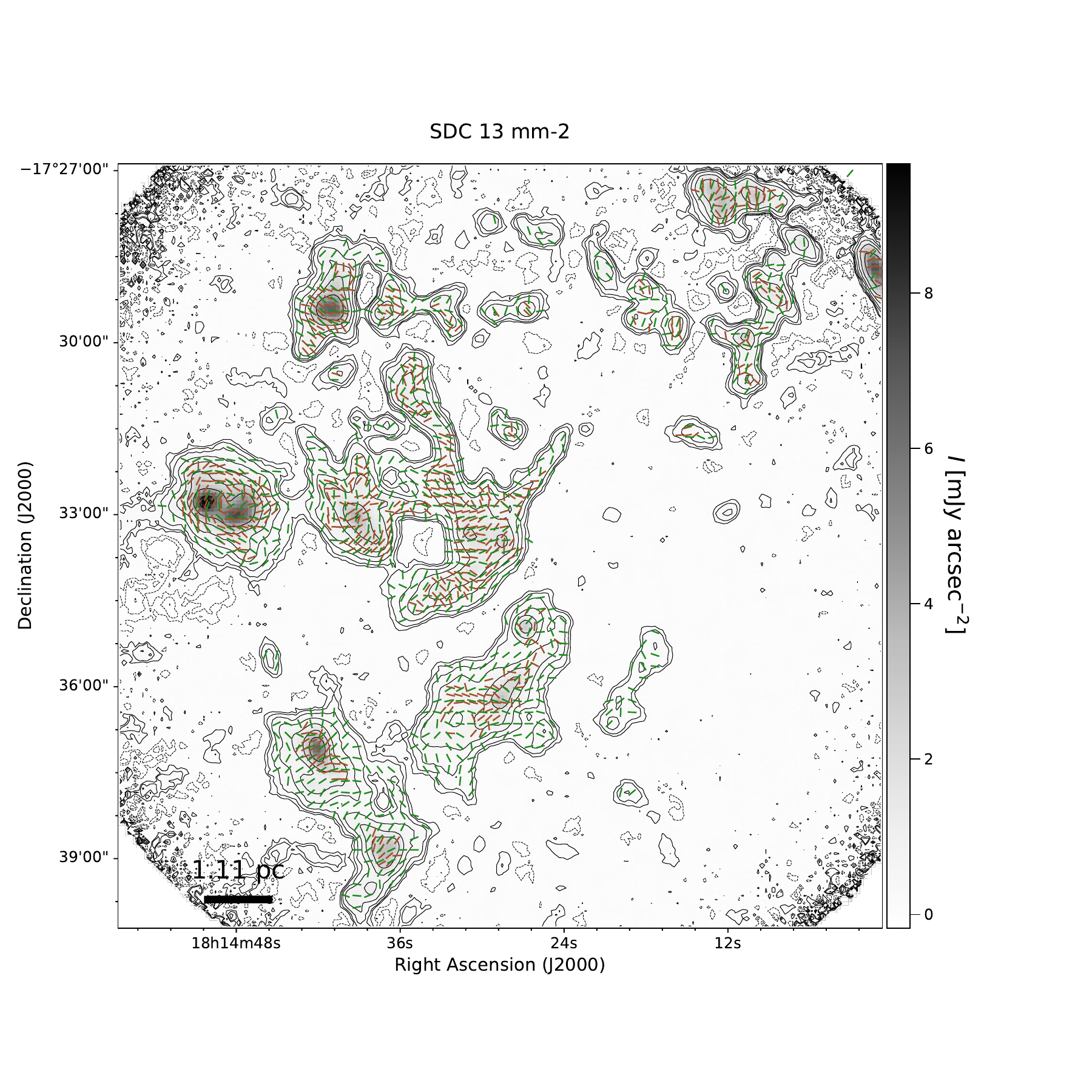}
\caption{Same as figure~\ref{afig:Bmap_L43L43}, but for SDC 13 mm-2.
The noise levels are $\sigma_{850} = 0.0039$\,mJy\,arcsec$^{-2}$ and $\sigma_{450} = 0.092$\,mJy\,arcsec$^{-2}$.
\\ {Alt text: Map with short line segments over contours.}
}
\label{afig:Bmap_SDC13M}
\end{figure}
\clearpage
\begin{figure}[H]
\centering
\includegraphics[width=0.95\textwidth]{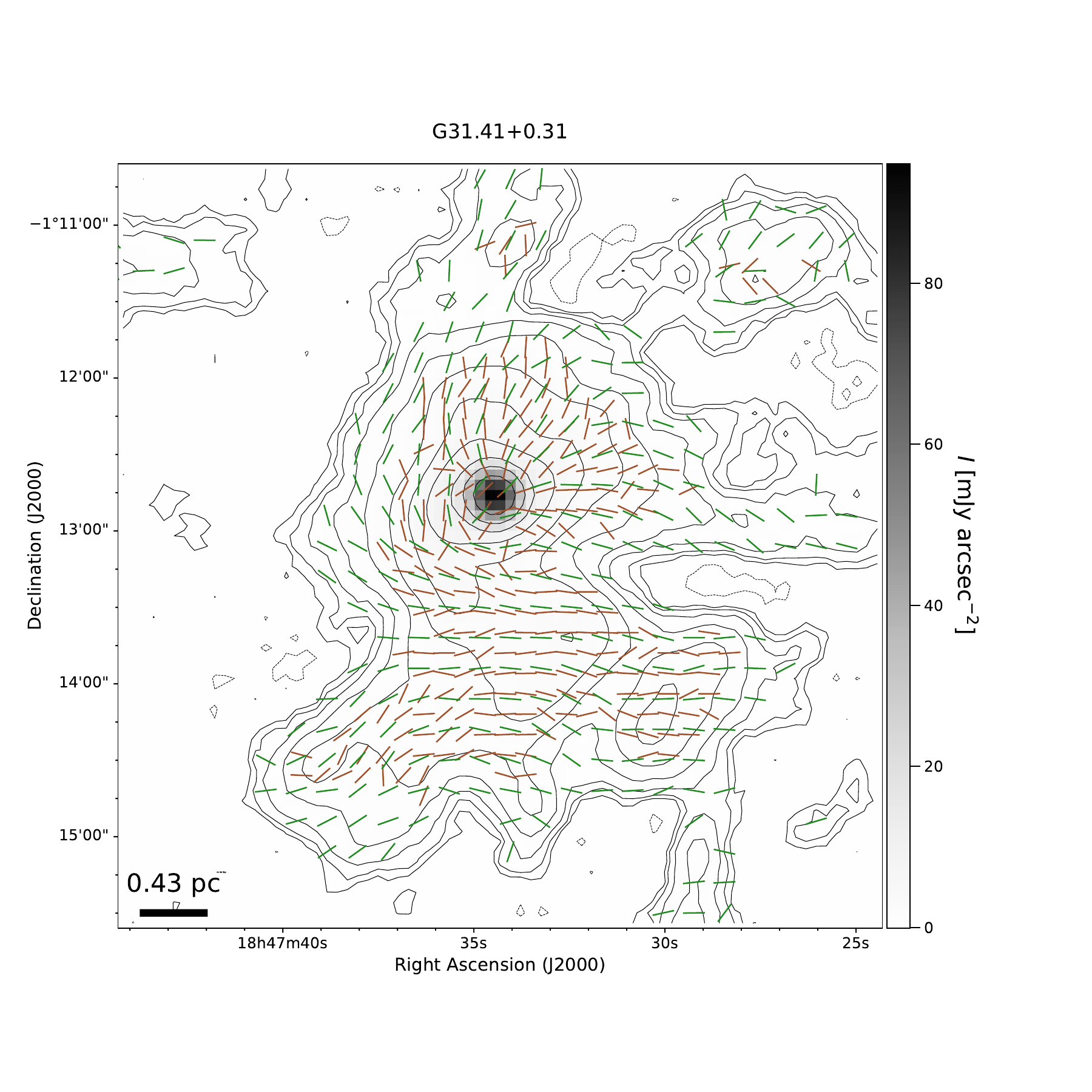}
\caption{Same as figure~\ref{afig:Bmap_L43L43}, but for G31.41+0.31.
The noise levels are $\sigma_{850} = 0.0032$\,mJy\,arcsec$^{-2}$ and $\sigma_{450} = 0.078$\,mJy\,arcsec$^{-2}$.
\\ {Alt text: Map with short line segments over contours.}
}
\label{afig:Bmap_G3141G}
\end{figure}
\clearpage
\begin{figure}[H]
\centering
\includegraphics[width=0.95\textwidth]{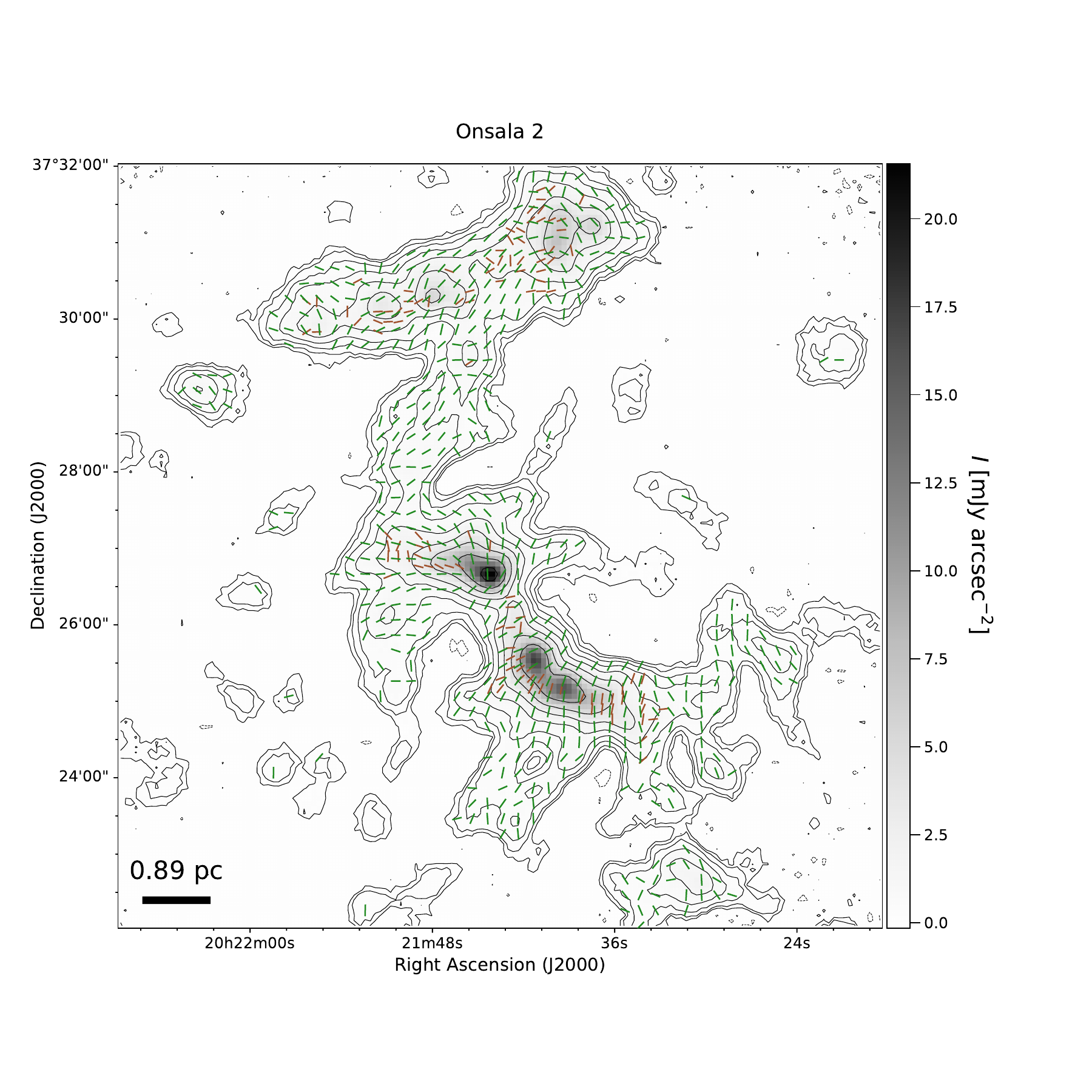}
\caption{Same as figure~\ref{afig:Bmap_L43L43}, but for Onsala 2.
The noise levels are $\sigma_{850} = 0.0046$\,mJy\,arcsec$^{-2}$ and $\sigma_{450} = 0.18$\,mJy\,arcsec$^{-2}$.
\\ {Alt text: Map with short line segments over contours.}
}
\label{afig:Bmap_DR15DR}
\end{figure}
\clearpage
\begin{figure}[H]
\centering
\includegraphics[width=0.95\textwidth]{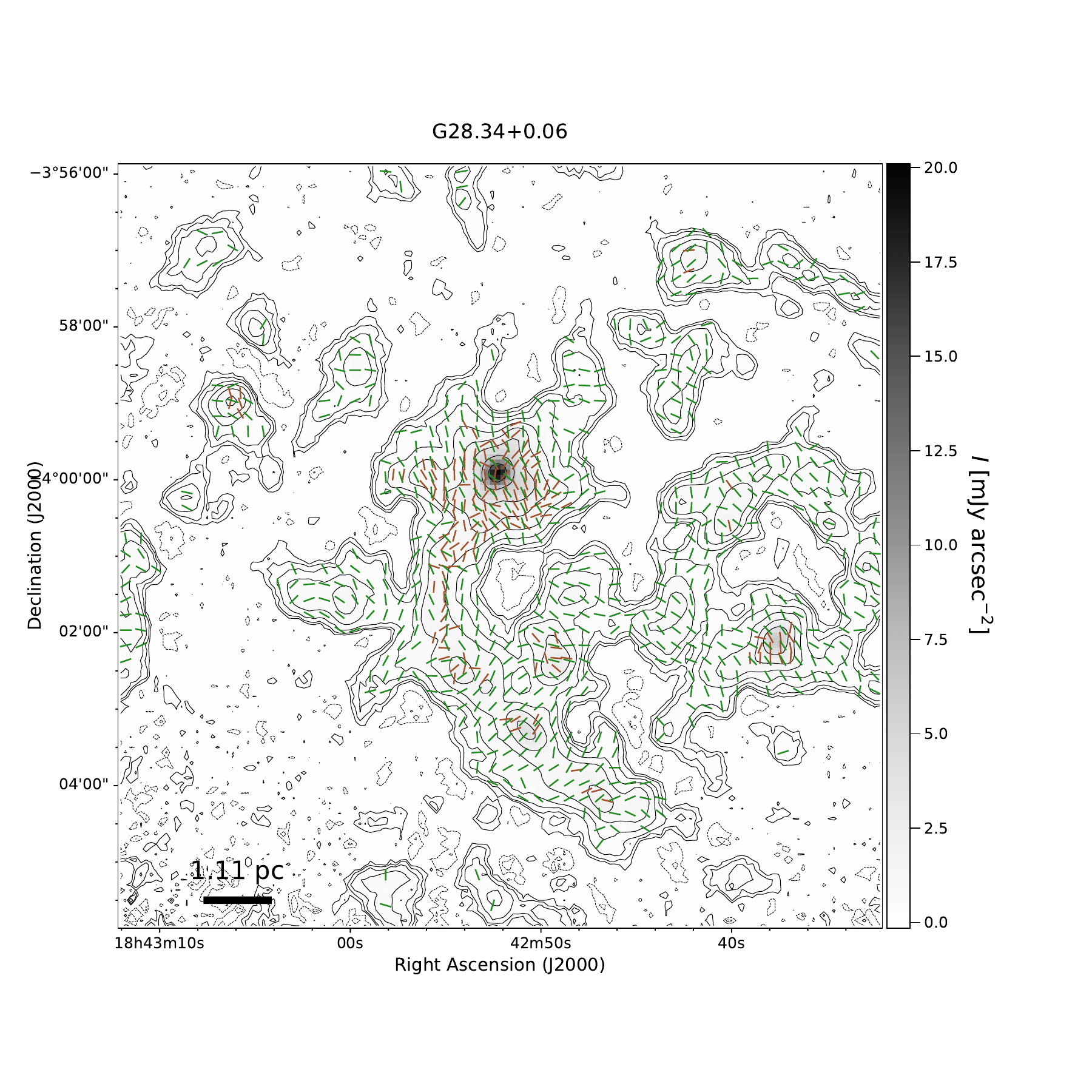}
\caption{Same as figure~\ref{afig:Bmap_L43L43}, but for G28.34+0.06.
The noise levels are $\sigma_{850} = 0.0034$\,mJy\,arcsec$^{-2}$ and $\sigma_{450} = 0.11$\,mJy\,arcsec$^{-2}$.
\\ {Alt text: Map with short line segments over contours.}
}
\label{afig:Bmap_G2834G}
\end{figure}
\clearpage
\begin{figure}[H]
\centering
\includegraphics[width=0.95\textwidth]{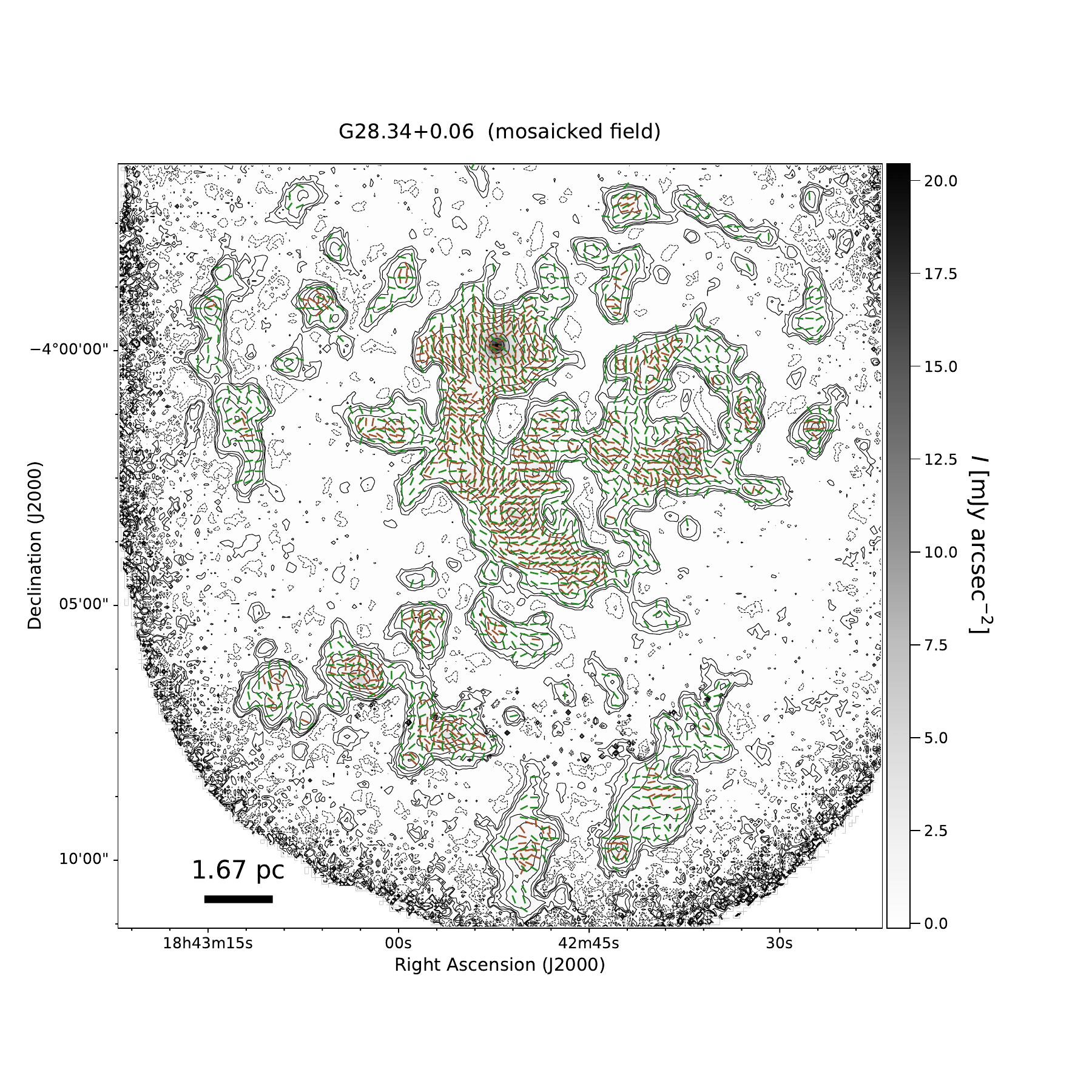}
\caption{Same as figure~\ref{afig:Bmap_L43L43}, but for G28.34+0.06  (mosaicked field).
The noise levels are $\sigma_{850} = 0.0030$\,mJy\,arcsec$^{-2}$ and $\sigma_{450} = 0.030$\,mJy\,arcsec$^{-2}$.
\\ {Alt text: Map with short line segments over contours.}
}
\label{afig:Bmap_G283MO}
\end{figure}
\clearpage
\begin{figure}[H]
\centering
\includegraphics[width=0.95\textwidth]{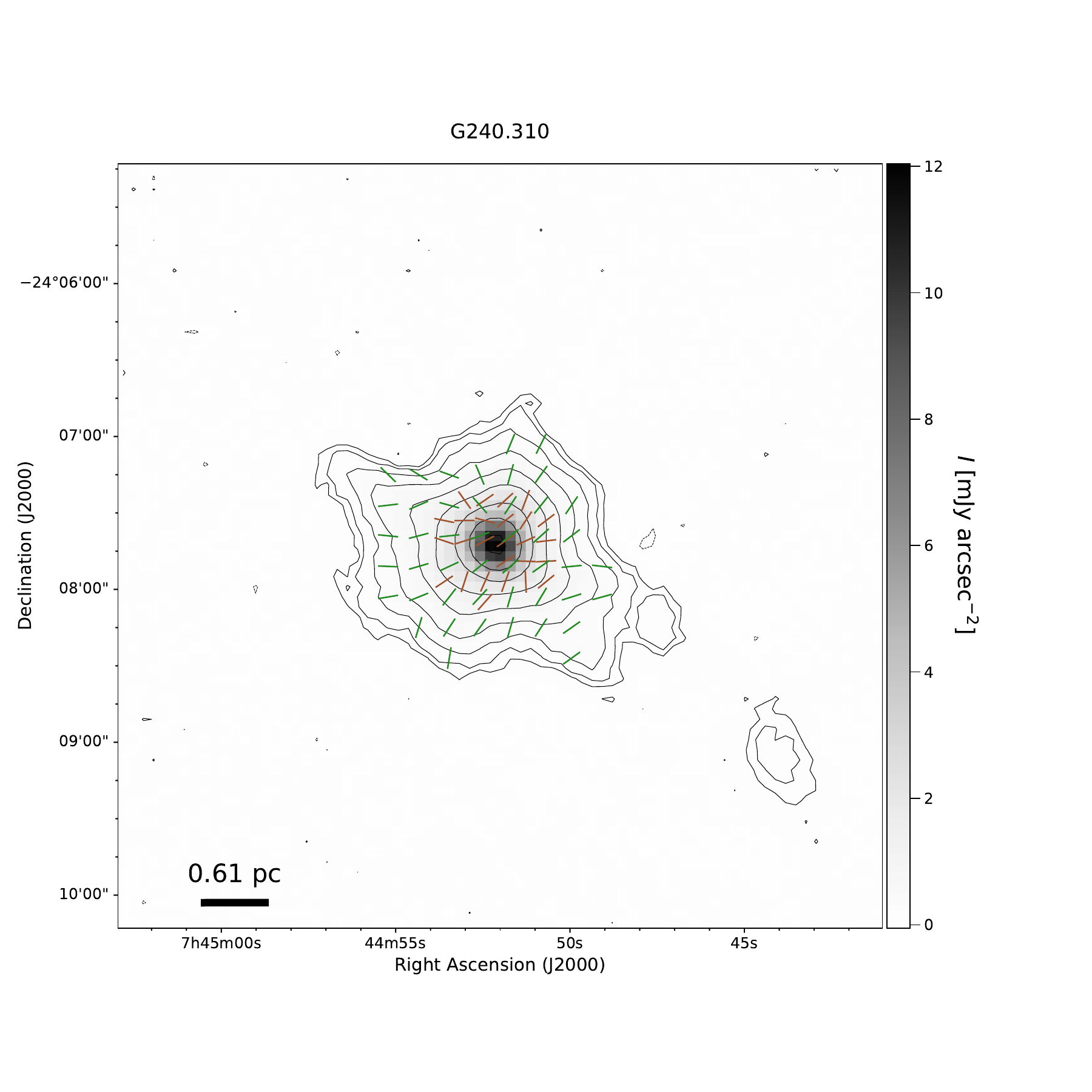}
\caption{Same as figure~\ref{afig:Bmap_L43L43}, but for G240.310.
The noise levels are $\sigma_{850} = 0.0028$\,mJy\,arcsec$^{-2}$ and $\sigma_{450} = 0.081$\,mJy\,arcsec$^{-2}$.
\\ {Alt text: Map with short line segments over contours.}
}
\label{afig:Bmap_G24031}
\end{figure}
\clearpage
\begin{figure}[H]
\centering
\includegraphics[width=0.95\textwidth]{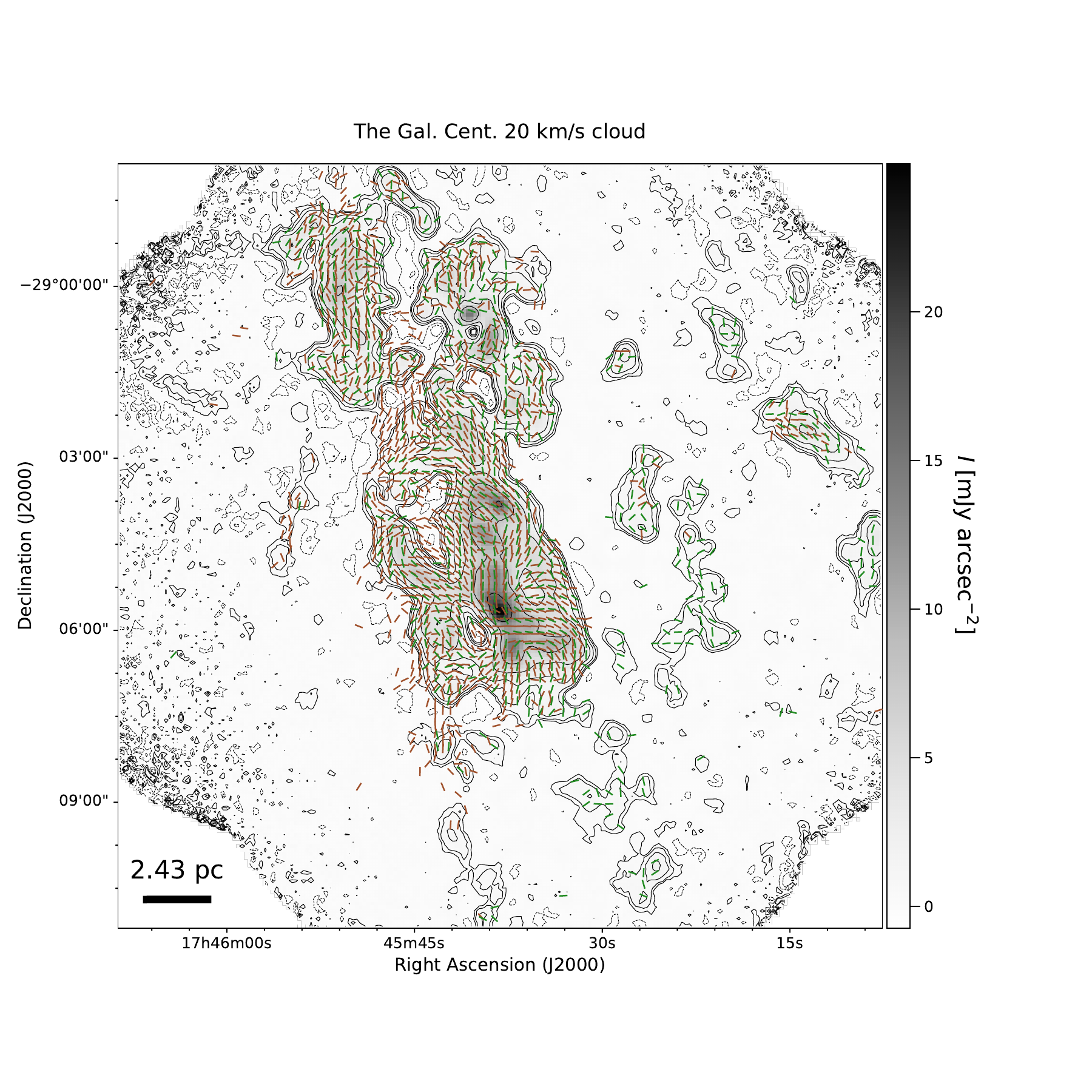}
\caption{Same as figure~\ref{afig:Bmap_L43L43}, but for the Galactic Center 20 km s$^{-1}$ cloud.
The noise levels are $\sigma_{850} = 0.016$\,mJy\,arcsec$^{-2}$ and $\sigma_{450} = 0.33$\,mJy\,arcsec$^{-2}$.
\\ {Alt text: Map with short line segments over contours.}
}
\label{afig:Bmap_GC20KC}
\end{figure}
\clearpage
\begin{figure}[H]
\centering
\includegraphics[width=0.95\textwidth]{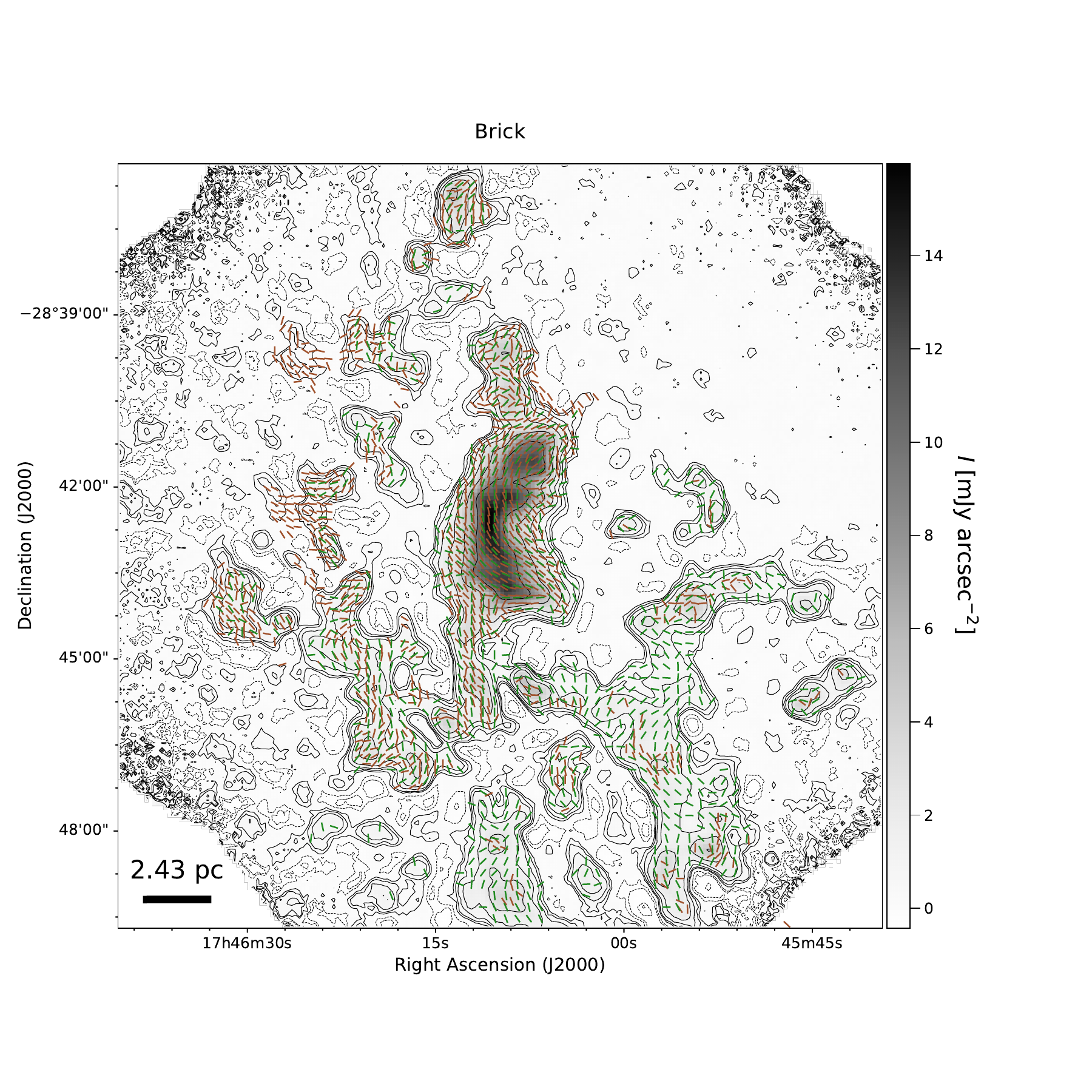}
\caption{Same as figure~\ref{afig:Bmap_L43L43}, but for the Brick.
The noise levels are $\sigma_{850} = 0.0095$\,mJy\,arcsec$^{-2}$ and $\sigma_{450} = 0.26$\,mJy\,arcsec$^{-2}$.
\\ {Alt text: Map with short line segments over contours.}
}
\label{afig:Bmap_GCBRCK}
\end{figure}
\clearpage
\begin{figure}[H]
\centering
\includegraphics[width=0.95\textwidth]{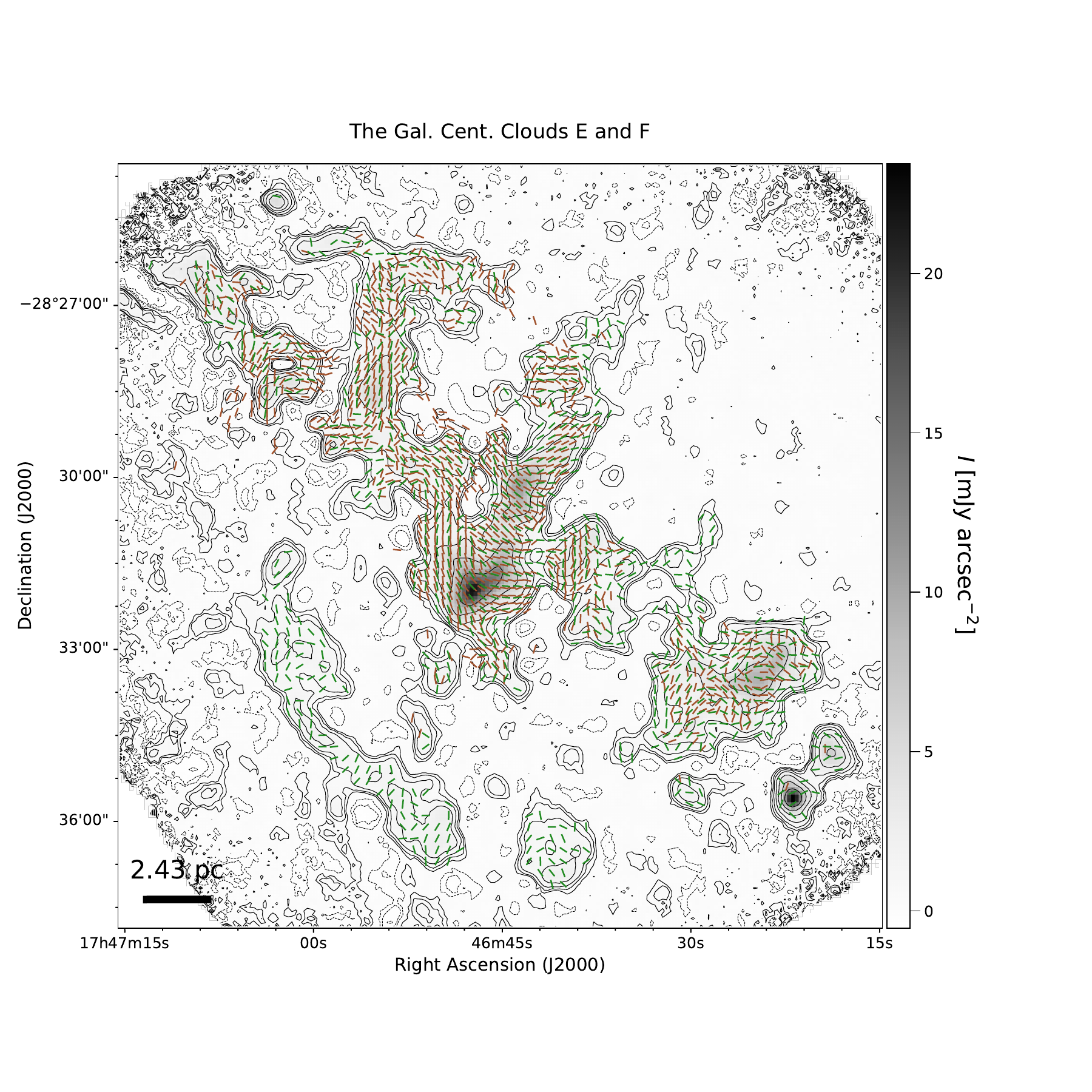}
\caption{Same as figure~\ref{afig:Bmap_L43L43}, but for Galactic Center Clouds E and F.
The noise levels are $\sigma_{850} = 0.012$\,mJy\,arcsec$^{-2}$ and $\sigma_{450} = 0.45$\,mJy\,arcsec$^{-2}$.
\\ {Alt text: Map with short line segments over contours.}
}
\label{afig:Bmap_GCCLEF}
\end{figure}
\clearpage
\begin{figure}[H]
\centering
\includegraphics[width=0.95\textwidth]{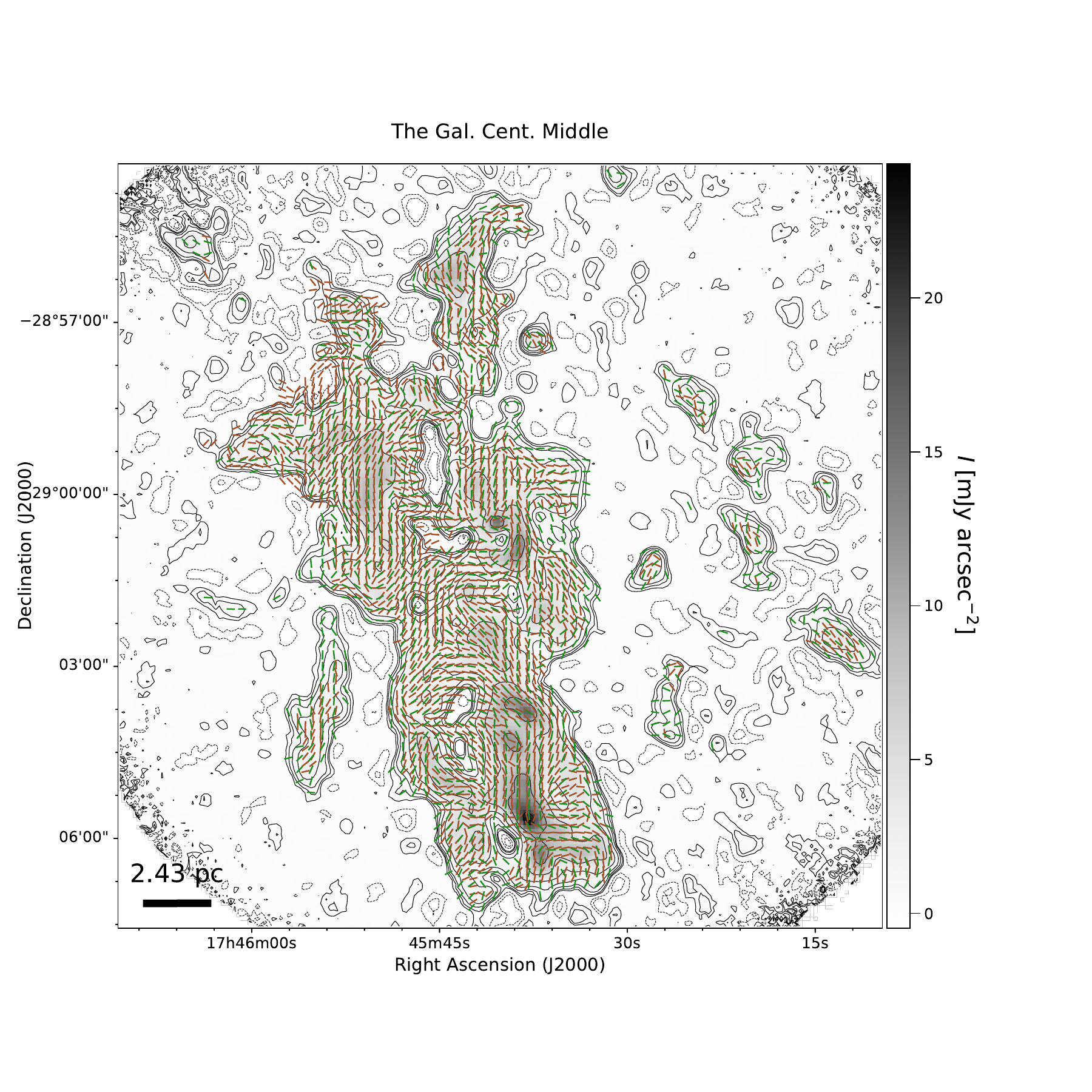}
\caption{Same as figure~\ref{afig:Bmap_L43L43}, but for Galactic Center Middle.
The noise levels are $\sigma_{850} = 0.011$\,mJy\,arcsec$^{-2}$ and $\sigma_{450} = 0.26$\,mJy\,arcsec$^{-2}$.
\\ {Alt text: Map with short line segments over contours.}
}
\label{afig:Bmap_GCMIDD}
\end{figure}
\clearpage
\begin{figure}[H]
\centering
\includegraphics[width=0.95\textwidth]{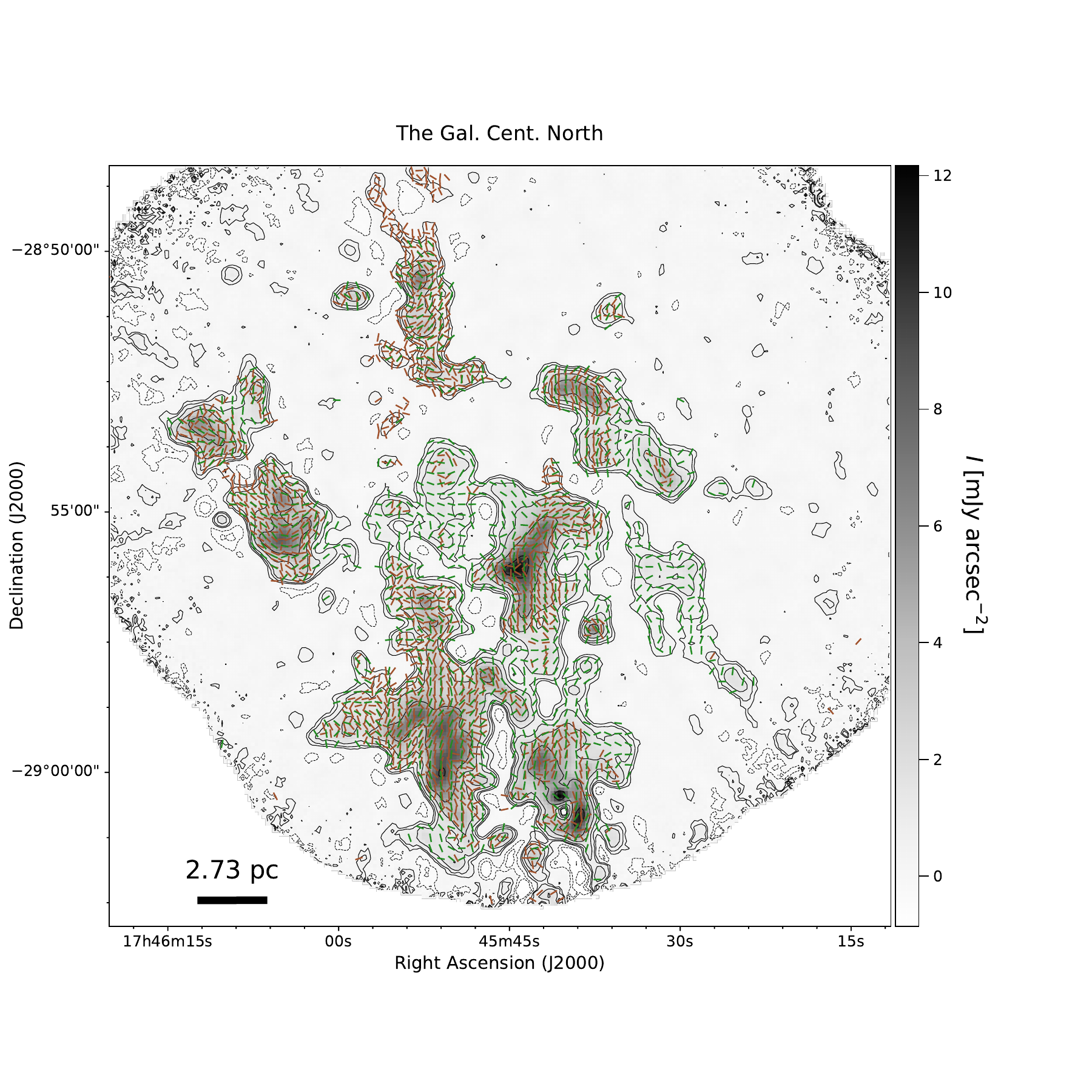}
\caption{Same as figure~\ref{afig:Bmap_L43L43}, but for Galactic Center North.
The noise levels are $\sigma_{850} = 0.019$\,mJy\,arcsec$^{-2}$ and $\sigma_{450} = 0.095$\,mJy\,arcsec$^{-2}$.
\\ {Alt text: Map with short line segments over contours.}
}
\label{afig:Bmap_GCNORT}
\end{figure}
\clearpage
\begin{figure}[H]
\centering
\includegraphics[width=0.95\textwidth]{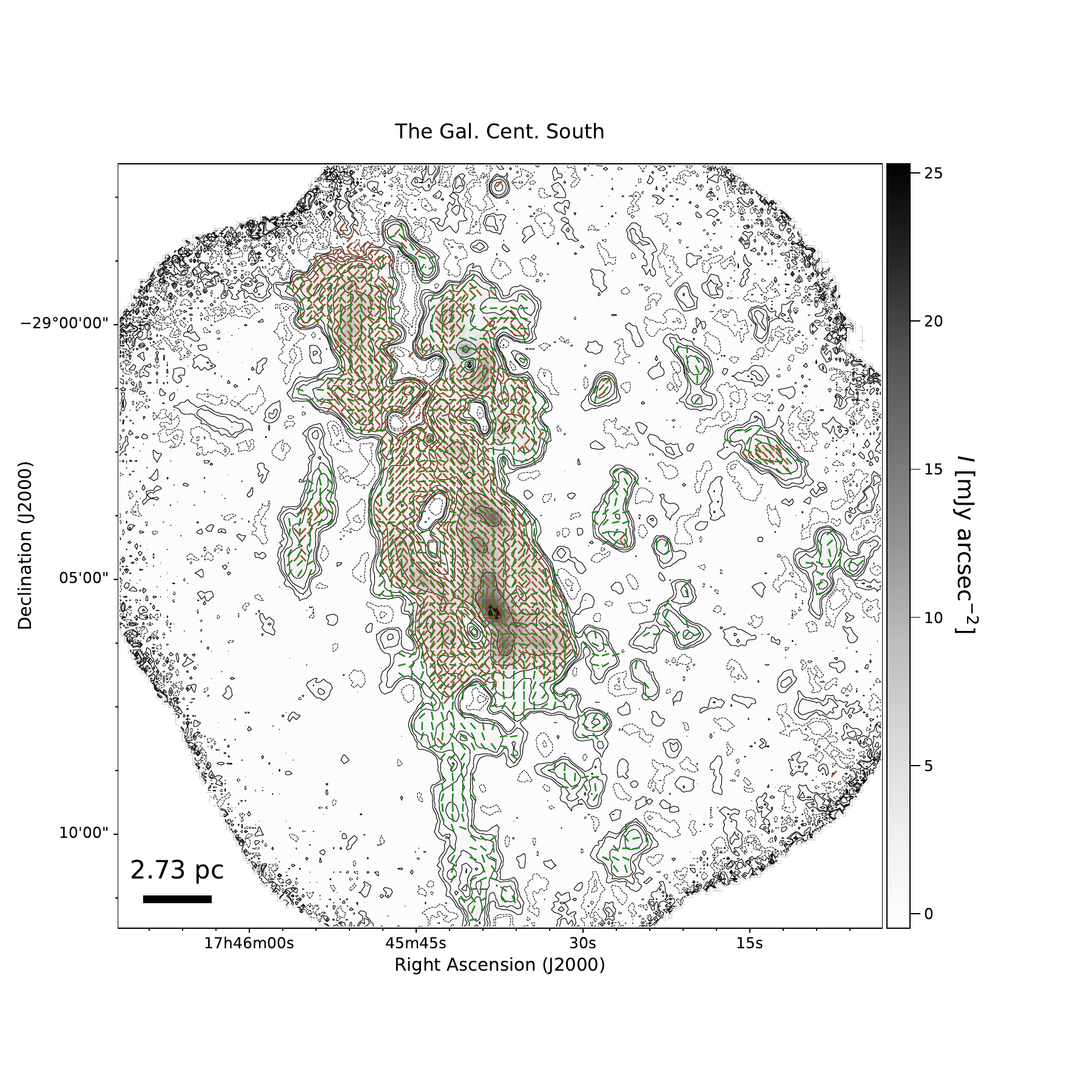}
\caption{Same as figure~\ref{afig:Bmap_L43L43}, but for Galactic Center South.
The noise levels are $\sigma_{850} = 0.011$\,mJy\,arcsec$^{-2}$ and $\sigma_{450} = 0.48$\,mJy\,arcsec$^{-2}$.
\\ {Alt text: Map with short line segments over contours.}
}
\label{afig:Bmap_GCSOUT}
\end{figure}
\clearpage
\begin{figure}[H]
\centering
\includegraphics[width=0.95\textwidth]{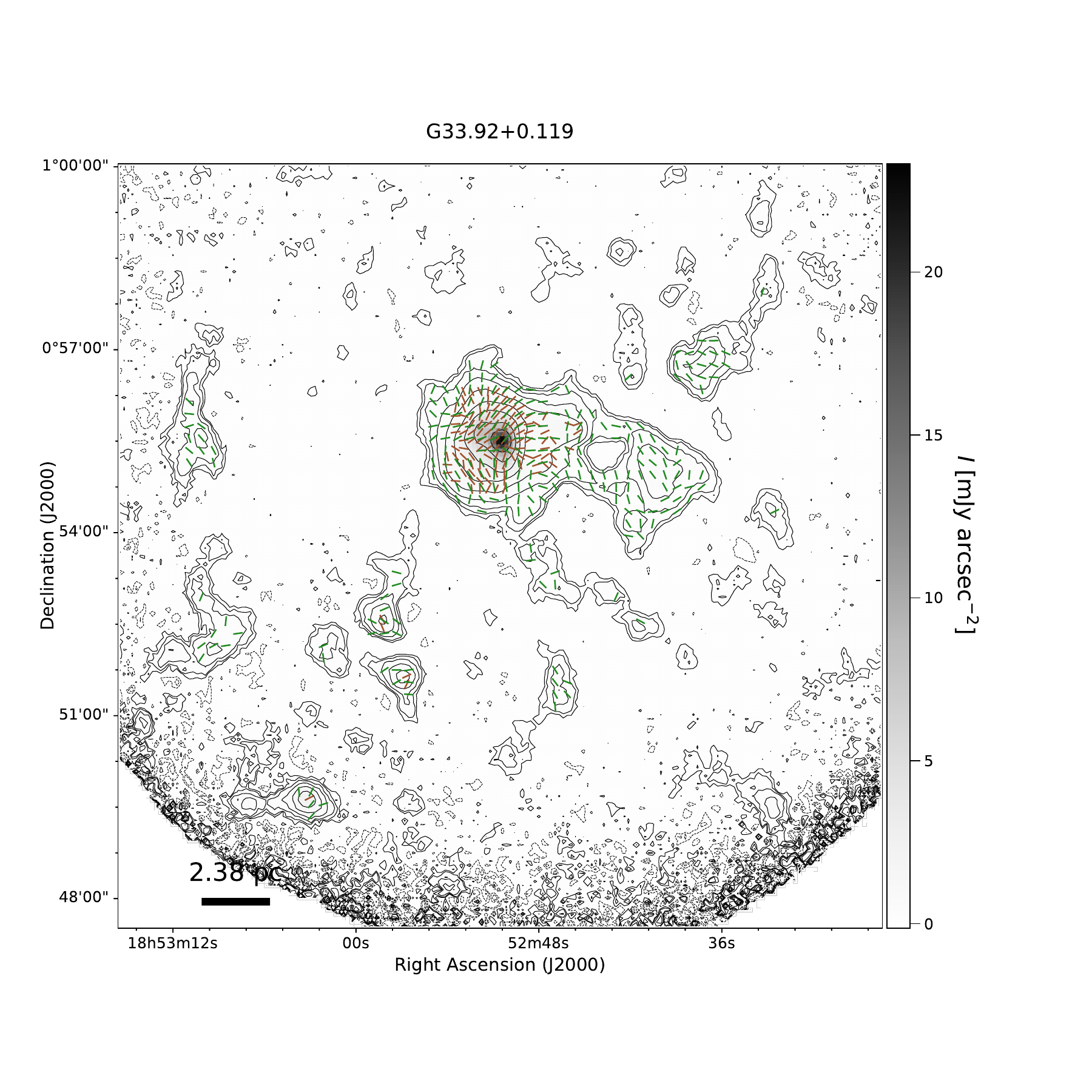}
\caption{Same as figure~\ref{afig:Bmap_L43L43}, but for G33.92+0.119.
The noise levels are $\sigma_{850} = 0.0031$\,mJy\,arcsec$^{-2}$ and $\sigma_{450} = 0.069$\,mJy\,arcsec$^{-2}$.
\\ {Alt text: Map with short line segments over contours.}
}
\label{afig:Bmap_G3392G}
\end{figure}
\clearpage
\begin{figure}[htbp]
\centering
\includegraphics[width=0.95\textwidth]{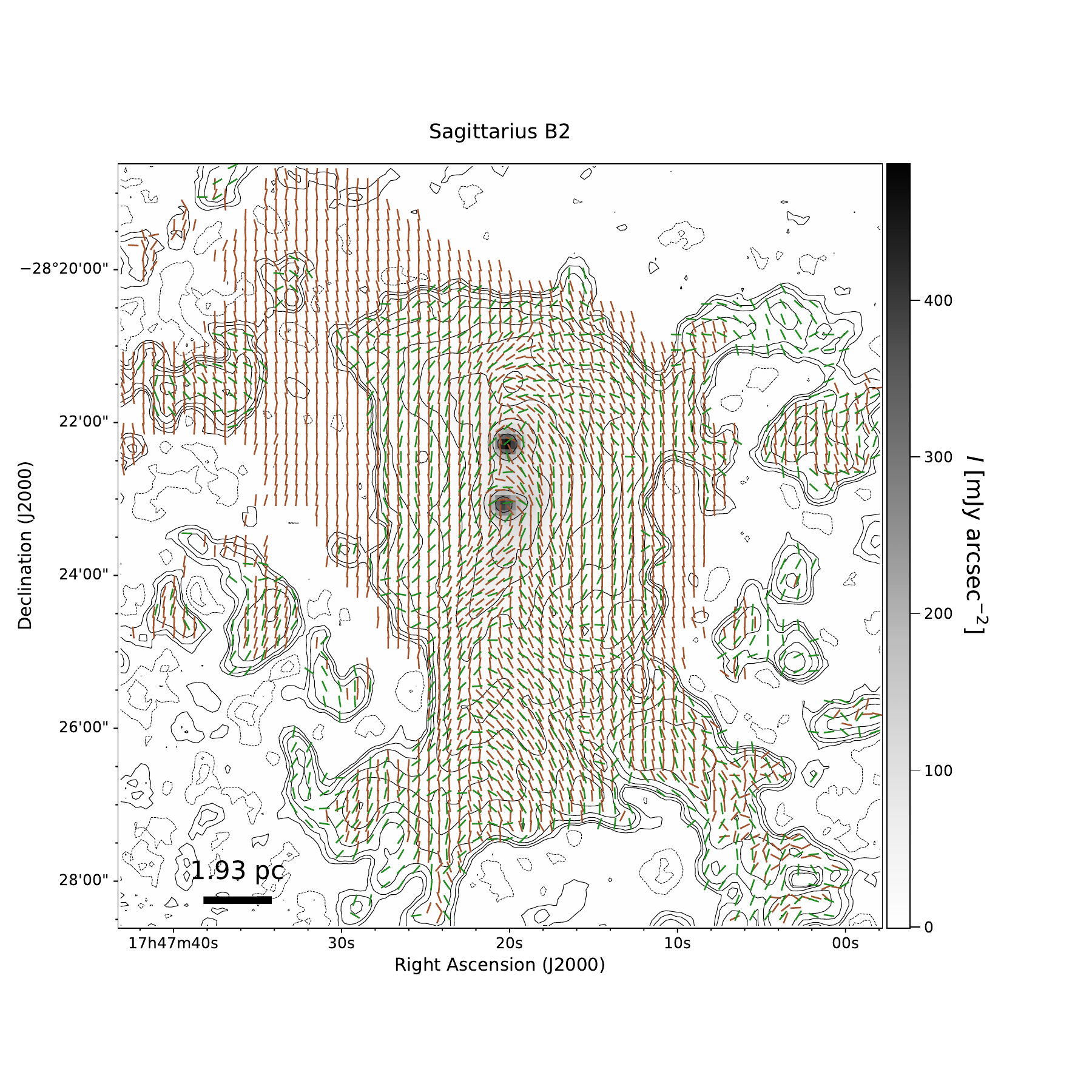}
\caption{
Same as figure~\ref{afig:Bmap_L43L43}, but for Sagittarius~B2.
The noise levels are $\sigma_{850} = 0.018$\,mJy\,arcsec$^{-2}$ and
$\sigma_{450} = 0.29$\,mJy\,arcsec$^{-2}$.
See appendix~\ref{ass:S2_SgrB2_450note} and figure~\ref{afig:SgrB2_450um_StokesI}
for a discussion of the extended 450\,$\mu$m polarization coverage.
\\ {Alt text: Map with short line segments over contours.}
}
\label{afig:Bmap_GSGRB2}
\end{figure}
 
\subsection{Note on the Extended 450\,$\mu$m Polarization Coverage of Sagittarius~B2}
\label{ass:S2_SgrB2_450note}
\begin{figure}[H]
\centering
\includegraphics[width=0.95\textwidth]{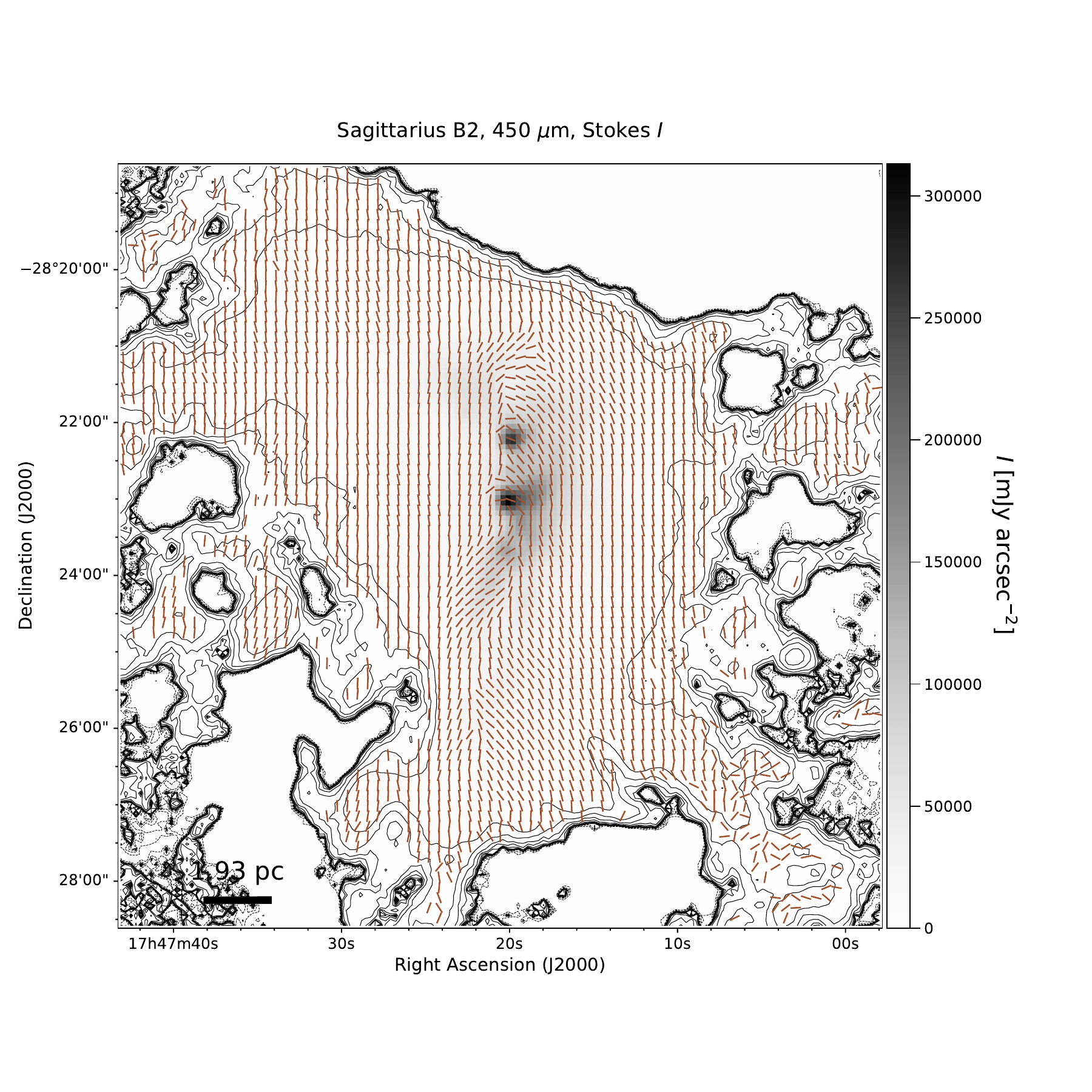}
\caption{
The 450\,$\mu$m $\vec{B}$-field orientation map of Sagittarius~B2.
Sienna polarization segments are overlaid on the 450\,$\mu$m Stokes~$I$ intensity (grayscale and contours).
We select these segments on Stokes~$I$ without an additional cut on the signal-to-noise ratio of polarized intensity, as in figure~\ref{afig:Bmap_L43L43}.
Contours are drawn at $10 \times 2^n\,\sigma_{450}$ ($n = 0$--10), where
$\sigma_{450} = 0.29$\,mJy\,arcsec$^{-2}$.
The 450\,$\mu$m segments extend beyond the 850\,$\mu$m contours shown in figure~\ref{afig:Bmap_GSGRB2} while overlapping the 450\,$\mu$m continuum emission displayed here.
This comparison describes their spatial coverage.
However, it does not by itself distinguish intrinsic wavelength-dependent structure from differences in sensitivity or recovered angular scales.
\\ {Alt text: Map with short line segments over contours.}
}
\label{afig:SgrB2_450um_StokesI}
\end{figure}

The apparent extent of polarized emission can vary between wavelengths if they weight different temperature components or sample different depths of a cloud.
The 450\,$\mu$m $\vec{B}$-field segments (sienna) toward Sagittarius~B2 extend beyond the region traced by the 850\,$\mu$m Stokes~$I$ contours in figure~\ref{afig:Bmap_GSGRB2}.
Figure~\ref{afig:SgrB2_450um_StokesI} overlays the same segments on the 450\,$\mu$m Stokes~$I$ map.
This overlap identifies the intensity structures on which the angles are displayed.
However, it does not establish a polarization detection at every displayed position.
The 450\,$\mu$m Stokes~$I$ intensity reaches values of
$\sim$1700\,mJy\,arcsec$^{-2}$ in the central region of Sagittarius~B2 and
remains above the detection threshold over a substantially larger area than the
850\,$\mu$m emission.
This difference could arise from the different optical depths and temperature weighting at the two wavelengths.
The 450\,$\mu$m band is more sensitive to warmer dust components, which can have a broader spatial distribution.
The criteria for displaying the polarization segments here are identical to those used throughout this appendix (see figure~\ref{afig:Bmap_L43L43}).
In particular, the displayed segments satisfy the adopted Stokes-$I$ selection.
We do not apply an additional cut on the signal-to-noise ratio of polarized intensity.
The extended 450\,$\mu$m segments are spatially associated with the 450\,$\mu$m continuum structure.
Interpreting their greater apparent coverage as a difference in the emitting material requires a comparison that also accounts for the sensitivities and selection functions of the two bands.
 
\subsection{Dataverse-Hosted Full Atlas}
\label{ass:S2_atlas_dataverse}
 
The Dataverse supplement includes the complete atlas of all map products alongside the 65-source figure atlas presented here.
It also contains, for each source, the Stokes $I$, $Q$, $U$, debiased $P_{450}$ and $P_{850}$, $\mathcal{S}_{850}$, and $\mathcal{S} \times \mathcal{P}^{\rm frac}$ maps, together with the tables of statistics.
The electronic supplement will be made available with the published version of this paper.

\section{450\,\micron\ Polarization Fraction Statistics for Individual Sources}
\label{app:S2_P450stats}
 
This appendix tabulates the statistics of the 450\,\micron\ polarization fraction for each source in the BISTRO {\sc PolSrc} sample (table~\ref{tbl:P450YYNYstats}).
It combines the MAS-debiased percentile statistics ($P_{25}$, $P_{50} \pm \mathrm{MAD}$, $P_{75}$), the results of the Ricean depolarization analysis ($P_{\rm ref}$, $\alpha_{\rm dep}$, $N_{\rm fit}$), and the quality flags (G/M/N/F).
The statistics provide the basis for the
discussions of individual sources in subsection~\ref{ss:S2_P_ratio} and the analysis of the stacked distribution in
appendix~\ref{app:S2_stacked}.
 
\begin{longrotatetable}
\footnotesize
\begin{deluxetable*}{lrcccccccc}
\tablecaption{450\,\micron\ Polarization Fraction Statistics at the $8\farcs6$ Beam Resolution \label{tbl:P450YYNYstats}}
\tablewidth{0pt}
\tablehead{
\colhead{Source} & \colhead{$N$} & \multicolumn{3}{c}{MAS-debiased Statistics} & & \multicolumn{3}{c}{Ricean Analysis} & \colhead{Q} \\
\cline{3-5} \cline{7-9}
\colhead{} & \colhead{} & \colhead{$P_{25}$} & \colhead{$P_{50} \pm \mathrm{MAD}$} & \colhead{$P_{75}$} & & \colhead{$P_{\mathrm{ref}}$} & \colhead{$\alpha_{\rm dep}$} & \colhead{$N_{\mathrm{fit}}$} & \colhead{} \\
\colhead{} & \colhead{} & \colhead{(\%)} & \colhead{(\%)} & \colhead{(\%)} & & \colhead{(\%)} & \colhead{} & \colhead{} & \colhead{}
}
\startdata
L1688 Ophiuchus A & 1736 & 0.88 & 1.73 $\pm$ 1.18 & 3.59 & & 5.26 & 0.51 & 197 & G \\
OMC 1 & 6348 & 1.72 & 3.34 $\pm$ 1.90 & 6.09 & & 7.30 & 0.35 & 334 & G \\
OMC 1, 2, and 3 (mosaicked field) & 16256 & 1.47 & 2.76 $\pm$ 1.53 & 4.76 & & 2.04 & 0.69 & 444 & G \\
OMC 2 & 5452 & 1.11 & 2.23 $\pm$ 1.29 & 4.03 & & 3.93 & 0.57 & 478 & G \\
OMC 3 & 3940 & 1.21 & 2.07 $\pm$ 1.00 & 3.42 & & 3.78 & 0.57 & 421 & G \\
Serpens Main (mosaicked field) & 2672 & 1.20 & 2.29 $\pm$ 1.26 & 4.07 & & 1.58 & 0.79 & 358 & G \\
SDC 13 mm-2 & 2840 & 1.11 & 2.17 $\pm$ 1.22 & 3.74 & & 0.68 & 0.97 & 155 & M \\
Monoceros R2 & 4464 & 1.46 & 2.43 $\pm$ 1.19 & 4.01 & & 7.77 & 0.40 & 306 & G \\
Onsala 2 & 556 & 1.86 & 3.05 $\pm$ 1.39 & 4.82 & & 3.88 & 0.60 & 161 & G \\
S106 (mosaic component field) & 1592 & 1.47 & 2.44 $\pm$ 1.18 & 4.03 & & 3.42 & 0.59 & 180 & G \\
DR21 & 5304 & 1.51 & 2.71 $\pm$ 1.44 & 4.71 & & 6.40 & 0.45 & 201 & G \\
NGC 6334 (mosaicked field) & 11732 & 1.17 & 2.09 $\pm$ 1.06 & 3.49 & & 3.32 & 0.60 & 612 & G \\
W3 HFS1 & 2460 & 1.25 & 2.37 $\pm$ 1.43 & 4.31 & & 7.34 & 0.42 & 236 & G \\
W3 HFS2 & 1264 & 0.74 & 1.90 $\pm$ 1.44 & 3.80 & & 3.02 & 0.64 & 237 & G \\
NGC 7538 IRS4 & 2288 & 1.35 & 2.28 $\pm$ 1.11 & 3.87 & & 4.09 & 0.57 & 330 & G \\
SDC 34.370 (mosaicked field) & 884 & 1.24 & 2.35 $\pm$ 1.44 & 4.48 & & 1.26 & 0.01 & 140 & G \\
NGC 1333 (mosaicked field) & 2172 & 1.69 & 3.05 $\pm$ 1.52 & 4.95 & & 0.21 & 1.07 & 332 & M \\
IC 5146 & 124 & 0.00 & 0.89 $\pm$ 0.89 & 2.47 & & 12.83 & 0.27 & 70 & M \\
IC 5146 Western Hub & 40 & \nodata & \nodata & \nodata & & 2.77 & 1.07 & 32 & M \\
NGC 2264 (mosaicked field) & 3708 & 0.91 & 1.73 $\pm$ 0.94 & 3.09 & & 4.58 & 0.53 & 91 & M \\
L43 & 648 & 1.60 & 2.58 $\pm$ 1.28 & 4.53 & & \nodata & \nodata & \nodata & N \\
L1688 Ophiuchus B & 792 & 1.80 & 2.87 $\pm$ 1.22 & 4.34 & & \nodata & \nodata & \nodata & N \\
L1689 North & 668 & 0.94 & 2.00 $\pm$ 1.01 & 2.91 & & \nodata & \nodata & \nodata & N \\
L1448 IRS2 & 300 & 0.92 & 2.26 $\pm$ 1.61 & 4.41 & & \nodata & \nodata & \nodata & N \\
Perseus IC348 & 1532 & 1.96 & 3.08 $\pm$ 1.38 & 4.82 & & \nodata & \nodata & \nodata & N \\
Orion-B NGC2071 & 332 & 0.91 & 1.71 $\pm$ 0.93 & 3.71 & & \nodata & \nodata & \nodata & N \\
SDC 24.013+0.488 & 344 & 1.49 & 2.63 $\pm$ 1.28 & 4.22 & & \nodata & \nodata & \nodata & N \\
SDC25.166 (mosaicked field) & 536 & 0.72 & 2.44 $\pm$ 1.69 & 4.12 & & \nodata & \nodata & \nodata & N \\
Auriga & 524 & 1.24 & 2.64 $\pm$ 1.74 & 5.01 & & \nodata & \nodata & \nodata & N \\
Serpens South & 468 & 1.46 & 2.19 $\pm$ 0.94 & 3.89 & & \nodata & \nodata & \nodata & N \\
Cepheus A & 712 & 0.49 & 1.37 $\pm$ 0.94 & 2.46 & & \nodata & \nodata & \nodata & N \\
IC 1396E & 264 & 1.16 & 1.77 $\pm$ 0.95 & 4.13 & & \nodata & \nodata & \nodata & N \\
IRAS 18360 & 376 & 1.03 & 2.34 $\pm$ 1.52 & 4.63 & & \nodata & \nodata & \nodata & N \\
L1688 Ophiuchus C & 8 & \nodata & \nodata & \nodata & & \nodata & \nodata & \nodata & F \\
L1689 South & 44 & \nodata & \nodata & \nodata & & \nodata & \nodata & \nodata & F \\
L1527 & 24 & \nodata & \nodata & \nodata & & \nodata & \nodata & \nodata & F \\
B335 & 84 & 0.00 & 0.93 $\pm$ 0.93 & 2.78 & & \nodata & \nodata & \nodata & F \\
L483 & 60 & 1.50 & 2.07 $\pm$ 1.31 & 5.55 & & \nodata & \nodata & \nodata & F \\
L1157 & 4 & \nodata & \nodata & \nodata & & \nodata & \nodata & \nodata & F \\
L1478 & 28 & \nodata & \nodata & \nodata & & \nodata & \nodata & \nodata & F \\
G16 & 12 & \nodata & \nodata & \nodata & & \nodata & \nodata & \nodata & F \\
IC 1396A & 172 & 1.96 & 3.07 $\pm$ 1.25 & 4.65 & & \nodata & \nodata & \nodata & F \\
G192.16--3.84 & 20 & \nodata & \nodata & \nodata & & \nodata & \nodata & \nodata & F \\
M16 & 8 & \nodata & \nodata & \nodata & & \nodata & \nodata & \nodata & F \\
Taurus B211/213 (mosaicked field) & 64 & 0.42 & 1.25 $\pm$ 1.08 & 2.27 & & \nodata & \nodata & \nodata & F \\
Perseus B1 & 56 & 1.33 & 1.87 $\pm$ 1.08 & 5.97 & & \nodata & \nodata & \nodata & F \\
PGCC G203.21-11.20 (mosaicked field) & 180 & 2.09 & 3.54 $\pm$ 1.59 & 5.13 & & \nodata & \nodata & \nodata & F \\
SDC 24.489 & 52 & 0.71 & 1.33 $\pm$ 0.91 & 2.84 & & \nodata & \nodata & \nodata & F \\
SDC 18.624-0.070 & 1324 & 0.64 & 1.84 $\pm$ 1.48 & 4.23 & & \nodata & \nodata & \nodata & F \\
Rosette Molecular Cloud (mosaicked field) & 384 & 2.20 & 3.51 $\pm$ 1.31 & 4.68 & & \nodata & \nodata & \nodata & F \\
Sharpless 2-201 & 92 & 1.42 & 2.79 $\pm$ 1.79 & 5.41 & & \nodata & \nodata & \nodata & F \\
\enddata
\tablecomments{
Column descriptions:
$N$ is the number of pixels satisfying $I/\sigma_I > 30$;
$P_{25}$, $P_{50}$, and $P_{75}$ are the 25th, 50th (median), and 75th percentiles of the distribution of observed polarization fractions;
MAD is the median absolute deviation of $P$;
$P_{\rm ref}$ is the intrinsic polarization fraction evaluated at the weighted median intensity $I_{\rm ref}$ of the fitted pixels, where $P_{\rm ref} = P_0 (\sigma_{QU}/I_{\rm ref})^{\alpha_{\rm dep}}$ (Paper~T appendix~\ref{T-app:ricean} describes the exact Bessel formulation of the Ricean expectation);
$\alpha_{\rm dep}$ is the depolarization index describing the polarization--intensity relationship, where $P \propto I^{-\alpha_{\rm dep}}$;
$N_{\rm fit}$ is the number of spatially contiguous pixels used in the Ricean fitting.
Here $P_0$ is the normalization of the intrinsic polarization--intensity relation, and $\sigma_{QU}$ is the common noise standard deviation of Stokes $Q$ and $U$ in the Ricean model.
The quality flag $Q$ indicates the reliability of the Ricean analysis.
Sources classified as G are considered good, with $N_{\rm fit} \geq 100$, $\alpha_{\rm dep} < 1.2$, and a fractional uncertainty of $P_0$ below 0.5.
The M flag denotes the remaining fitted sources (marginal cases), which may indicate a breakdown of the single-population assumption.
The N flag identifies sources with $\alpha_{\rm dep} \geq 1.4$, which are dominated by noise and for which we do not detect a meaningful polarization--intensity relationship.
We assign the F flag when the Ricean fit fails or uses fewer than 30 pixels.
We report observed statistics only for sources with $N \geq 50$ pixels.
We report the results of the Ricean analysis only for sources with quality flags G or M.
Measurements use a pixel spacing of $8\farcs0$, comparable to the native effective beam size of $8\farcs6\pm1\farcs3$ at 450\,\micron.
We fit the non-debiased polarization fractions directly, because the Rice distribution accounts for the positive noise bias.
}
\end{deluxetable*}
\end{longrotatetable}
 
\section{$\mathcal{S} \times \mathcal{P}^{\rm frac}$ Statistics for Individual Sources}
\label{app:S2_SP450stats}
 
This appendix tabulates the statistics of the grain-alignment efficiency proxy $\mathcal{S} \times \mathcal{P}^{\rm frac}$ at 450\,\micron\ for each of the 39 sources satisfying $N \geq 50$ (table~\ref{tbl:SP450stats}).
We use these statistics in the analyses of individual sources in subsection~\ref{ss:S2_SP} and the synthesis of the environmental results in subsection~\ref{ss:S2_EnvStrat}.
We group the sources by star-forming environment (LMSF / HMSF / CMZ), with a heading for each group.
The final block lists the pooled results for clusters covered by multiple fields (Orion\,A, NGC\,2264, NGC\,6334, CMZ).
 
\startlongtable
\begin{deluxetable*}{lcccccc}
\tablecaption{Statistics of 450\,$\mu$m Grain-Alignment Efficiency Proxy $\mathcal{S} \times \mathcal{P}^{\rm frac}$\label{tbl:SP450stats}}
\tablewidth{0pt}
\tablehead{
  \colhead{Source} &
  \colhead{$N$} &
  \multicolumn{3}{c}{$\mathcal{S} \times \mathcal{P}^{\rm frac}$ (\arcdeg)} &
  \multicolumn{2}{c}{$3\sigma$ upper limits} \\
  \cline{3-5} \cline{6-7}
  \colhead{} &
  \colhead{} &
  \colhead{25th pct.} &
  \colhead{median $\pm \sigma_{\rm MAD}$} &
  \colhead{75th pct.} &
  \colhead{median range (\arcdeg)} &
  \colhead{$f_{\rm lim}$ (\%)}
}
\startdata
\cutinhead{LMSF --- low-mass star-forming}
  Auriga & 106 & 0.519 & $0.921 \pm 0.642$ & 1.388 & 0.000--1.397 & 59.4 \\
  G209.29-19.65 & 59 & 0.262 & $0.621 \pm 0.547$ & 1.078 & 0.000--2.606 & 93.2 \\
  IC 5146 & 180 & 0.479 & $0.982 \pm 0.840$ & 1.618 & 0.000--2.684 & 87.2 \\
  L1448 IRS2 & 88 & 0.297 & $0.619 \pm 0.586$ & 1.100 & 0.000--2.232 & 86.4 \\
  L1688 Ophiuchus A & 298 & 0.180 & $0.364 \pm 0.324$ & 0.713 & 0.000--0.818 & 72.8 \\
  L1688 Ophiuchus B & 121 & 0.220 & $0.455 \pm 0.400$ & 0.905 & 0.000--1.675 & 93.4 \\
  L1689 North & 125 & 0.127 & $0.287 \pm 0.319$ & 0.639 & 0.000--0.889 & 84.0 \\
  L43 & 107 & 0.315 & $0.613 \pm 0.455$ & 0.949 & 0.000--1.400 & 85.0 \\
  NGC1333 North & 166 & 0.328 & $0.706 \pm 0.570$ & 1.114 & 0.000--1.656 & 85.5 \\
  NGC1333 South & 206 & 0.287 & $0.618 \pm 0.561$ & 1.159 & 0.000--1.960 & 92.2 \\
  Orion A Field 5 & 437 & 0.355 & $0.616 \pm 0.463$ & 1.036 & 0.000--1.198 & 72.1 \\
  Orion A Field 7 & 57 & 0.282 & $0.566 \pm 0.642$ & 1.041 & 0.000--2.457 & 93.0 \\
  Orion-B NGC2071 & 55 & 0.272 & $0.488 \pm 0.430$ & 0.930 & 0.000--1.245 & 80.0 \\
  Perseus IC348 & 304 & 0.302 & $0.544 \pm 0.408$ & 0.930 & 0.000--1.445 & 76.3 \\
  Serpens South & 190 & 0.305 & $0.540 \pm 0.430$ & 0.926 & 0.000--1.358 & 82.6 \\
\cutinhead{HMSF --- high-mass star-forming}
  Cepheus A & 135 & 0.190 & $0.325 \pm 0.296$ & 0.715 & 0.000--0.744 & 71.1 \\
  DR21 & 740 & 0.169 & $0.317 \pm 0.260$ & 0.579 & 0.129--0.441 & 40.9 \\
  G31.41+0.31 & 202 & 0.180 & $0.354 \pm 0.267$ & 0.558 & 0.000--0.672 & 66.3 \\
  Monoceros R2 & 714 & 0.168 & $0.279 \pm 0.235$ & 0.630 & 0.000--0.563 & 52.5 \\
  NGC 6334 North & 812 & 0.093 & $0.197 \pm 0.196$ & 0.421 & 0.000--0.727 & 72.3 \\
  NGC 6334 South & 737 & 0.237 & $0.380 \pm 0.257$ & 0.657 & 0.000--0.610 & 55.9 \\
  NGC 7538 IRS4 & 393 & 0.207 & $0.406 \pm 0.366$ & 0.782 & 0.000--1.217 & 75.3 \\
  NGC2264 Single Field\#1 & 294 & 0.159 & $0.377 \pm 0.388$ & 0.778 & 0.000--1.278 & 82.7 \\
  NGC2264 Single Field\#2 & 378 & 0.102 & $0.218 \pm 0.216$ & 0.461 & 0.000--0.693 & 86.0 \\
  OMC 1 & 935 & 0.208 & $0.374 \pm 0.288$ & 0.670 & 0.108--0.591 & 46.6 \\
  OMC 2 & 873 & 0.217 & $0.473 \pm 0.472$ & 0.969 & 0.000--1.002 & 70.6 \\
  OMC 3 & 637 & 0.162 & $0.345 \pm 0.327$ & 0.676 & 0.000--1.001 & 78.0 \\
  Onsala 2 & 331 & 0.236 & $0.556 \pm 0.573$ & 1.210 & 0.000--1.524 & 78.2 \\
  Orion A Field 4 & 799 & 0.273 & $0.442 \pm 0.320$ & 0.797 & 0.214--0.633 & 41.9 \\
  S106 (mosaic component field) & 270 & 0.225 & $0.447 \pm 0.393$ & 0.871 & 0.000--1.002 & 75.2 \\
  SDC 13 mm-2 & 564 & 0.232 & $0.561 \pm 0.532$ & 1.027 & 0.000--1.444 & 85.1 \\
  W3 HFS2 & 215 & 0.197 & $0.390 \pm 0.373$ & 0.880 & 0.000--0.770 & 56.3 \\
\cutinhead{CMZ --- Central Molecular Zone}
  Brick & 774 & 0.240 & $0.553 \pm 0.517$ & 0.988 & 0.000--1.300 & 74.5 \\
  Sagittarius B2 & 1007 & 0.173 & $0.311 \pm 0.255$ & 0.597 & 0.132--0.451 & 40.3 \\
  The Gal. Cent. 20 km s$^{-1}$ cloud & 843 & 0.251 & $0.543 \pm 0.559$ & 1.175 & 0.000--1.243 & 73.8 \\
  The Gal. Cent. Clouds E and F & 884 & 0.267 & $0.498 \pm 0.438$ & 1.002 & 0.000--1.063 & 63.2 \\
  The Gal. Cent. Middle & 986 & 0.182 & $0.306 \pm 0.224$ & 0.505 & 0.000--0.590 & 65.6 \\
  The Gal. Cent. North & 1166 & 0.270 & $0.526 \pm 0.464$ & 0.983 & 0.000--1.180 & 70.9 \\
  The Gal. Cent. South & 962 & 0.228 & $0.420 \pm 0.348$ & 0.774 & 0.000--0.672 & 59.4 \\
\cutinhead{Stacked fields}
  Orion\,A (stacked, 3 fields) & 2445 & 0.198 & $0.394 \pm 0.356$ & 0.745 & 0.000--0.836 & 63.4 \\
  NGC\,2264 (stacked, 2 fields) & 672 & 0.125 & $0.274 \pm 0.275$ & 0.577 & 0.000--0.976 & 84.5 \\
  NGC\,6334 (stacked, 2 fields) & 1549 & 0.141 & $0.288 \pm 0.259$ & 0.531 & 0.000--0.645 & 64.5 \\
  CMZ (stacked, 7 fields) & 6622 & 0.221 & $0.420 \pm 0.374$ & 0.845 & 0.000--0.866 & 63.6 \\
\enddata
\tablecomments{
Statistics of the grain-alignment efficiency proxy $\mathcal{S} \times \mathcal{P}^{\rm frac}$ at 450\,$\mu$m,
where $\mathcal{S}$ is the angle dispersion and $\mathcal{P}^{\rm frac}$ is the polarization fraction.
Columns show: source name, number of valid pixels $N$,
the 25th, 50th (median, with MAD-based dispersion), and 75th percentiles of
$\mathcal{S} \times \mathcal{P}^{\rm frac}$ within each source,
and the median range and percentage $f_{\rm lim}$ of the $3\sigma$ upper limits defined below.
We group the sources by star-forming environment (LMSF / HMSF / CMZ; subsection~\ref{ss:S2_envcls}) under printed block headings and order them alphabetically within each block.
The HMSF block collects regions hosting intermediate- to high-mass YSOs.
The grouping is by environment and not by distance.
The HMSF block spans $d \approx 400$\,pc (the OMC fields) to $d \approx 3.8$\,kpc (G31.41$+$0.31), and only the CMZ block lies at the Galactic Center.
Table~\ref{tab:surv} lists the distances separately.
The stacked entries in the final block combine the individual fields already listed above.
We exclude these entries from the totals quoted for each environment in table~\ref{tbl:env_4metric}.
The number of constituent fields is given with each entry.
We include only sources with $N \geq 50$ in this statistical sample.
We compute $\mathcal{S}$ with equation~(\ref{eq:S_def}) at a lag of one pixel and debias it as described in sub-subsection~\ref{sss:S_definition}.
We debias $\mathcal{P}^{\rm frac}$ with the MAS estimator \citep{Plaszczynski2014}.
We compute both quantities from the Method~3 maps, selecting pixels with $I_{450}/\sigma_{I} > 30$.
Where Paper~T requires a PCA threshold optimized for the field, we use the map obtained with that threshold.
Otherwise, we use the default map with the correction for the 450\,\micron\ Stokes-$Q/U$ zero-level offset described in Paper~T.
We treat a pixel as a $3\sigma$ upper limit when $\mathcal{S}$ is not detected at $3\sigma_{\mathcal{S}}$ or $P < 3\sigma_P$ (subsection~\ref{ss:S2_noise_floor}).
We obtain the median range by setting the values for these pixels to zero at one extreme and to their limits at the other, with the quantity below its floor set to $3\sigma$.
The percentage of these pixels is $f_{\rm lim}$.
The percentiles and the median are point estimates calculated from the debiased values of all selected pixels, including upper limits.
However, differences between these estimates do not by themselves establish differences in the underlying proxy (sub-subsection~\ref{sss:SP_global}).
}
\end{deluxetable*}
 
\bibliography{450um1G}{}
\bibliographystyle{aasjournal}
 
\end{document}